\documentclass[11pt,epsfig]{report}

\usepackage{graphicx}
\usepackage[usenames,dvipsnames]{color}
\usepackage{amssymb}
\usepackage{amsmath}
\usepackage{enumerate}
\usepackage{amsfonts}
\usepackage{epsfig}
\usepackage[section]{placeins}
\usepackage{cite}
\usepackage[colorlinks=true,linktocpage=true,linkcolor=blue,citecolor=blue]{hyperref}

\usepackage{ulem}

\def\Fig#1{Fig.~\ref{#1}}
\def\Eq#1{eq.~(\ref{#1})}
\def\be{\begin{equation}}
\def\ee{\end{equation}}

\def\del{\partial}
\def\llangle{\left <}
\def\rrangle{\right >}
\newcommand{\<}{\langle}
\renewcommand{\>}{\rangle}
\def\op{\mathcal{O}}
\def\N{\mathcal{N}}
\def\C{\mathcal{C}}
\def\T{\mathcal{T}}
\def\xtr{x_1}
\def\xipt{\zeta'^2(z)}
\def\xip{\zeta'(z)}
\def\tpa{2\pi\alpha'}
\def\pixz{\Pi^1_z}
\def\Ct{\frac{\tpa}{R^2}\pixz}
\def\arctanh{{\rm arctanh}}
\def\Efield{\mathcal{E}}
\def\rperp{{\bf r}_\perp}
\def\x{{\bf x}}
\def\tr{{\rm tr}}
\def\zh{\hat{z}}
\def\th{\hat{t}}

\def\wf{{\mathfrak{w}}}
\def\qf{{\mathfrak{q}}}

\def\rv{{\bf r}}

\newcommand{\ap}{\alpha'}
\newcommand{\ls}{\ell_s}
\newcommand{\gs}{g_s}
\newcommand{\tf}{T_\mt{str}}
\newcommand{\tc}{T_\mt{c}}
\newcommand{\tdec}{T_\mt{c}}
\newcommand{\mmes}{M_\mt{mes}}
\newcommand{\caln}{{\cal N}}
\newcommand{\bea}{\begin{eqnarray}}
\newcommand{\eea}{\end{eqnarray}}
\newcommand{\nn}{\nonumber \\}

\newcommand{\labell}[1]{\label{#1}} 
\newcommand{\beq}{\begin{equation}}
\newcommand{\eeq}{\end{equation}}
\newcommand{\beqa}{\begin{eqnarray}}
\newcommand{\eeqa}{\end{eqnarray}}

\newcommand{\reef}[1]{(\ref{#1})}
\newcommand{\eqn}[1]{(\ref{#1})}

\newcommand{\mt}[1]{\textrm{\tiny #1}}

\def\nc {N_\mt{c}}
\def\nf {N_\mt{f}}

\newcommand{\gten}{G}
\newcommand{\gym}{g}
\newcommand{\ads}{\mbox{$AdS_5 \times S^5$ }}
\newcommand{\mq}{M_\mt{q}}
\newcommand{\mth}{M_\mt{th}}

\newcommand{\sac}{\, , \qquad}
\newcommand{\eg}{{e.g. }}
\newcommand{\ie}{{i.e. }}
\def\vp {\varphi}

\newcommand{\fc}{\frac}
\newcommand{\ra}{\rightarrow}
\newcommand{\te}{t_\mt{E}}

\newcommand{\bbr}[1]{\mbox{${\mathbb R}^{#1}$}}
\newcommand{\nfour}{{\cal N} = 4}
\newcommand{\qc}{\langle \bar{\psi} \psi \rangle} 
\newcommand{\se}{S_\mt{E}}
\newcommand{\mbar}{\bar{M}}
\newcommand{\prt}{\partial}
\def\U{{\mathcal{U}}}
\newcommand{\vlim}{v_\mt{lim}}
\newcommand{\uem}{U(1)_\mt{EM}}
\newcommand{\jem}{J^\mt{EM}}

\newcommand{\td}{T_\mt{diss}}

\newcommand{\ccbar}{c\bar{c}}

\newcommand{\opeak}{\omega_\mt{peak}}
\newcommand{\jpsi}{J/\psi}

\newcommand{\eym}{E_\mt{YM}}
\newcommand{\dym}{d_\mt{YM}}

\newcommand{\dmes}{d_\mt{mes}}

\def\le{\left}
\def\ri{\right}
\def\ha{{1\over 2}}

\def\vev#1{\langle#1\rangle}
\def\det{{\rm det}}
\def\tr{{\rm tr}}
\def\Tr{{\rm Tr}}

\def\th{{\theta}}

\def\apr{{\alpha'}}
\newcommand{\p}{\partial}

\newcommand\ep{\epsilon}
\newcommand\sig{\sigma}
\newcommand\Sig{\Sigma}
\newcommand\lam{\lambda}
\newcommand\Lam{\Lambda}
\newcommand\om{\omega}
\newcommand\Om{\Omega}

\newcommand\Ga{{\ensuremath{{\Gamma}}}}

\newcommand\da{{\dagger}}
\newcommand\ov{\over}

\newcommand\De{{\Delta}}

\newcommand\de{{\delta}}

\newcommand{\field}[1]{\mathbb{#1}} 

\newcommand{\RR}{\field{R}}

\newcommand\sA{{\ensuremath{{\mathcal A}}}}

\newcommand\sG{{\ensuremath{{\mathcal G}}}}

\newcommand\sL{{\ensuremath{{\mathcal L}}}}

\newcommand\sO{{\ensuremath{{\mathcal O}}}}

\newcommand\sN{{\mathcal N}}
\newcommand\sC{{\mathcal C}}
\newcommand\sR{{\ensuremath{{\mathcal R}}}}
\newcommand\sT{{\ensuremath{{\mathcal T}}}}

\newcommand{\bln}{\begin{align}}
\newcommand{\eln}{\end{align}}
\newcommand{\bst}{\begin{split}}
\newcommand{\est}{\end{split}}
\newcommand{\bi}{\begin{itemize}}
\newcommand{\ei}{\end{itemize}}
\newcommand{\ben}{\begin{enumerate}}
\newcommand{\een}{\end{enumerate}}

\begin{document}
\bibliographystyle{utphys}

\pagestyle{plain}
\setcounter{page}{1}

\begin{titlepage}

\begin{center}
\vspace*{-1cm} \today \hfill CERN-PH-TH/2010-316 \\
{}~{} \hfill MIT-CTP-4198 \\ 
{}~{} \hfill  ICCUB-10-202 \\

\vskip 1cm

{\LARGE {\bf Gauge/String Duality, Hot QCD}}

\vskip 2mm

{\LARGE {\bf and Heavy Ion Collisions}}

\vskip 1 cm

{\large {\bf Jorge Casalderrey-Solana,$^1$ Hong Liu,$^2$
David Mateos,$^{3,4}$}}

\vspace{1.5mm}

{\large \bf Krishna Rajagopal,$^{2}$ and Urs Achim Wiedemann$^1$}

\vskip .8cm
   ${}^1$ {\it Department of Physics, CERN, Theory Unit, CH-1211 Geneva}

\medskip

  ${}^2$ {\it Center for Theoretical Physics, MIT, Cambridge, MA 02139, USA}

\medskip

  ${}^3$ {\it Instituci\'o Catalana de Recerca i Estudis Avan\c cats (ICREA), Passeig Llu\'\i s Companys 23, E-08010, Barcelona, Spain}

\medskip

  ${}^4$ {\it Departament de F\'\i sica Fonamental (FFN) \&  Institut de Ci\`encies del Cosmos (ICC), Universitat de Barcelona (UB), Mart\'{\i}  i Franqu\`es 1, E-08028 Barcelona, Spain}

\vskip 0.5cm

{\tt  jorge.casalderrey@cern.ch, hong$\_$liu@mit.edu, dmateos@icrea.cat, krishna@mit.edu, urs.wiedemann@cern.ch}

\vspace{5mm}

{\bf Abstract}
\end{center}
 
\noindent
Over the last decade, both experimental and theoretical advances have brought the need for strong coupling techniques in the analysis of deconfined QCD matter and heavy ion collisions to
the forefront.  As a consequence, a fruitful interplay has developed between analyses of strongly-coupled
non-abelian plasmas via the gauge/string duality (also referred to as the AdS/CFT correspondence)
and the phenomenology of heavy ion collisions.
We review some of the main insights gained from this interplay to date. To establish a common language, 
we start with an introduction to heavy ion phenomenology and finite-temperature QCD, and a 
corresponding introduction to important concepts and techniques in the gauge/string duality. These introductory sections are written for nonspecialists, with the goal of bringing readers ranging from beginning graduate students to experienced practitioners of either QCD or gauge/string duality to the point that they understand enough about both fields that they can then appreciate their interplay in all appropriate contexts.   We then review  the current state-of-the art in the application of the duality to the description of the dynamics of strongly coupled plasmas, with emphases that include: its thermodynamic, hydrodynamic and transport properties; the way it both modifies the dynamics of, and is perturbed by, high-energy or heavy quarks passing through it; and the physics of quarkonium mesons within it.   We seek throughout to stress the lessons that can be extracted from these 
computations for heavy ion physics as well as to
discuss future directions and open problems for the field.

\end{titlepage}

\tableofcontents

 \chapter{Preface}

The discovery in the late 1990s of the AdS/CFT correspondence, as well as its subsequent generalizations now referred to as the gauge/string duality, have provided a novel approach for studying the strong coupling limit of a large class of non-abelian quantum field theories. In recent years, there has  been a surge of interest in exploiting this approach to study properties of  the plasma  phase of such theories at non-zero temperature, including the transport properties of the plasma and the propagation and relaxation of
plasma perturbations. Besides the generic theoretical motivation of such studies, many of the recent  developments have 
been inspired by the phenomenology of ultra-relativistic heavy ion collisions.  Inspiration has acted in the other direction too, as properties of 
non-abelian plasmas that were determined via the gauge/string duality have helped to identify new avenues in heavy ion phenomenology.  There are many reasons for this at-first-glance surprising interplay among string theory, finite-temperature field theory, and heavy ion phenomenology, as we shall see throughout this review.
Here, 
we anticipate only that the analysis of data from the Relativistic Heavy Ion Collider (RHIC) had emphasized
the importance, indeed the necessity, of developing strong coupling techniques for heavy ion phenomenology.  For instance, in the calculation of an experimentally-accessible transport property, the dimensionless ratio of the shear viscosity to the entropy density, weak- and strong coupling results turn out to differ not only quantitatively but parametrically, and data favor the strong coupling result. Strong coupling presents no difficulty for lattice-regularized calculations of QCD thermodynamics, but the generalization of these methods beyond static observables to characterizing transport properties has well-known limitations. Moreover, these methods are quite unsuited to the study of the many and varied time-dependent problems that heavy ion collisions are making experimentally accessible.
It is in this context that the very different suite of opportunities provided by gauge/string calculations 
of strongly-coupled plasmas have started to provide a complementary source of insights for heavy ion phenomenology. Although these new methods come with limitations of their own, the results are obtained from first-principle calculations in non-abelian field theories at non-zero temperature.

The present review aims at providing an 
overview of the results obtained from this interplay between gauge/string duality, lattice QCD and heavy ion phenomenology within the last decade in a form such that readers from either community can appreciate and understand the emerging synthesis.   For the benefit of string theory practitioners, we begin in 
Sections~\ref{intro} and \ref{sec:latticeQCD} with targeted overviews of the data from and theory of heavy ion collisions, as well as of state-of-the art lattice calculations for QCD at non-zero temperature.
We do not provide a complete overview of all aspects of these subjects that are of current interest in their own terms, aiming instead for a self-contained, but targeted, overview of those aspects that are keys to understanding the impacts of gauge/string calculations.
In turn, Sections \ref{sec:Section3} and \ref{AdS/CFT} are mainly written for the benefit
of heavy ion phenomenologists and QCD practitioners. They provide a targeted introduction to the principles behind the gauge/string duality with a focus on those aspects relevant for calculations at non-zero temperature.  With the groundwork on both sides in place, we then provide an in-depth review of the gauge/string calculations of 
bulk thermodynamic and hydrodynamic properties (Section \ref{sec:BulkDynamicalProperties}); 
the propagation of probes (heavy or energetic quarks, and quark-antiquark pairs) through a strongly-coupled non-abelian plasma and the excitations of the plasma that result (Section~\ref{sec:Section6});
and a detailed analysis of mesonic bound states and spectral functions in a deconfined plasma (Section~\ref{mesons}). 

The interplay between hot QCD, heavy ion phenomenology and the gauge/string duality has been a very active field of research in recent years, and there are already a number of other reviews that cover various aspects of these developments. In particular, there are reviews  focussing on the techniques for calculating finite-temperature correlation functions of local operators from the gauge/string duality~\cite{Son:2007vk}; 
on the phenomenological aspects of perfect fluidity 
and its manifestation in different systems, including the quark-gluon plasma produced in heavy ion collisions and strongly coupled fluids made of trapped fermionic atoms that are more than twenty orders of magnitude colder~\cite{Schafer:2009dj} (we shall not review the connection to ultracold atoms here); and also shorter topical reviews~\cite{Peeters:2007ab,Mateos:2007ay,Erdmenger:2007cm,Gubser:2009md,Gubser:2009sn} that provide basic
discussions of the duality and its most prominent applications in the context of heavy ion phenomenology. The present review aims at covering a broader range of applications while at the same time providing readers from either the string theory or the QCD communities with the opportunity to start from square one on `the other side' and build an understanding of the needed context.  

Since any attempt at covering everything risks uncovering nothing, even given its length this review does remain limited in scope.   Important examples of significant advances in this rapidly-developing field that we have not touched upon include: (i) the physics of saturation in QCD and its application to understanding the 
initial conditions for heavy ion collisions \cite{Gelis:2010nm,Hatta:2007he,Hatta:2007cs,Mueller:2008bt,Bayona:2009qe,Albacete:2008ze,Cornalba:2008sp}; (ii) various field-theoretical approaches to understanding equilibration in heavy ion collisions \cite{Baier:2000sb,Romatschke:2003ms,Mrowczynski:2005ki,Arnold:2004ti,Arnold:2007pg}; 
and (iii) work from the string side on developing dual gravitational descriptions not just of the strongly-coupled plasma and its properties, as we do review, but of the dynamics of  how the plasma forms, equilibrates, expands, and cools 
after a collision. 
This last body of literature includes the early work of Refs.~\cite{Kang:2004jd,Nastase:2005rp,Nastase:2006eb,Nastase:2008hw}, continues in modern form with
the discovery of a dual gravitational description of a fluid that is cooling via boost-invariant longitudinal expansion ~\cite{Janik:2005zt} extends to the very recent discovery of the dual description of the collision of two finite-thickness sheets of energy density and the ensuing plasma formation~\cite{Chesler:2010bi}, and includes much in between ~\cite{Lin:2006rf,Kovchegov:2007pq,Lin:2007fa,Kajantie:2008rx,Grumiller:2008va,Gubser:2008pc,Albacete:2008vs,Lin:2008rw,AlvarezGaume:2008fx,Chesler:2008hg,Balasubramanian:2010ce,Lin:2009pn,Gubser:2009sx,Beuf:2009cx,Avsar:2009xf,Chesler:2009cy,LIn:2010cb,Aref'eva:2009kw}.
%
%
Also, we focus throughout on insights that have been obtained from calculations that are directly rooted in quantum field theories analyzed via the gauge/string duality. Consequently, we have omitted the so-called AdS/QCD approach that aims to optimize ans\"atze for gravity duals that do not correspond to known field theories, in order to best incorporate known features of QCD in the gravitational description 
\cite{Erlich:2005qh,DaRold:2005zs,Karch:2006pv,Andreev:2006ct,Brodsky:2007hb,Gursoy:2007cb,Gursoy:2007er,Gursoy:2008bu,Kajantie:2006hv,BallonBayona:2007vp,Polchinski:2001tt,BoschiFilho:2002vd,BoschiFilho:2002ta}.
So, while we have high hopes that the material reviewed here 
will serve in coming years as a pathway into the field for new practitioners with experience in QCD or string theory, we have not aimed for completeness. Indeed, we can already see now that this very active field continues to broaden and that adding new chapters to a review like this will soon be warranted.

\chapter{A heavy ion phenomenology primer}
\label{intro}
 
What macroscopic properties of matter emerge from the fundamental
constituents and interactions of a non-abelian gauge theory?
The study of ultra-relativistic heavy ion collisions addresses this question for the 
theory of the strong interaction, Quantum Chromodynamics, in the regime of extreme
energy density. To do this, heavy ion phenomenologists employ tools 
developed to identify and quantify 
collective phenomena in collisions that have many thousands of particles
in their final states.
Generically speaking, these tools quantify deviations with respect to
benchmark measurements (for example in proton-proton and
proton-nucleus collisions) in which collective effects are absent.
In this Section, we provide details for three
cases of current interest: (i) the characterization of elliptic flow, 
which teaches us how soon after the collision matter
moving collectively is formed and which allows us to
constrain the value of the shear viscosity of this matter; (ii) the
characterization of jet quenching via single inclusive hadron
spectra and angular correlations, which teaches us how this
matter affects and is affected by a high-velocity colored particle
plowing through it; and (iii) the characterization of 
the suppression of quarkonium production, which has the
potential to teach us about the temperature of the matter 
and of the degree to which it screens the interaction between
colored particles.

\section{General characteristics of heavy ion collisions}
\label{sec:generalcharacteristics}

In a heavy ion collision experiment, 
 large nuclei, such as gold  (at RHIC) or lead (at SPS and LHC), are collided at an 
ultra-relativistic center of mass energy $\sqrt{s}$.   The reason for 
using large nuclei
is to create as large a volume as possible of matter at a high energy density, to
have the best chance of discerning phenomena or properties that characterize
macroscopic amounts of strongly interacting matter.  In contrast, in energetic
elementary collisions (say electron-positron collisions but to a good approximation
also in proton-proton collisions) one may find many hadrons in the final
state but these are understood to result from a few initial 
partons that each fragment rather than from a macroscopic volume of
interacting matter.   Many years ago Phil Anderson coined the phrase ``more is
different'' to emphasize that macroscopic volumes of (in his case condensed)
matter manifest qualitatively new phenomena, distinct from those that can
be discerned in interactions among few elementary constituents and requiring
distinct theoretical methods and insights for their elucidation~\cite{Anderson:1972}. 
Heavy ion physicists do not have the luxury of studying systems
containing a mole of quarks, but by using the heaviest ions that nature provides
they go as far in this direction as is possible.

The purpose of building accelerators
that achieve heavy ion collisions  at higher and higher $\sqrt{s}$ is simply to create matter
at higher and higher energy density.   A simple argument 
to see why this may
be so arises upon noticing that in the center-of-mass frame we have the collision
of two Lorentz contracted nuclei, pancake shaped, and increasing the collision energy
makes these pancakes thinner.  Thus, at $t=0$ when these pancakes are coincident
the entire energy of the two incident nuclei is found within a smaller volume for higher
$\sqrt{s}$.   This argument is overly simple, however, because not all of the energy
of the collision is transformed into the creation of matter; much of it is carried by
the debris of the two colliding nuclei that spray almost along the beam directions.


\begin{figure} 
\begin{center}
\includegraphics[width=0.49\textwidth]{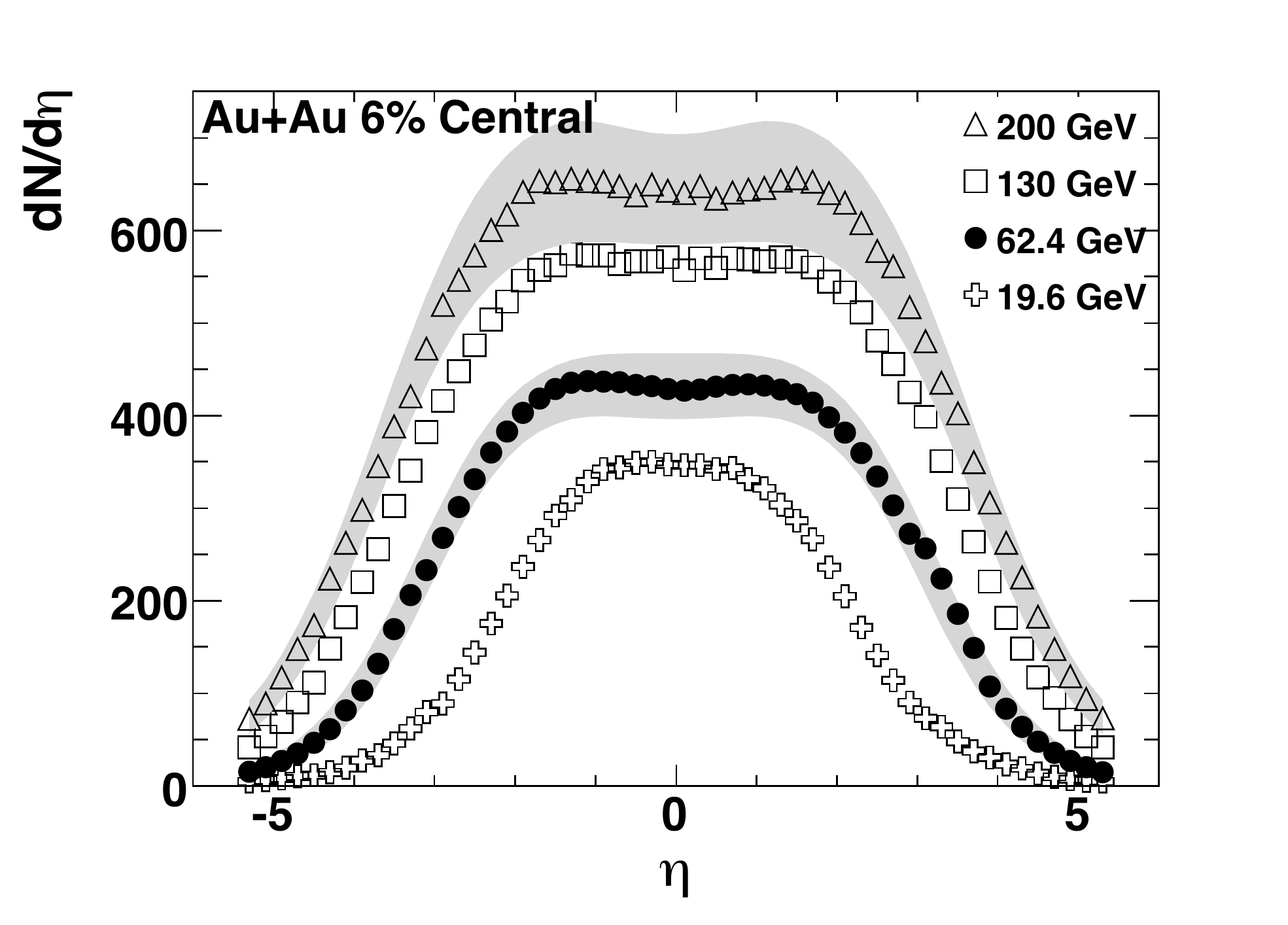}
\includegraphics[width=0.49\textwidth]{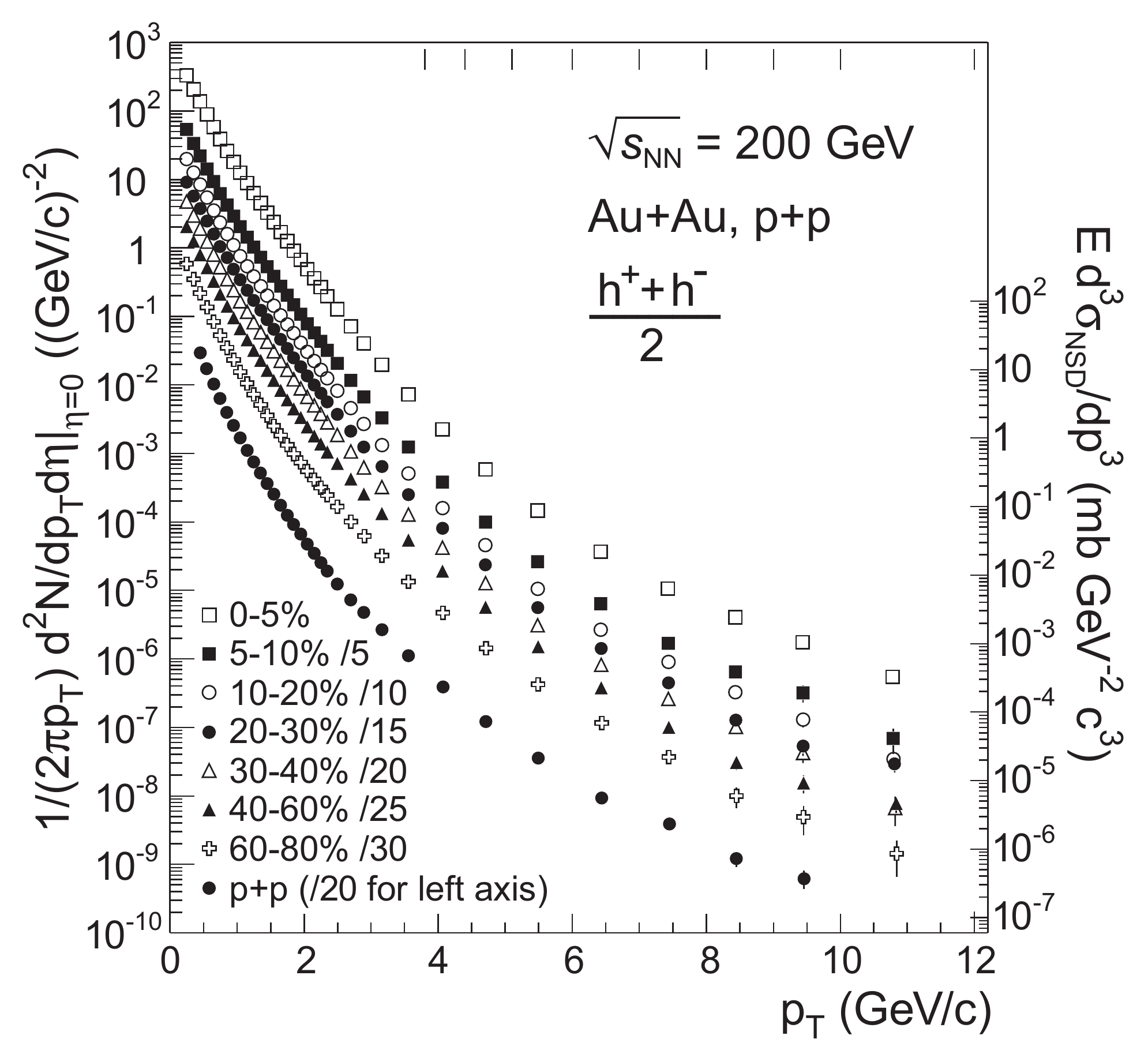}
\end{center}
\caption{\label{fig:general} \small
Left: Charged particle multiplicity distributions for collisions at RHIC with four different energies as a function of pseudorapidity \cite{Back:2005hs}.  Very recently, a charged 
particle multiplicity $dN_{\rm ch}/d\eta = 1584\pm 4 ({\rm stat}) \pm 76 ({\rm sys})$ has been measured at $\eta=0$
in heavy ion collisions with $\sqrt{s}=2.76$~TeV by the ALICE detector at the 
LHC~\cite{Aamodt:2010pb}.
Right: Charged particle spectrum as function of $p_T$ in gold-gold collisions at top
RHIC energy at different values of the impact parameter, with the
top (second-from-bottom) curve corresponding to nearly head-on (grazing)
collisions~\cite{Adams:2003kv}.  The bottom curve is data from proton-proton collisions at the same energy.}
\end{figure}

The question of how the initial state wave function of the colliding nuclei determines
precisely how much matter, containing how much entropy, 
is produced soon after the collision, and consequently determines the number
of particles in the final state, is a subject of intense theoretical interest.  At present, however,
calculations using gravity duals have not shed much light on the attendant issues and
so we shall not describe this branch of heavy ion phenomenology in any detail.
It is, however, worth having a quantitative sense of just how many particles are produced in a typical
heavy ion collision.   In 
the left panel of \Fig{fig:general} 
 we show the multiplicity of charged particles per unit pseudorapidity 
for RHIC collisions at four different values of $\sqrt{s}$.  Recall that the pseudorapidity
$\eta$ is related to the polar angle $\theta$ measured with respect to the 
beam direction by $\eta=-\log \tan (\theta/2)$.  Note also that, by convention,
$\sqrt{s}=200$~GeV (which is the top energy achieved at RHIC) 
means that the incident ions in these collisions have
a velocity such that individual nucleons colliding with
that velocity would collide with a
center of mass energy of 200 GeV.
Since each gold nucleus 
has 197 nucleons, the total center of mass energy in a heavy ion collision
at the top RHIC energy is about 40 TeV.  By integrating under the curve in 
the left panel of \Fig{fig:general}, 
 one finds that such a collision yields $5060 \pm 250$ charged
 particles~\cite{Back:2002wb,Back:2004je}. 
 The multiplicity measurement is made by counting tracks, meaning that
 neutral particles (like $\pi^0$'s and the photons they decay into) are not 
 counted.  So, the total number of hadrons is greater than the total
 number of charged particles.
 If all the hadrons in the final state were pions, and if the small isospin
 breaking introduced by the different number of protons and neutrons in
 a gold nucleus can be neglected, there would be equal numbers of
 $\pi^+$, $\pi^-$ and $\pi^0$ meaning that the total multiplicity would be
 $3/2$ times the charged multiplicity.  In reality, this factor turns out 
 to be about 1.6~\cite{Back:2005hs}, 
  meaning that heavy ion collisions at the top RHIC energy each produce about
 8000 hadrons in the final state.  
We see from the left panel of \Fig{fig:general} 
that this multiplicity grows with increasing
collision energy, and we see that the multiplicity per unit pseudorapidity is largest in a 
range of angles centered around $\eta=0$, meaning $\theta=\pi/2$. 
Heavy ion collisions at the LHC will have center of mass energies
up to $\sqrt{s}=5.5$~TeV per nucleon. 
Estimates for the charged particle multiplicity per unit pseudorapidity at this energy that were made before LHC heavy ion collisions began typically range from
$dN_{\rm ch}/d\eta\approx 1000$
to $dN_{\rm ch}/d\eta\approx 3000$~\cite{Abreu:2007kv}. 
  This range can now be considerably tightened, now that we know that $dN_{\rm ch}/d\eta = 1584 \pm 4 ({\rm stat}) \pm 76({\rm sys})$ at $\eta=0$ in most central collisions (5 \%) with $\sqrt{s}=2.76$~TeV~\cite{Aamodt:2010pb}.

The large multiplicities in heavy ion collisions indicate large
energy densities, since each of these particles carries a typical
(mean) transverse momentum of several hundred MeV.  There
is a simple geometric method due to Bjorken~\cite{Bjorken:1982qr},
that can be used to estimate the energy density
at a fiducial early time, conventionally chosen to be 
$\tau_0=1$~fm.  The smallest reasonable choice of $\tau_0$ would
be the thickness of the Lorentz-contracted pancake-shaped nuclei,
namely $\sim(14~{\rm fm})/100$ since gold nuclei have a radius
of about 7 fm.  But, at these early times $\sim 0.1~{\rm fm}$ the matter
whose energy density one would be estimating would still be far from
equilibrium.  We shall see below that elliptic flow data indicate
that by $\sim 1$~fm after the collision, matter is flowing collectively
like a fluid in local equilibrium.  The geometric estimate of the energy
density is agnostic about whether the matter in question is initial state partons
that have not yet interacted and are far from equilibrium
or matter in local equilibrium behaving collectively; because
we are interested in the latter, we choose 
$\tau_0=1$~fm.  Bjorken's geometric estimate can be written as
\be
\label{eq:bjestimate}
\varepsilon_{Bj}=\left.\frac{dE_T}{d\eta}\right|_{\eta=0}\frac{1}{\tau_0\pi R^2}\, ,
\ee
where $dE_T/d\eta$ is the transverse energy $\sqrt{m^2+p_T^2}$ of all the
particles per unit rapidity and $R\approx 7$~fm is the radius of the nuclei.
The logic is simply that at time $\tau_0$ the energy within a volume 
$2 \tau_0$ in longitudinal extent between the two receding  pancakes
and $\pi R^2$ in transverse area must be at least $2 dE_T/d\eta$,
the total transverse energy between $\eta=-1$ and $\eta=+1$.  
At RHIC with 
$dE_T/d\eta \approx 800\,   {\rm GeV}$~\cite{Back:2004je},
we obtain $\varepsilon_{Bj}\approx 5\,  {\rm GeV/fm^3}$.
In choosing the volume in the denominator in the estimate 
(\ref{eq:bjestimate}) we neglected transverse expansion because
$\tau_0 \ll R$.  But, there 
is clearly an arbitrariness in the range of $\eta$ used; if we
had included particles produced at higher pseudorapidity (closer
to the beam directions) we would have obtained a larger estimate
of the energy density.  Note also that there is another sense in
which (\ref{eq:bjestimate}) is conservative. If there is an epoch 
after the time $\tau_0$ during which the matter expands as a 
hydrodynamic fluid, and we shall later see evidence that this is so,
then during this epoch  its energy density drops more rapidly than $1/\tau$
because as it expands (particularly longitudinally) it is doing work.  This
means that by using $1/\tau$ to run the
clock backwards from the measured final state transverse energy
to that at $\tau_0$ we have significantly underestimated the energy density
at $\tau_0$.  It is striking that even though we have 
deliberately been conservative in making this underestimate, we have
found an energy density that is about five times larger than the QCD
critical energy density 
$\varepsilon_c\approx 1\, {\rm GeV/fm^3}$, 
where the crossover from hadronic matter to quark-gluon plasma
occurs, according to lattice calculations of QCD thermodynamics~\cite{Bazavov:2009zn}. 

As shown in 
the right panel of \Fig{fig:general}, 
the spectrum in a  nucleus-nucleus collision 
extends to very high momentum, much larger than the
mean. However, the multiplicity of high momentum particles
drops very fast with momentum, as a large power of $p_T$. 
We may separate the spectrum 
into two sectors. In the soft sector, spectra drop exponentially
with $\sqrt{m^2+p_T^2}$ as in thermal equilibrium.  In the hard sector, spectra
drop like power laws in $p_T$ as is the case for hard particles produced
by high momentum-transfer parton-parton collisions at $\tau=0$. 
The bulk of the particles have momenta in the soft sector; hard particles
are rare in comparison.
The separation between 
the hard and the soft sectors, which is by no means  sharp, lies
in the range of a few (say 2-5) GeV.

\begin{figure} 
\begin{center}
\includegraphics[width=0.45\textwidth]{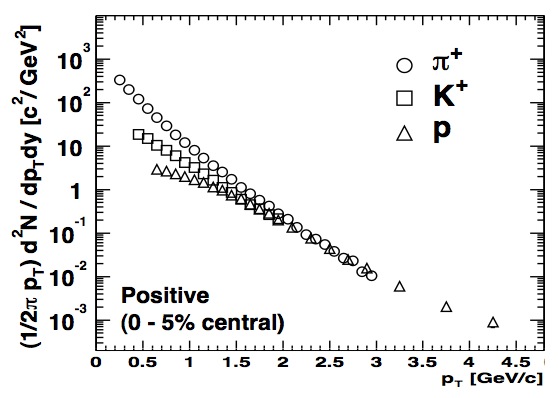}
\includegraphics[width=0.45\textwidth]{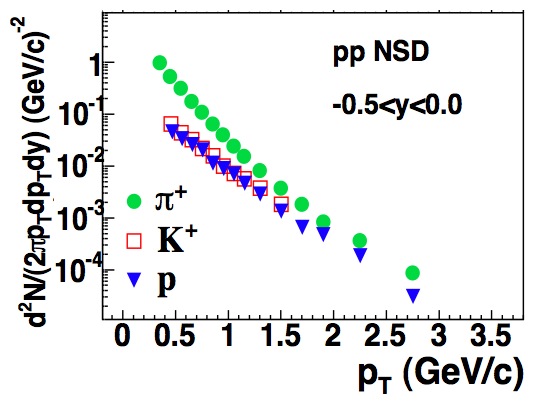}
\end{center}
\caption{\small \label{fig:generalb}
Left: Spectra for identified pions, kaons and protons as a function of $p_T$ in 
head-on gold-gold collisions at top RHIC energy \cite{Adler:2003cb}.
Right: Spectra for identified pions, kaons and protons as a function of $p_T$ 
in (non single diffractive) proton-proton collisions at
the same energy $\sqrt{s}=200~{\rm GeV}$~\cite{Adams:2003qm}.
 }
\end{figure}
%
There are several lines of evidence that indicate that the soft
particles in a heavy ion collision, which are the bulk of all the hadrons
in the final state, have interacted many times and come into local
thermal equilibrium.  The most direct approach comes via the analysis
of the exponentially falling spectra of identified hadrons.  Fitting a slope
to these exponential spectra and then extracting an ``effective temperature" for
each species of hadron yields different  ``effective temperatures" for each
species.  This species-dependence arises because the matter produced in a heavy ion
collision expands radially in the directions
transverse to the beam axis; perhaps explodes radially is a better phrase.
This means that we should expect the $p_T$ spectra
to be a thermal distribution boosted by
some radial velocity.  If all hadrons are boosted by the same {\it velocity},
the heavier the hadron the more its {\it momentum} is increased by the
radial boost.  
Indeed, what is found in data is that the effective temperature
increases with the mass of the hadron species.  This can be seen
at a qualitative level in 
the left panel of \Fig{fig:generalb}:
 in the soft regime, the proton, kaon and pion
spectra are ordered by mass, with the protons falling off most slowly with $p_T$,
indicating that they have the highest effective temperature.  Quantitatively, one
uses the data for hadron species with varying masses
species to first extract the mass-dependence of the effective temperature,
and thus the radial expansion velocity, and then to
extrapolate
the effective ``temperatures" to the  mass $\rightarrow$ zero limit, 
and in this way obtain a measurement
of the actual temperature of the final state hadrons.  This ``kinetic freezeout temperature''
is the temperature
at the (very late) time at which the gas of hadrons becomes so dilute that
elastic collisions between the hadrons cease,
and the momentum distributions therefore stop changing as the system expands.
In heavy ion collisions at the top RHIC energy, models of the kinetic freezeout
account for the data with freeze-out temperatures of $\approx 90$ MeV and radial
expansion velocities of $0.6\, c$ for collisions with the smallest impact 
parameters~\cite{Adams:2005dq}.
With increasing impact parameter, the radial velocity decreases and the freeze-out 
temperature increases. This is consistent with the picture that a smaller system builds
up less transverse flow and that during its expansion it cannot cool down as much
as a bigger system, since it falls apart earlier. 

The analysis just described is unique to heavy ion collisions: in elementary
electron-positron or proton-(anti)proton collisions, spectra at low transverse
momentum may also be fit by exponentials, but the ``temperatures" extracted
in this way do not have a systematic dependence on the hadron mass. In fact,
they are close to the same for all hadron species, as can
be seen qualitatively in 
the right panel \Fig{fig:generalb}.  
Simply seeing
exponential spectra and fitting a  ``temperature" therefore
does not in itself provide evidence for multiple interactions and equilibration.  
Making that case in the context of heavy ion collisions relies crucially
on the existence of a collective radial expansion with a common velocity
for all hadron species.

Demonstrating that the final state of a heavy ion collision at the time
of kinetic freezeout is a gas of hadrons in local thermal equilibrium 
is by itself not particularly interesting.  It does, however, embolden
us to ask whether the material produced in these collisions reaches
local thermal equilibrium at an earlier time, and thus at a higher temperature.
The best evidence for an affirmative answer to this question comes
from the analysis of  ``elliptic flow" in collisions with nonzero impact
parameter.  We shall discuss this at length in the next subsection.

\begin{figure} 
\begin{center}
\includegraphics[width=0.85\textwidth]{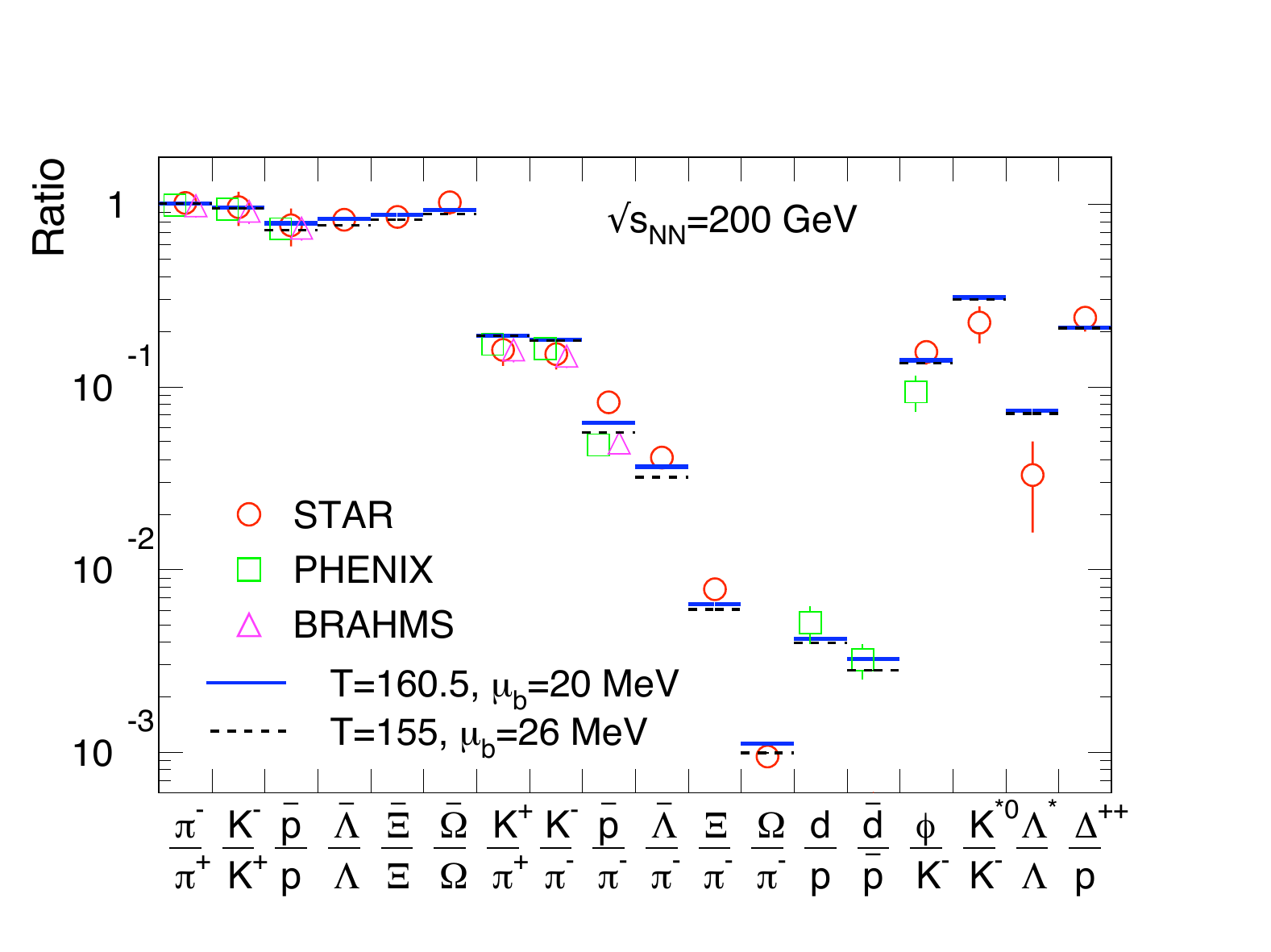}
\end{center}
\caption{\small \label{fig:general2a}
So-called thermal fit to different particle species. 
The relative abundance of different hadron species is well-described
by a two-parameter grand canonical ensemble in terms of temperature $T$
and baryon chemical potential $\mu_b$~\cite{Andronic:2005yp}.
}
\end{figure}

\begin{figure} 
\begin{center}
\includegraphics[width=0.45\textwidth]{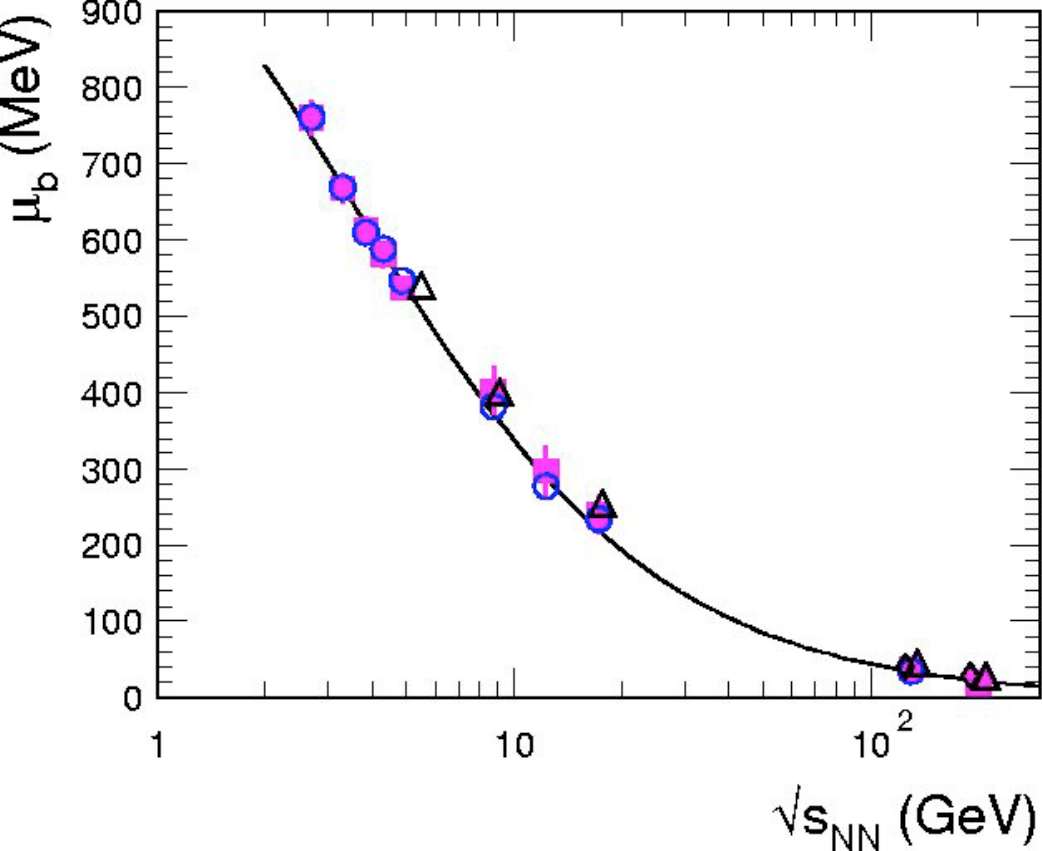}
\includegraphics[width=0.53\textwidth]{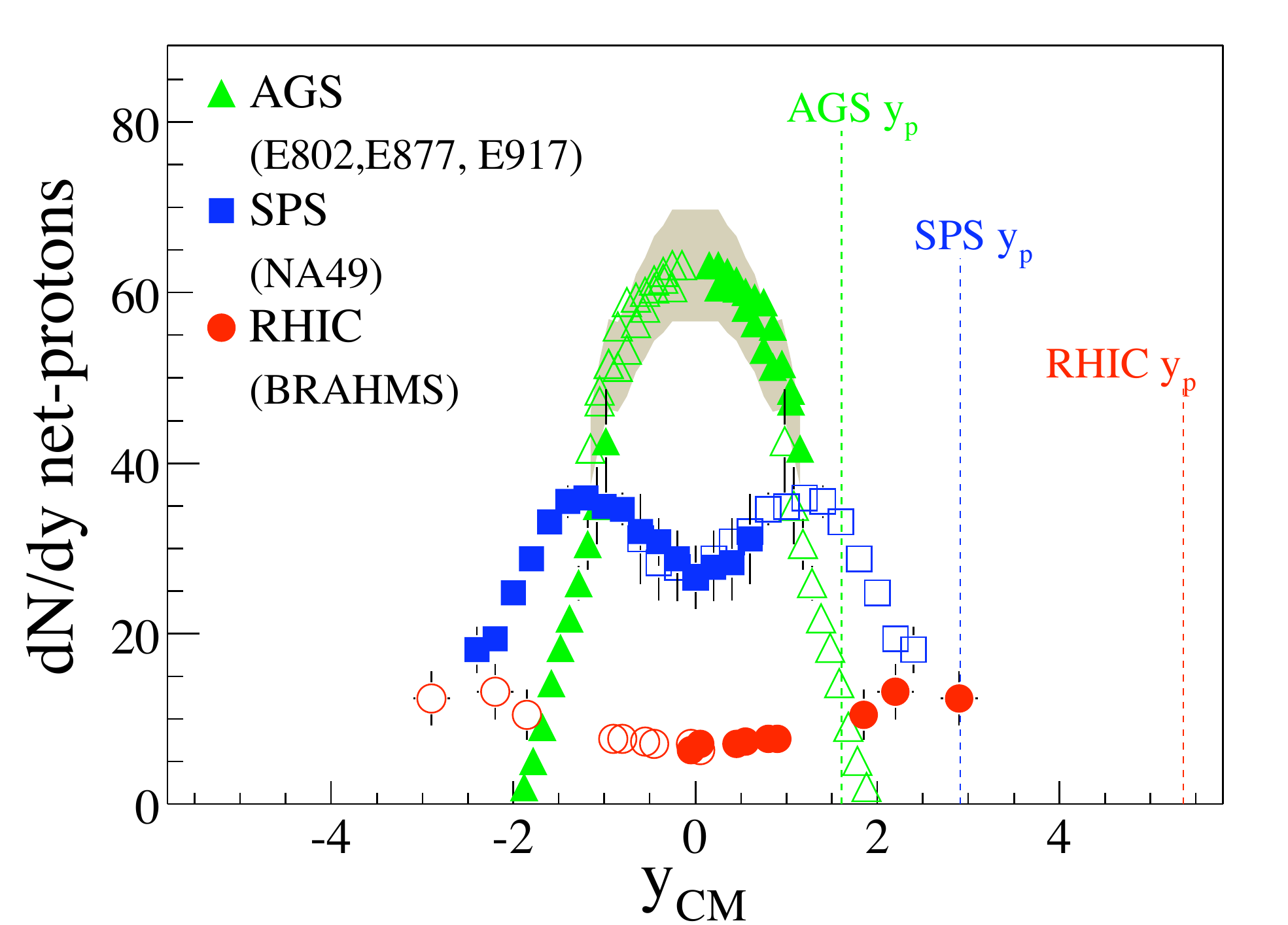}
\end{center}
\caption{\label{fig:general2} \small
Left: Chemical potential extracted from thermal fits at different center of mass 
energies \cite{Andronic:2005yp}. 
Right: The number of protons minus number of antiprotons
per unit rapidity for central heavy ion collisions~\cite{Bearden:2003hx}. 
This net proton number decreases with increasing center of mass energy
from $\sqrt{s}=5~{\rm GeV}$ (at the AGS collider
at BNL), via $\sqrt{s}=17~{\rm GeV}$ (at the SPS collider at CERN) to
$\sqrt{s}=200~{\rm GeV}$ (at RHIC).
(For each collision energy, $y_p$ indicates the rapidity of a hypothetical proton that
has the same velocity after the collision as it did before.)
}
\end{figure}

We close this subsection with a simpler analysis that lays further
groundwork by allowing us to see back to a somewhat earlier
epoch than that of kinetic freezeout.   If we think of a heavy ion
collision as a ``little bang", replaying the history of the big bang
in a small volume and with a vastly accelerated expansion rate, then
kinetic freezeout is the analogue of the (late) cosmological time at which 
photons and electrons no longer scatter off each other.  We now turn
to the analogue of the (earlier) cosmological epoch of nucleosynthesis,
namely the time at which the composition of the final state hadron
gas stops changing.  Experimentalists can measure the abundance
of more than a dozen hadron species, and it turns out that all the
ratios among these abundances can be fit upon assuming thermal
distributions 
with some temperature $T$ and some baryon number
chemical potential $\mu$, as shown in \Fig{fig:general2a}
This is a two parameter fit to about a dozen
ratios.  The temperature extracted in this way is called the chemical
freezeout temperature, since one interpretation is that it is the temperature
at which the hadronic matter becomes dilute
enough that inelastic hadron-hadron collisions cease to modify the
abundance ratios.  The chemical freezeout temperature in heavy
ion collisions at top RHIC energies is about 
155-180 MeV~\cite{BraunMunzinger:2003zd,Andronic:2005yp}.  
This is interesting for several reasons.
First, it is not far below the QCD phase transition temperature, which means
that the appropriateness of  a hadron gas description of this epoch may
be questioned.  Second, within error bars it is the same temperature that
is extracted by doing a thermal model fit to hadron production in electron-positron
collisions, in which final state rescattering, elastic or inelastic, can surely be
neglected.  So, by itself the success of the thermal fits to abundance ratios
in heavy ion collisions could be interpreted as telling us about the 
statistical nature of hadronization, as must be the case in electron-positron
collisions.  However, given that we know that in heavy ion collisions (and
not in electron-positron collisions) kinetic equilbrium is maintained down
to a {\it lower} kinetic freezeout temperature, and given that as we shall see
in the next subsection approximate local thermal equilbrium is achieved
at a {\it higher} temperature, it does seem most natural to interpret the
chemical freezeout temperature in heavy ion collisions as reflecting
the  temperature of the matter produced at the time when species-changing
processes cease.


We have not yet talked about the baryon number chemical potential extracted from
the thermal fit to abundance ratios.  As illustrated in 
the left panel of \Fig{fig:general2}, 
this $\mu$ decreases with increasing collision energy 
$\sqrt{s}$.  This energy dependence has two origins.  The dominant effect
is simply that at higher and higher collision energies more and more entropy
is produced while the total net baryon number in the collision is always
197+197.  At top RHIC energies, these baryons are diluted among the
8000 or so hadrons in the final state, making the baryon chemical potential
much smaller than it is in lower energy collisions where the final state
multiplicity is much lower.  The second effect is that, in the
highest energy collisions, most of the net baryon number from the two
incident nuclei stays at large pseudorapidity (meaning small angles near
the incident beam directions).  These two effects can be seen directly in
the data shown in 
the left panel of \Fig{fig:general}
and
 the right panel of \Fig{fig:general2}: 
 as the collision
energy increases, the total number of hadrons in the final state grows while
the net baryon number at mid-rapidity 
drops.\footnote{The data in the right panel of \Fig{fig:general2} is plotted relative
to rapidity 
\begin{equation}
y\equiv\frac{1}{2}\ln\left(\frac{E+p_L}{E-p_L}\right)\ ,
\end{equation}
where $E$ and $p_L$ are the energy and longitudinal momentum of
a proton in the final state.  Recall that rapidity and pseudorapidity $\eta$
(used in the plot in 
the left panel of \Fig{fig:general}) 
become the same in the limit
in which $E$ and $p_L$ are much greater than the proton mass.
When one plots data for all charged hadrons, as in the left panel of \Fig{fig:general}, only 
pseudorapidity can be defined since the rapidity of a hadron with a given
polar angle $\theta$ depends on the hadron mass.  When one plots data
for identified protons, pseudorapidity can be converted into rapidity.} 
This experimental fact that baryon number
is not ``fully stopped" teaches us about the dynamics of the earliest moments
of a hadron-hadron collision.  (In this respect, heavy ion collisions are not
qualitatively different than proton-proton collisions.)  In a high energy proton
proton collision, particle production at mid-rapidity is dominated by the
partons in the initial state that carry a small fraction of the momentum of
an individual nucleon --- small Bjorken $x$.  And, the small-$x$ parton
distributions functions that describe the initial state of the incident
nucleons or nuclei are dominated by gluons and to a lesser
extent by quark-antiquark pairs; the net baryon number is at larger-$x$.  
As we have already said, we shall not focus on the many interesting
questions related to the early-time dynamics in heavy ion collisions,
since gravity dual calculations have not had much to say about them.
Instead, in the next subsection we turn to the evidence that local
thermal equilibrium is established quickly, therefore at a high temperature.
This means that heavy ion collisions can teach us about properties of the
high temperature phase of QCD, namely the quark-gluon plasma.
And, we shall see later, so can gravity dual calculations.  
We shall henceforth always work at $\mu=0$. This is a good approximation
as long as $\mu/3$, the quark chemical potential, is much less than the
temperature $T$. The results in 
the left panel \Fig{fig:general2} 
show that this is a very good approximation
at top RHIC energies, and will be an even better approximation for heavy ion collisions
at the LHC.  Using lower energy heavy ion collisions to scan the QCD phase
diagram by varying $\mu$ is a very interesting ongoing research program,
but we shall not address it in this review.



\section{Elliptic flow}
\label{sec:EllipticFlow}
\subsection{Introduction and motivation}
\label{s:EllipticIntro}

\begin{figure}
\vspace{0.10in}
\centerline{\includegraphics[width=0.49\textwidth]
{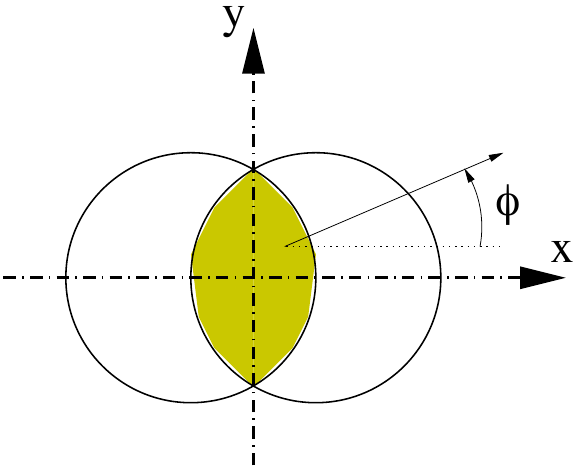}}
\caption{\small Sketch of the collision of two nuclei, shown in the transverse plane perpendicular
to the beam. The collision region is limited to the interaction almond in the center of
the transverse plane. Spectator nucleons located in the white regions of the nuclei
do not participate in the collision.  
Figure taken from Ref.~\cite{Ollitrault:2006va}. }
\label{fig1a}
\end{figure}

The phrase ``elliptic flow'' refers to a suite of experimental
observables in heavy ion physics that utilize the experimentalists' ability
to select events in which the impact parameter of the collision lies within
some specified range and use these events to study how the matter produced
in the collision flows collectively.  The basic idea is simple. Suppose we select
events in which the impact parameter is comparable to the nuclear radius. 
Now, imagine taking
a beam's eye view of one of these collisions. The
two Lorentz-contracted nuclei (think circular ``pancakes'') collide only in
an ``almond-shaped'' region, see Fig.~\ref{fig1a}
The fragments of the nuclei outside the almond
that did not collide (``spectator
nucleons") fly down the beam pipes.  All the few thousand particles at mid-rapidity
in the final state must have come from the few hundred 
nucleon-nucleon collisions that occurred within the almond.  
If these few thousand hadrons came instead from a few hundred independent
nucleon-nucleon collisions, just by the central limit theorem the few thousand final state
hadrons would be distributed uniformly in azimuthal angle $\phi$ (angle around
the beam direction).  This null hypothesis, which we shall make
quantitative below, is ruled out by the data as we shall see. 
If, on the other hand, the collisions
within the almond yield particles that interact, reach local equilibrium, and thus
produce some kind of fluid, our expectations for the ``shape'' of the 
azimuthal distribution of the final state hadrons is quite different.  The hypothesis
that is logically the opposite extreme to pretending that the thousands of partons
produced in the hundreds of nucleon-nucleon collisions do not see each other
is to pretend that what is produced is a fluid that flows according to the laws
of ideal, zero viscosity, hydrodynamics, since this extreme is achieved in the
limit of zero mean free path.  In hydrodynamics, the almond is thought
of as a drop of fluid, with zero pressure at its edges and a high pressure
at its center. This droplet of course explodes. And, since the pressure gradients
are greater across the short extent of the almond than they are across its
long direction, the explosion is azimuthally asymmetric.  
The first big news from the RHIC experimental
program was the discovery that, at RHIC energies, these azimuthal 
asymmetries can be large: the explosions can blast with summed
transverse momenta of the hadrons
that are twice as large in the short direction of the almond
as they are in the long direction.  And, 
as we shall see, it turns out that
ideal hydrodynamics does
a surprisingly good job of describing these asymmetric explosions of the matter
produced in heavy ion collisions with nonzero impact parameter. This has
implications which are sufficiently interesting that they motivate
our describing this story in considerable detail over the course
of this entire Section.   We close this introduction with a sketch
of these implications.

%

First, the agreement between data and ideal hydrodynamics
teaches us that the viscosity $\eta$ of the fluid produced in heavy ion collisions
must be low.   $\eta$ enters in the dimensionless ratio $\eta/s$, with
$s$ the entropy density, and it is $\eta/s$ that is constrained to be small.
A fluid that is close
to the ideal hydrodynamic limit, with small $\eta/s$,
requires strong coupling between 
the fluid constituents.  Small $\eta/s$ means that momentum is
not easily transported over distances that are long compared
to $\sim s^{-1/3}$, which means that there can be no 
well-defined quasiparticles with long
mean free paths in a low viscosity fluid since if they existed, they would transport
momentum and damp out shear flows.  No particles with long mean free paths
means strongly coupled constituents.

Second, we learn that the strong coupling between partons that results in
approximate local equilibration and fluid flow close to that described by
ideal hydrodynamics must set in very soon after the initial collision.  If partons
moved with significant mean free paths for many fm of time after the collision,
delaying equilibration for many fm, the almond would circularize to a significant
degree during this initial period of time and the azimuthal momentum
asymmetry generated
by any later period of hydrodynamic behavior would be less than observed.
When this argument is made quantitative, the conclusion is that RHIC collisions
produce strongly coupled fluid in approximate local thermal equilibrium within
close to or even somewhat less than 1 fm after the 
collision~\cite{Kolb:2003dz}.\footnote{Reaching approximate local thermal equilibrium and
hence hydrodynamic behavior within less than 1 fm after a heavy ion collision has been thought of as ``rapid equilibration'', since it is rapid compared to weak coupling 
estimates~\cite{Baier:2000sb}.  This observation has launched a large effort (that we shall not review) towards explaining equilibration as originating from weakly-coupled processes that arise in the presence of
the strong color fields that are present in the initial instants of a heavy ion collision.  A very recent calculation indicates, however, that the observed equilibration time may not be so rapid after all.  In a strongly coupled field theory with a dual gravitational description, when two sheets of energy density with a finite thickness collide at the speed of light a hydrodynamic description of the plasma that results becomes reliable only $\sim 3$ sheet-thicknesses after 
the collision~\cite{Chesler:2010bi}.  And, a Lorentz-contracted incident gold nucleus at RHIC has a maximum thickness of only 0.14 fm.  So, if the equilibration processes in heavy ion collisions could be thought of as  strongly coupled throughout, perhaps local thermal equilibrium and hydrodynamic behavior would set in even more rapidly than is indicated by the data.}

We can begin to see that the circle of ideas that emerge from the analysis
of elliptic flow data are what  make heavy ion collisions of interest to
the broader community of theoretical physicists for whom we are writing
this review.  These analyses justify the conclusion that only 1 fm after the
collision the matter produced can be described using the language
of thermodynamics and hydrodynamics. And, we have already seen
that at this early time the energy density is well above the hadron-QGP crossover in
QCD thermodynamics which is well-characterized
in lattice calculations.  This justifies the claim that heavy ion collisions
produce quark-gluon plasma.  Furthermore,  the same analyses teach
us that this quark-gluon plasma is a strongly coupled, low viscosity, fluid
with no quasiparticles having any significant mean free path.  Lattice
calculations have recently begun to cast some light on these transport properties
of quark-gluon plasma, but these lattice calculations that go beyond 
Euclidean thermodynamics
are still in their pioneering epoch.  Perturbative calculations of quark-gluon
plasma properties are built upon the existence of quasiparticles.  The analyses
of elliptic flow data
thus cast doubt upon their utility.  
And, we are motivated to study  the strongly coupled plasmas with
similar properties that can be analyzed via gauge theory / gravity
duality, since these calculational methods allow many questions
that go beyond thermodynamics to be probed rigorously
at strong coupling.  

In Section~\ref{s:EllipticData} we  first describe how to select classes of events all of whose impact
parameters lie in some narrow range.  Next, we define the experimental
observable used to measure azimuthal asymmetry, and describe how to falsify the null
hypothesis that the asymmetry is due only to statistical fluctuations. 
We then take a first look at the RHIC data and convince ourselves that the asymmetries
that are seen reflect collective flow, not statistical fluctuations. In Section~\ref{s:EllipticHydro} we
describe how to do a hydrodynamical calculation of the azimuthal asymmetry,
and in Section~\ref{s:EllipticCompare} we compare hydrodynamic calculations to RHIC data and
describe the conclusions that we have sketched above in more quantitative terms.
The reader interested only in the bottom line should skip directly to Section~\ref{s:EllipticCompare}.

\subsection{The elliptic flow observable $v_2$ at RHIC}
\label{s:EllipticData}

We want to study the dependence of collective flow on the size and anisotropy of
the almond-shaped region of the transverse plane, as seen in the qualitative beam's 
eye view sketch in Figure~\ref{fig1a}. To this end, it is obviously 
necessary to bin heavy ion collisions as a function of this impact parameter. 
This is possible in heavy ion collisions, since the number of hadrons produced 
in a heavy ion collision is anticorrelated with the impact parameter of the collision.  
For head-on collisions (conventionally
referred to as ``central collisions'') the multiplicity is high;
the multiplicity is much lower in collisions with impact parameters comparable
to the radii of the incident ions (often referred to as ``semi-peripheral collisions'');
the multiplicity is lower still in grazing (``peripheral'') collisions.   
Experimentalists therefore bin their events by multiplicity, using that
as a proxy for impact parameter.  The terminology used refers
to the ``0-5\% centrality bin'' and the ``5-10\%'' centrality bin and $\ldots$,
meaning the 5\% of events with the highest multiplicities, the next 5\% of
events with the next highest multiplicity, $\ldots$.  The correlation between
event multiplicity and impact parameter is described well by the Glauber
theory of multiple scattering, which we shall not review here.  Suffice to
say that even though the absolute value of the event multiplicities is the subject
of much ongoing research, the question of what distribution of impact parameters
corresponds to the 0-5\% centrality bin (namely the most head-on collisions) is 
well established.  Although experimentalists cannot literally pick a class
of events with a single value of the impact parameter, by binning their
data in multiplicity they can select a class of events with a reasonably
narrow distribution of impact parameters centered around any desired
value.  This is possible only because nuclei are big enough: in proton-proton
collisions, which in principle have impact parameters
since protons are not pointlike, there is no operational way to separate variations in 
impact parameter from event-by-event fluctuations in the multiplicity at
a given impact parameter.

Suppose that we have selected a class of semi-peripheral collisions. 
Since these collisions have a nonzero impact parameter,
the impact parameter vector together with the beam direction define a plane,
conventionally called the reaction plane.  Directions
within the transverse plane of Figure ~\ref{fig1a} specified by the azimuthal
angle $\phi$ now need not be equivalent.  We can ask to what extent
the multiplicity and momentum of hadrons flying across the short direction
of the collision almond (in the reaction plane) differs from that of 
the hadrons flying along the long direction of the collision almond (perpendicular
to the reaction plane).

Let us characterize this dependence on the reaction plane for the case of
the single inclusive particle spectrum $\frac{dN}{d^3{\bf p}}$
of a particular species of hadron.
The three-momentum ${\bf p}$ of a particle of mass $m$ is parametrized conveniently in terms 
of its transverse momentum $p_T$, its azimuthal angle $\phi$, and its
rapidity $y$ which specifies its longitudinal momentum. Specifically, 
\be
{\bf p} = \left(p_T\cos\phi, p_T\sin\phi, \sqrt{p_T^2+m^2} \sinh y\right)\ .
\ee 
The energy of the particle is  $E=\sqrt{p_T^2+m^2} \cosh y$. The single particle
spectrum can then be written as
\begin{equation}
 \frac{{\rm d}N}{{\rm d}^2{\bf p}_t\,{\rm d} y} = 
\frac{1}{2\pi p_T}\frac{{\rm d}N}{{\rm d}p_T\,{\rm d}y}
\left[ 1 + 2v_1\cos(\phi-\Phi_R) + 2v_2\cos 2(\phi-\Phi_R) + \cdots \right]\, ,
\label{eq3.9}
\end{equation}
where $\Phi_R$ denotes explicitly the azimuthal orientation of the reaction plane,
which we do not know {\it a priori}. 
Thus,  the azimuthal dependence of particle production is characterized by the harmonic
coefficients
\begin{equation}
	v_n \equiv \langle \exp\left[ i\, n\, (\phi-\Phi_R)\right] \rangle = 
	\frac{\int \frac{dN}{d^3{\bf p}}\, e^{i\, n\, (\phi-\Phi_R)}\, d^3p}{\int \frac{dN}{d^3{\bf p}}\, d^3p}\, .
	\label{eq3.1}
\end{equation}
The coefficients $v_n$ are referred to generically as $n$-th order flow. 
In particular, $v_1$ is referred to as
``directed flow'' and $v_2$ as ``elliptic flow''. 
In general, the $v_n$ can depend on the
transverse momentum $p_T$, the rapidity $y$, the impact
parameter of the collision, and they can differ for different particle species.
In the collision of identical nuclei at mid-rapidity,
the collision region is symmetric under $\phi \to \phi+\pi$ and all odd harmonics vanish. In
this case, the elliptic flow $v_2$ is the first non-vanishing coefficient. 
We shall focus only on mid-rapidity, $y=0$.  The most common observables
used are $v_2(p_T)$ for collisions with varying centrality (i.e. impact parameter)
and $v_2$ integrated over all $p_T$, which is just a function of centrality.
A nonzero $v_2$ (or $v_2(p_T)$) means that there are more particles (with
a given transverse momentum) going left and right than up and down.

%
\begin{figure*}[t]
\centering
\includegraphics[height=0.29\textheight,width=0.50\textwidth,origin=c,clip]
{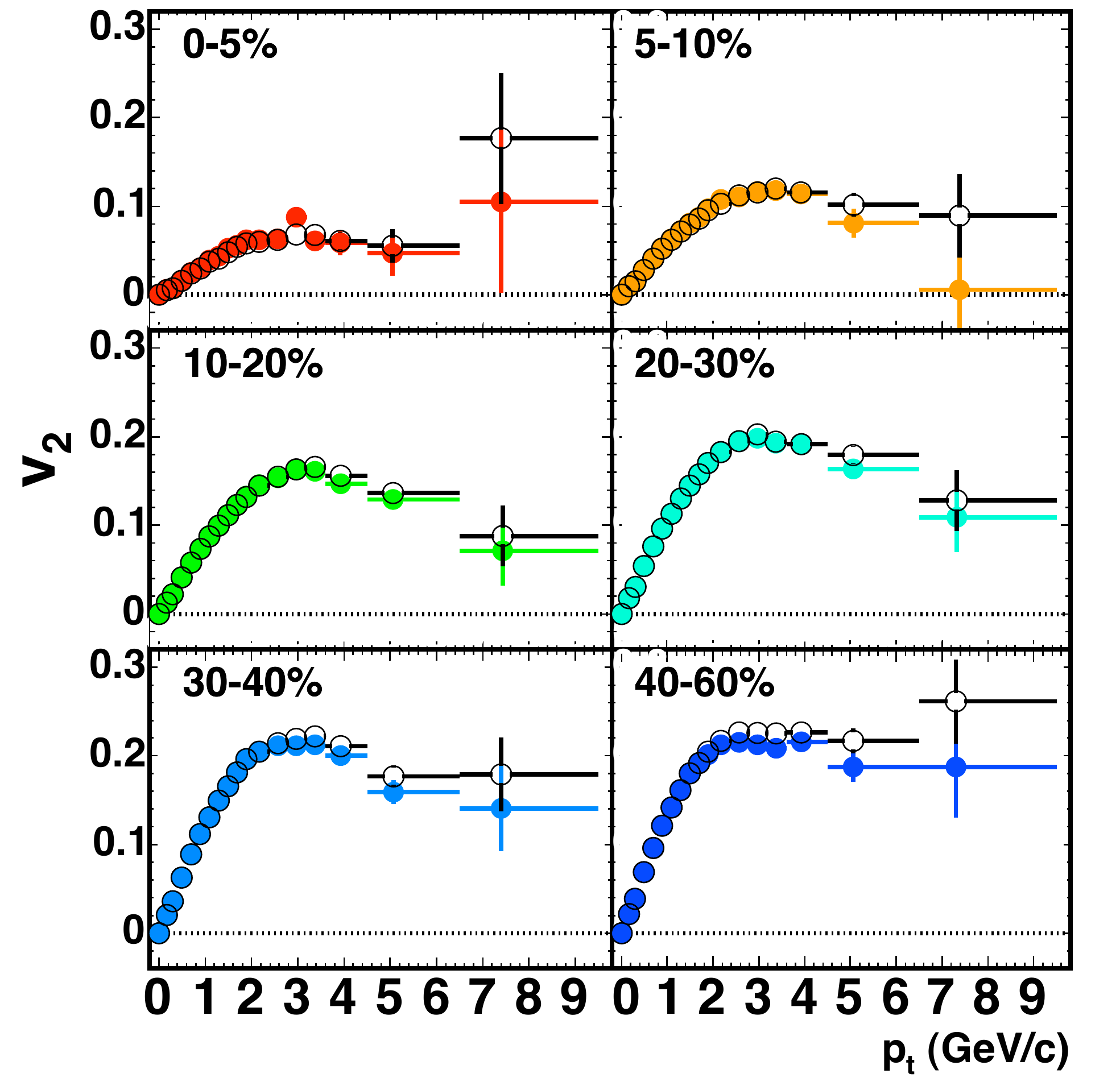}
\hfill
\includegraphics[height=0.25\textheight,width=0.46\textwidth,clip,origin=c]
{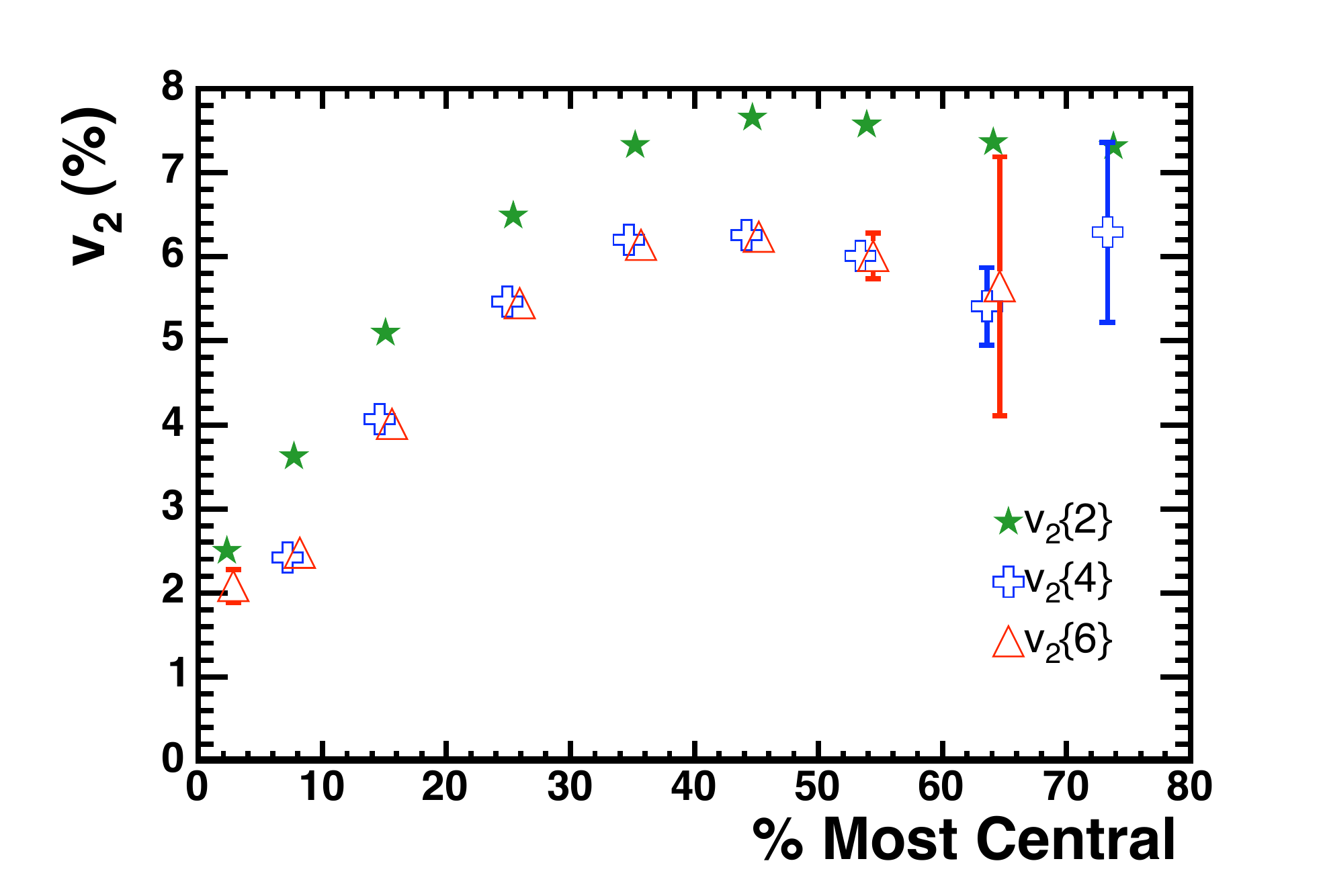}
\caption{\small 
Left: Transverse momentum dependence of the elliptic flow $v_2(p_T)$ for different 
centrality bins.  We see $v_2$ increasing as one goes from nearly head-on
collisions to semi-peripheral collisions.
Right: the $p_T$-integrated elliptic flow $v_2$ as a function of 
centrality bins, reconstructed with 2nd, 4th and 6th order cumulants. 
The most head-on collisions are on the left.
Figures taken from Ref.~\cite{Adams:2004bi}.}
\label{fig4}
\end{figure*}

In Fig.~\ref{fig4}, we show data for the transverse momentum 
dependence of the elliptic flow $v_2(p_T)$, and the transverse
momentum integrated quantity,  measured for different centrality classes in 
Au+Au collisions at RHIC.
The azimuthal asymmetry $v_2$ of the final
state single inclusive hadron spectrum is maximal in semi-peripheral collisions.
$v_2$ is less for more central collisions, 
since the initial geometric asymmetry is less --- the almond-shaped collision
region becomes closer to circular as the impact parameter is reduced.
In the idealized case of zero impact parameter, $v_2$ due to collective
effects should vanish. The reason that $v_2$ is not smaller 
even in the sample of the 5\% most head-on collisions
is that this sample includes events with a distribution of impact parameters
in the range $0< b < 3.5\, {\rm fm}$. 
In principle, $v_2$ should also vanish in the limit of grazing collisions.
Some decrease is visible in the most peripheral collisions shown in
the right panel of Fig.~\ref{fig4}, but the error bars become large long
before one reaches collisions which are literally grazing.

It is worth pausing to appreciate the size of the elliptic flow signal shown
in Fig.~\ref{fig4}. According to (\ref{eq3.9}), the ratio of $\frac{dN}{d^3{\bf p}}$
in whatever azimuthal direction it is largest to $\frac{dN}{d^3{\bf p}}$ ninety degrees
in azimuth away is $(1+2 v_2)/(1-2 v_2)$, which is a factor of 2 for $v_2=1/6$.
Thus, a $v_2$ of the order of magnitude shown in semi-peripheral collisions
at RHIC for $p_T\sim 2\, {\rm GeV}$ corresponds to a factor of more than 2 
azimuthal asymmetry.
It is natural to expect a nonzero quadrupolar
asymmetry $v_2$ at nonzero impact parameter 
due to collective effects occuring in the almond-shaped collision zone, given that
the almond shape itself has a quadrupole.
Of course, $v_4$, $v_6$, $v_8$, $\ldots$ will generically be nonzero as well.

There are two basic experimental problems in determining the coefficients
 (\ref{eq3.1}). The orientation $\Phi_R$ of the reaction plane is not known {\it a priori},
 and {\it a priori} it is unclear whether an azimuthal asymmetry in an event arises 
 as a consequence of collective motion, or whether it is the result of a 
 statistical fluctuation. In practice, both problems are connected. There are two methods in common use that each address both problems.
 One option is to use data at high rapidity to
determine $\Phi_R$, and then to measure the $v_n$ by applying (\ref{eq3.1})
to mid-rapidity data.  This eliminates the contribution of all statistical fluctuations  
uncorrelated with the reaction plane unless these fluctuations introduce
correlations between particles in the mid-rapidity and high rapidity regions
of the detector. 
The second option uses only the mid-rapidity data.  This means that
in each event the azimuthal direction of the maximal particle yield depends both on 
the orientation of the reaction plane  and on the statistical fluctuations
in that event. If it just so happens that there were 3 jets going up
and 3 jets going down, that would certainly skew the determination
of  the reaction plane $\Phi_R$ from the data. However, there is a 
systematic procedure, known as the cumulant method~\cite{Borghini:2000sa}, 
which makes an unambiguous separation of flow effects from non-flow 
effects possible. The basic idea is that if two particles are correlated in the azimuthal
angle because of a non-flow effect such as resonance decay or back-to-back jets or some other fluctuation,
then they are not correlated with all the other particles in the event. In contrast,
a flow effect leaves a signature in which all particles know
about the reaction plane, and thus are correlated. 
By a suitable analysis of 2-, 4-, and 6- particle correlations,
it is in practice possible to disentangle these two effects~\cite{Borghini:2000sa}. 
The right-hand side of Fig.~\ref{fig4} shows such an analysis. Subtracting  non-flow
effects in a more accurate 4-th order cumulant analysis, the elliptic flow signal is
reduced a bit. Further increasing the accuracy to 6th order, the signal does not
change, indicating that all non-flow effects have been corrected. 
In short, this demonstrates that the large elliptic flow signal cannot 
arise from statistical fluctuations in an incoherent superposition of nucleon-nucleon
collisions or from non-flow effects. It must characterize a {\it collective} phenomenon, reflecting
interactions among the debris produced by all the nucleon-nucleon collisions
occurring within the almond-shaped collision region. For more details,
we refer the reader to the original literature~\cite{Borghini:2000sa}.

\subsection{Calculating elliptic flow using (ideal) hydrodynamics}
\label{s:EllipticHydro}

We have now seen that the azimuthal asymmetry of the little explosions
created by heavy ion collisions with nonzero impact parameter,
parametrized by the second harmonic $v_2$, characterizes a 
strong {\it collective} phenomenon.   At time zero, the 
almond-shaped collision is azimuthally asymmetric {\it in space};  
subsequent collective dynamics must convert
that spatial asymmetry into an asymmetry {\it in momentum space}. After all,
$v_2$ measures the azimuthal asymmetry of the momentum distribution of the
final state hadrons.    This conversion of initial spatial anisotropy into final
momentum anisotropy is characteristic of any explosion --- one shapes the
explosive  in order to design a charge that blasts with greater force
in some directions than others.   Hydrodynamics provides the natural
language for describing such processes: the initial spatial anisotropy
corresponds to anisotropic pressure gradients.  Assuming that the pressure
is maximal at the center of the almond and zero at its edge, the gradient
is greater across the almond (in the reaction plane) than along it (perpendicular
to the reaction plane).
The elliptic flow $v_2$ measures the extent
to which these pressure gradients lead to an anisotropic explosion
with greater momentum flow in the reaction plane.
The size of $v_2$ characterizes the efficiency of translating initial pressure gradients
into collective flow.   By doing hydrodynamic calculations and comparing the
calculated $v_2$ to that in the data (and also comparing the final state radial
flow velocity to that determined from the single-particle spectra as described
in the previous section)  one can constrain the
input quantities that go into a hydrodynamic description.

The starting point in any hydrodynamic analysis is to consider
the limit of  ideal,
zero viscosity, hydrodynamics.
In this limit, the hydrodynamic description is specified entirely by an 
equation of state, which relates the pressure and the energy density, and by the
initial spatial distribution of energy density and fluid velocity.  
In particular, in ideal hydrodynamics
one is setting all dissipative coefficients (shear viscosity, bulk viscosity, and their
many higher order cousins) to zero.    If the equation of state is held fixed and
viscosity is turned on, $v_2$ must decrease:  turning on viscosity introduces 
dissipation that has the effect of turning some of the initial anisotropy
in pressure gradients into entropy production, rather than into directed collective flow.
So, upon making some assumption for the equation of state and for
the initial energy density distribution, setting the viscosities to zero yields
an upper bound on the $v_2$ in the final state. 
Ideal, inviscid, hydrodynamics has therefore long
been used as a calculational benchmark in heavy ion physics.  
As we shall see below, in heavy ion collisions at RHIC energies ideal hydrodynamics
does a good job of describing $v_2(p_T)$ for pions, kaons and protons
for transverse momenta $p_T$ below about $1-2$ GeV.    This motivates 
an ongoing research program in which one begins by comparing data
to the limiting case of ideal, inviscid, hydrodynamics and then turns
to a characterization of  dissipative effects, asking how large a viscosity
will spoil the agreement with data.  
In this subsection, we sketch how practitioners
determine the equation of state and initial energy density
profile, and we recall the basic principles behind
the hydrodynamic calculations on which all these studies rest. In the next
subsection, we summarize the current constraints on the shear viscosity
that are obtained by comparing to RHIC $v_2$ data.

The equation of state relates the pressure $P$ to the energy density $\varepsilon$.
$P$ is a thermodynamic quantity, and therefore can be calculated using the methods
of lattice quantum field theory.  Lattice calculations (or fits to them) 
of $P(\varepsilon)$  in the quark-gluon
plasma and in the crossover regime between QGP and hadron gas are often used as inputs
to hydrodynamic calculations.  At lower energy densities, practitioners either
use a hadron resonance gas model equation of state, or match the hydrodynamic
calculation onto a hadron cascade model.  One of the advantages of focussing on
the $v_2$ observable is that it is insensitive to the late time epoch of the collision,
when all the details of these choices matter.   This insensitivity is easy to
understand. $v_2$ describes the conversion of a spatial anisotropy into
anisotropic collective flow.  As this conversion begins, the initial almond-shaped
collision explodes with greater momentum across the short direction of the almond,
and therefore circularizes.  Once it has circularized, no further $v_2$ can develop.
Thus, $v_2$ is generated early in the collision, say over the first $\sim 5$~fm of time.
By the late times when a hadron gas description is needed, $v_2$ has already
been generated.  In contrast, the final state radial flow velocity reflects a time
integral over the pressure built up during all epochs of the collision: it is 
therefore sensitive
to how the hadron gas phase is described, and so is a less useful observable.

The discussion above reminds us of  a second sense in which the ideal hydrodynamic
calculation of $v_2$ is a benchmark:  ideal hydrodynamics requires local
equilibrium.  It therefore cannot be valid from time $t=0$.
By using an ideal hydrodynamic description beginning at $t=0$ we must again
be overestimating $v_2$, and so we can ask how long an initial phase during
which partons stream freely without, starting to circularize the almond-shaped region
without generating any momentum anisotropy, can be tolerated without spoiling
the agreement between calculations and data.

After choosing an equation of state, an initialization time, 
and viscosities (zero in the benchmark calculation), the only thing
that remains to be specified is the distribution of energy density as a function
of position in the almond-shaped collision region.  (The transverse
velocities are assumed to be zero initially.) In the simplest approach,
called Glauber theory, this energy density is proportional to the product
of the thickness of the two  nuclei at a given point in the transverse plane.
It is thus zero at the edge of the almond, where the thickness of one
nucleus goes to zero, and maximum at the center of the almond.
The proportionality constant is determined by fitting to data other than $v_2$,
see e.g.~\cite{Kolb:2001qz}. The assumptions behind this Glauber approach to
estimating how much energy density is created at a given location as a function of
the nuclear thickness at that location are assumptions
about physics of the collision at $t=0$. There is in particular one alternative
model parametrization, for which the energy density rises towards the center of the
almond more rapidly than the product of the nuclear thicknesses. This 
parametrization is referred to as the CGC initial condition, since it was first
motivated by ideas of parton saturation (called ``color glass condensate'')~\cite{Hirano:2005xf}.
The Glauber and CGC models for the initial energy density distribution are often
used as benchmarks in the hope that they bracket Nature's choice.

We turn now to a review of the formulation of the hydrodynamic equations of motion.
Hydrodynamics is an effective theory which describes the small frequency and 
long wavelength limit of an underlying interacting dynamical theory~\cite{Forster}. 
It is a classical field theory, where the fields can be understood as the 
expectation values of certain quantum operators in the underlying theory. 
In the hydrodynamic limit, since the length scales under consideration are longer than 
any correlation length  in the underlying theory, by virtue of the central limit theorem all 
$n$-point correlators of the underlying theory can be factorized into 
two point correlators (Gaussian approximation).
The fluctuations on these average values are small, 
and a description in terms of expectation values is meaningful.
If the underlying theory admits a (quasi)particle description, this statement is 
equivalent to saying that the hydrodynamic description involves averages over 
many of these fundamental degrees of freedom.

The hydrodynamic degrees of freedom include the expectation values of conserved currents such as the stress tensor $T^{\mu\nu}$ or the currents of conserved charges $J_B$. As a consequence of the conservation laws
\bea
\label{eq:emcons}
\del_\mu T^{\mu\nu} &=& 0 \,,
\\
\label{eq:ccons}
\del_\mu J_B^\mu &=& 0 \,,
\eea
long wavelength excitations of these fields can only relax on long 
timescales, since their relaxation must involve moving
stress-energy or charges over distances of order
the wavelength of the excitation. As a consequence, these conservation laws
lead to excitations whose lifetime diverges with their wavelength.  Such
excitations are called hydrodynamic modes.

If the only long-lived modes are those from conserved currents, then
hydrodynamics describes a normal fluid. However, there can be 
other degrees of freedom that lead to long lived modes in the long 
wave length limit. For example, in a phase
of matter in which some global symmetry is spontaneusly broken, the 
phase(s) of the symmetry breaking expectation value(s) is (are) also
hydrodynamic modes \cite{Forster}.   The classic example of this
is a superfluid, in which a global $U(1)$ symmetry is spontaneously
broken.  Chiral symmetry is spontaneously broken in QCD, but there
are two reasons why we can neglect the potential 
hydrodynamic modes associated with the chiral order
parameter~\cite{Son:1999pa}.  First, explicit chiral symmetry breaking
gives these modes a mass (the pion mass) and we are interested in
the hydrodynamic description of physics on length scales longer
than the inverse pion mass.  Second, 
we are interested in temperatures above the QCD crossover, 
at which the chiral order parameter is disordered, the
symmetry is restored, and this question does not arise.
So, we need only consider normal fluid hydrodynamics.
Furthermore, as we have discussed in Section~\ref{intro}, at RHIC energies the
matter produced in heavy ion collisions has only a very small 
baryon number density, and it is a good approximation to neglect
$J_B^\mu$.  The only hydrodynamic degrees of freedom are therefore
those described by $T^{\mu\nu}$.

At the length scales at which the hydrodynamic approximation is valid, each point of space can be 
regarded as a macroscopic {\it fluid cell}, characterized by its energy density $\varepsilon$, 
pressure $P$, and a velocity $u^{\mu}$. The velocity field can be defined by the energy flow
together with the constraint $u^2=-1$. 
In the so-called Landau frame, 
the four equations 
\be
\label{eq:landauframe}
u_\mu T^{\mu\nu}=- \varepsilon u^{\nu} \, .
\ee
determine $\varepsilon$ and  $u$ from the stress tensor.

Hydrodynamics can be viewed as a gradient expansion of the stress-tensor (and 
any other
hydrodynamic fields).  In general, the stress tensor can be separated into a term with no
gradients ({\it ideal}) and a term which contains all the gradients:
\be
T^{\mu\nu}=T_{\rm ideal}^{\mu\nu} + \Pi^{\mu \nu} \, .
\ee
In the rest frame of each fluid cell ($u^i=0$),  the ideal piece is diagonal 
and isotropic $T_{\rm ideal}^{\mu\nu}={\rm diag}\left(\varepsilon,P,P,P\right)$.
Thus, in any frame,
\be
\label{eq:stideal}
T_{\rm ideal}^{\mu \nu}=(\varepsilon+P) u^{\mu} u^{\nu}+P g^{\mu \nu} \, ,
\ee 
where $g^{\mu \nu}$ is the metric of the space.  

If there were a nonzero
density of some conserved charge $n$, the velocity field could either be defined
in the Landau frame as above or may
instead be  defined in the so-called Eckart frame,
with $J^\mu=n u^{\mu}$.  In the Landau frame,
the definition (\ref{eq:landauframe}) of $u^{\mu}$
implies $\Pi^{\mu\nu}u_\nu=0$ (transversality).  Hence,
there is no heat flow but there can be currents of the conserved
charge.  In the Eckart frame, the velocity field is comoving with
the conserved charge, but there can be heat flow.

\underline{A. Ideal fluid}\\
Ideal hydrodynamics is 
the limit in which all gradient terms in $T^{\mu\nu}$ are neglected. 
It can be used to describe motions of the fluid that occur on 
macroscopic length scales and
time scales associated with how the fluid is ``stirred''
that are long compared to any internal scales associated
with the fluid itself. Corrections to ideal hydrodynamics --- 
namely the gradient terms in $\Pi^{\mu\nu}$
that we shall discuss shortly --- introduce internal length and
time scales. For example, time scales for relaxation of perturbations
away from local thermal equilibrium, length scales
associated with mean free paths, etc. Ideal hydrodynamics works on
longer length scales than these.
At long time scales, the fluid cells are 
in equilibrium, $P$ is the
pressure in the rest frame of each fluid cell, 
and $\varepsilon$ and $P$ are related by the equation of 
state. 
This equation of state can be determined by studying 
a homogeneous system at rest with no gradients, for
example via a lattice calculation. 

The range of applicability of ideal fluid hydrodynamics can be characterized in terms of the
isotropization scale $\tau_{\rm iso}$ and the thermalization scale $\tau_{\rm eq}$.
The isotropization scale
measures the characteristic time over  which an initially
anisotropic stress tensor acquires a diagonal form in the local
rest frame. The thermalization scale sets the time over which entropy production ceases. 
These scales do not need to be the same, but $\tau_{\rm eq} \geq \tau_{\rm iso}$
since thermalization implies isotropy. 
$P$ and $\varepsilon$ are 
related by the equilibrium equation of state only at times $\tau>\tau_{\rm eq}$.
If there is a large separation of scales 
$\tau_{\rm iso}\ll \tau_{\rm eq}$, 
then between these timescales the relationship between $P$ and $\varepsilon$
can be different and is generically time-dependent. Ideal hydrodynamics
can nevertheless be used, as long as $P(\varepsilon)$ is replaced by
some suitable non-equilibrium equation
of state that will depend on exactly how the system
is out of thermal equilibrium.  It is worth noting that in the
case of a conformal theory, since the trace of the stress tensor vanishes, 
$P=\frac{1}{3}\varepsilon$ whether or not
the system is in equilibrium.  In this special case, as long as there
are no charge densities the equation of state is unmodified even if the
system is isotropic but out of equilibrium, and ideal hydrodynamics with 
the standard equation of state can be used as long as $\tau\gg \tau_{\rm iso}$.


\underline{B. 1st order dissipative fluid dynamics}

Going beyond the infinite wavelength limit requires the introduction of viscosities. 
To first order in gradients, the requirement that $\Pi^{\mu\nu}$ be transverse
means that it must take the form
\be
\label{eq:pilinear}
\Pi^{\mu\nu}= -\eta(\varepsilon) \sigma^{\mu\nu}-\zeta(\varepsilon) \Delta^{\mu\nu} \,\nabla \cdot u \, ,
\ee
where $\eta$ and $\zeta$ are the shear and bulk viscosities, 
$\nabla^\mu=\Delta^{\mu\nu} d_\nu$, with $d_\nu$  the covariant derivative
and
\bea
&&\Delta^{\mu\nu}=g^{\mu \nu} + u^{\mu} u^{\nu} \, ,\\
&&
\label{eq:shear}
\sigma^{\mu\nu}=\Delta^{\mu\alpha}\Delta^{\nu\beta} \left( \nabla_\alpha u_\beta + \nabla_\beta u_\alpha\right)
  -\frac{2}{3}\Delta_{\alpha\beta}\, \nabla \cdot u \, .
\eea
The operator $\Delta^{\mu\nu}$ is the projector into the space components of the fluid rest frame. 
Note that in this frame the only spatial gradients that appear in \Eq{eq:pilinear} are gradients
of the velocity fields. The reason that the derivative of $\varepsilon$ does not appear is that
it can be eliminated in the first order equations by using the zeroth order equation of motion
\be
D \varepsilon=-(\varepsilon+P)\,\nabla_\mu u^\mu\ ,
\label{entrop}
\ee 
where $D=u^\mu d_\mu$ is the time derivative in the fluid rest frame.  (Similarly, time derivatives of the energy density can be eliminated in the second order equations that we shall give below using the first order equations of motion.)

It is worth pausing to explain why we have introduced 
a covariant derivative, even though
we will
only ever be interested in heavy ion collisions --- and thus hydrodynamics --- occurring
in flat spacetime.  But, it is often convenient to use curvilinear coordinates with a nontrivial
metric. For example,
the longitudinal dynamics is more conveniently described using
proper time $\tau=\sqrt{t^2-z^2}$ and spacetime rapidity $\xi={\rm arctanh}(z/t)$
as coordinates rather than $t$ and $z$.   In these ``Milne coordinates'', 
the metric
is given by $g_{\mu\nu}={\rm diag}(g_{\tau\tau},g_{xx},g_{yy},g_{\xi\xi}) = (-1,1,1,\tau^2)$.
These coordinates are useful because boost
invariance simply translates into the requirement that $\varepsilon$, $u^\mu$ and $\Pi^{\mu\nu}$
be independent of $\xi$, depending on $\tau$ only. In particular, if the initial conditions
are boost invariant, then the fluid dynamic evolution will preserve this boost invariance,
and the numerical calculation reduces in Milne coordinates to a $2+1$-dimensional
problem. In the high energy limit, one expects that hadronic collisions distribute 
energy density approximately uniformly over a wide range rapidity. Accordingly,
boost-invariant initial conditions are often taken to be a good 
approximation~\cite{Bjorken:1982qr}. But, 
fully $3+1$-dimensional calculations that do not assume boost invariance can
also be found in the 
literature \cite{Hirano:2002ds,Hama:2005dz,Nonaka:2006yn,Renk:2006sx}.

It is often convenient to phrase the hydrodynamic equations in terms of the entropy density $s$. In the absence of conserved charges, i.e. with baryon chemical potential $\mu_B=0$, the entropy density is $s=(\varepsilon +P)/T$.  Using this and another fundamental thermodynamic relation, 
$DE = T\, DS -P\, DV$ (where $E/V\equiv\varepsilon$) the zeroth order equation of motion
(\ref{entrop}) becomes exactly the equation of entropy
flow for an ideal isentropic fluid
\be
D\, s=- s\,\nabla_\mu u^\mu
\label{entrop2}
\ee 
Repeating this analysis at first order, including the viscous terms,
one easily derives from $u_\mu\, \nabla_\nu\, T^{\mu\nu} = 0$
that
\be
\frac{D\, s}{s}=- \,\nabla_\mu u^\mu -  \frac{1}{s\, T} \, \Pi^{\mu\nu}\, \nabla_\nu u_\mu\, .
\label{entrop3}
\ee 
A similar analysis of the other three hydrodynamic equations then shows that they take the form
\be
D u_\alpha = -\frac{1}{Ts} \Delta_{\alpha\nu}\Biggl( \nabla^\nu P  + \nabla_\mu \Pi^{\mu\nu}\Biggr)\ .
\ee
It then follows from the structure of the shear tensor $\Pi^{\mu\nu}$ 
(\ref{eq:pilinear}) that shear viscosity and bulk viscosity always appear in the hydrodynamic
equations of motion in the dimensionless combinations $\eta/s$ and
$\zeta/s$. The net entropy increase is proportional to these dimensionless
quantities. Gradients of the velocity field are measured in units of $1/T$.

In a conformal theory, $\zeta=0$ since $\Pi^{\mu\nu}$ must be traceless.
There are a number of indications from lattice calculations that as the
temperature is increased above $(1.5-2)T_c$, 
with $T_c$ the crossover temperature, the quark-gluon
plasma becomes more and more conformal.  The equation of state
approaches $P=\frac{1}{3}\varepsilon$ \cite{Cheng:2007jq,Borsanyi:2010cj}. The bulk viscosity drops
rapidly~\cite{Meyer:2007dy}. So, we shall set $\zeta=0$ throughout the following, in
so doing neglecting
temperatures close to $T_c$.  One of the things that makes heavy ion
collisions at the LHC interesting is that in these collisions the plasma that
is created is expected to be better approximated as conformal
than is the case at RHIC, where the temperature at $\tau=1$~fm
is thought to be between $1.5 T_c$ and $2 T_c$.

Just like the equation of state $P(\varepsilon)$,
the shear viscosity $\eta(\varepsilon)$ is an input to the hydrodynamic
description that must be obtained either from experiment
or from the underlying microscopic theory.
We shall discuss in Section \ref{sec:latticeTransport} how transport coefficients like $\eta$
are obtained from 
correlation functions of the underlying microscopic theory
via Kubo formulae.

\underline{C. Second order dissipative Hydrodynamics}

Even though hydrodynamics is a controlled expansion in gradients,
the first order expression for the tensor 
$\Pi^{\mu\nu}$, \Eq{eq:pilinear},
is unsuitable for numerical computations. The problem is that
the set of equations (\ref{eq:emcons}) with the approximations (\ref{eq:pilinear})
leads to acausal propagation. 
Even though this problem only arises for modes outside of the region 
of validity of hydrodynamics (namely high momentum
modes with short wavelengths of the order of the microscopic length scale defined
by $\eta$),
the numerical evaluation of the first order equations of motion is sensitive to 
the acausality in
these hard modes. This problem is solved by going to one higher order in the 
gradient expansion. This is known as second order hydrodynamics.

There is a phenomenological approach to second
order hydrodynamics  due to M{\"u}ller, 
Israel and Stewart aimed at explicitly removing the acausal 
propagation~\cite{Mueller2od,Israel:1976tn,Israel:1979wp}.   
In this approach, the tensor $\Pi^{\mu\nu}$ is treated as a 
new hydrodynamic 
variable and a new dynamical equation is introduced. In its simplest form 
this equation is 
 \be
 \label{eq:rtaprox}
 \tau_{\Pi} D \Pi^{\mu\nu}=-\Pi^{\mu\nu}-\eta \sigma^{\mu\nu} \, ,
 \ee
where $\tau_{\Pi}$ is a new (second order) coefficient. Note that as $\tau_{\Pi} \rightarrow 0$,
\Eq{eq:rtaprox} coincides with \Eq{eq:pilinear} with the bulk viscosity $\zeta$ set to zero. Eq.~(
\ref{eq:rtaprox}) is such that $\Pi^{\mu\nu}$ relaxes
to its first order form in a (proper) time $\tau_{\Pi}$.
There are several variants of this equation in the literature, all of which follow the same 
philosophy. They all introduce the relaxation time as the characteristic time in which the 
tensor $\Pi^{\mu\nu}$ relaxes to its first order value. The variations arise from different
ways of  
fixing some pathologies of  \Eq{eq:rtaprox}, since as written
\Eq {eq:rtaprox} does not lead to 
a transverse stress tensor (although this is a higher order effect)
and is not conformally invariant. 
Since in this approach the relaxation time is introduced {\it ad hoc}, it
may not be possible to give a prescription for 
extracting  it from the underlying microscopic theory. 

The systematic extraction of second order coefficients demands
a similar analysis of the second order gradients as was done at first order.
 The strategy is, once again, to 
write all possible terms with two derivatives which are transverse and consistent with the
symmetries of the theory. As before, only spatial gradients (in the fluid rest frame) are considered, since
time gradients can be related to the former via the zeroth order equations of motion. 

In a conformal theory, second order hydrodynamics simplifies.
First, only terms such that $\Pi^{\mu}_\mu=0$ are allowed. Furthermore, the theory 
must be invariant under Weyl transformations 
\be
g_{\mu\nu} \rightarrow e^{-2\omega (x)} g_{\mu\nu}
\ee
which implies 
\be
T \rightarrow e^{\omega (x)} T\ ,\quad
u^\mu \rightarrow  e^{\omega (x)} u^\mu\ ,\quad
T^{\mu\nu} \rightarrow e^{(d+2)\omega (x)} T^{\mu\nu}\ ,
\ee
where $T$ is the temperature and $d$ the number of spacetime dimensions. 
It turns out that there are only five operators
that respect these constraints~\cite{Baier:2007ix}.
The second order contributions to the
tensor $\Pi^{\mu\nu}$  are linear combinations of these operators, and can be cast in the form~\cite{Baier:2007ix,Bhattacharyya:2008jc} 
\bea
\label{ISlikePi}
\Pi^{\mu\nu} & =& -\eta \sigma^{\mu\nu}- \tau_\Pi \left[ {}^\<D\Pi^{\mu\nu}{}^\> 
 + \frac d{d-1} \Pi^{\mu\nu}
    (\nabla{\cdot}u) \right] \nonumber\\
  &\quad &
  + \kappa\left[R^{\<\mu\nu\>}-(d-2) u_\alpha R^{\alpha\<\mu\nu\>\beta} 
      u_\beta\right]\nonumber\\
  & +& \frac{\lambda_1}{\eta^2} {\Pi^{\<\mu}}_\lambda \Pi^{\nu\>\lambda}
  - \frac{\lambda_2}\eta {\Pi^{\<\mu}}_\lambda \Omega^{\nu\>\lambda}
  + \lambda_3 {\Omega^{\<\mu}}_\lambda \Omega^{\nu\>\lambda}\, .
\eea
Here, $R^{\mu\nu}$ is the Ricci tensor,
the indices in brackets are the symmetrized traceless projectors onto the space components in
the fluid rest frame, namely
\begin{equation}
  {}^\<A^{\mu\nu}{}^\>
 \equiv \frac12 \Delta^{\mu\alpha} \Delta^{\nu\beta}
     (A_{\alpha\beta} + A_{\beta\alpha}) 
  - \frac1{d-1} \Delta^{\mu\nu} \Delta^{\alpha\beta} A_{\alpha\beta}\, 
\equiv A^{\<\mu\nu\>} \, ,
\end{equation}
and the vorticity tensor is defined as 
\begin{equation}
  \Omega^{\mu\nu} \equiv
  \frac12   
    \Delta^{\mu\alpha}\Delta^{\nu\beta}
    (\nabla_\alpha u_\beta - \nabla_\beta u_\alpha)\, .
\end{equation}
In deriving (\ref{ISlikePi}), we have replaced $\eta \sigma^{\mu\nu}$ by $\Pi^{\mu\nu}$
on the right-hand side in places where doing so makes no change at second order.
We see from (\ref{ISlikePi}) that five new coefficients 
$ \tau_{\Pi}$,  $\kappa$, $\lambda_1$,   $\lambda_2$, and  $\lambda_3$  arise at second order in the hydrodynamic description of a conformal fluid, in addition to $\eta$ and the equation of state which arise at first and zeroth order respectively. 
The coefficient
$\kappa$ is not relevant for hydrodynamics in flat spacetime. 
The $\lambda_i$ coefficients
involve nonlinear combinations of fields in the rest frame and, thus, are invisible in 
linearized hydrodynamics. Thus, these three coefficients cannot be extracted from 
linear response. 
Of these three, only $\lambda_1$ is relevant in the
absence of vorticity, as in the numerical simulations 
that we will describe in the
next Section. These simulations have also shown that, for physically 
motivated choices of $\lambda_1$, the results are insensitive to its precise value, 
leaving  $\tau_\pi$ as the only phenomenologically relevant second order parameter in the
hydrodynamic description of a conformal fluid.
In a generic, nonconformal fluid, there are nine additional transport 
coefficients~\cite{Romatschke:2009kr}.

For more in-depth discussions of  2nd order viscous hydrodynamics and its applications to heavy ion collisions,  
see e.g. Refs.~\cite{Muronga:2001zk,Teaney:2003kp,Muronga:2003ta,Muronga:2004sf,Baier:2006um,Baier:2006gy,Romatschke:2007mq,Song:2007fn,Dusling:2007gi,Song:2007ux,Luzum:2008cw,Song:2008si,Molnar:2008xj,Song:2008hj,Luzum:2009sb,Song:2009je,Song:2009rh,Heinz:2009xj,Romatschke:2009im,Teaney:2009qa,Shen:2010uy,Song:2010mg}.

\subsection{Comparing elliptic flow in heavy ion collisions and hydrodynamic calculations}
\label{s:EllipticCompare}

For the case of ideal hydrodynamics, the hydrodynamic equations of motion
are fully specified once the equation of state $P=P(\varepsilon)$ is given.
A second-order dissipative hydrodynamic calculation also requires knowledge
of the transport coefficients $\eta(\varepsilon)$ and $\zeta(\varepsilon)$ (although
in practice the latter is typically set to zero) and the relaxation time $\tau_\Pi$
and the second-order coefficient $\lambda_1$
entering Eq.~(\ref{ISlikePi}). ($\kappa$ would enter in curved
space time, and $\lambda_2$ and $\lambda_3$ would enter in the presence of
vorticity.)  All these parameters are well-defined in terms of correlation functions
in the underlying quantum field theory.  In this sense, the hydrodynamic 
evolution equations are model-independent.

 The output of any hydrodynamic  calculation depends on more than the evolution
 equations.  One must make model assumptions about the initial energy density
 distribution. As we have discussed in Section~\ref{s:EllipticData}, there are two benchmark
 models for the energy distribution across the almond-shaped collision region, which
 serve to give us a sense of the degree to which results are sensitive to our
 lack of knowledge of the details of this initial profile.  
Often, the initial transverse velocity fields are set to zero and 
boost invariance is assumed for the longitudinal velocity field and the evolution. 
For dissipative 
hydrodynamic simulations, the off-diagonal elements of the energy-momentum tensor 
are additional hydrodynamic fields which must be initialized.  
The initialization 
time $\tau_0$, at which these initial conditions are fixed, is an additional
model parameter. It can be viewed as characterizing the isotropization time, at which
hydrodynamics starts to apply but collective flow has not yet developed. 
In addition to initial state
 sensitivity, results depend on assumptions made about how the system stops
 behaving hydrodynamically and {\it freezes out}.  
 In practice, 
 freezeout is often assumed to happen as a rapid decoupling:  when a 
 specified criterion is satisfied (e.g. when a fluid cell drops below a critical
 energy or entropy density) then the hydrodynamic fields in the unit cell
 are mapped onto hadronic equilibrium Bose/Fermi distributions.  This treatment
 assumes that hydrodynamics is valid all the way down to the kinetic freezeout
 temperature, below which one has noninteracting
 hadrons.  Alternatively, at a higher temperature close to
 the crossover where hadrons are formed, one can map the hydrodynamic
 fields onto a hadron cascade which accounts for the effects of rescattering
 in the interacting hadronic phase without assuming that its behavior is 
 hydrodynamic~\cite{Bass:2000ib,Teaney:2001av}.  Indeed, recent work  suggests that hadronization may be triggered by cavitation induced by the large bulk viscosity in the vicinity of the crossover temperature~\cite{Rajagopal:2009yw}.
As we have discussed, $v_2$
 is insensitive to details of how the late-time evolution is treated because $v_2$
 is generated during the epoch when the collision region is azimuthally
 anisotropic. Nevertheless, these late-time issues do matter when one
 does a global  fit to $v_2$ and the single-particle spectra, since the latter
 are affected by the radial flow which is built up over the entire history
 of the collision.
 Finally, the validity of results from any hydrodynamic 
 calculation depends on the assumption that a hydrodynamic description
 is applicable.  This assumption can be checked at late times by checking the
 sensitivity to how freezeout is modelled and can be checked at early times
 by confirming the insensitivity of results to the values of the second order
 hydrodynamic coefficients and to the initialization
 of the higher order off-diagonal elements of the energy momentum tensor: 
 if hydrodynamics  is valid, the gradients
 must be small enough at all times that second order effects are small
 compared to first order effects.
  
 In practice, the dependence of physics conclusions on all these model assumptions has to
 be established by systematically varying the initial conditions and freeze-out prescriptions
 within a wide physically motivated parameter range, and comparing to data on
 both the single particle spectra (i.e. the radial velocity)  and the azimuthal
 flow anisotropy coefficient $v_2$. At the current time, several generic observations
 have emerged from pursuing this program in comparison to data from RHIC:
%
\begin{figure}
\centerline{\includegraphics[height=6.3cm,width=7.10cm,origin=c,clip]
{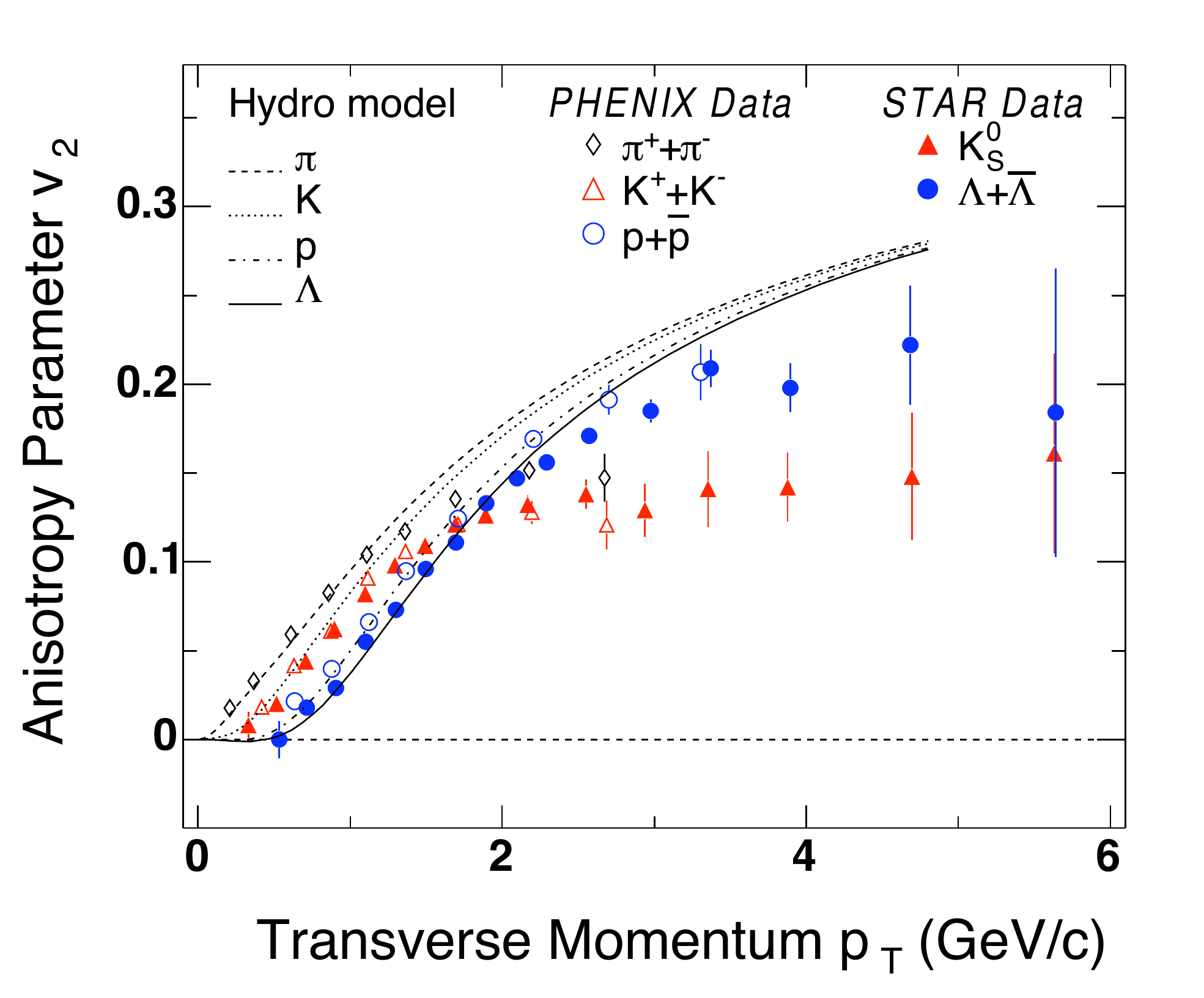}}
\caption{\small The elliptic flow $v_2$ versus $p_T$ for a large number of identified
hadrons (pions, kaons, protons, $\Lambda$'s) showing
the comparison between an {\it ideal} hydrodynamic calculation to data from RHIC.  
Figure taken from~\cite{Heinz:2005zg}.}
\label{v2hadchem}
\end{figure}

 \begin{enumerate}
   \item \emph{Perfect fluid dynamics reproduces the size and centrality dependence of $v_2$}  \\
   The RHIC data on single inclusive hadronic spectra $dN/ p_T\, dp_T\, dy$ and their
   leading azimuthal dependence $v_2(p_T)$ can be reproduced in magnitude and shape
   by ideal hydrodynamic calculations, for particles with $p_T<1 \, \rm{GeV}$. 
   The hydrodynamic picture is expected to break down for sufficiently small wavelength,
   i.e. high momenta, consistent with the observation that significant deviations 
   occur for $p_T> 2 \, \rm{GeV}$ (see left panel of Fig.~\ref{fig4}). The initialization time for these
   calculations is $\tau_0=0.6-1 \, \rm{fm}$. If $\tau_0$ is chosen larger, the agreement between
   ideal hydrodynamics and data is spoiled.
   This gives significant support to a picture 
   in which thermalization is achieved fast. It also provides the 
   first indication that the shear viscosity of the fluid produced at RHIC
   must be small.
   \item \emph{The mass-ordering of identified hadron spectra}\\
   The $p_T$-differential azimuthal asymmetry $v_2(p_T)$ of identified single inclusive hadron
   spectra shows a characteristic mass ordering in the range of $p_T < 2\, \rm{GeV}$: at small
   $p_T$, the azimuthal asymmetry of light hadrons is significantly more pronounced than that
   of heavier hadrons, see Fig.~\ref{v2hadchem}. This qualitative agreement of hydrodynamic 
   simulations with data supports the picture that all hadron species emerge from a 
   single fluid moving with a common flow field. 
   %
\begin{figure}
\vspace{0.10in}
\includegraphics[width=0.48\textwidth]
{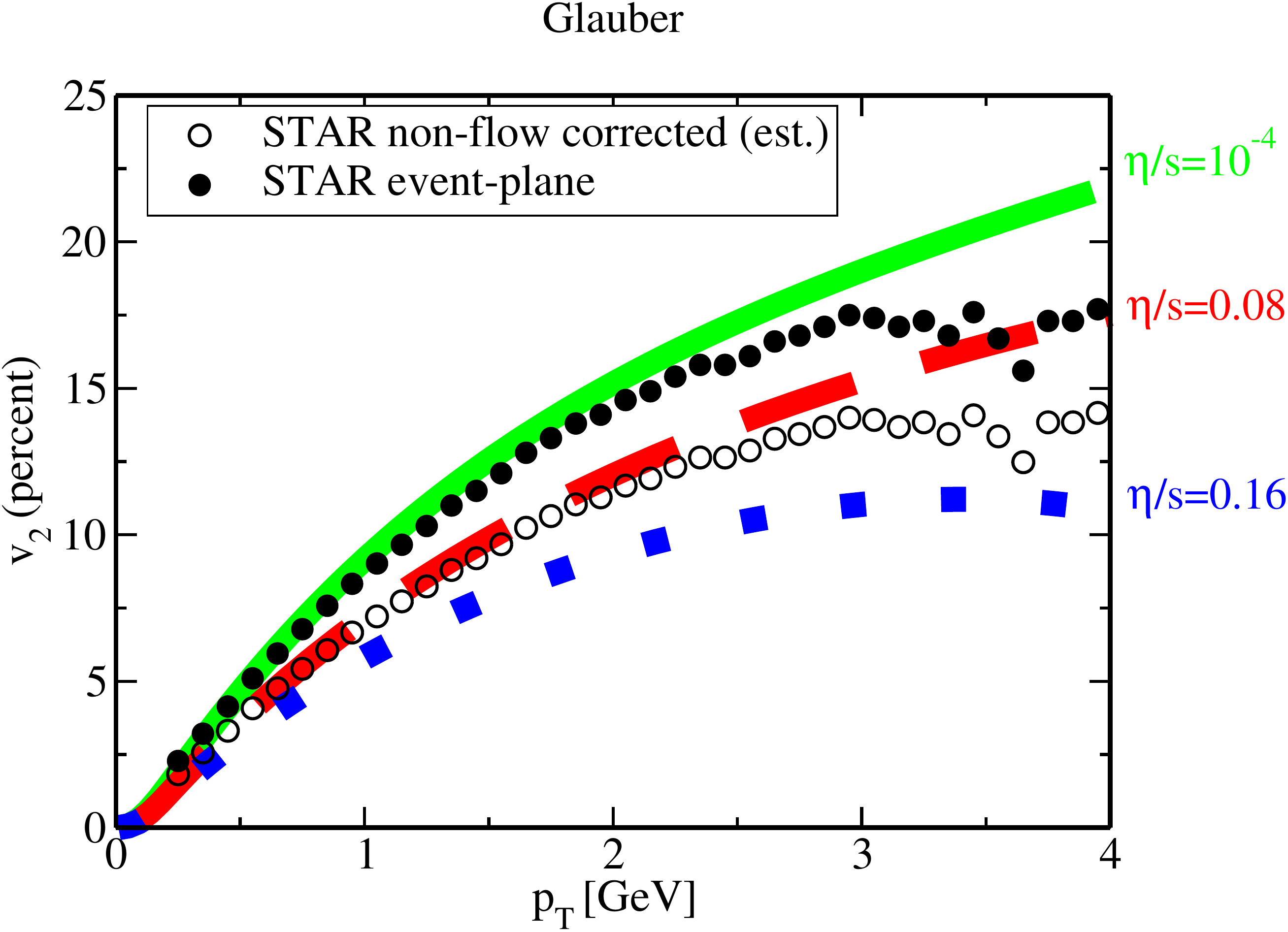}
\hfill
\includegraphics[width=0.48\textwidth]
{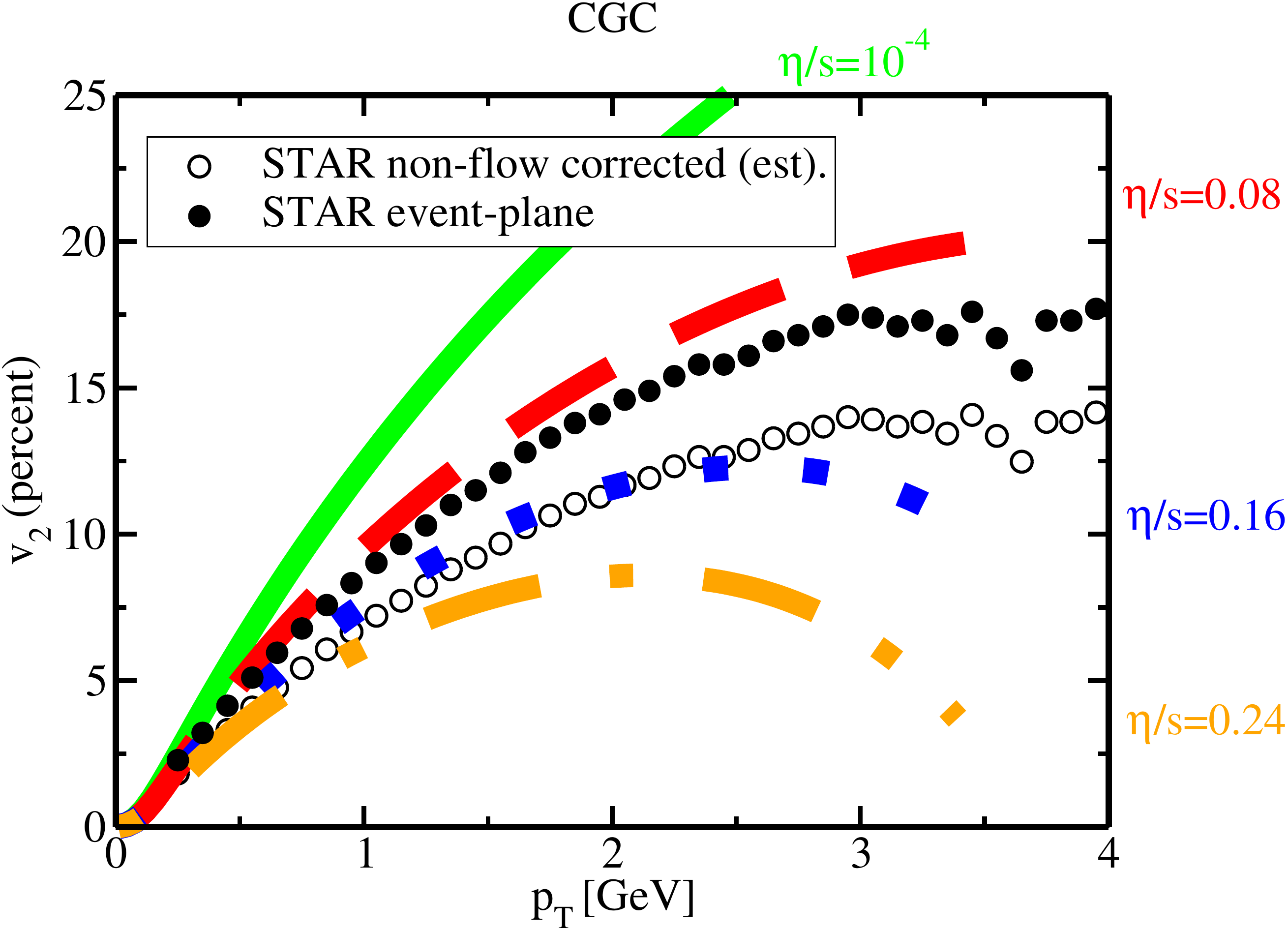}
\caption{\small The elliptic flow $v_2$ versus $p_T$ for 
charged hadrons showing the comparison between hydrodynamic calculations
with varying shear viscosity $\eta$ and data from RHIC.  The two data sets show
the reduction in $v_2$ when one uses the four-particle cumulant analysis that
strips out effects of statistical two-particle correlations.  The initial profile
for the energy density across the almond-shaped collision region is obtained
from (a) the Glauber model and (b) the color-glass condensate model. 
Figure taken from Ref.~\cite{Luzum:2008cw}.}
\label{v2viscosity}
\end{figure}
   \item \emph{Data support small dissipative coefficients such as shear viscosity}\\
   Above the crossover temperature, the largest dissipative correction is expected to arise from 
   shear viscosity $\eta$, which enters the equations of motion of second order dissipative
   hydrodynamics in the combination $\eta/s$, where $s$ is the entropy density.
   (Bulk viscosity may be sizable close to the 
   crossover temperature~\cite{Meyer:2007dy}.)
   As seen in Fig.~\ref{v2viscosity}, shear viscous corrections
   decrease collectivity, and so $v_2$ decreases with increasing   $\eta/s$.  The
   data are seen to favor small values $\eta/s < 0.2$, with $\eta/s > 0.5$ 
   strongly disfavored.  To give an example of the way this conclusion is stated in one recent review, it would be surprising if $\eta/s$ turned out to be larger than about $(3-5)/(4\pi)$ once all physical effects are properly included~\cite{Heinz:2009xj}.  The figure also shows that
   sensitivity to our lack of knowledge of the initial energy density profile precludes
   a precise determination of $\eta/s$ at present.\footnote{Recent analyses that include equations of state based upon parametrizations of recent lattice calculations~\cite{Huovinen:2009yb} and include dissipative effects from the late-time hadron gas epoch as well as estimates of the effects of bulk 
   viscosity~\cite{Song:2009je,Song:2009rh}, all of which turn out to be small, have maintained these conclusions: $\eta/s$ seems to lie within 
   the range $(1-2.5)/(4\pi)$ with the greatest remaining source of uncertainty coming from our lack of knowledge of the initial energy density profile~\cite{Song:2010mg}.}
   The smallness of $\eta/s$ is 
   remarkable, since almost all other known liquids have $\eta/s>1$ and most
   have $\eta/s\gg 1$.  The one liquid that is comparably close to ideal is an
   ultracold gas of strongly coupled fermionic atoms, whose $\eta/s$ is also
   $\ll 1$ and may be comparably small to that of the quark-gluon plasma 
   produced at RHIC~\cite{JohnThomasQM09}.  Both these fluids are much better described by ideal
   hydrodynamics than water is.   Both have $\eta/s$ comparable to the
   value $1/4\pi$ that, as 
   we shall see in Section \ref{sec:TransportProperties}, characterizes any strongly coupled gauge theory
   plasma with a gravity dual in the large number of colors limit.
 \end{enumerate}

We note that while hydrodynamic calculations reproduce elliptic flow,  
 a treatment in which the Boltzmann equation for quark and gluon
 (quasi)particles is solved, including all $2\rightarrow 2$ scattering processes
 with the cross-sections as calculated in perturbative QCD, fails dramatically.
 It results in values of $v_2$ that are much smaller than in the data.  Agreement
 with data can only be achieved if the parton scattering cross-sections are
 increased {\it ad hoc} by more than a factor of 10 \cite{Molnar:2001ux} .  With such large
 cross-sections, a Boltzmann description cannot be reliable since
 the mean free path of the particles becomes comparable to or smaller than
 the interparticle spacing.   Another way of reaching the same conclusion
 is to note that if a perturbative description of the QGP as a gas of interacting
 quasiparticles is valid, the effective QCD coupling $\alpha_s$ describing
 the interaction among these quasiparticles must be small, and for small 
 $\alpha_s$ perturbative calculations of $\eta/s$ are controlled and yield
 parametrically large values
$\propto 1/ \alpha_s^2 \ln \alpha_s$.  It is not possible to get as small a value
of $\eta/s$ as the data requires from the perturbative calculation without increasing
$\alpha_s$ to the point that the calculation is invalid.  In contrast, as 
we shall see in Section  \ref{sec:TransportProperties}, any gauge theory with
a gravity dual must have $\eta/s=1/4\pi$ in the large-$N_c$ 
and strong coupling limit and, 
furthermore, the plasma fluids described by these theories in 
this limit do not have any well-defined quasiparticles.  This calculational
framework thus seems to do a much better job of capturing the
qualitative features needed for a successful phenomenology of collective
flow in heavy ion collisions.


As we completed the preparation of this review, the first measurement of the elliptic flow $v_2$ in heavy ion collisions with $\sqrt{s}=2.76$~TeV was reported by the ALICE collaboration at the LHC~\cite{Aamodt:2010pa}.    It is striking that the $v_2(p_T)$ that they find for charged particles in three different impact parameter bins agrees within error bars at all values of $p_T$ out to beyond 4 GeV with that measured at $\sqrt{s}=0.2$~TeV by the STAR collaboration at RHIC.  Quantitative comparison of these data to hydrodynamic calculations is expected from various authors in the coming months, but at a qualitative level these data indicate that the quark-gluon plasma produced at the LHC is comparably strongly coupled, with comparably small $\eta/s$, to that produced and studied at RHIC.

\section{Jet quenching}
\label{sec:JetQuenching}

Having learned that heavy ion collisions produce a low viscosity, strongly coupled, fluid we now turn to experimental observables with which we may study properties of the fluid beyond just how it flows.   There are many such observables available.  In this subsection and the next we shall describe two classes of observables, selected because in both cases there is (the promise of) a substantive interplay between data from RHIC (and the LHC) and qualitative insights gained from the analysis of strongly coupled plasmas with dual gravity descriptions.

Jet quenching refers to a suite of experimental observables that together reveal what happens when a very energetic quark or gluon (with momentum much greater than the temperature) plows through the strongly coupled plasma.  Some measurements focus on how rapidly the energetic parton loses its energy; other measurements give access to how the strongly coupled fluid responds to the energetic parton passing through it.  These energetic partons are not external probes; they are produced within the same collision that produces the strongly coupled plasma itself.   

In a small fraction of {\it proton-proton} collisions at $\sqrt{s}=200$~GeV, partons from the incident protons scatter with a large momentum transfer, producing back-to-back partons in the final state with transverse momenta of order ten or a few tens of GeV.  These ``hard'' processes are rare, but data samples are large enough that they are nevertheless well studied.  The high transverse momentum partons in the final state manifest themselves in the detector as jets.  Individual high-$p_T$ hadrons in the final state come from such hard processes and are typically found within jets.  In addition to copious data from proton-(anti)proton collisions, there is a highly developed quantitatively controlled calculational framework built upon perturbative QCD that is used to calculate the rates for hard processes in high energy hadron-hadron collisions.  These calculations are built upon factorization theorems.  
Consider as an example the single inclusive charged hadron spectrum at 
high-$p_T$, Fig.~\ref{fig:general} (right). That is, the production cross-section 
 for a single charged hadron with a given high transverse momentum 
 $p_T$, regardless of what else is produced in the hadron-hadron 
 collision.  This quantity is calculated as a convolution of separate 
 (factorized) functions that describe different aspects of the process: 
 (i) the process-independent parton distribution function gives the 
 probability of finding partons with a given momentum fraction in the 
 incident hadrons; (ii) the process-dependent hard scattering 
 cross-section gives the probability that those partons scatter into 
 final state partons with specified momenta; and (iii) the 
 process-independent parton fragmentation functions that describe the 
 probability that a final state parton fragments into a jet that includes 
 a charged hadron with transverse momentum $p_T$.  Functions (i) and 
 (iii) are well-measured and at high transverse momentum function (ii) is 
 both systematically calculated and well-measured.
 This body of knowledge provides a firm foundation, a well-defined 
 baseline with respect to which we can measure changes if such a hard 
 scattering process occurs instead in an ultrarelativistic heavy ion 
 collision.


In hard scattering processes in which the momentum transfer $Q$ is high enough, the partonic hard scattering cross section (function (ii) above) is expected to be the same in an ultrarelativistic heavy ion collision as in a proton-proton collision.  This is so because the hard interaction occurs on a timescale and length scale $\propto 1/Q$ which is too short to resolve any aspects of the hot and dense strongly interacting medium that is created in the same collision. The parton distribution functions (function (i) above) are different in nuclei than in nucleons, but they may be measured  in proton-nucleus, deuteron-nucleus, and electron-nucleus collisions.  The key phenomenon that is unique to ultrarelativistic nucleus-nucleus collisions is that after a very energetic parton is produced, unless it is produced at the edge of the fireball heading outwards it must propagate through as much as 5-10 fm of the hot and dense medium produced in the collision.  These hard partons therefore serve as well-calibrated probes of the strongly coupled plasma whose properties we are interested in.  The presence of the medium results in the hard parton losing energy and changing the direction of its momentum. 
The change in the direction of its momentum is often referred to as ``transverse momentum broadening'', a phrase which needs explanation.  ``Transverse'' here means perpendicular to the original direction of the hard parton.  (This is different from $p_T$, the component of the (original) momentum of the parton that is perpendicular to the beam direction.)  ``Broadening'' refers to the effect on a jet when the directions of the momenta of many hard partons within it are kicked;  averaged over many partons in one jet, or perhaps in an ensemble of jets, there is no change in the mean momentum but the spread of the momenta of the individual  partons broadens.

Because the rates for hard scattering processes drop rapidly with increasing $p_T$, energy loss translates into a reduction in the number of partons produced with a given $p_T$.  (Partons with the given $p_T$ must have been produced with a higher $p_T$, and are therefore rarer than they would be in proton-proton collisions.)  
Transverse momentum broadening is expected to lead to more subtle modifications of jet properties.  Furthermore, the hard parton dumps energy into the medium, which motivates the use of observables involving correlations between soft final state hadrons and a high momentum hadron. Most generally, ``jet quenching'' refers to the whole suite of medium-induced modifications of high-$p_T$ processes in heavy ion collisions and modifications of the medium in heavy ion collisions in which a high-$p_T$ process occurs, all of which have their origin in the propagation of a highly energetic parton through the strongly coupled plasma.

%
\begin{figure}
\begin{center}
\includegraphics[width=1.0\textwidth]{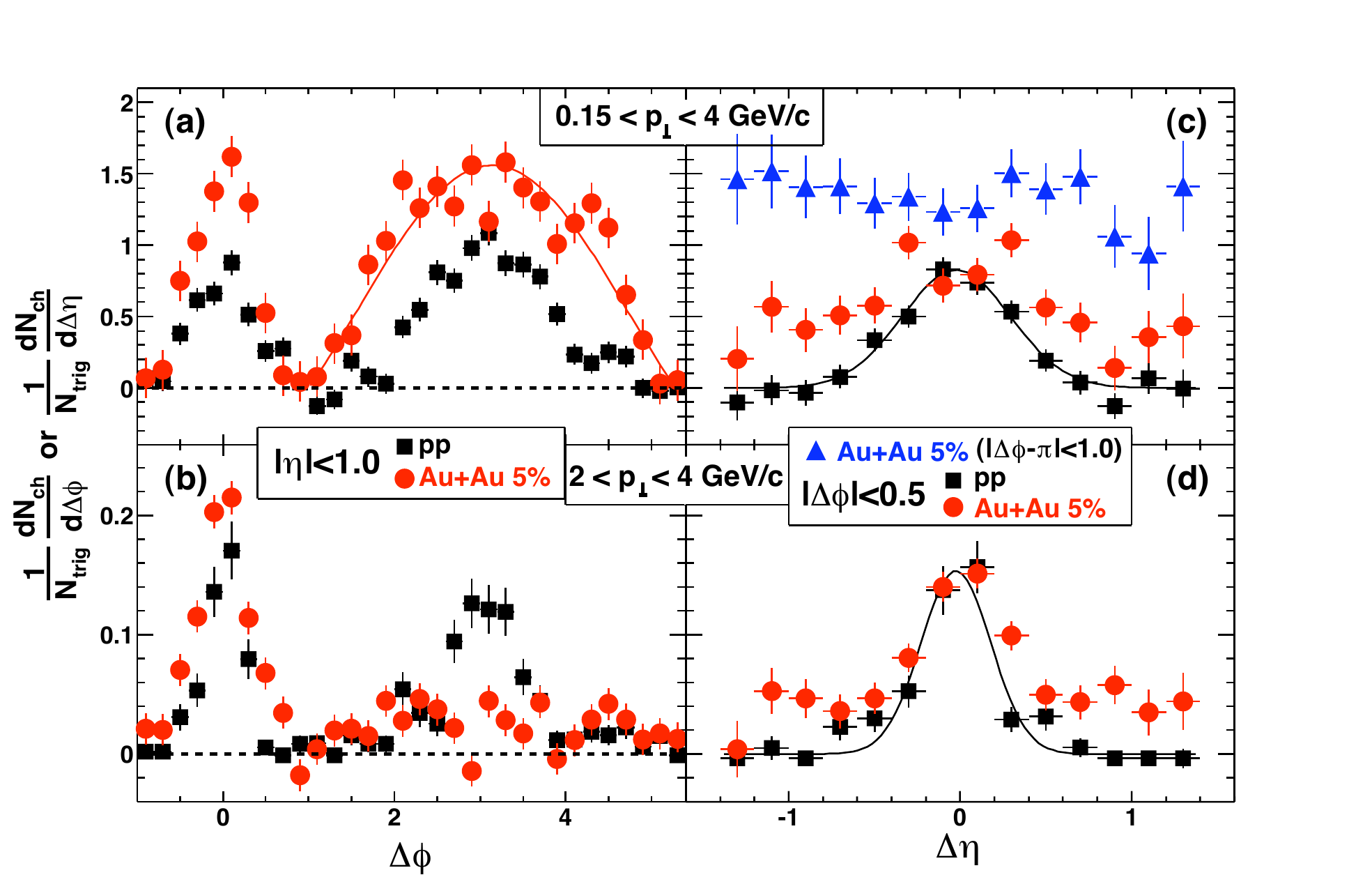}
\caption{\small 
Dihadron azimuthal and longitudinal two-particle correlations in RHIC collisions 
with $\sqrt{s}=200~GeV$ in the STAR detector 
for high trigger 
$p_T$, $4 < p_T^{\rm trig} < 6$~GeV,
and low (panels (a) and (c)) and intermediate (panels (b) and (d)) 
associated transverse momenta.  In each panel, data from nearly head-on heavy ion collisions (red) are compared to data from proton-proton collisions (black).  Panels (a) and (b) show the correlation in $\Delta \phi$, the difference in azimuthal angle of the two particles.  Panels (c) and (d) show the correlation in pseudorapidity for those particles that are close to each other in azimuth (red and black points) or close to back to back (blue points).  Figure taken from ~\cite{Adams:2005ph}.  
}
\label{fig:disappear}
\end{center}
\end{figure}
%

The most pictorial, although not the most generic, manifestation of jet quenching in heavy ion collisions at RHIC is provided by the data in Fig.~\ref{fig:disappear}.  In this analysis, one selects events in which there is a hadron in the final state with $p_T>4$~GeV, the trigger hadron.  The distribution of all other hadrons with $p_T>2$~GeV in azimuthal angle relative to the trigger hadron is shown in panel (b).  In proton-proton 
collisions, we see peaks at 0 and $\pi$ radians, corresponding to the jet in which the trigger hadron was found and the jet produced back to back with it.  In contrast, in gold-gold collisions the away-side jet is missing.  In panel (a), this analysis is repeated for {\it all} hadrons, not just those with $p_T>2$~GeV.  Here, we see an enhancement in gold-gold collisions on the away side.  The picture that these data convey is that we have triggered on events in which one jet escapes the medium while the parton going the other way loses so much energy that it produces no hadrons with $p_T>2$~GeV, dumping its energy instead into soft particles.  The jet has been quenched and the medium has been modified.  This picture is further supported by the longitudinal 
rapidity distribution shown in Fig.~\ref{fig:disappear} (c) and (d). On the near-side,
the rapidity distribution shows the shape of a jet-like structure. On the away side,
however, one finds an enhanced multiplicity distribution whose shape over many
units in pseudo-rapidity is consistent with background. 
This measurement is illustrative, but not generic in the sense that it depends on the choice 
of $p_T$ cuts used in the analysis.    A more generic consequence of jet quenching, equally 
direct but less pictorial than that in Fig.~\ref{fig:disappear}, is provided by simply measuring the diminution of the number of high-$p_T$ hadrons observed in heavy ion collisions. We shall turn to this momentarily. 

At the time of writing, the first results on jet quenching in heavy ion collisions at $\sqrt{s}=2.76$~TeV at the LHC are just becoming available~\cite{Collaboration:2010bu,Aamodt:2010jd}.  It is too soon to offer a quantitative analysis in a review such as this one, but it is clear from these early results that jet quenching remains strong even for the much higher energy jets that are produced in hard parton-parton scattering at these much higher collision energies.   And, in LHC collisions evidence of highly asymmetric dijets~\cite{Collaboration:2010bu}, as if one parton escaped relatively unscathed while its back-to-back partner was 
very significantly degraded by the presence of the medium, can now be obtained from single events rather than statistically as in Fig.~\ref{fig:disappear}.

\subsection{Single inclusive high-pt spectra and ``jet'' measurements}
\label{s:high_pT}

The RHIC heavy ion program has established that the measurement of single inclusive hadronic spectra yields a generic manifestation of jet quenching. 
Because the spectra in hadron-hadron collisions are steeply falling functions of $p_T$, if the hard partons produced in a heavy ion collision lose energy as they propagate through the strongly coupled plasma shifting the spectra leftward --- to lower energy --- is equivalent to depressing them.  This effect is quantified via the measurement of the nuclear modification factor $R^h_{AB}$, which 
characterizes how the number of hadrons $h$ produced in a collision between nucleus $A$ and nucleus $B$ differs from the number produced in an equivalent number of proton-proton collisions:
\begin{equation}
R^h_{AB}(p_T,\eta,{\rm centrality})= {  {{\rm d}N^{AB\to h}_{\rm medium}\over 
{\rm d}p_T\, {\rm d}\eta} \over
\langle N^{AB}_{\rm coll}\rangle {{\rm d}N^{pp\to h}_{\rm vacuum}\over 
{\rm d}p_T\,{\rm d}\eta}}\, .
\label{nucmod}
\end{equation}
Here,  $\langle N^{AB}_{\rm coll}\rangle$ is the average number of inelastic nucleon-nucleon 
collisions in $A$-$B$ collisions within a specified range of centralities.
This number is typically determined by inferring the transverse density distribution of
nucleons in a nucleus from the known radial density profile of nuclei, and then
calculating the average number of collisions with the help of the inelastic 
nucleon-nucleon cross section. This so-called Glauber calculation
can be checked experimentally by independent means, for instance via the measurement
of the nuclear modification factor for photons discussed below. 

 The nuclear modification factor depends in general
on the transverse momentum $p_T$ and pseudo-rapidity $\eta$ of the particle, the particle
identity $h$, the centrality of the collision and the orientation of the particle trajectory with respect
to the reaction plane (which is often averaged over).  If $R_{AB}$ deviates from 1 this reflects either medium effects or initial state effects --- the parton distributions in $A$ and $B$ need not be simply related to those in correspondingly many protons.  Measurements of $R_{dA}$ in deuteron-$A$ collisions --- which is a good proxy for $R_{pA}$ --- are used to determine whether an observed deviation of $R_{AA}$ from 1 is due to initial state effects or the effects of parton energy loss in medium.

 %
\begin{figure}
\begin{center}
\includegraphics[angle=0,width=0.49\textwidth]{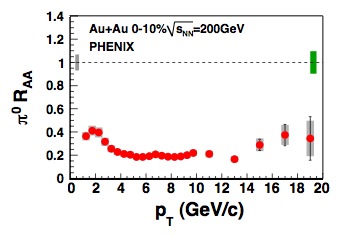}
\includegraphics[angle=0,width=0.49\textwidth]{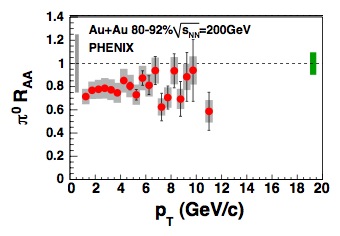}
\caption{\small 
$R_{AA}$ for neutral pions as a function of $p_T$ for central (left) and peripheral (right) 
collisions. Data taken from  \cite{:2008qa}.  
}
\label{fig:RAA}
\end{center}
\end{figure}
%

%
At mid-rapidity, RHIC data on $R_{AuAu}$ (which is often written as $R_{AA}$) 
show the following generic features:
\begin{enumerate}
	\item \emph{Characteristic strong centrality dependence of $R_{AA}$.}\\
     By varying the centrality of a heavy ion collision, one changes the typical in-medium path length
     over which hard partons produced in these collisions must  propagate through the dense matter. For the most central head-on collisions 
     (e.g., 0-10 \% centrality), the average $L$ is large, for a peripheral collision 
     (e.g., 80-92 \% centrality), the average $L$ is small.
     RHIC data (see Fig.~\ref{fig:RAA}) show that for the most peripheral centrality bin, the 
     nuclear modification factors are consistent with the absence of medium-effects, while
     $R_{AA}$ decreases monotonically with increasing centrality and reaches about 0.2 --- suppression by a factor of five! --- for the most central 
     collisions~\cite{Arsene:2003yk,Adler:2003ii,Back:2003ns,Adams:2003im}.   The left panel in Fig.~\ref{fig:RAA} is a direct manifestation of jet quenching:  80\% of the hard $\pi^0$'s that would be seen in the absence of a medium are gone.
\item \emph{Jet quenching is not observed in $R_{\rm dAu}$.}\\
In deuteron-gold collisions, $R_{\rm dAu}$ is consistent with or greater than 1 for all centralities and all transverse momenta.\footnote{While their physical origin is not unambiguously  constrained, 
the small deviations from unity observed in  $R_{\rm dAu}$ reflect the presence of initial state effects in the colliding nuclei. For instance any transverse momentum broadening of partons prior to hard processes is expected to deplete $R_{dAu}$ at low $p_T$ and enhance it at higher $p_T$. Alternatively, modifications in the nuclear parton distribution functions could also lead to such enhancement.}
  Jet quenching is not observed.  In fact, the centrality dependence is opposite to that seen in gold-gold collisions, with $R_{\rm dAu}$ reaching maximal values of around 1.5 for $p_T=3-5~$GeV/$c$ in the most central collisions~\cite{Adler:2003ii,Adams:2003im}. The high-$p_T$ hadrons are measured at or near mid-rapidity, meaning that they are well separated from the fragments of the struck gold nucleus.  And, d-Au collisions do not produce a medium in the final state.  In these collisions, therefore, the partons produced in hard scattering processes and tallied in $R_{dAu}$ do not have to propagate through any matter after they are produced.  The fact that $R_{dAu}$ is consistent with or greater than 1 in these collisions therefore demonstrates that the jet quenching measured in $R_{AuAu}$ is attributable to the propagation of the hard partons produced in heavy ion collisions through the medium that is present only in those collisions.
\item \emph{Photons are not quenched.}\\
     For single inclusive photon spectra, the nuclear modification factor shows only mild
     deviations from $R_{\rm AuAu}^\gamma \approx 1$~\cite{Isobe:2007ku}. 
  Within errors, these are consistent with perturbative predictions that  take into 
  account the nuclear modifications of parton distribution functions (mainly 
  the isospin difference between protons and nuclei)~\cite{Arleo:2006xb}. 
  Since photons, unlike partons or hadrons, do not interact strongly with the medium,
  this gives independent support that the jet quenching observed in 
	heavy ion collisions is a final state effect. And, it provides experimental evidence in support of the Glauber-type calculation of the factor $\langle N^{AA}_{\rm coll}\rangle$ in (\ref{nucmod}) discussed above.
\item \emph{Apparently $p_T$-independent and species-independent 
     suppression of $R_{AA}$ at high $p_T$.}\\
    At high transverse momentum, $p_T> 7 \, \rm{GeV}$, the suppression factor is 
    approximately independent of $p_T$ (see Fig.~\ref{fig:RAA}) in all centrality bins
    ~\cite{Adams:2003kv,Back:2004ra,Adler:2006hu}.  
    $R_{AuAu}$ is suppressed all the way out to the highest $p_T$ in Fig.~\ref{fig:RAA}, $p_T$ that are high enough that hadronization of the parton must occur only far outside the 
    medium.\footnote{We shall see later that the fact that the suppression of $R_{AA}$ is almost $p_T$-independent arises from a combination of various effects.  Here we stress only the fact that it is small all the way out to the highest $p_T$.}
    Moreover,  $R^h_{AuAu}$ is
    independent of the species of the hadron $h$~\cite{Adler:2003au}.  Either of these observations eliminates the possibility that hadrons are formed within the medium and then lose energy upon propagating through the medium, since different hadrons would have different cross sections for interaction with the medium.  This data supports the picture that the origin of the observed suppression is energy loss by a parton propagating through the medium prior to its hadronization.

     \item\emph{$R_{AA}$ for heavy-flavored and light-flavored hadrons is comparable.}\\
The current experiments at RHIC measure charm and bottom quarks via their decay products, typically via the measurement of single electrons produced in the weak decays of those 
quarks \cite{Adare:2006nq}.  These analyses cannot separate the contributions from 
charm and bottom quarks, since both have decay modes containing single electrons. The interest
in the dependence of parton energy loss on parton mass arises from the fact that models predict 
a significant dependence. It is a matter of ongoing discussion
to what extent the uncertainties in the existing data are already small enough to put interesting
contraints on models of parton energy loss.  More progress can be expected in the near
future, once detector upgrades at RHIC and measurements at the LHC allow for differentiation of bottom quarks via their displaced decay vertices.   
\end{enumerate}

In short, these observations support a picture in which highly energetic partons are produced
in high momentum transfer processes within the medium as if they were in the vacuum, but
where these partons subsequently lose a significant fraction of their initial energy due to
interactions with the medium.  Jet quenching is a  partonic final state effect that depends on the length of the medium through which the parton must propagate. It is expected to have many consequences in addition to the strong suppression of single inclusive hadron 
spectra, which tend to be dominated by the most energetic hadronic fragments of parent
partons. Rather, the entire parton fragmentation process is expected to be modified,
with consequences for observables including multi-particle jet-like correlations and calorimetric jet measurements.  (The early results from heavy ion collisions at the 
LHC~\cite{Collaboration:2010bu,Aamodt:2010jd} confirm the expectation  that at high enough jet energy calorimetric  jet observables are modified by the presence of the 
medium~\cite{Collaboration:2010bu}.)
 We shall not
review recent experimental and theoretical efforts to characterize multi-particle effects of jet
quenching here, but we shall touch upon them briefly in Section~\ref{sec:Synchrotron}.
Furthermore, the energy deposited into the medium by the energetic parton also has interesting and potentially observable effects. We shall discuss these in 
Section~\ref{sec:AdSCFTDragWaves}; the interested reader is also referred to 
the literature~\cite{Stoecker:2004qu,CasalderreySolana:2004qm}.

\subsection{Analyzing jet quenching}
\label{sec:AnalyzingJetQuenching}

For concreteness, we shall focus in this section on those aspects of the analysis of jet quenching that bear upon the calculation of the nuclear modification factor $R_{AA}$ defined in (\ref{nucmod}).    We shall describe other aspects of the analysis of jet quenching more briefly, as needed, in subsequent sections.  The single inclusive hadron spectra which define $R_{AA}$ are typically calculated upon assuming that the modification of the spectra in nucleus-nucleus collisions relative to that in proton-proton collisions arises due to parton energy loss.  This assumption is well supported by data, as we have described above.  But, from a theoretical point of view it is an assumption, not backed up by any formal factorization theorem.  Upon making this assumption, we write
\begin{equation}
   d\sigma^{AA\to h+{\rm rest}}_{(\rm med)}
      =\sum_f d\sigma^{AA\to f+X}_{\rm (vac)}
       \otimes P_f(\Delta E,L,\hat q,...)
       \otimes D^{\rm (vac)}_{f\to h}(z,\mu^2_F)\, .
      \label{jetquenansatz}
\end{equation}
Here,  
\begin{equation}
    d\sigma^{AA\to f+X}_{\rm (vac)}
     =\sum_{ijk}f_{i/A}(x_1,Q^2)
     \otimes f_{j/A}(x_2,Q^2)\otimes\hat\sigma_{ij\to f+k}
     \label{eq3}
\end{equation}
and $f_{i/A}(x,Q^2)$ are the nuclear parton distribution functions and
$\sigma_{ij\to f+k}$ are the perturbatively calculable partonic cross sections. 
The medium dependence enters via the function $P_f(\Delta E,L,\hat q,...)$, which
characterizes the probability that a parton $f$ produced with cross section 
$\sigma_{ij\to f+k}$ loses energy $\Delta E$ while propagating over a path length
$L$ in a medium.   
This probability depends of course on properties of the medium,
which are represented schematically in this formula by the symbol $\hat q$, the jet quenching parameter.  We shall see below that in the high parton energy limit, the properties of the medium enter $P_f$ only through one parameter, and in that limit $\hat q$ can be defined precisely.  
At nonasymptotic parton energies, $\hat q$ in (\ref{jetquenansatz}) is a place-holder, representing all relevant attributes of the medium.
It is often conventional to refer to the combination of $P_f$ and $D_{f\to h}^{\rm (vac)}$ together as a modified fragmentation function.   It is only in the limit of high parton energy where one can be sure that the parton emerges from the medium before fragmenting into hadrons in vacuum that these two functions can be cleanly separated as we have done in (\ref{jetquenansatz}).  This aspect of the ansatz (\ref{jetquenansatz}) is supported by the data: as we have described above, all hadrons exhibit the same suppression factor indicating that $R_{AA}$ is due to partonic energy loss, before hadronization.

The dynamics of
how parton energy is lost to the medium is specified in terms of the probability 
$P_f(\Delta E,L,\hat q,...)$.   In the high parton energy limit, the parton loses energy dominantly by inelastic processes that are the QCD analogue of bremsstrahlung: the parton radiates gluons as it interacts with the medium.  It is a familiar fact from electromagnetism that bremsstrahlung dominates the loss of energy of an electron moving through matter in the high energy limit.
The same is true in calculations of QCD parton energy loss in the high-energy 
limit, as established first in Refs.~\cite{Gyulassy:1993hr,Baier:1996sk,Zakharov:1997uu}.
The hard parton undergoes multiple inelastic interactions with the spatially extended medium, and this induces gluon bremsstrahlung. Here and throughout, by the high parton energy limit we mean the combined set of limits that can be summarized as:
\be
    E \gg \omega \gg \vert {\bf k}\vert, \vert {\bf q} \vert \equiv \vert \sum_i {\bf q}_i\vert \gg
                    T\, , \Lambda_{\rm QCD}\ ,
    \label{2.12}
\ee
where $E$ is the energy of the high energy projectile parton, where $\omega$ and ${\bf k}$ are 
the typical energy and momentum of the gluons radiated in
the elementary radiative processes $q\rightarrow q g$ or $g\rightarrow g g$, and where ${\bf q}$ is the transverse momentum (transverse to its initial direction) accumulated by the projectile parton due to many radiative interactions in the medium, and where  $T$ and $\Lambda_{\rm QCD}$ represent any energy scales that characterize the properties of the medium itself. This set of approximations underlies all analytical calculations of radiative parton energy loss to date~\cite{Baier:1996sk,Zakharov:1997uu,Wiedemann:2000za,Gyulassy:2000er,Guo:2000nz,Wang:2001ifa}. The premise of the analysis is the assumption that QCD at  scales of order $\vert {\bf k}\vert$ and $\vert{\bf q}\vert$ is weakly coupled, even if the medium (with its lower characteristic energy scales of order $T$ and $\Lambda_{\rm QCD}$) is strongly coupled.  
We shall spend most of this section on the analysis valid in this high parton energy limit, in which case all we need ask of analyses of strongly coupled gauge theories with gravity duals is insight into those properties of the strongly coupled medium that enter into the calculation of jet quenching in QCD.
However, this analysis based upon the limits (\ref{2.12}) may not be under quantitative control when applied to RHIC data, since at RHIC the partons in question have energies of at most a few tens of GeV, meaning that one can question whether all the scales separated by $\gg$ in (\ref{2.12}) are in fact well-separated.   We shall close this section with a brief look at elastic scattering as an example of an energy loss process that is relevant when (\ref{2.12}) is not satisfied.  And, in Section \ref{sec:HQDrag} we shall see that at low enough energies that physics at all scales in the problem all the way up to $E$ is strongly coupled (or in a conformal theory like ${\cal N}=4$ SYM in which physics is strongly coupled at all scales) new approaches are needed.  The analysis based upon (\ref{2.12}) that we focus on in this section will be under better control when applied to LHC data, since 
in heavy ion collisions at the LHC partons with $p_T$ all the way up to a few hundred GeV will be produced and therefore available as probes.

The analysis of radiative energy loss starts from (and extends)  the eikonal formalism, so we must begin with a few ideas and some notation from this approach
(for a self-contained introduction and references to earlier work, see e.g.
Ref.~\cite{Kovchegov:1998bi,Kovner:2001vi}).
As seen by a  high energy parton, a target that is spatially extended but of finite thickness appears Lorentz contracted, 
so in the projectile rest frame the parton propagates through the target in a short period of time and the transverse position of the projectile does not change during the propagation.
So, at ultra-relativistic energies, the main effect of the target on the projectile is a 
``rotation'' of the parton's color due to the color field of the target. 
These rotation phases are given by Wilson lines along the (straight line) trajectories 
of the propagating projectile:
\begin{equation}
  W({\bf x})={\cal P}\exp\{i\int dz^-T^aA^+_a({\bf x}, z^-)\}\, .
  \label{eq5.4}
\end{equation}
Here,  ${\bf x}$ is the {\it transverse} position of the projectile --- which does not change as the parton propagates at the speed of light along the $z^-\equiv (z-t)/\sqrt{2}$ lightlike direction.
$A^+$ is the large component of the target color field and $T^a$ is the generator of 
$SU(N)$ in the representation corresponding to the given projectile --- fundamental if the hard parton is a quark and adjoint if it is a gluon. 
The eikonal approach to scattering treats the (unphysical, in the case of colored projectiles) setting in which the projectile impinges on the target from outside, after propagating for an arbitrarily long time and building up a fully developed Weizs\"acker-Williams field proportional to $g{{\bf x}_i\over {\bf x}^2}$ (a coherent state cloud of gluons dressing the bare projectile).  The interaction of this dressed projectile with the target results in an eikonal phase (Wilson line) for the projectile itself and for each gluon in the cloud.  Gluon radiation then corresponds to the decoherence of components of the dressed projectile that pick up different phases.  Analysis of this problem yields a calculation of $N_{\rm prod}({\bf k})$, the number of radiated gluons with momentum ${\bf k}$, with the result:~\cite{Kovchegov:1998bi,Kovner:2001vi} 
\begin{eqnarray}
  && N_{\rm prod}({\bf k}) = 
  \nonumber\\
  && \quad 
  \frac{\alpha_s\, C_F}{2\pi}\, 
  \int d{\bf x}\, d{\bf y}\, e^{i{\bf k}\cdot({\bf x}-{\bf y})}
  \frac{ {\bf x}\cdot {\bf y}}{ {\bf x}^2\, {\bf y}^2}
  \Bigg[ 1 - \frac{1}{N^2 - 1}\, 
  \langle
  {\rm Tr}\left[ W^{A\, \dagger}({\bf x})\, W^A({\bf 0}) \right]
  \rangle
  \nonumber\\
&& \qquad \qquad \qquad \qquad \qquad \qquad  - \frac{1}{N^2 - 1}\, 
  \langle
  {\rm Tr}\left[ W^{A\, \dagger}({\bf y})\, W^A({\bf 0}) \right]
  \rangle
  \nonumber \\
&& \qquad \qquad \qquad \qquad \qquad \qquad
  +\frac{1}{N^2 - 1}\, 
  \langle
  {\rm Tr}\left[ W^{A\, \dagger}({\bf y})\, W^A({\bf x}) \right]
  \rangle\Bigg]\, ,
  \label{eq5.10}
\end{eqnarray}
where the $C_F$ prefactor is for the case where the projectile is a quark in the fundamental representation, where the projectile is located at transverse position ${\bf 0}$, and where the $\langle \ldots \rangle$ denotes averaging over the gluon fields of the target.  If the target is in thermal equilibrium, these are thermal averages.  

Although the simple result (\ref{eq5.10}) is not applicable to the physically relevant case as we shall describe in detail below, we can nevertheless glean insights from it that will prove relevant.
We note that the entire medium-dependence of the gluon number spectrum
(\ref{eq5.10}) is determined by target expectation values of the form  $\langle
  {\rm Tr}\left[ W^{A\, \dagger}({\bf x})\, W^A({\bf y}) \right]
  \rangle$ of two eikonal Wilson lines. The jet
  quenching parameter $\hat{q}$ that will appear below defines the fall-off properties of this correlation function
  in the transverse direction $L \equiv \vert {\bf x}-{\bf y}\vert$:
\begin{equation}
	 \langle
  {\rm Tr}\left[ W^{A}({\cal C}) \right]
  \rangle \approx \exp \left[ - \frac{1}{4\sqrt{2}} \hat{q}\, L^-\, L^2 \right]\, 
  \label{quenching}
\end{equation}
in the limit of small $L$, with $L^-$ (the extent of the target along the $z^-$ direction) assumed large but finite~\cite{Liu:2006ug,Liu:2006he}. 
Here, the contour ${\cal C}$ traverses a distance $L^-$ 
along the light cone at transverse position ${\bf x}$, and it returns at transverse position ${\bf y}$. These 
two long straight lightlike lines are connected by short transverse segments 
located at $z^- = \pm L^- / 2$, far outside the target.   We see from the form of (\ref{eq5.10}) that $|{\bf k}|$ and $L$ are conjugate: the radiation of gluons with momentum $|{\bf k}|$ is determined by Wilson loops with transverse extent $L\sim 1/|{\bf k}|$. This means that in the limit (\ref{2.12}), the only property of the medium that enters (\ref{eq5.10}) is $\hat q$.  Furthermore, inserting (\ref{quenching}) into (\ref{eq5.10}) yields the result that the gluons that are produced have a typical ${\bf k}^2$ that is of order $\hat q  L^-$.  This suggests that $\hat q$ can be interpreted as the transverse momentum squared picked up by the hard parton per distance $L^-$ that it travels, an interpretation that can be validated more rigorously via other 
calculations~\cite{CasalderreySolana:2007zz,D'Eramo:2010xk}.

The reason that the eikonal formalism cannot be applied verbatim to the problem of parton energy loss in heavy ion collisions is that the high energy partons we wish to study do not impinge on the target from some distant production site.  They are produced within the same collision that produces the medium whose properties they subsequently probe.  As a consequence,
they are produced with significant virtuality.  
This means that even if there were no medium present, they would radiate copiously.
They would fragment in what is known in QCD as a parton shower.  
The analysis of medium-induced parton energy loss
then requires understanding the interference between radiation 
in vacuum and the medium-induced bremsstrahlung radiation.  
It turns out that the resulting interference
resolves longitudinal distances in the
target~\cite{Baier:1996sk,Zakharov:1997uu,Wiedemann:2000za,Gyulassy:2000er}, 
meaning that its description goes beyond the eikonal approximation. 
The analysis of parton energy loss
in the high energy limit (\ref{2.12}) must include terms that are subleading in $1/E$, and therefore not present in the eikonal approximation, that describe the leading interference effects.  
To keep these ${\cal O}(1/E)$ effects, one must replace eikonal Wilson lines by retarded Green's functions that describe the propagation of a particle with energy $E$ from position $z^-_1,{\bf x}_1$ to position $z^-_2,{\bf x}_2$ without assuming ${\bf x}_1={\bf x}_2$~\cite{Zakharov:1997uu,Kopeliovich:1998nw,Wiedemann:2000ez}.   (In the $E\rightarrow \infty$ limit, ${\bf x}_1={\bf x}_2$ and the eikonal Wilson line is recovered.)  
It nevertheless turns out that even after Wilson lines are replaced by Green's functions the only attribute of the medium that arises in the  analysis, in the limit (\ref{2.12}), is the jet quenching parameter $\hat q$ defined in (\ref{quenching}) that already arose in the eikonal approximation.

We shall not present the derivation, but it is worth giving the complete (albeit somewhat formal) result for the distribution of gluons with energy $\omega$ and 
transverse momentum ${\bf k}$ that a high energy parton produced within a medium 
radiates:
\begin{eqnarray}
  \omega\frac{dI}{d\omega\, d{\bf k}}
  &=& \frac{\alpha_s\,  C_R}{ (2\pi)^2\, \omega^2}\,
    2{\rm Re} \int_{\xi_0}^{\infty}\hspace{-0.3cm} dy_l
  \int_{y_l}^{\infty} \hspace{-0.3cm} d\bar{y}_l\,
   \int d{\bf u}\,
  e^{-i{\bf k}\cdot{\bf u}}   \,
  \exp\left[ { -\frac{1}{4} \int_{\bar{y}_l}^{\infty} d\xi\, \hat{q}(\xi)\,
    {\bf u}^2 }\right] \,
  \nonumber \\
  && \times \frac{\partial }{ \partial {\bf x}}\cdot
  \frac{\partial }{ \partial {\bf u}}\,
  \int_{{\bf x}\equiv {\bf r}(y_l)\equiv {\bf 0} }^{{\bf u}\equiv{\bf r}(\bar{y}_l)}
  \hspace{-0.5cm} {\cal D}{\bf r}
   \exp\left[  \int_{y_l}^{\bar{y}_l} d\xi
        \left( \frac{i\, \omega}{2} \dot{\bf r}^2
          - \frac{1}{4} \hat{q}(\xi) {\bf r}^2  \right)
                      \right]\, .
    \label{2.14}
\end{eqnarray}
We now walk through the notation in this expression.
The Casimir operator $C_R$ is in the representation of the
projectile parton. The integration variables $\xi$, $y_l$ and $\bar{y}_l$ are all positions along the $z^-$ lightcone direction.  $\xi_0$ is the $z^-$ at which the projectile parton was created in a hard scattering process.  
Since we are not taking this to $-\infty$, the projectile is not assumed on shell. The projectile parton was created at the transverse position ${\bf x}={\bf 0}$. 
The integration variable ${\bf u}$ is also a transverse position variable, conjugate to ${\bf k}$.  The path integral is over all possible paths ${\bf r}(\xi)$ going from ${\bf r}(y_l)={\bf 0}$ 
to ${\bf r}(\bar{y}_l)={\bf u}$.  The derivation of (\ref{2.14}) proceeds by writing $dI/d\omega\,d{\bf k}$ in terms of a pair of retarded Green's functions in their path-integral representations, one of which describes the radiated gluon in the amplitude, radiated at $y_l$, and the other of which describes the radiated gluon in the conjugate amplitude, radiated at $\bar{y}_l$.  The expression (\ref{2.14}) then follows after a lengthy but purely technical calculation~\cite{Wiedemann:2000za}.  The properties of the medium enter (\ref{2.14}) only through the jet quenching parameter $\hat q(\xi)$.
The compact expression (\ref{2.14}) has been derived in the so-called path-integral
approach~\cite{Zakharov:1997uu}. For other, related, formulations of
QCD parton energy loss, we refer the reader to Refs.~\cite{Baier:2000mf,Guo:2000nz,Wang:2001ifa,Arnold:2002ja,Baier:2002tc,Kovner:2003zj,Gyulassy:2003mc,Jacobs:2004qv,Wicks:2005gt,Renk:2006sx,Zhang:2007ja,Majumder:2007ae,Qin:2007rn,CasalderreySolana:2007zz,Zapp:2008gi,Bass:2008rv,d'Enterria:2009am,Armesto:2009zi,Accardi:2009qv,Wiedemann:2009sh,Vitev:2009rd,Horowitz:2009eb,Chen:2010te,Majumder:2010qh,CaronHuot:2010zt}.  

The notation $\hat q(\xi)$ allows for the possibility that the nature of the medium and 
thus its $\hat q$ changes with time as the hard parton propagates through it.  
If we approximate the medium as unchanging, $\hat q(\xi)$ is just a constant.
In the strongly expanding medium of a heavy ion collision, however, this is not a 
good approximation. It turns out that the path-integral in (\ref{2.14}) can be solved
analytically in the saddle point approximation for quenching parameters of the
form $\hat{q}(\xi) = \hat{q}_0 \left(\xi_0/\xi \right)^\alpha$~\cite{Baier:1998yf}. 
Here the free exponent $\alpha$ may be chosen in model calculations to take 
values characterizing a one-dimensional
longitudinal expansion with so-called Bjorken scaling ($\alpha = 1$), or scenarios
which also account for the transverse expansion $1 < \alpha < 3$. Remarkably, one finds
that irrespective of the value of $\alpha$,
for fixed in-medium path length 
$L^-/\sqrt{2}$ 
the transverse momentum
integrated gluon energy distribution (\ref{2.14}) has 
the same $\omega$-dependence  if $\hat q(\xi)$ is simply replaced by a constant given by
the linear line-averaged transport coefficient~\cite{Salgado:2002cd}
\begin{equation}
	\langle \hat{q} \rangle \equiv \frac{1}{2\, {L^-}^2}\, 
		\int_{\xi_0}^{\xi_0 + L^-}\, d\xi\, \left(\xi - \xi_0\right)\, \hat{q}(\xi)\, .
	\label{qhataverage}
\end{equation}
In practice, this means that comparisons of different parton energy loss calculations to data
can be performed as if the medium were static. The line-averaged transport coefficient
$\langle \hat{q} \rangle$ determined in this way can then be related via (\ref{qhataverage})
to the transport coefficient at a given time, once a model for the expansion of the
medium is specified. Hence, we can continue 
our discussion for the case $\hat{q}(\xi) = \hat{q}$ without loss of generality. 

The result (\ref{2.14}) is both formal and complicated.  However, its central qualitative consequences can be characterized almost by dimensional analysis.  All dimensionful quantities can be scaled out of (\ref{2.14}) if $\omega$ is measured in units of the 
so-called characteristic gluon energy
\begin{equation}
	\omega_c \equiv \hat q (L^-)^2 
	\label{characteristico}
\end{equation}
and the transverse momentum ${\bf k}^2$ in units of $\hat q L^-$~\cite{Salgado:2003gb}. 
In a numerical analysis of (\ref{2.14}), one
finds that the transverse momentum distribution
of radiated gluons scales indeed with $\hat q L^-$, as expected for the  transverse momentum 
due to the Brownian motion in momentum space that is induced by multiple small angle scatterings. 
If one integrates the gluon distribution (\ref{2.14}) over transverse momentum and takes the upper limit of the ${\bf k}$-integration to infinity, one recovers~\cite{Salgado:2003gb}
an analytical expression first derived by Baier, Dokshitzer, Mueller, Peign\'e and 
Schiff~\cite{Baier:1996sk}:
\begin{equation}
   \omega \frac{dI_{\rm BDMPS}}{d\omega} =
   \frac{2\alpha_s C_R}{\pi}\, 
   \ln \Bigg \vert
   {\cos\left[\,(1+i)\sqrt{\frac{\omega_c}{2\omega}}\,\right]}
   \Bigg \vert \, ,
   \label{2.9}
\end{equation}
which yields the limiting cases
\begin{eqnarray}
   \omega \frac{dI_{\rm BDMPS}}{d\omega} \simeq 
           \frac{2\alpha_s C_R}{\pi} 
          \left\{ \begin{array} 
                  {r@{\qquad  \hbox{for}\quad}l}                 
                  \sqrt{\frac{\omega_c}{2\, \omega}}
                  & \omega \ll \omega_c\, , \\ 
                  \frac{1}{12} 
                  \left(\frac{\omega_c}{\omega}\right)^2
                  & \omega \gg \omega_c \, ,
                  \end{array} \right.
  \label{2.10}
\end{eqnarray}
for small and large gluon energies.
In the soft gluon limit, the BDMPS spectrum (\ref{2.9}) displays the characteristic
$1/\sqrt{\omega}$ dependence, which persists up to a gluon energy of the order of
the characteristic gluon energy (\ref{characteristico}). Hence, $\omega_c$ can be 
viewed as an effective energy cut-off, above which the contribution of 
medium-induced gluon radiation is negligible. These analytical limits provide
a rather accurate characterization of the full numerical result. In particular, 
one expects from the above expressions that 
the average parton energy loss $\langle \Delta E \rangle$, obtained by 
integrating (\ref{2.14}) over ${\bf k}$ and ${\omega}$, is proportional to
$\propto \int_0^{\omega_c} d\omega\, \sqrt{\omega_c}/\omega \propto \omega_c$.
One finds indeed
\begin{equation}
 \langle \Delta E \rangle_{\rm BDMPS} \equiv
 \int_0^\infty d\omega\, 
   \omega \frac{dI_{\rm BDMPS}}{d\omega}
 = \frac{\alpha_s C_R}{2}\, \omega_c\, . 
 \label{2.11}
\end{equation}
This is the well-known ${(L^-)}^2$-dependence of the average radiative parton energy 
loss~\cite{Baier:1996kr,Baier:1996sk,Zakharov:1997uu}.
In summary, the main qualitative properties of the medium-induced gluon energy
distribution (\ref{2.14}) are the scaling of ${\bf k}^2$ with $\hat{q}\, L^-$ dictated by
Brownian motion in transverse momentum space, the  $1/\sqrt{\omega}$ dependence of the 
${\bf k}$-integrated distribution characteristic of the non-abelian 
Landau-Pomerantschuk-Migdal (LPM) effect, and the resulting ${(L^-)}^2$-dependence
of the average parton energy loss. 

Medium-induced gluon radiation modifies the correspondence
between the initial parton momentum and the final hadron momenta. 
We now sketch how the resulting $P(\Delta E)$ in (\ref{jetquenansatz}) can be estimated.
If gluons are emitted independently, $P(\Delta E)$
is the normalized sum of the emission probabilities
for an arbitrary number of $n$ gluons which carry away the total
energy $\Delta E$:\cite{Baier:2001yt}
\begin{eqnarray}
  P(\Delta E) = 
      \exp\left[ - \int_0^\infty \hspace{-0.2cm}
      d\omega \frac{dI}{d\omega}\right]
  \sum_{n=0}^\infty \frac{1}{n!}
  \left[ \prod_{i=1}^n \int d\omega_i \frac{dI(\omega_i)}{d\omega}
    \right]
    \delta\left(\Delta E - \sum_{i=1}^n \omega_i\right)\, .
   \label{3.1}
\end{eqnarray}
Here, the factor $\exp\left[ - \int_0^\infty \hspace{-0.2cm} d\omega \frac{dI}{d\omega}\right]$
denotes the probability that no energy loss occurs. This factor ensures that $P(\Delta E)$
is properly normalized, $\int d\Delta E\, P(\Delta E) = 1$.
The mean energy loss is then
\begin{equation}
 \langle \Delta E\rangle  = \int d\Delta E\,  (\Delta E)\,  P(\Delta E) = \int d\omega \omega
 \frac{dI}{d\omega}\ ,
\end{equation}
as above, but it turns out that $P(\Delta E)$ is a very broad distribution, not peaked around its mean.  In particular, as seen from Fig.~\ref{fig:general}, single inclusive hadron spectra
are distributions which fall steeply with $p_T$. The modifications which parton energy loss
induce on such distributions cannot be characterized by an average energy loss.
Rather, what matters for a steeply falling distribution is not how much energy a parton
loses on average, but rather which fraction of all the partons gets away with much less
than the average energy loss~\cite{Baier:2001yt}. This so-called trigger bias effect
is quantitatively very important, and can be accounted for by the probability distribution
(\ref{3.1}).

\begin{figure}
\begin{center}
\includegraphics[width=1.0\textwidth]{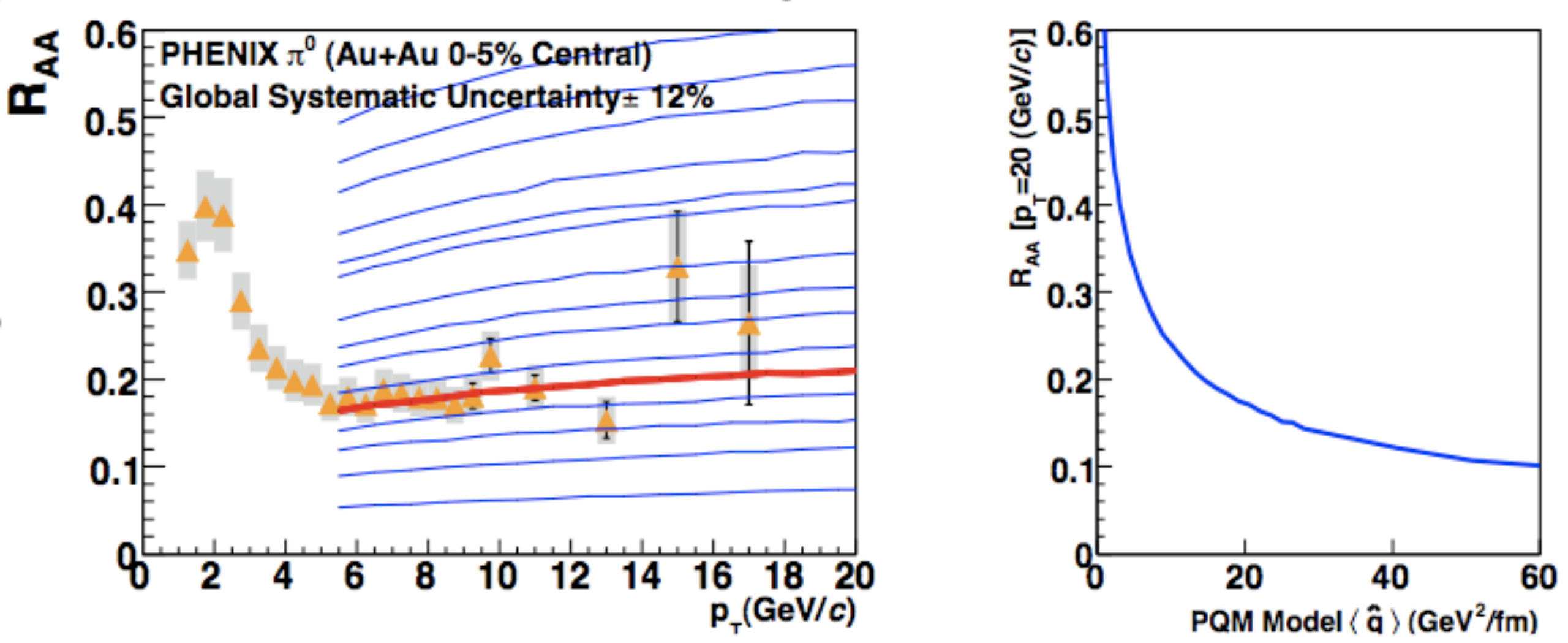}
\caption{\small 
Data for the nuclear modification factor of $\pi^0$ in central $0-5\%$ Au+Au collisions,
compared to a model calculation of jet quenching with $\langle \hat{q} \rangle$
values of 0.3, 0.9, 1.2, 1.5, 2.1, 2.9, 4.4, 5.9, 7.4, 10.3, 13.2, 17.7. 25.0, 40.5 and 101.4
${\rm GeV}^2/{\rm fm}$. Data taken from  \cite{Adare:2008cg}.  
}
\label{fig:RAAphenix}
\end{center}
\end{figure}

Making contact with phenomenology requires the implementation of $P(\Delta E)$ in a model in which  hard scattering events are distributed with suitable probability at locations in the transverse plane, with the hard partons produced going in random back-to-back directions and in which a hydrodynamic calculation is used to model how the medium in which the hard partons find themselves evolves subsequently, as the partons propagate through it.  
One such model is the PQM model~\cite{Dainese:2004te}, which 
calculates the hadronic spectrum (\ref{jetquenansatz}) by interfacing the
probability distributions $P(\Delta E)$ of Ref.~\cite{Salgado:2003gb} with a model of the
geometrical distribution of matter in the collision region. 
Although the PQM model is built upon the gluon energy distribution (\ref{2.10}), which gives the leading contribution to parton energy 
loss in the high energy limit (\ref{2.12}), it is nevertheless a model in several senses beyond its treatment of the distribution of energy density in space and time. 
For example, the
factorized ansatz of the inclusive cross section (\ref{jetquenansatz}) is not justified
from first principles but is an experimentally testable assumption. Also, the form (\ref{3.1})
of the probability distribution $P(\Delta E)$ assumes that in the case of multiple gluon 
emissions, the consecutive energy degradation of the leading parton is negligible. 
For this to be a good approximation, the total energy loss must be much smaller than
the projectile energy, $\Delta E \ll E$.   All these aspects of the model are expected to become reliable in the high energy limit (\ref{2.12}).  
The results of a comparison of this model to data, done by the PHENIX collaboration, 
are shown in Fig.~\ref{fig:RAAphenix}. 
We see that for a suitable choice of the line-averaged transport coefficient
$\langle \hat q \rangle$, defined in (\ref{qhataverage}), the calculation does a reasonable job of describing the data.
In this analysis, the approximate $p_T$-independence of $R_{AuAu}$ seen in the data arises from an interplay between the rapidly dropping parton production cross-section at $p_T$'s exceeding $\sim \sqrt{s}/10$, the fact that there are always some partons produced near the edge of the fireball that emerge relatively unscathed and the fact that $P(\Delta E)$ is significant even for $\Delta E=0$ even though the mean $\Delta E$ is large. 
The PHENIX authors quote for this PQM model study a jet quenching parameter
which is constrained by the experimental data as $13.2^{+ 2.1}_{-3.2}$ and
$13.2^{+ 6.3}_{-5.2}$ ${\rm GeV}^2/{\rm fm}$ at the one and two standard deviation
levels, respectively. 

More recently, the quantitative comparison between data from RHIC and the jet quenching calculations that we have sketched in this section has been revisited in Ref.~\cite{Armesto:2009zi}.  These authors use data on $R_{AA}$ as in Fig.~\ref{fig:RAAphenix} but in addition they use data on dihadron correlations as well as data on heavy quark suppression.  And, they include a hydrodynamical modelling of the expanding and cooling plasma.   They parametrize the jet quenching parameter as 
\begin{equation}
\hat q = 2 K \varepsilon^{3/4}
\label{qhatKrelation}
\end{equation}
where $\varepsilon$ is the energy density of the medium through which the energetic partons are propagating and where $K$ is a parameter to be obtained via comparing jet quenching calculations to data.  (In a weakly coupled quark-gluon plasma, $K \approx 1$~\cite{Baier:2002tc}.) $\varepsilon$ varies with position and decreases with time according to a hydrodynamical calculation.    These authors find that they are able to obtain a more stable fit to the value of $K$ than to a time-averaged $\hat q$.  Fitting to data from both PHENIX and STAR from RHIC collisions at $\sqrt{s}=200$~GeV, they obtain
\begin{equation}
K=4.1\pm 0.6\ .
\label{qhatKresult}
\end{equation}
We will compare this result to calculations done for strongly coupled plasmas in gauge theories with dual gravitational descriptions in Section~\ref{sec:AdSCFTJetQuenching}.

There is currently a vigorous debate about whether hard partons produced (and then 
quenched) in RHIC collisions have a high enough energy $E$ for the approximation 
(\ref{2.12}) to be under control, and for the above-mentioned model assumptions to be valid. 
The question of 
whether the comparison of RHIC data with different model
implementations of parton energy loss yield numerically consistent results is also still 
debated, see e.g.
\cite{Adare:2008cg}.  Here, we solely remark that while radiative energy loss is known
to dominate in the high energy limit, other mechanisms and in particular elastic
energy loss may contribute significantly at lower energies, and this may not be negligible
at RHIC~\cite{Wicks:2005gt,Peshier:2006mp,Kolevatov:2008bg}. In radiative energy loss, in the high energy limit, the longitudinal momentum of the incident parton is shared between the longitudinal momentum of the radiated gluon and that of the outgoing high energy parton.  Longitudinal momentum transfer to the medium is negligible in comparison to that carried by the radiated gluon.  In elastic scattering, however, the only energy lost by the projectile parton is that due to transferring longitudinal momentum to the medium.    
 What is clear from the outset is that if elastic mechanisms contribute, then parton energy loss must depend on more attributes of the medium than just $\hat q$, since $\hat q$ knows nothing about how ``bits'' of the medium recoil longitudinally when struck.
Note that this discussion goes through unchanged even in the context of a medium that is a strongly coupled fluid with no identifiable scattering centers.  If the medium is weakly coupled, then the conclusion can be stated more precisely: $\hat q$ does not differentiate between a medium made of heavy scatterers that hardly recoil and one made of light scatterers that recoil more easily.

\section{Quarkonia in hot matter}
\label{quarkonium}
 
One way of thinking about the operational meaning of the statement that quark-gluon plasma is deconfined is to ask what prevents the formation of a meson within quark-gluon plasma.
The answer is that the attractive force between a quark and an antiquark which are separated by a distance of order the size of a meson is screened by the presence of the quark-gluon plasma between them.   This poses a quantitative question: how close together do the quark  and antiquark have to be in order for their attraction not to be screened?  How close together do they have to be in order for them to feel the same attraction that they would feel if they were in vacuum?  It was first suggested by Matsui and Satz~\cite{Matsui:1986dk} in 1986 that measurements of  how many quarkonia --- mesons made of a heavy quark-antiquark pair --- are produced in heavy ion collisions could be used as a tool with which to answer this question, because they are significantly smaller than typical mesons or baryons.   

The generic term quarkonium refers to the charm-anticharm or charmonium, mesons 
($J/\Psi$, $\Psi'$, $\chi_c$, ...) and the bottom-antibottom, or bottomonium, mesons 
($\Upsilon$, $\Upsilon'$, ...).   The first quarkonium state that was discovered was the $1s$ state of 
the $c\, \bar{c}$ bound system, the $J/\Psi$.  It is roughly half the size of a typical meson like the $\rho$.  The bottomonium $1s$ state, the $\Upsilon$ is smaller again by roughly another factor of two.  It is therefore expected that if one can study quark-gluon plasma in a series of experiments with steadily increasing temperature, $J/\psi$ mesons survive as bound states in the quark-gluon plasma up to some dissociation temperature that is higher than the crossover temperature (at which generic mesons and baryons made of light quarks fall apart.)   More realistically, what Matsui and Satz suggested is that if high energy heavy ion collisions create deconfined quark-gluon plasma that is hot enough, 
then color screening would prevent 
charm and anticharm quarks from binding to each other in the deconfined interior of the 
droplet of matter produced in the collision, and as a result the number of $J/\Psi$ mesons produced in the collisions would be suppressed.
However, $\Upsilon$s should survive as bound states to even higher temperatures, until the quark-antiquark attraction is screened even on the short length scale corresponding to their size.

To study this effect, Matsui and Satz suggested comparing the temperature dependence of the screeing length for the quark-antiquark force, which can be obtained from lattice QCD calculations, with the $J/\Psi$ meson radius calculated in charmonium models.  They
then discussed the feasibility to detect this effect clearly in the mass spectrum 
of $e^+\, e^-$ dilepton pairs. 
Between 1986, when Matsui and Satz launched this line of investigation, suggesting it as a quantitative means of characterizing the formation and properties of deconfined matter, and today we know of no other measurement that has been advocated as a more direct experimental signature for the deconfinement transition. And, there is hardly any other measurement whose phenomenological analysis has turned out to be more involved.  In this subsection, we shall describe both the appeal of studying quarkonia in the hot matter produced in heavy ion collisions and the practical difficulties.
%
%
The theoretical basis for the argument 
of Matsui and Satz has evolved considerably within the last two 
decades~\cite{Satz:2005hx}. Moreover, the debate over how to interpret these measurements  is by now informed by 
data on $J/\Psi$-suppression in nucleus-nucleus collisions at the CERN 
SPS~\cite{Alessandro:2004ap,Arnaldi:2007zz},
 at RHIC~\cite{Adare:2006ns} and at the LHC~\cite{Collaboration:2010px}.  There is also a good possibility that qualitatively novel
information will become accessible in future high statistics runs at RHIC and in heavy ion collisions at the
LHC. 


%
\begin{figure}
\begin{center}
\includegraphics[width=0.75\textwidth]{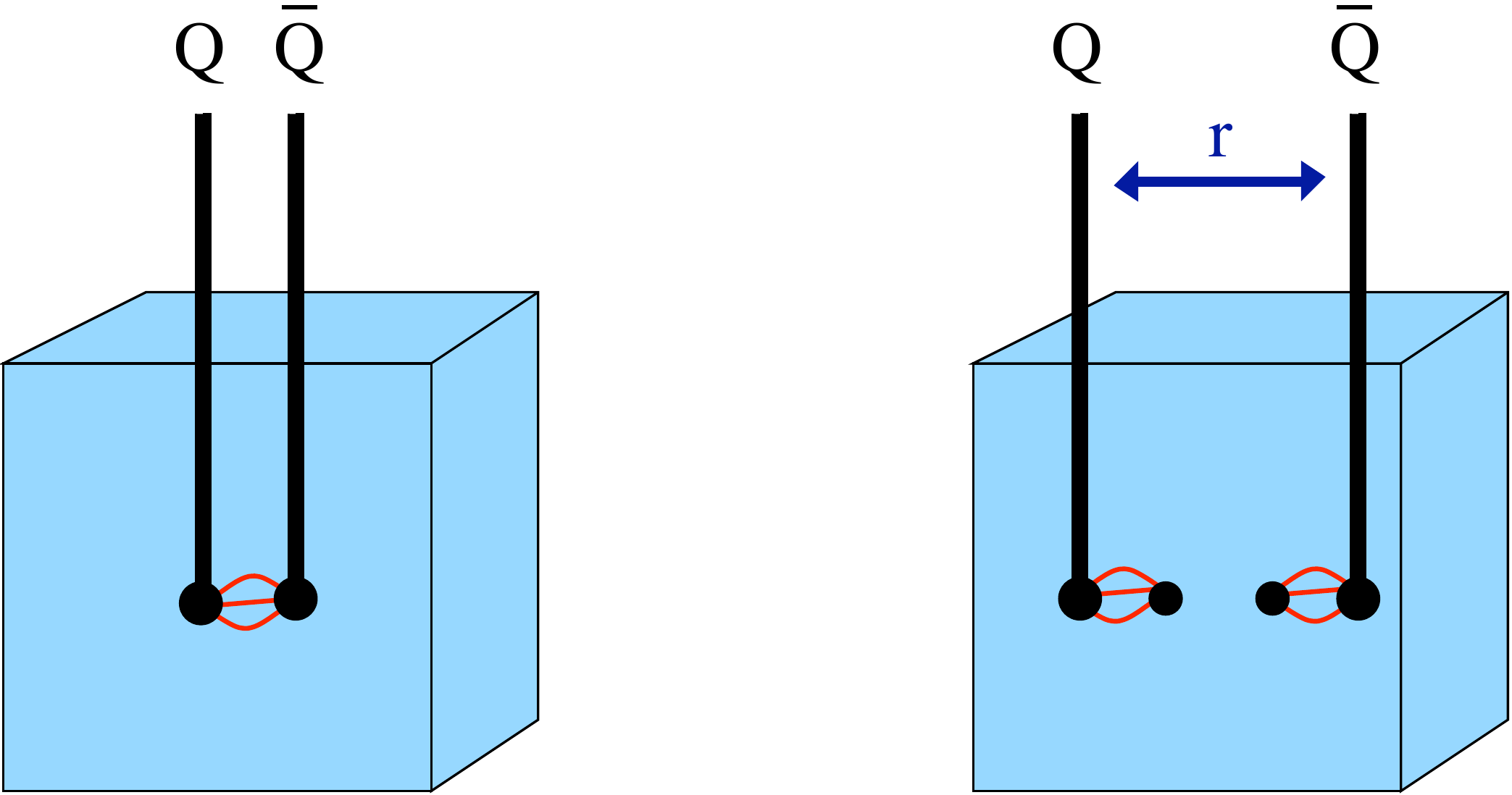}
\caption{\small 
Schematic picture of the dissociation of a $Q\, \bar{Q}$-pair in hot QCD
matter due to color screening. Figure taken from Ref.~\cite{Satz:2005hx}.
The straight black lines attached to the heavy $Q$ and $\bar{Q}$ indicate,
that these quarks are external probes, in contrast to the dynamical quarks
within the quark-gluon plasma.
}
\label{fig:QQbarDissoc}
\end{center}
\end{figure}
%

A sketch of the basic idea of Matsui and Satz is shown in Fig.~\ref{fig:QQbarDissoc}.
In very general terms, one expects that  the attractive interaction between the heavy quark 
and anti-quark in a putative bound state is sensitive to the medium in which the 
heavy particles are embedded, and that this attraction weakens with increasing temperature.
If the distance between the heavy quark and anti-quark is much smaller than $1/T$, there will not be much quark-gluon plasma between them.  Equivalently, typical momentum scales in the medium are of order the temperature $T$, and so the medium cannot resolve the separation between the quark-antiquark pair if they are much closer together than $1/T$.
However, if the distance is  larger, then the bound state
is resolved, and the color charges of the heavy quarks are screened by the 
medium, see Fig.~\ref{fig:QQbarDissoc}. 
In a heavy ion collision in which quark-gluon plasma with a temperature $T$ is created,
only those quarkonia with radii that are smaller than some length scale of order $1/T$ can form.
These basic arguments support the 
idea that quarkonium production rates are  an indicator of 
whether quark-gluon plasma is produced, and at what temperature.

In section~\ref{secmeson}, we review lattice calculations of the 
heavy quark static free energy $F_{Q\bar{Q}} (r)$. This static potential is typically 
defined via how the correlation function of a pair of
Polyakov loops, namely test quarks at fixed spatial positions whose worldlines wrap around the periodic Euclidean time direction, falls off
as the separation between the test quarks is increased.  This static potential is
renormalized such that it matches the zero temperature result at small distances.
Calculations of $F_{Q\bar{Q}} (r)$ were the earliest lattice results which substantiated 
the core idea that a quarkonium bound state, placed in hot QCD matter, 
dissociates (``melts'') above a critical temperature. As we now discuss, 
phenomenological models of quarkonium in matter are based upon interpreting
$F_{Q\bar{Q}} (r)$ as the potential in a Schr\"odinger equation whose eigenvalues and eigenfunctions describe the
heavy $Q$-$\bar{Q}$
bound states.  There is no rigorous basis for this line of reasoning, and if pushed too far it faces various conceptual challenges as we shall discuss in Section~\ref{secmeson}.  However, these models remain valuable as a source of semi-quantitative intuition.


At zero temperature, lattice results for $F_{Q\bar{Q}} (r)$ in QCD without dynamical quarks are well approximated
by the  ansatz $F_{Q\bar{Q}} (r) = \sigma\, r - \frac{\alpha}{r}$, where the linear 
term that dominates at long distance
 is characterized by the string tension $\sigma \simeq 0.2\, {\rm GeV}^2$
and the perturbative Coulomb term $\alpha/r$  is dominant at short distances. In QCD with dynamical quarks, beyond some radius $r_c$ the potential flattens because as the distance between the external
$Q$ and $\bar{Q}$ is increased, it becomes energetically favorable to break the
color flux tube  connecting them by producing a light quark-antiquark pair from the vacuum which, in a sense, screens the potential.  
With increasing temperature, the distance $r_c$ 
decreases, that is, the colors of $Q$ and $\bar{Q}$ are screened from each other
at increasingly shorter distances. This is seen clearly in  Fig.~\ref{fig:QQbarDissoc}
in Section~\ref{secmeson}. These lattice results are well parametrized by a 
screened potential of the form~\cite{Satz:2005hx,Karsch:2005nk}
\begin{equation}
 F_{Q\bar{Q}} (r) =  - \frac{\alpha}{r} +\sigma\, r \left(\frac{1-e^{-\mu\, r}}{\mu\, r} \right)\, ,
 \label{ScreenedPotentialParametrization}
\end{equation}
where $\mu \equiv \mu(T)$ can be interpreted at high temperatures 
as a temperature-dependent Debye screeing
mass. For suitably chosen $\mu(T)$, this ansatz reproduces the flattening of the potential found in lattice calculations  at the finite large distance 
value $F_{Q\bar{Q}} (\infty) = \sigma/\mu(T)$.
Taking this $Q\, \bar{Q}$ free energy $F_{Q\bar{Q}}(r,T)$ as the potential
in a Schr\"odinger equation, one may try to determine which
bound states in this potential remain, as the potential is weakened as the temperature increases.
Such potential model studies have led to predictions of the dissociation
temperatures $T_d$ of the charmonium family, which range from 
$T_d({\rm J}/\Psi) \simeq 2.1\, T_c$ to $T_d(\Psi') \simeq 1.1\, T_c$ for the 
more loosely bound and therefore larger $2s$ state. The deeply bound, small, $1s$ state of the bottomonium 
family is estimated to have a dissociation temperature $T_d(\Upsilon(1S))) > 4\, T_c$,
while dissociation temperatures for the corresponding $2s$ and $3s$ states were
estimated to lie at $1.6\, T_c$ and $1.2\, T_c$ respectively~\cite{Satz:2005hx,Karsch:2005nk}. 
Because the leap from the static quark-antiquark potential to a Schr\"odinger equation is not rigorously justified, the
uncertainties in quantitative results obtained from  these potential models are difficult to estimate.
(For more details on why this is so, see Section~\ref{secmeson}.)
However, these models with their inputs from lattice QCD calculations  do provide strong qualitative support for the central
idea of Matsui and Satz that quarkonia melt in hot QCD matter, and they 
provide strong support for the qualitative expectation that this melting
proceeds {\it sequentially}, with smaller bound states dissociating at a higher temperature.

Experimental data on the yield of $J/\Psi$ mesons in nucleus-nucleus collisions
have been reported by the NA50~\cite{Alessandro:2004ap} and NA60~\cite{Arnaldi:2007zz} 
collaborations at the CERN SPS and by the PHENIX and STAR collaborations at RHIC. 
As one increases the size of the collision system (either by varying the impact parameter selection or by changing the nuclear species used in the experiments) these yields
turn out to be increasingly suppressed, relative to the yields
measured in proton-proton or proton-nucleus collisions at the same center of mass energy. 
The measurements made in proton-nucleus collisions are 
particularly important to use as a baseline, since 
without them one would not know how much 
charmonium suppression arises due to the interaction between the charm-anticharm pair and ordinary confined hadronic matter~\cite{Brodsky:1988xz}. 
The operational procedure for separating such hadronic phase effects is to
measure them separately in proton-nucleus collisions~\cite{Kharzeev:1995id}, and to establish
then to what extent the number of $J/\Psi$ mesons produced in
nucleus-nucleus collisions drops below the yield
extrapolated from proton-nucleus collisions \cite{Adler:2005ph}.
We note that for nucleus-nucleus collisions, experimental data exist so far solely for the $J/\Psi$. 
The other bound states of the
charmonium family ($\Psi'$, $\chi_c$, ...) have not been characterized due to the lack
of statistics and resolution. Also, a characterization of bottomonium states in nuclear matter
is missing. Due to the much higher production rates for heavy ion collisions at the TeV scale,
this situation is expected to change at the LHC. Moreover, the higher luminosity of future
RHIC runs may give some access to the bottomonium system. Thus, measurements expected in the near future may  provide evidence for the sequential quarkonium suppression pattern,
which is a generic prediction of all models of quarkonium suppression and which has not
yet been tested.

The analysis of data on charmonium has to take into account a significant number of
important confounding effects. Here, we cannot discuss the phenomenology of these
effects in detail, but we provide a list of the most important ones:
\begin{enumerate}	
\item {\it Contributions from the decays of excited states}\\
In proton-proton collisions, a significant fraction of the observed yield of $J/\Psi$ mesons is known to arise from the production of excited states like the 
$\Psi'$ and $\chi_c$, which subsequently decay to $J/\Psi$.
In a nucleus-nucleus collision,
the suppression of the excited states is expected to set in at a lower temperature since these states are larger in size than the ground state $J/\Psi$.  
In particular, it has been
proposed that the observed suppression of the $J/\Psi$ mesons at RHIC
and at the SPS may arise solely from the dissociation of the more loosely bound
$\Psi'$ and $\chi_c$ states~\cite{Karsch:2005nk}, with the $J/\Psi$s themselves remaining bound in the quark-gluon plasma produced in all heavy ion collisions to date.
Regardless of whether this conclusion turns out to be quantitatively correct, it is certainly apparent that in the absence of
separate measurements of the production of the excited states, any conclusions about the
observed $J/\Psi$-suppression require careful modeling of, and inferences about, the contribution of the decays of excited states.
\item {\it Collective dynamics of the heavy ion collision: ``explosive expansion"}\\
Lattice calculations are done for heavy quark bound states which are at rest in a hot, static, medium.
In heavy ion collisions, however, even if the droplet of hot matter 
equilibrates rapidly, its temperature drops quickly during the subsequent explosive expansion. The observed quarkonium suppression must therefore result from a suitable time average over a dynamical medium. This is challenging in many ways.  One issue that arises is the question of
how long a bound state must be immersed in a sufficiently hot heat
bath in order to melt.   Or, phrased better, how long must the temperature be above the dissociation temperature $T_d$ in order to prevent an heavy quark and antiquark produced at the initial moment of the collision from binding to each other and forming a quarkonium meson?
\item {\it Collective dynamics of the heavy ion collision: ``hot wind''}\\
Another issue that faces any data analysis is that quarkonium mesons 
may be produced moving with significant transverse momentum through the hot medium.
In their own reference frame, the putative quarkonium meson sees a hot wind.
Phenomenologically, the question 
arises whether this leads to a stronger suppression since the bound state sees 
some kind of blue-shifted heat bath (an idea, which we will refine in Section \ref{sec:HotWind}), or whether
the bound state is less suppressed since it can escape the heat bath more quickly. 
\item {\it Formation of quarkonium bound states}\\
Neither quarkonia nor equilibrated quark-gluon plasma are produced at time zero in a heavy ion collision.
Quarkonia
have to form, for instance by a colored $c\, \bar{c}$-pair
radiating a gluon to turn into a color-singlet quarkonium state. This formation process is
not understood in elementary interactions or in heavy ion collisions. However, since
the formation process takes time, it is {\it a priori} unclear whether any observed quarkonium suppression is due to the effects of the hot QCD matter on a formed quarkonium bound state or on the precursor of such a bound state, which may have different attenuation properties in the hot medium.  And, it is unclear whether the suppression is due to processes occurring after the liquid-like strongly coupled plasma in approximate local thermal equilibrium is formed or earlier, before equilibration.
\item {\it Recombination as novel mechanism of quarkonium formation}\\
QCD is flavor neutral and thus charm is produced in $c\, \bar{c}$ pairs in primary interactions. 
If the average number of pairs produced per heavy ion collision is $\lesssim 1$, then all charmonium mesons produced in heavy ion collisions must be made from a $c$ and a $\bar c$ produced in the same primary interaction.
At RHIC and even more so at the LHC, however, more than one $c\, \bar c$ pair is produced per collision, raising the possibility of a new charmonium production mechanism 
in which a $c$ and a $\bar{c}$ from different 
primary  $c\, \bar{c}$-pairs meet and combine 
to form a charmonium meson~\cite{Thews:2000rj}. 
If this novel quarkonium production mechanism were to become significant, it could reduce the quarkonium suppression or even turn it into quarkonium enhancement. However, early LHC data on $J/\Psi$ production do not  show signs of any enhancement~\cite{Collaboration:2010px}.
\end{enumerate}

We have included this section about quarkonium in the present review, since calculations
based on the AdS/CFT correspondence and reviewed in Sections \ref{sec:HotWind} and
\ref{mesons} have provided
complementary information for phenomenological modeling, in particular by calculating
heavy quark potentials within a moving heat bath and by determining mesonic dispersion
relations. The above discussion illustrates the context in which such information is useful,
but it also emphasizes that such information is not sufficient. An understanding of quarkonium
production in heavy ion collisions relies on phenomeological modelling as the bridge between
experimental observations and the theoretical analysis of the underlying properties of hot QCD matter.

\newpage
\chapter{Results from lattice QCD}
\label{sec:latticeQCD}

At very high temperature, where the QCD coupling constant $g(T)$ is 
perturbatively small, hard thermal loop resummed perturbation theory
provides a quantitatively controlled approach to QCD thermodynamics.
However, in a wide temperature range around the QCD phase 
transition which encompasses the experimentally accessible regime, 
perturbative techniques become unreliable. Nonperturbative
lattice-regularized calculations provide the only known, quantitatively 
reliable, technique for the determination of thermodynamic properties of QCD matter
within this regime. 

We shall not review the techniques by which lattice-regularized calculations are implemented. We 
merely recall that the starting point of lattice-regularized calculations at nonzero
temperature is  the imaginary time formalism, which allows one to write
the QCD partition function  in Euclidean space-time with a periodic imaginary time
direction of length $1/T$.  Any thermodynamic quantity can be obtained via suitable 
differentiation of the partition function.
At zero baryon chemical potential, the QCD partition 
function is given by the exponent of a real action, integrated over all
field configurations in the Euclidean space-time. Since the action is real, the QCD partition function can then
be evaluated using standard
Monte Carlo techniques, which require the discretization of the field configurations
and the evaluation of the 
action on a finite lattice of space-time points. Physical results are obtained by
extrapolating calculated results to the limit of infinite volume and vanishing 
lattice spacing. In principle, this is a quantitatively reliable approach.
In practice, lattice-regularized calculations are CPU-expensive: the size of 
lattices in modern calculations does not exceed $48^3 \times 64$ \cite{Durr:2008zz}
and these calculations nevertheless require
the most powerful computing devices (currently at the multi-teraflop scale). In the continuum
limit, such lattices correspond typically to small volumes of $\approx (4\, {\rm fm})^3$ \cite{Durr:2008zz}. This means that
properties of QCD matter which are dominated by long-wavelength modes
are difficult to calculate with the currently available
computing resources, and there are only first exploratory studies.  For the same
reason, it is in practice difficult to carry out calculations using light quark masses
that yield realistically light pion masses.
Light quarks are also challenging
because of the  CPU-expensive complications which arise from the formulation of fermions
on the lattice. 

In addition to the practical challenges above, 
conceptual questions arise in two important domains.
First, at nonzero baryon chemical potential, 
the Euclidean action is no longer real, meaning that the so-called fermion sign problem
precludes the use of standard Monte Carlo techniques. Techniques have been found that evade this problem, but only
in the regime where the quark chemical potential $\mu_B/3$ is sufficiently small compared to $T$.
Second, conceptual questions arise
in the calculation of any physical quantities that cannot be written as derivatives of the partition function.  Calculating many such quantities that are of considerable interest requires the
analytic continuation of lattice results from Euclidean to Minkowski space (see below)
which is always under-constrained since the Euclidean calculations can only be done at finitely many values of the Euclidean time.  This means that lattice-regularized calculations, at least as currently formulated, are not
optimized for 
calculating transport coefficients and answering questions about, say, far-from-equilibrium dynamics or jet quenching.

We allude to these practical and conceptual difficulties to illustrate why alternative
strong coupling techniques, including the use of the AdS/CFT correspondence,  are 
and will remain of great interest for the study of QCD thermodynamics and quark-gluon plasma in heavy ion collisions, even though
lattice techniques can be expected to make steady progress in the coming years.
In the remainder of this Section, we discuss the current status of lattice calculations of some quantities of interest in QCD at nonzero temperature.  We shall begin 
in Section~\ref{sec:latticeQCDEoS} with quantities whose calculation does not run into any of
the conceptual difficulties we have mentioned, before turning to those that do.

\section{The QCD Equation of State from the lattice}
\label{sec:latticeQCDEoS}

The QCD equation of state at zero baryon chemical potential, namely the relation between the pressure and the energy density of hot QCD matter, is an example of a quantity that is well-suited to lattice-regularized calculation since, as a thermodynamic quantity, it can be obtained via suitable differentiations of the Euclidean partition function.  And, the phenomenological motivation for determining this quantity from QCD from first principles is great since, as we have seen in Section~\ref{sec:EllipticFlow}, it is the most important microphysical input for hydrodynamic calculations.
It illustrates the practical challenges of doing lattice-regularized calculations with light quarks that
we have mentioned above that 
while accurate calculations of the thermodynamics of pure glue QCD ($N_f=0$) have existed for a long time~\cite{Boyd:1996bx}, the extraction of the equation of state of quark-gluon plasma with light quarks with their physical masses from calculations that have been
extrapolated to the continuum limit has become 
possible only recently~\cite{Aoki:2006we,Bazavov:2009zn,Borsanyi:2010cj}. 
 
\begin{figure}[t]
\centering
\includegraphics[scale=0.40]{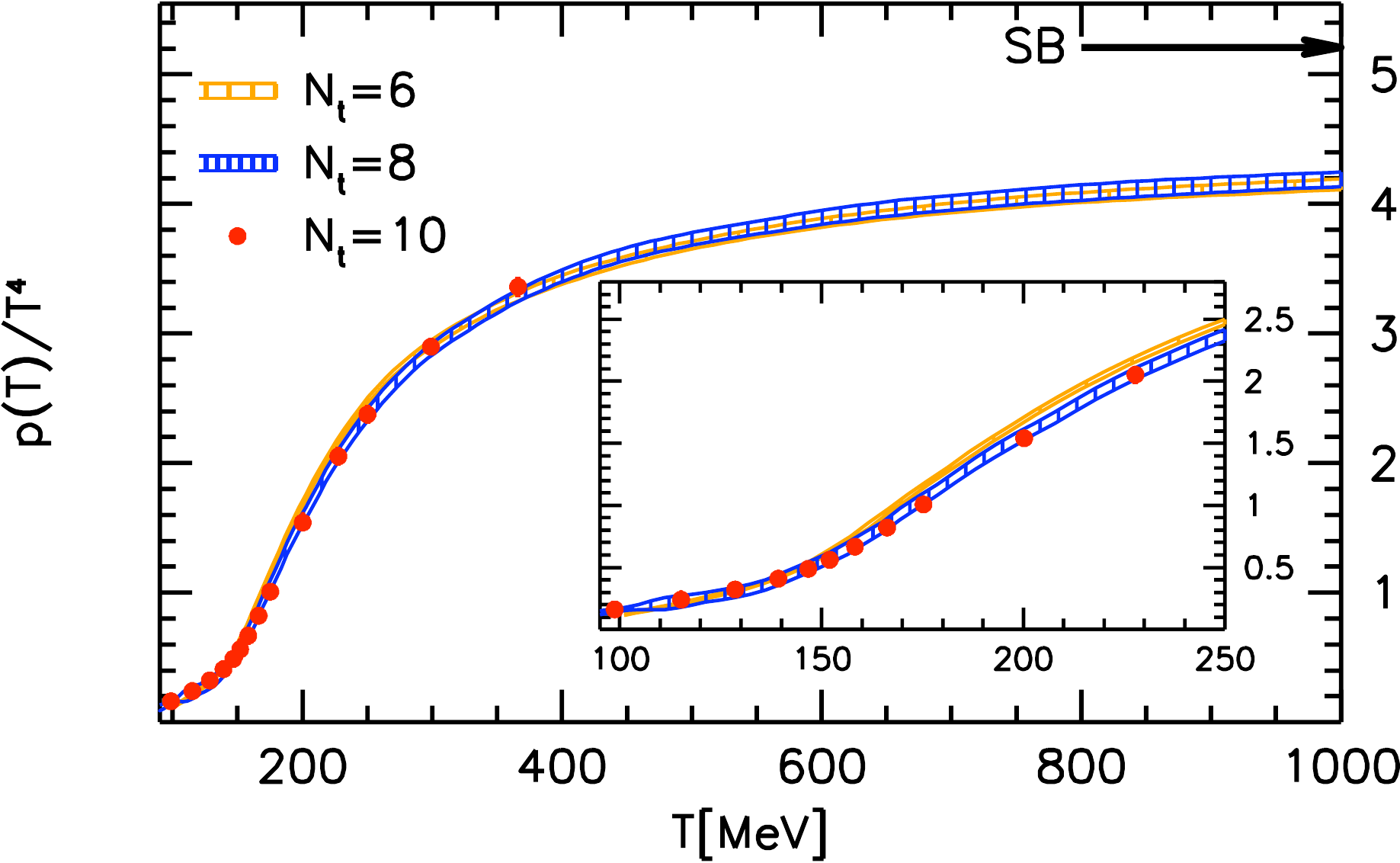}	
\includegraphics[scale=0.40]{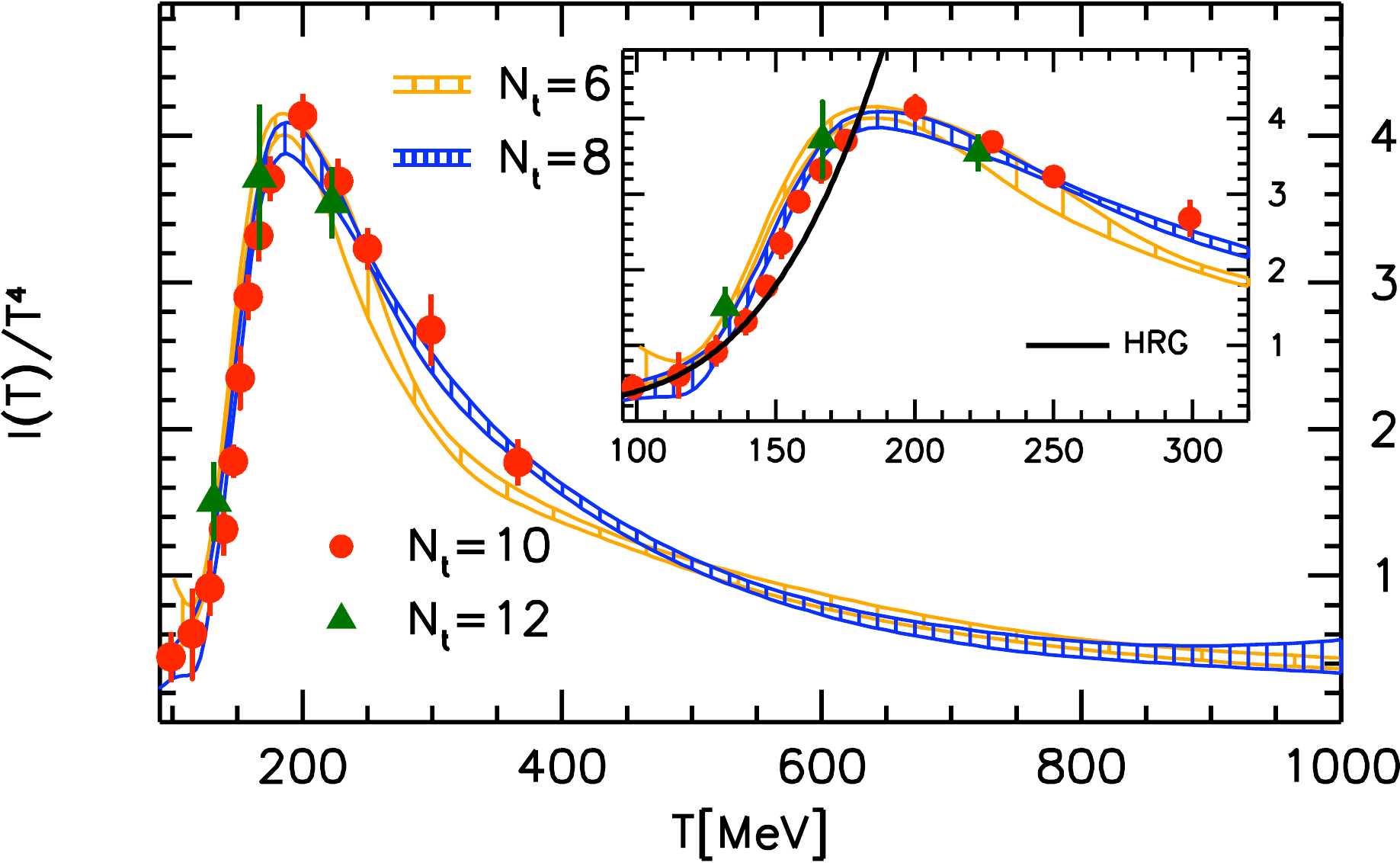}
\caption[]{\label{latticeresults} \small Results from a lattice calculation of QCD thermodynamics with physical quark masses ($N_f=3$, with appropriate light and strange masses).
 Left panel: Temperature
dependence of  the pressure in units of 
$T^4$. Right panel: The trace anomaly $(\varepsilon - 3P)$ in units of $T^4$. Data are for
lattices with the same temporal extent, meaning the same temperature, but with varying numbers of points in the Euclidean time direction $N_\tau$.  The continuum 
limit corresponds to taking $N_\tau\rightarrow\infty$.
Figures
taken from Ref.~\cite{Borsanyi:2010cj}.}
\end{figure}

The current understanding of QCD thermodynamics at the physical point ~\cite{Borsanyi:2010cj} is summarized in 
Fig.~\ref{latticeresults}. In the left panel, the pressure of QCD matter (in thermal equilibrium, with zero baryon chemical potential)
is plotted as a function of its temperature. In order to provide a physically meaningful reference,  it is customary to compare this quantity to the Stefan-Boltzmann result
\be
P_{SB}=\frac{8 \pi^2}{45} \left(1+ \frac{21}{32} N_f\right) T^4\,,
\ee
for a free gas of noninteracting gluons and massless quarks.  This benchmark is indicated by the arrow in the figure.
As illustrated by this plot, the number of degrees of freedom  rises rapidly above 
a temperature $T_c \sim 170$~MeV; at higher temperatures, the pressure takes an almost constant value which deviates from that of a noninteracting gas of quarks and gluons 
by approximately 20\%. This deviation is still present at temperatures as high as $1$~GeV, and convergence to the noninteracting limit is only observed at asymptotically high temperatures ($T>10^8$~GeV~\cite{Fodor:2007sy})  which are far from the reach of any collider experiment.  The right panel shows the trace anomaly, $\varepsilon-3P$, in units of $T^4$
in the same range of temperatures. $\varepsilon-3P$ is often called the ``interaction measure'', but this terminology is quite misleading since both noninteracting quarks and gluons 
on the one hand and very strongly interacting conformal matter on the other have
$\varepsilon-3P=0$, with $\varepsilon/T^4$ 
and $P/T^4$ both independent of temperature.  Large values of $(\varepsilon-3P)/T^4$ necessarily indicate significant interactions among the constituents of the plasma, but small values of this quantity should in no way be seen as indicating a lack of such interactions.
We see in the figure that $(\varepsilon-3P)/T^4$ rises rapidly in the vicinity of $T_c$.  This rapid rise corresponds to the fact that $\varepsilon/T^4$ rises more rapidly than $3P/T^4$, approaching roughly 80\% of its value in an noninteracting gas of quarks and gluons at a lower temperature, between 200 and 250~MeV.  At higher temperatures, as $3P/T^4$ rises 
toward roughly 80\% of {\it its} noninteracting value, $(\varepsilon-3P)/T^4$ falls off with increasing temperature and the quark-gluon plasma becomes more and more conformal.
Remarkably, after a proper re-scaling of the number of degrees of freedom and $T_c$, all the features described above remain the same when  the number of 
colors of the gauge group is increased and extrapolated to the large $N_c\rightarrow \infty $ limit \cite{Bringoltz:2005rr,Datta:2009tj,Panero:2009tv}. 

The central message for us from these lattice calculations of the QCD equation of state is that at high enough temperatures the thermodynamics of the QCD plasma becomes conformal while deviations from conformality are most severe at and just above $T_c$.   This suggests that the use of conformal theories (in which calculations can be done via gauge/gravity duality as described in much of this review) as vehicles by which to gain insights into real-world quark-gluon plasma may turn out to become (even?) more quantitatively reliable when applied to data from heavy ion collisions at the LHC than when applied to those at RHIC.  
In this respect, it is also quite encouraging that the charged particle elliptic flow $v_2(p_T)$ measured very recently in heavy ion collisions 
at $\sqrt{s}=2.76$~TeV at the LHC~\cite{Aamodt:2010pa} is, within error bars, the same as that measured at RHIC.  Quantitative hydrodynamic analyses of these data will come soon, but at a qualitative level they indicate that the quark-gluon plasma produced at the LHC is comparably strongly coupled to that at RHIC.

One of the first questions to answer with a calculation of the equation of state in hand
is whether the observed rapid rise in $\varepsilon/T^4$ and $P/T^4$ 
corresponds to a phase transition or to a continuous crossover. 
In QCD without quarks,
a first order deconfining phase transition is expected due to the breaking of the $Z_N$ center symmetry.  This symmetry is unbroken in the confined phase and broken above $T_c$ by a nonzero expectation value for the Polyakov loop.  The expected first order phase transition  is indeed seen in lattice calculations~\cite{Boyd:1996bx}.  The introduction of quarks introduces a small explicit breaking of the $Z_N$ symmetry even at low temperatures, removing this argument for a first order phase transition.  However, in QCD with massless quarks there must be a sharp phase transition (first order with three flavors of massless quarks, second order with two) since chiral symmetry is spontaneously broken at low temperatures and unbroken at high temperatures.
{\it This} argument for the necessity of a transition vanishes for quarks with nonzero masses, which break chiral symmetry explicitly even at high temperatures.
So, the question of what happens in QCD with physical quark masses, two light and one strange, cannot be answered by any symmetry argument.  Since both the center and chiral symmetries are explicitly broken at all temperatures,
it is possible for the transition from a hadron gas to quark-gluon plasma as a function of increasing temperature to occur with no sharp discontinuities.  And, in fact, lattice calculations have shown that this is what happens:  the dramatic increase in $\varepsilon/T^4$ and $P/T^4$ occurs continuously~\cite{Aoki:2006we}.  This is shown most reliably via the fact that the peaks in the chiral and Polyakov loop susceptibilities are unchanging as one increases the physical spatial volume $V$ of the lattice on which the calculation is done.  If there were a first order phase transition, the heights of the peaks of these susceptibilities should grow $\propto V$ in the large $V$ limit; for a second order phase transition, they should grow proportional to some fractional power of $V$. But, for a continuous crossover no correlation length diverges at $T_c$ and all physical quantities, including the heights of these susceptibilities, should be independent of $V$ 
once $V^{1/3}$ is larger than the longest correlation length. This is  indeed 
what is found~\cite{Aoki:2006we}.  The fact that the transition is a continuous crossover
means that there is no sharp definition of $T_c$, and different operational definitions can give different values.  However, the analysis performed in~\cite{Borsanyi:2010bp}
indicates that the 
chiral susceptibility and the Polyakov loop susceptibility 
peak in the range of $T=150-170$~MeV.

 Despite the absence of a phase transition in the mathematical sense, well above $T_c$ QCD matter is deconfined, since the Polyakov loop takes on large nonzero values. In this high temperature regime, the matter that QCD describes is
best understood in terms of quarks and gluons. This does not, however,
imply that the interactions amongst the plasma constituents is negligible. 
Indeed, we have already seen in Section~\ref{sec:EllipticFlow} that in the temperature regime accessible in heavy ion collisions at RHIC, the quark-gluon plasma behaves like a liquid, not at all like a gas of weakly coupled quasiparticles.
And, as we will discuss in Section~\ref{sec:BulkDynamicalProperties}, explicit calculations 
done via the AdS/CFT correspondence show that in the large-$N_c$ limit in gauge theories with gravity duals which are conformal, and whose coupling can therefore be chosen, the thermodynamic quantities change by only 25\% when the coupling is varied from zero (noninteracting gas) to infinite (arbitrarily strongly coupled liquid).  This shows that thermodynamic quantities are rather insensitive to the strength of the interactions among the constituents (or volume elements) of quark-gluon plasma.

Finally, we note that calculations of QCD thermodynamics done via perturbative methods have been compared to the results obtained from lattice-regularized calculations.
As is well known (see for example 
\cite{Kraemmer:2003gd} and references therein), the expansion of the pressure in powers of the coupling constant $g$ is a badly convergent series and, what is more, cannot be extended beyond  order $g^6 \log (1/g)$, where nonperturbative input is required. This means that perturbative calculations must 
resort to resummations
 and indeed different resummation schemes have been developed over the years~\cite{Braaten:1995jr,Kajantie:2002wa,Hietanen:2008tv,Blaizot:1999ip,Andersen:1999fw,Andersen:2003zk,Andersen:2010wu}. 
 The effective field theory techniques  
developed in \cite{Braaten:1995jr,Kajantie:2002wa}, in particular,  exploit a fundamental feature of any perturbative picture of the plasma:
at weak coupling, $\mu_D \propto gT$ and these methods all exploit the smallness of $\mu_D$ relative to $T$ since the basis of their formulation is that physics at these two energy scales is well separated.
As we will see in Section~\ref{sec:Quasi-particlesSpectralFunctions}, this characteristic is in fact essential for any description of the plasma in terms of quasiparticles.  The analysis 
performed in Ref.~\cite{Hietanen:2008tv} showed that in the region of $T=1-3 \, T_c$ these effective field theory calculations of the QCD pressure
become very sensitive to the matching between the scales $\mu_D$ and $T$, which indicates that there is  no separation of these scales.   This was foreshadowed much earlier by calculations of various different correlation lengths in the plasma phase which showed that at $T=2T_c$ some correlation lengths that are $\propto 1/(g^2 T)$ at weak coupling are in fact significantly 
{\it shorter} than others that are $\propto 1/(g T)$ at weak coupling~\cite{Hart:2000ha}, and showed that the perturbative ordering of these length scales is only achieved for $T>10^2 T_c$.
Despite the success of other 
resummation techniques~\cite{Blaizot:1999ip} in reproducing the main features of QCD thermodynamics, the absence of any separation of scales indicates that there are very significant interactions among constituents and casts doubt upon any approach based upon the existence of well-defined quasiparticles.

\subsection{\label{susceplat} Flavor susceptibilities}

The previous discussion focussed on thermodynamics in the absence of expectation values for any of the conserved (flavor) charges of QCD.
As is well known, 
 these charges are a consequence  of the three flavor symmetries that QCD possesses: 
 the U(1) symmetries  generated by electric charge, $Q$, and baryon number, $B$, and a 
 global SU(3) flavor symmetry. Within SU(3) there are two U(1) subgroups which can be chosen as those generated by $Q$ and by strangeness $S$.  
  Conservation of $Q$ is fundamental to the standard model, since the U(1) symmetry generated by $A$ is  gauge symmetry.  Conservation of $S$ is violated explicitly by the weak interactions and conservation of $B$ is violated by exceedingly small nonperturbative weak interactions, and perhaps by yet to be discovered beyond standard model physics.  As we are interested only in physics on QCD timescales, we can safely treat $S$ and $B$ as conserved.  
 (Instead of taking $B$, $Q$ and $S$ as the conserved quantities, we could just as well have chosen the linear combinations of them corresponding to the numbers of up, down and strange quarks.) With three  conserved quantities, we can introduce three independent chemical potentials. 
In spite of the difficulties in studying QCD at nonzero chemical potential on the lattice, derivatives of the pressure with respect to these chemical potentials at zero chemical potential can be calculated.  These derivatives describe moments of the distribution of these conserved quantities in an ensemble of volumes of quark-gluon plasma, and hence can be related to event-by-event fluctuations in heavy ion collision experiments.

When all three chemical potentials vanish, the lowest nonzero moments are the 
quadratic charge fluctuations, \ie the diagonal and off-diagonal susceptibilities defined as
 \bea
 \chi^X_2&=&\frac{1}{VT}\frac{\del^2}{\del\mu_X \del \mu_X}\log Z(T,\mu_X,\ldots)=\frac{1}{V T^3} \langle N^2_X\rangle \, ,
 \\
 \chi^{XY}_{12}&=&\frac{1}{VT}\frac{\del^2}{\del\mu_X \del \mu_Y}\log Z(T,\mu_X,\mu_Y,\ldots)=\frac{1}{V T^3} \langle N_X N_Y\rangle
 \,, 
 \eea
 where $Z$ is the partition function and the $N_X$ are the numbers of $B$, $Q$ or $S$ charge present in the volume $V$.
The diagonal susceptibilities 
quantify the fluctuations of the conserved quantum numbers in the 
plasma and the off-diagonal susceptibilities measure the  correlations among the conserved quantum numbers, and are more sensitive to the 
nature of the charge carriers~\cite{Koch:2005vg}.
 
\begin{figure}[t]
\centering
\includegraphics[scale=0.59]{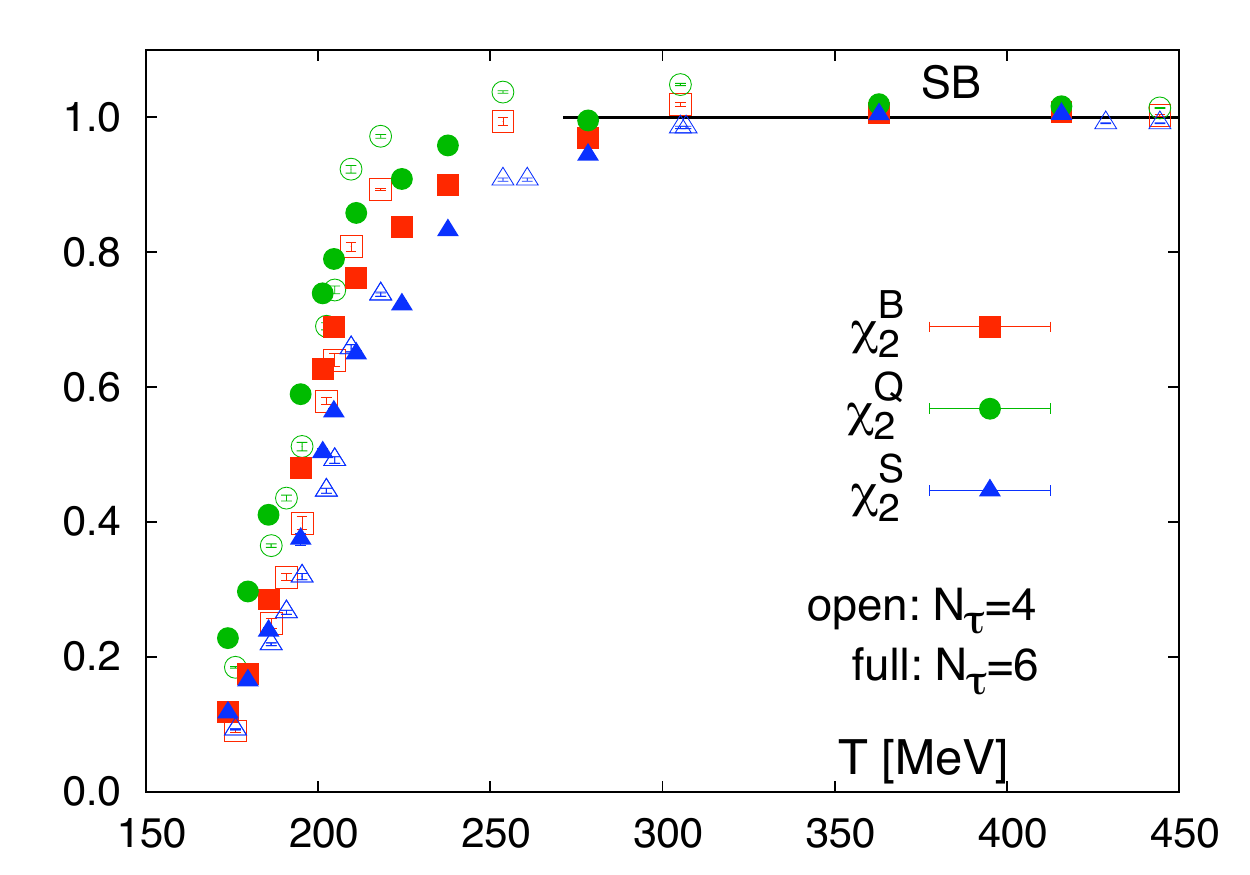}	
\includegraphics[scale=0.59]{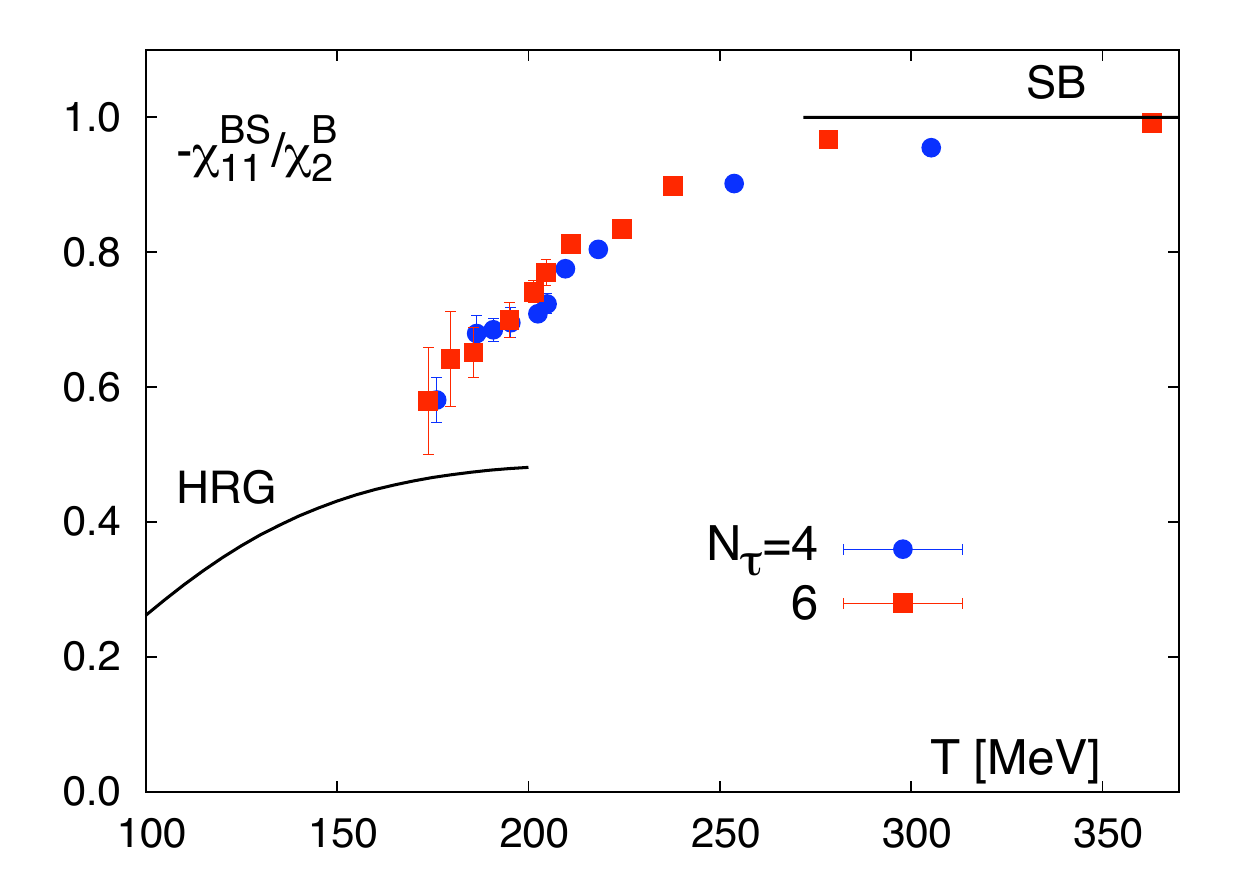}	       
\caption[]{\small 
Left: quadratic fluctuations of the baryon number, electric charge and strangeness normalized to their 
respective Stefan-Boltzmann limit. Right: off-diagonal susceptibility $\chi^{BS}_{11}$  normalized to the 
diagonal baryon charge fluctuations. The upper line corresponds to the Stefan-Boltzmann limits and the lower line corresponds to the hadron resonance gas. Figures taken from 
\cite{Cheng:2008zh}.}
\label{chisus}
\end{figure}
 
Lattice results for these quantities \cite{Cheng:2008zh} are shown in \Fig{chisus}; unlike the calculation of the pressure described in the previous section, these calculations have not been performed with physical quarks, but with larger quark masses ($m_\pi =220$ MeV) for which $T_c=200$ MeV is higher. In the left panel,
the diagonal susceptibilities are shown as a function of temperature. 
The susceptibilities are divided by their values in a noninteracting gas of gluons and massless quarks.
In the right panel, one of the off diagonal susceptibilities, $\chi^{SB}_{11}$, is shown; this result has been divided by the diagonal baryon susceptibility to ensure that this quantity approaches unity in the noninteracting limit. 
In both cases, a rapid rise is observed and above a 
temperature as low as $T=1.5\, T_c$ they are  compatible with a noninteracting gas. 

This is a puzzling result. On the one hand, both lattice calculations of $(\varepsilon - 3P)/T^4$ and experimental measurements of elliptic flow at RHIC indicate that significant interactions among the plasma constituents must be present above $T_c$. On the other hand, 
the lattice calculations shown in \Fig{chisus} suggest that
the flavor degrees of freedom appear to be 
uncorrelated above about  $1.5 T_c$, which would appear to 
favor a quasiparticle interpretation of the plasma already at these rather low temperatures. 
This observation further illustrates the fact that 
thermodynamic properties alone do not resolve the structure of the QGP, since they do not yield an answer to as simple a question as whether it behaves like a liquid or like a gas of quasiparticles.  
We will come back to the apparently contradictory pictures suggested by the  different static properties of the plasma in Section \ref{holsuscep} and
 we move on now to further determine the structure of the plasma via studying dynamical quantities rather than just static ones.  The lattice calculation of dynamical quantities, which require time and therefore Minkowski spacetime in their formulation, are subject to the conceptual challenges that we described above, meaning that the lattice results that we have described in this section are at present significantly more reliable than those to which we now turn.

\section{ Transport coefficients from the lattice}
\label{sec:latticeTransport}
Transport coefficients, such as the shear viscosity, are essential in the description of the real time 
dynamics of a system, since they describe how small deviations away from equilibrium relax toward equilibrium.   As we have discussed in Section \ref{sec:EllipticFlow}, the shear viscosity plays a particularly important role as it provides the connection between experimental data on elliptic flow and conclusions about the strongly coupled nature of the quark-gluon plasma produced in RHIC collisions.     In this section we describe how transport coefficients can be determined via lattice gauge theory calculations.

Transport coefficients 
can be extracted from the low momentum and low frequency limits of the Green's functions of a suitable conserved current of the theory, see Appendix \ref{sec:Green-Kubo}.
To illustrate this point, we concentrate on two examples: the stress tensor components
$T^{xy}$, and the longitudinal component of some conserved $U(1)$ current 
$J^{i}(\omega,{\bf k})$ which can be written $J (\omega, k) \, {\hat  k}$, with $\omega, \,k$ the Fourier modes.  The stress tensor correlator determines the shear viscosity; the current-current correlator determines the diffusion constant for the conserved charge associated with the current.  (The conserved charge could be baryon number, strangeness or electric charge in QCD or could be some $R$-charge in a supersymmetric theory.)
The retarded correlators of these operators are defined by
\bea
G^{xy\, xy}_R(t, x)&=&-i \theta (t) \llangle
                                                  \left[
                                                   T^{xy}(t,x) T^{xy} (0,0)
                                                  \right]
                                                   \rrangle \,, 
                                                   \\
G^{J\,J}_R(t, x)&=&-i \theta (t) \llangle
                                                  \left[
                                                   J(t,x) J (0,0)
                                                  \right]
                                                   \rrangle \,. 
\eea
And, according to the Green-Kubo relation (\ref{etaco})  the low momentum and low frequency limits of these correlators yield
\bea
\eta&=& - \lim_{\omega \rightarrow 0} \frac{{\rm Im}\, G^{xy\,xy}_R(\omega, k=0)}{\omega} \,  , 
\\
D \chi &=& - \lim_{\omega \rightarrow 0} \frac{{\rm Im}\, G^{J\,J}_R(\omega, k=0)}{\omega} \, ,
\eea
where $\eta$ is the shear viscosity, $D$ is the diffusion constant of the conserved charge, and 
$\chi$ is the charge susceptibility. Note that $\chi$ is a thermodynamic quantity which 
can be extracted from the partition function by suitable differentiation and so is straightforward to calculate on the lattice, 
while $\eta$ and $D$ are transport properties
which describe small deviations from equilibrium.          
In general, for any conserved current operator $\mathcal{O}$ whose
retarded correlator is given by
\be
G_R(t, x)=-i \theta (t) \llangle
                                                  \left[
                                                   \mathcal{O}(t,x) \mathcal{O} (0,0)
                                                  \right]
                                                   \rrangle \, ,
                                                   \ee
if we define a quantity $\mu$ by
\be
\mu= - \lim_{\omega \rightarrow 0} \frac{{\rm Im}\, G_R(\omega, k=0)}{\omega} \, ,
\ee
then $\mu$ is a transport coefficient, possibly multiplied by a thermodynamic quantity.

  Transport coefficients can be computed in perturbation theory. However, since 
  the quark-gluon plasma not too far above $T_c$ is strongly coupled, it is preferable to 
  extract information about the values of the transport coefficents from lattice calculations. Doing so is, however, quite challenging.  The difficulty arises from the fact that lattice quantum field theory is formulated in such a way that real time correlators cannot be calculated directly. Instead, these calculations determine 
    the thermal or 
  Euclidean correlator
  \be
  G_E (\tau,x)= \llangle \mathcal{O}_E (\tau, x) \mathcal{O}_E(0,0)\rrangle
  \ee 
  where the Euclidean operator is defined from its Minkowski counterpart by
  \be
  \mathcal{O}_M^{\mu_1 ... \mu_n}{}_{\nu_1.. \nu_m } (-i\tau, x) = (-i) ^r (i )^s 
                                                                                         \mathcal{O}_E^{\mu_1 .. \mu_n}{}_{\nu_1.. \nu_m}(\tau, x)
  \ee
  where $r$ and $s$ are the number of time indices in $\{\mu_1 \, ... \,  \mu_n \}$ and $\{ \nu_1 \, ... \, \nu_m \}$ 
  respectively. 
  Using the Kubo-Martin-Schwinger relation
 \be
 \llangle \op(t,x) \op(0,0)  \rrangle=\llangle \op(0,0) \op(t-i \beta,x)  \rrangle \, ,
 \ee
  the Euclidean correlator $G_E$
  can be related to the imaginary 
 part of the retarded correlator, 
 \be
\rho(\omega, k)\equiv -2 \, {\rm Im} \, G_R (\omega, k) \ ,
\label{sptfuncdef}
 \ee
 which is referred to as the spectral density.
The relation between  $G_E$ (which can be calculated on the lattice) and $\rho$ (which determines the transport coefficient) takes the form of a convolution with a known kernel:
 \be
 \label{eq:heatk}
 G_E(\tau, k)=(-1)^{r+s} \int^\infty_0 \frac{d\omega}{2\pi}\, \frac{\cosh\left(\omega \left(\tau -\frac{1}{2T}\right) \right)}{\sinh \left(\frac{\omega}{2T}\right)} \,\rho(\omega, k ) \, .
 \ee
A typical lattice computation provides values (with errors) for the Euclidean correlator at a set of values  of the Euclidean time, namely $\{\tau_i,  G_E\left(\tau_i,k\right)\}$. In general, it is not possible to extract a continuous 
 function $\rho(\omega)$ from a limited number of points on $G_E(\tau)$ without making assumptions about the functional form
 of either the spectral function or the Euclidean correlator. 
Note also that the Euclidean correlator at any one value of $\tau$ receives contributions from the spectral function at all frequencies. This makes  it hard to disentangle the low frequency behavior of the spectral function from a measurement of the Euclidean correlator at a limited number of values of $\tau$.   

 
 The extraction of the transport coefficient is also complicated by the fact that the high frequency part of the spectral function $\rho$ typically makes a large contribution to the measured $G_E$.
 At large $\omega$, the spectral function is 
  the same at nonzero temperature as at zero temperature and is given by
 \be
\rho (\omega,k=0)=A \, \omega^{2\Delta -d} \, , 
 \ee
 where $\Delta$ is the dimension of the operator ${\cal O}$ and $d$ is the dimension of spacetime. In QCD, the
 constant A can be computed in perturbation theory.  For the two examples that we introduced explicitly above, the spectral functions are given at $k=0$ to leading order 
 in perturbation theory by
\bea
  \rho^{JJ}_R (\omega,k=0)& =&\frac{N_c}{6\pi}  \, \omega^2 \, , \\
  \rho^{xy,xy}_R (\omega,k=0)& =&\frac{\pi (N_c^2-1)}{5 (4 \pi)^2} \,   \omega^4  \, ,
 \eea
where $N_c$ is the number of colors.
These results are valid at any $\omega$ to leading order in perturbation theory; because QCD is asymptotically free, they are the dominant contribution at large $\omega$.  This asymptotic domain of the spectral function does not contain any information about the transport coefficients, but it makes a large
contribution
to the Euclidean correlator.  
This means that the extraction of the contribution of the transport coefficient, which is small in comparison and $\tau$-independent, requires very precise lattice calculations.

 %
\begin{figure}
\begin{center}
\begin{minipage}{0.53\linewidth}
\includegraphics[angle=0,width=0.99\textwidth]{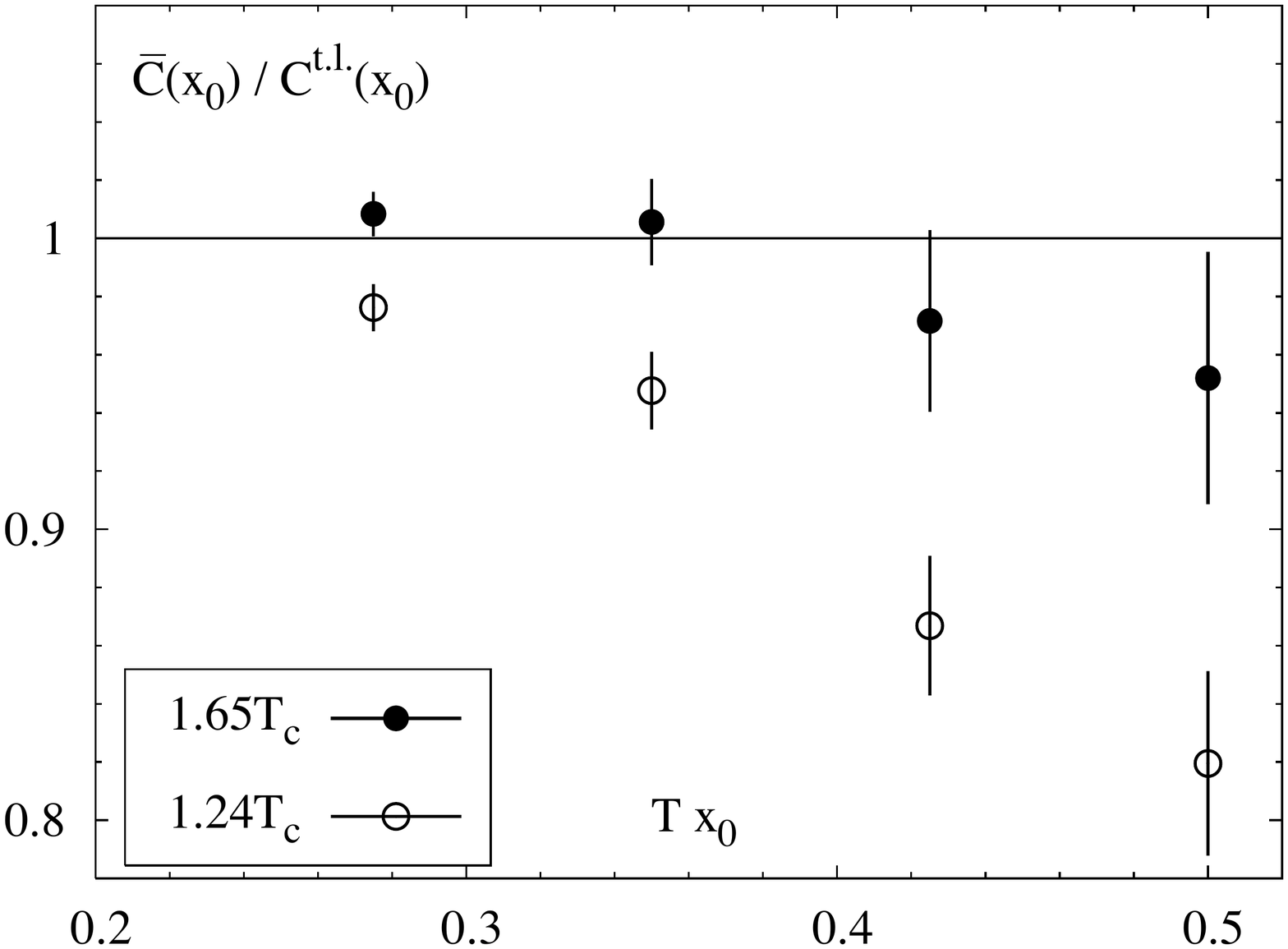}\hfill
\end{minipage}
\begin{minipage}{0.45\linewidth}
\includegraphics[angle=0,width=0.99\textwidth]{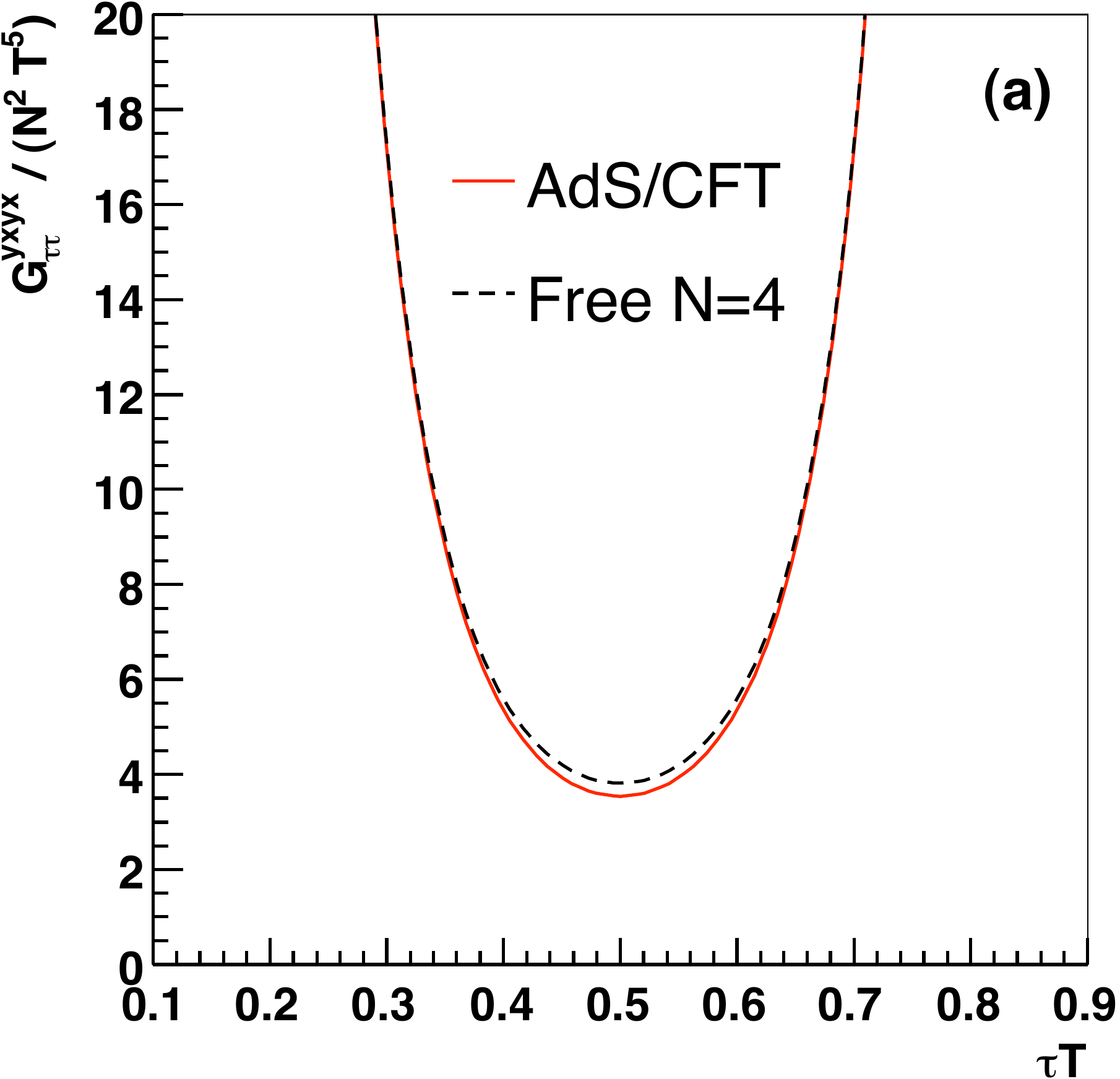}
\end{minipage}
\caption{\small 
\label{fig:euclidcor}
Left panel: Ratio of the stress tensor Euclidean correlator calculated 
on the lattice in Ref.~\cite{Meyer:2007ic} 
to that in the free theory 
for QCD with three colors and zero flavors at four values of the Euclidean time $x_0=\tau$ and two temperatures $T$.  This theory has a first order deconfinement transition, and $T$ is given in units of the critical temperature $T_c$ for this transition.  
Right panel: Stress tensor Euclidean correlator for  $\N=4$ 
SYM from Ref.~\cite{Teaney:2006nc}. The solid line corresponds to infinite coupling and dashed lines to the free theory. 
}
\end{center}
\end{figure}
%

The results of lattice computations for the shear correlator are shown in  the left panel of \Fig{fig:euclidcor}. The finite
temperature Euclidean correlator is normalized to the  free theory correlator at the same temperature. 
The measured correlator deviates from the free one
only by about $10-20 \% $. The errors of the numerical computation illustrate that it is
hard to distinguish the computed correlator from the free one, specially at the higher temperature. 
It is important to stress that the fact that the measured correlator is close to the free one comes from the fact that both receive a large contribution from the large $\omega$ region of the spectral function, and therefore cannot be interpreted as a signature
of large viscosity. 
To illustrate this point, it is illuminating to study 
$\mathcal{N}=4$ SYM as a concrete example in which we can compare weak and strong coupling behavior with both determined analytically. As we will discuss in Section \ref{sec:TransportProperties}, the AdS/CFT correspondence
allows us to compute $\rho$ in the limit of infinite coupling, where the viscosity is small.  
From this AdS/CFT result, we can then compute the Euclidean correlator
via \Eq{eq:heatk}. The result is shown in the right panel of
\Fig{fig:euclidcor}. In the same figure, we show the Euclidean correlator at zero coupling ---  noting that in the zero coupling limit the viscosity diverges as does the length scale above which hydrodynamics is valid.  As in the lattice computation in the left panel of the figure, the difference between the weak coupling and strong coupling Euclidean correlator is small and is only significant around $\tau=1/(2T)$, where $G_E$ is smallest and the contributions from the small-$\omega$ region of $\rho$ are most visible against the ``background'' from the large-$\omega$ region of $\rho$.
For this correlator in this theory,
  the difference between the infinite coupling and zero coupling
limits is only at most $10 \%$. Thus, the $\mathcal{N}=4$ SYM calculation gives us the perspective to realize that
the small deviation between the lattice and free  correlators in QCD
 must not be taken as an indication that the QGP at these temperatures behaves as a 
free gas. It merely reflect the lack of sensitivity of the Euclidean correlator to the low frequency part of the 
spectral function.   


The extraction of transport information from the 4 points in the left panel
of \Fig{fig:euclidcor}, as done in Ref.~\cite{Meyer:2007ic}, requires assumptions
about the spectral density. Since the high frequency behavior of the 
spectral function  is fixed due to asymptotic freedom, a first attempt can be made by writing 
\be
\frac{\rho(\omega)}{\omega}= \frac{\rho_{LF}(\omega)}{\omega} + \theta(\omega-\Lambda) \frac{\rho_{HF}(\omega)}{\omega}\ ,
\ee
where
\be
\label{eq:rhohf}
\rho_{HF}(\omega)=
 \frac{\pi (N^2-1)}{5(4\pi)^2}
\frac{\omega^4}{\tanh \omega/4 T}
\ee
is the free theory result at the high frequencies where this result is valid.  
In the analysis
performed in \cite{Meyer:2007ic}, the parameter $\Lambda$ is always chosen  to be $\ge 5T$. 
The functional form of the low frequency part $\rho_{LF}$ should be chosen such that $\rho_{LF}$ vanishes at high frequency.
A Breit-Wigner ansatz
 \be
 \rho_{LF}/\omega=\frac{\eta }{\pi (1+b^2 \omega^2)}=\rho_{BW}/\omega
 \ee
provides a simple example with which to start (and is in fact the form that arises in perturbation theory \cite{Aarts:2002cc}.)
This ansatz does not provide a good fit, but it nevertheless yields an important lesson. Fitting the parameters in this ansatz to the lattice results for $G_E$ at four values of $\tau$ 
favours large values (larger than $T$) for the width $\Gamma=2/b$
of the low frequency Breit-Wigner structure. This result provides quantitative
motivation for the assumption that the width of any peak or other structure at low frequency must
be larger than $T$.  From this assumption, a bound may be derived on the viscosity as follows. 
Since a wider function
than a  Breit-Wigner peak of width $\Gamma=T$ would lead to larger value of $\rho_{LF}$ for 
$\omega<\sqrt{2}T$ and since the spectral function is positive definite, we have
\be
G_E\left(\frac{1}{2T},k=0\right)\ge \frac{1}{T^5}\left[	
							\int^{2T}_0 \rho_{BW}(\omega) + \int^{\infty}_\Lambda \rho_{HF}(\omega)
							\right]\frac{d\omega}{\sinh \omega/2T}\ .
\ee
From this condition and the measured value of $G_E(\frac{1}{2T},k=0)$, an upper bound on the shear viscosity $\eta$ can be obtained, resulting in
\be
\eta/s < \left\{ \begin{array}{l@{~~~}l}
 0.96 & (T=1.65T_c) \\
 1.08 & (T=1.24T_c)    ,
          \end{array} \right.\label{conservativebound}
          \ee
          with $s$ the entropy density~\cite{Meyer:2007ic}.
The idea here is: (i) we know how much the $\omega>\Lambda$ region contributes to the integral $\int d\omega \rho(\omega) /\sinh \omega/2T$ which is what the lattice calculation determines, and (ii) we make the motivated assumption that the narrowest a peak at $\omega=0$ can be is $T$, and (iii) we can therefore put an upper bound on $\rho(0)$ by assuming that the entire contribution to the integral that does not come from $\omega>\Lambda$ comes from a peak at $\omega=0$ with width $T$.  The bound is conservative because it comes from assuming that $\rho$ is zero at intermediate $\omega$'s between $T$ and $\Lambda$.  Surely $\rho$ receives some contribution from this intermediate range of $\omega$, meaning that the bounds on $\eta/s$ obtained from this analysis are conservative.


Going beyond the conservative bound (\ref{conservativebound}) and making an estimate of $\eta$ is challenging, given the finite number of points at which $G_E(\tau)$ is measured,
and
relies on physically motivated parameterizations of the spectral function. 
 A sophisticated parameterization was introduced in Ref.~\cite{Meyer:2007ic} under the basic assumption 
that there are no narrow structures in the spectral function, which is supported by the 
Breit-Wigner analysis discussed above.
In Ref.~\cite{Meyer:2007ic}, the spectral function was expanded in an ordered basis of orthonormal
functions with an increasing number of nodes, defined and ordered such that
the first few functions are those that make the largest contribution to
the Euclidean correlator; in other words the latter is most sensitive to the
contribution of these functions. Due to the finite number of data points and
their finite accuracy, the basis has to be truncated to the first few
functions, which is a way of formalizing the assumption that there are no narrow structures
in the spectral function.
The analysis based on such parameterization leads to small values of the ratio of the shear viscosity to the entropy density. In particular, 
\be
\eta/s = \left\{ \begin{array}{l@{~~~}l}
 0.134(33) & (T=1.65T_c) \\
 0.102(56) & (T=1.24T_c)    \ .
          \end{array} \right.\label{latticeetaovers}
          \ee
The errors include both statistical errors and an estimate of those systematic errors due to the truncation of the basis of functions used in the extraction. The results of this study are compelling since, as discussed in Section \ref{sec:EllipticFlow}, they are consistent with the experimentally extracted
bounds on the shear viscosity of the QGP via hydrodynamical fits to data on elliptic flow in heavy ion collisions.
These results are also remarkably close to 
$\eta/s=1/4\pi \approx 0.08$, which is obtained in the infinite coupling limit of $\N=4$ SYM theory and 
which we will discuss extensively in Section \ref{sec:TransportProperties}. 

The lattice studies to date
 must be taken as
exploratory, given the various difficulties that we have described.  
As explained in Ref.~\cite{Meyer:2009jp}, there are ways to do better (in addition to using finer lattices and thus obtaining $G_E$ at more values of $\tau$).  For example,
a significant improvement may be achieved by analyzing the spectral function at varying nonzero values of the momentum $k$. 
One can then exploit energy and momentum conservation to relate different
Euclidean correlators to the same spectral function, in some cases constraining the same spectral function with 50-100 quantities calculated on the lattice rather than just 4.  Furthermore, the functional form of the spectral function is predicted order by order in the hydrodynamic expnasion and this provides guidance in interpreting the Euclidean data.  These analyses are still in progress, but results reported to date \cite{Meyer:2009jp} are consistent with (\ref{latticeetaovers}), given the error estimate therein. 

Let us conclude the discussion by remarking on the main points. The Euclidean correlators calculated on the lattice
are dominated by the contribution of the temperature-independent high frequency part of the spectral function, reducing their sensitivity to the transport properties that we wish to extract.
This fact, together with the finite number of points on the Euclidean correlators that are available from lattice computations, complicates the extraction of the shear viscosity from the lattice. Under the mild assumption that there are no narrow structures in the spectral function, an assumption that is supported by the lattice data themselves as we discussed, current lattice computations yield a conservative upper 
bound $\eta/s<1$ on the shear viscosity of the QGP at $T=(1.2-1.7) T_c$. A compelling but exploratory analysis of the lattice data has also been performed, yielding values of $\eta/s\approx 0.1$ for this range of
temperatures.   In order to determine $\eta/s$ with quantitative control over all systematic errors, however, further investigation is needed --- integrating information obtained from many Euclidean correlators at nonzero $k$ as well as pushing to finer lattices.

\section{Quarkonium Spectrum from the Lattice}
\label{secmeson}

Above the critical temperature, quarks and gluons are not confined. As we have discussed at length in  Section~\ref{intro}, experiments at RHIC have taught us that in this regime QCD describes a quark-gluon plasma in which the interactions among the quarks and gluons are strong enough to yield a strongly coupled liquid.
It is also possible that these interactions can result in the
formation of bound states within the deconfined fluid~\cite{Shuryak:2003ty}.  
This observation is of particular relevance for quarkonium mesons formed from
 heavy quarks in the plasma, namely quarks with 
$M\gg T$. For these quarks, $\alpha_s(M)$ is small and the zero temperature mesons are, to  a first approximation, described by a Coulomb-like potential between a $Q-\bar Q$ pair. Thus, the typical radius of the quarkonium meson is 
$r_M\sim 1/\alpha_s  M \ll 1/T$. As a consequence, the properties 
of these quarkonium mesons cannot be strongly modified in
the plasma.  Quarkonia are therefore expected to survive as bound states 
up to a temperature that is high enough that the screening length of the plasma has decreased 
to the point that it is of order the quarkonium radius~\cite{Matsui:1986dk}.

The actual masses  of the heavy quarks that can be accessed in heavy ion collisions, the charm and the bottom, are large enough that charmonium and bottomonium mesons are expected above the deconfinement transition, but they are not so large that these mesons are expected to be unmodified by the quark-gluon plasma produced in ultrarelativistic heavy ion collisions.  Heavy ion collisions at RHIC (at the LHC) may reach temperatures high enough to dissociate all the charmonium (bottomonium) states, and charmonia are certainly not expected to survive in the quark-gluon plasma produced at the LHC.  It is a non-trivial challenge to determine what QCD predicts for the temperatures up to which a particular quarkonium meson survives as a bound state, and above which it dissociates. In this Section, we review lattice QCD calculations done with this goal in mind.
This is a subject of on-going research, and definitive results for the dissociation temperatures of various quarkonia are not yet in hand.
For an example of a recent review on this subject, see Ref.~\cite{Bazavov:2009us}.

Some of the earliest~\cite{Matsui:1986dk,Karsch:1987pv}
attempts to describe the in-medium heavy mesons are based on solving the Schr\"odinger 
equation for a pair of heavy quarks in a potential determined from a lattice calculation. These 
approaches are known generically as potential models.
In this approach, it is assumed that  the 
  interactions between the quark anti-quark pairs and the medium
  can be expressed in the form of a temperature dependent potential. The mesons are identified as the bound states of quarks in this potential.  Such an approach has been very successful at zero temperature~\cite{Eichten:1979ms} and in this context it can be put in firm theoretical grounds by
means of  a non-relativistic effective theory for QCD~\cite{Pineda:1997bj,Brambilla:1999xf}.
However, at nonzero temperature it is not clear how to determine this
potential from first principles. (For some attempts in this direction, 
see Ref.~\cite{Brambilla:2008cx}.) 

\begin{figure}[t]
\begin{center}
\includegraphics[width=0.45\textwidth]{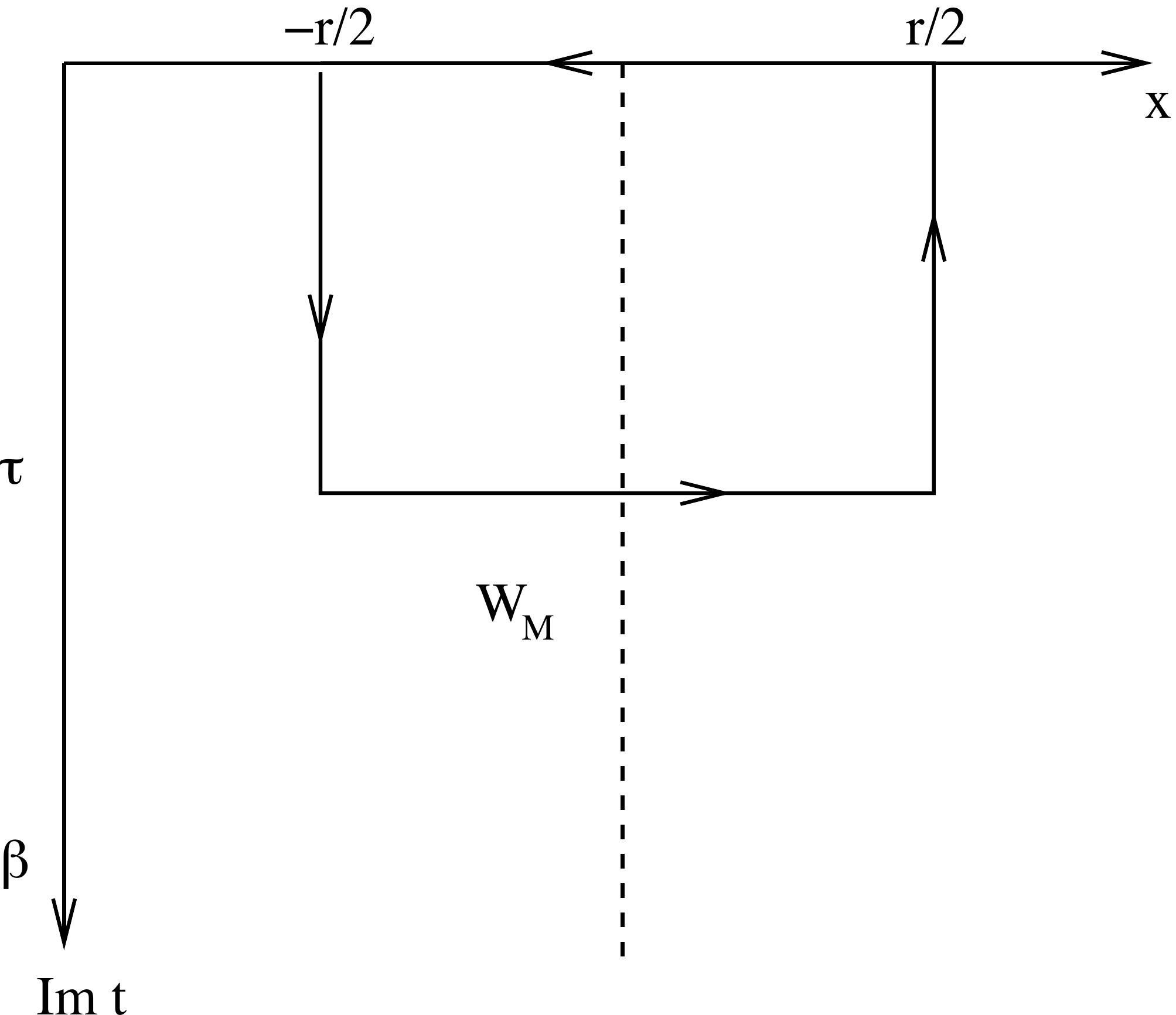}
\caption{\small Wilson line representing the propagation of a heavy quark-antiquark pair. The line at $-r/2$ is the 
heavy quark propagator in imaginary time while the line at $r/2$ is the antiquark. 
The space links ensure gauge invariance. The singlet free energy is obtained by setting $\tau=\beta$.
\label{plotWilsonC}
}
\end{center}
\end{figure}

If the binding energy of the quarkonium meson 
is small compared to the temperature and to any other energy scale that
characterizes the medium,
the potential can be extracted by analyzing a static (infinitely massive) $Q -\bar Q$ pair, in
the color singlet representation,
separated by a distance $r$.
In this limit,  both the quark and the antiquark remain 
static on the time-scale over which the medium fluctuates,  and their propagators in the medium 
reduce to Wilson lines along the time axis. In the
imaginary time formalism, these two Wilson lines wind around the periodic imaginary time direction and they are separated in space by the distance $r$.
These quark and anti-quark Wilson lines are connected by spatial links to 
ensure gauge invariance. These spatial 
links can be thought of as arising via applying a point-splitting procedure at the point where the 
quark and antiquark pair are produced by a local color singlet operator. A sketch of this Wilson line is shown in~\Fig{plotWilsonC}.

At zero temperature, the extension of the Wilson lines in the imaginary time direction $\tau$ can be taken to 
infinity; this limit yields Wilson's definition of the heavy quark potential~\cite{Wilson:1974sk}.
 In contrast, at nonzero temperature the 
imaginary time direction is compact and the imaginary time
$\tau$ is bounded by $1/T$. 
Nevertheless, inspired by the zero temperature case the early studies {\it postulated} that the potential should be obtained from the Wilson line with $\tau=\beta=1/T$.
This  Wilson line can be interpreted as the singlet free energy of the heavy quark pair, 
 i. e.~the energy 
change in the plasma due to the presence of a pair of quarks at a fixed distance and at fixed temperature~\cite{Philipsen:2002az,Nadkarni:1986cz}.

\begin{figure}[t]
\begin{center}
\includegraphics[width=0.47\textwidth]{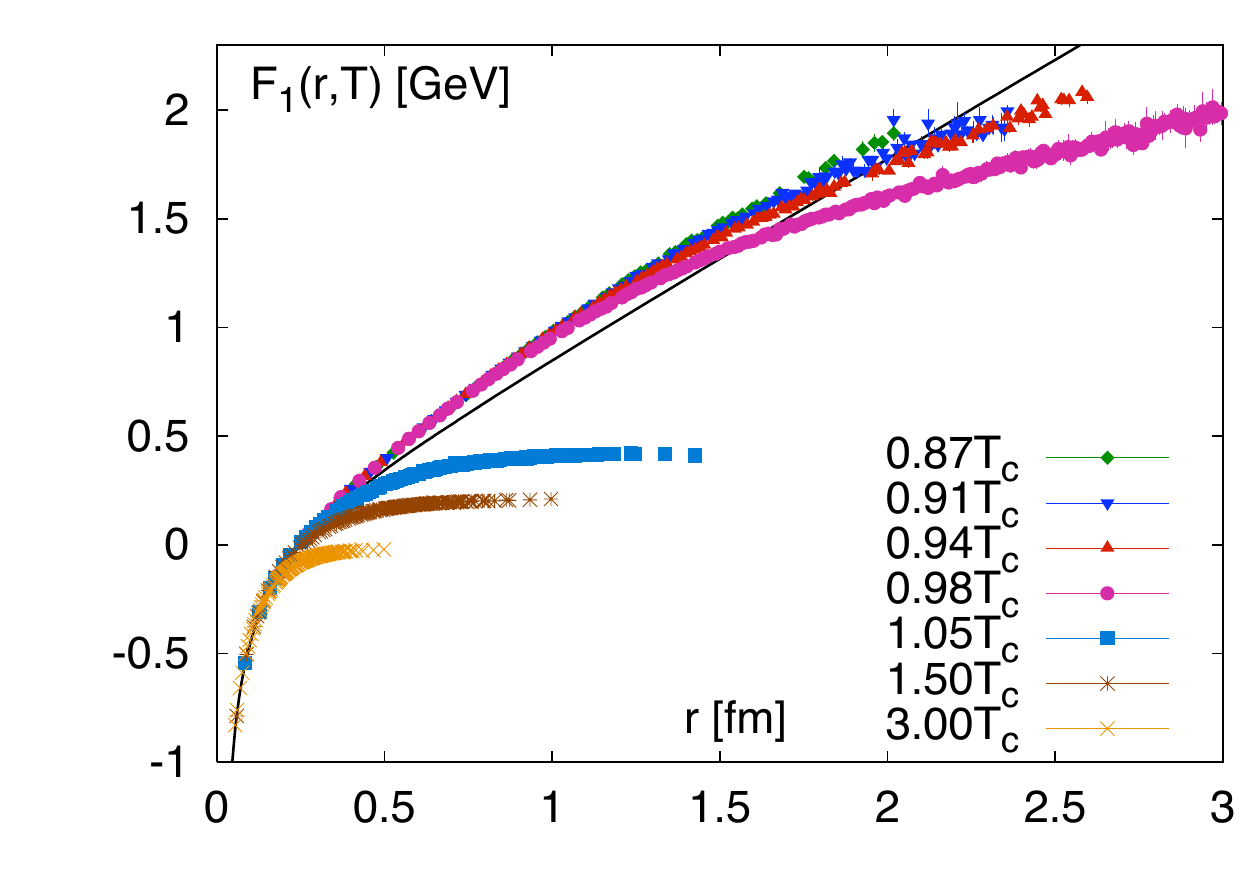}
\includegraphics[width=0.47\textwidth]{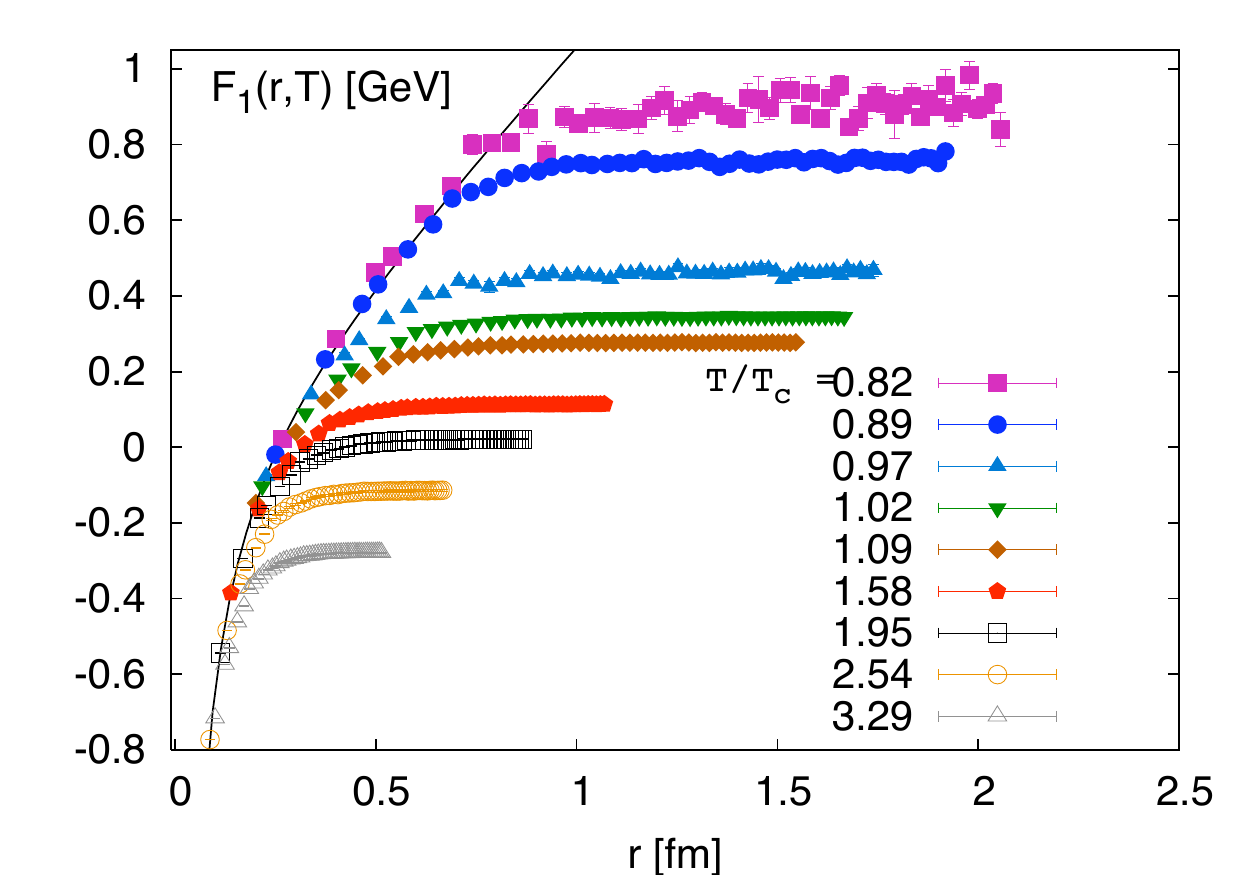}
\caption{\small 
Lattice results  for the singlet free energy $F_1(\rv,T)$ as 
 as a function of the
distance $r$ for different temperatures $T$, quoted as fractions of the
critical temperature $T_c$ at which the crossover  from hadron gas to
quark-gluon plasma occurs. 
 The left panel shows results for 
QCD without 
quarks~\cite{Kaczmarek:2002mc,Kaczmarek:2003dp,Kaczmarek:2004gv} 
and the right panel for 2+1 flavor QCD~\cite{Kaczmarek:2007pb}.
The fact that below $T_c$ the free energy 
goes above the zero temperature result is a lattice artifact \cite{Brambilla:2008cx}.
}
\label{fig:QQbarDissoc2}
\end{center}
\end{figure}

Lattice results for the singlet free energy  are shown 
in \Fig{fig:QQbarDissoc2}. In the left panel we show results for the gluon plasma described by QCD without any quarks~\cite{Kaczmarek:2002mc,Kaczmarek:2003dp,Kaczmarek:2004gv}.
The solid line in this figure denotes the 
 $T=0$ result, which rises linearly with the separation $r$  at large $r$,
as expected due to confinement. The potential is well approximated by the 
ansatz 
\begin{equation}
F_1 (r) = \sigma\, r - \frac{\alpha}{r}\ ,
\label{CornellPotential}
\end{equation}
where the linear 
long-distance part is characterized by the string 
tension $\sqrt{\sigma} = 420\, {\rm MeV}$~\cite{Necco:2001xg}
and the perturbative $1/r$ piece describes the short-distance regime.
Below $T_c$, as the temperature increases the theory remains confined
but the string tension decreases.
For temperatures larger 
than $T_c$, the theory is not confined and the free energy flattens to a finite value
in the large-$r$ limit. 
At these temperatures, the color charge in the plasma screens the interaction
between the heavy $Q$ and $\bar Q$.
In QCD with light dynamical quarks,
as in
 \Fig{fig:QQbarDissoc2} b)  from Ref.~\cite{Kaczmarek:2007pb}, the situation is more complicated.
In this case, the free energy flattens to a finite limit at large distance even at zero temperature, since once the heavy quark and antiquark have been pulled far enough apart it becomes favorable to produce a light $q-\bar q$ pair from the medium (in this case the vacuum)
which results in the formation of 
$Q\bar q$ and $\bar Q q$ mesons that can then be moved far apart without any further expenditure of energy.  In vacuum this process is usually referred to as ``string-breaking''.
In vacuum, at distances that are small enough that string-breaking does not occur the potential can be approximated by (\ref{CornellPotential}), 
but with a reduced string tension $\sqrt{\sigma} \approx 200\, {\rm MeV}$~\cite{Cheng:2007jq}.
Above $T_c$, the potential is screened at large distances by the presence of the colored
fluid, with the screening length beyond which the potential flattens shrinking with increasing temperature, just as in the absence of quarks.
As a consequence, in the right panel of the figure the potential evolves relatively smoothly with increasing temperature, with string-breaking below $T_c$ becoming screening at shorter
and shorter distance scales above $T_c$.
The decrease in the screening length with increasing temperature is a generic result, and it leads
us to expect that quarkonium mesons are expected to dissociate when the temperature is high enough that the vacuum quarkonium size corresponds to a quark-antiquark separation at which the potential between the quark and antiquark is screened~\cite{Matsui:1986dk}.

After precise lattice data for the singlet free energy became available, several authors have used them to solve the
Schr\"odinger equation. Since the expectation value of the Wilson loop in \Fig{plotWilsonC} leads to the singlet {\it free} energy
and not to the singlet {\it internal} energy, 
it has been argued that the potential to be used 
in the Schr\"odinger equation
should be that obtained after first subtracting the entropy contribution to the free energy, namely
\begin{equation}
\label{internalEdef}
U(r,T)=F(r,T)-T\frac{d F(r,T)}{d T} \,.
\end{equation}
Analyses performed with this potential 
indicate that the $J/\psi$ meson survives deconfinement, existing as a bound
state up to a dissociation temperature that lies in the range $T_{\rm diss} \sim(1.5-2.5)\,T_c$ \cite{Wong:2004zr,Alberico:2005xw,Shuryak:2004tx,Mocsy:2005qw,Alberico:2007rg}.
It is also a generic feature of these potential models that, since they are larger in size, other less bound charmonium
states like the $\chi_c$ and $\psi'$ dissociate at a lower temperature~\cite{Karsch:1987pv}, typically at temperatures as low as $T=1.1 \,T_c$. Let us state once more that these calculations are based on two key model assumptions: first, that the charm and bottom quarks are heavy enough for a potential model to apply and, second, that the potential is given by
\Eq{internalEdef}. Neither assumption has been demonstrated from first principles

Given that potential models {\it are} models, there has also been a lot of effort to extract model-independent information about the properties of quarkonium mesons in the medium at nonzero temperature by using lattice techniques to calculate 
 the Euclidean correlation function of 
a color singlet operator of the type
\begin{equation}
J_\Gamma(\tau, \x)=\bar \psi(\tau,\x) \Gamma \psi(\tau,\x) \,,
\end{equation}
where $\psi(\tau,\x)$ is the heavy quark operator and 
$\Gamma=1, \, \gamma_\mu, \, \gamma_5, \, \gamma_5 \gamma_\mu, \, \gamma_\mu\gamma_\nu$ correspond to 
the scalar, vector, pseudo-scalar, pseudo-vector and tensor channels. 
As in the case of the transport coefficients whose analysis we described in 
Section~\ref{sec:latticeTransport}, 
in order to obtain information about the in-medium mesons 
we are interested in extracting the spectral functions of these operators.
As in Section~\ref{sec:latticeTransport}, the Minkowski-space spectral function cannot be 
calculated directly on the lattice; it must instead be inferred from lattice calculations
of the Euclidean correlator
\begin{equation}
G_E(\tau,\x)=\left<
			J_\Gamma (\tau,\x) J_\Gamma (0,{\bf 0}) 
			\right> \, .
\end{equation}
which is related to the spectral function as in \Eq{eq:heatk}. 

\begin{figure}
\begin{center}
\includegraphics[width=0.5\textwidth]{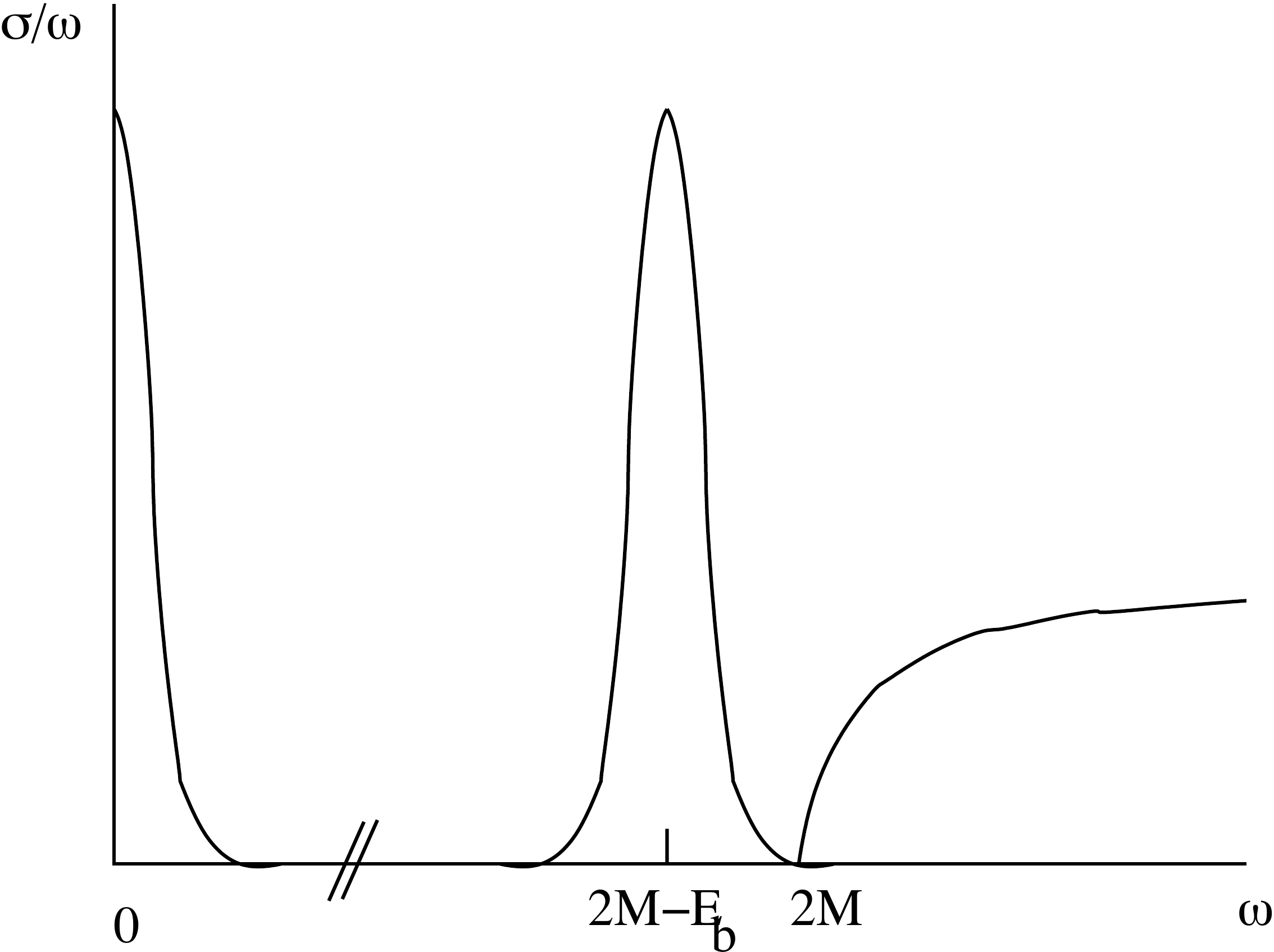}
\caption{\small Schematic view of the current-current spectral function as a function of frequency for 
heavy quarks. The structure at small frequency $\omega\sim 0$ is the transport peak which is due to
the interaction of the external current with heavy 
quarks and antiquarks from the plasma. At $\omega=2M$ there is a threshold for 
pair production. An in-medium bound state, like a quarkonium meson, 
appears as a peak below the threshold.}
\label{toyspf}
\end{center}
\end{figure}

The current-current correlator can be understood as describing
 the interaction of an external vector meson
which couples  only to heavy quarks 
with the plasma.  This interaction can proceed by scattering with 
the heavy quarks and antiquarks present in the plasma or by mixing
between the singlet quarkonium meson with (light quark) states from within the 
plasma that have the same quantum numbers as the external quarkonium.
The first physical process leads to a large absorption of those vector mesons in which the 
ratio $\omega/q$ matches the velocity 
of heavy quarks in the medium, yielding the so-called transport peak at small $\omega$. 
The second 
physical process populates the near-threshold region of $\omega \sim 2M$. Since the thermal distribution of the velocity of heavy quarks and anti-quarks is Maxwellian with a mean velocity 
$v\sim \sqrt{T/M}$, the transport peak is well-separated from the threshold region.  
Thus, 
the spectral function contains information not only 
about 
the properties of mesons in the medium, but also about 
the transport properties of the heavy quarks in 
the plasma.  A sketch of the general expectation for this spectral function is shown in \Fig{toyspf}.
Given these expectations, 
the extraction of properties of quarkonium mesons in the plasma
from the Euclidean correlator 
must take into account the presence of the transport peak. It is worth 
mentioning that for the 
particular case of pseudo-scalar quarkonia, 
the transport peak is suppressed 
by mass~\cite{Aarts:2005hg}; thus, the
extraction of meson properties is simplest in this channel. All other channels, including 
in particular the 
vector channel, include contributions from the transport peak.

From the relation (\ref{eq:heatk}) 
between the Euclidean correlator and the spectral function,
it is clear that the Euclidean correlator has two 
sources of temperature dependence: the temperature dependence of the spectral function itself which is of interest to us and the temperature dependence of the 
kernel in the relation (\ref{eq:heatk}). Since the latter is a trivial kinematical factor, lattice calculations of the Euclidean correlator are often presented compared to
\begin{equation}
\label{grecon}
G_{\rm recon}(\tau, T)=\int^\infty_0 d\omega    \frac{\cosh(\omega (\tau-1/2T))}{\sinh(\omega/2T)} \sigma(\omega,T=0) 
\, ,
\end{equation}
which takes into account the modification of the heat kernel. Any further  temperature
dependence of $G_E$ relative to that in 
$G_{recon}$ is due to the temperature dependence of the spectral function.

\begin{figure} 
\begin{center}
\includegraphics[width=0.49\textwidth]{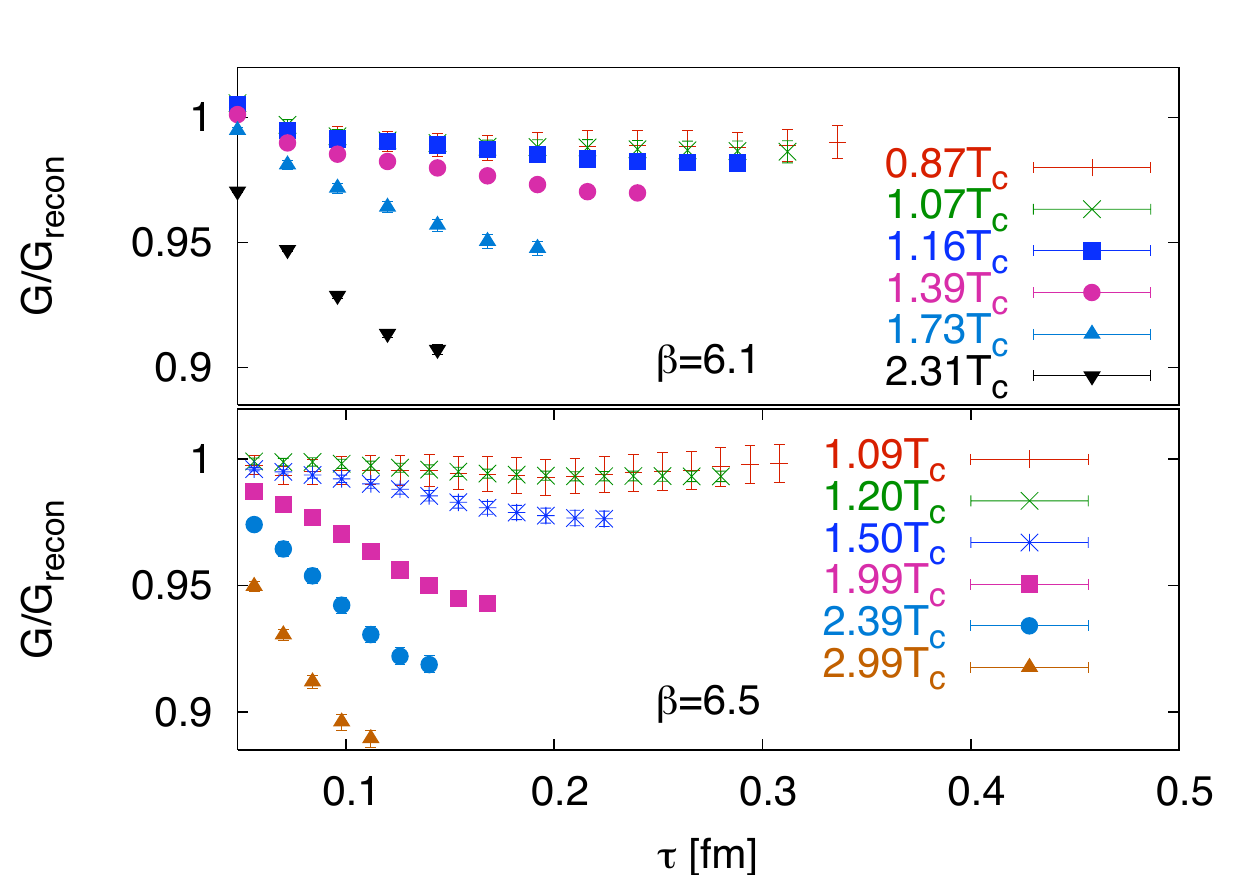}
\includegraphics[width=0.49\textwidth]{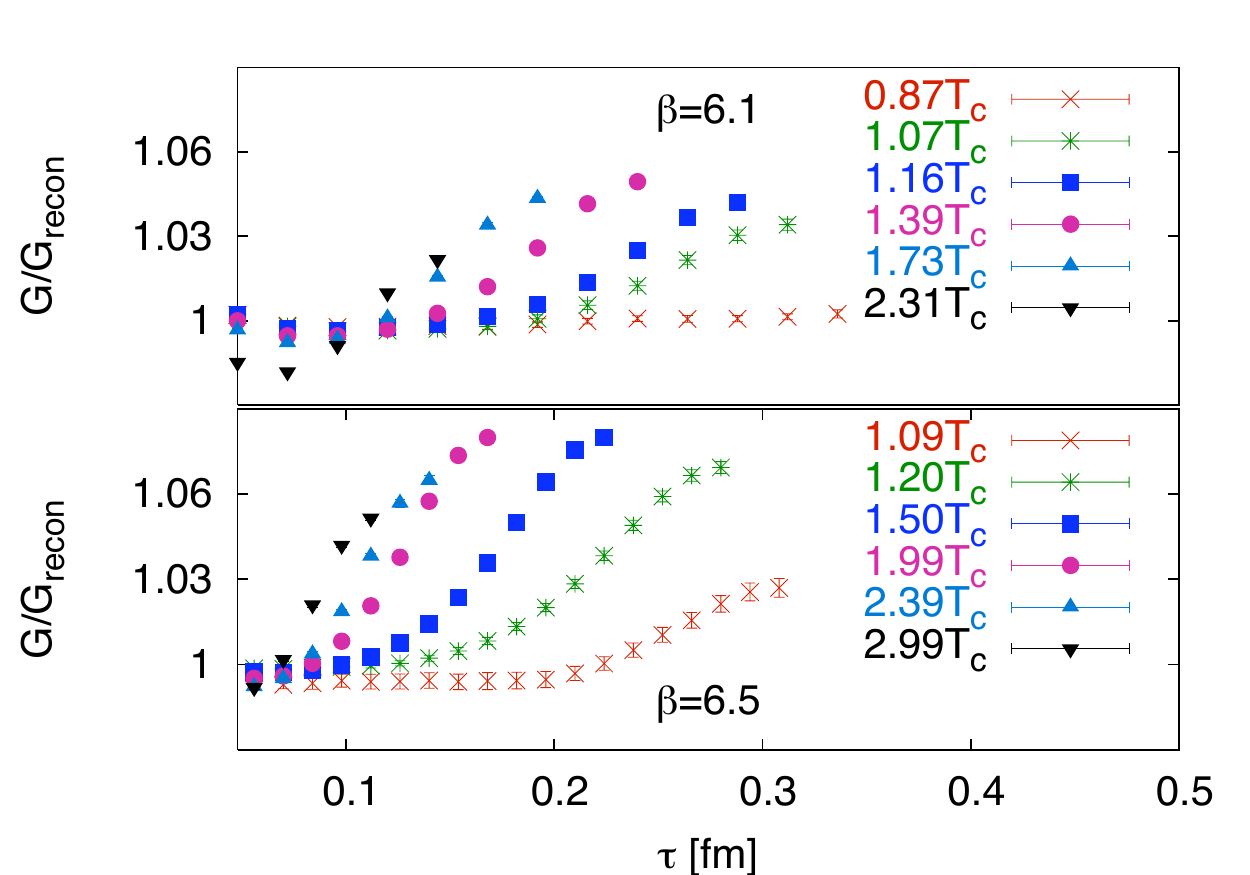}
\end{center}
\caption{\small The ratio of the Euclidean correlator $G_E$ to $G_{\rm recon}$ defined 
in  (\ref{grecon}) in the pseudo-scalar (left) and vector (right) channels for charm quarks versus the imaginary time $\tau$~\cite{Jakovac:2006sf}. Note that   the transport contribution is 
suppressed by the mass of the charmed quark   only in the pseudo-scalar  channel.
\label{fig:latcor}
}
\end{figure}

In \Fig{fig:latcor} we show the ratio of the computed lattice correlator $G_E$  to $G_{recon}$
defined in (\ref{grecon})  in the pseudoscalar  and vector channels for 
charm quarks~\cite{Jakovac:2006sf}.  The temperature dependence of this ratio is due only to the temperature dependence of the spectral function. The 
pseudoscalar  correlator shows little temperature dependence up to temperatures 
as high as $T=1.5 \, T_c$ while the vector correlator varies 
significantly in that range of temperatures.
Since, as we have already mentioned,  the transport peak is suppressed in the pseudoscalar channel, the lack of 
temperature dependence of the Euclidean correlator in this channel
can be interpreted as a signal of the  survival of pseudoscalar charmed mesons
(the $\eta_c$) above deconfinement.  However, the Euclidean correlator is a convolution integral over the spectral density and the thermal kernel, and in principle 
the spectral density can change radically while leaving the convolution integral relatively unchanged.
So, the spectral function must be extracted before definitive conclusions can
be drawn.

There has been a lot of effort towards extracting
these spectral densities in a model-independent way directly from the Euclidean correlators.  The method that has been developed the furthest is called the 
Maximum Entropy Method (MEM)~\cite{Asakawa:2000tr}.
It is an algorithm designed to find the 
most probable spectral function compatible with the lattice data on the Euclidean correlator.  This problem is underconstrained, since one has available lattice calculations of the Euclidean correlator only at finitely many values of the Euclidean time $\tau$, each with error bars, and one is seeking to determine a function of $\omega$.  This means that 
the algorithm must take advantage of prior information about the spectral function that is known absent any calculations of the value of the Euclidean correlator in the form of a default model for the spectral function.  Examples of priors that
are taken into account include
information about asymptotic behavior and sum rules. The MEM method has been very successful in extracting the spectral functions at zero temperature, where it turns out that the
extracted functions have little dependence on the details of how the priors are implemented in a default model for the spectral function.  The application of the same MEM procedure at nonzero temperature is complicated by two facts:  the number of 
data points is smaller at finite temperature than at zero temperature and the temporal extent of the correlators is limited
to $1/T$, which is reduced as the temperature increases. The first problem is a computational problem, which can be ameliorated over time as computing power grows by reducing the temporal lattice spacing and thus increasing the number of lattice points within the extent $1/T$. 
The second problem is intrinsic to the nonzero temperature calculation; 
all the structure in the Minkowski space spectral function, as in the sketch in Fig.~\ref{toyspf}, gets mapped onto fine details of the Euclidean correlator within a small interval of $\tau$
meaning that at nonzero temperature  it takes 
much greater precision in the calculation of the Euclidean correlator in order to disentangle even the main features of the spectral function.

To date, extractions of the pseudoscalar spectral density at nonzero temperature via 
the MEM indicate, perhaps not surprisingly, that the spectral function including its $\eta_c$ peak remains almost unchanged
 up to $T\approx1.5 \,T_c$~\cite{Asakawa:2003re,Datta:2003ww,Jakovac:2006sf,Aarts:2007pk}, 
especially when the comparison that is made is with the zero temperature spectral function extracted from only a reduced number of points on the Euclidean correlator.
The application of the MEM to the vector channel also indicates survival of 
the $J/\psi$ up 
to $T\approx 1.5-2 \,T_c$~\cite{Asakawa:2003re,Datta:2003ww,Jakovac:2006sf,Aarts:2007pk,Ding:2009ie},
but it fails to reproduce the transport peak that must be present in this correlator 
near $\omega=0$. It has been argued that most of the temperature dependence of the 
vector correlator seen in \Fig{fig:latcor} is due to the temperature dependence of the transport 
peak~\cite{Umeda:2007hy}. (Note that since the transport peak is a narrow structure centered at zero frequency, 
it corresponds to a temperature-dependent contribution to the Euclidean correlator that is approximately $\tau$-independent.)
This is supported by the fact that
the $\tau$-derivative of the ratio of correlators 
 is much less dependent on $T$~\cite{Datta:2008dq} and by the analysis  of the spectral functions extracted after 
 introducing a transport peak in the default model of the MEM~\cite{Ding:2009ie}.
  When the transport peak is taken into account
the MEM also shows that $J/\psi$ may survive at least up to $T=1.5 \,T_c$~\cite{Ding:2009ie}.
 However, above $T_c$ both the vector and the pseudoscalar 
 channels show  strong dependence (much stronger than at zero temperature) on the default model via which prior information is incorporated in the MEM~\cite{Jakovac:2006sf,Ding:2009ie}, which makes it difficult to extract solid conclusions on the survival of charmonium states 
  from this method. Despite these uncertainties,
   the conclusions of the MEM analyses agree with those reached via analyses of potential 
  models  in which the internal energy (\ref{internalEdef}) is used as the potential.  However, 
  before this agreement can be taken as firm evidence for the survival of charmonium states well above 
  the phase transition, it must be shown that the potential models and the lattice calculations are 
  compatible in other respects.  To this we now turn.

Potential models can be used for more than determining whether a temperature-dependent
potential admits bound states: they provide a prediction for the entire spectral density.  It is then
straightforward to start with such a predicted spectral density and compute the Euclidean correlator that would be obtained in a lattice calculation if the potential model correctly described all aspects of the physics.  One can then compare the Euclidean correlator predicted by the potential model
with that obtained in lattice computations like those illustrated in Fig.~\ref{fig:latcor}.  Following this
approach, the authors of Refs.~\cite{Mocsy:2007yj,Mocsy:2007jz} have shown that neither the spectral function obtained via identifying the singlet internal energy as the potential nor the one obtained via identifying the singlet free energy as the potential correctly reproduce the Euclidean correlator found in lattice calculations. This means that conclusions drawn based upon either of these potentials cannot be quantitatively reliable in all respects.
These authors then proposed a more phenomenological approach, constructing 
 a phenomenological potential 
(containing many of the qualitative 
features of the singlet free energy but differing from it)
that reproduces the Euclidean correlator obtained in lattice calculations
at the percent level~\cite{Mocsy:2007yj,Mocsy:2007jz}. 
These authors also point out that at nonzero temperature all putative bound
states must have some nonzero thermal width,
and states whose binding energy is smaller than this width should not be considered bound. 
These considerations lead the authors of  Refs.~\cite{Mocsy:2007yj,Mocsy:2007jz} 
to conclude that  the $J/\psi$ and $\eta_c$ dissociate by $T\sim 1.2 T_c$ while less bound states like the $\chi_c$ or $\psi'$ do not survive
the transition at all.  These conclusions differ from those obtained via the MEM. 
Although these conclusions are dependent on the potential used, an important 
and lasting lesson from this work is that the spectral function above $T_c$ can 
be very different from that at zero temperature even if  
the Euclidean correlator computed on the lattice does not show any strong temperature dependence.  
This lesson highlights the challenge, and the need for precision, in trying to extract the spectral
function from lattice calculations of the Euclidean correlator in a model independent fashion.

Finally, we note once again that there is no argument from first principles for 
using the Schr\"odinger equation with either the 
phenomenological potential of Refs.~\cite{Mocsy:2007yj,Mocsy:2007jz} or the internal energy or the free energy as the potential. 
To conclude this Section, we would like to add some remarks on why the identification of the potential with the singlet
internal or singlet free energy cannot be correct~\cite{Laine:2006ns,Beraudo:2007ky}. 
If the quarkonium states can be described
by a Schr\"odinger equation, the current-current correlator must reduce  to the propagation of a   quark-antiquark pair at a given distance $r$ from each other. The correlator must then satisfy 
\be
\label{Gfschr}
\left(-\del_\tau +\frac{\nabla^2}{2M} -2 M - V(\tau, \rv) \right)G_M(\tau,\rv)=0 
\ee
where we have added the subscript $M$ to remind the reader that this expression is only valid in the near threshold region. From this expression, it is clear that the potential can be extracted from the infinitely massive limit, where the 
propagation of the pair is given by the Wilson  line  $W_M$ in \Fig{plotWilsonC} (up to a trivial phase factor proportional to $2Mt$). In this limit, the potential in the 
Euclidean equation (\ref{Gfschr}) is then defined by 
\be
\label{potdef}
-\del_\tau W_M(\tau,\rv)  =  V(\tau,\rv) W_M(\tau,\rv)\ ,
\ee
where $\tau$ and $r$ are the sides of the Wilson loop in \Fig{plotWilsonC}.
In principle, the correct real time potential $V(t,\rv)$ should then be obtained via 
analytic continuation of $V(\tau,\rv)$.
And, for bound states with sufficiently low binding energy
it would then suffice to consider the long time limit of the potential, 
$V_\infty(\rv)\equiv V(t=\infty,\rv)$.

  The difficulty of extracting the correct potential resides in the analytic  continuation from $V(\tau,\rv)$ to $V(t,\rv)$. At zero temperature, $\tau$ is not periodic and we can take 
  the $\tau\rightarrow \infty$ limit and relate what we obtain to  $V_\infty$.
   At nonzero temperature, $\tau$ is periodic and so there is no $\tau\rightarrow\infty$ limit.
   It is also apparent that $V_\infty $ need not coincide with the value 
   of $V(\beta,r)$ as postulated in some  potential models; in fact,
  due to the periodicity of of $\tau$ a lot of information is lost by 
  setting $\tau=\beta$~\cite{Beraudo:2007ky}. Explicit calculations within perturbative thermal field theory, where the analytic continuation can be performed, show that
  $V_\infty$ does not coincide with the internal energy (\ref{internalEdef}) and, what is more, 
  the in-medium potential develops an imaginary 
  part which can be interpreted as the collisional width of the state in the 
  plasma~\cite{Laine:2006ns}. If the log of the 
  Wilson loop is a quadratic function of the gauge potential, as in QED or in QCD to leading order in perturbation theory, then it is possible to show
  that the real part of the potential agrees with the singlet free energy~\cite{Beraudo:2007ky}; however this is not the case in general. So, the correct potential to describe shallow bound states is not 
  given by \Eq{internalEdef} and care must be
  taken in drawing conclusions from
  potential model calculations.

\chapter{Introducing the gauge/string duality}
\label{sec:Section3}

Sections~\ref{sec:Section3} and \ref{AdS/CFT} together constitute a primer on gauge/string duality, written for a QCD audience.

Our goal in this section is to state what we mean by gauge/string duality, via a clear statement  of the original example of such a duality \cite{Maldacena:1997re,Gubser:1998bc,Witten:1998qj}, namely the conjectured equivalence between a certain conformal gauge theory and a certain gravitational theory in Anti de Sitter spacetime.  We shall do this in section~\ref{sec:conj}. In order to get there, in section~\ref{sec:motivating} we will first motivate from a gauge theory perspective why there must be such a duality.  Then, in section~\ref{sec:allyouneed}, we will give the reader a look at all that one needs to know about string theory itself in order to understand section~\ref{sec:conj}, and indeed to read this review.

Since some of the contents in this section are by now standard textbook material, in some cases we will not give specific references. The reader interested in a more detailed review of string theory may consult the many textbooks available such as \cite{Green:1987sp,Lust:1989tj,Polchinski:1998rq,Johnson:2003gi,Zwiebach:2004tj,Kiritsis:2007zz,Becker:2007zj,Dine:2007zp}. The reader interested in complementary aspects or extra details about the gauge/string duality may consult some of the many existing reviews, e.g.~\cite{Aharony:1999ti,D'Hoker:2002aw,Maldacena:2003nj,Son:2007vk,Peeters:2007ab,Mateos:2007ay,Erdmenger:2007cm,Gubser:2009md,Schafer:2009dj,Polchinski:2010hw}.

\section{Motivating the duality}
\label{sec:motivating}

Although the AdS/CFT correspondence was originally discovered \cite{Maldacena:1997re,Gubser:1998bc,Witten:1998qj} by studying D-branes and black holes in string theory, the fact that such an equivalence may exist can be directly motivated from certain aspects of gauge theories and gravity.\footnote{Since string theory is a quantum theory of gravity and the standard Einstein gravity arises as the 
low-energy limit of string theory, we will use the terms gravity and string theory interchangeably below.} In this subsection we motivate such a direct path from gauge theory to string theory without going into any details about string theory and D-branes.

\subsection{An intuitive picture: Geometrizing the renormalization group flow} \label{geom}

Consider a quantum field theory (more generally, a many-body system) in $d$-dimensional Minkowski spacetime with coordinates $(t, \vec x)$, possibly defined with a short-distance cutoff $\ep$.
From the work of Kadanoff, Wilson and others in the sixties, a good way to describe the system is to organize the physics in terms of length (or energy) scales, since degrees of freedom at widely separated scales
are largely decoupled  from each other. If one is interested in properties of the system at a length scale $z \gg \ep$, instead of using the bare theory defined at 
scale $\ep$, it is more convenient to integrate out short-distance degrees of freedom
and obtain an effective theory at length scale $z$. Similarly, for physics at an even longer length scale $z' \gg z$, it is more convenient to use the effective theory
at scale $z'$. This procedure defines a renormalization group (RG) flow and gives rise to a continuous family of effective theories in $d$-dimensional Minkowski spacetime labeled by the RG scale $z$. One may now visualize this continuous family of $d$-dimensional theories as a single theory in a $(d+1)$-dimensional spacetime with the RG scale $z$ now becoming a spatial coordinate.\footnote{Arguments suggesting that the string dual of a Yang-Mills  theory must involve an extra dimension were put forward in \cite{Polyakov:1998ju}.} By construction, this $(d+1)$-dimensional theory should have the following properties:
 
\ben
\item 
The theory should be intrinsically non-local, since an effective theory at a scale $z$ should only describe physics at scales longer than $z$. However, there should still be some degree of locality in the $z$-direction, since degrees of freedom of the original theory at different scales are not strongly correlated with each other. For example, the renormalization group equations governing the evolution of the couplings are local with respect to length scales.

\item 
The theory should be invariant under reparametrizations of the $z$-coordinate, since the physics of the original theory is invariant under reparametrizations of the RG scale.

\begin{figure}
    \begin{center}
    	\includegraphics[width=.8\textwidth]{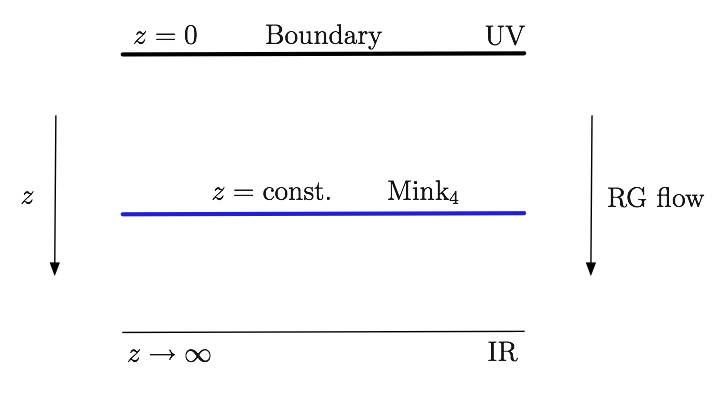}
    \end{center}
    \caption{\small A geometric picture of $AdS_5$.}
\label{AdSpicture}
\end{figure}

\item 
All the physics in the region below the Minkowski plane at $z$ (see fig.~\ref{AdSpicture}) should be describable
by the effective theory of the original system defined at a RG scale $z$. In particular, this $(d+1)$-dimensional description has only the number of degrees of freedom of a $d$-dimensional theory. \label{foire}

\een

In practice, it is not yet clear how to `merge' this continuous family of $d$-dimensional theories into a coherent description of a $(d+1)$-dimensional system, or whether this way of rewriting the renormalization group gives rise to something sensible or useful. Property number~(\ref{foire}) above, however,  suggests that if such a description is indeed sensible, the result may be a theory of quantum gravity. The clue comes from the holographic 
 principle~\cite{tHooft:1993gx,Susskind:1994vu} (for a review, see \cite{Bousso:2002ju}), which says that a theory of quantum gravity in a region of space should be described by a {\it non-gravitational} theory living at the boundary of that region. In particular, one may think of the quantum field theory as living on the $z=0$ slice, the boundary of the entire space. 

We now see that the gauge/gravity duality, when interpreted as a geometrization of the RG evolution of a quantum field theory, appears to provide  a specific realization of the holographic principle. An important organizing principle which follows from this discussion is the UV/IR connection \cite{Susskind:1998dq,Peet:1998wn} between the physics of the boundary and the bulk systems. From the viewpoint of the bulk, physics near the $z=0$ slice corresponds to physics near the boundary of the space, \ie to large-volume or IR physics. In contrast, from the viewpoint of the quantum field theory, physics at small $z$ corresponds to short-distance physics, \ie UV physics.


\subsection{The large-$\nc$ expansion of a non-Abelian gauge theory vs. the string theory expansion} \label{sec:nce}

The heuristic picture of the previous section does not tell us for which many-body system such a gravity description is more likely to exist, or what kind of properties such a gravity system should have. A more concrete indication that a many-body theory may indeed have a gravitational description comes from the large-$\nc$ expansion of a non-Abelian gauge theory.

That it ought to be possible to reformulate a non-Abelian gauge theory as a string theory can be motivated at different levels. After all, string theory was first invented to describe strong interactions. Different vibration modes of a string provided an economical way to explain many resonances discovered in the sixties which obey the so-called Regge behavior, \ie the relation $M^2 \propto J$ between the mass and the angular momentum of a particle.  After the formulation of QCD as the microscopic theory for the strong interactions, confinement provided a physical picture for possible stringy degrees of freedom in QCD. Due to confinement, gluons at low energies behave to some extent like flux tubes which can close on themselves or connect a quark-antiquark pair, which naturally suggests a possible string formulation. Such a 
low-energy effective description, however, does not extend to high energies if the theory becomes weakly coupled, or to non-confining gauge theories.

A strong indication that a fundamental (as opposed to effective) string theory description may exist for any non-Abelian gauge theory (confining or not) comes from  't Hooft's large-$\nc$ expansion~\cite{'tHooft:1973jz}. Due to space limitations, here we will not give a self-contained review of the expansion (see e.g.~\cite{Witten:1979kh,Coleman,Manohar:1998xv} for reviews) and will only summarize the most important features. The basic idea of 't Hooft was to treat the number of colors $\nc$ for a non-Abelian gauge theory as a parameter, take it to be large, and expand physical quantities in $1/\nc$. For example, consider the Euclidean partition function for a $U(\nc)$ pure gauge theory with gauge coupling $\gym$:
 \begin{equation}
 Z = \int D A_\mu \; \exp \left( - {1 \ov 4 \gym^2} \int d^4 x\,  \Tr F^2 \right) \ .
 \end{equation}
Introducing the 't Hooft coupling
\be
\lam = \gym^2 \nc \ ,
\ee
one finds that the vacuum-to-vacuum amplitude $\log Z$ can be expanded in $1/\nc$ as
\be \label{laN}
\log Z = \sum_{h=0}^\infty \nc^{2-2h} f_h (\lam)
 = \nc^2 f_0 (\lam) + f_1 (\lam) + {1 \ov \nc^2} f_2 (\lam) + \cdots \,,
\ee
where $f_h (\lam), h=0,1, \ldots$ are functions of the 't Hooft coupling
$\lam$ only. What is remarkable about the large-$\nc$  expansion~\eqref{laN} is that, at a fixed $\lam$, Feynman diagrams are organized by their topologies. For example, diagrams which can be drawn on a plane without crossing any lines (``planar diagrams'') all have the same $\nc$ dependence, proportional to $\nc^2$, and are included in $f_0 (\lam)$. Similarly, $f_h (\lam)$ includes the contributions of all diagrams which can be drawn on a two-dimensional surface with $h$ holes without crossing any lines. Given that the topology a two-dimensional compact orientable surface is classified by its number of holes, the large-$\nc$ expansion~\eqref{laN} can be considered as an expansion in terms of the topology of two-dimensional compact surfaces.

This is in remarkable parallel with the perturbative
expansion of a closed string theory, which expresses physical quantities in terms of the propagation of a string in spacetime. The worldsheet of a {\it closed} string is a two-dimensional compact surface and the string perturbative expansion is given by a sum over the topologies  of two-dimensional surfaces. For example the vacuum-to-vacuum amplitude in a string theory can be written as
 \be \label{strE}
 \sA = \sum_{h=0}^\infty g_s^{2h-2} F_h (\apr) = 
 \frac{1}{g_s^2} F_0 (\apr) + F_1 (\apr)
 + g_s^2 F_2 (\apr) + \cdots \,,
 \ee
where $g_s$ is the string coupling, $2 \pi \apr$ is the inverse string tension, and $F_h (\apr)$ is the contribution of 2d surfaces with $h$ holes.

Comparing~\eqref{laN} and~\eqn{strE}, it is tempting to identify~\eqn{laN} as the perturbative expansion of some string theory with
 \be \label{couR}
 g_s \sim {1 \ov \nc}
 \ee
and the string tension given as some function of the 't Hooft coupling $\lam$.
Note that the identification of~\eqref{laN} and~\eqn{strE} is more than just a mathematical analogy.  Consider for example $f_0 (\lam)$, which is given by the sum over all Feynman diagrams which can be drawn on a plane (which is topologically a sphere).
Each planar Feynman diagram can be thought of as a discrete triangulation of the sphere. Summing all planar diagrams can then be thought of as summing over all possible discrete triangulations of a sphere, which in turn can be considered as summing over all possible embeddings of a two-dimensional surface with the topology of a sphere
in some ambient spacetime. This motivates the conjecture of identifying $f_0 (\lam)$ with $F_0$ for some closed string theory, but leaves open what the specific string theory is.

One can also include quarks, or more generally matter in the fundamental representation. Since quarks have $\nc$ degrees of freedom, in contrast with the $\nc^2$ carried by gluons, including quark loops in the Feynman diagrams will lead to $1/\nc$ suppressions. For example, in a theory with $\nf$ flavours, the single-quark loop planar-diagram contribution to the vacuum amplitude scales as $\log Z \sim \nf \nc$ rather than as $\nc^2$ as in~\eqref{laN}. In the large-$\nc$ limit with finite $\nf$, the contribution from
quark loops is thus suppressed by powers of $\nf/\nc$.  Feynman diagrams with quark loops can also be classified using topologies of two-dimensional surfaces, now with inclusion of surfaces with boundaries. Each boundary can be identified with a quark loop.
On the string side, two-dimensional surfaces with boundaries describe worldsheets of a string theory containing both closed and open strings, with boundaries corresponding to the worldlines of the endpoints of the open strings.

\subsection{Why AdS?}

Assuming that a $d$-dimensional field theory can be described by
a $(d+1)$-dimensional string or gravity theory, we can try to derive
some properties of the $(d+1)$-dimensional spacetime.
The most general metric in $d+1$ dimensions consistent with 
$d$-dimensional Poincare symmetry can be written as
 \be \label{teme}
 ds^2 = \Om^2 (z) \le( - dt^2 + d \vec x^2 + dz^2 \ri) \,,
 \ee
 with $z$ is the extra spatial direction. Note that in order to have translational symmetries in the $t, \vec x$ directions, the warp factor $\Om(z)$ can depend on $z$ only. At this stage not much can be said of the form of $\Om(z)$ for a general quantum field theory. However if we consider field theories which are conformal (CFTs), then we can determine $\Om (z)$ using the additional symmetry constraints!
A conformally invariant theory is invariant under the scaling
 \be \label{scaO}
 (t, \vec x ) \to C (t, \vec x) 
 \ee
with $C$ a constant. For the gravity theory formulated in~\eqn{teme} to describe such a field theory, the metric~\eqn{teme} should respect the scaling symmetry~\eqn{scaO} with the simultaneous scaling of the $z$ coordinate $z \to C z$, since $z$ represents a length scale in the
boundary theory. For this to be the case we need  $\Om(z)$ to scale as
 \be
  \Om (z) \to C^{-1} \Om (z) \quad {\rm under} \quad z \to C z \,.
 \ee
 This uniquely determines
 \be \label{Wads}
 \Om (z) = {R \ov z} \,,
 \ee
 where $R$ is a constant. The metric~\eqref{teme} can now be written as
 \be \label{Pads}
 ds^2 = {R^2 \ov z^2} \le( - dt^2 + d \vec x^2 + dz^2 \ri),
 \ee
which is precisely the line element of $(d+1)$-dimensional anti-de Sitter  spacetime, $AdS_{d+1}$. This is a maximally symmetric spacetime with curvature radius $R$ and constant negative curvature proportional to $1/R^2$. See e.g.~\cite{Hawking:1973uf} for a detailed discussion of the properties of AdS space.

In addition to Poincare symmetry and the scaling~\eqref{scaO}, a conformal field theory in $d$-dimension is also invariant under $d$ special conformal transformations, which altogether form the $d$-dimensional conformal group $SO(2,d)$. It turns out that the isometry group\footnote{Namely the spacetime coordinate transformations which leave the metric invariant.} of~\eqref{Pads} is also $SO(2,d)$, precisely matching that of the field theory. Thus one expects that a conformal field theory should have a string theory description in AdS spacetime!

Note that it is not possible to use the discussion of this section to deduce the precise string theory dual of a given field theory, nor the precise relations between their parameters. In next section we will give a brief review of some essential aspects of string theory which will then enable us to arrive at a precise formulation of the duality, at least for some gauge theories.

\section{All you need to know about string theory}
\label{sec:allyouneed}

Here we will review some basic concepts of string theory and D-branes, which will enable us to establish an equivalence between IIB string theory in $AdS_5 \times S^5$ and $\sN=4$ SYM theory. Although some of the contents of this section are not indispensable to understand some of the subsequent chapters, they are important for building the reader's intuition about the AdS/CFT correspondence.

\subsection{Strings}

Unlike quantum field theory, which describes the dynamics of point particles, string theory is a quantum theory of interacting, relativistic one-dimensional objects. It is characterized by the string tension, $T_\mt{str}$, and by a dimensionless coupling constant, $\gs$, that controls the strength of interactions. It is customary to write the string tension in terms of a fundamental length scale $\ell_s$, called string length, as
\be
T_\mt{str} \equiv \frac{1}{2\pi \ap}  \, \qquad {\rm with} \qquad \ap \equiv \ls^2 \ .
\label{tension}
\ee
We  now describe the conceptual steps involved in the definition of the theory, in a first-quantized formulation, \ie we consider the dynamics of a single string propagating in a fixed spacetime. Although perhaps less familiar, an analogous first-quantized formulation also exists for point particles~\cite{Polyakov:1987ez,Casalbuoni:1974pj}, whose
second-quantized formulation is a quantum field theory. In the case of string theory, the corresponding second-quantized formulation is provided by string field theory, which contains an infinite number of quantum fields, one for each of the vibration modes of a single string. In this review we will not need to consider such
a formulation. For the moment we also restrict ourselves to closed strings --- we will discuss open strings in the next section.

A string will sweep out a two-dimensional worldsheet which, in the case of a closed string, has no boundary. We postulate that the action that governs the dynamics of the string is simply the area of this worldsheet. This is a natural generalization of the action for a relativistic particle, which is simply the length of its worldline. In order to write down the string action explicitly, we parametrize the worldsheet with local coordinates $\sigma^\alpha$, with $\alpha = 0, 1$. For fixed worldsheet time, $\sigma^0 = \mbox{const.}$, the coordinate $\sigma^1$ parametrizes the length of the string. Let $x^M$, with $M=0, \ldots, D-1$, be spacetime coordinates. The trajectory of the string is then described by specifying $x^M$ as a function of $\sigma^\alpha$. In terms of these functions, the two-dimensional metric $g_{\alpha\beta}$ induced on the string worldsheet has components
\be
g_{\alpha \beta} = \partial_\alpha x^M \partial_\beta x^N \, g_{MN} \,,
\label{induced}
\ee
where $g_{MN}$ is the spacetime metric. (If $x^M$ are Cartesian coordinates in flat spacetime then $g_{MN} = \eta_{MN} = \mbox{diag}(-+ \cdots +)$.) The action of the string is then given by
\be
S_\mt{str} = - T_\mt{str} \int d^2 \sigma \sqrt{- \det g}\,.
\label{string-action}
\ee

In order to construct the quantum states of a  single string, one needs to quantize this action. It turns out that the quantization imposes strong constraints on the spacetime one started with; not all spacetimes allow a consistent string propagation 
--- see e.g.~\cite{Green:1987sp}. For example, if we start with a $D$-dimensional Minkowski spacetime, then a consistent string theory exists only for $D=26$. Otherwise the spacetime Lorentz group becomes anomalous at the quantum level and the theory contains negative norm states.

Physically, different states in the spectrum of the two-dimensional worldsheet theory correspond to different vibration modes of the string. From the spacetime viewpoint, each of these modes appears as a particle of a given mass and spin. The spectrum typically contains a finite number of massless modes and an infinite tower of massive modes with masses of order $m_s \equiv \ls^{-1}$. A crucial fact about 
a closed string theory is that one of the massless modes is a particle of spin two, \ie a graviton. This is the reason why string theory is, in particular, a theory of quantum gravity. The graviton describes small fluctuations of the spacetime metric, 
implying that the fixed spacetime that we started with is actually dynamical. 

One can construct other string theories by adding degrees of freedom to the string worldsheet. The theory that will be of interest here is a supersymmetric theory of strings, the so-called type IIB superstring theory \cite{Green:1981yb,Schwarz:1982jn}, which can be obtained by adding two-dimensional worldsheet fermions to the action \eqn{string-action}. Although we will of course be interested in eventually breaking supersymmetry in order to obtain a dual description of QCD, it will be important that the underlying theory be supersymmetric, since this will guarantee the stability of our constructions. For a superstring, absence of negative-norm states requires the dimension of spacetime to be $D=10$. In addition to the graviton, the massless spectrum of IIB superstring theory includes
two scalars, a number of antisymmetric tensor fields, and
various fermionic partners as required by supersymmetry. One of the scalars, the so-called dilaton $\Phi$, will play an important role here. 

\begin{figure}
    \begin{center}
    	\includegraphics[width=0.7\textwidth]{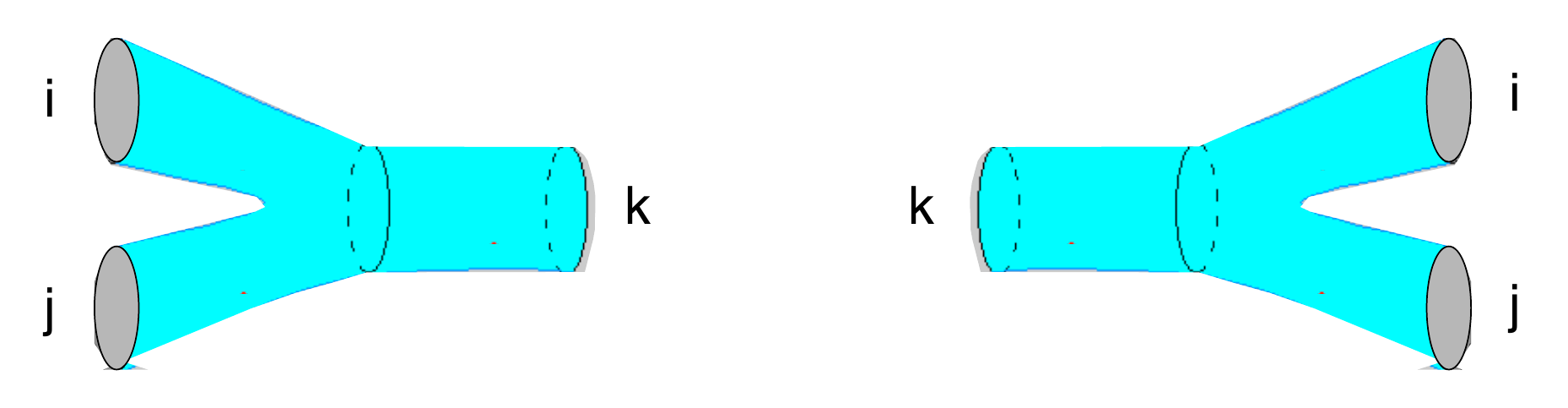}
    \end{center}
    \caption{\small `Geometrical' interactions in string theory: two strings
            in initial states $i$ and $j$ can join into one string in
            a state $k$ (left) or vice-versa (right).}
\label{strings}
\end{figure}

\begin{figure}
    \begin{center}
    	\includegraphics[width=0.9\textwidth]{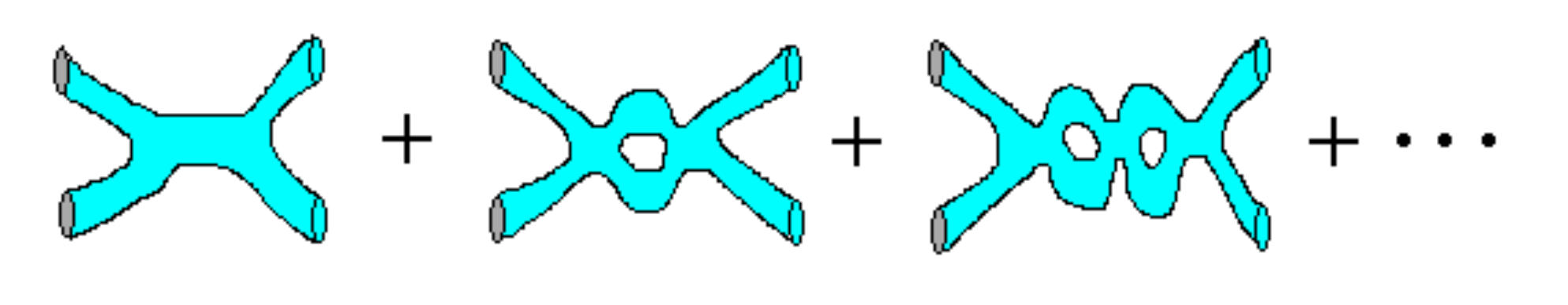}
    \end{center}
    \caption{\small Sum over topologies contributing to the two-to-two amplitude.}
\label{genus}
\end{figure}

Interactions can be introduced geometrically by postulating that two strings can join together and that one string can
split into two through a vertex of strength 
$g_s$ --- see Fig.~\ref{strings}.
Physical observables like scattering amplitudes can be found by summing over string propagations (including all possible splittings and joinings) between initial and final states. After fixing all the gauge symmetries on the string worldsheets, 
such a sum reduces to a sum over the topologies of two-dimensional surfaces, with contributions from surfaces of $h$ holes weighted by a factor $g_s^{2h-2}$. This is illustrated in Fig.~\ref{genus} for the two-to-two amplitude.

At low energies $E \ll m_s$, one can integrate out the massive string modes and obtain a low-energy effective theory for the massless modes. Since the massless spectrum of a closed string theory always contains a graviton, to second order in derivatives, the low energy effective action has the form of Einstein gravity coupled to other (massless) matter fields, \ie
 \be
S = \frac{1}{16 \pi \gten} \int d^{D} x \sqrt{-g} {\cal R} + \cdots \,,
\label{sugra}
\ee
where ${\cal R}$ is the Ricci scalar for the spacetime with $D$  spacetime dimensions and where 
the dots stand for additional terms associated with the rest of massless modes. For type IIB superstring theory, the full low-energy effective action at the level of two derivatives is given by the so-called IIB supergravity \cite{Schwarz:1983wa,Schwarz:1983qr}, a supersymmetric generalization of~\eqref{sugra}~(with $D=10$). The higher-order corrections to~\eqref{sugra} take the form of a double expansion, in powers of 
$\ap E^2$ from integrating out the massive stringy modes, and in powers of the string coupling $g_s$ from loop corrections.

\begin{figure}
\begin{center}
    \begin{tabular}{c@{\hspace{2cm}}c}
    	\includegraphics[width=0.30\textwidth]{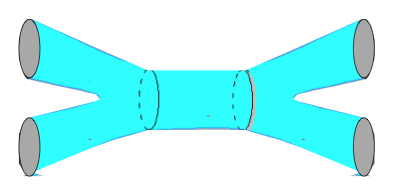} &
	\includegraphics[width=0.30\textwidth]{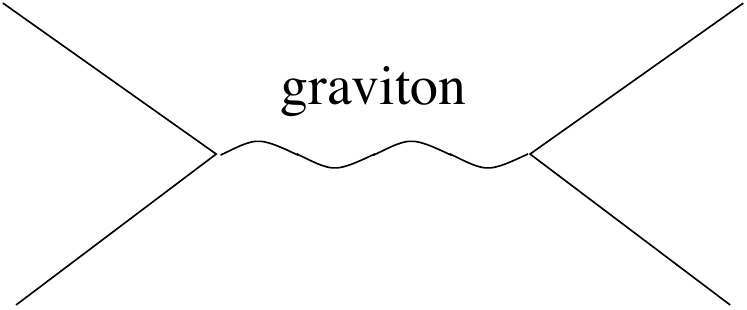} \\
        (a) & (b)
    \end{tabular}
\end{center}
\caption{\small (a) Tree-level contribution, of order $\gs^2$, to a two-to-two scattering process in string theory. The low-energy limit of this tree-level diagram must coincide with the corresponding field theory diagram depicted in (b), which is proportional to Newton's constant, $G$.}
\label{string-tree-graviton}
\end{figure}

We conclude this section by making two important comments. First, we note that the ten-dimensional Newton's constant $G$
in type IIB supergravity can be expressed in terms of the string coupling and the string length as
\be
16 \pi G = (2\pi)^7 \gs^2  \ls^8 \,.
\label{G}
\ee
The dependence on $\ls$ follows from dimensional analysis, since in $D$ dimensions Newton's constant has dimension (length)$^{D-2}$. The dependence on $\gs$ follows from considering two-to-two string scattering. The leading string theory diagram, depicted in 
Fig.~\ref{string-tree-graviton}(a), is proportional to $\gs^2$, since it is obtained by joining together the two diagrams of
Fig.~\ref{strings}. The corresponding diagram in supergravity is drawn in Fig.~\ref{string-tree-graviton}(b), and is proportional to $G$. The
requirement that the two amplitudes yield the same result at energies much lower than the string scale implies $G \propto \gs^2$.

Second, the string coupling constant $g_s$ is not a free parameter, but is given by the expectation value of the dilaton field $\Phi$ as $g_s = e^{\Phi}$.
As a result, $\gs$ may actually vary over space and time. Under these circumstances we may still speak of the string coupling constant, \eg in formulas like \eqn{G} or \eqn{tdp}, meaning the asymptotic value of the dilaton at infinity, $g_s = e^{\Phi_\infty}$.

\subsection{D-branes and gauge theories} \label{sec:dbran1}

\begin{figure}
    \begin{center}
	    \includegraphics[width=0.3\textwidth]{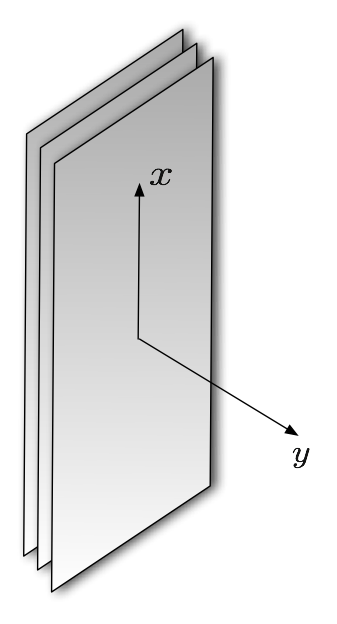}
    \end{center}
    \caption{\small Stack of D-branes.}
\label{brane}
\end{figure}

Perturbatively, string theory is a theory of strings. Non-perturbatively, the theory also contains a variety of higher-dimensional solitonic objects. D-branes \cite{Polchinski:1995mt} are a particularly important class of solitons. To be definite, let us consider a superstring theory (\eg~type IIA or IIB theory) in a 10-dimensional flat Minkowski spacetime, labeled by time 
$t \equiv x_0$ and spatial coordinates $x_1, \ldots, x_9$.  A D$p$-brane is then a ``defect'' where closed strings can break and open strings can end that occupies a $p$-dimensional subspace --- see Fig.~\ref{brane}, where the $x$-directions are parallel to the branes and the $y$-directions are transverse to them.  When closed strings break, they become open strings. The end points of the open strings can move freely along the directions of the D-brane, but cannot leave it by moving in the transverse directions. Just like domain walls or cosmic strings in a quantum field theory, D-branes are dynamical objects which can move around. A D$p$-brane then sweeps out a $(p+1)$-dimensional worldvolume in spacetime. D0-branes are particle-like objects, D1-branes are string-like, D2-branes are membrane-like, etc. Stable D$p$-branes in Type IIA superstring theory exist for $p=0,2,4,6,8$, whereas those in Type IIB have $p=1,3,5,7$  \cite{Polchinski:1995mt}.\footnote{D9-branes also exist, but additional consistency conditions must be imposed in their presence. We will not consider them here.} This can be seen in a variety of (not unrelated) ways, two of which are: (i) In these cases the corresponding D$p$-branes preserve a fraction of the supersymmetry of the underlying theory; (ii) In these cases the  corresponding D$p$-branes are the lightest states that carry a conserved charge. 

Introducing a D-brane adds an entirely new sector to the theory of closed strings, consisting of open strings whose endpoints must satisfy the boundary condition that they lie on the D-brane. Recall that in the case of closed strings we started with a fixed spacetime and discovered, after quantization, that the close string spectrum corresponds to dynamical fluctuations of the spacetime. An analogous situation holds for open strings on a D-brane. Suppose we start with a D$p$-brane extending in the $x^\mu = (x^0, x^1, \ldots, x^p)$ directions, with transverse directions labelled as 
$y^i = (x^{p+1}, \ldots, x^9)$. Then, after quantization, one obtains an open string spectrum which can be identified with fluctuations of the D-brane.

More explicitly, the open string spectrum consists of a finite number of massless modes and an infinite tower of massive modes with masses of order $m_s = 1/\ell_s$. For a single D$p$-brane, the massless spectrum consists of an Abelian gauge field 
$A_\mu (x)$, $\mu =0,1,\ldots, p$, $9-p$ scalar fields $\phi^i (x)$, $i=1,\ldots, 9-p$, and their superpartners. Since these fields are supported on the D-brane, they depend only on the $x^\mu$ coordinates along the worldvolume, but not on the transverse coordinates. The $9-p$ scalar excitations $\phi^i$ describe fluctuations of the D-brane in the transverse directions $y^i$, including deformations of the brane's shape and linear motions. They are the exact parallel of familiar collective coordinates for a domain wall or a cosmic string in a quantum field theory, and can be understood as the Goldstone bosons associated to the subset of translational symmetries spontaneously broken by the brane. The presence of a $U(1)$ gauge field $A_\mu (x)$ as part of collective excitations  lies at the origin of many fascinating properties of D-branes, which (as we will discuss below) ultimately lead to the gauge/string duality. Although this gauge field is less familiar in the context of quantum field theory solitons (see e.g.~\cite{Coleman:1975qj,Jackiw:1977yn,Rajaraman:1982is}), it can nevertheless be understood as a Goldstone mode associated to large gauge transformations spontaneously broken by the brane \cite{Callan:1991ky,Kaplan:1995cp,Adawi:1998ta}.

\begin{figure}[t!]
    \begin{center}
	    \includegraphics[width=0.7\textwidth]{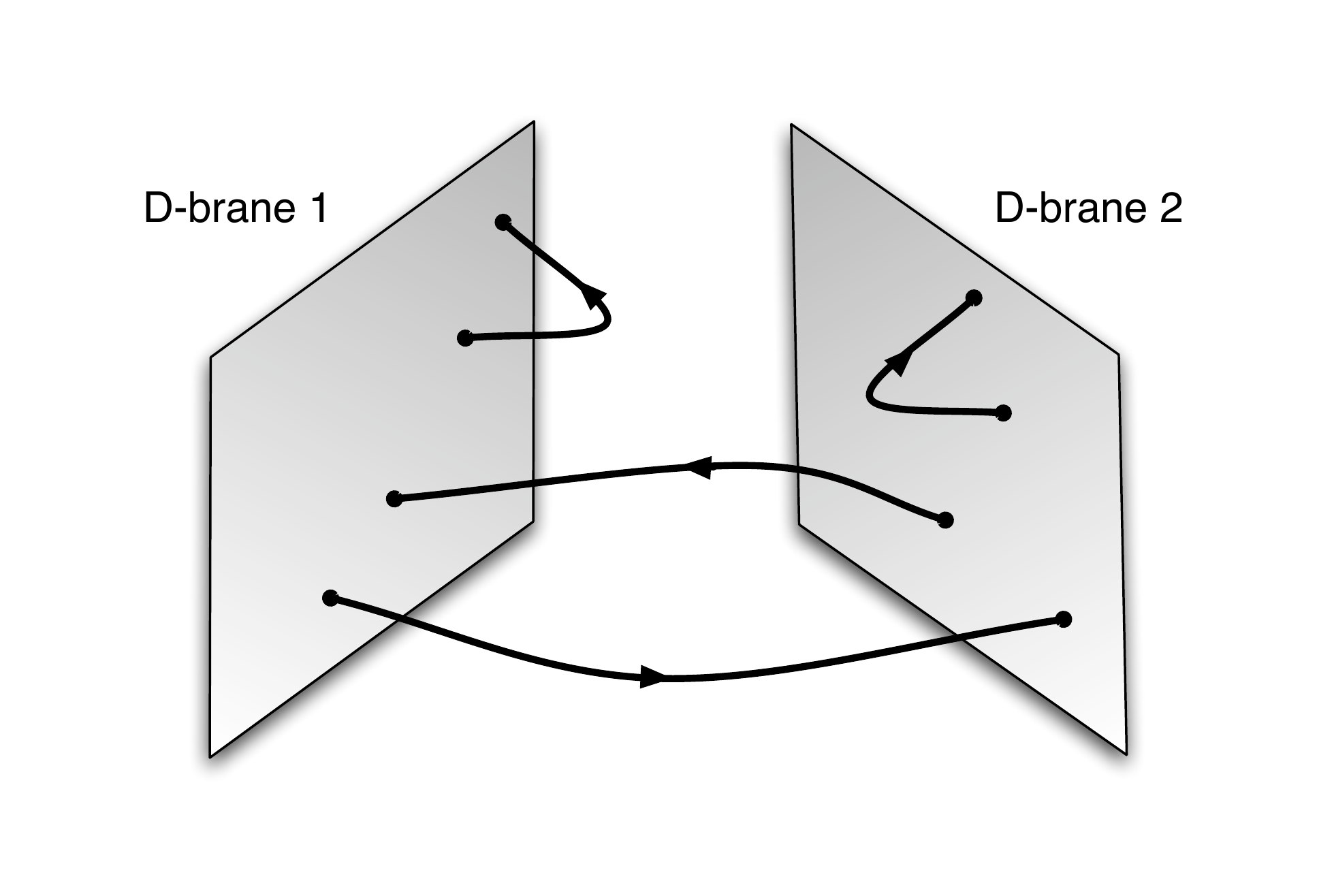}
    \end{center}
    \caption{\small Strings stretching between two D-branes.}
\label{d-branes}
\end{figure}

Another striking new feature of D-branes, which has no parallel in field theory, is the appearance of a non-Abelian gauge theory when multiple D-branes become close to one another \cite{Witten:1995im}. In addition to the degrees of freedom pertaining to each D-brane, now there are new sectors corresponding to open strings stretched between different branes. For example, consider two parallel branes separated from each other by a distance $r$, as shown in Fig.~\ref{d-branes}. Now there are four types of open strings, depending on which brane their endpoints lie on. The strings with both endpoints on the same brane give rise, as before, to two massless gauge vectors, which can be denoted by $(A_\mu)^1{_1}$ and  $(A_\mu)^2{_2}$,
where the upper (lower) numeric index labels the brane on which the string starts (ends). Open strings stretching between different branes give rise to two additional vector fields $(A_\mu)^1{_2}$ and  $(A_\mu)^2{_1}$, which have a mass given by the tension of the string times the distance between the branes, 
\ie~$m = r / 2 \pi \apr$.  These become massless when the branes lie on top of each other, $r =0$. In this case there are four massless vector fields altogether, $(A_\mu)^a{_b}$ with $a,b =1,2$, which precisely correspond to  the gauge fields of a non-Abelian $U(2)$ gauge group. Similarly, one finds that the $9-p$ massless scalar fields also become $2 \times 2$ matrices $(\phi^{i})^{a}{_b}$, which transform in the adjoint representation of the $U(2)$ gauge group. In the general case of $\nc$ parallel coinciding branes one finds a $U(\nc)$ multiplet of non-Abelian gauge fields with $9-p$ scalar fields in the adjoint representation of $U(\nc)$.
The low-energy dynamics of these modes
can be determined by integrating out the massive open string modes, and it turns out to be governed by a non-Abelian gauge theory \cite{Witten:1995im}.
To be more specific, let us consider $\nc$ D3-branes in type IIB theory.
The massless spectrum consists of a gauge field $A_\mu$, six scalar fields $\phi^i$, $i=1, \ldots, 6$ and four Weyl fermions, all of which are in the adjoint representation of $U(\nc)$ and can be written as 
$\nc \times \nc$ matrices. At the two-derivative level the low-energy effective action for these modes turns out to be precisely \cite{Witten:1995im} the $\sN=4$ super-Yang-Mills theory with gauge group $U(\nc)$ in (3+1) dimensions \cite{Brink:1976bc,Gliozzi:1976qd} (for reviews see e.g.~\cite{Kovacs:1999fx,D'Hoker:2002aw}), the bosonic part of whose Lagrangian can be written as
\be \label{n=4l}
\sL = -{1 \ov  \gym^2 } \Tr \left( {1 \ov 4} F^{\mu \nu}F_{\mu \nu}
 + \ha D_{\mu} \phi^i D^{\mu} \phi^i + [\phi^i, \phi^j]^2 \right) \,,
 \ee
with the Yang-Mills coupling constant given by
\be
\gym^2 = 4 \pi g_s  \, .
\label{YM}
\ee
Equation~\eqref{n=4l} is in fact the (bosonic part of the) most general renormalizable Lagrangian consistent with $\sN=4$ global supersymmetry. Due to the large number of supersymmetries the theory has many interesting properties, including the fact that the beta function vanishes exactly \cite{Avdeev:1980bh,Grisaru:1980nk,Sohnius:1981sn,Caswell:1980ru,Howe:1983sr,Mandelstam:1982cb,Brink:1982wv} (see section 4.1 of \cite{Polchinski:2010hw} for a one-paragraph proof). Consequently, the coupling constant is scale-independent and the theory is conformally invariant. 

Note that the $U(1)$ part of~\eqref{n=4l} is free and can be decoupled. Physically, the reason for this is as follows. Excitations of the overall, diagonal $U(1)$ subgroup of $U(\nc)$ describe motion of the branes' centre of mass, i.e.~rigid motion of the entire system of branes as a whole. Because of the overall translation invariance, this mode decouples from the remaining 
$SU(\nc) \subset U(\nc)$ modes that describe motion of the branes relative to one another. This is the reason why, as we will see, IIB strings in $AdS_5 \times S^5$ are dual to $\sN=4$ super-Yang-Mills theory with gauge group $SU(\nc)$.

\begin{figure}
    \begin{center}
    	\includegraphics[width=0.35\textwidth]{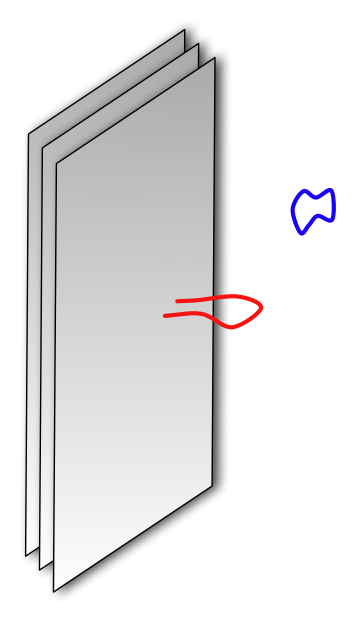}
    \end{center}
    \caption{\small Open and closed string excitations of the full system.}
\label{D3excitations}
\end{figure}

The Lagrangian \eqref{n=4l} receives higher-derivative corrections suppressed by $\apr E^2$. The full system also contains closed string modes (\eg~gravitons) which propagate in the bulk of the ten-dimensional spacetime (see Fig.~\ref{D3excitations}) and the full theory contains interactions between closed and open strings.  The strength of interactions of closed string modes with each other is controlled by Newton's constant $G$, so the dimensionless coupling constant at an energy $E$ is $GE^8$. This vanishes at low energies and so in this limit closed strings become non-interacting, which is essentially the statement that gravity is infrared-free. Interactions between closed and open strings are also controlled by the same parameter, since gravity couples universally to all forms of matter. Therefore at low energies closed strings decouple from open strings. We thus conclude that at low energies the interacting sector reduces to an $SU(\nc)$ $\caln =4$ SYM theory in four dimensions.

Before closing this section we note that, for a single D$p$-brane with constant $F_{\mu \nu}$ and $\partial_\mu \phi^i$, all higher-order 
$\apr$-corrections to~\eqref{n=4l} (or its $p$-dimensional generalizations) can be resummed exactly into the so-called Dirac-Born-Infeld (DBI) action \cite{Leigh:1989jq}
 \be
S_\mt{DBI} = - T_\mt{Dp} \int d^{p+1} x \, e^{-\Phi}
\sqrt{-\det \left( g_{\mu \nu} +  2\pi \ell_s^2 \, F_{\mu \nu} \right)} \,,
\label{DBI}
\ee
where
\be
T_\mt{Dp} = \frac{1}{\left( 2\pi  \right)^p g_s \ls^{p+1}} \,
\label{tdp}
\ee
is the tension of the brane, namely its mass per unit spatial volume. In this action, $\Phi$ is the dilaton and $g_{\mu \nu}$ denotes the induced metric on the brane. In flat space, the latter can be written more explicitly as
 \be \label{indM}
g_{\mu\nu} = \eta_{\mu\nu} + (2\pi \ell_s^2)^2
\partial_\mu \phi^i \partial_\nu \phi^i \,.
\ee
The first term in~\eqref{indM} comes from the flat spacetime metric along the worldvolume directions, and the second term arises from fluctuations in the transverse directions. Expanding the action \eqn{DBI} to quadratic order in $F$ and $\partial \phi$ one recovers the Abelian version of eqn.~\eqref{n=4l}. The non-Abelian generalization of the DBI action \eqn{DBI} is not known in closed form --- see e.g.~\cite{Tseytlin:1999dj} for a review. 
Corrections to \eqn{DBI} beyond the approximation of slowly-varying fields have been considered in \cite{Andreev:1988cb,Kitazawa:1987xj,Bachas:1999um,Green:2000ke}.

\subsection{D-branes as spacetime geometry}
\label{sec:dbran2}

Due to their infinite extent along the $x$-directions, the total mass of a D$p$-brane is infinite. However, the mass per unit $p$-volume, known as the tension, is finite and is given in terms of fundamental parameters by equation~\eqref{tdp}. The dependence of the tension on the string length is dictated by dimensional analysis. The inverse dependence on the coupling $g_s$ is familiar from solitons in quantum field theory (see e.g.~\cite{Coleman:1975qj,Jackiw:1977yn,Rajaraman:1982is}) and signals the non-perturbative nature of D-branes, since it implies that they become infinitely massive (even per unit volume) and hence decouple from the spectrum in the perturbative limit $g_s \to 0$. The crucial difference is that the D-branes'  tension scales as $1/g_s$ instead of the $1/g^2$ scaling that is typical of field theory solitons. This dependence can be anticipated based on the divergences of string perturbation theory \cite{Shenker:1990uf} and, as we will see, is of great importance for the gauge/string duality.

In a theory with gravity, all forms of matter gravitate. D-branes are no exception, and their presence deforms the spacetime metric around them. The spacetime metric sourced by $\nc$ D$p$-branes can be found by explicitly solving the supergravity equations of motion~\cite{Gibbons:1987ps,Garfinkle:1990qj,Horowitz:1991cd}. For illustration we again use  the example of D3-branes in type IIB theory, for which one finds:
  \be \label{d3bG}
ds^2 = H^{-1/2} \left( -dt^2 + dx_1^2 + dx_2^2 + dx_3^2 \right) +
H^{1/2} \left( dr^2 + r^2 d\Omega_5^2 \right) \,.
\ee
The metric inside the parentheses in the second term is just the flat metric in the $y$-directions transverse to the D3-branes written in spherical coordinates, with radial coordinate $r^2 =  y_1^2 + \cdots + y_{6}^2 $.
The function $H(r)$ is given by 
\be
H = 1 + \frac{R^4}{r^4} \,,
\label{defH}
\ee
where 
\be
R^4  =  4\pi g_s \nc \ell_s^4 \,.
\label{defR}
\ee

Let us gain some physical intuition regarding this solution. Since 
D3-branes extend along three spatial directions, their gravitational effect is similar to that of a point particle with mass $M \propto \nc T_{\rm D3}$ in the six transverse directions.
Thus the metric~\eqref{d3bG} only depends on the radial coordinate 
$r$ of the transverse directions. For $r \gg R$ we have $H \simeq 1$ and the metric reduces to that of flat space with a small correction proportional to
 \be \label{newtP}
 {R^4 \ov r^4} \sim {\nc g_s \ell_s^4 \ov r^4} \sim {G M \ov r^4} ,
 \ee
which can be interpreted as the gravitational potential due to a massive object of mass $M$ in six spatial dimensions.\footnote{Recall that a massive object of mass $M$ in $D$ spatial dimensions generates a gravitational potential $G M / r^{D-2}$ at a distance $r$ from its position.} Note that in the last step in eqn.~\eqref{newtP} we have used the fact that $G \propto g_s^2 \ell_s^8 $ and 
$M \propto \nc T_{\rm D3} \propto \nc /g_s \ell_s^{4}$ --- see~\eqref{G} and~\eqref{tdp}.

The parameter $R$ can thus be considered as the length scale characteristic of the range of the gravitational effects of $\nc$ 
D3-branes. These effects are weak for $r \gg R$, but become strong for 
$r \ll R$. In the latter limit, we may neglect the `1' in eqn.~\eqn{defR}, in which case the metric \eqn{d3bG} reduces to
\be
ds^2 = ds^2_{AdS_5} + R^2 d\Omega_{\it 5}^2 \,,
\label{metric10DK}
\ee
where
\be
ds^2_{AdS_5} = \frac{r^2}{R^2} \left( -dt^2 + dx_1^2 + dx_2^2 + dx_3^2 \right) +
\frac{R^2}{r^2} dr^2 \,
\label{AdSmetricK}
\ee
is the metric \eqn{Pads} of five-dimensional anti-de Sitter spacetime written in terms of $r=R^2/z$. We thus see that in the strong gravity region the ten-dimensional metric factorizes into $AdS_5 \times S^5$.

\begin{figure}
    \begin{center}
    	\includegraphics[width=0.7\textwidth]{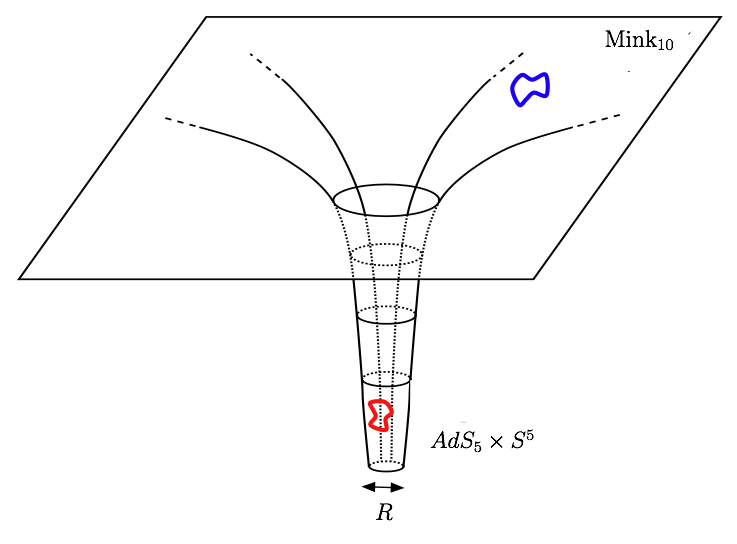}
    \end{center}
    \caption{\small Excitations of the system in the closed string  description.}
\label{AdSexcitations}
\end{figure}

We conclude that the geometry sourced by the D3-branes takes the form displayed 
in Fig.~\ref{AdSexcitations}: far away from the branes the spacetime is flat, ten-dimensional Minkowski space, whereas close to the branes a `throat' geometry of the form $AdS_5 \times S^5$ develops. The size of the throat is set by the length-scale $R$, given by~\eqref{defR}. As we will see, the spacetime geometry~\eqref{d3bG} can be considered as providing an {\it alternative} description of the D3-branes.

\section{The AdS/CFT conjecture} 
\label{sec:conj}

In the last two Sections we have seen two descriptions of D3-branes. In the description of 
Section~\ref{sec:dbran1}, which we will refer to as the open string description, D-branes correspond to a hyperplane in a {\it flat } spacetime. In this description the D-branes' excitations are open strings living on the branes, and closed strings propagate outside the branes --- see 
Fig.~\ref{D3excitations}.  

In contrast, in the description of Section~\ref{sec:dbran2}, which we will call the closed string description, D-branes correspond to a spacetime geometry in which only {\it closed} strings propagate, as displayed in fig.~\ref{AdSexcitations}. In this description there are no open strings. In this case the low-energy limit consists of focusing on excitations that have arbitrarily low energy with respect to an observer in the asymptotically flat Minkowski region. We have here two distinct sets of degrees of freedom, those propagating in the Minkowski region and those propagating in the throat --- see fig.~\ref{AdSexcitations}. In the Minkowski region the only modes that remain  are those of the massless ten-dimensional graviton supermultiplet. Moreover, at low energies these modes decouple from each other, since their interactions are proportional to $GE^8$. In the throat region, however, the whole tower of massive string excitations survives. This is because a mode in the throat must climb up a gravitational potential in order to reach the asymptotically flat region. Consequently, a closed string of arbitrarily high proper energy in the throat region may have an arbitrarily low energy as seen by an observer at asymptotic infinity, provided the string is located sufficiently deep down the throat. As we focus on lower and lower energies these modes become supported deeper and deeper in the throat, and so they decouple from those in the asymptotic region. We thus conclude that in the closed string description, the interacting sector of the system at low energies reduces to closed strings in \ads.

These two representations are tractable in different parameter regimes. For $g_s \nc \ll 1$, we see from eqn.~\eqref{defR} that $R \ll \ell_s$, \ie the radius characterizing the gravitational effect of the D-branes becomes small in string units, and closed strings feel a flat spacetime everywhere except very close to the hyperplane where the D-branes are located. In this regime the closed string description is not useful since one would need to understand sub-string-scale geometry. In the opposite regime, $g_s N \gg 1$, we find that $R \gg \ell_s$ and the geometry becomes weakly curved. In this limit the closed string description simplifies and essentially reduces to classical gravity. Instead, the open string description becomes intractable, since  
$g_s \nc$ controls the loop expansion of the theory and one would need to deal with strongly coupled open strings.  Note that both representations exist in both limiting regimes, and in between.

To summarize, the two descriptions of $\nc$ D3-branes that we have discussed and their low-energy limits are: 
\ben

\item  A hyperplane in a {\it flat } spacetime with open strings attached.
The low-energy limit is described by $\sN=4$ SYM theory~\eqref{n=4l} with a gauge group $SU(\nc)$.

\item A curved spacetime geometry~\eqref{d3bG} where only closed strings propagate. The low-energy limit is described by closed IIB string theory in $AdS_5 \times S^5$.

\een

It is natural to conjecture that these two descriptions are equivalent. 
Equating in particular their low-energy limits, we are led to conjecture that 
\be
\Big\{ \sN = 4 \; SU(\nc) \; {\rm SYM \; theory} \Big\} =
\Big\{  {\rm IIB \; string \; theory \; in} \; AdS_5 \times S_5 \Big\} .
\label{n=4Du}
 \ee
From equations~\eqref{YM} and~\eqref{defR} we find how the parameters of the two theories are related to one another:
\be  \label{impP}
g_s  = {\gym^2 \ov 4 \pi } , \qquad 
{R \ov \ell_s} = (\gym^2 \nc)^{1/4}  \ .
 \ee
One can also use the ten-dimensional Newton constant~\eqref{G} in place of $g_s$ in the first equation above and obtain equivalently
\be \label{NewtC1}
{G  \ov R^8}=  {\pi^4 \ov 2 \nc^2} , \qquad {R \ov \ell_s} = (\gym^2 \nc)^{1/4}  \,.
 \ee
Note that, in particular, the first equation in \eqn{impP} implies that the criterion that  $g_s N_c$ be large or small translates into the criterion that the gauge theory 't Hooft coupling $\lambda = \gym^2 \nc$ be large or small. Therefore the question of which representation of the D-branes is tractable becomes the question of whether the gauge theory is strongly or weakly coupled. We will come back to this in section~\ref{AdS/CFT}.

The discussion above relates string theory to $\sN =4$ SYM theory at zero temperature, as we were considering the ground state of the $\nc$ D3-branes. On the supergravity side this corresponds to the so-called extremal solution. The above discussion can easily be generalized to a nonzero temperature $T$ by exciting the degrees of freedom on the D3-branes to a finite temperature, which corresponds \cite{Gubser:1996de,Witten:1998zw} to the so-called non-extremal solution \cite{Horowitz:1991cd}. It turns out that the net effect of this is solely to modify the AdS part of the metric, replacing~\eqn{AdSmetricK} by
\be
ds^2 = \frac{r^2}{R^2} \left( - f dt^2 + dx_1^2 + dx_2^2 + dx_3^2 \right) + \frac{R^2}{r^2 f} dr^2 \,,
\label{AdSfinite1}
\ee
where
\be
f(r) =  1 - \frac{r_0^4}{r^4} \,.
\ee
Equivalently, in terms of the $z$-coordinate of section~\ref{sec:motivating} we replace~\eqn{Pads} by
\be
ds^2 = \frac{R^2}{z^2} \left( - f dt^2 + dx_1^2 + dx_2^2 + dx_3^2 + \frac{dz^2}{f} \right) \,,
\label{AdSfinite1BIS}
\ee
where 
\be
f(z) =  1 - \frac{z^4}{z_0^4} \,.
\ee
These two metrics are related by the simple coordinate transformation $z=R^2/r$, and represent a black brane in AdS spacetime with a horizon located at $r=r_0$ or $z=z_0$ which extends in all three spatial directions of the original brane. As we will discuss in more detail in the next section, $r_0$ and $z_0$ are related to the temperature of the $\sN=4$ SYM theory as 
\be
r_0 \propto \frac{1}{z_0} \propto T \,.
\label{prop}
\ee
Thus we conclude that $\sN=4$ SYM theory at finite temperature is described by string theory in an AdS black brane geometry.

To summarize this section, we have arrived at a duality \eqn{n=4Du} of the type anticipated in section~\ref{sec:motivating}, that is, an equivalence between a conformal field theory with zero $\beta$-function and trivial RG-flow and string theory on a scale-invariant metric that looks the same at any $z$.  In the finite-temperature case, Eqn.~\eqn{prop} provides an example of the expected relationship between energy scale in the gauge theory, set in this case by the temperature, and position in the fifth dimension, set by the location of the horizon.

\chapter{General aspects of the duality}
\label{AdS/CFT}

\section{Gauge/gravity duality}
\label{sec:GaugeGravityDuality}

In the last section we outlined the string theory reasoning behind the equivalence \eqn{n=4Du} between $\sN = 4 \ SU(\nc)$ SYM  theory and type IIB string theory on $AdS_5 \times S^5$. $\sN=4$ SYM theory is the unique maximally supersymmetric gauge theory in $(3+1)$ dimensions, whose field content includes a gauge field $A_\mu$, six real scalars $\phi^i, i=1,\ldots 6$, and four Weyl fermions $\chi_a, a =1,\ldots 4$, all them in the adjoint representation of the gauge group.
The metric of $AdS_5 \times S^5$ is given by
 \be
ds^2 = ds^2_{AdS_5} + R^2 d\Omega_{\it 5}^2 \,,
\label{metric10D}
\ee
with
\be
ds^2_{AdS_5} = \frac{r^2}{R^2}\eta_{\mu \nu} dx^\mu d x^\nu  +
\frac{R^2}{r^2} dr^2 \sac r \in (0, \infty) \,.
\label{AdSmetric}
\ee
In the above equation $x^\mu = (t, \vec x)$, $\eta_{\mu \nu}$ is the Minkowski metric in four spacetime dimensions, and $d \Om_5^2$ is the metric on a 
unit five-sphere. The metric~\eqref{AdSmetric} covers the so-called `Poincar\'e patch' of a global AdS spacetime, and it is sometimes convenient to rewrite~\eqref{AdSmetric} using a new radial coordinate $z = R^2/r \in (0,\infty)$, in terms of which we have
\be
ds^2_{AdS_5} =  g_{MN} dx^M dx^N = \frac{R^2}{z^2} \left( \eta_{\mu \nu} dx^\mu d x^\nu  + dz^2\right)
,  \quad  x^M = (z, x^\mu) \,,
\label{poincare-z}
\ee
as used earlier in~\eqref{Pads}.

In~\eqref{poincare-z}, each constant-$z$ slice of $AdS_5$ is isometric to four-dimensional Minkwoski spacetime with $x^\mu$ identified as the coordinates of the gauge theory~(see also fig.~\ref{AdSpicture}). As $z \ra 0$ we approach the  `boundary' of $AdS_5$. This is a boundary in the conformal sense of the word but not in the topological sense, since the prefactor ${R^2 /z^2}$ in~\eqref{poincare-z} approaches infinity there. Although this concept can be given a precise mathematical meaning, we will not need these details here. As motivated in Section~\ref{geom} it is natural to imagine that the Yang-Mills theory lives at the boundary of $AdS_5$. For this reason, below we will often refer to it as the boundary theory.  As $z \to \infty$, we approach the so-called Poincar\'e horizon, at which the prefactor ${R^2 / z^2}$ and the determinant of the metric go to zero.

\subsection{UV/IR connection} \label{sec:IRUV}

Due to the warp factor ${R^2 / z^2}$ in front of the Minkowski metric in~\eqref{poincare-z}, energy and length scales along  Minkowski directions in $AdS_5$ are related to those in the gauge theory by a $z$-dependent rescaling. More explicitly, consider an object with energy 
$\eym$ and size $\dym$ in the gauge theory. These are the energy and the size of the object measured in units of the coordinates $t$ and $\vec x$. From~\eqref{poincare-z} we see that the corresponding proper energy $E$ and proper size $d$ of this object in the bulk are 
\be \label{relbd}
d = {R \ov z} \, \dym \sac E = {z \ov R} \, \eym \,,
\ee
where the second relation follows from the fact that the energy is conjugate to time, and so it scales with the opposite scale factor than $d$. We thus see that physical processes in the bulk with identical proper energies but occurring at different radial positions correspond to different gauge theory processes with energies that scale as $E_\mt{YM} \sim 1/z$. In other words, a gauge theory process with a characteristic energy $\eym$ is associated with a bulk process localized at $z \sim 1/\eym$~\cite{Maldacena:1997re,Susskind:1998dq,Peet:1998wn}. 
This relation between the radial direction $z$ in the bulk and the energy scale of the boundary theory makes concrete the heuristic discussion of Section~\ref{geom} that led us to identify the $z$-direction with the direction along the renormalization group flow of the gauge theory. In particular the high-energy (UV) limit  $E_\mt{YM} \to \infty$ corresponds to $z \to 0$, i.e. to the near-boundary region, while the low-energy (IR) limit $E_\mt{YM} \to 0$ corresponds to $z \to \infty$, i.e. to the near-horizon region. 

In a conformal theory, there exist excitations of arbitrarily low energies. This is reflected in the bulk in the fact that the geometry extends all the way to $z \to \infty$. As we will see in Section~\ref{sec:conf}, for a confining theory with a mass gap $m$, the geometry ends smoothly at a finite value $z_0 \sim 1/m$. Similarly, at a finite temperature $T$, which provides an effective IR cutoff, the spacetime will be cut off by an event horizon at a finite $z_0 \sim {1 /T}$ (see Section~\ref{sec4.1.4}).

\subsection{Strong coupling from gravity} \label{sec:coupling}

$\sN=4$ SYM theory is a scale-invariant theory characterized by two parameters: the Yang-Mills coupling $\gym$ and the number of color $\nc$. The theory on the right hand side of~\eqref{n=4Du} is a quantum gravity theory in a maximally symmetric spacetime which is characterized by the Newton constant $G$ and the string scale $\ell_s$
in units of the curvature radius $R$. The relations between these parameters are given by~\eqref{NewtC1}. Recalling that 
$G \sim \ell_p^8$, with $\ell_p$ the Planck length, these relations imply
 \be  \label{NewtC}
{\ell_p^8  \ov R^8} \propto  {1 \ov \nc^2} \sac 
{\ell_s^2 \ov R^2} \propto \frac{1}{\sqrt{\lambda}} \,,
\ee
where $\lam=\gym^2 \nc$ is the 't Hooft coupling and we have only omitted purely numerical factors. 

The full IIB string theory on $AdS_5 \times S^5$ is rather complicated and right now a systematic treatment of it is not available. However, as we will explain momentarily, in the limit
\be \label{larN}
 {\ell_p^8  \ov R^8} \ll 1  \sac  {\ell_s^2 \ov R^2} \ll 1
\ee
the theory dramatically simplifies and can be approximated by classical supergravity, which is essentially Einstein's general relativity coupled to various matter fields. An immediate consequence of the relations~\eqref{NewtC} is that the limit~\eqref{larN} corresponds to
\be
\nc \gg 1 \sac \lam \gg 1 \,.
\ee
Equation~\eqref{n=4Du} then implies that the planar, strongly coupled limit of the SYM theory can be described using just classical supergravity.

Let us return to why string theory simplifies in the limit~\eqref{larN}. Consider first the requirement $\ell_s^2 \ll R^2$. This can be equivalently rewritten as 
$m_s^2 \gg  {\cal R}$ or as $\tf \gg  {\cal R}$, where ${\cal R} \sim 1/R^2$ is the typical curvature scale of the space where the string is propagating. The condition $m_s^2 \gg  {\cal R}$ means that one can omit the contribution of all the massive states of {\it microscopic} strings in low-energy processes. In other words, only the massless modes of microscopic strings, i.e.~the supergravity modes, need to be kept in this limit. This is tantamount to treating these strings as point-like particles and ignoring their extended nature, as one would expect from the fact that their typical size, $\ell_s$, is much smaller than the typical size of the space where they propagate, $R$. The so-called $\ap$-expansion on the string side (with $\apr = \ell_s^2$), which incorporates stringy effects associated with the finite length of the string in a derivative expansion, corresponds on the gauge theory side to an expansion around infinite coupling  in powers of $1/\sqrt{\lambda}$.

The extended nature of the string, however, cannot be ignored in all cases. As we will see in the context of the Wilson loop calculations of Section \ref{sec:Wilson} and in many other examples in Section~\ref{sec:Section6}, the description of certain physical observables requires one to consider long, {\it macroscopic} strings whose typical size is much larger than $R$ --- for example, this happens when the string description of such observables involves non-trivial boundary conditions on the string. In this case the full content of the second condition in \eqref{larN} is easily understood by rewriting it as $\tf \gg  {\cal R}$. This condition says that the tension of the string is very large compared to the typical curvature scale of the space where it is embedded, and therefore implies that fluctuations around the classical shape of the string can be neglected. These long strings can still break and reconnect, but in between such processes their dynamics is completely determined by the Nambu-Goto equations of motion. In these cases, the $\ap$-expansion (that is, the expansion in powers of $1/\sqrt{\lambda}$) incorporates stringy effects associated with fluctuations of the string that are suppressed at $\lambda\rightarrow\infty$ by the tension of the string becoming infinite in this limit. From this viewpoint, the fact that the massive modes of microscopic strings can be omitted in this limit is just the statement that string fluctuations around a point-like string can be neglected. 


Consider now the requirement $\ell_p^8 \ll R^8$. Since the ratio $\ell_p^8/R^8$ controls the strength of quantum gravitational fluctuations, in this regime we can ignore quantum fluctuations of the spacetime metric and talk about a fixed spacetime like $AdS_5 \times S^5$. The quantum gravitational corrections can be incorporated in a power series in $\ell_p^8/R^8$, which corresponds to the $1/\nc^2$ expansion in the gauge theory.  Note from (\ref{impP}) that taking the $N_c\rightarrow\infty$ limit at fixed $\lambda$ corresponds to taking the string coupling $g_s\rightarrow 0$, meaning that quantum corrections corresponding to loops of string breaking off or reconnecting are suppressed in this limit.


In summary, we conclude that the strong-coupling limit in the gauge theory suppresses the stringy nature of the dual string theory, whereas the large-$\nc$ limit suppresses its quantum nature. When both limits are taken simultaneously the full string theory reduces to a classical gravity theory with a finite number of fields.

Given that the $S^5$ factor in~\eqref{metric10D} is compact, it is often convenient to express a 10-dimensional field in terms of a tower of fields in $AdS_5$ by expanding it in terms of  harmonics on $S^5$. For example, the expansion of a scalar field $\phi (x, \Om)$ can be written schematically as
 \be
 \phi (x, \Om) = \sum_\ell \phi_\ell (x) Y_\ell (\Om) \,,
 \ee
 where $x$  and $\Om$ denote coordinates in $AdS_5$ and $S^5$ respectively, and $Y_l (\Om)$ denote the spherical harmonics on $S^5$.  Thus, for many purposes (but not all) the original duality~\eqref{n=4Du} can also be considered as the equivalence of $\sN=4$ SYM theory (at strong coupling) with a gravity theory in $AdS_5$ only. This perspective is very useful in two important aspects. First, it makes manifest that the duality~\eqref{n=4Du} can be viewed as an explicit
realization of the holographic principle mentioned in Section~\ref{geom}, with the bulk spacetime being $AdS_5$ and the boundary being four-dimensional Minkowski spacetime. Second, as we will mention in Section~\ref{sec:othgen}, this helps to give a unified treatment of many different examples of the gauge/gravity duality. In most of this review we will adopt this five-dimensional perspective and work only with fields in $AdS_5$.

After dimensional reduction on $S^5$, the supergravity action can be written as 
\be 
S = {1 \ov 16 \pi G_5} \int d^5 x \, \le[\sL_{\rm grav} + \sL_{\rm matt} \ri] \,,
\label{effective}
\ee
where 
\be
\sL_{\rm grav} = \sqrt{-g} \le(\sR + {12 \ov R^2} \ri) \
\ee
is the Einstein-Hilbert Lagrangian with a negative cosmological constant \mbox{$\Lam = - {6 / R^2}$} and $\sL_{\rm matt}$ is the Lagrangian for matter fields. In the general case, the latter would include the infinite towers $\phi_\ell(x)$ coming from the expansion on the $S^5$. The metric~\eqref{poincare-z} is a maximally-symmetric solution of the equations of motion derived from the action~\eqn{effective} with all matter fields set to zero.

The relation between the effective five-dimensional Newton's constant $G_5$  and its ten-dimensional counterpart $G$ can be read off from the reduction of the Einstein-Hilbert term,
\be
{1 \ov 16 \pi G} \int d^5 x d^5 \Omega \, \sqrt{-g_\mt{10}} \, \sR_\mt{10} = 
{R^5 \Omega_5 \ov 16 \pi G_5} \int d^5 x  \, \sqrt{-g_\mt{5}} \, \sR_\mt{5} + \cdots \,,
\ee
where $\Omega_5 = \pi^3$ is the volume of a unit $S^5$. This implies
\be
G_5 = {G \ov \Omega_5} = {G \ov \pi^3 R^5}  \sac \mbox{i.e.} \quad
{G_5 \ov R^3}  =  {\pi \ov 2 \nc^2} \,,
\label{G5}
\ee
where in the last equation we made use of~\eqref{NewtC1}.

\subsection{Symmetries} \label{sec:symm}

Let us now examine the symmetries on both sides of the correspondence. The $\caln =4$ SYM theory is invariant not only under dilatations but under $\mbox{Conf}(1,3) \times SO(6)$. The first factor is the conformal group of
four-dimensional Minkowski space, which contains the Poincar\'e group, the dilatation symmetry generated by $D$, and four special conformal transformations whose generators we will denote by $K_\mu$. The second factor is the R-symmetry of the theory under which the $\phi^i$ in~\eqref{n=4l} transform as a vector. In addition, the theory is invariant under sixteen ordinary or `Poincar\'e' supersymmetries, the fermionic superpartners of the translation generators $P_\mu$, as well as under sixteen special conformal supersymmetries, the fermionic superpartners of the special conformal symmetry generators $K_\mu$.

The string side of the correspondence is of course invariant under the group of  diffeomorphisms, which are gauge transformations. The subgroup of these consisting of large gauge transformations that leave the asymptotic (i.e.~near the boundary) form of the metric invariant is precisely $SO(2,4) \times SO(6)$. The first factor, which is isomorphic to $\mbox{Conf}(1,3)$, corresponds to the isometry group of $AdS_5$, and the second factor corresponds to the isometry group of $S^5$. As usual, large gauge transformations must be thought of as global symmetries, so we see that the bosonic global symmetry groups on both sides of the correspondence agree. In more detail, the Poincar\'e group of four-dimensional Minkowski spacetime is realized inside $SO(2,4)$ as transformations that act separately on each of the 
constant-$z$ slices in~\eqref{poincare-z} in an obvious manner. The dilation symmetry of Minkowski spacetime is realized in $AdS_5$ as the transformation $(t, \vec x) \to \Lambda (t, \vec x)$, $z \to \Lambda z$ (with $\Lambda$ a positive constant), which indeed leaves the metric~\eqref{poincare-z} invariant. The four special conformal transformations of Minkowski spacetime are realized in a slightly more involved way as isometries of $AdS_5$. 
 
An analogous statement can be made for the fermionic symmetries. $AdS_5 \times S^5$ is a maximally supersymmetric solution of type IIB string theory, and so it possesses thirty-two Killing spinors which generate fermionic isometries. These can be split into two groups that match those of the gauge theory.\footnote{In both boundary and bulk, bosonic and fermionic symmetries combine together to form a supergroup $SU(2,2|4)$.}

We therefore conclude that the global symmetries are the same on both sides of the duality. It is important to note, however, that on the gravity side the global symmetries arise as large gauge transformations. In this sense there is a correspondence between global symmetries in the gauge theory and gauge symmetries in the dual string theory. This is an important general feature of all known gauge/gravity dualities, to which we will return below after discussing the field/operator correspondence. It is also consistent with the general belief that the only conserved charges in a theory of quantum gravity are those associated with global symmetries that arise as large gauge transformations.

\subsection{Matching the spectrum: the field/operator correspondence}
\label{sec:MatchingSpectrum}

We now consider the mapping between the spectra of two theories. To motivate the main idea, we begin by recalling that the SYM coupling constant $\gym^2$ is identified (up to a constant) with the string coupling constant $g_s$. As discussed below~\eqref{G}, in string theory this is given by $g_s=e^{\Phi_\infty}$, where $\Phi_\infty$ is the value of the dilaton at infinity, in this case at the AdS boundary ($\partial$AdS). This suggests that deforming the gauge theory by changing the value of a coupling constant corresponds to changing the value of a bulk field at $\partial$AdS. More generally, one may imagine deforming the gauge theory action as
\be \label{source}
S \ra S + \int d^4 x \, \phi(x) {\cal O} (x) \,,
\ee
where ${\cal O} (x)$ is a local, gauge-invariant operator and $\phi(x)$ is a possibly point-dependent coupling, namely a source. If $\phi(x)$ is constant, then the deformation above corresponds to simply changing the coupling for the operator ${\cal O} (x)$. The example of $\gym$ suggests that to each possible source $\phi(x)$ for each possible local, gauge-invariant operator ${\cal O} (x)$ there must correspond a dual bulk field $\Phi (x,z)$ (and vice-versa) such that its value at the AdS boundary may be identified with the source, namely:
\be
\phi(x) = \left. \Phi \right|_{\partial AdS} (x) =  \lim_{z \ra 0} \Phi (x,z)  \,.
\label{identification}
\ee
As will be discussed around Eqn.~\eqref{genid}, Eqn.~\eqref{identification} is correct for a massless field $\Phi$, but it must be generalized for a massive field.

This one-to-one map between bulk fields in AdS and local, gauge-invariant operators in the gauge theory is known as the field/operator correspondence. The field and the operator must have the same quantum numbers under the global symmetries of the theory, but there is no completely general and systematic recipe to identify the field dual to a given operator. Fortunately, an additional restriction is known for a very important set of operators in any gauge theory: conserved currents associated to global symmetries, such as the  $SO(6)$ symmetry in the case of the $\sN =4 $ SYM theory. The source $A_\mu (x) $ coupling to a conserved current $J^\mu (x)$ as
\be
\int d^4 x \, A_\mu(x) J^\mu (x) 
\ee
may be thought of as an external background gauge field, and from~\eqref{identification} we can view it as the boundary value of a dynamical gauge field $A_\mu (x,z)$ in AdS. This identification is very natural given the discussion in Section~\ref{sec:symm} that continuous global symmetries in the boundary theory should correspond to large gauge transformations in the bulk. 

An especially important set of conserved currents in any 
translationally-invariant theory are those encapsulated in the energy-momentum tensor operator ${\cal T}^{\mu \nu}(x)$. The source $g_{\mu\nu} (x)$ coupling to ${\cal T}^{\mu \nu}(x)$ as
\be
\int d^4 x \, g_{\mu\nu} (x) {\cal T}^{\mu\nu} (x) 
\label{density}
\ee
can be interpreted as an external spacetime metric deformation. According to~\eqref{identification} we can then associate it with the boundary value of the bulk metric $g_{\mu \nu} (x,z)$. We thus reach the important conclusion that the dual of a translationally-invariant gauge theory, in which the energy-momentum tensor is conserved,  must involve dynamical gravity.

\subsection{Normalizable vs. non-normalizable modes and mass-dimension relation }
\label{sec:norma}
Having motivated the field/operator correspondence, we now elaborate on two important aspects of this correspondence: how the conformal dimension of an operator is related to properties of the dual bulk field~\cite{Gubser:1998bc,Witten:1998qj}, and how to interpret normalizable and non-normalizable modes of a bulk field in the boundary theory~\cite{Balasubramanian:1998sn,Balasubramanian:1998de}.

For illustration we will consider a massive bulk scalar field $\Phi$, dual to some scalar operator $\sO$ in the boundary theory.  Although our main interest is the case in which the boundary theory is four-dimensional, it is convenient to present the equations for a general boundary spacetime dimension $d$. For this reason we will work with a generalization of the AdS metric~\eqref{poincare-z} in which 
$x^\mu = (t, \vec x)$ denote coordinates of a $d$-dimensional Minkowski spacetime. 

The bulk action for $\Phi$ can be written as
 \be \label{quaDcBIS}
S  =   - {1 \ov 2 } \int dz \, d^{d} x \, \sqrt{-g} \,  \left[  g^{MN} \p_M \Phi \p_N \Phi + m^2 \Phi^2 \right] + \cdots \ .
 \ee
We have canonically normalized $\Phi$, and the dots stand for terms of order higher than quadratic. We have omitted these terms because they are proportional to positive powers Newton's constant, and are therefore suppressed by positive powers of $1/\nc$. 

Since the bulk spacetime is translationally invariant along the 
$x^\mu$-directions, it is convenient to introduce a Fourier decomposition in these directions by writing\footnote{For notational simplicity we will use the same symbol to denote a function and its Fourier transform, distinguishing them only through their arguments.}
  \be \label{FF1}
 \Phi (z,x^\mu) = \int {d^d k \ov (2 \pi)^d}  \, e^{i k \cdot x} \, 
 \Phi (z,k^\mu)  \,,
 \ee
where $k \cdot x \equiv \eta_{\mu \nu }k^\mu x^\mu$ and 
$k^\mu \equiv (\om, \vec k)$, with $\omega$ and $\vec k$ the energy and the spatial momentum, respectively. In terms of these Fourier modes the equation of motion for $\Phi$ derived from the action~\eqn{quaDcBIS} is 
 \be \label{eom1A}
z^{d+1} \p_z (z^{1-d} \p_z \Phi) - {k^2 z^2}  \Phi - m^2 R^2 \Phi = 0 \sac k^2 = -\omega^2 + {\vec k}^2  \,.
 \ee
Near the boundary $z \to 0$, the second term in~\eqref{eom1A} can be neglected and the equation can be readily solved to find the general asymptotic form of the solution:
 \be \label{asumA}
 \Phi (z,k) \approx A (k) \, z^{d-\Delta} + B(k) \, z^{ \Delta} 
 \qquad \mbox{as } z \to 0 \,,
 \ee
 where 
  \be \label{dimObis}
  \Delta = {d \ov 2} + \nu \sac
 \qquad \nu = \sqrt{m^2 R^2+ {d^2 \ov 4}} \,.
 \ee
The integration `constants' $A$ and $B$ actually acquire a dependence on $k$ through the requirement that the solution be regular in the interior  of AdS, i.e.~for all $z>0$. Fourier-transforming~\eqref{asumA} back into coordinate space, we then find
 \be \label{asumC}
 \Phi (z,x) \approx A (x) \, z^{d-\Delta} + B(x) \, z^{\Delta} 
 \qquad \mbox{as } z \to 0 \,.
 \ee
The exponents in~\eqref{asumC} are real provided
\be \label{BFB}
m^2 R^2 \geq -{d^2 \ov 4}  \,.
\ee
In fact, one can show that the theory is stable for any $m^2$ in the range~\eqref{BFB}, whereas for $m^2 R^2 < -{d^2 / 4}$ there exist modes that grow exponentially in time and the theory is unstable~\cite{Breitenlohner:1982bm,Breitenlohner:1982jf,Mezincescu:1984ev}. 
In other words, in AdS space a field with a negative mass-squared does not lead to an instability provided the mass-squared is not `too negative'. 
Equation~\eqref{BFB} is often called the Breitenlohner-Freedman (BF)  bound. In the stable region \eqn{BFB} one must still distinguish between the finite interval $-{d^2 / 4} \leq m^2 R^2 <  -{d^2 / 4} +1$ and the rest of the region, $m^2 R^2 \geq -{d^2 / 4} +1$. In the first case both terms in~\eqref{asumC} are normalizable with respect to the inner product 
\be
(\Phi_1, \Phi_2) = - i \int_{\Sig_t} dz d\vec x  \, 
\sqrt{-g} \, g^{tt} (\Phi_1^* \p_t \Phi_2 - \Phi_2 \p_t \Phi_1^* ) \,,
\ee
where $\Sig_t$ is a constant-$t$ slice. We will comment on this case at the end of this section. For the moment let us assume that 
$m^2 R^2 \geq -{d^2 / 4} +1$. In this case the first term in~\eqref{asumC}  is non-normalizable and the second term, which is normalizable, does not affect the leading boundary behavior. As motivated in the last subsection, the boundary value of a bulk field
 $\Phi$ should be identified with the source for the corresponding boundary operator $\sO$.
Since in~\eqref{asumC} the boundary behavior of $\Phi$ is controlled by $A(x)$, the presence of such a non-normalizable term should correspond to a deformation of the boundary theory of the form 
\be \label{source1}
S_\mt{bdry} \ra S_\mt{bdry} + \int d^d x \, \phi(x) {\cal O} (x) \sac \mbox{with } \phi (x) = A(x) \,. 
\ee
In other words, {\it the non-normalizable term determines the boundary theory Lagrangian}. In particular, we see that in order to obtain a finite source $\phi(x)$, equation~\eqref{identification} should be generalized for $\De \neq d$ to 
\be \label{genid}
\phi (x) = \left. \Phi \right|_{\partial AdS} (x) \equiv 
\lim_{z \to 0} z^{\De-d} \Phi (z,x)  \,.
\ee
In contrast, the normalizable modes are elements of the bulk Hilbert space. More explicitly, in the canonical quantization one expands $\Phi$ in terms of a basis of normalizable solutions of~\eqref{eom1A}, from which one can then build the Fock space and compute the bulk Green's  functions, etc. The equivalence between the bulk and boundary theories implies that their respective Hilbert spaces should be identified. Thus we conclude that {\it normalizable modes should be identified with states of the boundary theory}. This identification gives an important tool for finding the spectrum of low-energy excitations of a strongly coupled gauge theory, and we shall verify it below by showing that normalizable solutions of the wave equation for $\Phi$ with well-defined momentum $k$ give rise to poles in the two-point function $\langle \sO(q) \sO(-q)\rangle$ of its dual operator at momentum $q=k$. In the particular example at hand, one can readily see from~\eqref{eom1A} that, for a given $\vec k$, there is a continuous spectrum of $\om$, consistent with the fact that the boundary theory is scale invariant. 

Furthermore, as will be discussed in Section~\ref{sec:CORR} (and in Appendix~\ref{app:A}), the coefficient $B(x)$ of the normalizable term in~\eqref{asumC} can be identified with the expectation value of 
$\sO$ in the presence of the source $\phi (x) = A(x)$, namely
\be \label{bvev}
\vev{\sO(x)}_\phi = 2 \nu B(x) \,.
\ee
In the particular case of a purely normalizable solution, i.e. one with $A(x)=0$, this equation yields the expectation value of the operator in the undeformed theory.  

Equations (\ref{asumC}), \eqref{source1} and~\eqref{bvev} imply that 
$\Delta$, introduced in~\eqref{dimObis}, should be identified as the conformal dimension of the boundary operator $\sO$ dual to $\Phi$. Indeed, recall that a scale transformation of the boundary coordinates $x^\mu \to \Lambda x^\mu$ corresponds to the isometry 
$x^\mu \to \Lambda x^\mu, \;\; z \to \Lambda z$ in the bulk. Since $\Phi$ is a scalar field, under such an isometry it transforms as 
$\Phi' (\Lambda z, \Lambda x^\mu) = \Phi (z, x^\mu)$, which implies that the corresponding functions in the asymptotic form~\eqref{asumC} must  transform as $A' (\Lambda x^\mu) = \Lambda^{\Delta - d} A (x^\mu)$ and $B' (\Lambda x^\mu) = \Lambda^{-\Delta} B (x^\mu)$. This means that $A(x)$ and $B(x)$ have mass scaling dimensions $d-\Delta$ and $\Delta$, respectively. Eqs.~\eqref{source1} and~\eqref{bvev} are then consistent with each other and imply that $\sO(x)$ has mass scaling dimension $\De$. 

The relations~\eqref{dimObis} and the near-boundary behaviour \eqn{asumC} also apply to a bulk spin-two field. In particular, they apply to the five-dimensional metric\footnote{Although in some cases additional logarithmic terms must be included --- see e.g.~\cite{Skenderis:2002wp}.} \cite{Witten:1998qj}, for which $m^2=0$. Instead, in the case of a bulk $p$-form the dimension $\De$ of the dual operator is the largest root of the equation
\be
m^2 R^2 = (\De-p)(\De + p - d)
\ee
and the near-boundary fall-off becomes \cite{Witten:1998qj}
\be 
\Phi_{\mu_1 \cdots \mu_p} \approx A_{\mu_1 \cdots \mu_p}  z^{d-p-\De} + B_{\mu_1 \cdots \mu_p}  z^{\De-p} 
 \qquad \mbox{as } z \to 0 \,.
 \label{below}
\ee
A gauge field $A_\mu$ corresponds to the particular case $p=1, m^2=0$. These results show that the metric and the gauge field are dual to operators of dimensions $\Delta=d$ and $\Delta=d-1$, as expected for their respective dual operators, the stress-energy tensor and a vector current --- see Section \ref{sec:MatchingSpectrum}. An intuitive way to understand the near-boundary behaviours above is to note that the transverse traceless part of metric fluctuations behaves like a minimally coupled massless scalar --- see Section \ref{universalshear}. Consequently, the boundary behavior of this part, and by covariance of the rest of the metric (in Fefferman-Graham coordinates), is the same as that of a massless scalar. Instead, for a $p$-form one finds a massless scalar but with a spacetime-dependent coupling.

Before closing this section, let us return to the range $-{d^2 / 4} \leq m^2 <  -{d^2 / 4} +1$. We shall be brief because this is not a case that arises in later Sections.  Since in this case both terms in~\eqref{asumC} are normalizable, either one can be used to build the Fock space of physical states of the theory \cite{Breitenlohner:1982bm,Breitenlohner:1982jf}. This gives rise to two different boundary CFTs in which the dimensions of the operator $\sO(x)$ are $\Delta$ or $d-\Delta$, respectively \cite{Klebanov:1999tb}. It was later realized \cite{Witten:2001ua,Berkooz:2002ug} that even more general quantizations are possible in which the modes used to build the physical states have both $A$ and $B$ nonzero, with a relation between the two that ensures that the asymptotic symmetries of AdS are preserved \cite{Hertog:2004dr,Henneaux:2004zi}.

\section{Generalizations}

\subsection{Nonzero temperature and nonzero chemical potential}
\label{sec4.1.4}
As discussed in Section~\ref{sec:conj}, the same string theory reasoning giving rise to the equivalence~\eqref{n=4Du} can be generalized to nonzero temperature by replacing the pure AdS metric~\eqn{AdSmetric} by that of a black brane in $AdS_5$~\cite{Witten:1998zw}, Eqn.~\eqn{AdSfinite1}, which we copy here for convenience: 
 \be
ds^2 = \frac{r^2}{R^2} \left( - f dt^2 +d \vec x^2 \right) +
\frac{R^2}{r^2 f} dr^2 \, \sac f(r) = 1 - \frac{r_0^4}{r^4} \ .
\label{AdSfiniteR}
\ee
Equivalently, in terms of the $z$-coordinate, we replace Eqn.~\eqn{AdSmetric} by Eqn.~\eqn{AdSfinite1BIS}, i.e.
 \be
ds^2 = \frac{R^2}{z^2} \left( - f dt^2 +d \vec x^2 \right) +
\frac{R^2}{z^2 f} dz^2 \, \sac f(z) = 1 - \frac{z^4}{z_0^4} \ .
\label{AdSfiniteZ}
\ee
The metrics above have an event horizon at $r=r_0$ and $z= z_0$, respectively, and the regions outside the horizon correspond to 
$r \in (r_0, \infty)$ and $z \in (0, z_0)$. This generalization can also be directly deduced from~\eqref{n=4Du} as the black brane~\eqref{AdSfiniteR}--\eqn{AdSfiniteZ} is the only metric on the gravity side that satisfies the following properties: (i) it is asymptotically $AdS_5$; (ii) it is translationally-invariant along all the boundary directions and rotationally-invariant along the boundary spatial directions; (iii) it has a temperature and satisfies all laws of thermodynamics. It is therefore natural to identify the temperature and other thermodynamical properties  of~\eqref{AdSfiniteR}--\eqn{AdSfiniteZ} with those of the SYM theory at nonzero temperature.

We mention in passing that there is also a nice connection between the black brane geometry~\eqref{AdSfiniteR}--\eqn{AdSfiniteZ} and the thermal-field formulation of finite-temperature field theory in terms of real time. Indeed, the fully extended spacetime of the black brane has two boundaries. Each of them supports an identical copy of the boundary field theory which can be identified with one of the two copies of the field theory in the Schwinger-Keldysh formulation. The thermal state can also be considered as the maximally-entangled state of the two field theories. For more details see~\cite{Maldacena:2001kr,Herzog:2002pc}.

The Hawking temperature of the black brane can be calculated via the standard method~\cite{Gibbons:1976ue} (see Appendix~\ref{app:HawT} for details) of demanding that the Euclidean continuation of the metric~\eqn{AdSfiniteZ} obtained by the replacement $t \ra - i \te$,
\be
ds_\mt{E}^2 = \frac{R^2}{z^2} \left( f d\te^2 + dx_1^2 + dx_2^2 + dx_3^2 \right) +
\frac{R^2}{z^2 f} dz^2 \,,
\label{AdSeuclidean}
\ee
be regular at $z=z_0$. This  requires that $\te$ be periodically identified with a period $\beta$ given by
\be
\beta = \frac{1}{T} = \pi z_0 \,.
\label{eq:haw}
\ee
The temperature $T$ is identified with the temperature of the boundary SYM theory since, since $\te$ corresponds precisely to the Euclidean time coordinate of the boundary theory.
We emphasize here that while the Lorentzian spacetime~\eqref{AdSfiniteZ} can be extended beyond the horizon $z = z_0$, the Euclidean metric~\eqref{AdSeuclidean} only exists for $z \in (0, z_0)$ as the spacetime ends at $z=z_0$, and ends smoothly once the choice (\ref{eq:haw}) is made. 

For a boundary theory with a $U(1)$ global symmetry, like $\sN=4$ SYM theory, one can furthermore turn on a  chemical potential $\mu$ for the corresponding $U(1)$ charge. From the discussion of Section~\ref{sec:MatchingSpectrum}, this requires that the bulk gauge field $A_\mu$ which is dual to a boundary current $J_\mu$ satisfy the boundary condition 
\be 
\lim_{z \to 0} A_t = \mu \ .
\label{AmuChemPotential}
\ee
The above condition along with the requirement that the field $A_\mu$ should be regular at the horizon implies that there should be a radial electric field in the bulk, i.e.~the black hole is now charged. 
We will not write the metric of a charged black hole explicitly, as we will not use it in this review. For more details see~\cite{Chamblin:1999tk,Chamblin:1999hg,Gubser:1998jb,Cai:1998ji,Cvetic:1999ne,Cvetic:1999rb}. Similarly, in the case of theories with fundamental flavour introduced as probe D-branes, a baryon number chemical potential corresponds to an electric field on the branes \cite{Kim:2006gp,Horigome:2006xu,Kobayashi:2006sb,Mateos:2007vc,Nakamura:2006xk,Karch:2007pd,Yamada:2007ys,Bergman:2007wp,Davis:2007ka,Rozali:2007rx,Nakamura:2007nx,Ghoroku:2007re,Karch:2007br}.

\subsection{A confining theory} \label{sec:conf}

Although our main interest is the deconfined phase of QCD, in this section we will briefly describe a simple example of a duality for which the field theory possesses a confining phase~\cite{Witten:1998zw}. For simplicity we have chosen a model in which the field theory is three-dimensional, but all the essential features of this model extend to the string duals of more realistic confining theories in four dimensions.

We start by considering $\sN=4$ SYM theory at finite temperature. In the Euclidean description the system lives on $\RR^3 \times S^1$. The circle direction corresponds to the Euclidean time, which is periodically identified with period $\beta=1/T$. As is well known, at length scales much larger than $\beta$ one can effectively think of this theory as the Euclidean version of pure three-dimensional Yang-Mills theory. The reasoning is that at these scales one can perform a Kaluza-Klein reduction along the circle. Since the fermions of the $\sN=4$ theory obey antiperiodic  boundary conditions around the circle, their zero-mode is projected out, which means that all fermionic modes acquire a tree-level mass of order $1/\beta$. The scalars of the $\sN=4$ theory are periodic around the circle, but they acquire masses at the quantum level thorough their couplings to the fermions. The only fields that cannot acquire masses are the gauge bosons of the $\sN=4$ theory, since masses for them are forbidden by gauge invariance. Thus, at long distances the theory reduces to a pure Yang-Mills theory in three dimensions, which is confining and has a map gap. The Lorentzian version of the theory is simply obtained by analytically continuing one of the $\RR^3$ directions into the Lorentzian time. Thus, in this construction the `finite temperature' of the original four-dimensional theory is a purely theoretical device. The effective Lorentzian theory in three dimensions is at zero temperature. 

In order to obtain the gravity description of this theory we just need to implement the above procedure on the gravity side. We start with the Lorentzian metric~\eqn{metric10D}-\eqn{AdSmetric} dual to $\sN=4$ SYM at zero-temperature. Then we introduce a nonzero temperature by going to Euclidean signature via $t \ra - i \te$ and periodically identifying the Euclidean time. This results in the metric~\eqn{AdSeuclidean}. Finally, we analytically continue one of the $\RR^3$ directions, say $x_3$, back into the new Lorentzian time: $x_3 \rightarrow it$. The final result is the metric
\be
ds^2 = \frac{R^2}{z^2} \left( -dt^2 + dx_1^2 + dx_2^2 + f d\te^2 \right) +
\frac{R^2}{z^2 f} dz^2 \,.
\label{3dcon}
\ee
In this metric the directions $t, x_1, x_2$ correspond to the directions in which the effective three-dimensional YM theory lives. The direction $\te$ is now a compact spatial direction. Note that since the original metric~\eqref{AdSeuclidean} smoothly ends at $z=z_0$, so does~\eqref{3dcon}. This leads to a dramatic difference between the gauge theory dual to \eqn{3dcon} and the original $\sN=4$ theory: the fact that the radial direction smoothly closes off at $z=z_0$ introduces a mass scale in the boundary theory. To see this, note that the warp factor 
${R^2 / z^2}$ has a lower bound. Thus, when applying the discussion of Section~\ref{sec:IRUV} to~\eqref{3dcon}, $\eym$ in Eqn.~\eqref{relbd} will have a lower limit of order $M\sim {1 / z_0}$, implying that the theory develops a mass gap of this order. This can also be explicitly verified by  solving the equation of motion of a classical bulk field (which is dual to some boundary theory operator)  in the metric~\eqref{3dcon}: for any fixed $\vec k$ one finds a discrete spectrum of normalizable modes  with a mass gap of order $M$. (Note that since the size of the circle parametrized by $\te$ is proportional to ${1 / z_0}$, the mass gap is in fact comparable to the energies of Kaluza-Klein excitations on  the circle.) As explained in Section~\ref{sec:norma}, these normalizable modes can  be identified with the glueball states of the boundary theory.  

The fact that the gauge theory dual to the geometry \eqn{3dcon} is a confining theory is further supported by several checks, including the following two. First, analysis of the expectation value of a Wilson loop reveals an area law, as will be discussed in 
Section~\ref{sec:Wilson}. Second, the gravitational description can be used to establish that the theory described by \eqn{3dcon} undergoes a deconfinement phase transition at a temperature $T_c \sim M$ set by the mass gap, above which the theory is again described by a geometry with a black hole horizon~\cite{Witten:1998zw} (see \cite{Mateos:2007ay,Peeters:2007ab} for reviews).  

The above construction resulted in an effective confining theory in three dimensions because we started with the theory on the worldvolume of D3-branes, which is a four-dimensional SYM theory. By starting instead with the near horizon solution of a large number of non-extremal D4-branes, which describes a SYM theory in five dimensions, the above procedure leads to the string dual of a Lorentzian confining theory that 
at long distance reduces to a four-dimensional pure Yang-Mills theory~\cite{Witten:1998zw}. This has been used as the starting point of the Sakai-Sugimoto model for 
QCD~\cite{Sakai:2004cn,Sakai:2005yt}, which incorporates spontaneous chiral symmetry breaking and its restoration at high temperatures \cite{Aharony:2006da,Parnachev:2006ev}. For reviews on some of these topics see for example \cite{Mateos:2007ay,Peeters:2007ab}.

\subsection{Other generalizations} \label{sec:othgen}

In addition to~\eqref{n=4Du}, many other examples of gauge/string dualities are known in different spacetime dimensions (see e.g.~\cite{Aharony:1999ti} and references therein). These include theories with fewer supersymmetries and theories which are not scale invariant, in particular confining theories \cite{Polchinski:2000uf,Klebanov:2000hb,Maldacena:2000yy} (see e.g.~\cite{Aharony:2002up,Strassler:2005qs} for reviews).

For a $d$-dimensional conformal theory, the dual geometry on the gravity side contains a factor of AdS$_{d+1}$ and some other compact manifold.\footnote{Not necessarily in a direct product; the product may be warped.} When expanded in terms of the harmonics of the compact manifold, the duality again reduces to that between a $d$-dimensional conformal theory and a gravity theory in AdS$_{d+1}$. In particular, in
the classical gravity limit, this reduces to Einstein gravity in AdS$_{d+1}$ coupled to various matter fields with the precise spectrum of matter fields depending on the specific theory under consideration.
 For a non-conformal theory the dual geometry is in general more complicated. Some simple, early examples were discussed in~\cite{Itzhaki:1998dd}. If a theory has a mass gap it is always the case that the dual bulk geometry closes off at some finite value of $z_0$ as in the example of Section~\ref{sec:conf}. 

All known examples of gauge/string dual pairs share the following common features with~\eqref{n=4Du}: (i) the field theory is described by elementary bosons and fermions coupled to non-Abelian gauge fields whose gauge group is specified by some $\nc$; (ii) the string description reduces to classical (super)gravity in the large-$\nc$, strong coupling limit of the field theory. In this review we will use~\eqref{n=4Du} as our prime example for illustration purposes, but the discussion can be immediately applied to other examples including non-conformal ones.

\section{Correlation functions of local operators} 
\label{sec:CORR}

In this section we will explain how to calculate correlation functions of local gauge-invariant operators of the boundary theory in terms of the dual  gravity description. In this review we are mostly concerned with the properties of strongly coupled gauge theories at nonzero temperature. For this reason we are particularly interested in real-time retarded correlators, since these are relevant for determining linear responses, transport coefficients, spectral functions, etc. Below, we will first describe the general prescription for computing $n$-point Euclidean correlation functions, and then turn to the computation of finite-temperature retarded two-point functions.

\subsection{Euclidean correlators} \label{sec:eucl}

In view of the field/operator correspondence discussed in Sections~\ref{sec:MatchingSpectrum} and~\ref{sec:norma}, it is natural to postulate that the Euclidean partition functions of the two theories must agree upon the identification \eqn{genid}, namely that~\cite{Gubser:1998bc,Witten:1998qj}
\be
Z_\mt{CFT} \left[ \phi(x) \right] =
Z_\mt{string} \le[\left. \Phi \right|_{\partial AdS} (x) \ri] \,.
\label{Z}
\ee
Note that this equation makes sense because both sides of the equation are functionals of the same variables. Indeed, the most general partition function on the CFT side would include a source for each gauge-invariant operator in the theory, so one should think of $\phi(x)$ in Eqn.~\eqn{Z} as succinctly indicating the collection of all such sources. Since AdS has a boundary, in order to define the path integral over bulk fields in AdS one needs to specify a boundary condition for each field. The collection of all such boundary conditions is indicated by
$\left. \Phi \right|_{\partial AdS} (x)$ in Eqn.~\eqn{Z}.

The right-hand side of \eqn{Z} is in general not easy to compute, but it simplifies dramatically in the classical gravity limit~\eqref{larN}, in which it can be obtained using the saddle point approximation as
\be
Z_\mt{string} [\phi] \simeq   \exp \left( S^{\rm (ren)}[\Phi_c^{(E)}] \right)
\,,
\label{equi}
\ee
where we have absorbed a conventional minus sign into the definition of the Euclidean action which avoids having some additional minus signs in various equations below and in the analytic continuation to Lorentzian signature.

In this equation  $S^{\rm (ren)} [\Phi_c^{(E)}]$ is the renormalized on-shell classical supergravity action \cite{Henningson:1998gx,Balasubramanian:1999re,Myers:1999psa,Emparan:1999pm,Kraus:1999di,deHaro:2000xn,Bianchi:2001kw}, namely the classical action evaluated on a solution $\Phi_c^{(E)}$ of the equations of motion determined by the boundary condition
\be
\lim_{z \to 0}  \, z^{\De-d} \, \Phi_c^{(E)} (z, x) = \phi(x)  
\label{bc}
\ee
and by the requirement that it be regular everywhere in the interior of the  spacetime. The on-shell action needs to be renormalized because it typically suffers from infinite-volume (i.e.~IR) divergences due to the integration region near the boundary of AdS \cite{Witten:1998qj}. These divergences are dual to UV divergences in the gauge theory, consistent with the UV/IR correspondence. The procedure to remove these divergences on the gravity side is well understood and goes under the name of `holographic renormalization'.  Although it is an important ingredient of the gauge/string duality, it is also somewhat technical.  In Appendix~\ref{app:A} we briefly review it in the context of a two-point function calculation. 
The interest reader may consult some of the references in this paragraph, as well as the review \cite{Skenderis:2002wp} for details.

Corrections to Eqn.~\eqn{equi} can be included as an expansion in $\ap$ and $\gs$, which correspond to $1/\sqrt{\lam}$ and $1/\nc$ corrections in the gauge theory, respectively. Note that since the classical action~\eqref{effective} on the gravity side is  proportional to $1/G_5$, from Eqn.~\eqn{G5} we see that $S^{\rm (ren)}[\Phi_c^{(E)}] \sim \nc^2$, as one would expect for the generating functional of an $SU(\nc)$ SYM theory in the 
large-$\nc$ limit.

We conclude from Eqs. (\ref{Z}) and (\ref{equi}) that, in the large-$\nc$ and large-$\lam$ limit, connected correlation functions of the gauge theory are given simply by functional derivatives of the on-shell, classical gravity action:
\be
\label{eepp}
\vev{\sO (x_1) \ldots \sO (x_n)}
 = {\delta^n S^{\rm (ren)} [\Phi_c^{(E)}] \ov \delta \phi (x_1) \ldots \delta \phi (x_n)} \biggr|_{\phi =0} \,.
\ee
It is often convenient to consider the one point function of $\sO$ in the presence of the source $\phi$, which is given by
\be
\label{defexpO}
\langle \sO(x) \rangle_{\phi} = {\delta S^{\rm (ren)} [\Phi_c^{(E)}] \ov \delta \phi (x)} 
= \lim_{z \to 0} z^{d-\De} {\delta S^{\rm (ren)} [\Phi_c^{(E)}] \ov \delta \Phi_c^{(E)} (z,x)} \ .
\ee
A very important application of this formula arises when the operator in question is the gauge theory stress-energy tensor. In this case the dual five-dimensional field is the bulk metric and, as we discussed below eqn.~\eqn{below}, $\Delta=d$, so the equation above becomes
\begin{equation}
\label{stgenBIS}
\langle {\cal T}^{\mu \nu}(x) \rangle = 
\lim_{z \rightarrow 0} \frac{\delta S^{\rm (ren)} [g^{(E)}]}{\delta g_{\mu\nu}^{(E)}(x,z)}\ .
\end{equation}
At this point we must distinguish between a tensor density and a tensor. From eqn. \eqn{density} it is clear that ${\cal T}^{\mu \nu}$ is a tensor density. In order to construct a tensor from it we define 
\begin{equation}
\label{stgenBISBIS}
\langle T^{\mu \nu}(x) \rangle = \lim_{z \rightarrow 0} \frac{2}{\sqrt{-g^{(E)}(x,z)}}
 \frac{\delta S^{\rm (ren)} [g^{(E)}]}{\delta g_{\mu\nu}^{(E)}(x,z)}\ .
\end{equation}
Except for the dependence on the holographic coordinate $z$, the reader will probably recognize the right-hand side as a familiar definition of the stress tensor in classical field theory (see e.g.~\cite{landau2}). The square root of the metric in the denominator turns the result into a true tensor. This and the factor of 2 above can be fixed in a variety of ways, including the requirements that $T_{\mu\nu}$: agree with the known results in simple cases such as a free scalar field; be the appropriately-normalized generator of translations; be the appropriately-normalized stress-tensor that appears on the right-hand side of Einstein equations. 
Eqn. \eqn{stgenBISBIS} will play an important role in Section \ref{sec:AdSCFTDragWaves}.

In classical mechanics, it is well known that the variation of the action with respect to the boundary value of a field results in the canonical momentum $\Pi$ conjugate to the field, where the boundary in that case is usually a constant-time surface (see e.g.~\cite{landau}). In the present case the boundary is a constant-$z$ surface, but it is still useful to proceed by analogy with classical mechanics and to think of the derivative in the last term in \eqn{defexpO} as the renormalized canonical momentum conjugate to $\Phi_c^{(E)}$ evaluated on the classical solution: 
\be
\Pi^{\rm (ren)}_c (z,x) = {\delta S^{\rm (ren)} [\Phi_c^{(E)}] \ov \delta \Phi_c^{(E)} (z,x)} \,.
\ee
With this definition eqn.~\eqn{defexpO} takes the form
\be \label{canjr}
\langle \sO(x) \rangle_{\phi} = \lim_{z \to 0} z^{d-\De} \Pi^{\rm (ren)}_c (z,x) \,.
\ee
More explicitly, for a bulk scalar field $\Phi$ with action~\eqref{quaDcBIS} the corresponding $\Pi$ is given by
\be 
\Pi = -  g^{zz} \sqrt{-g} \, \p_z \Phi \,. 
\ee
We show in Appendix~\ref{app:A}  that this implies that
\be \label{imeq}
\langle \sO(x) \rangle_{\phi}  = 2  \nu B (x) 
\ee
when identifying $A (x)$ in~\eqref{asumC} with $\phi(x)$. In linear response theory the response in momentum space, i.e~the expectation value of an operator, is proportional to the corresponding source, and the constant of proportionality (for each momentum) is the two-point function of the operator: 
\be 
\vev{\sO( \om_E , \vec k)}_\phi = G_E (\om_E, \vec k) \phi (\om_E, \vec k) \,.
\ee
Thus equation~\eqref{imeq} immediately yields 
\be \label{GEex}
G_E (\om_E, \vec k) = {\vev{\sO( \om_E , \vec k)}_\phi \ov \phi (\om_E, \vec k)} = 
\lim_{z \to 0} z^{2 (d-\De)} {\Pi^{\rm (ren)}_c \ov \Phi_c^{(E)}} =
2 \nu {B (\om_E, \vec k) \ov A (\om_E, \vec k)} \,,
\ee
where $\om_E$ denotes Euclidean frequency. In Appendix~\ref{app:A}, we give an explicit evaluation of~\eqref{GEex}. For some early work on the evaluation of higher-point functions see Refs.~\cite{Freedman:1998tz,Liu:1998bu,Chalmers:1998xr}. Incidentally, Eqn.~\eqref{GEex} shows that the Euclidean correlator possesses a pole precisely at those frequencies 
for which $A (\om_E, \vec k)$ vanishes. In other words, the poles of the two-point function are in one-to-one correspondence with normalizable solutions of the equations of motion, as anticipated in the paragraph below Eqn.~\eqn{genid}.

\subsection{Real-time thermal correlators}
\label{sec:rtthc}

We now proceed to the prescription for calculating real-time correlation functions. 
We will focus our discussion on {\it retarded} two-point functions because of their important role in 
characterizing linear response.  Also, once the retarded function is known one can then use standard relations to obtain the other Green's functions. 

We start with the relation between the retarded and the Euclidean two-point
functions in momentum space, 
\be 
G_E (\om_E, \vec k) = G_R (i \om_E, \vec k) , \qquad \om_E > 0 \,,
\ee
or inversely 
\be \label{grana}
G_R (\om, \vec k) = G_E (- i (\om + i \ep), \vec k) \,.
\ee
If the Euclidean  correlation functions $G_E$ are known exactly, the retarded functions $G_R$ can then be obtained though the simple analytic continuation~\eqref{grana}. In most examples of interest, however, the Euclidean correlation functions can only be found numerically and analytic continuation to Lorentzian signature becomes difficult. Thus, it is important to develop techniques to calculate real-time correlation functions directly.
Based on an educated guess that passed several consistency checks, a prescription for calculating retarded two-point functions in Lorentzian signature was first proposed by Son and Starinets in Ref.~\cite{Son:2002sd}. Ref.~\cite{Herzog:2002pc} later justified the prescription and extended it to $n$-point functions. Here we will follow the treatment given 
in Refs.~\cite{Iqbal:2008by,Iqbal:2009fd}. For illustration we consider the retarded two-point function for a scalar operator $\sO$ at nonzero temperature, which can be obtained from the propagation of the dual scalar field $\Phi$ in the geometry of an AdS black hole. The action for $\Phi$ again takes the form~\eqref{quaDcBIS} with $g_{MN}$ now given by the black brane metric~(\ref{AdSfiniteZ}).

Before giving the prescription, we note that in Lorentzian signature one cannot directly apply the procedure summarized by Eqn.~\eqn{eepp} to obtain retarded functions. There are two immediate complications/difficulties. First, the Lorentzian black hole spacetime contains an event horizon and one also needs to impose appropriate boundary conditions there when solving the classical equation of motion for $\Phi$. Second, since partition functions are defined in terms of path integrals, the resulting correlation functions should be  time-ordered.\footnote{While it is possible to obtain Feynman functions this way, the procedure is quite subtle, since Feynman functions require imposing different boundary conditions for positive- and negative-frequency modes at the horizon and
the choices of positive-frequency modes are not unique in a black hole spacetime. The correct choice corresponds to specifying the so-called Hartle-Hawking vacuum. For details see~\cite{Herzog:2002pc}. In contrast, the retarded function does not depend on the choice of the bulk vacuum in the classical limit as the corresponding bulk retarded function is given by the commutator of the corresponding bulk field --- see Eqn.~\eqref{relreT}.} As we now review, both complications can be dealt with in a simple manner.

The idea is to analytically continue the Euclidean classical solution $\Phi_c^{(E)} (\om_E, \vec k)$, as well as Eqs.~\eqref{canjr}--\eqref{GEex}, to Lorentzian signature according to~\eqref{grana}. 
Clearly the analytic continuation of $\Phi_c^{(E)} (\om_E, \vec k)$,
\be \label{anasol}
\Phi_c (\om, \vec k) = \Phi_c^{(E)} (- i (\om + i \ep), \vec k) \,,
\ee 
solves the Lorentzian equation of motion. In addition, this solution obeys the in-falling boundary condition at the future event horizon of the black-brane metric~(\ref{AdSfiniteZ}). This property is important as it ensures that the retarded correlator is causal  and only propagates information forward in time. This is intuitive since we expect that, classically, information can fall into the black hole horizon but not come out, so the retarded correlator should have no out-going component.
Although it is intuitive, given its importance let us briefly verify that the in-falling boundary condition is satisfied. The Lorentzian equation of motion in momentum space for $\Phi_c$ in the black brane metric~(\ref{AdSfiniteZ}) takes the form
\be
\label{eeom2}
z^5 \p_z \Big[ z^{-3} f (z) \p_z \Phi \Big] + {\om^2 z^2 \ov f(z)} \Phi
- {\vec k^2 z^2}  \Phi - m^2 R^2 \Phi = 0 \,, 
\ee
where $\vec k^2 = \delta^{ij} k_i k_j$. The corresponding Euclidean equation is obtained by setting $\om = i \om_E$. Near the horizon $z \to z_0$, since $f \to 0$ the last two terms in (\ref{eeom2}) become negligible compared with the second term and can be dropped.
The resulting equation (with only the first two terms of (\ref{eeom2})) then takes the simple form
\bea
& {\rm Lorentzian:} \quad & \p_\xi^2 \Phi + \om^2 \Phi = 0 \,,\nonumber \\
& {\rm Euclidean:} \quad & \p_\xi^2 \Phi - \om^2_E \Phi = 0 \,,
\eea
in terms of a new coordinate 
\be
\xi \equiv \int^z {dz' \ov f (z')} \,.
\ee
Since $\xi \to + \infty$ as $z \to z_0$, in order for the Euclidean solution to be regular at the horizon we must choose the solution with the decaying exponential, i.e.~$\Phi_c^{(E)} (\om_E, \xi) \sim e^{- \om_E \xi}$. The prescription~\eqref{anasol} then yields $\Phi_c (\om, \xi) \sim e^{ i \om \xi}$. Going back to coordinate space we find that near the horizon
\be
\label{plane1}
\Phi_c  (t, \xi) \sim  e^{-i \om \le(t  - \xi \ri)} \,. 
\ee
As anticipated, this describes a wave propagating towards the direction in which 
$\xi$ increases, i.e.~falling into the horizon. (Had we chosen the opposite sign in the prescription \eqn{anasol} we would have obtained an out-going wave, as appropriate for the advanced correlator which obeys an out-going boundary condition at the past event horizon of the
metric (\ref{AdSfiniteZ}), and has no in-falling component.)

Given a Lorentzian solution satisfying the in-falling boundary condition at the horizon which can be expanded near the boundary according to~\eqref{asumA},  Eqs.~\eqref{canjr}--\eqref{GEex} can be analytically continued to Lorentzian signature to obtain an intrinsic Lorentzian prescription for the expectation value and the retarded two-point function of an operator. The former is given by 
\be 
\vev{\sO(\om, \vec k)}_{\phi} = \lim_{z \to 0} z^{d-\De} \Pi^{\rm (ren)}_c  = 2 \nu B (\om, \vec k)  \,,
\ee
with $\phi (\om, \vec k) = A(\om, \vec k)$, and the latter takes the form
\be
\label{RetB}
G_R (\om,\vec k) =  \lim_{z \to 0} z^{2 (d-\De)}{ \Pi^{\rm (ren)}_c  \ov \Phi_c (\om, \vec k)} =2 \nu {B (\om,\vec k) \ov A (\om,\vec k)} \,.
\ee 

For practical purposes, let us recapitulate here the main result of this section, namely the algorithmic procedure for computing the real-time, finite-temperature retarded two-point function of a local, gauge-invariant operator $\sO(x)$. This consists of the following steps: 
\ben
\item Identify the bulk mode $\Phi(x,z)$ dual to $\sO(x)$.
\item Find the Lorentzian-signature bulk effective action for $\Phi$  to quadratic order, and the corresponding linearized equation of motion in momentum space. 
\item Find a solution $\Phi_c (k, z)$ to this equation with the boundary conditions that the solution is in-falling at the horizon and behaves as 
\be
 \Phi_c (z,k) \approx A (k) \, z^{d-\Delta} + B(k) \, z^{ \Delta} 
 \ee
near the boundary ($z \to 0$), where $\De$ is the dimension of $\sO(x)$, $d$ is the spacetime dimension of the boundary theory and $A (k)$ should be thought of as an arbitrary source for $\sO(k)$. $B(k)$ is not an independent quantity but is determined by the boundary condition at the horizon and $A(k)$. 
\label{poi:3}
\item The retarded Green's function for $\sO$ is then given by
  \be \label{RestB}
 G_R (k) = 2 \nu {B (k) \ov A (k)}  \,,
 \ee
where $\nu$ is defined in Eqn.~\eqn{dimObis}.
\een
In Section~\ref{sec:BHMesonSpectrum} we will discuss in detail an example of a retarded correlator of two electromagnetic currents. 

Before closing this section, we note that an alternative way to compute boundary correlation functions which works in both Euclidean and Lorentzian signature is~\cite{Banks:1998dd}
\be
\vev{\sO (x_1) \cdots \sO (x_n)} = 
\lim_{z_i\to 0} (2 \nu z_1^\De) \cdots  (2 \nu z_n^\De)
\vev{\Phi(z_1,x_1) \cdots \Phi (z_n,x_n)}
\label{bbre}
\ee
where the correlator on the right-hand side is a correlation function in the bulk theory. 
In~\eqref{bbre} it should be understood that whatever ordering one wants to consider, it should be same on both sides.  For example, for the retarded two-point function $G_R$ of $\sO$
\be \label{relreT}
G_R (x_1-x_2) =  \lim_{z_1, z_2 \to 0} (2 \nu z_1^\De)  (2 \nu z_2^\De)\, 
\sG_R (z_1,x_1;z_2,x_2) \,,
\ee
where $\sG_R$ denotes the retarded Green's function of the bulk field $\Phi$.

\section{Wilson loops} 
\label{sec:Wilson}


The expectation values of Wilson loops
\begin{equation}
   W^r({\cal C}) ={\rm Tr}\, {\cal P} \exp \left[ i \int_{\cal C} dx^\mu\, A_\mu(x) \right]\, ,
   \label{2.1}
\end{equation}
are an important class of non-local observables in any gauge theory.
Here, $\int_{\cal C}$ denotes a line integral along the closed path ${\cal C}$,
$W^r({\cal C})$ is the trace of an $SU(N)$-matrix
in the representation $r$ (one often considers fundamental or adjoint representations, i.e. $r=F,A$), the vector potential $A_{\mu}(x) = A_{\mu}^a(x)\, T^a$ can be expressed in terms of the generators $T^a$ of the corresponding representation, and ${\cal P}$ denotes path ordering. The expectation values of Wilson loops contain information about the non-perturbative physics of non-Abelian gauge field theories and have applications to many physical phenomena  such as confinement, thermal phase transitions, quark screening, etc. For many of these applications it is useful to think of the path ${\cal C}$ as that traversed by a quark. We will discuss some of these applications in Section~\ref{sec:Section6}. Here, we describe how to compute expectation values of Wilson loops in a strongly coupled gauge theory using its gravity description.

We again use $\sN=4$ SYM theory as an example. Now recall that the field content of this theory includes six scalar fields $\vec \phi = (\phi^1, \cdots \phi^6)$ in the adjoint representation of the gauge group. This means that in this theory one can write down the following generalization of \eqn{2.1}~\cite{Maldacena:1998im,Rey:1998ik}:
\be
\label{Wils}
W(\sC) = {1 \ov \nc} \Tr P \exp \le[i \oint_{\sC} ds \, \le( A_\mu \dot x^\mu +
\vec n \cdot \vec \phi \sqrt{\dot x^2} \ri) \ri] \,,
\ee
where $\vec{n}(s)$ is a unit vector in $\bbr{6}$ that parametrizes a path in this space (or, more precisely, in $S^5$), just like $x^\mu(s)$ parametrizes a path in $\bbr{(1,3)}$. The factor of $\sqrt{\dot x^2}$ is necessary to make $\vec n \cdot \vec \phi \sqrt{\dot x^2}$ a density under worldline reparametrizations.
Note that the operators (\ref{2.1}) and (\ref{Wils}) are equivalent in the case of a light-like loop (as will be discussed in Section~\ref{sec:AdSCFTJetQuenching}) for which $\dot x^2=0$. 

An important difference between the operators  (\ref{2.1}) and (\ref{Wils}) is that (\ref{2.1})  breaks supersymmetry, whereas (\ref{Wils}) is locally 1/2-supersymmetric, meaning that for a straight-line  contour (that is time-like in Lorentzian signature) the operator is invariant under half of the supercharges of the $\mathcal {N}=4$ theory.

\begin{figure}
    \begin{center}
    	\includegraphics[width=0.7\textwidth]{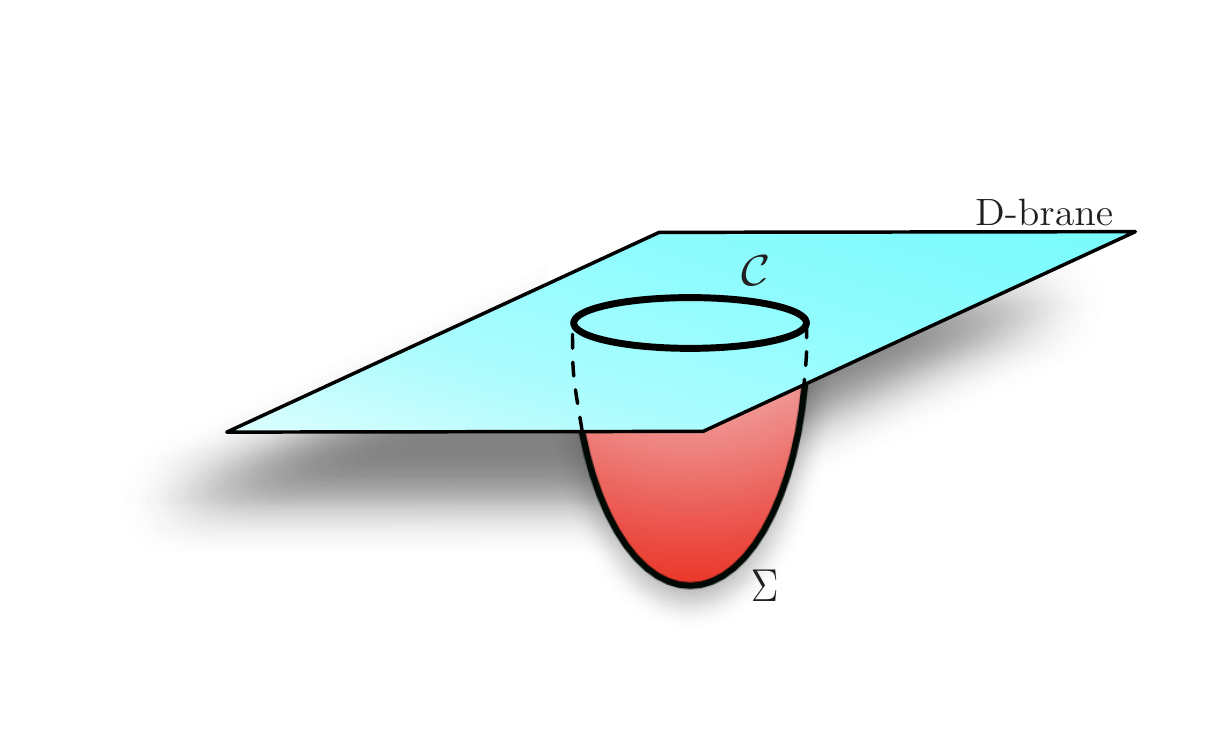}
    \end{center}
    \caption{\small String worldsheet associated with a Wilson loop.}
\label{wilsonloop}
\end{figure}

We will now argue that the generalized operator \eqn{Wils} 
has a  dual description in terms of a string worldsheet. For this purpose it is useful to think of the loop ${\cal C}$ as the path traversed by a quark. Although the $\sN=4$ SYM theory has no quarks, we will see below that these can be simply included by introducing in the gravity description open strings attached to a D-brane sitting at some radial position proportional to the quark mass. The endpoint of the open string on the D-brane is dual to the quark, so the boundary $\partial \Sigma$ of the string worldsheet $\Sigma$ must coincide with the path ${\cal C}$ traversed by the quark --- see Fig.~\ref{wilsonloop}.
This suggests that we must identify the expectation value of the Wilson loop operator, which gives the partition function (or amplitude) of the quark traversing $\sC$, with the partition function of the dual string worldsheet $\Sigma$~\cite{Maldacena:1998im,Rey:1998ik}:
\be
\vev{W (\sC)} = Z_{\rm string} [\p \Sig = \sC] \,.
\ee
For simplicity, we will focus on the case of an infinitely heavy (non-dynamical) quark. This means that we imagine that we have pushed the D-brane all the way to the AdS boundary. Under these circumstances the boundary $\p \Sig = \sC$ of the string worldsheet also lies within the boundary of AdS.

The key point to recall now is that the string endpoint couples both to the gauge field and to the scalar fields on the D-brane. This is intuitive since, after all, we obtained these fields as the massless modes of a quantized open string with endpoints attached to the D-brane. Physically, the coupling to the scalar fields is just a reflection of the fact that a string ending on a D-brane `pulls' on it and deforms its shape, thus exciting the scalar fields which parametrize this shape. The direction orthogonal to the D-brane in which the string pulls is specified by $\vec{n}$. The coupling to the gauge field reflects the fact that the string endpoint behaves as a point-like particle charged under this gauge field. 
We thus conclude that an open string ending on a D-brane with a fixed $\vec{n}$ excites both the gauge and the scalar fields, which suggests that the correct Wilson loop operator dual to the string worldsheet must include both types of fields and must therefore be given by 
\eqn{Wils}. 

The dual description of the operator (\ref{2.1}) is the same as that of (\ref{Wils}) except that the Dirichlet boundary conditions on the string worldsheet along the $S^5$ directions must be replaced by Neumann boundary conditions \cite{Alday:2007he} (see also \cite{Drukker:1999zq}). One immediate consequence is that, to leading order, the strong coupling results for the Wilson loop  (\ref{Wils}) with constant $\vec n$ and for the Wilson loop (5.62) are the same. However, the two results differ at the next order in the $1/\sqrt{\lambda}$ expansion, since in the case of (\ref{2.1}) we would have to integrate over the point on the sphere where the string is sitting. More precisely, at the one-loop level in the $\alpha'$-expansion one finds that the determinants for quadratic fluctuations are different in the two cases \cite{Drukker:2000ep}.

 In the large-$\nc$, large-$\lam$ limit, the string partition function
 $Z_{\rm string} [\p \Sig = \sC]$ greatly simplifies and is given by the exponential of the classical string action, i.e.
 \be \label{calss}
 Z_{\rm string} [\p \Sig = \sC] = e^{i S (\sC)} \quad \to \quad \vev{W (\sC)}
 = e^{i S (\sC)} \ .
 \ee
The classical action $S (\sC)$ can in turn be obtained by extremizing the Nambu-Goto action for the string worldsheet with the boundary condition that the string worldsheet ends on the curve ${\cal C}$.  More explicitly, parameterizing the two-dimensional world sheet by the coordinates 
 $\sigma^{\alpha}=(\tau,\sigma)$, the location of the string world sheet in the five-dimensional spacetime with coordinates $x^M$ is given by the Nambu-Goto action \eqn{string-action}. The fact that the action is invariant under coordinate changes of $\sigma^\alpha$ will allow us to pick the most convenient worldsheet coordinates $(\tau,\sigma)$ for each occasion.

Note that the large-$\nc$ and large-$\lambda$ limits are both crucial for~\eqref{calss} to hold.  Taking $\nc\to\infty$ at fixed $\lambda$ corresponds to taking the string coupling to zero, meaning that we can ignore the possibility of loops of string breaking off from the string world sheet.  Additionally taking 
$\lambda\to\infty$ corresponds to sending the string tension to infinity, which implies that we can neglect fluctuations of the string world sheet. Under these circumstances the string worldsheet `hanging down' from the contour ${\cal C}$ takes on its classical configuration, without fluctuating or splitting off loops.

As a simple example let us first consider a contour $\sC$ given by a straightline along the time direction with length $\sT$ which describes an isolated static quark at rest. On the field theory side we expect that the expectation value of the Wilson line should be given by
 \be \label{rrp}
 \vev{W(\sC)} = e^{- i M \sT} \,,
 \ee
where $M$ is the mass of the quark. From the symmetry of the problem, the corresponding bulk string worldsheet should be that of a straight string connecting the boundary and the Poincar\'e horizon and translated along the time direction by $\sT$. The action of such a string worldsheet is infinite since the proper distance from the boundary to the center of AdS is infinite. This is consistent with the fact that the external quark has an infinite mass.
A finite answer can nevertheless be obtained if we introduce an IR regulator in the bulk, putting the boundary at $z = \ep$ instead of $z=0$. From the IR/UV connection this corresponds to introducing a short-distance (UV) cutoff in the boundary theory. Choosing $\tau = t$ and $\sig = z$ the string worldsheet is given by $x^i (\sig, \tau) = {\rm const.}$, and the induced metric on the worldsheet is then given by
 \be
 ds^2 = {R^2 \ov \sig^2} (-d\tau^2 + d \sig^2) \,.
 \ee
Evaluating the Nambu-Goto action on this solution yields
 \be \label{statM}
 S=S_0  \equiv -{\sT R^2 \ov 2 \pi \apr} \int_\ep^\infty {dz  \ov z^2}
 =  -{\sqrt{\lam} \ov 2 \pi \ep} \, \sT \,,
  \ee
where we have used the fact that $R^2/\alpha' = \sqrt{\lambda}$. Using~\eqref{calss} and~\eqref{rrp} we then find that
\be \label{quarM}
 M = {\sqrt{\lam} \ov 2 \pi \ep}  \ .
\ee

\subsection{Rectangular loop: vacuum}

Now let us consider a rectangular loop sitting at a constant position on the $S^5$ \cite{Rey:1998ik,Maldacena:1998im}. The long side of the loop extends along the time direction with length $\sT$, and the short side extends along the $x_1$-direction with length $L$.
We will assume that $\sT \gg L$. Such a configuration can be though of as consisting of a static quark-antiquark pair separated by a distance $L$. Therefore we expect that the expectation value of the Wilson loop (with suitable renormalization) gives the potential energy between the pair, i.e. we expect that 
 \be
 \vev{W (\sC)} = e^{-i E_{\rm tot} \sT} =  e^{-i (2 M + V (L)) \sT} = e^{ i S (\sC)} \,,
 \ee
where $E_{\rm tot}$ is the total energy for the whole system and $V(L)$ is the potential energy between the pair. In the last equality we have used~\eqref{calss}. We will now proceed to calculate $S(\sC)$ for a rectangular loop.

\begin{figure}[t]
    \begin{center}
    	\includegraphics[width=0.4\textwidth]{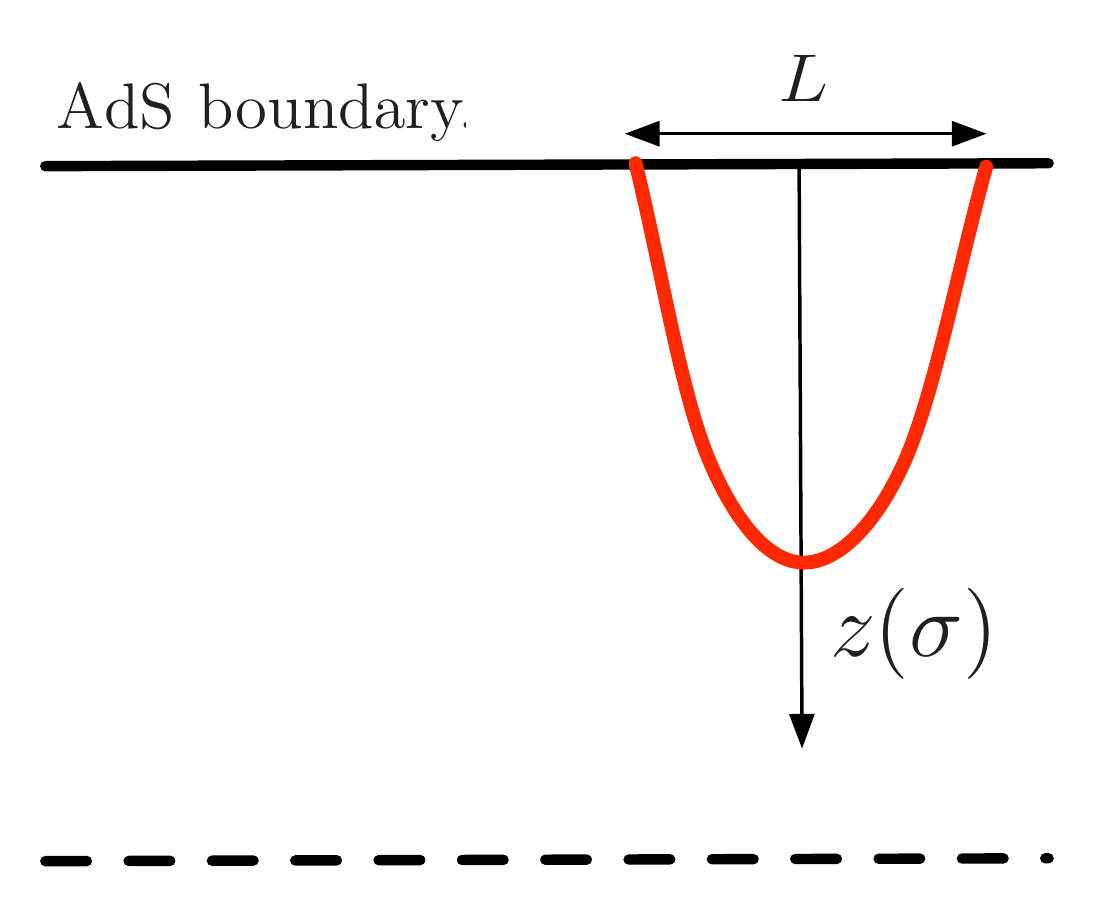}
    \end{center}
    \caption{\small String (red) associated with a quark-antiquark pair.}
\label{wilsonlooprectangle}
\end{figure}

It is convenient to choose the worldsheet coordinates to be
\be \label{wsC}
\tau = t, \qquad \sig = x_1 \ .
\ee
Since $\sT \gg L$, we can assume that the surface is translationally invariant along the $\tau$-direction, i.e. the extremal surface should have nontrivial dependence only on $\sig$. Given the symmetries of the problem we can also set
\be \label{othF}
x_3 (\sig) = \mbox{const.} \sac x_2 (\sig) = \mbox{const.} 
\ee
Thus the only nontrivial function to solve for is $z=z (\sig)$ (see Fig.~\ref{wilsonlooprectangle}), subject to the  boundary condition
\be \label{bdcs}
z \le(\pm  {L \ov 2} \ri) = 0 \,.
\ee

Using the form (\ref{poincare-z})  of the spacetime metric and Eqs.~(\ref{wsC})--(\ref{othF}), the induced metric on the worldsheet is given by
\be
ds^2_{ws} = {R^2 \ov z^2} \le(-d\tau^2 + (1 + z'^2) d\sig^2 \ri) \,,
\ee
giving rise to the Nambu-Goto action
\be \label{vacac}
S_{\rm NG}  = -{R^2 \sT \ov 2 \pi \apr} \int^{{L \ov 2}}_{-{L \ov 2}} d\sig {1 \ov z^2} \sqrt{1 + z'^2} \,,
\ee
where $z' = {dz / d \sig}$.
Since the action and the boundary condition are symmetric under 
$\sig \to -\sig$, $z(\sig)$ should be an even function of $\sig$. Introducing dimensionless coordinates via
\be
\sig = L \, \xi \sac z(\sig) = L \, y(\xi) 
\ee
we then have
\be \label{fulac}
S_{\rm NG}  =  -{2 R^2 \ov 2 \pi \apr} {\sT \ov L} \, Q \sac {\rm with} \ 
Q = \int_{0}^{\ha} {d \xi \ov y^2} \sqrt{ 1 + y'^2} \,.
\ee
Note that $Q$ is a numerical constant. As we will see momentarily, it is in fact divergent and therefore it should be defined more carefully. The equation of motion for $y$ is given by
\be \label{eomY}
y'^2 = {y_0^4 - y^4 \ov y^4} 
 \ee
 with $y_0$ the turning point at which $y' =0$, which by symmetry should happen at $\xi=0$.  $y_0$ can thus be determined by the condition 
 \be
 \ha = \int_0^\ha d \xi = \int_0^{y_0} {dy \ov y'} = \int_0^{y_0} dy \, {y^2 \ov \sqrt{y_0^4 - y^4}}
 \quad \to \quad y_0 = {\Ga ({1 \ov 4}) \ov 2 \sqrt{\pi} \Ga ({3 \ov 4})} \,.
 \ee
It is then convenient to change integration variable in $Q$ from $\xi$ to $y$ to get
  \be \label{QE}
Q = y_0^2 \int_{0}^{y_0} {d y \ov y^2 \sqrt{ y_0^4 - y^4}} \,.
\ee
This is manifestly divergent at $y=0$, but the divergence can be interpreted as coming from the infinite rest masses of the quark and the antiquark. As in the discussion after~\eqref{rrp}, we can obtain a finite answer by introducing an IR cutoff in the bulk by putting the boundary at $z = \ep$, i.e. by replacing the lower integration limit in~\eqref{QE} by $\ep$. The potential $V(L)$ between the quarks is then obtained by subtracting $2 M \sT$ from~\eqref{fulac} (with $M$ given by~\eqref{quarM}) and then taking $\ep \to 0$ at the end of the calculation. One then finds the finite answer
\be  \label{potB}
V(L) = -{ 4 \pi^2 \ov \Ga^4 ({1 \ov 4})} {\sqrt{\lam} \ov L} \,,
\ee
where again we used the fact that ${R^2 / \apr} = \sqrt{\lam}$ to translate from gravity to gauge theory variables. Note that the $1/L$ dependence is simply a consequence of conformal invariance. The non-analytic dependence on the coupling, i.e. the $\sqrt{\lam}$ factor, could not be obtained at any finite order in perturbation theory. From the gravity viewpoint, however, it is a rather generic result, since it is due the fact that the tension of the string is proportional to $1/\apr$.
The above result is valid at large $\lam$. At small $\lam$, the potential between a quark and an anti-quark in an $\sN=4$ theory is given by~\cite{Erickson:1999qv}
\be \label{coul}
E = -{\pi \lam \ov L} 
\ee
to lowest order in the weak-coupling expansion.

It is remarkable that the calculation of a Wilson loop in a strongly interacting gauge theory has been simplified to a classical mechanics
problem no more difficult than finding the catenary curve describing a string suspended from two points, hanging in a gravitational field --- in this case the gravitational field of the AdS spacetime.

Note that given~\eqref{potB}, the boundary short-distance cutoff $\ep$ in~\eqref{quarM} can be interpreted as the size of the external quark. One might have expected (incorrectly) that a short distance cutoff on the size of the quark should be given by the Compton wavelength 
${1 / M} \sim {\ep / \sqrt{\lam}}$, which is much smaller than $\ep$.
Note that the size of a quark should be defined by either its Compton wavelength or by the distance between a quark and an anti-quark at which the potential is of the order of the quark mass, whichever is bigger. In a weakly coupled theory, the Compton wavelength is bigger, while in a strongly coupled theory with potential (\ref{potB}), the latter is bigger and is of order $\ep$.

\subsection{Rectangular loop: nonzero temperature}
\label{Loop-nonzero}
\begin{figure}
    \begin{center}
    	\includegraphics[width=0.5\textwidth]{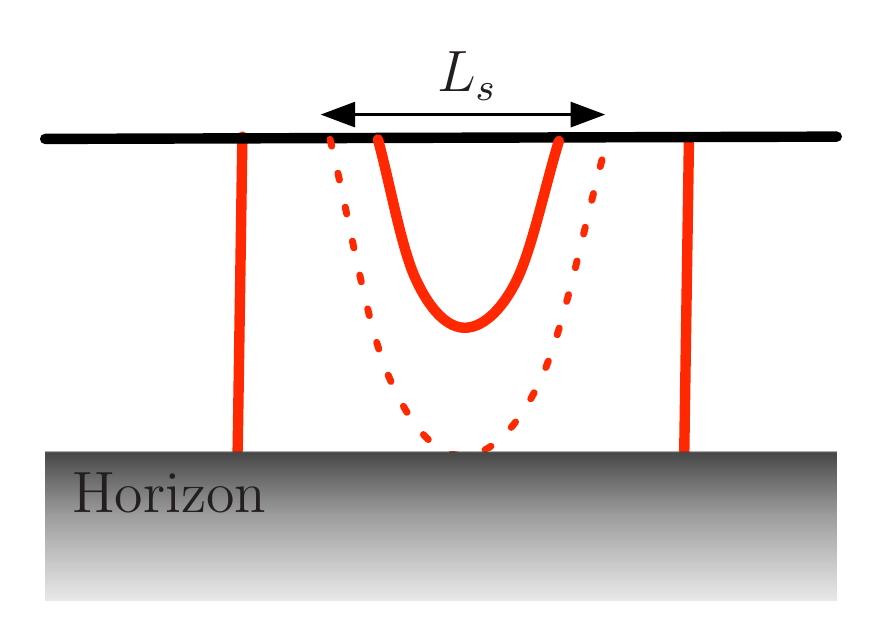}
    \end{center}
    \caption{\small String (red) associated with a quark-antiquark pair in a plasma with temperature $T>0$. The preferred configuration beyond a certain separation $L_s$ consists of two independent strings.}
\label{wilsonloopfiniteT}
\end{figure}

We now consider the expectation of the rectangular loop at nonzero temperature \cite{Rey:1998bq,Brandhuber:1998bs}. In this case the bulk gravity geometry is given by that of the black brane~\eqref{AdSfiniteZ}. The set-up of the calculation is exactly the same as in  Eqs.~\eqref{wsC}--\eqref{bdcs} for the vacuum. The induced worldsheet metric is now given by
 \be
ds^2_{ws} = {R^2 \ov z^2} \le(- f (z) d\tau^2 + 
\le(1 + {z'^2 \ov f} \ri) d\sig^2 \ri) \,,
\ee
which yields the Nambu-Goto action
\be \label{sceac}
S_{\rm NG}  = -{R^2 \sT \ov 2 \pi \apr} 
\int^{{L \ov 2}}_{-{L \ov 2}} d\sig {1 \ov z^2} \sqrt{f(z) + z'^2} \,.
\ee
The crucial difference between the equation of motion following from~\eqref{sceac} and that following from~\eqref{vacac} is that in the present case there exists a maximal value $L_s \sim 1/T$ beyond which no nontrivial solutions exist \cite{Rey:1998bq,Brandhuber:1998bs} --- see Fig.~\ref{wilsonloopfiniteT}.
Instead, the solution beyond this maximal separation consists of two disjoint vertical strings ending at the black hole horizon. The physical reason can be easily understood qualitatively from the figure. At some separation, the
lowest point on the string touches the horizon.   
Surely at and beyond
this separation the string can minimize its energy by splitting into two independent strings, each of which falls through the horizon.  The precise value of $L_s$
is defined as the quark-antiquark separation at which the free energy of the disconnected configuration becomes smaller than that of the connected configuration. This happens at a value of $L$ for which the lowest point of the connected configuration is close to but still somewhat above the horizon.
Once  $L>L_s$, the quark-antiquark separation can then be increased further at no additional energy cost,
so the potential becomes constant  and the quark and the antiquark are perfectly screened from each other by the plasma between them.  (See, for example, Ref.~\cite{Bak:2007fk} for a careful discussion of the corrections to this 
large-$\nc$, large-$\lambda$ result.)

\subsection{Rectangular loop: a confining theory}

\begin{figure}
    \begin{center}
    	\includegraphics[width=0.6\textwidth]{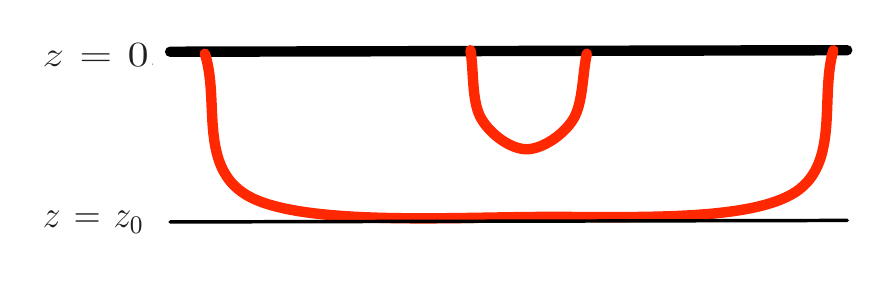}
    \end{center}
    \caption{\small String (red) associated with a quark-antiquark pair in a confining theory.}
\label{wilsonloopconfining}
\end{figure}

For comparison, let us consider the expectation value of  a rectangular
loop in the $2+1$-dimensional confining theory \cite{Witten:1998zw} (for a review see \cite{Sonnenschein:2000qm}) whose metric is given by~\eqref{3dcon}, which we reproduce here for convenience:
 \be \label{3dcon1}
ds^2 = \frac{R^2}{z^2} \left(  -dt^2 + dx_1^2 + dx_2^2 + f dt_\mt{E} \right) +
\frac{R^2}{z^2 f} dz^2, \qquad  \, f = 1 - {z^4 \ov z_0^4} \ .
\ee
As discussed earlier the crucial difference between~\eqref{3dcon1} and  AdS is that the spacetime \eqref{3dcon1} ends smoothly at a finite value $z=z_0$, which introduces a scale in the theory. The difference as compared to the finite-temperature case is that in the confining geometry the string has no place to end, so in order to to minimize its energy it tends  to drop down to $z_0$ and to run parallel there --- see Fig.~\ref{wilsonloopconfining}.

Again the set-up of the calculation is completely analogous to the cases above. The induced worldsheet metric is now given by
 \be
ds^2_{ws} = {R^2 \ov z^2} \le(-  d\tau^2 + \le(1 + {z'^2 \ov f} \ri) d\sig^2 \ri),
\ee
and the corresponding the Nambu-Goto action is
\be \label{sceac1}
S_{\rm NG}  =- {R^2 \sT \ov 2 \pi \apr} \int^{{L \ov 2}}_{-{L \ov 2}} d\sig {1 \ov z^2}
\sqrt{1 + {z'^2 \ov f(z)}} \,.
\ee
When $L$ is large, the string quickly drops to $z=z_0$ and runs parallel there. We thus find that the action can be approximated by (after subtracting the vertical parts which can be interpreted as due to the static quark masses)
\be \label{areaL}
-S(\sC) - 2 M \sT \approx  {R^2 \sT \ov 2 \pi \apr}  {L \ov z^2_0} \,,
\ee
 which  gives rise to a confining potential 
 \be
 V(L) = \sig_s L \sac \sig_s = {\sqrt{\lam} \ov 2 \pi z_0^2} \,.
 \ee
The constant $\sig_s$ can be interpreted as the effective string tension. As mentioned in Sec.~\ref{sec:conf} 
the mass gap for this theory is $M \sim {1 / z_0}$, so we find that 
$\sig_s \sim \sqrt{\lam} M^2$. 
Although we have described the calculation only for one example of a confining gauge theory, the qualitative features of Fig.~\ref{wilsonloopconfining} generalize.  In any confining gauge theory with a dual gravity description, as a quark-antiquark pair are separated the string hanging beneath them sags down to some `depth' $z_0$ and then as the separation is further increased it sags no further.  Further increasing the separation means adding more and more string at the same depth $z_0$, which costs an energy that increases linearly with separation.  Clearly, any metric in which a suspended string behaves like this cannot be conformal; it has a length scale $z_0$ built into it in some way.  This length scale $z_0$ in the gravitational description corresponds via the IR/UV correspondence to the mass gap $M\sim 1/z_0$ for the gauge theory and to the size of the `glueballs' in the gauge theory, which is of order $z_0$. 

To summarize, we note that  the qualitative behavior of  the Wilson loop discussed  in various examples above is only determined by  gross features of the bulk geometry. The $1/L$ behavior~\eqref{coul} in the conformal vacuum follows directly from the scaling symmetry of the bulk geometry;  the area law~\eqref{areaL} in the confining case has to do with the fact that a string has no place to end in the bulk when the geometry smoothly closes off; and the screening behavior at finite temperature is a consequence of the fact that a string can fall through the black hole horizon. The difference between Figs. \ref{wilsonlooprectangle} and \ref{wilsonloopconfining} highlights the fact that 
${\cal N}=4$ SYM theory is not a good model for the vacuum of a confining theory like QCD.  However, as we will discuss in Section 7.6, the potential obtained from Fig.~\ref{wilsonloopfiniteT} is not a bad caricature of what happens in the deconfined phase of QCD.  This is one of many ways of seeing that ${\cal N}=4$ SYM at $T \neq 0$ is more similar to QCD above $T_c$ than ${\cal N}=4$ SYM at $T=0$ is to QCD at $T=0$.  A heuristic way of thinking about this is to note that at low temperatures the putative horizon would be 
at a $z_\mt{hor} >  z_0$, i.e. it is far below the bottom of Fig.~\ref{wilsonloopconfining}, and therefore it plays no role while
at large temperatures, the horizon is far above $z_0$ and it is $z_0$ that plays no role.
At some intermediate temperature, the theory has undergone a phase transition from a confined phase
described by Fig.~\ref{wilsonloopconfining} into a deconfined phase described by 
Fig. \ref{wilsonloopfiniteT}.\footnote{The way we have described the transition is a crude way of thinking about the so-called Hawking-Page phase transition between a spacetime without and with a black hole \cite{Hawking:1982dh,Witten:1998qj}.} Unlike in QCD, this deconfinement phase transition is a 1st order phase transition in the large-$\nc$, strong coupling limit under consideration, and the theory in the deconfined phase looses all memory about the confinement scale $z_0$. Presumably corrections away from this limit, in particular finite-$\nc$ corrections, could turn the transition into a higher-order phase transition or even a cross-over. 


\section{Introducing fundamental matter}
\label{fundamental}

All the matter degrees of freedom of ${\cal N}=4$ SYM, the fermions and the scalars, transform in the adjoint representation of the gauge group. In QCD, however, the quarks transform in the fundamental representation. Moreover, most of what we know about QCD phenomenologically comes from the study of quarks and their bound states. Therefore, in order to construct holographic models more closely related to QCD, we must introduce degrees of freedom in the fundamental representation. It turns out that there is a rather simple way to do this in the limit in which the number of quark species, or flavours, is much smaller than the number of colours, \ie when $\nf \ll \nc$. Indeed, in this limit the introduction of $\nf$ flavours in the gauge theory corresponds to the introduction of $\nf$ D-brane probes in the AdS geometry sourced by the D3-branes \cite{Aharony:1998xz,Karch:2000gx,Karch:2002sh}. This is perfectly consistent with the well-known fact that the topological representation of the large-$\nc$ expansion of a gauge theory with quarks involves Riemann surfaces with boundaries --- see Section~\ref{sec:nce}. In the string description, these surfaces correspond to the worldsheets of open strings whose endpoints must be attached to D-branes. In the context of the gauge/string duality, the intuitive idea is that closed strings living in AdS are dual to gauge-invariant operators constructed solely out of gauge fields and adjoint matter, \eg ${\cal O} =\Tr F^2$, whereas open strings are dual to meson-like operators, \eg ${\cal O} =\bar{q} q$.
In particular, the two end-points of an open string, which are forced to lie on the D-brane probes, are dual to a quark and an antiquark, respectively.

\subsection{The decoupling limit with fundamental matter}
\label{fundec}
\begin{figure}[t]
    \begin{center}
    	\includegraphics[width=0.65\textwidth]{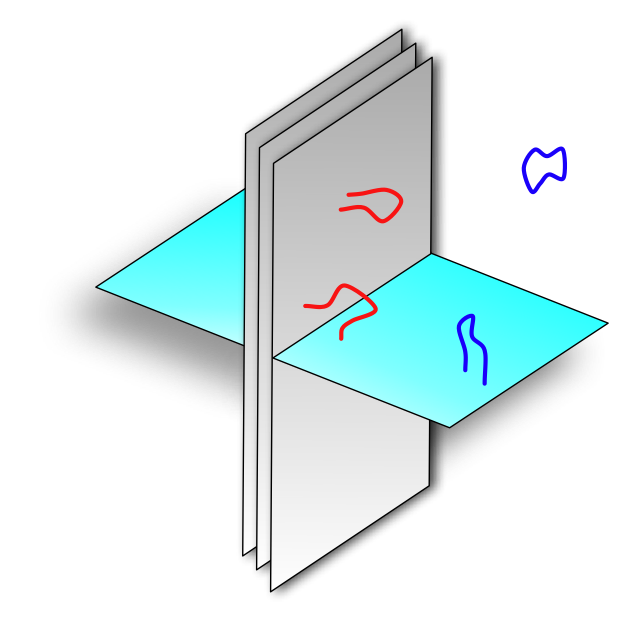}
    \end{center}
    \caption{\small Excitations of the system in the open string description.}
\label{D3D7excitations}
\end{figure}

The fact that the introduction of gauge theory quarks corresponds to the introduction of D-brane probes in the string description can be more `rigorously' motivated by repeating the arguments of Sections~\ref{sec:dbran1}, \ref{sec:dbran1} and \ref{sec:conj} in the presence of $\nf$ D$p$-branes, as indicated in Fig.~\ref{D3D7excitations}.
We shall be more precise about the value of $p$ and the precise orientation of the branes later; for the moment we simply assume $p > 3$.

As in Section~\ref{sec:dbran1}, when $g_s \nc \ll 1$ the excitations of this system are accurately described by interacting closed and open strings living in flat space. In this case, however, the open string sector is richer. As before, open strings with both end-points on the D3-branes give rise, at low energies, to the ${\cal N}= 4$ SYM multiplet in the adjoint of $SU(\nc)$.  We see from Eqn.~\eqn{YM} that the coupling constant for these degrees of freedom  is dimensionless, and therefore these degrees of freedom remain interacting at low energies. The coupling constant for the open strings with both end-points on the D$p$-branes, instead, has dimensions of (length)$^{p-3}$. Therefore the effective dimensionless coupling constant at an energy $E$ scales as
$g_\mt{Dp} \propto E^{p-3}$. Since we assume that $p > 3$, this implies that, just like the closed strings, the $p$-$p$ strings become non-interacting at low energies. Finally, consider the sector of open strings with one end-point on the
D3-branes and one end-point on the D$p$-branes. These degrees of freedom transform in the fundamental of the gauge group on the D3-branes and in the fundamental of the gauge group on the D$p$-branes, namely in the bifundamental of $SU(\nc) \times SU(\nf)$. Consistently, these 3-$p$ strings interact with the 3-3 and the $p$-$p$ strings with strengths given by the corresponding coupling constants on the D3-branes and on the D$p$-branes. At low energies, therefore, only the interactions with the 3-3 strings survive. In addition, since the effective coupling on the D$p$-branes vanishes, the corresponding gauge group $SU(\nf)$ becomes a global symmetry group. This is the origin of the flavour symmetry expected in the presence of $\nf$ (equal mass) quark species in the gauge theory.

To summarize, when $g_s \nc \ll 1$ the low-energy limit of the D3/D$p$ system yields two decoupled sectors. The first sector is free and consists of closed strings in ten-dimensional flat space and $p$-$p$ strings propagating on the worldvolume of $\nf$ D$p$-branes. The second sector is interacting and consists of a four-dimensional ${\cal N}= 4$ SYM multiplet in the adjoint of $SU(\nc)$, coupled to the light degrees of freedom coming from the 3-$p$ strings. We will be more precise about the exact nature of these degrees of freedom later, but for the moment we emphasize that they transform in the fundamental representation of the $SU(\nc)$ gauge group, and in the fundamental representation of a global, flavour symmetry group $SU(\nf)$.

\begin{figure}
    \begin{center}
    	\includegraphics[width=0.7\textwidth]{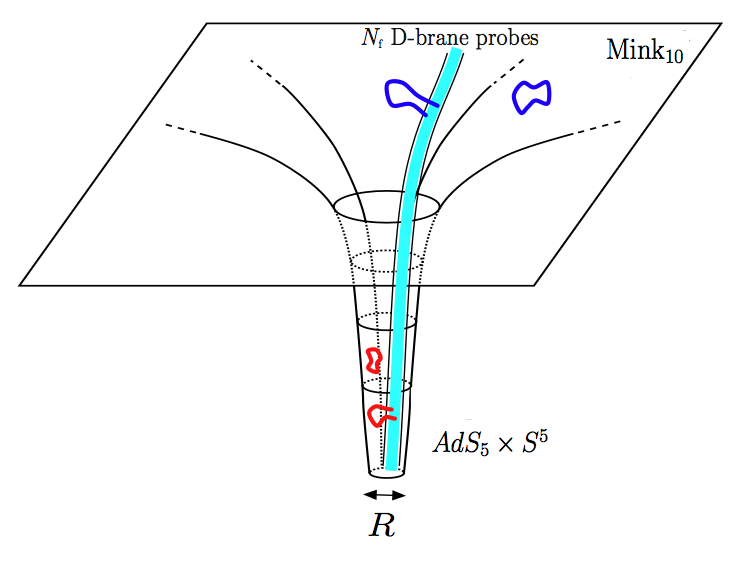}
    \end{center}
    \caption{\small Excitations of the system in the second description.}
\label{D3D7AdSexcitations}
\end{figure}

Consider now the closed string description  at $g_s \nc \gg 1$. In this case, as in Section~\ref{sec:dbran2}, the D3-branes may be replaced by their backreaction on spacetime. If we assume that $g_s \nf \ll 1$, which is consistent with $\nf \ll \nc$, we may still neglect the backreaction of the D$p$-branes. In other words, we may treat the D$p$-branes as probes living in the geometry sourced by the D3-branes, with the D$p$-branes not modifying this geometry. The excitations of the system in this limit consist of closed strings and open $p$-$p$ strings that propagate in two different regions, the asymptotically flat region and the $AdS_5 \times S^5$ throat -- see Fig.~\ref{D3D7AdSexcitations}.
As in section \ref{sec:dbran2}, these two regions decouple from each other in the low-energy limit. Also as in section \ref{sec:dbran2}, in this limit the strings in the asymptotically flat region become non-interacting, whereas those in the throat region remain interacting because of the gravitational redshift.

Comparing the two descriptions above, we see that the low-energy limit at both small and large values of $g_s \nc$ contains a free sector of closed and open $p$-$p$ strings. As in section \ref{sec:conj} we identify these free sectors, and we conjecture that the interacting sectors on each side provide dual descriptions of the same physics. In other words, we conjecture that the ${\cal N}= 4$ SYM coupled to $\nf$ flavours of fundamental degrees of freedom is dual to type IIB closed strings in $AdS_5 \times S^5$, coupled to open strings propagating on the worldvolume of $\nf$ D$p$-brane probes.

It is worth clarifying the following conceptual point before closing this section. It is sometimes stated that, in the 't Hooft limit in which $\nc \rightarrow \infty$ with $\nf$ fixed, the dynamics is completely dominated by the gluons, and therefore that the quarks can be completely ignored. One may then wonder what the interest of introducing fundamental degrees of freedom in a large-$\nc$ theory may be. There are several answers to this. First of all, in the presence of fundamental matter, it is more convenient to think of the large-$\nc$ limit {\it a la} Veneziano, in which $\nf/\nc$ is kept small but finite. Any observable can then be expanded in powers of $1/\nc^2$ and $\nf/\nc$. As we will see, this is precisely the limit that is captured by the dual description in terms of $\nf$ D-brane probes in
$AdS_5 \times S^5$. The leading D-brane contribution will give us the leading contribution of the fundamental matter, of relative order $\nf/\nc$. The Veneziano limit is richer than the 't Hooft limit, since setting $\nf/\nc = 0$ one recovers the 't Hooft limit. The second point is that, even in the 't Hooft limit, the quarks should not be regarded as irrelevant, but rather as valuable probes of the gluon-dominated dynamics. It is their very presence in the theory that allows one to ask questions about heavy quarks in the plasma, jet quenching, meson physics, photon emission, etc. The answers to these questions are of course dominated by the gluon dynamics, but without dynamical quarks in the theory such questions cannot even be posed. There is a completely analogous statement in the dual gravity description: To leading order the geometry is not modified by the presence of
D-brane probes, but one needs to introduce these probes in order to pose questions about heavy quarks in the plasma, parton energy loss, mesons, photon production, etc. In this sense, the D-brane probes allow one to decode information already contained in the geometry.

\subsection{Models with fundamental matter}
Above, we motivated the inclusion of fundamental matter via the introduction of
$\nf$ `flavour' D$p$-brane probes in the background sourced by $\nc$ `colour' D3-branes. However, we were deliberately vague about the value of $p$, about the relative orientation between the flavour and the colour branes, and about the precise nature of the flavour degrees of freedom in the gauge theory. Here we will address these points. Since we assumed $p>3$ in order to decouple the $p$-$p$ strings, and since we wish to consider stable D$p$-branes in type IIB string theory, we must have $p=5$ or $p=7$ --- see Section \ref{sec:dbran1}. In other words, we must consider D5- and D7-brane probes.

\begin{figure}
    \begin{center}
    	\includegraphics[width=0.5\textwidth]{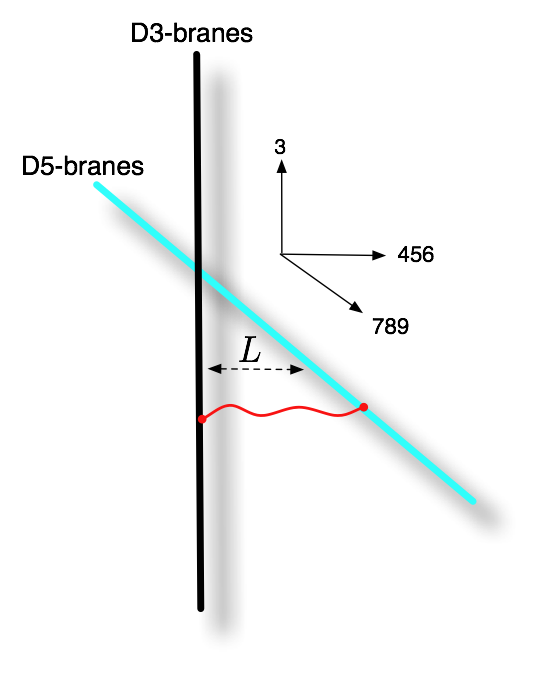}
    \end{center}
    \caption{\small D3-D5 configuration \eqn{D3D5on2} with a string stretching between them. The 12-directions common to both branes are suppressed.}
\label{D3D5orientation}
\end{figure}

Consider first adding flavour D5-branes. We will indicate the relative orientation between these and the colour D3-branes by an array like, for example,
\be
\begin{array}{rcccccccccl}
\mbox{D3:}\,\,\, & 1 & 2 & 3 & \_ & \_ & \_ & \_ & \_ & \_ & \, \\
\mbox{D5:}\,\,\, & 1 & 2 & \_ & 4 & 5 & 6 & \_ & \_ & \_ & \,.
\end{array}
\label{D3D5on2}
\ee
This indicates that the D3- and the D5-branes share the 12-directions. The 3-direction is transverse to the D5-branes, the 456-directions are transverse to the D3-branes, and the 789-directions are transverse to both sets of branes. This means that the two sets of branes can be separated along the 789-directions, and therefore they do not necessarily intersect, as 
indicated in Fig.~\ref{D3D5orientation}.
It turns out that the lightest states of a D3-D5 string have a minimum mass given by what one would have expected on classical grounds, namely $M=\tf L = L/2\pi \ell_s^2$, where $\tf$ is the string tension \eqn{tension} and $L$ is the minimum distance between the D3- and the D5-branes.\footnote{In order to really establish this formula one must quantise the D3-D5 strings and compute the ground state energy. In the case at hand, the result coincides with the classical expectation. The underlying reason is that, because the configuration \eqn{D3D5on2} preserves supersymmetry, corrections to the classical ground state energy coming from bosonic and fermionic quantum fluctuations cancel each other out exactly. For other brane configurations like \eqn{D3D5on3} this does not happen.} These states can therefore be arbitrarily light, even massless, provided $L$ is sufficiently small.
Generic excited states, as usual, have an additional mass set by the string scale alone, $m_s$. The only exception are excitations in which the string moves rigidly with momentum $\vec{p}$ in the 12-directions, in which case the energy squared is just $M^2 + \vec{p}\,^2$. This is an important observation because it means that in the decoupling limit, in which one focuses on energies $E \ll m_s$, only a finite set of modes of the D3-D5 strings survive, and moreover these modes can only propagate along the directions common to both branes. From the viewpoint of the dual gauge theory, this translates into the statement that the degrees of freedom in the fundamental representation are localised on a defect --- in the example at hand, on a plane that extends along the 12-directions and lies at a constant position in the 3-direction. As an additional example, the configuration
\be
\begin{array}{rcccccccccl}
\mbox{D3:}\,\,\, & 1 & 2 & 3 & \_ & \_ & \_ & \_ & \_ & \_ & \, \\
\mbox{D5:}\,\,\, & 1 & \_ & \_ & 4 & 5 & 6 & 7 & \_ & \_ &
\end{array}
\ee
corresponds to a dual gauge theory in which the fundamental matter is localised on a line --- the 1-direction.

We thus conclude that, if we are interested in adding to the $\nfour$ SYM theory fundamental matter degrees of freedom that propagate in 3+1 dimensions (just like the gluons and the adjoint matter), then we must orient the flavour D-branes so that they extend along the 123-directions. This condition leaves us with two possibilities:
\be
\begin{array}{rcccccccccl}
\mbox{D3:}\,\,\, & 1 & 2 & 3 & \_ & \_ & \_ & \_ & \_ & \_ & \, \\
\mbox{D5:}\,\,\, & 1 & 2 & 3 & 4 & 5 & \_ & \_ & \_ & \_ &
\end{array}
\label{D3D5on3}
\ee
and
\be
\begin{array}{rcccccccccl}
\mbox{D3:}\,\,\, & 1 & 2 & 3 & \_ & \_ & \_ & \_ & \_ & \_ & \, \\
\mbox{D7:}\,\,\, & 1 & 2 & 3 & 4 & 5 & 6 & 7 & \_ & \_ & \,.
\end{array}
\label{D3D7}
\ee

So far we have not been specific about the precise nature of the fundamental matter degrees of freedom --- for example, whether they are fermions or bosons, etc. This also depends on the relative orientation of the branes. It turns out that for the configuration \eqn{D3D5on3}, the ground state energy of the D3-D5 strings is (for sufficiently small $L$) negative, that is, the ground state is tachyonic, signaling an instability in the system. This conclusion is valid at weak string coupling, where the string spectrum can be calculated perturbatively. While it is possible that the instability is absent at strong coupling, we will not consider this configuration further in this review.

We are therefore left with the D3-D7 system \eqn{D3D7}. Quantisation of the D3-D7 strings shows that the fundamental degrees of freedom in this case consist of $\nf$ complex scalars and $\nf$ Dirac fermions, all of them with equal masses given by
\be
M_\mt{q} = \frac{L}{2\pi \alpha'} \,.
\label{quarkmass}
\ee
In a slight abuse of language, we will collectively refer to all these degrees of freedom as `quarks'. The fact that they all have exactly equal masses is a reflection of the fact that the addition of the $\nf$ D7-branes preserves a fraction of the original supersymmetry of the SYM theory. More precisely, the original
$\nfour$ is broken down to ${\cal N}=2$, under which the fundamental scalars and fermions transform as part of a single supermultiplet. In the rest of the review, especially in Section \ref{mesons}, we will focus our attention on this system as a model for gauge theories with fundamental matter.




\newpage
\chapter{Bulk  properties of strongly coupled plasma}
\label{sec:BulkDynamicalProperties}

Up to this point in this review, we have laid the groundwork needed for what is to come in two halves.  In
Sections \ref{intro} and \ref{sec:latticeQCD} 
we have introduced the theoretical, phenomenological and experimental challenges posed by  the study of the deconfined phase of QCD and in Sections \ref{sec:Section3} and \ref{AdS/CFT} we have motivated and described gauge/string duality, providing the reader with most of the conceptual and computational machinery necessary to perform many calculations.   Although we have foreshadowed their interplay at various points, these two long introductions have to a large degree been separately self-contained.  In the next three sections, we weave these strands together.  In these sections, we shall review applications of gauge/gravity duality to the study of the strongly coupled plasma of ${\cal N}=4$ SYM theory at nonzero temperature, focussing on the ways in which these calculations can guide us toward the resolution of the challenges described in Sections \ref{intro} 
and \ref{sec:latticeQCD}.

The study of the zero temperature vacuum  of strongly coupled  $\N=4$ SYM theory
is a rich subject with numerous physical insights into the dynamics of gauge theories. 
Given our goal of gaining insights into the deconfined phase of QCD,
we will largely concentrate on the description of  strongly coupled $\N=4$ SYM theory at nonzero temperature, where it describes a strongly coupled non-abelian plasma with
$\mathcal{O} (N^2_c)$ degrees of freedom. 
The vacua of QCD and $\N=4$ SYM theory have very different properties.
However, when we compare $\N=4$ SYM at $T\neq 0$ with QCD at a temperature above the temperature $T_c$ of the crossover from a hadron gas to quark-gluon plasma, many of the
qualitative distinctions disappear or become 
unimportant. In particular:  
\begin{enumerate}
\item 
QCD confines, while $\N=4$ does not.   This is a profound difference in vacuum.  But, above its $T_c$ QCD is no longer confining.  
The fact that its $T=0$ quasiparticles are hadrons within which quarks are confined is not particularly relevant at temperatures above $T_c$.  
\item
In QCD, chiral symmetry is broken by a chiral condensate which sets a scale that is certainly not present in $\N=4$ SYM theory.  However, in QCD above its $T_c$ the chiral condensate melts away and this distinction between the vacua of the two theories also ceases to be relevant.
\item 
$\N=4$ is a scale invariant theory while in QCD scale invariance is broken by the confinement scale, the chiral condensate, and just by the running of the coupling constant.  Above $T_c$, we have already dispensed with the first two scales.  
Also, as we have described in Section~\ref{sec:latticeQCD}, QCD thermodynamics is significantly 
nonconformal just 
above $T_c\sim 170$~MeV, but at higher temperatures the quark-gluon plasma becomes more and more scale invariant, at least in its thermodynamics.  (Thermodynamic quantities converge to their values in the noninteracting limit, due to the running of the coupling towards zero, only at vastly higher temperatures which are far from the reach of any collider experiment.)  So, here again, 
QCD above (but not asymptotically far above) its $T_c$  is much more 
similar to $\N=4$ SYM theory at $T\neq 0$ than the vacua of the two theories are.
\item 
$\N=4$ SYM theory is supersymmetric. However, supersymmetry is explicitly broken at nonzero temperature.  In a thermodynamic context, this can be seen by noting that fermions have antiperiodic boundary conditions along the Euclidean time circle while bosons are periodic.  
For this reason, 
supersymmetry does not play a major role in the characterization of properties of the $\N=4$ SYM plasma at nonzero temperatures.
\item 
QCD is an asymptotically free theory and, thus, high energy processes are weakly coupled. However, as we have described in Section~\ref{intro}, in the regime of temperatures above $T_c$ that are accessible to heavy ion collision experiments the QCD plasma is strongly coupled,
which opens a window of applicability for strong coupling techniques. 
\end{enumerate}
For these and other reasons, the strongly coupled plasma of $\N=4$ SYM theory has been studied by many authors with the aim of gaining insights into the dynamics of deconfined QCD plasma.  

In fairness, we should also mention the significant differences between the two theories that remain at nonzero temperature:
\begin{enumerate}
\item
$\N=4$ SYM theory with $N_c=3$ has more degrees of freedom than QCD with $N_c=3$.  
To seek guidance for QCD from results in $\N=4$ SYM, the challenge is to
evaluate how an observable of interest depends on the number of degrees of freedom, as we do at several points in Section~\ref{sec:Section6}.  The best case scenario is that there is no such dependence, as for example arises for the ratio $\eta/s$ between the shear viscosity and the entropy density that we introduced in Section~\ref{intro} and 
that we shall discuss in Section~\ref{sec:TransportProperties} below. 
\item
Most of the calculations that we shall report are done in the strong 
coupling ($\lambda\rightarrow\infty$) limit.  This is of course a feature not a bug.  The ability to do these calculations in the strong coupling regime is a key part of the motivation for all this work.  But, although in the temperature regime of interest $g^2(T) N_c = 4\pi N_c \alpha_s(T)$ is large, it is not infinite.  This motivates the calculation of corrections to various results that we shall discuss that are proportional to powers of $1/\lambda$, for the purpose of testing the robustness of conclusions drawn from calculations done with $\lambda\rightarrow\infty$.
\item
QCD has $N_c=3$ colors, while all the calculations that we shall report are done in the $N_c\rightarrow \infty$ limit.  Although the large-$N_c$ approximation is familiar in QCD, the standard way of judging whether it is reliable in a particular context is to compute corrections suppressed by powers of $1/N_c$.
And, determining the $1/N_c^2$ corrections to the  calculations done via the gauge/string duality that we review remains an outstanding challenge.
\item
Although we have argued above that the distinction between bosons and fermions is not important at nonzero temperature, the distinction between degrees of freedom in the adjoint or fundamental representation of $SU(N_c)$ is important.
QCD has $N_f=3$ flavors in the fundamental representation, namely $N_f=N_c$.  These fundamental degrees of freedom contribute significantly to its thermodynamics at temperatures above $T_c$.  And, the calculations that we shall report are either done with $N_f=0$ or with $0<N_f\ll N_c$.  Extending  methods based upon gauge/string duality to the regime in which $N_f\sim N_c$ remains an outstanding challenge.
\end{enumerate}
The plasmas of QCD and strongly coupled $\N=4$ SYM theory certainly differ.  
At the least, using one to gain insight into the other follows in the long tradition of modelling, in which a theoretical physicist employs the simplest instance of a theory that captures the essence of a suite of phenomena that are of interest in order to gain insights. $\N=4$ SYM theory may not seem simple from a field theory perspective, but its gravitational description makes it clear that it is in fact the simplest, most symmetric, strongly coupled non-Abelian plasma.  The question then becomes whether there are quantities or phenomena that are universal across many different strongly coupled plasmas.  The qualitative, and in some instances even semi-quantitative, successes that we shall review that have been achieved in comparing results or insights obtained in $\N=4$ SYM theory to those in QCD suggest a positive answer to this question, but no precise definition of this new kind of universality has yet been conjectured.  Absent a precise understanding of such a universality, we can hope for reliable insights into QCD but not for controlled calculations.


We begin our description of the $\N=4$ strongly coupled plasma in this Section by characterizing its macroscopic properties,  i.e., those that involve temporal and spatial scales 
much larger than the microscopic scale  $1/T$.
In Section~\ref{sec5.1} we briefly review the determination of the thermodynamics of $\N=4$ SYM theory.  The quantities that we calculate are accessible in QCD, via lattice calculations as we have reviewed in Section~\ref{sec:latticeQCD}, meaning that in Section \ref{sec5.1} we will be able to compare calculations done in $\N=4$ SYM theory via gauge/string duality to reliable information about QCD.
In Section~\ref{sec:TransportProperties} we turn to transport coefficients like the shear viscosity $\eta$, which govern the relaxation of small deviations away from thermodynamic equilibrium.  Lattice calculations of such quantities are in their infancy but, as we have seen 
in Section~\ref{sec:EllipticFlow}, phenomenological analyses of collective effects in heavy ion collisions in comparison to relativistic, viscous, hydrodynamic calculations are yielding information about $\eta/s$ in QCD.
Finally, Section \ref{sec:Quasi-particlesSpectralFunctions} will be devoted to illustrating one of the most important qualitative differences between the strongly coupled $\N=4$ plasma and  any weakly coupled plasma: the absence of quasiparticles. As we will argue in this Section, this is a generic feature of strong coupling which, at least at a qualitative level, provides a strong motivation in the context of the physics of QCD above $T_c$
for performing studies 
within the framework of gauge/string duality.

\bigskip

\section{Thermodynamic properties}

\label{sec5.1}
\subsection{Entropy, energy and free energy}
As discussed in Section~\ref{sec4.1.4},  $\N=4$ SYM theory in equilibrium at nonzero temperature
is described in the gravity theory by introducing black-branes  which change the AdS$_5$ metric to the  black-brane metric \eqn{AdSfiniteZ} and lead to  the formation of an event-horizon at position $z_0$.  As in standard black hole physics, the presence of the horizon allows us to compute the entropy in the gravity description, which is given by  
the Bekenstein-Hawking formula 
\be
S_{\lam=\infty} = S_\mt{BH} = {A_3 \ov 4 G_5} \,,
\label{five}
\ee
where $A_3$ is the area of the 3-dimensional event horizon of the non-compact part of the metric and  $G_5$ is the five-dimensional Newton constant. This entropy is identified as the entropy in the strong-coupling limit
 \cite{Gubser:1996de}. 
 The area $A_3$ is determined from the horizon metric, obtained by setting 
$t=\mbox{const}, z=z_0$ in eqn.~\eqn{AdSfiniteZ}, \ie
\be
ds^2_\mt{Hor} = 
\frac{R^2}{z_0^2} \left( dx_1^2 + dx_2^2 + dx_3^2 \right) \,. 
\ee
The total horizon area is, then, 
\be
A_3 = {R^3 \ov z_0^3} \int dx_1 dx_2 dx_3 \,, 
\ee
 where $\int dx_1 dx_2 dx_3$ is the volume in the gauge theory. While the total entropy is infinite, the
 entropy density per unit gauge-theory volume is finite and is given by 
\be
s_{\lam=\infty} = {S_\mt{BH} \ov \int dx_1 dx_2 dx_3} = 
{R^3 \ov 4 G_5 z_0^3} =  {\pi^2 \ov 2} \nc^2 T^3 \,,
\label{enDs}
\ee
where in the last equality we have used Eqs.~\eqn{G5} and \eqn{eq:haw} to translate the gravity parameters $z_0$, $R$ and $G_5$ into the gauge-theory parameters $T$ and $\nc$. Note that  we would have obtained the same result if we had used the full ten-dimensional geometry, which includes the $S^5$. In this case the horizon would have been 8-dimensional, of the form $A_8 = A_3 \times S^5$, and the entropy would have taken the form
\be
S_\mt{BH} = {A_8 \ov 4 G} = {A_3 V_{S^5}  \ov 4 G} \,,
\ee
which equals \eqn{five} by virtue of the relation \eqn{G5} between the 10- and the 5-dimensional Newton constants.

Once the entropy density is known, the rest of the thermodynamic potentials  are obtained through  standard thermodynamic  relations. In particular, the pressure  $P$ obeys  $s=\partial P/\partial T$, and the energy density is given by $\varepsilon = -P + T s$. Thus we find:
  \be
  \varepsilon_{\lam=\infty} = {3 \pi^2 \ov 8} \nc^2 T^4, \qquad 
  P_{\lam=\infty} =  { \pi^2 \ov 8} \nc^2 T^4 \,.
  \ee
The $\nc$ and temperature dependence of these results could have been anticipated.
The former follows from the fact that the number of degrees of freedom in an $SU(\nc)$ gauge theory in its deconfined phase grows as $\nc^2$, whereas the latter follows from dimensional analysis, since the temperature is the only scale in the $\caln=4$ theory. What is remarkable about these results is that they show that the prefactors in front of the $N_c$ and temperature dependence in these thermodynamic quantities attain finite values in the limit of infinite coupling, 
$\lambda \ra \infty$, which is the limit in which the gravity description becomes strictly applicable. 

It is instructive to compare the above expressions at infinite coupling with those for the free $\sN=4$ SYM theory, \ie at $\lam =0$. 
Since $\sN=4$ SYM has $8$ bosonic and $8$ fermonic adjoint degrees of freedom and since 
the contribution of each boson to the free entropy is  
 $2 \pi^2 T^3/45$ whereas the contribution of each fermion is $7/8$ of that of a boson, the zero coupling entropy is given by
\be
\label{s:free}
s_{\lam =0} = \le(8 + 8 \times {7 \ov 8} \ri) {2 \pi^2 \ov 45} 8(\nc^2-1) T^3
\simeq {2 \pi^2\ov 3 } \nc^2 T^3 \,,
\ee
where in the last equality we have used the fact that $\nc \gg 1$. 
As before, the $\nc$ and $T$ dependences are set by general arguments. The only difference between
the infinite and zero coupling entropies is an overall numerical factor: comparing Eqs.~(\ref{enDs}) and (\ref{s:free}) we find \cite{Gubser:1996de}
\be \label{ratio}
{s_{\lam = \infty} \ov  s_{\lam = 0} } = 
{P_{\lam = \infty} \ov  P_{\lam = 0} } = 
{\varepsilon_{\lam = \infty} \ov  \varepsilon_{\lam = 0} } = 
{3 \ov 4} \,.
\ee
This is a very interesting result: while the coupling of $\N=4$ changes radically between the two limits, the thermodynamic potentials vary very mildly. 
This observation is, in fact, not unique to the special case of $\N=4$ SYM theory, but seems to be a generic phenomenon for field theories with a gravity dual. In fact, in 
Ref.~\cite{Nishioka:2007zz} it was found that for several different classes of theories, each encompassing infinitely many instances, the change in entropy 
between the infinitely strong and infinitely weak coupling limit is 
\be
{s_{\rm strong} \ov s_{\rm free}} = {3 \ov 4} h  \,,
\label{DefnOfh}
\ee
with $h$ a factor of order one, ${8 \ov 9} \leq h \leq 1.09662$. These explicit calculation strongly suggest that the thermodynamic potentials  of non-Abelian gauge-theory plasmas (at least for near-conformal ones) are quite insensitive to the particular value of the gauge coupling. This is particularly striking since, as we will see in Sections~\ref{sec:TransportProperties} and \ref{sec:Quasi-particlesSpectralFunctions},  the transport properties of these gauge theories change dramatically 
as a function of coupling, going from a nearly ideal gas-like plasma of quasiparticles at weak coupling to a nearly ideal liquid with no quasiparticles at strong coupling.   So, we learn an important lesson from the calculations of thermodynamics at strong coupling via gauge/string duality:  thermodynamic quantities are not good observables for distinguishing a weakly coupled gas of quasiparticles from a strongly coupled liquid;  transport properties and the physical picture of the composition of the plasma are completely different in these two limits, but no thermodynamic quantity changes much.

Returning to the specific case of ${\cal N}=4$ SYM theory, in this case the leading finite-$\lambda$ correction to (\ref{ratio}) has been calculated, yielding~\cite{Gubser:1998nz}
\be \label{ratioCorrection}
{s_{\lam = \infty} \ov  s_{\lam = 0} } = 
{P_{\lam = \infty} \ov  P_{\lam = 0} } = 
{\varepsilon_{\lam = \infty} \ov  \varepsilon_{\lam = 0} } = 
\frac{3}{4}+\frac{1.69}{\lambda^{3/2}}+\ldots \,,
\ee
suggesting that this ratio increases from 3/4 to 7/8 as $\lambda$ drops from infinity 
down to $\lambda\sim 6$, corresponding to $\alpha_{\rm SYM}\sim 0.5 /N_c$.  This reminds us that the control parameter for the strong-coupling approximation is $1/\lambda$, meaning that it can be under control down to small values of $\alpha_{\rm SYM}$.  The ratio (\ref{ratio}) is also expected to receive corrections of order $1/N_c^2$, but these have not been computed.

It is also interesting to compare (\ref{ratio})  to what we know about QCD thermodynamics from lattice calculations like those described in Section~\ref{sec:latticeQCDEoS}.  
The ratio (\ref{ratio}) has the advantage that the leading dependence on the number of degrees of freedom 
drops out, making it meaningful to compare directly to QCD.
While theories which have been analyzed in Ref.~\cite{Nishioka:2007zz}  are rather different from
QCD, the regularity observed in these theories compel us to evaluate the 
ratio of the entropy density computed in the lattice calculations to that which would 
be obtained for free quarks and gluons.
Remarkably, Fig.~\ref{latticeresults} shows that  for  $T=(2-3) \, T_c$, the coefficient defined in (\ref{DefnOfh}) is $h\simeq 1.07$, which is 
in the ballpark of what the calculations done via gauge/gravity duality have taught us to expect for a strongly coupled gauge theory.  While this observation is interesting, by itself it is not strong evidence that the QCD plasma at these temperatures is strongly coupled.  The central lesson is, in fact, that the ratio (\ref{ratio}) is quite insensitive to the coupling.  The proximity of the lattice results to the value for free quarks and gluons should never have been taken as indicating that the quark-gluon plasma at these temperatures is a weakly coupled gas of quasiparticles. And, now that experiments at RHIC that we described in Section~\ref{sec:EllipticFlow} combined with calculations that we shall review in Section~\ref{sec:TransportProperties} have shown us a strongly coupled QCD plasma, the even closer proximity of the lattice results for QCD thermodynamics to that expected for a strongly coupled gauge theory plasma should also not be overinterpreted.




\subsection{Holographic susceptibilities}
\label{holsuscep}

The previous discussion focussed on a plasma at zero chemical potential $\mu$. While gauge/gravity duality allows us to explore the phase diagram of the theory at nonzero values of $\mu$, 
in order to parallel our discussion of QCD thermodynamics  in Section~\ref{sec:latticeQCD}, in our analysis of strongly coupled
$\N=4$ SYM theory here  we will concentrate on the calculation of susceptibilities. 
 As explained in Section~\ref{susceplat}, their study requires the introduction of $U(1)$ conserved charges. In $\N$=4 SYM, there is an $SU(4)$ global symmetry, the $R$-symmetry, which in the 
 dual gravity theory corresponds to rotations in the 5-sphere. Chemical potential is introduced by studying
 rotating black-holes  in these 
 coordinates~\cite{Russo:1998mm,Cvetic:1999ne,Kraus:1998hv,Gubser:1998jb}; 
 these solutions demand non-vanishing values of an abelian vector potential $A_\mu$ in the gravitational theory which, in turn, lead  to a non-vanishing 
 charge density $n$ in the gauge theory proportional to the angular momentum density of the black hole. The chemical potential can be extracted from the boundary value of the temporal component of the  Maxwell field as in (\ref{AmuChemPotential}) and 
 is also a function of the angular momentum of the black hole. 
 The explicit calculation performed in Ref.~\cite{Son:2006em}
 leads to 
 \be
 \label{nofmuAdS}
 n=\frac{N_c^2 T^2}{8} \mu \, ,
 \ee
 in the small chemical potential limit.
 Note that, unlike in QCD, the susceptibility $dn/d\mu$ inferred from  Eq.~(\ref{nofmuAdS}) is proportional to $N^2_c$
instead of $N_c$. This is a trivial consequence of the fact that R-symmetry operates over adjoint degrees of freedom. 

As in the case of the entropy, the different number of degrees of freedom can be taken into account  by comparing the susceptibility at strong coupling to the free theory result, which yields
 \be
 \frac{\chi_{\lam=\infty}}{\chi_{\lam = 0}}=\frac{1}{2} \, ,
 \ee  
 where $\chi_{\lam=0}=N_c^2 T^2/4$ \cite{Teaney:2006nc}.  
 Similarly to the case of the entropy density, the ratios of susceptibilities between these two extreme limits saturates into an  order one constant. In spite of the radical change in the dynamics of the degrees of freedom in the two systems, all the variation in this observable is a 50\% reduction, comparable to the 25\% reduction of the energy density in the same limit. 
On the other hand, the lattice calculations in Section~\ref{susceplat} show
that above $T_c$ the susceptibilities approach closer to their values
in the free limit more quickly than in the case of the energy density.
 In fact, the rapid approach of the susceptibilities to the free limit has been interpreted by some as a sign that the QCD quark-gluon plasma is not strongly 
 coupled~\cite{Koch:2005vg,Sasaki:2006ww,Bluhm:2008sc,Petreczky:2009at}. 
Another possible interpretation of this fact could be that while the gluonic (adjoint) degrees of freedom in the plasma are strongly coupled, the fermionic (fundamental) ones remain quasi-free. This interpretation is based on the observation that for fermions 
the lowest Matsubara frequency, $\omega_F=\pi T$, is different than that for gluons, $\omega_G=0$,
and thus the thermodynamics of fermions are insensitive to the 
softer components of the gluon fields. 
This possible interpretation of the discrepancy between the $\N=4$ SYM susceptibilities and the QCD susceptibilities would hinge on the fermions in the QCD plasma remaining weakly coupled at scales $\sim \pi T$ while softer gluonic modes become strongly coupled.  We note that if this interpretation is taken literally, the quarks in the QCD plasma should increase its $\eta/s$ which does not seem to be indicated by the elliptic flow data.   
Nevertheless, it is not clear whether this deviation poses a serious challenge to the interpretation of the QCD plasma as strongly coupled, given the manifest insensitivity of thermodynamic quantities to the coupling. Furthermore, the value of this deviation is not universal and
it can be different for other holographic models which are closer to QCD than $\mathcal{N}=4$ SYM.
It is in fact conceivable that if it were possible to use gauge/string duality to analyze strongly coupled theories 
with $N_f\sim N_c$ and compute the susceptibility for a $U(1)$ charge carried by the fundamental degrees of freedom in such a theory, it would turn out to be close to that at weak coupling as in QCD, even when all degrees of freedom are strongly coupled.  Were this speculation to prove correct, it would be an example of a result from QCD leading to insight into strongly coupled gauge theories with a gravitational description, i.e. it would be an example of using the duality in the opposite direction from that in which we apply it throughout most of this review. 
We shall not speculate further on this discrepancy, simply acknowledging that its interpretation is an open question.


\section{Transport properties}
\label{sec:TransportProperties}

We now turn to the calculation of the transport coefficients
of a strongly coupled plasma with a dual gravitational description, which control how
such a plasma responds to small deviations from 
equilibrium.
 As we have reviewed in Section \ref{sec:latticeTransport}, since the relaxation of these perturbations is intrinsically a real time process, the lattice determination of transport coefficients is very challenging. While initial steps toward determining them in QCD have been taken, definitive results are not in hand. As a consequence, the determination of transport coefficients via gauge/string duality is extremely valuable since it  opens up their analysis in a regime which is not tractable otherwise. 
A remarkable consequence of this analysis,  which we review in Section \ref{universalshear}, is a universal relation between the shear viscosity and the entropy density for the plasmas in all strongly  coupled large-$N_c$ gauge  theories with a gravity dual~\cite{Kovtun:2003wp,Buchel:2003tz,Kovtun:2004de,Buchel:2004qq}. This finding, together with the comparison of the universal result $\eta/s=1/(4\pi)$ with values extracted by comparing data on elliptic flow in heavy ion collisions to analyses in terms of viscous hydrodynamics as we have reviewed in Section~\ref{sec:EllipticFlow},  
has been one of the most influential results obtained via the gauge/string duality.

\subsection{A general formula for transport coefficients}
\label{generaltransportads}

The most straightforward way in which transport coefficients can be determined using the gauge/gravity correspondence is via Green-Kubo formulas, see Appendix \ref{sec:Green-Kubo}, which rely on the analysis of the retarded correlators in the field theory at small four-momentum. The procedure for determining these correlators using the correspondence has been outlined in 
Section \ref{sec:CORR}.  In this section we will try to keep our analysis as general as possible so that it can be used for the transport coefficent that describes the relaxation of 
any conserved current in the theory. In addition, we will not restrict ourselves to the particular form of the metric (\ref{AdSfiniteZ}) so that our discussion can be applied to any theory with a gravity dual. Our discussion will closely follow the formalism developed 
in \cite{Iqbal:2008by}, which builds upon earlier analyses in Refs.~\cite{Policastro:2001yc,Kovtun:2003wp,Buchel:2003tz,Kovtun:2004de,Buchel:2004qq,Saremi:2007dn,Starinets:2008fb,Fujita:2007fg}.

In general, if the field theory at nonzero temperature  is invariant under  translations and rotations, the gravitational theory will be described by a $(4+1)$-dimensional metric of the form
  \be \label{bkG}
 ds^2 =  - g_{tt} dt^2 + g_{zz} dz^2 +g_{xx} \delta_{ij} d x^i dx^j
 = g_{MN} dx^M dx^N
 \ee
 with all the metric components solely dependent on $z$. Since a nonzero  temperature is characterized in the dual theory by the presence of an event horizon, we will assume that 
  $g_{tt}$ has a first order zero and $g_{zz}$ has a first order pole at a particular value $z=z_0$.

We are interested in computing the transport coefficient $\chi$ associated with  
some operator ${\cal O}$ in this theory, namely
\be \label{trcotext}
\chi = -\lim_{\om \to 0} \lim_{\vec k \to 0} {1 \ov \om} \,{\rm Im} \,G_R (\om, {\bf k}) \, .
\ee
(See Appendix \ref{sec:Green-Kubo} for the exact definition of $G_R$ and for a derivation of this formula.) For concreteness, 
we assume that the quadratic effective action for the bulk mode $\phi$ dual to ${\cal O}$ has the form of a massless scalar field\footnote{Note that restricting to a massless mode
does not result in much loss of generality, since almost all transport coefficients calculated to date
are associated with operators whose gravity duals are massless fields. The only exception is the bulk viscosity.}
  \begin{equation} \label{ree1}
S = - \frac{1}{2} \int d^{d+1}x \sqrt{-g}\, {1 \ov q (z)} \, g^{MN} \p_M \phi \p_N \phi,
\end{equation}
where $q(z)$ is a function of $z$ and can be considered a spacetime-dependent coupling constant. As we will see below, Eqs.~\eqref{bkG} and~\eqref{ree1} apply to various examples of interest including the shear viscosity and the momentum broadening for the motion of a heavy quark in the plasma.
Since transport coefficients are given by the  Green-Kubo formula  Eq.~(\ref{trcotext}),  the general expression for the retarded correlator  (\ref{RetB}) with $\De=d$ since $m=0$ leads to
 \be \label{coe}
 \chi =  -\lim_{k_\mu \to 0} \lim_{z \to 0} {\rm Im} \left\{{\Pi (z, k_\mu) \ov  \om \phi (z, k_\mu)} \right\}
= -\lim_{k_\mu \to 0} \lim_{z \to 0} {\Pi (z, k_\mu) \ov i \om \phi (z, k_\mu)} \, ,
\ee
where $\Pi$ is the canonical momentum of the field $\phi$:
\be
\label{eom1}
  \Pi  = \frac{\delta S}{\delta \del_z \phi} =-{\sqrt{-g} \ov q(z)} \, g^{zz}\partial_z\phi \,.
\ee
The last equality in (\ref{coe}) follows from the fact that the real part of $G_R(k)$ vanishes faster than linearly in $\omega$ as $k \to 0$, as is proven by the fact that the final result that we will obtain, Eq.~\eqn{coe1}, is finite and real.

In (\ref{coe}) both $\Pi$ and $\phi$ must be solutions of the 
 classical equations of motion which, in the Hamiltonian formalism, are given by 
 (\ref{eom1}) together with 
 \be
  \p_z \Pi  =-{\sqrt{-g} \ov q(z)} \, g^{\mu \nu} k_\mu k_\nu \phi \, .
   \label{eom2}
 \ee
 The evaluation of $\chi$, following 
 Eq.~(\ref{coe}), requires the determination of both $\om \phi$ and $\Pi$ 
 in the small four momentum $k_\mu \to 0$ limit . Remarkably, in this limit the  equations of motion (\ref{eom1}) and (\ref{eom2}) are trivial
\begin{equation} \label{flow_scalar}
\partial_z \Pi = 0 + \mathcal{O}(k_\mu \om \phi)\, , \qquad \partial_z(\om \phi) = 0 + \mathcal{O}(\omega\Pi) \ ,
\end{equation}
and both quantities become independent of $z$, which allows their evaluation at any $z$. For simplicity, and since the only restriction on the general metric (\ref{bkG})  is that it possesses a horizon,
we will evaluate them at $z_0$ where the in-falling boundary condition should be imposed. 
Our assumptions on the metric imply that in the vicinity of the horizon   $z \to z_0$
 \be
 g_{tt} = -c_0 (z_0-z), \qquad g_{zz} = {c_z \ov z_0-z} \, ,
 \ee
 and eliminating $\Pi$ from (\ref{eom1}) and (\ref{eom2}) we find an equation for $\phi$ given by
 \be
 \sqrt{c_0 \ov c_z} (z_0-z) \p_z \le(\sqrt{c_0 \ov c_z} (z_0-z) \p_z \phi \ri)
 + \om^2 \phi =0 \,.
 \ee
  The two general solutions for this equation are  
 \be
 \label{torN}
 \phi\propto e^{-i\om t} \left(z_0-z\right) ^{\pm i \om \sqrt{c_z/c_0} }\, .
 \ee
 Imposing  in-falling boundary condition implies that we should take the negative sign in the exponent. Therefore, from  Eq.~(\ref{torN}) we find that at the horizon
\be
\p_z \phi = \sqrt{{g_{zz} \ov -g_{tt}}} (i \om \phi) \, ,
\ee
and using Eqs.~(\ref{eom1}) and (\ref{torN}) we obtain
  \be \label{coe1}
 \chi = -\lim_{k_\mu \to 0} \lim_{z \to 0} {\Pi (z, k_\mu) \ov i \om \phi (z, k_\mu)} =-\lim_{k_\mu \to 0} \lim_{z \to z_0} {\Pi (z, k_\mu) \ov i \om \phi (z, k_\mu)}
 = {1 \ov q(z_0)} \sqrt{\frac{-g} {-g_{zz}g_{tt}}} \biggr|_{z_0} \, . 
\ee
Note that the last equality in (\ref{coe1}) can also be written as 
\be \label{ero}
\chi =  {1 \ov q (z_0)} {A \ov V} \, ,
\ee
where $A$ is the area of the horizon and $V$ is the spatial volume of the boundary theory. From our  analysis of the thermodynamic properties of the plasma in Section~\ref{sec:BulkDynamicalProperties}, the area of the event horizon is related to the 
entropy of the boundary theory via
 \be
 s = {A \ov V}{1 \ov 4 G_N} \, .
 \ee
 From this analysis we  conclude that for any theory with a gravity dual the ratio of any transport coefficient to the 
 entropy density depends solely on the properties of the dual fields at the horizon, 
  \be  \label{eyao}
  {\chi \ov s} = {4 G_N \ov q(z_0)} \ .
  \ee
  In the next subsection we will use this general expression to compute the shear viscosity of the AdS plasma.

Finally, we would like to remark  that the above discussion applies to more general effective actions of the form
\be \label{moreG}
 S = -\frac{1}{2}\int \frac{d\omega d^{d-1}k}{(2\pi)^d} dz\sqrt{-g}\left[\frac{g^{zz}(\partial_z\phi)^2}{Q(z; \omega,k)} + P(z; \omega,k)\phi^2\right],
 \ee
provided that the equations of motion (\ref{flow_scalar}) remain trivial in the zero-momentum limit. This implies that $Q$ should go to a nonzero constant at zero momentum  and $P$ must be at least quadratic in momenta. For (\ref{moreG}) the corresponding transport coefficient $\chi$ is given by
\be \label{ratio1}
 \chi =  {1 \ov Q (z_0, k_\mu=0)} {A \ov V}  \qquad {\rm and} \qquad {\chi \ov s} = {4  G_N \ov Q (z_0, k_\mu=0)} \ .
 \ee

\subsection{Universality of the Shear Viscosity}
\label{universalshear}

We now apply the result of last subsection to the calculation of the shear
viscosity $\eta$ of a strongly coupled plasma described by the metric
(\ref{bkG}). As in Appendix \ref{sec:Green-Kubo}  we must compute the correlation function of the operator ${\cal O} = T_{xy}$, where the coordinates $x$ and $y$ are orthogonal to the momentum vector. The bulk field $\phi$ dual to ${\cal O}$ should have a metric perturbation $h_{xy}$ as its boundary value. It then follows that $\phi =  (\delta g)^x_y = g_{zz} h_{xy}$, where $\delta g$ is the perturbation of the bulk metric. For Einstein gravity in a geometry with no off-diagonal components in the background metric, as in (\ref{bkG}), 
a standard  analysis of the Einstein equations to linear order in the perturbation upon assuming that the momentum vector is perpendicular to the $(x,y)$ plane shows that
the effective action for $\phi$ is simply that of a minimally coupled massless scalar field, namely
 \be
 \label{Sclphi}
 S = - {1 \ov 16 \pi G_N } \int d^{d+1} x \, \sqrt{-g} \,  \left[ \ha g^{MN} \p_M \phi \p_N \phi \ri] \,.
 \ee
The prefactor $1 / 16 \pi G_N $ descends from that of the Einstein-Hilbert action.
This action has the form of Eq.~(\ref{ree1}) with
\be \label{eei}
q (z) = 16 \pi G_N = {\rm const} \ ,
\ee
which, together with  Eq.~(\ref{eyao}), leads to 
the celebrated result
  \be \label{etas}
  {\eta_{\lambda=\infty} \ov s_{\lambda=\infty}} = {1 \ov 4 \pi}\,
  \ee
that was first obtained in 2001 by Policastro, Son and Starinets~\cite{Policastro:2001yc}.
In (\ref{eei}), we have added the subscript $\lambda=\infty$ to stress that the numerator and denominator are both
computed in the strict
infinite coupling limit. Remarkably, this ratio converges to a constant at strong coupling. And, 
 this is not 
only a feature of $\N=4$ SYM theory because 
 this 
 derivation applies to any gauge theory  with a gravity dual given by
Einstein gravity coupled to matter fields, since in Einstein gravity the coupling constant for gravity is always given by Eq.~(\ref{eei}). In this sense, this result is universal \cite{Kovtun:2003wp,Buchel:2003tz,Kovtun:2004de,Buchel:2004qq} since it applies in the strong-coupling and large-$N_c$ limits  to the large class of theories with a gravity dual, regardless of whether the theories  are conformal or not, confining or not, supersymmetric or not and with or without chemical potential. In particular, if large-$N_c$ QCD has a gravity dual, 
its $\eta/s$ should also be given by $1/(4\pi)$ up to corrections due to the finiteness of the coupling.  Even if large-$N_c$ QCD does not have a gravity dual, Eq.~\ref{etas} may still apply since
the universality of this result may be due to generic properties of strongly coupled theories (for example the absence of quasiparticles, see Section~\ref{sec:Quasi-particlesSpectralFunctions}) 
which may not depend on whether they are dual to a gravitational theory. 

The finite-coupling corrections to Eq.~(\ref{etas}) in ${\cal N}=4$ SYM theory are given 
by\cite{Buchel:2004di,Buchel:2008ac,Buchel:2008sh,Myers:2008yi}
\be
\label{etasfc}
  {\eta_{\lambda\rightarrow \infty } \ov s_{\lambda \rightarrow \infty}} =
  \frac{1}{4\pi} \left(1 + \frac{15\, \zeta(3)}{\lambda^{3/2}} + ...\right) \, ,
\ee
where $\zeta(3)=1.20...$ is Ap\'ery's constant. 
While the coefficient of the leading finite-$\lambda$ correction proportional to $1/\lambda^{3/2}$ 
is universal 
 when expressed in terms
of the parameters $R$ and $l_s$ in the gravity theory, 
this coefficient is different for different gauge theories~\cite{Buchel:2008ae}, since the
relation between $R$ and $l_s$ and the gauge theory parameters $\lambda$ and $N_c$ changes. Thus, expression  (\ref{etasfc}) is only valid for $\N=4$ SYM theory. 
It is interesting to notice that, according to  Eq.~(\ref{etasfc}), $\eta/s$ increases 
to $\sim 2/(4\pi)$ once $\lambda$ decreases to $\lambda\sim 7$, meaning 
$\alpha_{\rm SYM}\sim 0.5/N_c$. This is the same range of couplings
at which the finite coupling corrections (\ref{ratioCorrection}) 
to thermodynamic quantities become significant.    These results together suggest that 
strongly coupled theories with gravity duals may yield insight into the quark-gluon plasma in QCD even down to apparently rather small values of $\alpha_s$, at which $\lambda$ is still large.  

To put the result (\ref{etas})  into further context, we can compare this strong coupling result to results for the same ratio $\eta/s$ at weak coupling in both $\N=4$ SYM theory and QCD.  These have been computed at next to leading 
log accuracy, and take the form
\be
\label{etaspert}
  {\eta_{\lambda\rightarrow 0 } \ov s_{\lambda \rightarrow 0}} =
  \frac{A}{\lambda^2 \log \left(B/\sqrt{\lambda}\right)} \, ,
\ee
with $A=6.174$ and $B=2.36$ in $\N=4$ SYM theory and $A=34.8$ (46.1) and $B=4.67$ (4.17) in QCD with $N_f=0$ ($N_f=3$)~ \cite{Arnold:2003zc,Huot:2006ys}, where we have defined $\lambda=g^2 N_c$ in QCD as in $\N=4$ SYM theory.    
Quite unlike the strong coupling result (\ref{etas}), these weak-coupling results show a strong dependence on $\lambda$, and in fact diverge in the weak-coupling limit.
The divergence reflects the fact that a weakly-coupled gauge-theory plasma is a gas of quasiparticles, with strong dissipative effects.  In a gas, $\eta/s$ is proportional to the ratio of the mean-free path of the quasiparticles to their average separation. A large mean-free path, and hence a large $\eta/s$, mean that momentum can easily be transported over distances that are long compared to the average spacing between particles.
In the $\lambda\rightarrow 0$ limit the mean-free path diverges.  The strong $1/\lambda^2$  dependence of $\eta/s$ can be traced to the fact that the two-particle scattering cross-section is proportional to $g^4$.
It is reasonable to guess that the $\lambda$-dependence of $\eta/s$ in $\N=4$ SYM theory 
is monotonic, increasing from $1/(4\pi)$ as in (\ref{etasfc}) as $\lambda$ decreases from $\infty$ and then continuing to increase until it diverges according to (\ref{etaspert}) as $\lambda\rightarrow 0$.
The weak-coupling result (\ref{etaspert}) also illustrates a further important point:  $\eta/s$ is not universal for weakly-coupled gauge theory plasmas.  The coefficients $A$ and $B$ can vary significantly from one theory to another, depending on their particle content.  It is only in the strong coupling limit that universality emerges, with all large-$N_c$ theories with a gravity dual having plasmas with $\eta/s=1/(4\pi)$.
And, we shall see in Section~\ref{sec:Quasi-particlesSpectralFunctions} that a strongly coupled gauge theory plasma does not have quasiparticles, which makes it less surprising that $\eta/s$ at strong coupling is independent of the particle content of the theory at weak coupling.


One lesson from the calculations of  $\eta/s$ is that this quantity changes significantly with the coupling constant, going from infinite at zero coupling to $1/(4\pi)$ at strong coupling, at least for large-$N_c$ theories with gravity duals.  This is in marked contrast to the behavior of the thermodynamic quantities described in Section~\ref{sec:BulkDynamicalProperties}, which change only by 25\% over the same large range of couplings.  Thermodynamic observables are insensitive to the coupling, whereas $\eta/s$ is a much better indicator of the strength of the coupling 
because it is a measure of whether the plasma is liquid-like or gaseous.


These observations 
prompt us to revisit the phenomenological extraction of the shear viscosity of quark-gluon plasma
in QCD 
from measurements of elliptic flow in heavy ion collisions, described in  
Section~\ref{sec:EllipticFlow}.
As we saw, the comparison between data and calculations done using relativistic viscous hydrodynamic yields a conservative conclusion that
$\eta/s <(3-5)/(4\pi)$, with a current estimate 
being that $\eta/s$ seems to lie within the range $(1-2.5)/(4\pi)$ in QCD, in the same ballpark as the strong-coupling result (\ref{etas}).
And, as we reviewed in Section~\ref{sec:latticeTransport}, current lattice calculations of $\eta/s$ in $N_f=0$ QCD come with caveats but also indicate a value that is in the ballpark of $1/(4\pi)$, likely somewhat above it.
Given the sensitivity of $\eta/s$ to the coupling, these comparisons constitute one of the main
lines of evidence that, in the temperature regime accessible at RHIC, the quark-gluon plasma is a strongly coupled fluid.   If we were to attempt to extrapolate the weak-coupling result (\ref{etaspert}) 
for $\eta/s$ in QCD with $N_f=3$ to the values of $\eta/s$  
favored by experiment, we would need $\lambda \sim (14-24)$, well beyond the regime of applicability of perturbation theory.  (To make this estimate we had to set the $\log$ in (\ref{etaspert}) to 1 to avoid negative numbers, which reflects the fact that the perturbative result is being applied outside its regime of validity.)

A central lesson from the strong-coupling calculation of $\eta/s$ via gauge/string duality,
arguably even more significant than the qualitative agreement between the result (\ref{etas}) and current extractions of $\eta/s$ from heavy ion collision data, is simply the fact that values of 
$\eta/s\ll 1$ are possible in non-abelian gauge theories, and in particular in non-abelian gauge theories whose thermodynamic observables are not far from weak-coupling expectations.
These calculations, done via gauge/string duality, provided theoretical support for considering a range of small values of $\eta/s$ that had not been regarded as justified before,
and inferences drawn from RHIC data have  pushed $\eta/s$ into this regime.
The computation of the shear viscosity that we have just reviewed is one of the most influential results supporting the notion that
the application of gauge/gravity duality can yield insights into the phenomenology of 
hot QCD matter.

It has  also been conjectured~\cite{Kovtun:2004de} that the value of $\eta/s$  in Eq.~(\ref{etas}) 
is, in fact, a lower bound for all systems in nature. This conjecture is supported by the finite-coupling corrections 
shown in Eq.~(\ref{etasfc}).   And, 
all substances known in the laboratory satisfy the bound.  Among conventional liquids, the 
lowest $\eta/s$ is achieved by liquid helium, but it is about an order of magnitude above $1/(4\pi)$;  water --- after which hydrodynamics is named --- has an $\eta/s$ that is larger still, by about another order of magnitude.  The best liquids known in the laboratory are the quark-gluon plasma produced in heavy ion collisions and an ultracold gas of fermionic atoms at the unitary point, 
at which the $s$-wave atom-atom scattering length has been dialed 
to infinity~\cite{Schafer:2009dj}, both of which have $\eta/s$ in the ballpark of $1/(4\pi)$ but, according to current estimates, somewhat larger.




However, in recent years the conjecture that (\ref{etas}) is a lower bound on $\eta/s$
has been questioned and counter-examples have been found among theories with 
gravity duals.
As emphasized in Section \ref{AdS/CFT},  Einstein gravity in the dual gravitational description
corresponds to the large-$\lam$ and
large-$N_c$ limit of the boundary gauge theory.
When higher order corrections to Einstein gravity are included, which correspond
to $1/\sqrt{\lam}$ or $1/N_c$ corrections in the boundary gauge theory, Eq.~(\ref{etas}) 
will no longer be universal. 
In particular, as pointed out in Refs.~\cite{Brigante:2007nu,Kats:2007mq} and generalized 
in Refs.~\cite{Brustein:2008cg,Brustein:2008xx,Cai:2008ph,Ge:2009ac,Pal:2009qg,Shu:2009ax,Amsel:2010aj,Brustein:2010ku,Camanho:2010ru}, generic higher derivative corrections to Einstein gravity can violate the proposed bound. 
Eq.~(\ref{ratio1}) indicates that $\eta/s$ is smaller than
Eq.~(\ref{etas}) if  the ``effective'' gravitational coupling for the $h^{x}_y$ polarization at the horizon is {\it stronger} than the universal value (\ref{eei}) for Einstein gravity. Gauss-Bonnet gravity as discussed in Refs.~\cite{Brigante:2007nu,Brigante:2008gz} is an example in which this occurs. There, the effective action for $h_x^y$ has the form of Eq.~(\ref{moreG}) with the effective coupling $Q(r)$ at the horizon satisfying~\cite{Brigante:2007nu}
 \be
 \frac{1}{Q(r_0)} = \frac{(1-4\,\lambda_{\rm GB})}{16\pi G_N} \, ,
 \ee
 leading to
 \be
 {\eta \ov s} = \frac{(1-4\,\lambda_{\rm GB})}{4\pi}\,,
 \ee
 where $\lam_{\rm GB}$ is the coupling for the Gauss-Bonnet higher-derivative term. Thus, 
 for $\lam_{\rm GB} > 0$ the graviton in this theory is more strongly coupled
than that of Einstein gravity and the value of $\eta/s$ is smaller than $1/4\pi$.
In Ref.~\cite{Kats:2007mq}, an explicit gauge theory has been proposed whose gravity dual
corresponds to $\lam_{\rm GB} > 0$. (See Refs.~\cite{Myers:2008yi,Buchel:2008vz} for generalizations.)
Despite not being a lower bound, the smallness of $\eta/s$, the qualitative agreement 
between Eq.~(\ref{etas}) and values obtained from heavy ion collisions, and the universality of the result (\ref{etas}) which applies to any gauge theory with a gravity dual in the large-$N_c$ and strong-coupling limits, are responsible for the great impact that this calculation done via gauge/string duality has had on our understanding of the properties of deconfined QCD matter.


\subsection{Bulk viscosity}
\label{bulkviscads}

As we have discussed in  Section~\ref{sec:EllipticFlow}, while the bulk viscosity $\zeta$ is very small in the QCD plasma at  temperatures larger than  $1.5 -2\, T_c$, with $\zeta/s$ much 
smaller than $1/4\pi$, $\zeta/s$ rises in the vicinity of  $T_c$, a feature which can be important for heavy ion collisions.  Since the plasma of a conformal theory has zero  bulk viscosity, 
${\cal N}=4$ SYM theory is not a useful example to study the bulk viscosity of a strongly coupled plasma. However, the bulk viscosity has been calculated both in more sophisticated examples of the gauge/string duality in which the gauge theory is not conformal \cite{Benincasa:2005iv,Buchel:2005cv,Buchel:2007mf,Mas:2007ng,Buchel:2008uu}, as well as in AdS/QCD models
that incorporate an increase in the bulk viscosity near a deconfinement phase
transition~\cite{Gubser:2008yx,Gubser:2008sz,Gursoy:2008bu}.

We will only briefly review what is possibly the simplest among the first type of examples, the so-called D$p$-brane theory. This is a $(p+1)$-dimensional cousin of ${\cal N}=4$ SYM, namely a
$(p + 1)$-dimensional SYM theory (with 16 supercharges) living at the boundary of the geometry describing a large number of non-extremal black D$p$-branes~\cite{Itzhaki:1998dd} with $p\neq 3$. The case 
$p = 3$ is $\N = 4$ SYM, while the cases $p = 2$ and $p = 4$ correspond to non-conformal theories in $(2 + 1)$- and
$(4 + 1)$-dimensions. We emphasize that we choose this example for its simplicity rather than because it is directly relevant for phenomenology.

The metric sourced by a stack of black D$p$-branes can be written
as
 \be \label{omet}
ds^2 = \apr { (d_p \tilde\lam z^{3-p})^{1 \ov 5-p}  \ov z^2} \le(-{\tilde f} dt^2
+ ds_p^2
  + \le({2 \ov 5-p}\ri)^2 {dz^2 \ov {\tilde f}} + z^2 d \Om_{8-p}^2
  \ri)\,,
 \ee
where
  \be
\tilde\lam = g^2 N, 
\qquad {\tilde f} = 1- \le({z \ov z_0} \ri)^{14-2p \ov
5-p } , \qquad d_p = 2^{7-2p} \pi^{9-3p \ov 2}\, \Gamma
 \le( 7 - p \ov 2\ri)
  \ee
and
\be
g^2 = (2 \pi)^{p-2} g_s \alpha'^{3-p \over 2} 
\ee
is the Yang-Mills coupling constant, which is dimensionful if $\p \neq 3$. For $p=2$ and $4$ there is also a nontrivial profile for the dilaton field but we shall not give its explicit form here. The metric above is dual to 
$(p + 1)$-dimensional SYM theory at finite temperature. 

The bulk viscosity can be computed from the dual gravitational theory via 
the Kubo formula (\ref{trco}). However this computation is more complicated in the bulk  channel than in the shear channel and we will not reproduce it here.
%
%
 An alternative and simpler way to compute the bulk viscosity is based on the fact that,
 in the hydrodynamic limit, the sound mode has the following dispersion relation:
  \be
  \om = c_s q - {i \ov \ep + p} \le({d-1 \ov d-2} \eta  + {\zeta \ov 2 } \ri) q^2 + \cdots \,, 
  \ee
   with $c_s$ the speed of sound. Thus, $\zeta$ contributes to the damping of  sound. 
   In the field theory, the dispersion relation for the sound mode can be found by examining the poles of the retarded Green's function for the stress tensor in the sound channel.
   As discussed in Section~\ref{sec:eucl}, on the gravity side these poles correspond to normalizable
   solutions to the equations of motion for metric perturbations.  Due to the in-falling boundary conditions at the horizon, see  Eq.~(\ref{plane1}), 
   the frequencies of these modes have nonzero and negative imaginary parts and they are therefore called quasi-normal modes. 
   The explicit computation of these modes for the AdS black hole metric can be used to extract the dispersion relation for sound waves in the 
strongly coupled $\N=4$ SYM plasma, yielding an alternative derivation of its shear viscosity~\cite{Policastro:2002se}.  
The explicit computation of these quasi-normal modes  for the metric (\ref{omet})  
performed in Ref.~\cite{Mas:2007ng} showed that the sound mode has the dispersion relation
  \be
  \om = \sqrt{5-p \ov 9-p} q - i \,{2 \ov 9-p }\, {q^2 \ov 2 \pi T}  + \cdots
  \ee
  from which one finds that (after using $\eta/s = 1 /( 4 \pi)$)
   \be
   c_s = \sqrt{5-p \ov 9-p}, \qquad {\zeta \ov s} = { (3-p)^2 \ov 2 \pi p (9-p)} \,.
   \ee
 The above expressions imply an interesting  relation~\cite{Buchel:2007mf}
   \be
   {\zeta \ov \eta} = 2 \le({1 \ov p} - c_s^2 \ri)
   = 2 \le(c_{s,{\rm CFT}}^2 - c_s^2 \ri) \, ,
   \ee
where we have used the fact that the sound speed for a CFT in $(p+1)$-dimension is
$c_{s,{\rm CFT}} = 1/ \sqrt{p}$. This result might not seem surprising since
the bulk viscosity of 
 a theory which is close to being conformal can be expanded in powers 
of $c_{s,{\rm CFT}}^2 - c_s^2$, which is a measure of deviation from conformality. 
The nontrivial result is that even though the Dp-brane gauge theories are {\it not} close to being conformal, their bulk viscosities are nevertheless linear in $c_{s,{\rm CFT}}^2 - c_s^2$. While this is an interesting observation, it is not clear to what extent it is particular to the  Dp-brane gauge theories or whether it is more generic.

\subsection{Relaxation times and other 2nd order transport coefficients}
\label{s:relaxationAdS}

As we have reviewed in Section \ref{s:EllipticHydro}, transport coefficients correspond to the leading order gradient expansion of an interacting theory  which corrects the hydrodynamic description. 
{\it A priori}, there is no reason to stop the extraction of these coefficients at first
order, and higher order ones can be (and have been) computed using gauge/string duality. Of particular importance is the determination of the five second order coefficients, $\tau_\pi$, $\kappa$, $\lambda_1$, $\lambda_2$, $\lambda_3$ defined in Eq.~(\ref{ISlikePi}).
Unlike for the first order coefficients, the gravitational computation of these second order 
coefficients is quite  technical and we shall not 
review it here. We shall only describe the main points and refer the reader 
to Refs.~\cite{Baier:2007ix,Bhattacharyya:2008jc} for details. 

The strategy for determining these coefficients is complicated by the fact that 
the three  coefficients $\lambda_i$ involve 
only non-linear combinations of the hydrodynamic fields. Thus, even though 
 formulae can be derived for the linear 
 coefficients $\tau_\pi$ and 
$\kappa$~\cite{Baier:2007ix,Moore:2010bu}, the non-linear coefficients cannot be determined from two-point correlators, since these coefficients are invisible in the linear perturbation analysis of the background. Their determination, thus, demands the small gradient analysis of non-linear solutions to the Einstein equations\footnote{
Kubo-like formulas involving three-point correlators (as opposed to two) can also be use to determine the coefficients $\lambda_i$ \cite{Moore:2010bu}.
 At the time of writing, this approach had not been explored within the gauge/gravity context.} as performed in Ref.~\cite{Bhattacharyya:2008jc} (see also Ref.~\cite{Baier:2007ix}) which yields
 \bea\label{lotsof2ndordercoeffs}
 &\tau_\pi=\frac{2-\ln 2}{2\pi T} \, ,
 \quad
 \kappa=\frac{\eta}{\pi T} \, ,
 & 
 \\
 \lambda_1=\frac{\eta}{2\pi T} \, , \quad \quad 
 &
 \lambda_2=-\frac{\eta \ln 2}{\pi T} \, ,  \quad \quad 
 &
 \lambda_3=0\,. 
 \nonumber
 \eea
These results are valid in
the large-$N_c$ and strong-coupling limit.
Finite coupling corrections to some of these coefficients can be found in Ref.~\cite{Buchel:2008bz}.   Additionally, the first and second order coefficients have been  studied in a large class of non-conformal theories with or without flavor in Refs.~\cite{Bigazzi:2009tc,Bigazzi:2010ku}.

 To put these results in perspective we will compare them to those extracted in the weakly coupled  limit of QCD  ($\lambda \ll 1$)~\cite{York:2008rr}. We shall not comment on the values of all the coefficients, since, as discussed in Section~\ref{s:EllipticHydro}, the only one with any impact on current phenomenological applications to heavy ion collisions is the shear relaxation time 
 $\tau_\pi$. In the weak-coupling limit, 
 %
 \begin{equation}
 \label{taupipert}
 	\lim_{\lambda \to 0} \tau_\pi \simeq \frac{5.9}{T} \frac{\eta}{s}\, ,
 \end{equation}
 where the result is expressed in such a way as to show that $\tau_\pi$ and $\eta$ have  the same
 leading-order dependence on the coupling $\lambda$ 
 (up to logarithmic corrections). For comparison, 
 the strong-coupling result from (\ref{lotsof2ndordercoeffs}) may be written as
 \be
 \lim_{\lambda \gg 1} \tau_{\pi} = \frac{7.2}{T} \frac{\eta}{s}\, ,
 \ee
 which is remarkably close to (\ref{taupipert}). But, of course, the value of $\eta/s$ is vastly different in the weak- and strong-coupling limits.
%
 %
 %
 %
On general grounds, one may expect that relaxation and equilibration processes are more efficient in the strong 
coupling limit, since they rely on the interactions between different modes in the medium. 
This general expectation 
is satisfied for the shear relaxation time of the ${\cal N}=4$ SYM plasma, with $\tau_\pi$ diverging at weak-coupling and taking on the small value
 \begin{equation}\label{taupiStrongCoupling}
 	\lim_{\lambda \to \infty} \tau_{\pi} \simeq \frac{0.208}{T}\, .
 \end{equation}
 in the strong-coupling limit.
%
 %
For the temperatures $T > 200$ MeV,   which are relevant for the quark-gluon 
plasma produced in heavy ion collisions,
 this relaxation time is of the order of $0.2$ fm/c, which is much smaller than perturbative expectations. We have recalled 
 already at other places in this review that caveats enter if one seeks quantitative guidance for heavy ion phenomenology
 on the basis of calculations made for ${\cal N}=4$ SYM plasma. However, the 
 qualitative (and even semi-quantitative)
 impact of the  result (\ref{taupiStrongCoupling}) on heavy ion phenomenology should not be underestimated: 
the computation of $\tau_\pi$
demonstrated for the first time that at least some excitations in a strongly coupled non-abelian plasma dissipate on timescales that are much shorter than $1/T$, \ie
on time scales much shorter than 1 fm/c. Such small values did not have a theoretical underpinning
 before, and they are clearly relevant for phenomenological 
studies based on viscous fluid dynamic simulations.   As we reviewed in Section~\ref{sec:EllipticFlow},  the success of the comparison of such simulations to heavy ion collision data implies that a hydrodynamic description of the matter produced in these collisions is valid only $\sim 1$ fm/c after the collision. 
Although this equilibration time is related to out-of-equilibrium dynamics, whereas $\tau_\pi$ is related to near-equilibrium dynamics (only to second order), the smallness of $\tau_\pi$ makes the rapid equilibration time seem less surprising.
 As in the case of $\eta/s$, the gauge/gravity calculation of $\tau_\pi$ has made it legitimate to consider values of an important parameter that had not been considered before by showing that this regime arises in the strongly coupled plasma of a quantum field theory that happens to be accessible to reliable calculation because it possesses a gravity dual.
 
 Let us also mention that the
 second order transport coefficients are known 
 for the same non-conformal gauge theories whose bulk viscosity we discussed in
 Section \ref{bulkviscads}. 
  Since conformal symmetry is broken in these models, 
  there are a total of  15 first- and second-order transport coefficients, 9 more than 
    in the conformal case (including both shear and bulk viscosities in the counting) \cite{Romatschke:2009kr}.
  In addition, the velocity of sound $c_s$ is a further independent
 parameter that characterizes  the zeroth-order hydrodynamics of non-conformal plasmas, whose equations of state are not given simply by $P=\varepsilon/3$. 
As for the case of the bulk viscosity,  the variable
 $\left( \frac{1}{3} - c_s^2\right)$ can be used to parametrize deviations from conformality, and all 
 transport coefficients can indeed be written 
 explicitly as functions of $\left( \frac{1}{3} - c_s^2\right)$ ~\cite{Kanitscheider:2009as}. 
 
Let us conclude this section with a curious remark.  As we saw in Section \ref{s:EllipticHydro}, the second order 
hydrodynamic equations are hyperbolic and, as such, they are causal.
 However, hydrodynamic equations are a small gradient expansion of the full theory and
 there is no {\it a priori} reason why hydrodynamic dispersion relations must provide physically sensible results for short 
 wavelength, highly energetic, excitations. In particular, 
  the group velocity $\frac{d\omega(k)}{dk}$ of a hydrodynamic excitation does not need to remain smaller than the velocity 
 of light in the limit $k \to \infty$.
And yet, the 
  structure of the second order viscous hydrodynamic equations ensures that the group velocities 
  of all hydrodynamic excitations approach finite values in the limit $k \to \infty$. 
  For example, the hydrodynamic propagation speeds for the shear mode and the sound mode
  depend on $(\epsilon + p) = T\, s$, the shear and bulk viscosities, as well as the corresponding relaxation times
  $\tau_\pi$ and $\tau_\Pi$. 
  In the limit $k\to\infty$,  their group velocities take the simple and general forms
  \begin{eqnarray}
  	\lim_{k\to\infty} \frac{d\omega_{\rm sound}(k)}{dk} &=& \sqrt{\frac{\eta}{\tau_\pi \left( \epsilon + p\right)}}\, , \\
	  	\lim_{k\to\infty} \frac{d\omega_{\rm shear}(k)}{dk} &=& \sqrt{ 
											c_s^2 + \frac{4}{3}\frac{\eta}{\tau_\pi \left( \epsilon + p\right)}
													+ \frac{\zeta}{\tau_\Pi \left( \epsilon + p\right)} }\, .
  \end{eqnarray}
 For a derivation, see Ref.~\cite{Romatschke:2009im}.
  In principle, one could find a field theory with relaxation times $\tau_\pi$ and/or $\tau_\Pi$ that are so short that
the limit $\lim_{k\to\infty} \frac{d\omega(k)}{dk}$ of some mode exceeds the velocity of light; in this case one would simply state
that this limit is outside the range of validity of hydrodynamics. But, for these two modes and
for all hydrodynamic modes studied so far, and for all field theories studied so far, when the values of all quantities are plugged into these expressions,
the resulting  
$\lim_{k\to\infty} \frac{d\omega(k)}{dk}$ describes propagation within the forward light cone. We emphasize that the current understanding of this finding admits the possibility that it is accidental.

\section{Quasiparticles and spectral functions}
\label{sec:Quasi-particlesSpectralFunctions}

In Sections~\ref{sec5.1} and \ref{sec:TransportProperties} 
we have illustrated the power of gauge/string duality  by performing, in a remarkably simple way, computations that via standard field theoretical methods either take Teraflop-years of computer time or are not accessible.
However, the simplicity of the calculations comes with a price.
Because we do the calculations in the dual gravitational description of the theory, all we get are reliable results; we do not get the kind of intuition of what is happening in the gauge theory that we would get automatically from a field theory calculation done with Feynman diagrams or could get with effort from one done on the lattice.  The gravitational calculation yields answers, and new kinds of intuition, but since by using it we are abandoning the 
description of the plasma in terms of
quark and gluon quasi-particles interacting with each other, we are 
losing our prior physical intuition about how the dynamics of the gauge theory works.
In particular, from the results that we extract alone it is difficult to understand whether the dynamics within a strongly coupled plasma differs in a qualitative way from those in a weakly coupled plasma, or merely differs quantitatively.   We have given up the description in terms of quasiparticles, but maybe the familiar quasiparticles or some new kind of quasiparticles are in fact nevertheless present.  We rule out this possibility in this Section, illustrating that a strongly coupled non-abelian gauge-theory plasma really is qualitatively different than a weakly coupled one: while in perturbation theory the degrees of freedom of the plasma are long-lived quasiparticle excitations 
which carry momentum, color and flavor, 
there are no quasiparticles in the strongly coupled plasma.  The pictures that we are used to using that frame how we think about a weakly coupled plasma are simply invalid for the strongly coupled case.

Determining whether a theory possesses quasiparticles with a given set of quantum numbers is
a conceptually well defined task: it suffices to analyze the spectral function of operators with that set of quantum numbers and look for narrow peaks in momentum space. 
In weakly coupled Yang-Mills theories, the quasiparticles (gluons and quarks in QCD)   are colored
and are identified by studying 
operators that are {\it not} gauge invariant. Within perturbation theory, it can be shown that the poles of these correlators, which determine the physical properties of the quasiparticles, are gauge invariant~\cite{Braaten:1989kk}. However, non-perturbative gauge-invariant operators corresponding to these excitations are not known, which complicates the search for these 
quasiparticles at strong coupling.
 Note, however, that even if such operators were known, 
 demonstrating the absence of quasiparticles with the same quantum numbers as in the perturbative limit 
 does not  guarantee the absence of quasiparticles, since at strong coupling the system could reorganize itself into quasiparticles with different quantum numbers. Thus, proving the absence of quasiparticles along these lines would require  
exploring all possible spectral functions in the theory. Fortunately, there is an indirect method
which can answer the question of whether any quasiparticles that carry some conserved `charge' (including momentum) exist, although this method cannot determine
the quantum numbers of the long-lived excitations if any are found to exist.
The method involves the analysis of the
small frequency structure of the spectral functions of those conserved currents of the
theory which do not describe a propagating hydrodynamic mode like sound.
As we will see, the presence of quasiparticles leads to a narrow structure (the transport peak)  in these spectral functions~\cite{Aarts:2002cc,Teaney:2006nc}. In what follows we will use this method to demonstrate that the strongly coupled ${\cal N}=4$ SYM
plasma does not possess any quasiparticles that carry momentum.   In order to understand how the method works, we first apply it at weak coupling where there are quasiparticles to find.

\subsection{Quasiparticles in perturbation theory}

We start our analysis by using kinetic theory to predict the general features of the low-frequency structure of correlators of conserved currents in a weakly coupled plasma. 
Kinetic theory is governed by the Boltzmann equation, 
which  describes excitations of a quasiparticle system at scales which are long compared to 
the inter-particle separation. The applicability of the kinetic description demands that there is a separation of scales such that
the duration of interactions among particles is short
compared to their mean free path ($\lambda_{\rm mfp}$) and 
that multiparticle distributions are consequently 
determined by the single particle distributions. 
In  Yang-Mills theories at
 nonzero temperature and weak coupling,
 kinetic theory is important since it coincides with the Hard Thermal 
 Loop description~\cite{Heinz:1983nx,Braaten:1989mz,Frenkel:1989br,Taylor:1990ia,Kelly:1994dh,Blaizot:1993zk}, which is the effective field theory for physics at momentum scales of order $gT$,  and   
  the Boltzmann equation can be derived from first principles~\cite{Blaizot:1993zk,Calzetta:1986cq,Calzetta:1999ps,Blaizot:1999xk,Blaizot:2001nr,Arnold:2002zm}.
 In Yang-Mills theory at weak coupling and nonzero temperature, 
 the necessary separation of scales arises by virtue of the small coupling constant
$g$, since $\lambda_{\rm mfp}\sim 1/(g^4 T)$ 
and the time-scale of interactions is
$1/\mu_D \sim 1/(gT)$, where $1/\mu_D$ is the Debye screening length of the 
plasma.\footnote{Strictly speaking, $\lambda_{\rm mfp}\sim 1/(g^4 T)$  is the  length-scale over which an order 1 change of the momentum-vector
of the quasiparticles occurs.  Over the shorter length-scale $1/\mu_D$, soft exchanges (of order $g T$; not enough to change the momenta which are $\sim T$ significantly) occur. These soft exchanges are not relevant for transport.} 
The small value of the coupling constant also leads to the factorization of higher-point correlation functions.

In the kinetic description,   the system is characterized by a distribution function
\be
f(x, {\bf p}) \, ,
\ee
which determines the number of particles of momentum ${\bf p}$ at space-time position $x$.
Note that this position should be understood as the center of a 
region in space-time with a typical size  much larger, at least,  than the de-Broglie wave length of the particles, as demanded by the uncertainty principle. As a consequence, the Fourier 
transform of $x$, which we shall 
denote by $K=(\omega,{\bf q})$, must be much smaller than 
the typical momentum scale of the particles, $K \ll p \sim T$. 
(Here and below, when we write a criterion like $K\ll p$ we mean that both $\omega$ and $|{\bf q}|$ must be $\ll |{\bf p}|$.)
Due to this separation  in momentum scales,
the $x$-dependence of the distribution functions is said to describe the soft modes of the 
gauge theory while the momenta ${\bf p}$ are those of the 
hard modes. If $K$ is sufficiently small (smaller than the inverse inter-particle separation
$\sim T$), the mode describes collective excitations which involve the motion of many particles, while ${\bf p}$ is the momentum of those particles. In this case, the 
Fourier-transformed distribution $f(K,{\bf p})$ can be 
interpreted approximately as the number of particles within the wavelength of the excitation.  At the long distances at which the kinetic-theory description is valid, 
particles are on mass shell, as determined by the position of the peaks in the correlation functions of the relevant operators  ($p^0=E_{\bf p}$), and these hard modes describe particles that follow 
classical trajectories, at least between the microscopic collisions.
All the properties of the system can be extracted from the distribution function. In particular, the stress tensor is given by 
\be
\label{Toff}
T^{\mu\nu}(x)=\int \frac{d^3p}{(2\pi)^3} \frac{p^\mu p^\nu}{E} f(x,{\bf p}) \,.
\ee

Since all quasiparticles carry energy and momentum, we will concentrate only on the kinetic theory 
description of stress-tensor correlators.  Our analysis is analogous to the one performed for the determination of the Green-Kubo formulae in Appendix~\ref{sec:Green-Kubo}, and proceeds by studying the response of the system to small metric fluctuations. The dynamics are, then, governed by the Boltzmann equation which states the continuity of the distribution function $f$ up to particle collisions~\cite{LandauTransport}:
\be\label{BoltzmannEquation}
E \frac{d}{dt} f(x, {\bf p})= p^\mu \del_{x^\mu} f(x,{\bf p}) + E_p  \frac{d {\bf p}}{dt} \frac{\del}{\del {\bf p}} f(x,{\bf p})= \mathcal{C}\left[ f\right] \ ,
\ee
where 
 $\mathcal{C}\left[ f \right]$ is the collision term which encodes the microscopic collisions among the
plasma constituents and vanishes for the equilibrium distribution $f_{\rm eq}(E_p)$  (which does not depend on $x$ and which does not depend on the direction of ${\bf p}$).
In writing (\ref{BoltzmannEquation}), we are assuming that ${\bf p}=E\, {\bf v}_p $, where ${\bf v}_p$ is the velocity of the particle. In curved space, in the absence of external forces, the Boltzmann equation becomes
\be
\label{Boltcs}
p^{\mu} \del_{x^\mu} f(x,{\bf p}) 
- \Gamma^\lambda_{\mu \nu} p^\mu p^\nu \del_{p^\lambda} f(x,{\bf p})
=
\mathcal{C}\left[ f\right] \, ,
\ee
where $\Gamma^\lambda_{\mu \nu}$ are the Christoffel symbols of the background metric. 
As in Appendix \ref{sec:Green-Kubo}, we shall determine the stress-tensor correlator by 
introducing 
a perturbation in which 
the metric deviates from flat space by a small amount, $g_{\mu \nu}= \eta_{\mu \nu} + h_{\mu\nu}$,
and studying the response of the system. 
Even though the analysis for a generic perturbation can be performed, it will suffice for our 
purposes to restrict ourselves to fluctuations which in Fourier space have only one non-vanishing component $h_{xy}(K)$. We choose the directions $x$ and $y$ perpendicular to the 
wave-vector ${\bf q}$, which  lies in the  $z$  direction. For this metric, the only Christoffel symbols that are non-vanishing at leading order in $h_{xy}$ are 
$\Gamma^t_{xy}=\Gamma^x_{ty}=\Gamma^y_{tx}= -i\omega \, h_{xy}/2$ and 
$\Gamma^x_{zy}=\Gamma^y_{zy}=-\Gamma^z_{xy}=i q \, h_{xy}/2$.
 
 We will assume that prior to the perturbation the system is in equilibrium. In response to the external disturbance the equilibrium distribution changes 
\be
f(x,{\bf p})=f_{\rm eq}(E_p) + \delta f(x,{\bf p}) \, .
\ee
In the limit of a small perturbation, the modified distribution function $\delta f(x,{\bf p})$ is linear in the perturbation $h_{xy}$.   We will also assume that the theory is rotationally invariant so that the energy  of the particle $E_p$ is only a function of the modulus of $p^2=g_{ij} p^i p^j$.
As a consequence,  the metric 
perturbation also changes  the on-shell relation, and the equilibrium distribution must also be expanded to first order  in the perturbation, yielding
\be
f_{\rm eq}= f_0 + f'_0 p^x p^y \frac{|v_p|}{p} h_{xy} \approx f_0 + f'_0 \frac{p^x p^y}{E} h_{xy} \, ,
\ee
where $f_0$ is the equilibrium distribution in flat space, $f'_0(E)=d f(E)/ dE$, 
and the velocity is given by $v_p=d E_p / dp$. In the last equality
we have again approximated $v_p\approx p/E$. 

The solution of the Boltzmann equation requires the computation of the collision term $\mathcal{C}$. In general this is a very complicated task since it takes into account the interactions among all the system constituents, which are responsible for maintaining equilibrium. However, since our only goal is to understand generic features of the spectral function, it will be sufficient to
employ the relaxation time approximation 
\be
\label{relC}
\mathcal{C}=-E \frac{f-f_{\rm eq}}{\tau_R} \, 
\ee
for the collision term, in which the parameter $\tau_R$ is 
referred to as the relaxation time.\footnote{In this approximation, this relaxation time coincides with the shear relaxation time: $\tau_R=\tau_\pi$ \cite{Baier:2006um}. However, since $\tau_\pi$ is a 
property of the theory itself (defined as the appropriate coefficient in the effective field theory, aka the hydrodynamic expansion)
whereas $\tau_R$ is a parameter specifying a simplified approximation to the collision kernel, which in general is not of the form \Eq{relC}, we will maintain the notational distinction between $\tau_\pi$ and $\tau_R$.
}
Since small perturbations away from equilibrium are driven back to
equilibrium by particle collisions, the relaxation time must be of the order of the mean free path 
$\lambda_{\rm mfp}$ (which is long compared to the interparticle distance). 
The relaxation-time approximation is a very significant simplification of the full dynamics, but it will allow us to illustrate the main points that we wish to make. A complete analysis 
of the collision term within perturbation theory for the purpose of extracting the 
transport coefficients of a weakly coupled plasma can be found in 
Refs.~\cite{Arnold:2000dr,Arnold:2003zc,Hong:2010at,York:2008rr}.

Within the approximation (\ref{relC}), upon taking into account that the distribution function does not depend independently on the energy of the particles in Eq.~(\ref{Boltcs}), the 
solution to the linearized Boltzmann equation is given by
\be
\delta f(K,{\bf p})=\frac
{- i \omega p^x p^y f'_0(p)
}{-i \omega + i{\bf v}_p {\bf q}+\frac{1}{\tau_R}}
\frac{h_{xy}(K)}{E} \, .
\ee
Substituting this into Eq.~(\ref{Toff})
we learn that the perturbation of the distribution function leads to a perturbation of the stress tensor given by
\be
\delta T^{\mu \nu} (K)= \int \frac{d^3p}{(2\pi)^3} \frac{p^\mu p^\nu}{E} \delta f(K,{\bf p}) 
                                = -G^{xy, xy}_R (K) h_{xy} (K) \,,
\ee
where the retarded correlator is given by
\be
G^{xy, xy}_R (K)= -  \int \frac{d^3 p }{(2\pi)^3}
 v^x v^y
\frac
{
\omega\, p^x p^y\, f'_0(p)
}{ \omega - {\bf q} {\bf v}_p + \frac{i}{\tau_R}} \,.
\ee
From the definition (\ref{sptfuncdef}), the spectral function associated with this correlator is
\bea
\label{rhotpeak}
\rho^{xy, xy}(K)&=& - \omega \int \frac{d^3 p }{(2\pi)^3} 
\frac{\left(p^x p^y\right)^2}{E^2}  f'_0(p) 
\frac{\frac{2}{\tau_{R}}}{(\omega-{\bf q v_p})^2 + \frac{1}{\tau^2_R}} \, .
\eea
Obtaining this spectral function was our goal, because as we shall now see it has qualitative features that indicate the presence (in this weakly coupled plasma) of quasiparticles.

To clarify the structure of the spectral function (\ref{rhotpeak}) we begin by describing the free theory limit, in which 
 $\tau_R\rightarrow \infty$ since the collision term vanishes. In this limit, the Lorentzian may be 
replaced by a $\delta$-function, yielding
\be
\rho^{xy, xy}(K)=-\omega \int \frac{d^3 p }{(2\pi)^3} 
\frac{\left(p^x p^y\right)^2}{E^2}  f'_0(p) 
\, 2 \pi\, \delta \left(\omega-{\bf q \cdot v}_p\right) \, .
\ee
The $\delta$-function arises because, in this limit, the external perturbation (the gravity wave)
interacts with free particles. 
The $\delta$-function selects those components of the
gravity wave whose phase velocity $\omega/|{\bf q}|$ 
coincides with the velocity of any of the free particles in the plasma. 
Thus, in the free-theory limit, this $\delta$-function encodes the existence of free particles 
in the plasma.  
For an isotropic distribution of particles, such as the thermal distribution,
at any ${\bf q} \neq 0$
the integration over angles washes out the $\delta$-function and one is left with some function of $\omega$ that is characterized by the typical momentum scale of the particles ($\sim T$) and that is not of interest to us here.\footnote{A distinct peak at in the spectral density at some $\omega\neq 0$ 
could be observed if the initial distribution
were very anisotropic.  This can arise if the theory has a (gauged) 
conserved charge and if the system is analyzed in the presence of a constant force that acts on this charge --- \ie an electric field.
}
On the other hand, at ${\bf q}=0$ we find that
$\frac{1}{\omega}\rho^{xy,xy}(\omega,0)$ is proportional to $\delta(\omega)$.
This $\delta$-function at $\omega=0$ in the low-momentum spectral function is a direct consequence of the presence of free particles in the plasma.  As we now discuss, the effect of weak interactions is to dress the particles into quasiparticles and to broaden the $\delta$-function into a narrow, tall, peak at $\omega=0$.

When the interactions do not  vanish, we can proceed by relating the relaxation time to the shear viscosity.
To do so, we work in the hydrodynamic limit in which all momenta must be smaller than any internal scale. This means that we can set ${\bf q}$ to zero, but 
 we must keep the relaxation time $\tau_R$ finite.
The spectral density at zero momentum is then given by 
\be
\label{sdkinth}
\rho^{xy, xy}(\omega,0)=-\omega \int \frac{d^3 p }{(2\pi)^3} 
\frac{\left(p^x p^y\right)^2}{E^2}  f'_0 (p)\, 
\frac{\frac{2}{\tau_{R}}}{\omega^2 + \frac{1}{\tau^2_R}}\ .
\ee
Note that the spectral density at zero momentum has a peak at $\omega=0$, and note in particular that the width in $\omega$ of this peak is $\sim 1/\tau_R \ll T$. The spectral density has vanishing strength for $\omega\gg 1/\tau_R$.  This low-frequency structure in the zero-momentum spectral function is called the `transport peak'.   It is clear that in the $\tau_R\rightarrow\infty$ limit it becomes the $\delta$-function that characterizes the spectral density of the free theory that we described above.  Here, in the presence of weak interactions, this peak at $\omega=0$ is a direct consequence of the presence of momentum-carrying quasiparticles whose 
mean free time is $\sim \tau_R$.

The expression (\ref{sdkinth})  is only valid for $\omega \ll T$ where the modes are 
correctly described by the Boltzmann equation. For $\omega \gg T$,  since the quasiparticles can be
resolved the structure of the spectral density is close to that in vacuum. The separation of scales 
in the spectral density is directly inherited from the separation of scales which allows the Boltzmann description. 
Finally, using the Green-Kubo formula for the shear viscosity (\ref{etaco}), we find
\be
\eta =  -\tau_R \int \frac{d^3 p }{(2\pi)^3}  \frac{\left(p^x p^y\right)^2}{E^2}  f'_0(p) \,.
\ee
Thus, since $\eta$ is determined by the collisions among the quasiparticles, we can understand
$1/\tau_R$ as the width that arises because the quasiparticles do not have well-defined momenta due to the collisions among them.
 In particular, in perturbation theory~\cite{Arnold:2000dr,Arnold:2003zc} 
\be
\frac{1}{\tau_R}\sim \frac{1}{T}  \,g^4 \ln \frac{1}{g} \sim \frac{1}{\lambda_{\rm mfp}} \, .
\ee

\begin{figure} [t]
\includegraphics[width=0.45\textwidth]{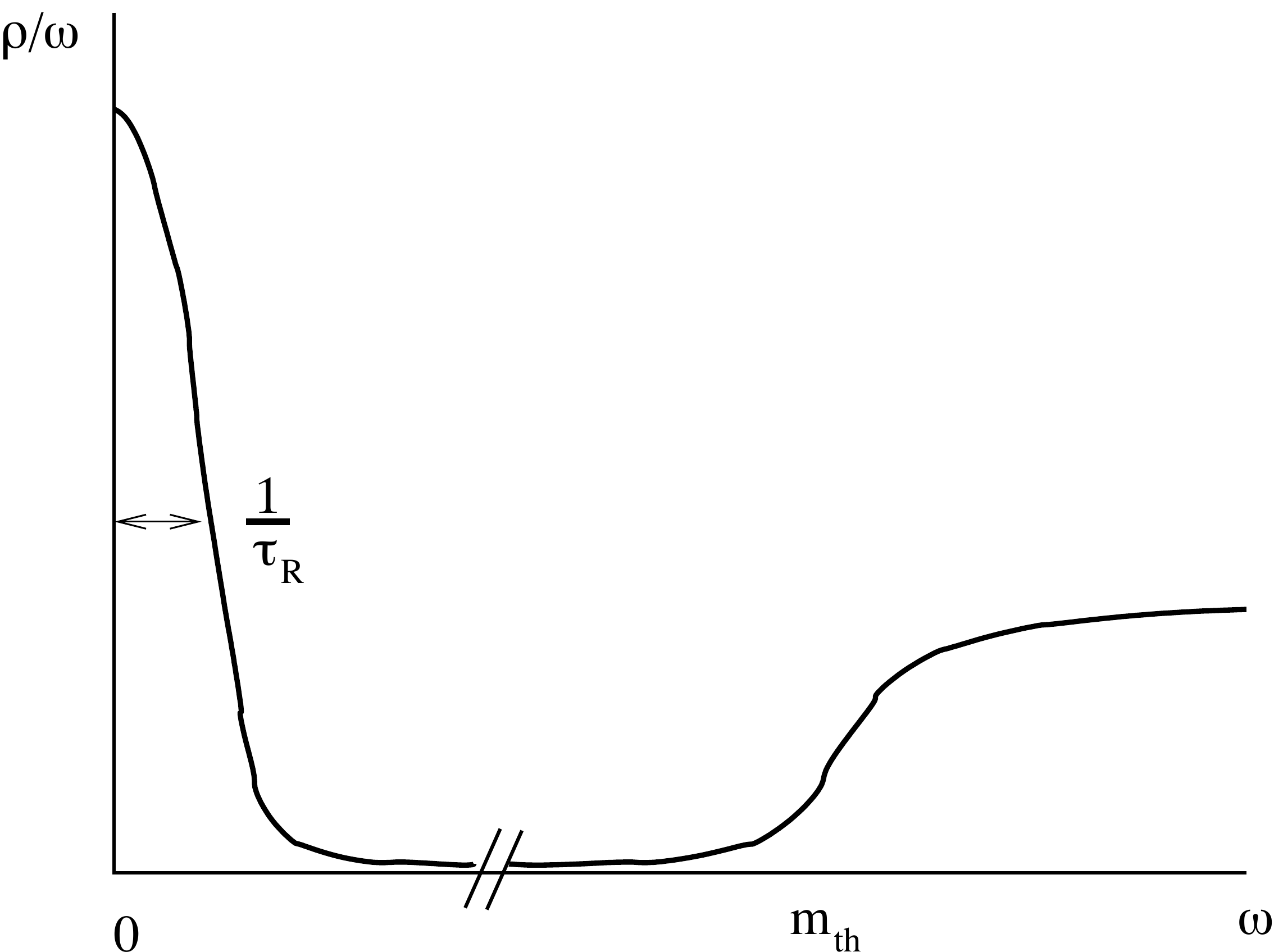}
\includegraphics[width=0.45\textwidth]{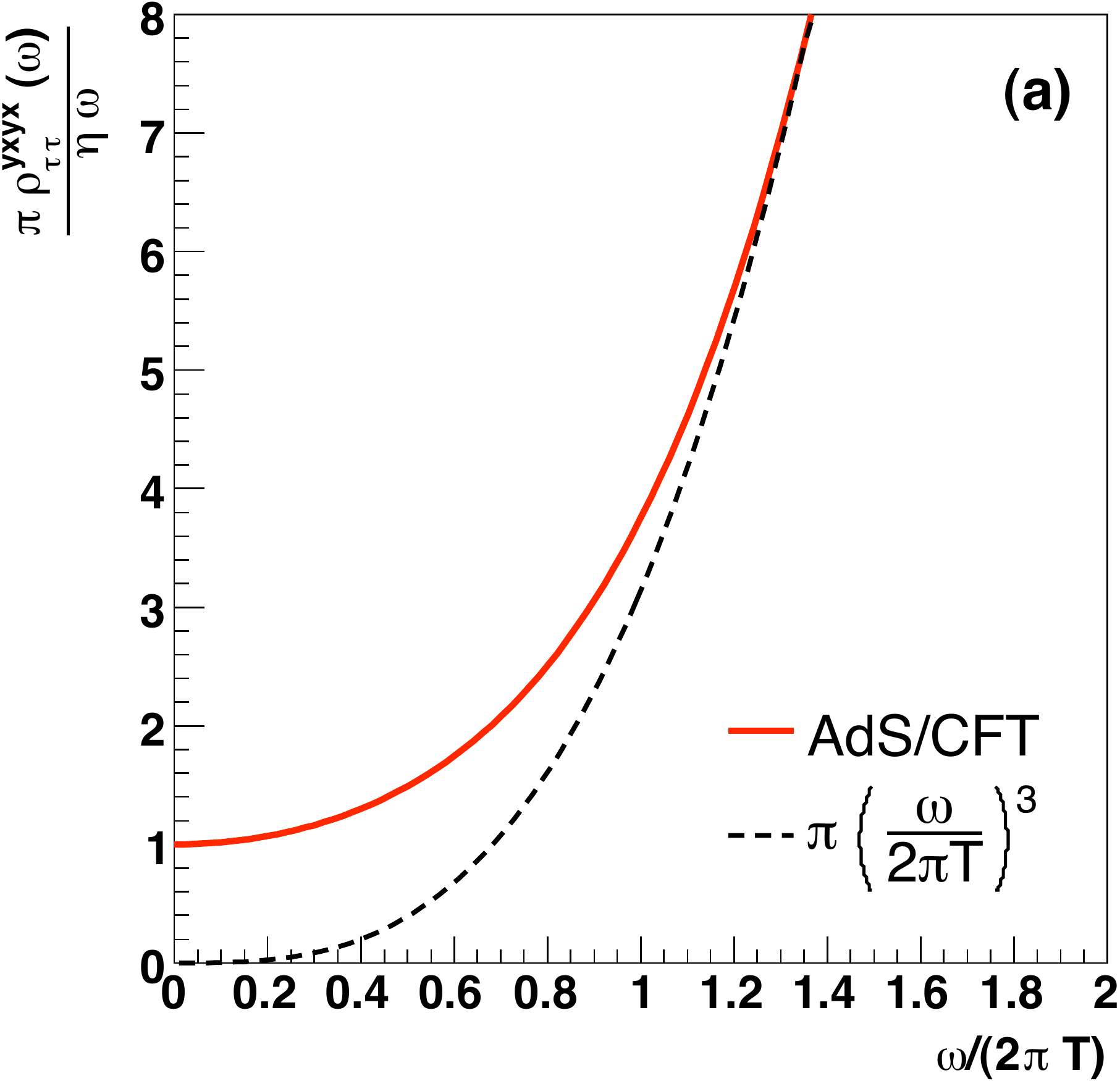}
\caption{\label{plotspf} \small
Left: Sketch of the spectral function at zero momentum as a function of frequency for a weakly coupled plasma, as obtained 
from kinetic theory. The narrow structure at small frequency is 
the transport peak with a width $1/\tau_R$ that is suppressed by the 
coupling ($1/\tau_R \sim g^4 T$). The
thermal mass is $m_{\rm th}\propto g T$.
Right: Spectral function for the shear channel in the strongly coupled plasma of ${\cal N}=4$ SYM theory computed via gauge/string duaity~\cite{Teaney:2006nc} (solid) and 
a comparison with the vacuum spectral function (dashed) which it approaches at high frequencies.
The vertical axis of this figure has been scaled by the shear viscosity
$\eta=s/4\pi$ of the strongly coupled plasma.
 Note that the definition of $\rho^{xyxy}=- {\rm Im} G_R/\pi$ used in Ref.~\cite{Teaney:2006nc} is different from that in Eq.~(\ref{sptfuncdef}) by a factor of $\pi/2$.
}
\end{figure}

Let us summarize the main points.  The zero-momentum spectral densities of a plasma with 
quasiparticles  have a completely distinctive structure: there is a separation of scales between the scale $T$ 
(the typical momentum of the quasiparticles in the plasma) and the much lower scale 
$1/\lambda_{\rm mfp}$. In particular, there is a narrow peak in $\rho(\omega,0)/\omega$ around
$\omega=0$ of width $\tau_R\sim 1/\lambda_{\rm mfp}$ and height $2\eta$. 
At larger frequencies, the strength of the spectral function is very small. At the scale of the mass of the quasiparticles, the spectral function grows again. For massless  particles or those with mass much smaller than any temperature-related scale, 
the role of the mass threshold is played by the thermal mass of the particles, $g T$,
which is  much higher than  the scale $1/\lambda_{\rm mfp} \sim g^4 T$ associated with the mean free path due to the weakness of the coupling.
Finally, above 
the scale $T$ the structure of the spectral function approaches what it would be in vacuum.
A sketch of this behavior can be found in 
Fig.~\ref{plotspf}.    These qualitative features are independent of any details of the theory, and do not even depend on its symmetries.  All that matters is the existence of momentum-carrying quasiparticles.  In the presence of quasiparticles, no matter what their  quantum numbers are, these qualitative features must be present in the spectral density.

\subsection{Absence of quasiparticles at strong coupling}
\label{AbsenceAdS}

We return now to the strongly coupled ${\cal N}=4$  SYM plasma, with its dual gravitational description, in order to compare the expectation (\ref{sdkinth}) for how the spectral density should look if the plasma contains any momentum-carrying quasiparticles to an explicit computation of the retarded correlator at strong coupling, of course done via gauge/string duality.
In this subsection we will benefit from the general analyses of 
Sections~\ref{generaltransportads} and \ref{universalshear}.
 As in the kinetic-theory computation, we study the response to a
metric fluctuation $h_{xy}(\omega,{\bf q})$ in the boundary theory,
with the same conventions as before. As in 
Section \ref{universalshear}, the fluctuation in the boundary leads to a metric perturbation in the bulk, $h^y_x=g^{yy} h_{yx}$ of the form
\be
h^y_x(\omega, q,z)=\phi(\omega,q,z) e^{-i \omega t+ i q z} \, .
\ee
The field $\phi$ is governed by the classical action (\ref{Sclphi}) which yields an equation of motion
for $\phi(\omega,q,z)$ that is given by
\be
\label{eqforphi}
\phi''(\omega, q,u) -\frac{1+u^2}{u f} \phi'(\omega, q, u) + \frac{\wf^2-\qf^2(1-u^2)}{u f^2} \phi(\omega, q, u)=0\, ,
\ee
where $u=z^2$, $\wf=\omega/(2\pi T)$ and $\qf=q/(2\pi T)$. We may now use the general program 
outlined in Section \ref {sec:rtthc} to determine the retarded correlator.  It is given by 
Eq.~(\ref{RestB}) which, together with Eqs.~(\ref{eom1}) and (\ref{eei}), leads to 
\be
\label{gxyxyads}
G^{xy,xy}_R=-\lim_{u \rightarrow 0} \frac{1}{16\pi G_N}\frac{\sqrt{-g} g^{uu} \del_u \phi(\wf,\qf,u)}{\phi_0(\wf,\qf,u)}\,,
\ee 
where $\phi(\wf,\qf,u)$ is the solution to the equation of motion (\ref{eqforphi}) 
with in-falling boundary conditions at the horizon.
For arbitrary values of $\wf$ and $\qf$, Eq.~(\ref{eqforphi}) must be solved 
numerically~\cite{Teaney:2006nc,Kovtun:2006pf}. 
From the correlator (\ref{gxyxyads}), 
the spectral function is evaluated using the definition (\ref{sptfuncdef}).  The result of this computation at zero spatial momentum ${\bf q}=0$ is shown in the right panel of Fig.~\ref{plotspf}, where we have plotted $\rho/\omega$ which should have a peak at $\omega=0$ if there are any quasiparticles present.

 In stark contrast to the kinetic-theory expectation, there is no transport peak in the spectral function at strong coupling. In fact, the spectral function has no interesting structure at all 
at small frequencies. The numerical computation whose results are plotted in the right panel of  
Fig.~\ref{plotspf} also shows that there is no separation of scales in the spectral function.  In the strong-coupling calculation, quite unlike in perturbation theory, the small and large frequency behaviors join smoothly and the 
spectral density is only a function of $\wf=\omega/2\pi T$.  This could perhaps have been expected in a conformal theory with no small coupling constant,  but note that a free massless theory is conformal and that theory does have a $\delta$-function peak in its spectral function at zero frequency.  So, having the explicit computation that gauge/string duality provides is necessary to give us confidence in the result that there is no transport peak in the strongly coupled plasma.
The absence of the 
transport peak shows unambiguously that there are no momentum-carrying quasiparticles   
in the strongly coupled plasma.  Thus, the physical picture of the system is completely different from that in perturbation theory. 

To conclude this section we would like to argue that the absence of quasiparticles is a generic property
of strong coupling and  is not specific to any particular theory with any particular symmetries or matter content.
To do so, let us recall that in the kinetic theory calculation the separation of scales required for its consistency are a consequence of the weak coupling; this is so in perturbative QCD or in 
perturbative $\mathcal{N}=4$ or in any weakly coupled plasma. 
Now, imagine increasing the coupling.
According to kinetic theory, independent of the symmetries or the matter content of the theory, the width of the transport peak grows and its height decreases as the coupling increases.  This reflects the fact that as the coupling grows so does the width of the quasiparticles.
Extrapolating this trend to larger and larger couplings leads to the
 disappearance of the transport peak which, at a qualitative level, agrees nicely with the strong-coupling result for the  $\N=4$ SYM plasma obtained by explicit computation and shown in the 
 right panel of Fig.~\ref{plotspf}.
 This observation is one of the most salient motivations 
 for the phenomenological applications of AdS-based techniques.  
 Since, as we have argued extensively
 in Sections~\ref{sec:EllipticFlow} and  \ref{sec:latticeQCD} and \ref{sec:TransportProperties}, the quark-gluon plasma of QCD at temperatures a few times its $T_c$ is strongly coupled, the quasiparticle picture that has conventionally been used to think about its dynamics is unlikely to be valid in this regime.   This makes it very important to have new techniques at our disposal that allow
 us to study strongly coupled plasmas with no quasiparticles, seeking 
 generic 
  consequences of the absence of quasiparticles for physical observables.
 Gauge/gravity duality is an excellent tool for these purposes, as we have already seen 
in Sections~\ref{sec5.1} and \ref{sec:TransportProperties} and as we will further see in the remaining Sections of this review.  Indeed, as we use gauge/gravity duality to calculate more, and more different, physical observables we will discover that the calculations done in the dual gravitational description begin to yield a new form of physical intuition, phrased in the dual language rather than in the gauge theory language, in addition to yielding reliable results.

\chapter{Probing strongly coupled plasma}
\label{sec:Section6}

As discussed in Sections~\ref{sec:JetQuenching} and~\ref{quarkonium}, two of the most informative probes of strongly coupled plasma that are available in heavy ion collisions are the rare highly energetic partons and quarkonium mesons produced in these collisions.  In this Section and in Section~\ref{mesons}, we review results obtained by employing the AdS/CFT correspondence that are shedding light on these classes of phenomena.   In Sections~\ref{sec:HQDrag} and~\ref{sec:HQBroadening}, we review how a test quark of mass $M$ moving through the strongly coupled ${\cal N}=4$ SYM plasma loses energy and picks up transverse momentum.  In Section~\ref{sec:AdSCFTDragWaves} we consider how the strongly coupled plasma responds to the hard parton plowing through it; that is, we describe the excitations of the medium which result.  In Section~\ref{sec:Stopping}, we review a calculation of the stopping distance of a single light quark moving through the strongly coupled plasma. 
Throughout Sections~\ref{sec:HQDrag}, \ref{sec:HQBroadening},  \ref{sec:AdSCFTDragWaves} and \ref{sec:Stopping} we assume that all aspects of the phenomena associated with an energetic parton moving through the plasma are strongly coupled.
In Section~\ref{sec:AdSCFTJetQuenching},  we review an alternative approach in which we assume that QCD is weakly coupled at the energy and momentum scales that characterize gluons radiated from the energetic parton, while the medium through which the energetic parton and the radiated gluons propagate is strongly coupled. In this case, one uses the AdS/CFT correspondence only in the calculation of those properties of the strongly coupled plasma that arise in the calculation of radiative parton energy loss and transverse momentum broadening.  
In Section~\ref{sec:Synchrotron}, we describe a calculation of synchrotron radiation in strongly coupled ${\cal N}=4$ SYM theory that allows the construction of  a narrowly collimated beam of gluons (and adjoint scalars), opening a new path toward analyzing jet quenching.

In Section~\ref{sec:HotWind},
we review those insights into the physics of quarkonium mesons in heavy ion collisions that have been obtained via AdS/CFT calculations of the temperature-dependent screening of the potential between a heavy quark and antiquark.  To go farther, we need to introduce a holographic description of quarkonium-like mesons themselves. In Section~\ref{mesons}, we first review this construction and then review the insights that it has yielded.  In addition to shedding light upon the physics of quarkonia in hot matter that we have introduced in Section~\ref{quarkonium},
as we review in 
Section~\ref{sec:Cherenkov} these calculations have also resulted in the discovery of a new and significant process by which a hard parton propagating through a strongly coupled plasma can lose energy: Cherenkov radiation of quarkonium mesons.


\section{Parton energy loss via a drag on heavy quarks}
\label{sec:HQDrag}

When a heavy quark moves through the strongly coupled plasma of a conformal theory, it feels a drag force and consequently loses energy~\cite{Herzog:2006gh,Gubser:2006bz}. We shall review the original calculation of this drag force in ${\cal N}=4$ SYM theory~\cite{Herzog:2006gh,Gubser:2006bz}; it has subsequently been done in many other gauge theories with dual gravitational 
descriptions~\cite{Herzog:2006se,Caceres:2006dj,Caceres:2006as,Matsuo:2006ws,Nakano:2006js,Talavera:2006tj,Gubser:2006qh,Gursoy:2009kk,HoyosBadajoz:2009pv,Bertoldi:2007sf,Bigazzi:2009bk}.
In calculations of the drag on heavy quarks, one determines the energy per unit time
needed to maintain the forced motion of the quark in the plasma. In these calculations
one regards the quark as an external source moving at fixed velocity, $v$, and
one performs thermal averages over the medium.  This picture can be justified if
the mass of the quark is assumed to be much larger than the typical momentum scale of the medium 
(temperature), and if the motion of the quark  is studied in a time window that is large 
compared with the relaxation scale of the medium but short compared to the time 
it takes the quark to change its trajectory. 
In this limit the heavy quark is described by a Wilson line 
along the worldline of the quark. 
%
%

The dual description of the Wilson line is given by a classical string hanging down from the
quark on the boundary of AdS. Since we are considering a single quark, the other end of the 
string hangs downs  into the bulk of the AdS space. We consider the stationary situation, 
in which the quark has been moving at a fixed velocity for a long time, meaning that the shape of the string trailing down and behind it is no longer changing with time. 
For concreteness, we will assume that the quark moves in the 
$\xtr$ direction, and we choose to parametrize the string 
world sheet by $\tau=t$ and $\sigma=z$. 
By symmetry, we can set two of the perpendicular coordinates, $x_2$ and $x_3$ to a constant. 
The problem of finding the string profile reduces, then, to finding a function
\begin{equation}
\xtr(\tau,\sigma) \, ,
\end{equation}
that fulfills the string equations of motion. The string solution must also satisfy the 
boundary condition
\begin{equation}
\xtr(t,z\rightarrow0)=vt \,. 
\end{equation}
Since we are interested in the stationary situation, the string solution takes the form 
\begin{equation}
\xtr(t,z)=v t +\zeta (z)\, ,
\end{equation}
with $\zeta(z\rightarrow 0)=0$. We work in an ${\cal N}=4$ plasma, whose dual gravitational description is the AdS black hole with the metric $G_{\mu\nu}$ given in (\ref{AdSfiniteZ}).  The induced metric on the string worldsheet $g_{\alpha\beta}=G_{\mu\nu}\partial_\alpha x^\mu \partial_\beta x^\nu$ is then given by
\begin{eqnarray}
ds^2_{ws}&=&\frac{R^2}{z^2}\Biggl(
                                                       -\left(f(z)-v^2\right) d\tau^2 +
                                                        \left(\frac{1}{f(z)}+\xipt \right) d\sigma^2\nonumber\\ 
                                                        & & \qquad \qquad+
                                                        v\, \xip v \left(d\tau d\sigma+d\sigma d\tau \right)
                                              \Biggr) \, ,
\end{eqnarray}
where, as before, $f(z)=1-z^4/z^4_0$ and $\xip$ denotes differentiation  with respect to $z$.

The Nambu-Goto action for this string reads
\be
\label{Sdrag}
S=-\frac{R^2}{\tpa}\mathcal{T} \int \frac{d z}{z^2}\sqrt{\frac{f(z)-v^2+f(z)^2\xipt}{f(z)}}
   = \mathcal{T} \int dz \mathcal{L} \, ,
\ee
with $\mathcal{\T}$ the total time traveled by the quark. Extremizing this action yields the equations 
 of motion that must be satisfied by $\zeta(z)$.
The action (\ref{Sdrag}) has a constant of motion given by the canonical momentum
\begin{equation}
\label{eq:fluxdef}
\pixz=\frac{\del \mathcal{L}}{\del x'_1}=-\frac{R^2}{\tpa} \frac{1}{z^2} \frac{f(z)^{3/2}\xip}{\sqrt{f(z)-v^2+f(z)^2\xipt}}
\, ,
\end{equation}
which coincides with the longitudinal momentum flux in the z direction.  In terms of $\pixz$, the equation of motion for $\zeta$ obtained from (\ref{Sdrag}) takes the form
\be
\label{eq:xipzeq}
\xipt=\left(\Ct \right)^2 \frac{z^4}{f(z)^2}\frac{f(z)-v^2}{f(z)-\left(\Ct\right)^2 z^4}\, .
\ee
The value of $\pixz$ can be fixed by inspection of this equation, as follows: both the numerator and the denominator of (\ref{eq:xipzeq}) are positive at the boundary $z=0$ and negative at the horizon $z=z_0$; since $\zeta'(z)$ is real, both the numerator and the denominator must change sign at the same $z$; this is only the case if
\begin{equation}
\label{eq:pixz}
\pixz=\pm \frac{R^2}{\tpa z^2_0} \gamma v \, ,
\end{equation}
with $\gamma=1/\sqrt{1-v^2}$ the Lorentz $\gamma$ factor. Thus, stationary solutions can only be found
for these values of the momentum flux. (Or, for $\pixz=0$, for which $\xi=$constant. This solution has real action only for $v=0$.) 


\begin{figure} 
\begin{center}
\includegraphics{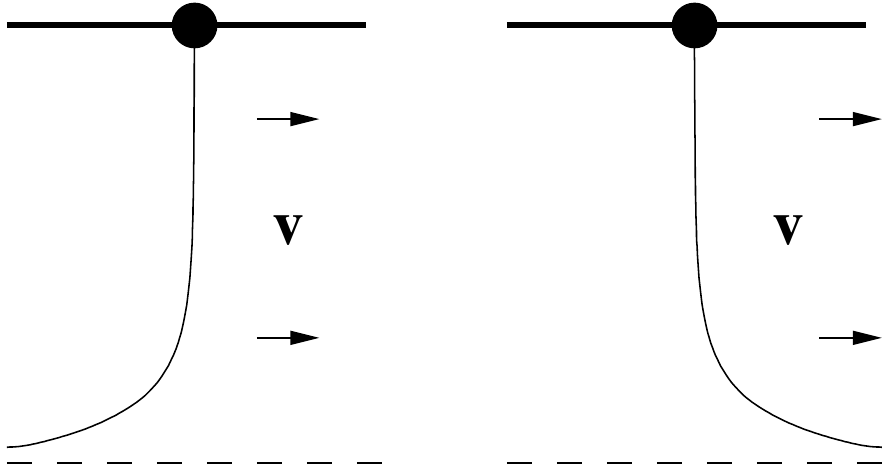}
\caption{\small 
\label{fig:xisol}
String solutions of \Eq{eq:pixz}. The physical (unphysical) solution in which  momentum flows into  (out of) the  horizon and the string trails behind (curves ahead) of the quark at the boundary is plotted inn the left (right) panel.
}
\end{center}
\end{figure}

The two solutions (\ref{eq:pixz}) correspond to different boundary choices of boundary conditions at the horizon.   Following Refs.~\cite{Herzog:2006gh,Gubser:2006bz}, we choose the solution for which the momentum flux along the string world sheet flows from the boundary into the horizon, corresponding to the physical case in which the energy provided by the external agent that is pulling the quark through the plasma at constant speed is dissipated into the medium.  
This solution to (\ref{eq:xipzeq}) is given by
\begin{equation}
\label{eq:solxi}
\zeta(z)=- \frac{v\, z_0}{2}\left(
                                                        \arctanh\left(\frac{z}{z_0}\right)-\arctan\left(\frac{z}{z_0}\right)
						\right) \, .
\end{equation}
As illustrated in Fig.~\ref{fig:xisol}, this solution describes a string that trails behind the moving quark as it hangs down from it into the bulk spacetime.


The momentum flux flowing down from the boundary, along the string world sheet (\ref{eq:solxi}), and towards the horizon
determines the amount
of momentum lost by the quark in its propagation through the plasma.  In terms of 
the field theory variables,
\begin{equation}
\label{eq:dpdt}
\frac{dp}{dt}=-\pixz=-\frac{\pi T^2 \sqrt{\lambda}}{2}  \gamma v\, .
\end{equation}
Note, however, that in the stationary situation we have described, there is 
by construction no change in the actual momentum of the quark at the boundary; instead, 
in order to keep the quark moving with constant speed $v$ against the force (\ref{eq:dpdt}) there must be some external agency pushing the quark through the strongly coupled plasma.
This force can be viewed as due to a constant electric field
acting on the string endpoint, with the magnitude of the field given by
\begin{equation}
\label{eq:Efield}
\Efield=\frac{\pi T^2 \sqrt{\lambda}}{2}  \gamma v\, .
\end{equation}
The physical setup described by the string (\ref{eq:solxi}) is thus that of forced motion of the quark
through the plasma at constant speed in the presence of a constant electric field. The external force on the quark balances the backward {\it drag force} (\ref{eq:dpdt}) on the quark exerted by the medium through which it is moving.  To make it explicit that the medium exerts a drag force, we can rewrite (\ref{eq:dpdt}) as
\begin{equation}
\label{eq:drag}
\frac{dp}{dt}=-\eta_D p \, ,
\end{equation}
with $p=M\gamma v$ the relativistic expression of the momentum of the quark and $M$ the 
mass of the heavy particle. The drag coefficient is then
\begin{equation}
\label{eq:etaads}
\eta_D=\frac{\pi \sqrt{\lambda} T^2 }{2 M} \, .
\end{equation}
For test quarks with $M\rightarrow\infty$, as in the derivation above, this result is valid for motion with arbitrarily relativistic speeds $v$.  It is remarkable that the energy loss of a  heavy quark moving through the quark-gluon plasma with constant speed is described so simply, as due to a drag force.
In contrast, in either a weakly 
coupled plasma~\cite{Moore:2004tg} or a strongly coupled plasma that is not conformal~\cite{Liu:2008tz}, $dp/dt$ is not proportional to $p$ even at low velocities.

We shall see in Section~\ref{sec:HQBroadening} that a heavy quark moving through the strongly coupled plasma of ${\cal N}=4$ SYM theory experiences transverse and longitudinal momentum broadening, in addition to losing energy via the drag that we have analyzed above.  We shall review the implications of the understanding of how the presence of the strongly coupled plasma affects the motion of heavy quarks for heavy ion collision phenomenology 
at the end of Section~\ref{sec:HQBroadening}.


\subsection{Regime of validity of the drag calculation}

In the derivation of the drag force above, we considered a test quark with $M\rightarrow\infty$.  The result is, however, valid for quarks with finite mass $M$, as long as $M$ is not too small.  As we now show, the criterion that must be satisfied by $M$ depends on the velocity of the quark $v$.  The closer $v$ is to 1, the larger $M$ must be in order for the energy loss of the quark to be correctly described via the drag force calculated above. In deriving the regime of validity of the drag calculation, we shall assume for simplicity that we are interested in large enough $\gamma=1/\sqrt{1-v^2}$ that the $M$ above which the calculation is valid satisfies $M\gg \sqrt{\lambda} T$.
We will understand the need for this condition in Section \ref{mesons}.

The introduction of quarks with finite mass $M$ in the fundamental representation of the gauge group corresponds in the dual gravitational description to the introduction of D7-branes~\cite{Karch:2002sh}, as we have reviewed in Section~\ref{fundamental} and as we will further pursue in Section~\ref{mesons}.  The D7-brane extends from the boundary at $z=0$ down to some $z_q$, related to the mass of the quarks it describes by 
\begin{equation}
M = \frac{\sqrt{\lambda}}{2\pi z_q}\ ,
\end{equation}
a result that we shall explain in Section~\ref{sec:AddingQuarks}.
The physical reason that the calculation of the drag force breaks down if $M$ is too small or $v$ is too large is that if the electric field $\Efield$ required to keep the quark moving at constant speed $v$ gets too large, one gets copious production of pairs of quarks and antiquarks with mass $M$, and the picture of dragging a single heavy quark 
through the medium breaks down completely~\cite{CasalderreySolana:2007qw}.
%
The parametric dependence of the critical field $\Efield_c$ at which pair production becomes copious can be estimated by inspection of how the Dirac-Born-Infeld action for the D7-brane depends on $\Efield$, namely
\begin{equation}
S_{DBI} \sim \sqrt{1-\left(\frac{\tpa}{R^2} \Efield \, z_q^2\right)^2}\, .
\end{equation}
The critical maximum field strength that the D7 brane can support is the $\Efield_c$ at which this action vanishes. This yields a criterion for the validity of the drag calculation, namely that $\Efield$ must be less than of order
 \begin{equation}
 \label{eq:Ecrit}
 \Efield_c=\frac{2 \pi M^2}{\sqrt{\lambda}} \, .
 \end{equation}
This maximum value of the electric field implies a maximum value of  $\gamma$ up to which 
the drag calculation can be applied for quarks with some finite mass $M$. From \Eq{eq:Efield} and \Eq{eq:Ecrit}, this criterion is
\begin{equation}
\label{eq:glimit1}
\gamma v < \left( \frac{2 M}{\sqrt{\lambda} T} \right)^2 \, .
\end{equation}
We shall assume that $M \gg \sqrt{\lambda} T$, meaning that in (\ref{eq:glimit1}) we can take $\gamma v \simeq \gamma$.  And, the estimate is only parametric, so the factor of two is not to be taken seriously.  Thus, the result to take away is that the drag calculation is valid as long as
\begin{equation}
\label{eq:glimit}
\gamma  \lesssim \left( \frac{M}{\sqrt{\lambda} T} \right)^2 \, .
\end{equation}

The argument in terms of pair production for the limit (\ref{eq:glimit}) on the quark velocity gives a nice physical understanding for its origin, but this limit arises in a variety of other ways.   For example, at  (\ref{eq:glimit}) the velocity of the quark $v$ becomes equal to the local speed of light in the bulk at $z=z_q$, where the trailing string joins onto the quark on the D7-brane.  For example, at (\ref{eq:glimit}) the screening length $L_s$ (described below in Section~\ref{sec:HotWind}) at which the potential between a quark and antiquark is screened becomes as short as the Compton wavelength of a quark of mass $M$, meaning that the calculation of Section~\ref{sec:HotWind} is also valid only in the regime~(\ref{eq:glimit})~\cite{Liu:2006he}. 

Yet further understanding of the meaning of the limit (\ref{eq:glimit}) can be gained by asking the question of what happens if the electric field is turned off, and the quark moving with speed $v$ begins to decelerate due to the drag force on it.    We would like to be able, at least initially, to calculate the energy loss of this now decelerating quark by assuming that this energy loss is due to the drag force, which from (\ref{eq:dpdt}) means
\begin{equation}
\label{EnergyLossLinearDrag}
\frac{dE}{dt}\Biggr|_{\rm drag} = - \frac{\pi}{2} \sqrt{\lambda} T^2 \gamma v^2
= -\frac{\pi}{2} \sqrt{\lambda} T^2 \,\frac{p v}{M}\ .
\end{equation}
However, once the quark is decelerating it is natural to expect that, due to its deceleration, it radiates and loses energy via this radiation also.   The energy lost by a quark in strongly 
coupled ${\cal N}=4$ SYM theory moving in vacuum along a trajectory with arbitrary acceleration has been calculated by Mikhailov~\cite{Mikhailov:2003er}.  For the case of a linear trajectory with deceleration $a$, his result takes the form
\begin{equation}
\label{EnergyLossLinearVacuumRadiation}
\frac{dE}{dt}\Biggr|_{\rm vacuum~radiation} = - \frac{\sqrt{\lambda}}{2\pi}\, a^2\gamma^6
= - \frac{\sqrt{\lambda}}{2\pi} \, \frac{1}{M^2} \, \left( \frac{dp}{dt} \right)^2\ .
\end{equation}
At least initially, $dp/dt$ will be that due to the drag force, namely (\ref{eq:dpdt}). We now see that the condition that $dE/dt$ due to the vacuum radiation (\ref{EnergyLossLinearVacuumRadiation}) caused by the drag-induced deceleration (\ref{eq:dpdt}) be less than $dE/dt$ due to the drag itself (\ref{EnergyLossLinearDrag}) simplifies considerably and becomes
\begin{equation}
\gamma  < \left( \frac{2 M}{\sqrt{\lambda} T} \right)^2 \, ,
\label{eq:glimit2}
\end{equation}
the same criterion that we have seen before.  This gives further physical intuition into the criterion for the validity of the drag calculation and at the same time demonstrates that this calculation cannot be used in the regime in which energy loss due to deceleration-induced radiation becomes dominant.

Motivated by the above considerations, the authors of Ref.~\cite{Fadafan:2008bq} considered the (academic) case of a test quark moving in a circle of radius $L$ with constant angular frequency $\omega$.  They showed that in this circumstance, $dE/dt$ is given 
by (\ref{EnergyLossLinearDrag}), as if due to drag with no radiation, as long 
as $\omega^2 \ll (\pi T)^2 \gamma^3$, with $\gamma$ the Lorentz factor for velocity $v=L\omega$.  But, for $\omega^2 \gg (\pi T)^2 \gamma^3$, the energy loss of the quark moving in a circle through the plasma is precisely what it would be in vacuum according to Mikhailov's result, which becomes
\begin{equation}
\label{EnergyLossCircularVacuumRadiation}
\frac{dE}{dt}\Biggr|_{\rm vacuum~radiation} = \frac{\sqrt{\lambda}}{2\pi}\, v^2 \omega^2 \gamma^4
=  \frac{\sqrt{\lambda}}{2\pi} \, a^2 \gamma^4 
\end{equation}
for circular motion.  Note that the radiative energy loss (\ref{EnergyLossCircularVacuumRadiation}) is greater than that due to drag, (\ref{EnergyLossLinearDrag}), 
for 
\begin{equation}\label{RadiationDominatedCriterion}
\omega^2 \gg (\pi T)^2 \gamma^3\ ,
\end{equation}
so the result of the calculation is that energy loss is dominated by that due to acceleration-induced radiation or that due to drag wherever each is larger.  (Where they are comparable in magnitude, the actual energy loss is somewhat less than their sum~\cite{Fadafan:2008bq}.)
This calculation shows that the {\it calculational method} that yields the result that a quark moving in a straight line with constant speed $v$ in the regime (\ref{eq:glimit2}) loses energy via drag can yield other results in other circumstances (see \cite{Chernicoff:2008sa,Chernicoff:2009re,Chernicoff:2009ff} for further examples).  In the case of circular motion, the criterion for the validity of the calculational method is again (\ref{eq:glimit2}), but there is a wide range of parameters for which this criterion and (\ref{RadiationDominatedCriterion}) are both 
satisfied~\cite{Fadafan:2008bq}. This means that for a quark in circular motion,  the calculation is reliable in a regime where energy loss is as if due to radiation in vacuum.

\section{Momentum broadening of a heavy quark}
\label{sec:HQBroadening}

In the same regime in which a heavy quark moving through the strongly coupled plasma of ${\cal N}=4$ SYM theory loses energy via drag, as reviewed in Section~\ref{sec:HQDrag}, it is also possible to use gauge/gravity duality to calculate the transverse (and, in fact, longitudinal) momentum broadening induced by motion through the 
plasma~\cite{CasalderreySolana:2006rq,Gubser:2006nz,CasalderreySolana:2007qw,CasalderreySolana:2009rm}.  We shall review these calculations in this section.  They have been further analyzed~\cite{Dominguez:2008vd,deBoer:2008gu,Giecold:2009cg}, and extended to study the effects of nonconformality~\cite{Liu:2008tz,Peng:2008ru,Gursoy:2010aa} and 
acceleration~\cite{Xiao:2008nr,Caceres:2010rm}.

For non-relativistic heavy quarks, the result (\ref{eq:drag}) is not surprising. The dynamics of this
particle is that of Brownian motion which can be described by the effective equation of motion
\begin{equation}
\label{eq:Langevine}
\frac{dp}{dt}=-\eta_D p + \xi(t)  \, ,
\end{equation}
where $\xi(t)$ is a random force that encodes the interaction of the medium with the heavy probe and that causes the momentum broadening that we describe in this Section. 
For heavy quarks, we have seen in (\ref{eq:etaads}) that
$\eta_D$ is suppressed by mass.   This reflects the obvious fact that 
the larger the mass the harder
it is to change the momentum of the particle. Thus, for a heavy quark the typical time for such a change, $1/\eta_D$, is long compared to any microscopic time scale of the medium 
$\tau_{\rm med}$.  
 This fact allows us to characterize the
force distribution by the two point correlators
\begin{eqnarray}
\llangle \xi_T(t) \xi_T (t') \rrangle &=& \kappa_T \delta(t-t') \, , \nonumber\\
\llangle \xi_L(t) \xi_L (t') \rrangle &=& \kappa_L \delta(t-t') \, ,  \label{eq:whitenoise}
\end{eqnarray}
where the subscripts $L$ and $T$ refer to the forces longitudinal and transverse to the direction of the particle's 
motion. 
Here, we are also assuming an isotropic plasma which leads to $\llangle \xi_L(t) \rrangle=\llangle \xi_T(t) \rrangle=0$.
 In general, the force
correlator would have a nontrivial dependence on the time difference (different from $\delta(t-t')$).
However, since the dynamics of the heavy quark happen on timescales that are much larger than 
$\tau_{\rm med}$, we can approximate all medium correlations as happening instantaneously. 
It is then easy to see that the coefficient  $\kappa_T$ ($\kappa_L$) 
 corresponds to the mean squared transverse (longitudinal) momentum transferred to the heavy quark  per unit time.  For example,  the transverse momentum broadening   is given by
 \begin{equation}
 \llangle {\bf p}^2_\perp \rrangle = 2 \int dt dt' \llangle \xi_T(t)_T \xi_T (t') \rrangle  =2 \kappa_T \T
 \, ,
 \end{equation}
 where $\T$ is the total time duration (which should be smaller than $1/\eta_D$) and where the 2 is the number of transverse dimensions. 
 It is clear from the correlator that $\kappa_{T}$ is a property of the medium, independent
 of any details of the heavy quark probe.   Our goal in this section is to calculate $\kappa_T$ and $\kappa_L$.  We shall do so first at low velocity, and then throughout the velocity regime in which the calculation of the drag force is valid.

Before we begin, we must show that
in the limit we are considering the noise distribution
is well characterized by its second moment.  Odd number correlators vanish because of symmetry, so the first higher moment  to consider is the fourth moment of the distribution of the  transverse momentum picked up by the heavy quark moving through the plasma  
\begin{equation}
\llangle {\bf p^4_\perp} \rrangle= \int dt_1 dt_2 dt_3 dt_4 \llangle
                                                                                                   \xi_T(t_1)                                                                                                                                                                                                                 
                                                                                                   \xi_T(t_2) 
                                                                                                   \xi_T(t_3) 
                                                                                                   \xi_T(t_4) 
                                                                                                   \rrangle \, .
\end{equation}
The four-point correlator may be decomposed as
\bea
\label{eq:fpcor}
\llangle
\xi_T(t_1)                                                                                                                                                                                                                 
\xi_T(t_2) 
\xi_T(t_3) 
\xi_T(t_4) 
\rrangle 
             &=&             \llangle
             \xi_T(t_1)                                                                                                                                                                                                                 
\xi_T(t_2) 
\xi_T(t_3) 
\xi_T(t_4) 
\rrangle_c      \\
             &&+ \llangle \xi_T(t_1) \xi_T(t_2) \rrangle      \llangle \xi_T(t_3) \xi_T(t_4) \rrangle       \nonumber \\                                                                                         
             &&+ \llangle \xi_T(t_1) \xi_T(t_3) \rrangle      \llangle \xi_T(t_2) \xi_T(t_4) \rrangle       \nonumber \\                                                                                        
             &&+ \llangle \xi_T(t_1) \xi_T(t_4) \rrangle      \llangle \xi_T(t_2) \xi_T(t_3) \rrangle   \,   ,\nonumber \\                                                                                     
\eea
which is the definition of the connected correlator. Due to time translational invariance,
the connected correlator is a function
\be
             \llangle
	\xi_T(t_1)                                                                                                                                                                                                                 
\xi_T(t_2) 
\xi_T(t_3) 
\xi_T(t_4) 
\rrangle_c      = f(t_4-t_1, t_3-t_1,t_2-t_1) \, .
\ee
As before, the  correlator has a characteristic scale of the order of the medium scale. As a consequence,
since the expectation value due to  the connected part has only one free integral,
we find
\be
\llangle {\bf p^4_\perp} \rrangle= \left(3 \, \left(2  \kappa_T  \right) ^2+ \mathcal{O} (\frac{\tau_{\rm med}}{\T})\right)\T^2  \, ,
\ee
where the dominant term comes from the disconnected parts in \Eq{eq:fpcor}.
Since we are interested in times parametrically long compared to $\tau_{\rm med}$, we can neglect the connected part of the correlator.   

\subsection{$\kappa_T$ and $\kappa_L$ in the $p\rightarrow 0$ limit}

The  dynamical equations (\ref{eq:Langevine})  together with (\ref{eq:whitenoise}) constitute
the Langevin description of heavy quarks in a medium.  In the $p\rightarrow 0$ limit, there is no distinction between transverse and longitudinal, meaning that both the fluctuations in (\ref{eq:whitenoise}) must be described by the same correlator with
$\kappa_L=\kappa_T\equiv\kappa$.   The Langevin equations (\ref{eq:Langevine}) and (\ref{eq:whitenoise}) describe the time evolution of the probability distribution for the momentum of an ensemble of heavy quarks in a medium.  
A standard analysis shows that, independent of the initial probability distribution, after sufficient time any solution to the Langevin equation yields the probability distribution
\begin{equation}
\mathcal{P} ({\bf p},t\rightarrow \infty) 
 =\left(\frac{1}{ \pi}  \frac{\eta_D}{
                                                                               \kappa 
                                                                               }
\right)^{3/2}
                                      \exp\left\{ -
                                                           {\bf p}^2
                                                           \frac{\eta_D}{
                                                                               \kappa 
                                                                               }
                                          \right\}\ ,
\end{equation}
which coincides with the equilibrium (i.e. Boltzmann) momentum distribution for the heavy quark provided that
\begin{equation}
\label{eq:Einsteinrel}
\eta_D=\frac{\kappa}{2 M T} \, .
\end{equation}
This expression is known as the Einstein relation.  
Thus,
the Langevin dynamics of non-relativistic heavy quarks is completely determined by the momentum
broadening $\kappa$, and the heavy quarks equilibrate at asymptotic times.

The Einstein relation (\ref{eq:Einsteinrel}) together with the computation 
of $\eta_D$ in (\ref{eq:etaads}) for strongly coupled ${\cal N}=4$ SYM theory allow us to infer the value of $\kappa$ for this strongly coupled conformal plasma, namely
\begin{equation}
\kappa=\pi \sqrt{\lambda} T^3 \, .
\label{KappaResult}
\end{equation}
The dynamical equation (\ref{eq:drag}) that we used in the previous section does not include the noise term simply because in that section we were describing the change in the
mean heavy quark momentum in the ensemble of this section. 


\subsection{Direct calculation of the noise term}

We would like to have a direct computation of the noise term 
in the description of a heavy quark
in a strongly coupled gauge theory  plasma.  There are two motivations for this: 1) to explicitly check
that the Einstein relation (\ref{eq:Einsteinrel}) 
is fulfilled and 2) to compute the momentum broadening for moving
heavy quarks, which are not in equilibrium with the plasma and to which the Einstein relation therefore does not apply.  This computation is somewhat technical; the reader interested only in the results for $\kappa_T$ and $\kappa_L$ for a moving heavy quark may skip 
to Section~\ref{sec:kappaCalculation}.

We need to express the momentum broadening in terms which are easily computed within
the gauge/gravity 
correspondence. To do so, we prepare a state of the quark at an initial time $t_0$ which
is moving at given velocity $v$ in the plasma.  In quantum mechanics, the state is characterized by 
a density matrix, which is a certain distribution of pure states
\begin{equation}
\rho(t_0)= \sum_n w(n) \left |n\rrangle \llangle n \right|
\end{equation}
where the sum is performed over a complete set of states and the weight $w(n)$ is the ensemble. 
For a thermal distribution, the states are eigenstates of the Hamiltonian and  $w(n)=\exp\{-E_n/T\}. $

 In the problem we are interested in, the density matrix includes not only the quark degrees of freedom
 but also the gauge degrees of freedom. However, we start our discussion using 
 a one particle system --- initially for illustrative purposes.  In this case, 
 the distribution function of the particle is
defined from the density matrix as
\begin{equation}\label{DensityMatrixCoordinateSpace}
\hat f (x, x'; t_0 )= \sum_n  w(n)  \llangle x \right | \left. n \rrangle \llangle n \right | \left. x' \rrangle \, ,
\end{equation}
where, as usual,  $\llangle x \right | \left. n \rrangle$ is the wave function of the particle in the state 
$ \left| n \rrangle$.
  It  is also common to call  $f(x,x')$ the density matrix. 
It is conventional to introduce the mean and relative coordinates and express the density 
matrix as 
\begin{equation}
f(X,r;t_0)=\hat f \left(X+\frac{r}{2}, X-\frac{r}{2} ; t_0\right) \, ,
\end{equation}
where $X=(x+x')/2$ and $r=x-x'$.
It is then easy to see that 
the mean position and mean momentum of the single particle with a given density matrix
are given by
\begin{eqnarray}
\llangle x \rrangle =
{\rm tr} \left\{\rho (t_0) \,x \right\}&=& 
 \int dx  x \hat f (x,x;t_0)
 =\int dX X  f(X,0;t_0)\nonumber\\
\llangle p \rrangle = 
{\rm tr} \left\{\rho (t_0) \,p \right\}&=& 
\int dx  \frac{-i}{2} \left( \del_x -\del_{x'}  
                                                      \right) \left. \hat f (x,x'; t_0) \right |_{x'=x}\nonumber\\
&=&   -i \int dX  \partial_r f(X,r;t_0)|_{r=0}\ ,                                                
\end{eqnarray}
meaning that $r$ is the conjugate variable to the momentum
and, of most interest to us, the mean squared momentum of the distribution is
\begin{equation}
\llangle p^2 \rrangle= -\int d X \del^2_r \left. f(X,r; t_0) \right |_{r=0}\, .
\end{equation}

Returning now to the problem of interest to us, we must consider an ensemble containing 
the heavy quark and also the 
gauge field degrees of freedom. Since we assume the mass of the quark to be much larger than
the temperature, we can described the pure states of the system as 
\begin{equation}
\left | A '\right>= Q_a^{\dagger} (x) \left | A\right> \, .
\end{equation}
where $|A\rangle$ is a state of the gauge fields only, $|A'\rangle$ denotes a state of the heavy quark plus the gauge fields, and
$Q_a^{\dagger}(x)$
is the creation operator (in the Schr\"odinger picture) of a heavy quark with color $a$ at position 
$x$. 
Corrections to this expression are (exponentially) suppressed by $T/M$. 
The Heisenberg representation of the operator $Q(x)$ satisfies the equation of motion
\begin{equation}
\label{eq:nreq}
\left( iu\cdot D - M \right) Q =0 \,  ,
\end{equation}
where $u$ is the four velocity of the quark and $D$ is the covariant derivative with 
respect to the gauge fields of the medium. This equation realizes the physical intuition
that the heavy quark trajectory is not modified by the interaction with the 
medium, which leads only to a modification of the quark's phase.\footnote{The expression
(\ref{eq:nreq}) can also be derived from the Dirac equation by performing a
Foldy-Wouthuysen transformation, which in the heavy quark rest frame is given by
$Q=\exp\{\gamma \cdot D /2 M\} \,\psi $, 
where $\gamma=1/\sqrt{1-v^2}$.}  

The full density matrix of the system, $\rho$, describes 
an ensemble of  all the degrees of freedom of the
system. Since we are only interested in the effects of the medium on the momentum of the heavy quark probe, 
we can define
a one-body density matrix from the full density matrix by integrating over the  gauge degrees of freedom
\begin{eqnarray}
\label{eq:trdef}
f(X,r; t_0)&=&\llangle Q^\dagger_a\left(X-\frac{r}{2};\right) U_{ab}
                                  Q_b\left(X+\frac{r}{2} \right) \rrangle 
                                  \nonumber
                                  \\
&=&
                                  {\rm Tr} \left[ \rho \, Q^\dagger_a\left(X-\frac{r}{2} \right)
                                   U_{ab} 
                                  Q_b\left(X+\frac{r}{2} \right) 
                                                 \right] \, ,
\end{eqnarray}
where the trace is taken over a complete set of states
\be
\sum_{A, a} \int d x \, Q_a^\dagger (x) \left | A\right> \left< A\right| Q_a(x) \, .
\ee
Note that the inclusion of the operators in the trace in (\ref{eq:trdef})
plays the same role as the projectors $\left | x \right\rangle$ in (\ref{DensityMatrixCoordinateSpace}).  The gauge link
$U_{ab}$  in (\ref{eq:trdef}) joins the points $X+r/2$ and $X-r/2$ to ensure gauge
invariance. In the long time limit,  the precise path is not important, and we 
will assume that $U_{ab}$ is a straight link.
To simplify our presentation, we shall explicitly treat only transverse 
momentum broadening, which means taking the separation $r$ to be in a direction 
perpendicular to the direction of motion of the heavy quark,  $\rperp$.

At a later time $t$, after the heavy quark has propagated through the plasma for a time $t-t_0$, the one-body density matrix has evolved from (\ref{eq:trdef}) to
\begin{eqnarray}
\label{eq:flater}
f(X,\rperp; t)&=
                                  {\rm Tr} & \Big[ \rho \,  
                                  e^{i H (t-t_0)} 
                                  Q^\dagger_a\left(X-\frac{\rperp}{2} \right)
                                  e^{-i H (t-t_0)}  \nonumber \\
                                 & &
                                  e^{i H (t-t_0)}  U_{ab} \,  e^{-i H (t-t_0)} 
                                  \nonumber \\
                                  &&
                                  e^{iH(t-t_0)}
                                  Q_b\left(X+\frac{\rperp}{2} \right)
                                  e^{-iH(t-t_0)} 
                                                 \Big] \, ,
\end{eqnarray}
where we have introduced evolution operators to express the result in 
the Heisenberg picture.
We then introduce a complete set of states, obtaining
\begin{eqnarray}
\label{eq:falmostdone}
f(X,\rperp; t)&=&
                      \int d\x_1 d\x_2\sum_{A_1, A_2, A_3, A_4} 
                       \rho_{a_1 a_2}[\x_1,\x_2; A_1, A_2]  \nonumber \\
                      & & \llangle A_2 \right |
                                 Q_{a_2}(\x_2) Q^{\dagger}_a \left(X-\frac{\rperp}{2}; t \right)
                             \left | A_3 \rrangle \nonumber \\
                      & &  \llangle A_3 \right |
                                U_{ab}(t)
	                      \left | A_4 \rrangle \nonumber \\
	              & & \llangle A_4 \right |
                                 Q_b \left(X+\frac{\rperp}{2}; t \right) Q^{\dagger}_{a_1}(x_1)
                             \left | A_1 \rrangle \,, 
\end{eqnarray}
where we have defined
\bea
\rho_{a_1 a_2}[\x_1,\x_2; A_1, A_2] &\equiv &
\left <A_1 \right|
Q_{a_1} \left(\x_1 \right) \, \rho \, Q^{\dagger} _{a_2} \left(\x_2 \right)
\left | A_2 \right> \, .
\eea

\begin{figure}
\setlength{\unitlength}{3947sp}%
\begingroup\makeatletter\ifx\SetFigFont\undefined%
\gdef\SetFigFont#1#2#3#4#5{%
  \reset@font\fontsize{#1}{#2pt}%
  \fontfamily{#3}\fontseries{#4}\fontshape{#5}%
  \selectfont}%
\fi\endgroup%
\begin{picture}(5790,906)(-200,-1774)
\put(1726,-1036){\makebox(0,0)[lb]{\smash{{\SetFigFont{14}{16.8}{\rmdefault}{\mddefault}{\updefault}{\color[rgb]{0,0,0}$t_{\scriptscriptstyle C}$}%
}}}}
{\color[rgb]{0,0,0}\thinlines
\put(1801,-1186){\circle*{150}}
}%
{\color[rgb]{0,0,0}\put(3000,-1441){\circle*{150}}
}%
\thicklines
{\color[rgb]{0,0,0}\put(0,-1198){\line( 1, 0){5447}}
}%
{\color[rgb]{0,0,0}\put(5463,-1444){\line(-1, 0){5447}}
}%
\put(4573,-1036){\makebox(0,0)[lb]{\smash{{\SetFigFont{14}{16.8}{\rmdefault}{\mddefault}{\updefault}{\color[rgb]{0,0,0}$X= (t, x_o + v\,\Delta t)$}%
}}}}
\put(3973,-1711){\makebox(0,0)[lb]{\smash{{\SetFigFont{14}{16.8}{\rmdefault}{\mddefault}{\updefault}{\color[rgb]{0,0,0}$X= (t, x_o + v\,\Delta t-i v \epsilon)$}%
}}}}
\put(0,-1711){\makebox(0,0)[lb]{\smash{{\SetFigFont{14}{16.8}{\rmdefault}{\mddefault}{\updefault}{\color[rgb]{0,0,0}$X_f=(t_o - i\epsilon , x_o - i v \epsilon)$}%
}}}}
\put(0,-1036){\makebox(0,0)[lb]{\smash{{\SetFigFont{14}{16.8}{\rmdefault}{\mddefault}{\updefault}{\color[rgb]{0,0,0}$X_o=(t_o , x_o)$}%
}}}}
\put(3000,-1711){\makebox(0,0)[lb]{\smash{{\SetFigFont{14}{16.8}{\rmdefault}{\mddefault}{\updefault}{\color[rgb]{0,0,0}$t'_{\scriptscriptstyle C}$}%
}}}}
{\color[rgb]{0,0,0}\put(5465,-1444){\oval(  4,  0)[bl]}
\put(5465,-1321){\oval(246,246)[br]}
\put(5465,-1321){\oval(246,246)[tr]}
\put(5465,-1198){\oval(  4,  0)[tl]}
}%
\end{picture}%
\caption{\small \label{ctour}
Time contour $C$ in the complex time plane for the path integral (\ref{eq:fwpi}). Here, 
$\Delta t\equiv  t-t_0$ and
the $i \, \epsilon$
prescription in time is translated to the longitudinal coordinate $x$ 
since the quark trajectory is $x=v t$.
The two-point functions computed from the partition function (\ref{eq:fwpi}) 
 are evaluated at two arbitrary points 
$t_C$ and $t'_C$ on the contour. }
\label{path}
\end{figure}
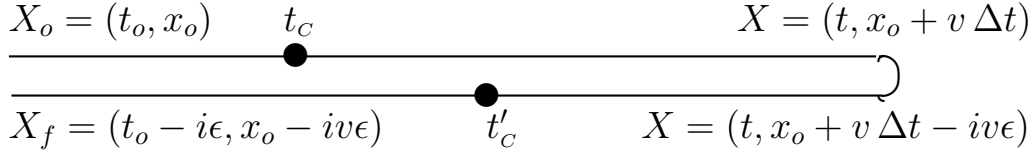

\begin{figure}
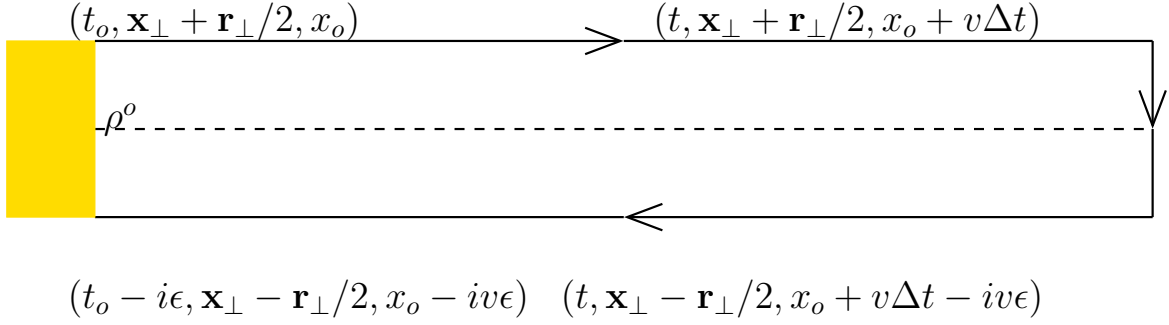
\caption{\small 
Graphical representation of \Eq{frperp}.  The Wilson line indicated by the black line is denoted $W_{\C}[\rperp/2, -\rperp/2]$. This Wilson line is traced with the initial 
density matrix, $\rho^{o}_{a_1 a_2}$. The horizontal axis is along the time direction and the vertical axis
is along one of the transverse coordinates, $x_\perp$. $\Delta t\equiv t-t_0$.
}
\label{path2}
\end{figure}

The expression (\ref{eq:falmostdone})  can be  expressed as a path integral. Note that the 
second line
in (\ref{eq:falmostdone}) is an anti-time ordered correlator; thus, its path integral representation involves
a time reversal of the usual path integral. Instead of introducing two separate path integrals corresponding to the
second and fourth lines of (\ref{eq:falmostdone}), we introduce 
the time 
contour shown in \Fig{ctour} and use this contour to define a single path integral. The fields $A_1$ and $A_2$
are the values at the endpoints of the contour.
The one-body density matrix then reads
\begin{eqnarray}
\label{eq:fwpi}
f(X,\rperp; t)&=&
                           \sum_{A_1, A_2} \int d\x_1 d\x_2 
                        \int \left[ \mathcal{D} A \right] 
                        \, \mathcal{D} Q \, \mathcal{D} Q^{\dagger} 
                        e^{i\int_C d^4x \left\{ \mathcal{L}_{YM}
                        +Q^\dagger \left( i u \cdot D -M \right) Q \right\} } \nonumber \\
                        &&
                       \rho_{a_1 a_2}[\x_1,\x_2; A_1, A_2] \,  U_{ab}(t)  \nonumber \\
&&                        Q_{a_2}\left(\x_2, t_0-i\epsilon  \right)\, 
                             Q^\dagger_{a}\left(X-\frac{\rperp}{2}, t-i\epsilon  \right) 
                             \nonumber \\
&&                        Q_{b}\left(X+\frac{\rperp}{2}, t  \right)  \,
			  Q^{\dagger}_{a_1}\left(\x_1, t_0\right)   \, .                                                                            
\end{eqnarray}
Integrating out the quark field and working to leading order in $T/M$ (neglecting the fermionic determinant)
yields
\begin{equation}
\label{frperp}
f\left(X,\rperp; t \right)=\llangle
                               {\rm  tr} \left[ 
                               \rho\left[X+\frac{\rperp}{2},X-\frac{\rperp}{2};A_1, A_2 \right]
                               W_\C\left[\frac{\rperp}{2}, -\frac{\rperp}{2}\right]
                               \right]
	  	             \rrangle_A \, ,
\end{equation}
where the subscript $A$ indicates averaging with respect to the gauge fields, and where the Wilson 
line 
$W_\C\left[\rperp/2, -\rperp/2\right]$ is defined in \Fig{path2}.  We have used the fact that the 
Green's function of \Eq{eq:nreq} is the (contour ordered) Wilson line.

Next, we perform a Taylor expansion of the time-evolved density matrix (\ref{frperp}) 
about $\rperp=0$, obtaining
\begin{eqnarray}
f\left(X,\rperp; t \right)& =& f\left(X,0;t  \right) +\nonumber \\
&&
                               \frac{\rperp^2}{2}
                                           \llangle
                               {\rm  tr} \left[ 
                               \frac{\del^2}{\del \rperp^2}\rho\left[X+\frac{\rperp}{2},X-\frac{\rperp}{2};A_1, A_2 \right]
                               W_\C\left[0\right]
                               \right]
	  	             \rrangle_A  + \nonumber \\
		             &&
		             \frac{\rperp^2}{2} \kappa_T  \Delta t
\llangle
                               {\rm  tr} \left[ 
                               \rho\left[X,X;A_1 ,A_2 \right]
                               W_\C\left[0\right]
                               \right]
	  	             \rrangle_A \, .	             
\end{eqnarray}
The second term in this expression involves only derivatives of the initial density matrix; thus, it is
the mean transverse momentum squared of the initial distribution (which may be supposed to be small).
In the last term, which scales with the elapsed time $\Delta t$, we have defined 
\begin{equation}
\label{eq:baswilson}
\kappa_{T} {\Delta t}  = \frac{1}{4} \, \frac{1}{\llangle  \tr \rho W_\C[0] \rrangle_{A}} 
\int_C d t_C d  t'_C \, \llangle \tr \, \rho[X,X; A_1, A_2] \,
\frac{ \delta^2 W_{\C}[\delta y] }{\delta y(t_C) \,\delta y(t'_C)}  \rrangle_{A} \, ,
\end{equation}
where $t_C$ denotes time along the contour depicted in \Fig{path}.  $\kappa_T \Delta t$ 
is the mean transverse momentum squared picked up by the heavy 
quark during the time $\Delta t$.
We have expressed the transverse derivatives of the Wilson line as functional
derivatives with respect to the path of the Wilson line. The path $\delta y$ denotes a small 
transverse fluctuation $\delta y(t)$ away from the path $X_1=vt$. 

The contour $\delta y$ may be split into two  pieces, $\delta y_1$ and $\delta y_2$, which run along the
time ordered and anti-time ordered part of the path. Thus, the fluctuation calculation defines four 
correlation functions  
\begin{eqnarray}
iG_{11}(t, t') &=& 
\frac{1}{\llangle  \tr \rho^o W_\C[0,0] \rrangle_{A}} 
\, \llangle \tr \, \rho^o  \; 
\frac{ \delta^2 W_{\C}[\delta y_1, 0] }{\delta y_1(t) \,\delta y_1(t')}  \rrangle_{A} \, ,  \\
iG_{22}(t, t') &=& 
\frac{1}{\llangle  \tr \rho^o W_\C[0,0] \rrangle_{A}} 
\, \llangle \tr \, \rho^o  \; 
\frac{ \delta^2 W_{\C}[0, \delta y_2] }{\delta y_1(t) \,\delta y_2(t')}  \rrangle_{A} \, , \\
iG_{12}(t, t') &=& 
\frac{1}{\llangle  \tr \rho^o W_\C[0,0] \rrangle_{A}} 
\, \llangle \tr \, \rho^o  \; 
\frac{ \delta^2 W_{\C}[\delta y_1, \delta y_2] }{\delta y_1(t) \,\delta y_2(t')}  \rrangle_{A} \, ,  \\
iG_{21}(t, t') &=& 
\frac{1}{\llangle  \tr \rho^o W_\C[0,0] \rrangle_{A}} 
\, \llangle \tr \, \rho^o  \; 
\frac{ \delta^2 W_{\C}[\delta y_2, \delta y_2] }{\delta y_1(t') \,\delta y_2(t)}  \rrangle_{A}  \, .
\end{eqnarray}
Note that the first two correlators correspond to time ordered and anti-time ordered
correlators, while the last two are unordered.
We can then divide the integration over $t_C$ and $t_{C'}$ in (\ref{eq:baswilson}) into four parts
corresponding to the cases where each of $t_C$ and $t_{C'}$ is on the upper or lower half of the contour in Fig.~\ref{ctour}.  
In the large $\Delta t$ limit we can then use time translational invariance  
to cast (\ref{eq:baswilson}) as
\begin{equation}
\label{hatqf}
\kappa_T=\lim_{\omega\rightarrow 0} \frac{1}{4}  \int dt e^{+i\omega t} 
        \left[ i G_{11}(t,0) + iG_{22}(t,0) + i G_{12}(t,0)+i G_{21}(t,0)\right]    \, .
\end{equation}
This admittedly rather formal 
expression for $\kappa_T$ is as far as we can go in general.  In Section~\ref{sec:kappaCalculation} we evaluate $\kappa_T$ (and $\kappa_L$) in the strongly coupled plasma of ${\cal N}=4$ SYM theory.

Although our purpose in deriving the expression (\ref{hatqf}) is to use it to analyze the case $v\neq 0$, it can be further simplified in the case that $v=0$.
On the time scales under consideration, the static quark is in 
equilibrium with the plasma, and the Kubo-Martin-Schwinger relation which takes
the form
\begin{equation}
\label{KMS}
i\left[ G_{11}(\omega)+G_{22}(\omega) + G_{12}(\omega) +G_{21}(\omega) \right]= -4 \coth\left(\frac{\omega}{2T}\right) \, {\rm Im} G_{R}(\omega)
\end{equation}
for $\epsilon\rightarrow 0$ applies~\cite{LeBellac}. Here, $G_R$ is the retarded correlator. Thus,
we find
\begin{equation}
\label{eq:kappacor}
\kappa_T(v=0)=\lim_{\omega\rightarrow 0}  \left(-\frac{2 T}{\omega} \right) {\rm Im } \,G_R (\omega) \, .
\end{equation}
If $v\ne0$, 
however, we must 
evaluate the four correlators in the expression (\ref{hatqf}).

\subsection{$\kappa_T$ and $\kappa_L$ for a moving heavy quark}
\label{sec:kappaCalculation}

We see from the expression (\ref{eq:baswilson}) that the transverse momentum broadening 
coefficient $\kappa_T$ is extracted by analyzing small fluctuations in 
the path of the Wilson line depicted in Fig.~\ref{path2}. In the strongly coupled plasma of ${\cal N}=4$ SYM theory, we can use gauge/gravity duality to evaluate $\kappa_T$ starting 
from (\ref{eq:baswilson}).
In the dual gravitational description, the small fluctuations in the path of the  Wilson line amount  to perturbing
the location on the boundary at which the classical string (whose unperturbed shape is given by (\ref{eq:solxi})) terminates according to
\begin{equation}
\left(x_1(t, z), 0, 0 \right) \rightarrow \left(x_1(t, z),  y (t,z), 0 \right) \, .
\label{PerturbationsOfWorldSheet}
\end{equation}
The perturbations of the Wilson line at the boundary yield fluctuations
on the  string world sheet dragging behind
the quark.   Because we wish to calculate $\kappa_T$, in (\ref{PerturbationsOfWorldSheet}) we have only introduced perturbations transverse to the direction of motion of the quark.  We shall quote the result for $\kappa_L$ at the end; calculating it requires extending (\ref{PerturbationsOfWorldSheet}) to include perturbations to the function $x_1(t,z)$.

In order to analyze fluctuations of the string world sheet, we begin by casting the
metric induced on the string world sheet in the absence of any perturbations
\begin{equation}
ds^2_{\rm ws}=\frac{R^2}{z^2}\left(
                                                       -\left(f(z)-v^2\right) d\tau^2 +
                                                        \frac{\hat f (z)}{f^2(z)} d\sigma^2-
                                                         v^2 \frac{z^2/z_0^2}{f(z)} \left(d\tau d\sigma+d\sigma d\tau \right)
                                              \right) \, ,
\label{WorldSheetMetric}
\end{equation}
in a simpler form.  In (\ref{WorldSheetMetric}),
we have defined $\hat{f}(z)\equiv 1-z^4/(z_0^4 \gamma^2)$. 
The induced metric (\ref{WorldSheetMetric}) is diagonalized
by the change of world sheet coordinates
\begin{eqnarray}
\label{eq:chstraight}
\hat t &=&\frac{t}{\sqrt{\gamma}} +\frac{z_0}{2\sqrt{\gamma}}\Biggl(
                                    \arctan\left(
                                                \frac{z}{z_0}
                                                \right)
                                     -\arctanh\left(
                                                    \frac{z}{z_0}
                                                     \right)  \nonumber\\
                                                     &&\qquad\qquad\qquad\qquad
                                                -\sqrt{\gamma}\arctan\left(
                                                \frac{\sqrt{\gamma}z}{z_0}
                                                \right)
                                    +\sqrt{\gamma}\arctanh\left(
                                                \frac{\sqrt{\gamma}z}{z_0}
                                                \right)
                                        \Biggr) \nonumber \, , \\
  \zh&=&\sqrt{\gamma}z \, ,
\end{eqnarray}
in terms of which the induced metric takes the simple form
\begin{equation}
ds^2_{\rm ws}=\frac{R^2}{\zh^2}
                     \left(
                     - f(\zh) d\hat t^2 + \frac{1}{f(\zh)} d\zh^2 
                      \right) \, .
    \label{DiagonalWorldSheetMetric}                  
\end{equation}
Note that this has the same form as the induced metric for the world sheet hanging below a motionless quark, upon making the replacement
$(\hat t,\zh)\rightarrow (t,z)$. In particular, the metric (\ref{DiagonalWorldSheetMetric}) has a horizon at 
$\hat z=z_0$, which means that the metric describing the world sheet of the string trailing
behind the moving quark has a world sheet horizon at 
$z=z_{\rm ws}\equiv z_0/\sqrt{\gamma}$.   For $v\rightarrow 0$, the location of the world sheet horizon drops down toward the spacetime horizon at $z=z_0$.  But, for $v\rightarrow 1$, the world sheet horizon moves closer and closer to the boundary at $z=0$, i.e. towards the ultraviolet.
As at any horizon, the singularity at $z=z_{\rm ws}$ (i.e. at $\hat z=z_0$) 
in (\ref{DiagonalWorldSheetMetric}) is just a coordinate singularity.  In the present case, this is manifest since (\ref{DiagonalWorldSheetMetric}) was obtained from (\ref{WorldSheetMetric}) which is regular at $z=z_{\rm ws}$ by a coordinate transformation (\ref{eq:chstraight}).
Nevertheless, the world sheet horizon has clear physical significance:  at $z=z_{\rm ws}$
the local speed of light at this depth in the bulk matches the speed $v$ with which the quark at the boundary is moving.
Furthermore, and of direct relevance to us here, because of the world 
sheet horizon at $z=z_{\rm ws}$ fluctuations of the string world sheet at 
 $z>z_{\rm ws}$, below --- to the infrared of ---  the world sheet horizon, 
are causally disconnected from fluctuations at $z<z_{\rm ws}$ above the world sheet horizon and in particular are causally disconnected from the boundary at $z=0$.  

The remarkable consequence of the picture that emerges from the above analysis of the unperturbed string world sheet trailing behind the quark at the boundary moving with speed $v$
is that the momentum fluctuations of this quark can be thought of as due to the Hawking radiation on the string world sheet, originating from the world sheet horizon 
at $z=z_{\rm ws}$~\cite{Gubser:2006nz,CasalderreySolana:2007qw}.  It is as if the force fluctuations that the quark in the boundary gauge theory feels are due to the fluctuations of the string world sheet to which it is attached, with these fluctuations arising due to the Hawking radiation originating from the world sheet horizon.
It will  therefore prove useful to calculate the Hawking temperature of the world sheet horizon, which we denote $T_{\rm ws}$.
As detailed in  Appendix  \ref{app:HawT} this can be done in the standard fashion, upon using a further coordinate transformation to write the 
metric (\ref{DiagonalWorldSheetMetric}) in the vicinity of the world sheet horizon in the 
form 
$ds^2_{\rm ws} = - b^2 \rho^2 d{\hat t}^2 + d\rho^2$ 
for some constant $b$, where the world sheet horizon is at $\rho=0$.    Then, it is a standard argument that in order to avoid having a conical singularity at $\rho=0$ in the Euclidean version of this metric, namely
$ds^2_{\rm ws} =  b^2 \rho^2 d\hat\theta^2 + d\rho^2$,
$b \hat\theta$ must be periodic with period $2\pi$. 
The periodicity of the variable $\hat \theta$, namely $2\pi/b$, is $1/T$. Since at the boundary, where $z=0$, Eq.~(\ref{eq:chstraight}) becomes
$\hat t = t/\sqrt{\gamma}$,  this argument yields 
\begin{equation}
\label{WorldSheetTemperature}
T_{\rm ws} = \frac{T}{\sqrt{\gamma}}\ ,
\end{equation}
a result that we shall use below.

We have gained significant physical intuition by analyzing the unperturbed string world sheet, but in order to obtain a quantitative result for $\kappa_T$ we must introduce the transverse fluctuations $y(t,z)$ defined in (\ref{PerturbationsOfWorldSheet}) explicitly.  We write the Nambu-Goto action for the string world sheet with $y(t,z) \neq 0$, and expand it to second order in $y$, obtaining the zeroth order action (\ref{Sdrag}) plus a second order contribution
\begin{equation}
\label{eq:NGquadratic}
S_T^{(2)}[y]=\frac{\gamma R^2 }{\tpa} \int \frac{d\hat t d\zh}{\zh^2} \frac{1}{2}  
                                                   \left(
                                                   \frac{\dot{y}^2}{f(\zh)}- f(\zh) y'^{2}
                                                   \right) \, 
\end{equation}
where $\dot{}$ and $'$ represent differentiation with respect to $\hat t$ and $\hat{z}$ respectively.
This action is conveniently expressed as
\begin{equation}
\label{eq:asgen}
S_T^{(2)}[y]=-\frac{\gamma R^2 }{\tpa} \int \frac{d\hat t d\zh}{\zh^2} \frac{1}{2} \sqrt{-h} h^{ab} \del_a y \del_b y
\end{equation}
with $h_{ab}$ the induced metric on the unperturbed world sheet that we have analyzed above.
The existence of the world sheet horizon means that we are only interested in solutions to the equations of motion for the transverse fluctuations $y$ obtained from (\ref{eq:asgen}) that satisfy infalling boundary conditions at the world sheet horizon.  This constraint in turn implies a relation among the correlators analogous to those in (\ref{hatqf}) that describe the transverse fluctuations of the world sheet, and in fact
the relation turns out to be analogous to the Kubo-Martin-Schwinger relation (\ref{KMS}) among the gauge theory correlators~\cite{CasalderreySolana:2007qw}.
%
Consequently, for a quark moving with velocity $v$ the transverse momentum broadening coefficient $\kappa_T(v)$ is given by the same expression (\ref{eq:kappacor}) that is valid at $v=0$ with $T$ replaced by the world sheet 
temperature $T_{\rm ws}$ 
of (\ref{WorldSheetTemperature})~\cite{Gubser:2006nz,CasalderreySolana:2007qw}.
That is,
\begin{equation}
\kappa_T(v)=\lim_{\omega\rightarrow 0}  \left(-\,\frac{2 \,T_{\rm ws}}{\omega} \, {\rm Im } \hat{G}_R (\omega) \right) \, ,
\end{equation}
where $\hat{G}_R$ denotes the retarded correlator at the world-sheet horizon.  
The fact that in the strongly coupled theory there is a KMS-like relation at $v\ne0$ after all
is a nontrivial consequence of the development of the world-sheet horizon.

The computation of the retarded correlator follows the general procedure of 
Ref.~\cite{Son:2002sd} described in Section~\ref{sec:CORR}. 
Since the action (\ref{eq:NGquadratic}) 
is a function of $\hat t$ which is given by $t/\sqrt{\gamma}$ at the boundary, the retarded correlator is a function 
of $\hat {\omega}=\sqrt{\gamma} \omega$ (with $\omega$ the frequency of oscillations at the boundary).
To avoid this complication, and in particular in order to be able to apply the general 
results for ${\rm Im}\,G_R$ that we derived in Section~\ref{sec:TransportProperties}, it is
convenient to define
\begin{equation}
\tilde t = \sqrt {\gamma} \, \hat t \ ,
\end{equation}
so that $\tilde t= t$ at the boundary. We now wish to apply the general expressions (\ref{trco}), (\ref{coe}) and (\ref{eom1}).  In order to do so, we identify the world sheet metric $h^{ab}$ and the field $y$ in the action (\ref{eq:asgen}) with the metric $g^{MN}$ and the field $\phi$ in the 
action (\ref{ree1}), meaning that in our problem the function $q$ in (\ref{ree1}) takes the specific form
\begin{equation}
\frac{1}{q(z)}=\frac{\gamma R^2}{\tpa} \frac{1}{\zh^2} = \frac{\sqrt{\lambda}}{2\pi z^2}\,.
\end{equation}
Furthermore, for the two-dimensional world sheet metric we have $-h=h_{\tilde t \tilde t} h_{zz}$, 
meaning that from the general result (\ref{coe1}) we find
\begin{equation}
- \lim_{\omega \rightarrow 0} \frac{{\rm Im} \, \hat G_R (\omega)}{\omega} = \frac{1}{q(z_{\rm ws})}=
\frac{\gamma \sqrt{\lambda}}{2\pi}  (\pi T)^2\ ,
\end{equation}
and thus
\begin{equation}
\kappa_T=
\sqrt{\lambda \gamma} \pi T^3 \, ,
\label{KappaTResult}
\end{equation}
which is our final result for the transverse momentum broadening coefficient.

The analysis of longitudinal fluctuations and the extraction of $\kappa_L$ proceed analogously to the analysis we have just reviewed, except that in (\ref{PerturbationsOfWorldSheet}) we introduce a perturbation to the function $x_1$ instead of a transverse perturbation $y$.  At quadratic order, there is no coupling between the transverse and longitudinal perturbations.
Remarkably,
the action for longitudinal fluctuations of the string is the same as that for transverse fluctuations 
\Eq{eq:NGquadratic} up to a constant:
\begin{equation}
S_L^{(2)}[x]=\gamma^2 S_T^{(2)}[x] \, ,
\end{equation}
with $\gamma$ the Lorentz factor.  Following the analogous derivation through, we 
conclude that 
\begin{equation}
\kappa_L=\gamma^2 \kappa_T =\gamma^{5/2} \sqrt{\lambda} \pi T^3 \, .
\label{KappaLResult}
\end{equation}
This result shows that $\kappa_L$ depends very strongly 
on the velocity of the heavy quark. Indeed, $\kappa_L$  grows faster with increasing velocity than 
the energy squared of the heavy quark, $\gamma^2 M^2$. Thus, the longitudinal momentum
acquired by a quark moving through a region of strongly coupled ${\cal N}=4$ SYM plasma 
of finite extent does not become a negligible fraction of the energy of the quark in the high energy limit.  This is very different from the behavior of a quark moving through a weakly coupled QCD plasma, in which the longitudinal momentum transferred to the quark can be neglected in the high energy limit.
However,
 we should keep in mind that due to the bound (\ref{eq:glimit}), for a given value of the
 mass $M$ and
 the coupling $\sqrt{\lambda}$
 the calculation of $\kappa_L$ (and of $\kappa_T$) is only valid for finite energy quarks, with $\gamma$ limited by (\ref{eq:glimit}). 
 
 We see from the expressions (\ref{KappaTResult}) and (\ref{KappaLResult})  
 for $\kappa_T$ and $\kappa_L$ derived by explicit analysis of the
  fluctuations that in the $v\rightarrow 0$ limit we have $\kappa_T=\kappa_L=\kappa$ with $\kappa$ given by (\ref{KappaResult}), as we obtained previously from the drag coefficient $\eta_D$ via the use of the 
 Einstein relation (\ref{eq:Einsteinrel}).  This is an example of the fluctuation-dissipation theorem.
 
 In the gauge theory, momentum broadening is due to the fluctuating force exerted on the heavy quark by the fluctuating plasma through which it is moving.  In the dual gravitational description, 
the quark at the boundary feels a fluctuating force due to the fluctuations of the world sheet that describes the profile of the string to which the quark is attached. 
These fluctuations have their origin in the Hawking radiation of fluctuations of the string world sheet originating from the world sheet horizon.  The explicit computation of this world sheet Hawking radiation for a quark at rest was performed in Refs.~\cite{deBoer:2008gu,Son:2009vu}, and these results nicely reproduce those we have obtained within a Langevin formalism.  This computation was extended to quarks moving at nonzero velocity in Refs.~\cite{Giecold:2009cg,CasalderreySolana:2009rm}.

 \subsection{Implications for heavy quarks in heavy ion collisions}
 
 Models based upon Langevin dynamics have been used by a number of authors to describe the motion of heavy quarks within the hot expanding fluid produced in heavy ion 
collisions~\cite{Moore:2004tg,vanHees:2005wb,Horowitz:2007su,vanHees:2007me,Akamatsu:2008ge,Beraudo:2009pe,Gubser:2009sn,Rapp:2009my,Noronha:2009vz,Noronha:2010zc,Alberico:2010tb}, 
with the goal of using data from RHIC~\cite{Adler:2005xv,Abelev:2006db,Adare:2006nq} to constrain the 
diffusion constant $D$ that describes the Brownian motion of a heavy quark in the hot fluid.  These analyses typically assume that the relative velocity of the heavy quark and the hot fluid is not relativistic, meaning that $\kappa_T=\kappa_L\equiv \kappa$.  In this regime, 
the diffusion constant is given by $D=2 T^2/\kappa$, 
meaning that the result (\ref{KappaResult}) translates into the statement that a heavy quark in the strongly coupled ${\cal N}=4$ SYM theory plasma obeys a Langevin equation with 
\begin{equation}
D=\frac{4}{\sqrt{\lambda} }\, \frac{1}{2\pi T} \approx \frac{1.1}{2\pi T} \, \sqrt{ \frac{1}{\alpha_{\rm SYM} N_c} }\ .
\label{StronglyCoupledDiffusionConstant}
\end{equation}
The diffusion constant $D$ parametrizes how strongly the heavy quark couples to the medium, with smaller $D$ corresponding to stronger coupling and shorter mean free path.  $D$ is well defined even if it is so small that it does not make sense to define a mean free path, as is the case in the plasma of strongly coupled ${\cal N}=4$ SYM theory, given 
(\ref{StronglyCoupledDiffusionConstant}).
In a weakly coupled QCD plasma, perturbative calculations suggest~\cite{Moore:2004tg}
\begin{equation}
D_{\rm weakly~coupled} \approx \frac{14}{2\pi T} \, \left( \frac{.33}{\alpha_s} \right)^2\ 
\label{WeaklyCoupledDiffusionConstant}
\end{equation}
or larger by a factor of more than three~\cite{vanHees:2007me},
although it should be noted that the perturbative expansion converges poorly meaning that such calculations can only become quantitatively reliable  at values of $\alpha_s$ that are much 
smaller than 0.33~\cite{CaronHuot:2007gq,CaronHuot:2008uh}.
Note that the result (\ref{StronglyCoupledDiffusionConstant}) is valid for ${\cal N}=4$ SYM theory when $\lambda = 4\pi \alpha_{\rm SYM} N_c$ is large, and $2\pi T D$ is small, while the result (\ref{WeaklyCoupledDiffusionConstant}) is valid for QCD when $\alpha_s$ is small, and $2\pi T D$ is large.

At present, the central lessons from RHIC data on heavy quarks are qualitative. This is because, so far, the best experimental constraints on the spectra of $c$ and $b$ quarks come mainly from single-electron spectra that are known to be dominated by the semileptonic decays of  heavy quarks. The experimental pointing resolution is not yet sufficient to separate $b$ from $c$ quark decays via their displaced vertices. Upgrades at PHENIX and STAR will soon allow one to make this separation. Moreover, with the onset of heavy ion collisions at the LHC, it will soon be possible to measure B- and D-meson spectra up to $p_T \sim 20$ GeV, e.g.~via the decay channel $D^0 \to K^+\, \pi^-$. Also, heavy quark energy loss can be characterized via muon spectra at the LHC. The experimental understanding of how heavy quarks behave in the matter produced in relativistic heavy ion collisions can be expected to become much more quantitative with these developments.

At present, there are two qualitative lessons from the RHIC 
data~\cite{Adler:2005xv,Abelev:2006db,Adare:2006nq}: (i) $R_{AA}$ for isolated electrons (some linear combination of $c$ and $b$ quarks) is small, not that much bigger than $R_{AA}$ for pions; and (ii) $v_2$ for isolated electrons is significant, not that much smaller than that for pions.  At a qualitative level, both these observations suggest that $c$ and $b$ quarks are strongly coupled to the medium, as they must initially be slowed by their relative motion through it and must then flow along with the collective hydrodynamic expansion.    Theoretical calculations involve calculating 
$b$ and $c$ quark production, using a Langevin model to describe their diffusion in the hot fluid produced in the collision, which is described hydrodynamically, and then modelling the freezeout and decay of the heavy $B$ and $D$ mesons, before finally comparing to data on $R_{AA}$ and $v_2$ for isolated electrons.  Two recent calculations along these lines are qualitatively consistent with RHIC data as long as the diffusion constant is in the 
range~\cite{Akamatsu:2008ge}
\begin{equation}
D_{\rm RHIC} \approx \frac{2-6}{2\pi T}\ ,
\label{RHICDiffusionConstant}
\end{equation}
or $(3-5)/(2\pi T)$~\cite{vanHees:2007me}.
Certainly the extraction of $D$, or equivalently $\kappa$, will become more quantitative once $b$'s and $c$'s can be measured separately.   On the theoretical side, it will be interesting to redo these analyses with $\kappa_T \neq \kappa_L$ and both velocity-dependent as in (\ref{KappaTResult}) and (\ref{KappaLResult}).  It will be interesting to see whether the explicit velocity 
dependence of $\kappa_T$ and $\kappa_L$ serves to improve the fit to $R_{AA}$ and $v_2$ for $b$ and $c$ quarks.   There is some tension in the present analyses of isolated electrons with a single fit parameter $D$, with the $R_{AA}$ data favoring somewhat larger values of $D$ while the $v_2$ data favors somewhat smaller values~\cite{Adare:2006nq}.

It is striking that $D$ for strongly coupled ${\cal N}=4$ SYM theory, 
given in (\ref{StronglyCoupledDiffusionConstant}), is even in the same ballpark as that extracted by fitting to present RHIC data, given in (\ref{RHICDiffusionConstant}).  To understand whether this  is a coincidence, we need to ask how $D$ would change if we could deform ${\cal N}=4$ SYM theory so as to turn it into QCD.   This is not a question to which the answer is known, but we can make two observations.  First, in a large class of conformal theories, at a given value of $T$, $N_c$ 
and $\lambda$  both $\kappa$ and  the drag coefficient $\eta_D$ scale with the square root of the entropy density.  (The argument is the same as that for the jet quenching parameter $\hat q$ 
of Section \ref{sec:AdSCFTJetQuenching}, and is given in Ref.~\cite{Liu:2006he}.)   
The number of degrees of freedom in QCD is smaller than that in ${\cal N}=4$ SYM theory by a factor of $47.5/120$ for $N_c=3$, suggesting that $\kappa$ and $\eta_D$ should be smaller in QCD by a factor of $\sqrt{47.5/120}=0.63$, making $D$ larger by 
a factor of $\sqrt{120/47.5}=1.59$.\footnote{Gubser has suggested that the difference in the numbers of degrees of freedom between QCD and ${\cal N}=4$ SYM theory be taken into account by comparing results in the two theories at differing temperatures, 
with $T_{\rm QCD} = (47.5/120)^{1/4} T_{\rm SYM}$~\cite{Gubser:2006qh,Gubser:2009sn}.  
This prescription is inconsistent with known results for a large class of 
conformal theories~\cite{Liu:2006he}, in which $\hat q$, $\kappa$, $\eta_D$ and $D$ {\it all} scale with the square root of the number of degrees of freedom when compared at fixed $T$, $N_c$ and $\lambda$. Perhaps coincidentally, since the drag coefficient $\eta_D$ is proportional to $T^2$ scaling it by the square root of the number of degrees of freedom is equivalent to scaling $T$ by the one-fourth power of the number of degrees of freedom as prescribed in Refs.~\cite{Gubser:2006qh,Gubser:2009sn}.  But, this prescription  is not correct when applied 
to $D\propto 1/T$ or to $\kappa\propto T^3$ or  $\hat q \propto T^3$.}
Second, 
${\cal N}=4$ SYM theory is of course conformal, while QCD is not.  Analysis of a toy model in which nonconformality can be introduced by hand suggests that turning on nonconformality to a degree suggested by lattice calculations of QCD thermodynamics reduces $D$, perhaps by as much as a factor of two~\cite{Liu:2008tz}.  Turning on nonconformality in ${\cal N}=2^*$ theory also reduces $D$. 
So, $D$ in a strongly coupled QCD plasma is not known reliably but the reduction in degrees of freedom and the nonconformality seem to push in opposite directions, suggesting 
that $D_{\rm QCD}$
may not be far from that given in (\ref{StronglyCoupledDiffusionConstant}).  It will be very interesting to watch how this story evolves once it is possible to measure $b$ and $c$ quarks separately in heavy ion collisions.



\section{Disturbance of the plasma induced by an energetic heavy quark}
\label{sec:AdSCFTDragWaves}

In Sections~\ref{sec:HQDrag} and \ref{sec:HQBroadening} we have analyzed the effects of the strongly coupled plasma of ${\cal N}=4$ SYM theory on an energetic heavy quark moving through it, focussing on how the heavy quark loses energy in Section~\ref{sec:HQDrag} and on the momentum broadening that it experiences in Section~\ref{sec:HQBroadening}.
In this Section, we turn the tables and analyze the effects of the energetic heavy quark on the 
medium through which it is 
propagating~\cite{Rischke:1990jy,Stoecker:2004qu,CasalderreySolana:2004qm,Ruppert:2005uz,Satarov:2005mv,Renk:2005si,CasalderreySolana:2005rf,CasalderreySolana:2006sq,Renk:2006mv,Friess:2006fk,Yarom:2007ap,Gubser:2007nd,Yarom:2007ni,Chesler:2007an,Gubser:2007xz,Gubser:2007ga,Gubser:2007ni,Chesler:2007sv,Noronha:2007xe,Neufeld:2008fi,Gubser:2008vz,Neufeld:2008hs,Noronha:2008un,Neufeld:2008dx,Gyulassy:2008fa,Betz:2008wy,Betz:2008ka,Neufeld:2009ep,Gubser:2009sn,Neufeld:2010tz,Betz:2010qh}.
From the point of view of QCD calculations and heavy ion collision phenomenology, the problem of understanding the response of the medium to an energetic probe is quite complicated.
An energetic particle passing through the medium can 
excite the medium on many different wavelengths. 
And, even if the medium was thermalized prior
to the interaction with the probe, the disturbance caused by the probe must drive
the medium out of equilibrium, at least close to the probe.  And, non-equilibrium processes are difficult to treat, especially at strong coupling.

It is natural to attempt to describe the disturbance of the medium using hydrodynamics, with the energetic particle treated as a source for the hydrodynamic equations.  This approach
is based on  two assumptions. First, one must assume that
 the medium itself can be described hydrodynamically.   Second, one has to assume that the non-equilibrium disturbance in the vicinity of the energetic particle relaxes to some locally equilibrated (but still excited) state after the energetic particle has passed on a timescale that is short compared to the lifetime of the hydrodynamic medium itself.
The first assumption is clearly supported by data from heavy ion collisions at RHIC, as discussed in Section~\ref{sec:EllipticFlow}. The second assumption is stronger, and less well justified.  Even though, as we saw in Section~\ref{sec:EllipticFlow},  there is evidence from the data that in heavy ion collisions at RHIC a hydrodynamic medium in local thermal equilibrium forms rapidly, after only a short initial thermalization time, it is not clear {\it a priori}  that the relaxation time for the disturbance caused by an energetic quark plowing through this medium is comparably short, particularly since the density of the medium drops with time.
Finally, even if a hydrodynamic approach to the dynamics of these disturbances is valid, 
the details of the functional form of this hydrodynamic source are unknown, 
since the relaxation process is not under theoretical control.

Keeping the above difficulties in mind, it is still possible to use 
the symmetries of the
problem and some physical considerations to 
make some progress toward understanding the
source for the hydrodynamic equations corresponding to the disturbance caused by an energetic quark.
If the propagating parton is sufficiently energetic, we may assume that it moves at a fixed velocity; this ansatz forces
the source to be a function of $x-vt$, with the parton moving in the $x$-direction. We may also assume that
the source has cylindrical symmetry around the parton direction. 
We may also constrain the source by the amount of energy and momentum that is fed into the plasma, which for the case of the plasma of strongly coupled ${\cal N}=4$ SYM theory we calculated in Section~\ref{sec:HQDrag}.
In an
infinite medium, at  late enough times, all the energy lost by the probe must thermalize and 
be incorporated into heating and/or
hydrodynamic motion.  (This may not  be a good approximation for a very energetic parton propagating through weakly coupled plasma of finite extent since, as we have discussed in Section~\ref{sec:JetQuenching}, in this setting the parton loses energy by the radiation of gluons whose energy and momentum are large relative to the temperature of the medium, which may escape from the medium without being thermalized.)


Although the caveats above caution against attempting to draw quantitative conclusions without further physical inputs, the success of the hydrodynamical description of the medium itself support the conclusion that there 
must be some hydrodynamic response to the passage of the energetic particle through it. 
In particular, there must be some coupling of the energetic particle to sound waves in the medium. 
Since an energetic particle is moving through the plasma with a velocity greater than the velocity of sound $c_s$, the coupling of the energetic particle to the sound mode 
must lead to the
formation of a Mach cone, namely a sound front moving away from the 
trajectory of the energetic particle
at the Mach angle
\begin{equation}
\cos \Theta_M=\frac{c_s}{v} \,,
\end{equation}
with $v$ the velocity of the energetic particle. However, absent further physical inputs it is hard to estimate how strong the Mach cone produced by a {\it point-like} particle shooting through the medium will be.

Several years ago, the interest in these phenomena intensified considerably when 
the formation of Mach cones was suggested as a candidate to explain  certain nontrivial particle correlations seen in heavy ion collisions at RHIC, illustrated in the top-left panel of 
Fig.~\ref{fig:disappear}.  Recall that in these experiments one triggers on (that is, selects events containing) a high energy hadron in the final state, and then measures the distribution of all the other soft hadrons in azimuthal angle $\Delta \phi$ relative to the direction of the trigger particle.
After subtracting the effects of elliptic  flow,
the correlation functions for nucleus-nucleus collisions show a near-side jet of hadrons at small $\Delta\phi$ and a very broad distribution of soft particles on the away-side, at angles around $\Delta\phi\sim \pi$, which may have low peaks at 
$\Delta \phi = \pi \pm \phi_v$ with $\phi_v\approx 1 - 1.2$~radians, on the 
shoulders of the away-side distribution rather than at its center at $\Delta\phi=\pi$.
This correlation
pattern is very different from that seen in proton-proton collisions, 
in which the away-side peak is broader than the near-side peak but is approximately Gaussian in shape, centered about a single peak at $\Delta\phi=\pi$.
Many authors have argued that this nontrivial structure is the debris resulting from the near-complete quenching (thermalization) of the energetic particle that was initially produced back-to-back with the hard parton that became the trigger hadron, with the Mach cone excited in the medium by this energetic particle leading to the peaks on the shoulders of the away-side distribution at $\Delta\phi=\pi \pm \phi_v$~\cite{Stoecker:2004qu,CasalderreySolana:2004qm,Renk:2005si,Betz:2008wy}.
This qualitative picture motivates the need for a good quantitative model for the coupling between an energetic parton and the collective modes of the strongly coupled plasma.

More recently, some doubt has been cast on the interpretation of the qualitative features seen in the top-left panel 
of Fig.~\ref{fig:disappear} in terms of a Mach cone.  Alver and Roland realized that even though $v_3=0$ on average, it is nonzero in individual heavy ion collisions~\cite{Alver:2010gr}.
On average the shape of the overlap between two nuclei in an ensemble of collisions all with the same nonzero impact parameter is almond-shaped, meaning that $v_n=0$ for all odd $n$.  But, individual nuclei are made of 200 nucleons, meaning that when they are Lorentz-contracted into pancakes these pancakes are not perfectly circular.  This means that the overlap between two individual nuclei will not be perfectly almond shaped.  In general, the shape of this overlap zone will have some ``triangularity'', which raises the possibility of a nonzero $v_3$ in the final state, since the hydrodynamic expansion of an initial state that is initially triangular in position space will result in a nonzero $v_3  = \langle \exp [ i\, 3(\phi - \Phi_R )] \rangle$ in momentum space, and hence in the observed hadrons.  (Recall that $\Phi_R$ is the angle of the reaction plane.)      In the analysis of data that goes into producing Fig.~\ref{fig:disappear}, $v_2$ has been subtracted but $v_3$ has not.  If $v_3$ is nonzero in individual events, the consequence in Fig.~\ref{fig:disappear} would be
an excess at angles $\Delta\phi=0$ and $\Delta\phi=\pi\pm\frac{\pi}{3}$.    And, Alver and Roland noted, whenever low peaks are seen at some $\phi_v$ as in Fig.~\ref{fig:disappear}, these peaks always seem to occur at $\phi_v \approx \pi/3$, independent of any cuts that are made that should change the velocity of the energetic particle on the away side and consequently should change the putative Mach angle.    The preliminary model calculations in Ref.~\cite{Alver:2010gr} suggest that the incident nuclei are sufficiently asymmetric that the ``triangular flow'', which is to say $v_3$, that is produced is sufficient to explain the features in Fig.~\ref{fig:disappear} that had previously been attributed to the excitation of a Mach cone.  As of this writing, it remains to be seen whether once the event-by-event $v_3$ is quantified and subtracted, there will be any remaining evidence for a Mach cone~\cite{Alver:2010gr}.

Notwithstanding these recent developments, it remains an interesting question of principle to understand in quantitative terms how strong a Mach cone is induced by the passage of an 
energetic point-like quark through strongly coupled plasma.  Remarkably, every one of the difficulties associated with answering this problem in QCD, sketched above, can be addressed for the case of an energetic heavy quark propagating through the strongly 
coupled plasma of ${\cal N}=4$ SYM theory.  In this section we shall review this calculation, which is done via the dual gravitational description of the phenomenon.  As in Sections~\ref{sec:HQDrag} and \ref{sec:HQBroadening}, we shall assume that the relevant physics is strongly coupled at all length scales, treating the problem entirely within strongly coupled ${\cal N}=4$ SYM theory.  
In this calculation, the AdS/CFT correspondence is used to determine the stress tensor of the medium, excited by the passing energetic quark, at all length scales.  This dynamical computation
will allow us to quantify to what extent hydrodynamics can be used to describe the response of the strongly coupled plasma of this theory to the disturbance produced by the energetic quark, as well as to study the relaxation of the initially far-from-equilibrium disturbance.
This calculation applies to quarks with mass $M$ whose velocity respects the 
bound (\ref{eq:glimit}).

\subsection{Hydrodynamic preliminaries}
\label{hdp}

 From the point of view of hydrodynamics, the disturbance of the medium induced by the passage of an energetic probe must be described by adding some source to the conservation equation:
 \begin{equation}
 \label{tmunuconsv}
 \del_\mu T^{\mu \nu}(x) = J^\nu (x) \ .
 \end{equation}
 As we have stressed above, we do not know the functional form of the source,
 since it not only involves the way in which energy is lost by the energetic particle
 but also how this energy is thermalized and how it is incorporated  into the medium. 
 The source will in general depend not only on the position of the quark but also on 
 its velocity.
In this subsection, we will use general considerations valid in any hydrodynamic medium to constrain the 
functional form of the source.  
 From \Eq{tmunuconsv} it is clear that the amount of energy-momentum
 deposited in the plasma is given by.\footnote{
 We note as an aside that if the source moves supersonically, one component of its energy loss is due to the emission of sound waves.  This is conventionally known as sonic drag, and is a part of the energy loss computed in Section~\ref{sec:HQDrag}.
 }
 \begin{equation}
 \frac{dP^\nu}{dt}= \int d^3 x J^\nu(x) \, .
 \label{EnergyDeposited}
 \end{equation}

 We now attempt to characterize the hydrodynamic modes that
 can be excited in the plasma due to the deposition of the energy (\ref{EnergyDeposited}). 
 We will assume, for simplicity, that the perturbation
 on the background plasma is small. We will also assume than the background plasma
 is static. The modification of the stress tensor
 \begin{equation}
 \delta T^{\mu \nu} \equiv T^{\mu\nu} - T_{\rm background}^{\mu\nu}
 \end{equation}
 satisfies a linear equation.  
 
 Since in the hydrodynamic limit the stress tensor is characterized by the local energy density, $\epsilon$, and the three components of the fluid spatial  velocity, $u^i$, there are only 4 independent fields, which can be chosen to be
 \begin{equation}
 \label{deflincomp}
 \mathcal{E}\equiv \delta T^{00} \qquad {\rm and} \qquad S^i\equiv\delta T^{0i}\ .
 \end{equation}
 Using the hydrodynamic form of the stress tensor, (\ref{eq:stideal} ), all other stress tensor components can 
 be expressed as a function of these variables. Since we have assumed that these perturbations are 
 small, all the stress-tensor components can be expanded to first order in the four independent fields (\ref{deflincomp}).
 
 In Fourier space, keeping the shear viscosity correction, 
 the linearized  form of the equations 
 (\ref{tmunuconsv})
 for the mode with a wave vector ${\bf q}$ that has the magnitude $q\equiv |{\bf q}|$ 
 take the form
 \begin{eqnarray}
 \del_t \mathcal{E} + i q S_L&=& J^0\,,
 \nonumber \\
 \del_t  S_L + i c^2_s q \mathcal{E} + \frac{4}{3}\frac{\eta}{\varepsilon_0+p_0} q^2 S_L&=& 
 J_L\,,
 \nonumber \\
 \del_t {\bf S}_T + \frac{\eta}{\varepsilon_0+ p_0} q^2 {\bf S}_T &=&{\bf J}_T\,,
 \label{LinearizedHydroEquations}
  \end{eqnarray}
 where ${\bf S}=S_L {\bf q}/q + {\bf S}_T$,
 ${\bf J}=J_L {\bf q}/q + {\bf J}_T$, $L$ and $T$ refer to longitudinal and transverse relative to the 
 hydrodynamical wave-vector {\bf q}
and $\varepsilon_0$, $p_0$, $c_s=\sqrt{d p/d\epsilon}$  and $\eta$ are the energy density, 
pressure, speed of sound and shear viscosity of the unperturbed background plasma.
 We observe that the longitudinal and transverse modes are independent.
 This
 decomposition is possible since the homogeneous equations have a $SO(2)$ symmetry 
 corresponding to rotation around
  the wave vector ${\bf q}$. The spin zero (longitudinal) and spin one (transverse) modes correspond
 to the sound and diffusion mode respectively. (The spin two mode is a subleading perturbation in the gradient expansion, since its leading contribution is proportional to velocity gradients.)
 After combining the first two equations of Eq.~(\ref{LinearizedHydroEquations}) and doing a Fourier transformation, we find
 \bea
 \left(\omega^2 - c_s^2 q^2 + i \,\frac{4}{3}\, \frac{\eta}{\epsilon_0 +p_0} q^2 \omega \right) S_L&=& i\,c^2_s  q J^0 + i\omega J_L\, ,\nonumber \\
 \left(i\omega  - \frac{\eta}{\epsilon_0+p_0} q^2\right){\bf S}_T&=&-{\bf J}_T \, .
 \eea 
 The sound mode ($S_L$) satisfies a wave equation and propagates 
 with the speed of sound while the diffusion mode $({\bf S}_T)$, which does not propagate,
describes the diffusion of transverse momentum as opposed to wave propagation. 
 We also note that only the sound mode results in fluctuations of the energy density, 
 while the diffusion mode involves only momentum densities (the $S^i$ of Eq.~(\ref{deflincomp})).
 In the linear approximation that we are using, the excitation of
 the diffusion mode produces fluid motion but does not affect the energy density.
 This result can be further illustrated by expressing the energy fluctuations in terms 
 of the velocity fields
 \begin{equation}
 \label{eq:kineticcontribution}
 \delta T^{00}=\delta \varepsilon + \frac{1}{2}(\varepsilon+P) \left(\delta v\right)^2 + \, . . .
 \end{equation}
The second term in this expression corresponds to the kinetic energy contribution of the fluid motion
which takes a non-relativistic form due to the small perturbation approximation. This expression is
quadratic in the velocity fluctuation, and thus is not described 
in the linearized approximation.    The sound mode corresponds to both compression/rarefaction of the fluid and motion of the fluid; sound waves result in fluctuations of the energy density as a consequence of  the associated compression and rarefaction.  But, the diffusion mode corresponds to fluid motion only and, to this order, does not affect the energy density.

Solving the linearized hydrodynamic equations (\ref{LinearizedHydroEquations}) yields
  hydrodynamic fields 
 given by
 \begin{eqnarray}
 \label{eq:ep}
 \mathcal{E}(t,{\bf x})&=& 
 \int \frac{d\omega}{2\pi} \frac{d^3 q}{(2\pi)^3} \frac
 									{i q J_L + i\omega J^0 - \Gamma_s q^2 J^0 }
 									{\omega^2 -c^2_s q^2+i\Gamma_s q^2 \omega}
									e^{-i \omega t + i {\bf q}\cdot {\bf x}}
\\
 \label{eq:gl}
{\bf S}_L(t,{\bf x})&=&
 \int \frac{d\omega}{2\pi} \frac{d^3 q}{(2\pi)^3} \frac{{\bf q}}{q}
 									\frac
 									{c_s^2 i q J^0 + i\omega J_L }
 									{\omega^2 -c^2_s q^2+i\Gamma_s q^2 \omega}
									e^{-i \omega t + i {\bf q}\cdot {\bf x}}
\\
\label{eq:gt}
{\bf S}_T(t,{\bf x})&=&
 \int \frac{d\omega}{2\pi} \frac{d^3 q}{(2\pi)^3}
 									\frac
 									{-{\bf J}_T }
 									{i\omega-D q^2}
									e^{-i \omega t + i {\bf q}\cdot {\bf x}}
 \end{eqnarray}
  where the sound attenuation length and the diffusion constant are
  \begin{eqnarray}
  \Gamma_s&\equiv&\frac{4}{3} \frac{\eta}{\varepsilon_0+p_0}
  \\
  D&\equiv& \frac{\eta}{\varepsilon_0+p_0}\ .
  \end{eqnarray}
  We note in passing that the integral of the longitudinal momentum density over all space vanishes. 
  
  
  The hydrodynamic solutions (\ref{eq:ep}), (\ref{eq:gl}) and (\ref{eq:gt}) are only of formal value without any information about the source.  And, as we have stressed above, a lot of nonlinear, non-equilibrium physics goes into determining the source as a function of the coordinates.  Still, we can make some further progress.  If we assume that the energetic quark moves at a constant velocity $v$ for a long time (as would be the case if the quark is either ultrarelativistic or very heavy) then we expect
   \begin{equation}
  J^{\mu}(\omega,k)=2 \pi \delta(\omega-{\bf v} \cdot {\bf q}) J_v^\mu ({\bf q})\ ,
  \end{equation}
 where the factor $\delta(\omega-{\bf v} \cdot {\bf q})$ comes from  Fourier transforming 
  $\delta({\bf x} -{\bf v} t)$. 
We also note that  far away from the source, and at sufficiently small $q$  that we can neglect any energy scales characteristic of the medium and any internal structure of the particle moving through the medium, 
  the only possible vectors from which to construct the source are ${\bf v}$ 
  and ${\bf q}$. 
In this regime, we may decompose the source as
  \begin{eqnarray}
  J_v^0({\bf q})&=&e_0({\bf q})\nonumber\\
  {\bf J}_v({\bf q})&=& {\bf v}\, g_0 ({\bf q}) + {\bf q} \,  g_1 ({\bf q})\ .
  \label{genHJ}
  \end{eqnarray}
  Then, inspection of the solutions
  (\ref{eq:ep}), (\ref{eq:gl}) and (\ref{eq:gt}) together with the observation that 
  a particle moving with a velocity close to the speed of light loses similar amounts of energy and momentum,
  shows that, at least for an ultrarelativistic probe, nonvanishing values of $e_0({\bf q})$ must be linked to nonvanishing values of $g_0({\bf q}$).
  We call this case scenario 1. 
 However, if the interaction of the probe with the plasma is such  that both $g_0$ and $e_0$ 
 are zero 
 (or parametrically small compared to $g_1$), from Eqs.~(\ref{EnergyDeposited}) and 
 (\ref{genHJ}) and 
 since ${\bf q}\,  g_1({\bf q})$ is a total derivative,
 one may mistakenly conclude that the energetic probe has created a disturbance carrying zero energy and momentum.  In this scenario, which we shall call scenario 2, the energy and momentum loss are actually quadratic in the fluctuations.
 %
%
 %
 These two scenarios  lead to disturbances with different characteristics.
 In scenario
 2, only the sound mode is excited while in scenario 1, both the sound and diffusion mode are 
 excited.   The correct answer for a given energetic probe may lie in between these two extreme cases.
 
 The phenomenological implications of this analysis depend critically on the degree to which the diffusion mode is excited. 
 This mode leads to an excess of momentum density along the direction of the source which does not propagate out of the region of deposition, but only diffuses away. Therefore, 
  the diffusion mode excited by an energetic quark moving through the plasma corresponds to a wake of moving fluid, trailing behind the quark and moving in the same direction as the quark.  In a heavy ion collision, therefore, the diffusion wake excited by the away-side energetic quark will become hadrons at $\Delta\phi \sim \pi$, whereas the Mach cone will become a cone of hadrons with moment at some angle away from $\Delta\phi = \pi$.   If most of the energy dumped into the medium goes into the diffusion wake, even if a Mach cone were produced it would be overwhelmed in the final state, and invisible in the data. 
 Only in the case in which the diffusion 
 mode is absent (or sufficiently small) is the formation of a Mach cone reflected in a visible nontrivial correlation of the type seen in Fig.~\ref{fig:disappear}.

 
 \subsection{AdS computation}
 
In Section~\ref{sec:HQDrag} we have computed the amount of energy lost by a heavy quark as it plows through the strongly coupled ${\cal N}=4$ SYM theory plasma. In order to address the fate of this energy, we must determine the stress tensor of the gauge theory fluid at the boundary that corresponds to the string (\ref{eq:solxi})  trailing behind the quark in the bulk.
In the dual gravitational theory, this string modifies the metric of the $(4+1)$-dimensional geometry.  That is, it produces gravitational waves.  The stress energy tensor of the gauge theory plasma at the boundary is determined by the asymptotic behavior of the bulk metric perturbations as they approach the boundary \cite{Friess:2006fk,Chesler:2007an,Chesler:2007sv,Gubser:2007xz}.   

The modifications of the $4+1$-dimensional metric due to the presence of the trailing string are obtained by solving
the Einstein equations
\begin{equation}
\mathcal{R}_{\mu\nu}-\frac{1}{2} G_{\mu\nu} (\mathcal{R} + 2 \Lambda) = \kappa^2_5 t_{\mu\nu}\ ,
\label{EinsteinEquations}
\end{equation}
where $\kappa^2_5=4\pi^2 R^3/N^2_c$ and $\Lambda=6/R^2$ with $R$ the AdS radius 
and where $t_{\mu\nu}$ is the five-dimensional string stress tensor, which
can be computed from the Nambu-Goto action:
\be
t^{\mu\nu}=-\frac{1}{2\pi\alpha'} \int d\tau d\sigma \frac{\sqrt{-h}}{\sqrt{-G}} h^{ab} \del_a X^\mu (\tau,\sigma)\del_b X^\nu(\tau,\sigma) \delta^{(5)}\left(x-X(\tau,\sigma)\right) \, ,
\ee 
where $h^{ab}$ is the induced metric on the string and $X(\tau,\sigma)$ is the string profile.
For the case of a trailing string (\ref{eq:solxi}), the stress tensor
is given by
\begin{eqnarray}
t_{00}&=&s(f+v^2 z^4/z^4_0) \, , \nonumber\\
t_{0i}&=&-s v_i \, , \nonumber\\
t_{0z}&=&-s v^2 z^2/z^2_0 f \, , \nonumber\\
t_{zz}&=&s(f-v^2)/f^2 \, , \nonumber \\
 t_{ij}&=&s v_i v_j \, , \nonumber \\
 t_{iz}&=&s v_i z^2/z^2_0 f \, ,
 \label{StringStressTensor}
\end{eqnarray}
where
\begin{equation}
s=\frac{z \gamma \sqrt{\lambda}}{2\pi R^3} 
\delta^3\left({\bf x} -{v}t - {\bf \zeta}(z) \right)
\, ,
\end{equation}
with $\zeta(z)$ the string profile  (\ref{eq:solxi}).
After solving the Einstein equations (\ref{EinsteinEquations}) with the string stress tensor (\ref{StringStressTensor}),
the expectation value of the boundary stress tensor is then obtained by following the 
prescription (\ref{stgenBISBIS}),
namely by
 performing functional derivatives of the Hilbert action evaluated on the classical solution 
with respect to the boundary metric.

We need to analyze the small fluctuations on top of the background AdS black hole metric.  Denoting these fluctuations by $h_{\mu\nu}$ and the background metric by $g_{\mu\nu}$, the left-hand side of the Einstein equations (\ref{EinsteinEquations}) 
are given to leading order in $h_{\mu\nu}$ by
\begin{eqnarray}
\label{eqhgen}
-D^2 h_{\mu\nu} + 2 D^\sigma D_{(\mu}  h_{\nu)\sigma}
-D_\mu D_\nu h + \frac{8}{R^2} h_{\mu \nu} 
&& \nonumber
\\
+\left(
D^2 h - D^\sigma D^{\delta} h_{\sigma \delta} -\frac{4}{R^2} h 
\right) g_{\mu\nu} =0 \, ,
\end{eqnarray}
with $D_\mu$ the covariant derivative with respect to the full metric, 
namely $g_{\mu\nu}+h_{\mu\nu}$. 
This equation has a gauge symmetry 
\begin{equation}
h_{\mu \nu}\rightarrow h_{\mu\nu} + D_\mu \xi_\nu + D_\nu \xi_\mu\ ,
\end{equation}
inherited from reparameterization  invariance,
that, together with 5 constraints from the linearized Einstein equations (\ref{eqhgen}), reduces the number of degrees of freedom from 15 to 5. 
It is therefore convenient 
to introduce gauge invariant combinations which describe the independent degrees of freedom. 
These can be found after Fourier transforming the $(3+1)$-dimensional coordinates.  
The gauge invariants can be classified by how they transform under SO(2) rotations around the wave vector ${\bf q}$.
Upon introducing $H_{\mu\nu}=z^2 h_{\mu\nu} /R^2$, one possible choice of gauge invariants is given by \cite{Chesler:2007an,Chesler:2007sv}
\begin{eqnarray}
Z_{(0)}&=&q^2 H_{00} + 2 \omega q H_{0q} + \omega^2 H_{qq} + \frac{1}{2}\left[(2-f)q^2 -\omega^2\right]H_{}\nonumber
\\
Z_{(1)\alpha}&=&\left(H'_{0\alpha} -i\omega H_{\alpha 5}\right)\nonumber 
\\
Z_{(2)\alpha\beta}&=&\left(H_{\alpha\beta} - \frac{1}{2} H  \delta_{\alpha\beta} \right)
\end{eqnarray}
where $q\equiv |{\bf q}|$,  $\hat{q}\equiv {\bf q}/q$, $H_{0 q}\equiv H_{0i} \hat{q}^i$, $H_{q q}\equiv  H_{ij} \hat{q}^i  \hat{q}^j$, $\alpha$ and $\beta$ (which are each either 1 or 2)
are space coordinates transverse to $\hat q$, $'$ means $\partial_z$, and $H\equiv H_{\alpha\alpha}$.
When written in terms of these gauge invariants, the Einstein equations (\ref{eqhgen}) become three independent equations for 
%
$Z_{(0)}$, $Z_{(1)\alpha}$ and $Z_{(2)\alpha \beta}$,
  which correspond to the spin zero, one and two 
fluctuations of the stress tensor.
We focus on the spin zero and spin one fluctuations, since these are the relevant modes in the hydrodynamic
limit. 
Their equations of motion are given by 
\begin{eqnarray}
\label{gieq}
Z_{(1)\alpha}''  +
 \frac{z f^{'} -3 f}{z f} Z_{(1)\alpha}'
+ \frac{3 f^2 - z (z q^2 +3 f^{'})f + z^2 \omega^2}{z^2 f^2} Z_{(1)\alpha}
&=& S_{(1)\alpha}
\\
Z_{(0)}''+\frac{1}{u} \left[
                                  1+ \frac{u f^{'}}{f} + \frac{24(q^2 f- \omega^2)}{q^2(u f^{'}-6 f) + 6 \omega^2}
                                  \right] Z_{(0)}' \quad 
                                  && \nonumber \\
+\frac{1}{f}
\left[
-q^2 + \frac{\omega^2}{f} - \frac{32 q^2 z^6/ z^8_0}{q^2(u f^{'} -6 f) + 6\omega^2}
\right]
Z_{(0)}                                  
&=&  S_{(0)}\ ,
\label{gieq2}
\end{eqnarray}
where the sources are combinations of the string stress tensor and its derivatives. Choosing
one of the transverse directions (which we shall denote by $\alpha=1$)
 to lie in the $({\bf v}, {\bf q})$ plane, the source for the 
trailing string is given explicitly by
\bea
S_{(1)1}&=&\frac{2 \kappa^2_5\gamma \sqrt{\lambda}}{ R^3}\, \frac{v q_\perp } {q f}
\, \delta (\omega -{\bf v} \cdot {\bf q}) e^{-i {\bf q} \cdot {\bf \zeta}}  \, ,
\nonumber\\
S_{(1)2}&=&0 \,,\\
S_{(0)} &=& 
\frac{\kappa^2_5 \gamma \sqrt{\lambda}}{3 R^3} 
\,\frac{q^2(v^2+2)-3 \omega^2}{q^2} 
\,\frac{z\left[q^4 z^8+ 48 i q^2 z^2_0 z^5 - 9(q^2-\omega^2)^2 z^8_0\right]}{f (f q^2 +2 q^2  -3 \omega^2) z^8_0}
\nonumber\\
&& 
\quad \times \delta (\omega - {\bf v} {\bf q})  e^{-i {\bf q}\cdot \zeta}\, ,
\eea
where $q_\perp$ is the magnitude of the component of ${\bf q}$ perpendicular to ${\bf v}$.
The boundary action can be expressed in terms of the gauge invariants 
$Z_{(1)\alpha}$ and $Z_{(0)}$
plus certain counterterms
(terms evaluated at the boundary). This procedure, which can be found in 
Ref.~\cite{Chesler:2007sv},
is  somewhat cumbersome but straightforward, and we shall not repeat it here.  Once this is achieved, the
 stress tensor components can be obtained from the classical solution to (\ref{gieq})-(\ref{gieq2}), following  
the prescription (\ref{stgenBISBIS}).

To find the classical solution to (\ref{gieq})-(\ref{gieq2})  we must specify boundary conditions.  
Since the quark propagates in flat space, the metric fluctuations must vanish at the boundary.
Also, 
since we are interested in the response of the 
medium, 
the solution must satisfy retarded boundary condition,  meaning that at the horizon
 it must  be composed only of infalling modes. Thus, we may construct the Green's function
\begin{equation}
G_s(z,z')=\frac{1}{W_s (z')}
\left(
\theta(z'-z) g^n_s(z) g^i_s(z') + \theta(z-z') g^i_s(z) g^n_s(z')
\right)
\end{equation}
where the subscript $s$ which can be $0$ or $1$ denotes the spin component 
and $g^n_s$ and  $g^i_s$ denote 
the normalizable and infalling solutions to the homogeneous equations obtained by setting the left-hand side of (\ref{gieq}) equal to zero.  $W_s$ is 
the Wronskian of the two homogeneous solutions. The full solution to (\ref{gieq}) may be then 
written as
\begin{equation}
Z_s(z)=\int^{z_h}_0 dz' G_s (z, z') S_s(z')
\end{equation}
Close to the boundary, these solutions behave as
\begin{eqnarray}
Z_{(0)}&=&z^3 Z^{[3]}_{(0)} + z^4 Z^{[4]}_{(0)} + \, . \, . \, .
\nonumber\\
\vec{Z}_{(1)}&=&z^2 \vec{Z}^{[2]}_{(1)}+ z^3 \vec{Z}^{[3]}_{(1)} + \, . \, . \, .
\end{eqnarray}
The components $Z^{[3]}_{(0)}$ and   $Z^{[2]}_{(1)}$ can be computed analytically and are
temperature independent. They yield a divergent  contribution to the 
boundary stress tensor. However, this contribution is analytic in $q$ and, thus, has 
$\delta$-function support at the position of the heavy quark. This divergent 
contribution is the contribution of the heavy quark mass to the boundary theory stress tensor.
The response of the boundary theory gauge fields to the disturbance induced by the passing energetic quark 
 is encoded in the components
$Z^{[4]}_{(0)}$ and $Z^{[3]}_{(1)}$, which must be computed numerically. 
After expressing the boundary actions in terms of gauge invariants, the nondivergent spin zero and one components of the boundary stress tensor are given by 
\begin{eqnarray}
\mathcal{T}_0&=& \frac{4 q^2}{3 \kappa^2_5 (q^2-\omega^2)^2} Z^{[4]}_{(0)}
 + D + \varepsilon_0
 \,,
\\
\mathcal{\vec{T}}_1&=&-\frac{L^3}{2 \kappa_5} \vec{Z}^{[3]}_{(1)}\ .
\end{eqnarray}
where $ \mathcal{T}_0=T^{00}$ and $\mathcal{\vec{T}}_1=T^{0a} \hat \epsilon_a$, with 
$\hat \epsilon_a$ the spatial unit vectors orthogonal to the spatial momentum ${\bf q}$, and where the counterterm $D$ is a complicated function of $\omega$ and $q$ that depends on the quark velocity and the plasma temperature and that is given in
Ref.~\cite{Chesler:2007sv}. 

\begin{figure} 
\includegraphics[width=\textwidth]{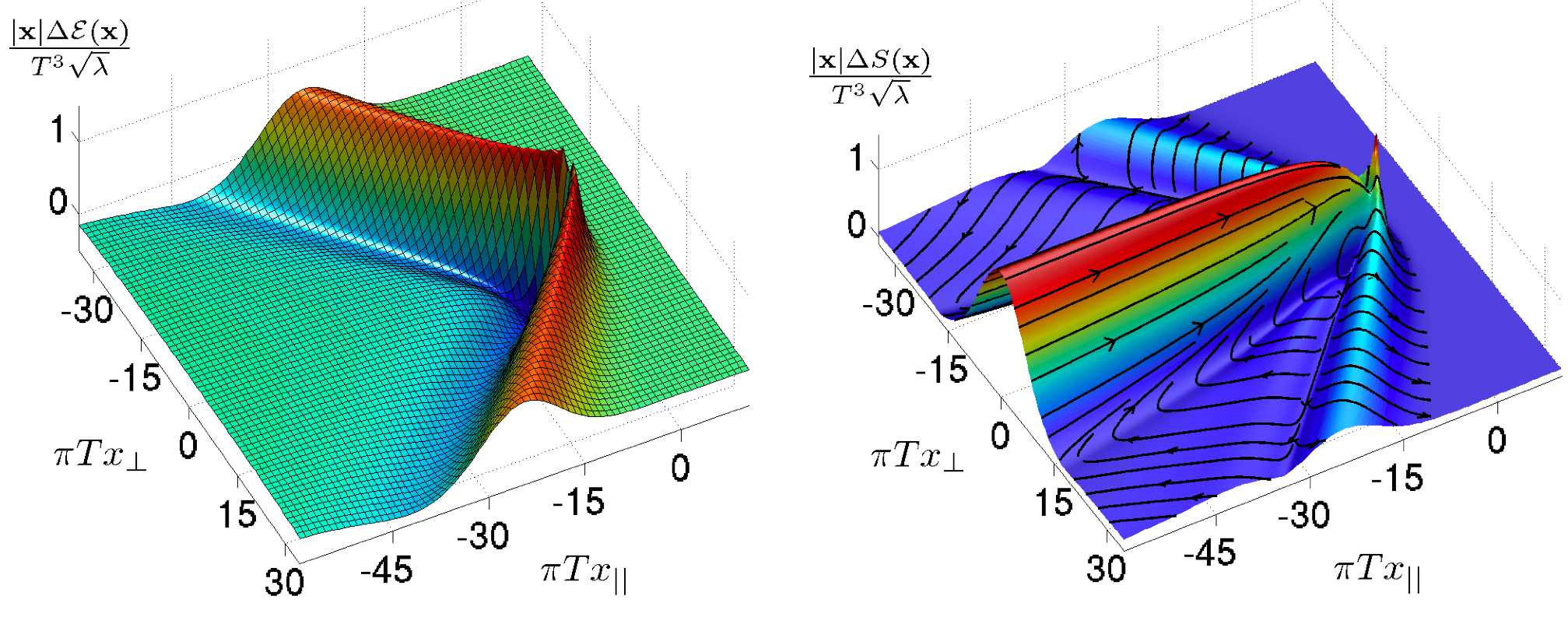}
\caption{\small 
\label{fig:sttsor}
Energy density (left) and momentum flux (right) induced by the passage of a
supersonic heavy quark moving through the strongly coupled ${\cal N}=4$ SYM theory plasma 
in the $x_\parallel$ direction with speed $v=0.75$.  ($\Delta\varepsilon({\bf x})$ is the difference between $\varepsilon({\bf x})$ and the equilibrium energy density; since ${\bf S}=0$ in equilibrium, $\Delta {\bf S}({\bf x})$ is simply ${\bf S}({\bf x})$.)
The flow lines on the surface are flow lines of $\Delta {\bf S}({\bf x})$.
These disturbances are small compared to the background energy density 
and pressure of the plasma (both of which are $\propto N_c^2$). 
The perturbation is small and it is well
described by linearized hydrodynamics everywhere except within a distance 
$R\approx 1.6/T$ from the quark.
Since the perturbation is small, the kinetic energy contribution of the diffusion mode to the
energy density is suppressed by $N^2_c$ and, thus, it does not
contribute in the left panel. 
}
\end{figure}

Results from Ref.~\cite{Chesler:2007sv} on the numerical computation of the disturbance in the gauge theory plasma created by a supersonic quark moving with speed $v=0.75$ are shown in \Fig{fig:sttsor}. 
The left panel shows the energy density of the disturbance and clearly demonstrates that a Mach cone has been excited by the supersonic quark.  The front is moving outwards at the Mach angle
$\Theta_M$, where $\cos \Theta_M=c_s/v = 4/ (3 \sqrt{3})$. Recall from our general discussion above that fluid motion is invisible in the energy density, to the linear order at which we are working; the energy density is nonzero wherever the fluid is compressed.  Thus, the Mach cone is made up of sound modes, as expected.  In the right panel of Fig.~\ref{fig:sttsor}, we see the density of fluid momentum 
 induced by the supersonic quark.
 This figure reveals the presence of a sizable wake of moving fluid behind the quark, a wake that is invisible in the energy density and is therefore made up of moving fluid without any associated compression, meaning that it is made up of diffusion modes.  We conclude that the supersonic quark passing through the strongly coupled plasma excites both the sound mode and the diffusion mode, meaning that the interaction of the quark with the plasma is as in what we called scenario 1 above.  Quantitatively, it turns out that the momentum carried by the sound waves is greater than that carried by the diffusion wake, but only by a factor of $1+v^2$~\cite{Gubser:2007ga}.

Since hydrodynamics describes the long wavelength  limit of
the stress tensor excitation, it is reasonable to find a Mach cone at long distances.
 And, since the gravitational equations whose solution we have reviewed are linear, the long distance behavior of the gauge theory fluid must be described by linearized hydrodynamics.
It is easy to justify the linearization from the point of view of the field theory: the background plasma has an energy density that is proportional to $N_c^2$ while that of the perturbation
is proportional to the number of flavors, which is just $N_f=1$ in the present case since we are considering only one quark. 
The strong coupling computation leads to 
a perturbation of magnitude $N_f \sqrt{\lambda}$. Thus, the energy density
of the fluctuations are suppressed by $\sqrt{\lambda}/N^2_c$ with respect to that of the background plasma, justifying the linearized 
treatment. 
%
%
Remarkably,  it turns out that disturbances 
like those in Fig.~\ref{fig:sttsor} are well described by hydrodynamics everywhere 
except within $\approx 1.6/T$ of the position of the quark~\cite{Chesler:2007sv}
%
%
So, the calculation that we have reviewed in this section is important for two reasons.
First, it demonstrates that a {\it point-like} probe passing through the strongly coupled plasma does indeed excite hydrodynamic modes.  And, second, it demonstrates that in the strongly coupled plasma, the resulting disturbance relaxes to a hydrodynamic excitation in local thermal 
equilibrium surprisingly close to the probe.


The observation that point particles moving through the strongly coupled fluid excite  
sound waves, which are collective excitations, is at odds with intuition based upon the interaction of, say, electrons with water. 
In this example, most of the energy lost by 
the electron is transferred to photons and not to the medium. These photons, in turn, have long 
mean free paths and  dissipate their energy
 far away from the electron (or escape the medium entirely). 
 Thus, the effective size of the region 
 where energy is dissipated is very  large, given by the photon mean free path. 
 Hydrodynamics will only describe the physics on longer length scales than this.
The reason that no Mach cone is formed is that the length scale over which the energy is deposited is long compared to the length scale over which the electron slows and stops. 
 The situation is similar in weakly coupled gauge theory plasmas; even though the 
 gauge modes in these theories do interact, they still have 
 long mean free paths proportional to $1/g^4$.
  In sharp contrast, in the strongly coupled plasma of
    $\N=4$ SYM theory there are no long-lived quasiparticle excitations (leave apart photons) that could transport the energy deposited by the point-like particle over long distances.  Instead, all the energy lost by the point-like probe is dumped into collective hydrodynamic modes over a characteristic length scale $\sim 1/T$, which is the only length scale in this conformal plasma.
    

\subsection{Implications for heavy ion collisions}

The calculation that we have reviewed in this section suggests that a high energy quark plowing through the strongly coupled plasma produced in heavy ion collisions at RHIC should excite a Mach cone.
As we argued just above, this phenomenon is not expected in a weakly coupled plasma.
%
The Mach cone should have consequences that are observable in the soft particles on the away-side of a high energy trigger hadron.
However, for a hydrodynamic solution like that in Fig.~\ref{fig:sttsor}, it turns out that the diffusion wake contains enough momentum flux along the direction of the energetic particle to ``fill in'' the center of the Mach cone, meaning that the Mach cone is not sufficiently prominent as to result in peaks in the particle distribution at $\Delta\phi = \pi \pm \Theta_M$~\cite{CasalderreySolana:2004qm,Noronha:2008un}.
As we discussed above, the observed peaks at $\Delta\phi=\pi \pm \phi_v$ receive a significant contribution from the event-by-event $v_3$ due to event-by-event fluctuations that introduce ``triangularity''.    Detecting evidence for  Mach cones in heavy ion collisions will require careful subtraction of these effects from the data on the one hand, and careful theoretical analysis of the effects of the rapid expansion of the fluid produced in heavy ion collisions on the putative Mach cones on the other.

\subsection{Disturbance excited by a moving quarkonium meson}

Strong coupling calculations like that of the disturbance excited by an energetic quark moving through the plasma of  ${\cal N}=4$ SYM theory can help guide the construction of more phenomenological models of the coupling of energetic particles to hydrodynamic modes.  To further that end, we close with an example which shows that not all probes behave in the same way.

As we shall describe in Section~\ref{sec:HotWind}, a simple way of modelling a ``quarkonium'' meson made from a heavy quark and antiquark embedded in the strongly coupled plasma of ${\cal N}=4$ SYM theory is to consider a string with both ends at the boundary --- the ends representing the quark and antiquark.  We shall see in Section~\ref{sec:HotWind} that even when this string is moving through the plasma, it hangs straight downward  into the AdS black hole metric, rather than trailing behind as happens for the string hanging downward from a single moving quark.  The fact that the ``U'' of string hangs straight down and does not trail behind the moving quark and antiquark implies that the heavy quarkonium meson moving through the strongly coupled plasma does not lose any energy, at least at leading order.  The energy loss of such a meson has been computed and is in fact nonzero but is suppressed by $1/N_c^2$~\cite{Dusling:2008tg}.


Despite the fact that the leading order quarkonium 
energy loss vanishes, the leading order disturbance of 
the fluid through which the meson is moving does not vanish~\cite{Gubser:2007zr}. 
Instead, the meson excites a Mach cone with no diffusion wake, providing an example
of what we called scenario 2 at the end of Section~\ref{hdp}.  It is as though the moving meson ``dresses itself'' with a Mach cone, and then the meson and its Mach cone propagate through the fluid without dissipation, to leading order.  
To illustrate this point, the metric fluctuation and consequent boundary stress tensor induced
by a semiclassical string with both ends on the boundary 
moving with a velocity ${\bf v}$ has been calculated~\cite{Gubser:2007zr}.
For a string with the
two endpoints aligned along the direction of motion and separated by a distance $l$ the long distance part (low momentum) part
of the associated stress tensor is given by
\bea
\label{mstts}
\delta T^{00}&=& \frac{\Pi }{q^2-3 \left({\bf} q \cdot {\bf v} \right)^2} 
                              \left( 
                               -q^2 \left(1+2 v^2 \sigma \right)
                               -
                               3 v^2 \left({\bf} q \cdot {\bf v}\right)^2 \left(1- 2 \sigma  \right)
                               \right)
                               \\
\delta T^{0i}&=&   
\frac{\Pi }{k^2-3 \left({\bf} q \cdot {\bf v} \right)^2} 
2 {\bf} q \cdot {\bf v} \left(1 -  (1-v^2) \sigma \right) q^i + 2 \, \Pi \, \sigma \,v^i                  
\\
\delta T^{i j}&=&  \frac{\Pi }{k^2-3 \left({\bf} q \cdot {\bf v} \right)^2}  2\left({\bf} q \cdot {\bf v} \right)^2 
\left( 
-1 + (1-v^2)\sigma
\right) \delta^{i j} -
\\ \nonumber 
&& 
\quad  -\frac{\Pi  }{v^2} \left(1 + 2 v^2 \sigma \right) \frac{v^i v^j}{v^2}
 \eea
where $\sigma=\sigma(l,T)$ is a dimensionless function of the length of the meson and the temperature and the prefactor takes the form
\begin{equation}
\Pi= \sqrt{\lambda} \frac{F(lT)}{l} \, ,
\end{equation}
with $F$ a dimensional function.

The expression (\ref{mstts}) clearly shows that all the spin 0,1 and 2 components of the stress tensor
are excited. The spin 0 components are multiplied by the sound propagator, signaling the 
emission of sound waves. (Note that in the low-$q$ limit the width of the sound pole vanishes.) 
The spin 1
component corresponds to the terms proportional to the velocity of the particle $v^i$. These terms are 
analytic in $q$; in particular it seems that there is no pole contribution from the diffusive mode. 
More careful analysis shows that the diffusive mode decays faster than that excited by a quark probe.
%
%

The magnitude of the disturbance in the strongly coupled plasma that is excited by a passing quarkonium meson is no smaller than that excited by a passing quark.
However, the total integral 
of the energy and momentum deposited is zero, as can be  seen by multiplying the momentum
densities by $\omega={\bf v \cdot \bf q}$ and taking the limit $q\rightarrow 0$. This is consistent with the
fact that, to the order at which this calculation has been done, the meson does not lose any energy.
This is an interesting example  since it 
indicates that the loss of energy and the excitation of hydrodynamic modes are distinct phenomena, controlled by different physics.  This example also illustrates the value of computations done at strong coupling in opening ones eyes to new possibilities: without these calculations it would have been very hard to guess or justify that such a separation in magnitude between the strength of the hydrodynamic fields excited by a probe and the energy lost by that probe could be possible.  It would be interesting to analyze the soft particles in heavy ion collisions in which a high transverse momentum quarkonium meson is detected, to see whether there is any hint of a Mach cone around the meson --- in this case without the complication of soft particles from a diffusion wake filling in the cone.



\section{Stopping light quarks}
\label{sec:Stopping}

As we have discussed extensively in Section~\ref{sec:JetQuenching}, the dominant energy loss process for a parton moving through the QCD plasma with energy $E$  in 
the limit in which $E\rightarrow\infty$ is gluon radiation, and in this limit much (but not all;
see Section~\ref{sec:AdSCFTJetQuenching}) of the calculation can be done at weak coupling.
However, since it is not clear how quantitatively reliable the $E\rightarrow \infty$ approximation is for the jets produced in RHIC collisions, it is also worth analyzing the entire problem of parton energy loss and jet quenching at strong coupling to the degree that is possible.  For the case of a heavy quark propagating through the strongly coupled plasma of ${\cal N}=4$ SYM theory, this approach has been pursued extensively, yielding the many results that we have reviewed in the previous three subsections.   Less work has been done on 
the energy loss of an energetic light quark or gluon in the ${\cal N}=4$ SYM plasma, in particular since they do not fragment into anything like a QCD jet.  (As we shall review in 
Section~\ref{sec:Synchrotron}, there are no true jets in ${\cal N}=4$ SYM theory, although it is possible to cook up a collimated beam of radiation.)  It is nevertheless worth asking how a light quark or gluon loses energy in the ${\cal N}=4$ SYM plasma in the hope that even if 
this is not a good model for jets and their quenching in QCD
some qualitative strong-coupling benchmarks against which to compare experimental results may be obtained. This program has been pursued in 
Refs.~\cite{Gubser:2008as,Chesler:2008wd,Chesler:2008uy}.

As introduced in Section~\ref{fundamental} and described extensively in Section~\ref{mesons}, dynamical quarks can be introduced into ${\cal N}=4$ SYM theory by introducing a D7-brane that fills the $3+1$ Minkowski dimensions and fills the fifth dimension from the boundary at $z=0$ down to $z=z_q$.  The mass of the (heavy) quarks that this procedure introduces in the gauge theory is $\sqrt{\lambda}/(2\pi z_q)$.  Light quarks are obtained by taking $z_q \rightarrow \infty$, meaning that the D7-brane fills all of the $z$-dimension.  At $T\neq 0$, what matters is that the D7-brane fills the $z$-dimension all the way down to, and below, the horizon.  In this set up, a light quark-antiquark pair moving back-to-back each with some initial high energy can be modelled as a quark and anti-quark located at some depth $z$ that are moving apart from each other in, say, the $x$ direction and that are connected by a string~\cite{Chesler:2008wd,Chesler:2008uy}.  The quark and antiquark must be within the D7 brane, but since this D7 brane fills all of $z$ there is nothing stopping them from falling to larger $z$ as they fly apart from each other, and ultimately there is nothing stopping them from falling into the horizon.   It should be evident from this description that there is an arbitrariness to the initial condition:  At what $z$ should the quark and antiquark be located initially?  What should the string profile be initially?  What should the initial profile of the velocity of the string be?  These choices correspond in the gauge theory to choices about the initial quantum state of the quark-antiquark pair and the gauge fields surrounding them. And, there is no known way to choose these initial conditions so as to obtain a QCD-like jet so the choices made end up being arbitrary.  (The analogous set up for a back-to-back pair of high energy gluons~\cite{Gubser:2008as} involves a doubled loop of string, rather than an open string with a quark and antiquark at its ends.)

Ambiguities about the initial conditions notwithstanding, several robust qualitative insights have been obtained from these calculations.  First, the quark and the antiquark always fall into the horizon after travelling some finite distance  $x_{\rm stopping}$.  (The string between them falls into the horizon also.) $x_{\rm stopping}$ corresponds in the gauge theory to the stopping distance for the initially energetic quark, namely the distance that it takes this quark to slow down, thermalize, and equilibrate with the bulk plasma --- the gauge theory analogue of falling into the horizon.  This is qualitatively reminiscent of the discovery at RHIC of events with a single jet (manifest as a high $p_T$ hadron that is triggered on) but no jet (no high $p_T$ hadrons) back-to-back with it.

Second, although $x_{\rm stopping}$ does depend on details of the initial conditions, the dominant dependence is that it scales like $E^{1/3}$, 
where $E$ is the initial energy of the quark~\cite{Gubser:2008as,Chesler:2008uy,Arnold:2010ir,Arnold:2011qi}.  
More precisely, upon analyzing varied initial conditions the maximum possible stopping distance is given by~\cite{Chesler:2008uy}
\begin{equation}
x_{\rm stopping} =  \frac{C}{T} \left( \frac{E}{T\sqrt{\lambda}} \right)^{1/3}\ ,
\label{StoppingScaling}
\end{equation}
with $C\approx 0.5$.   If there is a regime of $E$ and $T$  in which it is reasonable to treat the entire problem of jet quenching at strong coupling, and if in this regime the droplet of plasma produced in a heavy ion collision is large enough and lives long enough that it can stop and thermalize an initial parton with energy $E$ that would in vacuum have become a jet, then the scaling 
(\ref{StoppingScaling}) has interesting qualitative consequences. For example, if this scaling applies to collisions with two different collision energies $\sqrt{s_1}$ and $\sqrt{s_2}$, yielding plasmas that form at different temperatures $T_1$ and $T_2$, then jets in these two experiments whose energies 
satisfy $E_1/E_2 \sim (T_1/T_2)^4$ should have similar observed phenomenology.   Turning this speculation into semi-quantitative expectations for experimental observables requires careful study of jet stopping in a realistic model of the dynamics in space and time of the expanding droplet of plasma produced in a heavy ion collision.

Third, a light quark with initial energy $E$ that loses this energy over a distance $x_{\rm stopping}$ loses most of its energy near the end of its trajectory, where it thermalizes (falls into the horizon)~\cite{Chesler:2008uy}.  This pattern of energy loss is reminiscent of the `Bragg peak' that characterizes the energy loss of a fast charged particle in ordinary matter, where the energy loss has a pronounced peak near the stopping point.  It is quite different from the behavior of a heavy quark in strongly coupled plasma which, as we saw in Section~\ref{sec:HQDrag}, loses energy at a rate 
proportional to its momentum, making it reasonable to expect that a heavy quark that slows from a high velocity to a stop loses more energy earlier in its trajectory than later. 

It will be very interesting to see how these insights fare when compared with results on jet quenching in heavy ion collisions at the LHC, and in particular to comparisons between such results and results at RHIC energies.

\section{Calculating the jet quenching parameter}
\label{sec:AdSCFTJetQuenching}

As we have described in  Section~\ref{sec:JetQuenching}, when a parton with large transverse momentum is produced in a hard scattering that occurs within a heavy ion collision, the presence of the medium in which the energetic parton finds itself has two significant effects: it causes the parton to lose energy and it changes the direction of the parton's momentum. The latter effect is referred to as ``transverse momentum broadening''.  
In the high parton energy limit, 
as established first in Refs.~\cite{Gyulassy:1993hr,Baier:1996sk,Zakharov:1997uu},
the parton loses energy dominantly by 
inelastic processes that are the QCD analogue of bremsstrahlung: the parton radiates gluons as it interacts with the medium.  
It is crucial to the calculation of this radiative energy loss process that the incident hard parton, the outgoing parton, and the radiated gluons are all continually being jostled by the medium in which they find themselves:  they are {\it all} subject to transverse momentum broadening.    
The transverse momentum broadening of a hard parton is described by $P(k_\perp)$, defined as the probability that after propagating through the medium for a distance $L$ the hard parton has acquired 
transverse momentum $k_\perp$.
For later convenience, we shall choose to normalize $P(k_\perp)$ as follows:
\begin{equation}
 \int \frac{d^2 k_{\perp}}{(2\pi)^2} P (k_{\perp}) = 1\ .
\label{eq:Pnormalization}
\end{equation}
{}From the probability density $P(k_\perp)$, it is straightforward to obtain the mean transverse momentum picked up by the hard parton per unit distance travelled (or, equivalently in the high parton energy limit, per unit time):
\begin{equation}
\hat q \equiv \frac{\langle k_\perp^2 \rangle}{L} =
\frac{1}{L} \int \frac{d^2 k_{\perp}}{(2\pi)^2} k_\perp^2 P (k_{\perp})\ .
\label{qhatFirstTime}
\end{equation}
$P(k_\perp)$, and consequently $\hat q$, can be evaluated for a hard quark or a hard gluon.
In the calculation of radiative parton energy 
loss~\cite{Baier:1996sk,Zakharov:1997uu,Wiedemann:2000za,Gyulassy:2000er,Guo:2000nz,Wang:2001ifa,Arnold:2002ja} 
that we have reviewed in Section \ref{sec:JetQuenching} and that is
also reviewed in Refs.~\cite{Baier:2000mf,Kovner:2003zj,Gyulassy:2003mc,Jacobs:2004qv,CasalderreySolana:2007zz,Accardi:2009qv,Wiedemann:2009sh,Majumder:2010qh},
$\hat q$ for the radiated gluon plays a central role, and this quantity is referred 
to as the ``jet quenching parameter''.
Consequently, $\hat q$ should be  thought of as a (or even the) property of the strongly coupled medium that is ``measured'' (perhaps constrained is a better phrase) by radiative parton energy loss and hence jet quenching.   But, it is important to note that $\hat q$ is {\it defined} via transverse momentum broadening only.  Radiation and energy loss do not arise in its definition, although they are central to its importance.

The BDMPS calculation of parton energy loss in QCD
involves a number of scales which must be well-separated in order for this 
calculation to be relevant.  The radiated gluons have energy of 
order $\omega_c\sim \hat q L^2$ and transverse momenta of order $\sqrt{ \hat q L}$.  Both these scales must be much less than $E$ and much greater than $T$.  And, $\alpha_S$ evaluated at both these scales must be small enough that physics at these scales is weakly coupled, even if physics at scales of order $T$ is strongly coupled.  
In heavy ion collisions at RHIC, with the highest energy partons having $E$ only of order many tens of GeV, this separation of scales can be questioned.  It is expected that in heavy ion collisions at the LHC it will be possible to study the interaction of partons with energies of order a few hundred GeV, improving the reliability of the calculations 
reviewed in this subsection.   In the RHIC regime, and particularly for heavy quarks, the alternative approach of the previous subsection in which the entire problem is assumed to be strongly coupled and parton energy loss occurs via drag is just as plausibly relevant.  In this subsection, we shall review the calculation of $\hat q$ --- the property of the plasma that describes transverse momentum broadening directly and, in the high parton energy limit in which the relevant scales are well separated, controls the transverse momentum and the energy of the gluon radiation that dominates parton energy loss.

It has been shown via several different calculations done via conventional field 
theoretical 
methods~\cite{CasalderreySolana:2007zz,Liang:2008vz,D'Eramo:2010xk} 
that the transverse momentum broadening of a hard parton in the $\mathcal{R}$ representation of $SU(N)$ takes the form
\begin{equation} \label{provb}
P(k_\perp) = \int d^2 x_{\perp} \, e^{-i k_{\perp} \cdot x_{\perp}}\,
 {\mathcal W}_\sR (x_\perp)
\end{equation}
with
 \begin{equation} \label{QuenchWils}
   {\mathcal W}_{\sR} (x_\perp)  = \frac{1}{d\left(\mathcal{R}\right)} 
   \left\langle\Tr \le[ W_{\sR}^\da [0,x_\perp] \, W_{\sR} [0,0] \ri]\right\rangle
  \end{equation}
where
\begin{equation}
W_\mathcal{R}\left[x^+, x_{\perp}\right] \equiv P \left\{ \exp \left[i g \int_{0}^{L^-} d x^{-}\, A^+_\mathcal{R} (x^+, x^-,x_{\perp})\right] \right\}
\label{SingleWils}
\end{equation}
is the representation-$\mathcal{R}$ Wilson line along the lightcone, $L^-=\sqrt{2}L$ is the distance along the lightcone corresponding to travelling a distance $L$ through the medium, and
where $d\left(\mathcal{R}\right)$ is  the dimension of the representation $\mathcal{R}$.  
The result (\ref{provb}) is similar to (\ref{frperp}) although the physical context in which it arises is different as is the path followed by the Wilson line, and one of the 
derivations~\cite{CasalderreySolana:2007zz} of (\ref{provb}) is analogous to the 
derivation of (\ref{frperp}) that we reviewed in Section~\ref{sec:HQBroadening}. Another 
derivation~\cite{D'Eramo:2010xk} proceeds via the use of the optical theorem to relate $P(k_\perp)$ to an appropriate forward scattering matrix element  which can then be calculated explicitly via formulating the calculation of transverse momentum broadening in the language of Soft Collinear Effective Theory~\cite{Bauer:2000ew,Bauer:2000yr,Bauer:2001ct,Bauer:2001yt,Bauer:2002nz}.
This derivation in particular makes it clear that (\ref{provb}) is valid whether the plasma through which the energetic quark is propagating, i.e. the plasma which is causing the transverse momentum broadening, is weakly coupled or strongly coupled.
The result (\ref{provb}) is an elegant expression saying that the probability for the 
quark to obtain transverse momentum $k_\perp$ is simply given by the two-dimensional Fourier 
transform in $x_\perp$ of the expectation value~\eqref{QuenchWils} of two light-like Wilson lines separated in the transverse plane by the vector $x_\perp$.    Note that the 
requirement (\ref{eq:Pnormalization}) that the probability distribution  $P(k_\perp)$ be normalized is equivalent to the requirement that ${\mathcal W}_{\mathcal R}(0)=1$.

\begin{figure}
 \begin{center}
\includegraphics[scale=0.55]{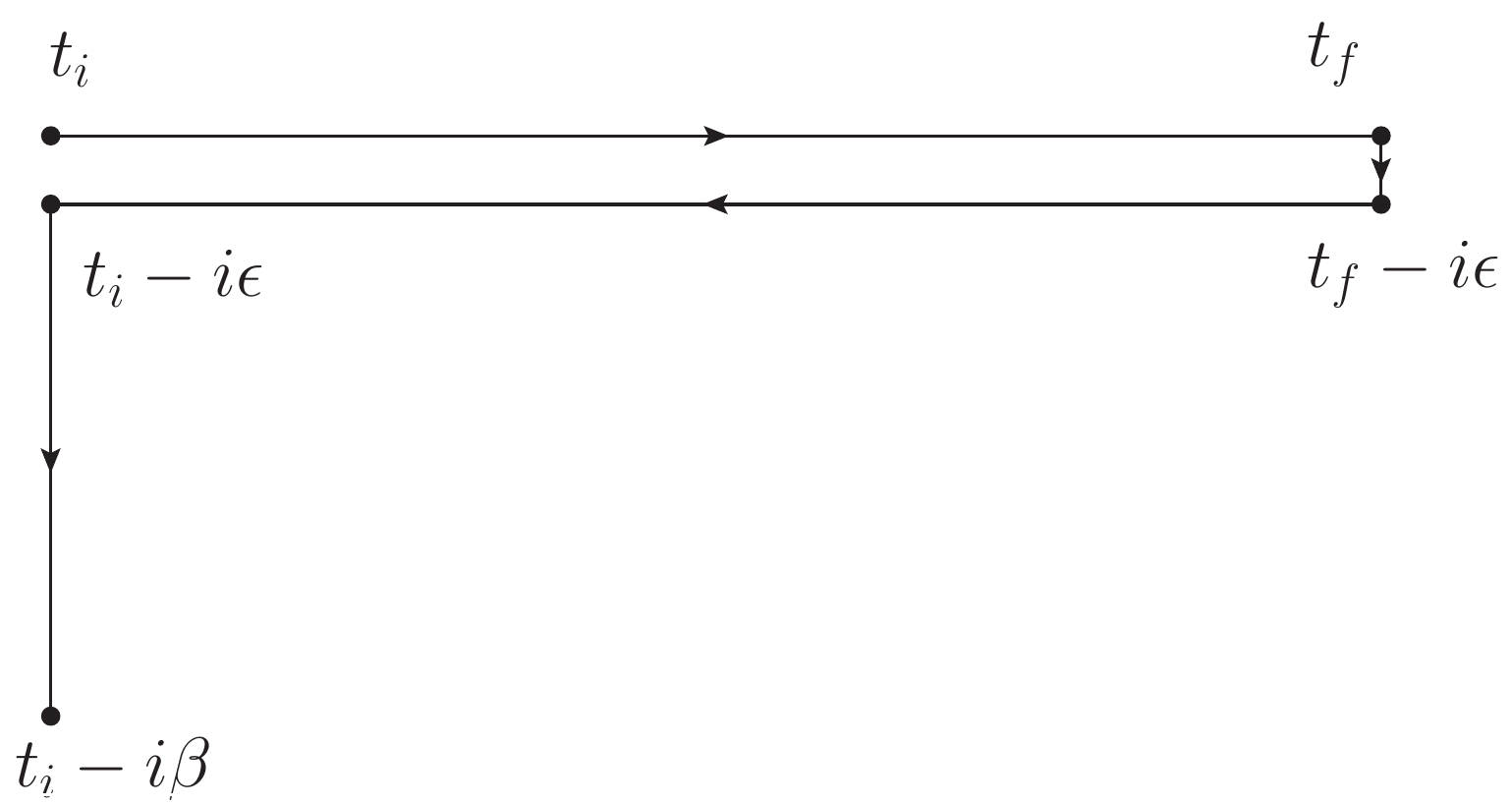}
\end{center}
\caption{\small The Schwinger-Keldysh contour that must be used in the evaluation 
of ${\mathcal W}_{\sR}(x_\perp)$. It is similar to that in Fig.~\ref{path}.}
\label{fig:skc}
\end{figure}

It is important to notice that the expectation value of the trace of the product of two light-like Wilson lines
that arises in $P(k_\perp)$ and hence in $\hat q$, namely ${\mathcal W}_{\sR}(x_\perp)$ of (\ref{QuenchWils}), has a different operator ordering from that in a standard Wilson loop.  Upon expanding the exponential, each of the $A^+$ that arise can be written as the product of an operator and a group matrix: $A^+=(A^+)^a t^a$.  It is clear (for example either by analogy with our discussion around (\ref{eq:falmostdone}) in the analysis of momentum broadening of heavy quarks or from the explicit derivation in Ref.~\cite{D'Eramo:2010xk})
 that in ${\mathcal W}_{\sR}(x_\perp)$ both the operators and the group matrices are path ordered.  In contrast, in a conventional Wilson loop the group matrices are path ordered but the operators are time ordered.  Because the operators in (\ref{QuenchWils}) are path ordered, the expectation value in (\ref{QuenchWils}) should be described using the 
 Schwinger-Keldysh contour in Fig.~\ref{fig:skc} with one of the light-like Wilson lines on the ${\rm Im}\, t=0$ segment of the contour and the other light-like Wilson line on the ${\rm Im}\, t=-i\epsilon$ segment of the contour.  The infinitesimal displacement of one Wilson line with respect to the other in Fig.~\ref{fig:skc} ensures that the operators from the two lines are ordered such that all operators from one line come before any operators from the other.
In contrast,  the loop $\sC$ for a standard Wilson loop operator
lies entirely at ${\rm Im} \, t =0$, and the operators for a standard Wilson loop are time ordered.   

The transverse momentum broadening of a hard parton with energy $E$ is due to repeated interactions with gluons from the medium which, if the medium is in equilibrium at temperature $T$, carry transverse momenta of order $T$ and lightcone momenta of order 
$T^2/E$~\cite{Idilbi:2008vm,D'Eramo:2010xk}.  The relation (\ref{provb}) between $P(k_\perp)$ (and hence $\hat q$) and the expectation value ${\mathcal W}$ of  (\ref{QuenchWils}) is valid 
as long as $E\gg {\hat q}L^2$  (which is to say $E$ must be much greater than the characteristic energy of the radiated gluons)
even if $\alpha_S(T)$ is in no way small, i.e. it is valid in the large-$E$ limit 
even if the hard parton is interacting with a strongly coupled plasma and even if the soft interactions that generate transverse momentum broadening are not suppressed by 
any weak coupling either~\cite{D'Eramo:2010xk}.   However, in this circumstance even 
though (\ref{provb}) is valid it was not particularly useful until recently because there is no known conventional field theoretical {\it evaluation} of ${\mathcal W}$ for a strongly coupled plasma.  (Since lattice quantum field theory is formulated in Euclidean space, it is not well-suited  for the evaluation of the expectation value of lightlike Wilson lines.)    In this subsection we review the evaluation of ${\mathcal W}$, and hence $\hat q$, in the strongly coupled plasma of ${\cal N}=4$ SYM theory with gauge group $SU(N_c)$ in the large $N$ and strong coupling limit using its 
gravitational dual, namely 
the AdS Schwarzschild black hole at nonzero 
temperature~\cite{Liu:2006ug,Buchel:2006bv,Caceres:2006as,Lin:2006au,Avramis:2006ip,Armesto:2006zv,Nakano:2006js,Liu:2006he,Gursoy:2009kk,HoyosBadajoz:2009pv,D'Eramo:2010xk}.   
The calculation is not simply an application of results reviewed in Section \ref{sec:Wilson} both because the operators are path ordered and because the Wilson lines are light-like.

We begin by sketching how the standard AdS/CFT procedure for computing a Wilson loop in the fundamental representation in the large $N_c$ and strong coupling limit, reviewed in Section~\ref{sec:Wilson}, applies to a light-like Wilson loop with standard operator ordering~\cite{Liu:2006ug,Liu:2006he}, and then below describe how the calculation (but not the result) changes when the operator ordering is as in~\eqref{QuenchWils}.
Consider a Wilson loop operator $W(\sC)$ specified by a closed loop $\sC$ in the $(3+1)$-dimensional field theory, and thus on the boundary of the $(4+1)$-dimensional AdS space.
$\langle W(\sC)\rangle$  is then given by the exponential of the classical action of an extremized string worldsheet $\Sig$ in AdS which ends on $\sC$. The contour ${\cal C}$ lives within the $(3+1)$-dimensional
Minkowski space boundary, but the string world sheet $\Sig$ attached to it hangs ``down'' into the bulk of the curved five-dimensional AdS$_5$ spacetime.
More explicitly, consider two long parallel light-like Wilson lines separated by a distance $x_\perp$ in a transverse direction.\footnote{Note that for a light-like contour ${\cal C}$, the Wilson 
line (\ref{Wils}) of ${\cal N}=4$ SYM theory reduces to the familiar (\ref{SingleWils}).}
(The string world sheet hanging down into the bulk from these two Wilson lines can be visualized as in Fig.~\ref{fig:horizon} below if one keeps everything in that figure at ${\rm Im} \, t =0$, i.e.~if one ignores the issue of operator ordering.)
Upon parameterizing the two-dimensional world sheet by the coordinates $\sigma^{\alpha}=(\tau,\sigma)$,
the location of the string world sheet in the five-dimensional spacetime with coordinates
$x^\mu$ is
\begin{equation}
    x^\mu = x^\mu(\tau,\sigma)\, 
    \label{para}
\end{equation}
and the Nambu-Goto action for the string world sheet is given
by
 \begin{equation}
S =- \frac{1 }{ 2 \pi \alpha'} \int d\sigma d \tau \, \sqrt{ - \det
g_{\alpha \beta}}\, . \label{ngac}
 \end{equation}
Here,
\begin{equation}
  g_{\alpha \beta} = G_{\mu\nu} \partial_\alpha x^\mu \partial_\beta x^\nu\,
  \label{inm}
\end{equation}
is the induced metric on the world sheet and $G_{\mu\nu}$ is the metric of the
$(4+1)$-dimensional AdS$_5$ spacetime. Denoting by $S (\sC)$ the classical action which  extremizes the Nambu-Goto action~\eqref{ngac} for the string worldsheet with the boundary condition that it ends on the curve ${\cal C}$, 
the expectation value of the Wilson loop operator is then given by
 \begin{equation}
\langle{W({\cal C})}\rangle = \exp\left[ i\, \lbrace S ({\cal C})  - S_0 \rbrace \right] \, ,\label{exewi}
 \end{equation}
where the subtraction
$S_0$ is the action of two disjoint strings hanging straight down from the two Wilson lines. In order to obtain the thermal expectation value at nonzero temperature, one takes 
the metric $G_{\mu\nu}$ in~\eqref{inm} to be that of an AdS Schwarzschild 
black hole~\eqref{AdSfiniteR}
with a horizon at $r=r_0$ and Hawking temperature $T=r_0/(\pi R^2)$.
The AdS curvature radius $R$ 
and the string tension $1/(2\pi\alpha')$ are related to the 't Hooft coupling
in the Yang-Mills theory $\lambda\equiv g^2 N_c$ by $\sqrt{\lambda} = R^2/\alpha'$.

We shall assume that the length of the two light-like lines $L^-=\sqrt{2}L$ is much greater than their transverse separation $x_\perp$, which can be justified after the fact by using the result 
for ${\cal W} (x_\perp)$ to show that the $x_\perp$-integral in (\ref{provb})  is dominated by 
values of $x_\perp$ that satisfy $x_\perp  \ll   1/\sqrt{\hat q L} \sim  1/\sqrt{\sqrt{\lambda}LT^3}$.  As long as we are interested in $L \gg 1/T$, then $x_\perp \ll 1/(T \lambda^{1/4}) \ll 1/T \ll L$.
%
%
With $L^-\gg x_\perp$, we can ignore the ends of the light-like Wilson lines and assume that the shape of the surface $\Sigma$ is translationally invariant along the light-like direction. The action (\ref{ngac}) now takes the form
\begin{equation}
S=i\frac{\sqrt{2}r_0^2 \sqrt{\lambda} L^-}{2\pi R^4}\int_0^{x_\perp/2} d\sigma \sqrt{1+\frac{r'^2R^4}{r^4-r_0^4}}
\end{equation}
where the shape of the worldsheet $\Sigma$ is described by the function $r(\sigma)$ that satisfies $r(\pm \frac{x_\perp}{2})=\infty$, which preserves the symmetry $r(\sigma)=r(-\sigma)$, and where $r'=\partial_\sigma r$.  The equation of motion for $r(\sigma)$ is then
\begin{equation}\label{qhatEoM}
r'^2=\frac{\gamma^2}{R^4} \left(r^4-r_0^4\right)
\end{equation}
with $\gamma$ an integration constant.  Eq.~(\ref{qhatEoM}) has two solutions.  One has $\gamma=0$ and hence $r'=0$, meaning that $r(\sigma)=\infty$ for all $\sigma$:  the surface $\Sigma$ stays at infinity.  Generalizations of this solution have also been studied~\cite{Argyres:2006yz,Argyres:2008eg}.  We shall see below that such solutions are not relevant.  The other solution has $\gamma>0$. It ``descends'' from $r(\pm \frac{x_\perp}{2})=\infty$ and has a turning point where $r'=0$ which, by symmetry, must occur at $\sigma=0$.  From (\ref{qhatEoM}), the turning point must occur at the horizon $r=r_0$.  Integrating (\ref{qhatEoM}) gives the condition that specifies the value of $\gamma$:
\begin{equation}
\frac{x_\perp}{2}=\frac{R^2}{\gamma}\int_{r_0}^\infty \frac{dr}{\sqrt{r^4-r_0^4}} = \frac{a R^2}{\gamma r_0}
\end{equation}
where we have defined 
\begin{equation}
a\equiv \sqrt{\pi}\Gamma(\frac{5}{4})/\Gamma(\frac{3}{4})\approx 1.311\ .
\end{equation}
Putting all the pieces together, we find~\cite{Liu:2006ug,Liu:2006he}
\begin{equation}
S=\frac{i a \sqrt{\lambda} T L^-}{\sqrt{2}}\sqrt{1+\frac{\pi^2T^2x_\perp^2}{4a^2}}\ .
\end{equation}
We see that $S$ is imaginary, because when the contour ${\mathcal C}$ at the boundary is lightlike the surface $\Sigma$ hanging down from it is spacelike.  It is worth noting that $S$ had to turn out to be imaginary, in order for $\langle W \rangle$ in (\ref{exewi}) to be real and the transverse momentum broadening $P(k_\perp)$ to be real, as it must be since it is a probability distribution.
The surface $\Sigma$ that we have used in this calculation descends from infinity, skims the horizon, and returns to infinity.  Note that the surface descends all the way to the horizon regardless of how small $x_\perp$ is.  This is reasonable on physical grounds, as we expect $P(k_\perp)$ to depend on the physics of the thermal medium~\cite{Liu:2006ug,Liu:2006he}.
We shall see below that it is also required on mathematical grounds: when we complete the calculation by taking into account the nonstandard operator ordering in (\ref{provb}), we shall see that only a worldsheet that touches the horizon is relevant~\cite{D'Eramo:2010xk}.

 \begin{figure*}[t]
 \begin{center}
\includegraphics[scale=0.55]{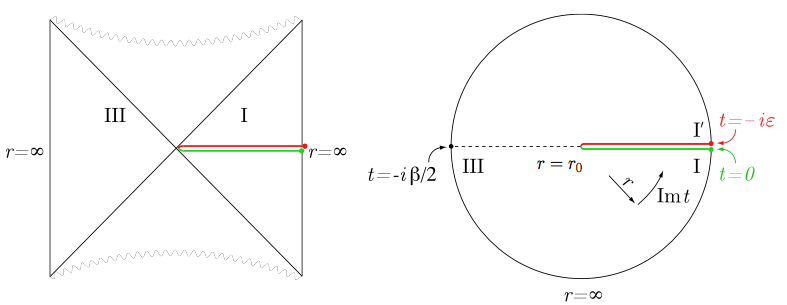}
\end{center}
\caption{\small Penrose diagrams for Lorentzian (${\rm Im}\,t=0$) and Euclidean (${\rm Re}\,t=0$) sections of an AdS black hole. In the right panel, the two light-like Wilson lines are points at $r=\infty$, indicated by colored dots.  These dots are the boundaries of a string world sheet that extends inward to $r=r_0$, which is at the origin of the Euclidean section of the black hole. In the left panel, the string world sheet and its endpoints at $r=\infty$ are shown at ${\rm Re}\, t=0$;  as ${\rm Re}\,t$ runs from $-\infty$ to $\infty$, the string worldsheet sweeps out the whole of quadrant $I$.  }
\label{fig:disk}
\end{figure*}


We now consider the computation of~\eqref{QuenchWils}, with its nonstandard operator ordering corresponding to putting one of the two light-like Wilson lines on the ${\rm Im} \, t =0$ contour in Fig.~\ref{fig:skc} and the other on the ${\rm Im}\, t=-i\epsilon$ contour.  The procedure we shall describe is a specific example of the
more general discussion of Lorentzian
AdS/CFT given recently 
 in Refs.~\cite{Skenderis:2008dh,Skenderis:2008dg,vanRees:2009rw,Barnes:2010jp}.
In order to compute~\eqref{QuenchWils} we first need to construct the bulk geometry corresponding to the ${\rm Im}\,t=-i\epsilon$ segment of the Schwinger-Keldysh contour 
in Fig.~\ref{fig:skc}. For this purpose it is natural to consider the black hole geometry with complex time.
In Fig.~\ref{fig:disk}, we show two slices of this complexified geometry. The left plot is the Penrose diagram for the fully extended black hole spacetime with quadrant $I$ and $III$ corresponding to the slice ${\rm Im} \, t =0$ and ${\rm Im} t = -{\beta \ov 2}$ respectively, while the right plot is for the Euclidean black hole geometry, i.e. corresponding to the slice ${\rm Re} t =0$. Note that because the black hole has a nonzero temperature, the imaginary part of $t$ is periodic with the period given by the inverse temperature $\beta$. 
In the left plot the imaginary time direction can be considered as a circular direction coming out of the paper at quadrant $I$, going a half circle to reach quadrant $III$ and then 
going into the paper for a half circle to end back at $I$. In the right plot the real time direction can be visualized as the direction perpendicular to the paper.

The first segment of the Schwinger-Keldysh contour in Fig.~\ref{fig:skc}, with ${\rm Im}\,t=0$, lies at the boundary ($r=\infty$) of quadrant $I$ in Fig.~\ref{fig:disk}. 
The second segment of the Schwinger-Keldysh contour, 
with ${\rm Im}\, t=-i\epsilon$, lies at the $r=\infty$ boundary of a copy of $I$ that in the left plot of Fig.~\ref{fig:disk} lies infinitesimally outside the paper and in the right plot of Fig.~\ref{fig:disk} lies at an infinitesimally different angle.  We shall denote this copy of $I$ by $I'$. 
The geometry and metric in $I'$ are identical to those of $I$. Note that $I'$ and $I$ are joined together at the horizon $r=r_0$, namely at the origin in the right plot of Fig.~\ref{fig:disk}. Now, the thermal expectation value~\eqref{QuenchWils} can be computed by putting the two parallel light-like Wilson lines at the boundaries of $I$ and $I'$, and finding the extremized string world sheet which ends on both of them.
Note that since $I$ and $I'$ 
meet only at the horizon, the only way for there to be a nontrivial (i.e.~connected) string world sheet whose boundary is the two Wilson lines in (\ref{QuenchWils}) is for such a string world sheet to touch the horizon.
Happily, this is precisely the feature of the string world sheet found in the explicit calculation 
that we reviewed above.
So, we can use that string world sheet in the present analysis, with the only difference being that half the string world sheet now lies on $I$ and half on $I'$, as 
illustrated in Fig.~\ref{fig:horizon}.\footnote{
The calculation of $\hat q$ in ${\cal N}=4$ SYM theory via~\eqref{QuenchWils} nicely resolves a subtlety. As we saw above, in addition to the extremized string configuration which touches the horizon, the string action also has another trivial solution 
which lies solely at the boundary, at $r=\infty$.   
Based on the connection between position in the $r$ dimension in the gravitational theory and energy scale in the quantum field theory, the authors of Ref.~\cite{Liu:2006ug,Liu:2006he} argued that physical considerations (namely the fact that $\hat q$ should reflect thermal physics at energy scales of order $T$) require selecting the extremized string configuration that touches the horizon.  Although this physical argument remains valid, we now see that it is not necessary. 
In (\ref{QuenchWils}), the two Wilson lines are at the boundaries of $I$ and $I'$, with different values of ${\rm Im}\,t$.  That means that there are no string world sheets that connect the two Wilson lines without touching the horizon.  So, once we have understood how the nonstandard operator ordering in (\ref{QuenchWils}) modifies the boundary conditions for the string world sheet, we see that the trivial world sheet of Refs.~\cite{Liu:2006ug,Liu:2006he} and all of its generalizations in Refs.~\cite{Argyres:2006yz,Argyres:2008eg} do not satisfy the correct boundary conditions.  The nontrivial world sheet illustrated in Fig.~\ref{fig:horizon}, which is sensitive to thermal 
physics~\cite{Liu:2006ug,Liu:2006he,Liu:2008tz}, is the only extremized world sheet bounded by the two lightlike Wilson lines in (\ref{QuenchWils})~\cite{D'Eramo:2010xk}.}

\begin{figure}
 \begin{center}
\includegraphics[scale=0.60]{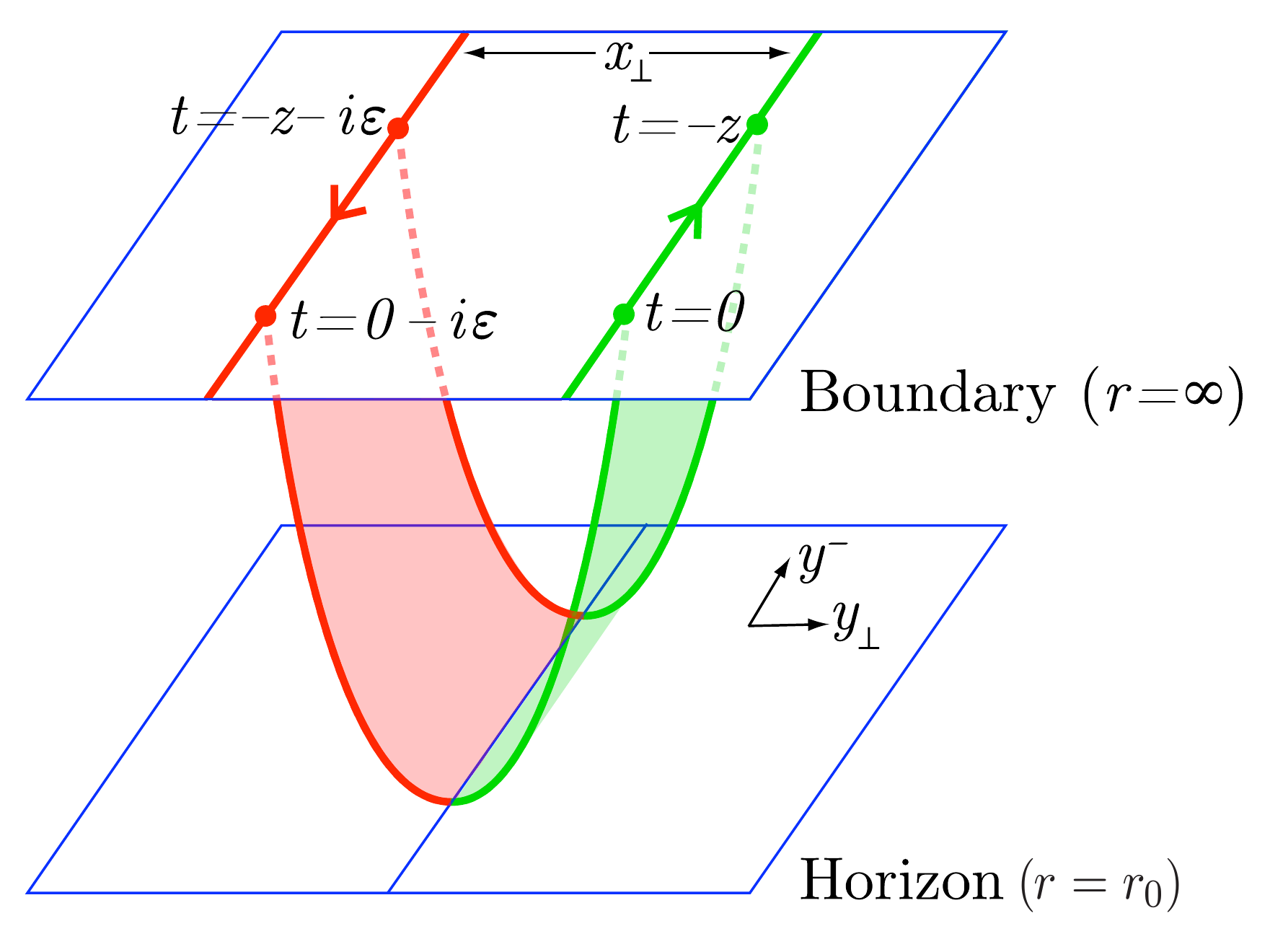}
\end{center}
\caption{\small String configuration for the thermal expectation value of~\eqref{QuenchWils}.}
\label{fig:horizon}
\end{figure}

We conclude that the result for the expectation value~\eqref{QuenchWils}, with its nonstandard path ordering of operators, is identical to that obtained in Refs.~\cite{Liu:2006ug,Liu:2006he} for a light-like Wilson loop with standard time ordering of operators~\cite{D'Eramo:2010xk}.  
That is, in strongly coupled ${\cal N}=4$ SYM theory ${\mathcal W}(x_\perp)$ in the adjoint representation
is given by 
\begin{equation}
\mathcal{W}_{\mathcal{A}}(x_\perp) = 
 \exp \left[ - \sqrt{2} a \sqrt{\lambda}\,  L^- T \left(  \sqrt{1 + \frac{\pi^2 T^2 x_\perp^2}{4a^2}} - 1\right)\right]\ .
\label{fullW}
\end{equation}
We have quoted the result for $\mathcal{W}_{\mathcal{A}}(x_\perp)$, which is given by  
$\mathcal{W}^2_{\mathcal{F}}(x_\perp)$ in the large-$N_c$ limit,
because that is what arises in the analysis of jet quenching, see Section~\ref{sec:AnalyzingJetQuenching}.  (Radiative parton energy loss depends on the medium through the transverse momentum broadening of the radiated gluons, which are of course in the adjoint representation.)  
The $x_\perp$-independent term in the exponent in (\ref{fullW}), namely ``the -1'', is the finite subtraction of $S_0$, which was identified in Ref.~\cite{Liu:2006ug} as the action of two disjoint strings hanging straight down from the two Wilson lines to the horizon of the AdS black hole.
Our calculation serves as a check of the value of $S_0$, since  
only with the correct $S_0$ do we obtain $\mathcal{W}_{\mathcal{A}}(0)=1$  and a correctly normalized probability distribution $P(k_\perp)$. 
Note that
our field theory set-up requires $L^- T \gg 1$, and our supergravity calculation requires $\lambda \gg 1$, meaning that our result (\ref{fullW})  is only valid for
\begin{equation}
\sqrt{\lambda} \, L^- T \gg 1 .
\label{StrongCouplingGaussian}
\end{equation}
In this regime, (\ref{fullW}) is very small unless $\pi x_\perp T/(2a)$ is small.  This means that when we take the Fourier transform of \eqref{fullW} to obtain the probability 
distribution $P(k_\perp)$, in the regime (\ref{StrongCouplingGaussian})
where the calculation is valid 
the Fourier transform is dominated by small values of $x_\perp$, for which
\begin{equation}
\mathcal{W}_{\mathcal{A}}(x_\perp) \simeq 
\exp \left[ - \frac{\pi^2} {4 \sqrt{2} a} \sqrt{\lambda} L^- T^3 x_\perp^2 \right]\ ,
\label{StrongCouplingGaussian2}
\end{equation}
and we therefore obtain
\begin{equation}
P(k_\perp) = \frac{4\sqrt{2}a}{\pi \sqrt{\lambda} T^3 L^-}\exp\left[ - \frac{\sqrt{2} a k_\perp^2}{\pi^2 \sqrt{\lambda} T^3 L^-} \right] \ .
\label{PkperpResult}
\end{equation}
Thus, the probability distribution $P(k_\perp)$ is a Gaussian  and the jet-quenching parameter (\ref{qhatFirstTime}) can easily be evaluated, yielding~\cite{Liu:2006ug}
\begin{equation}
\hat q=\frac{\pi^{3/2}\Gamma(\frac{3}{4})}{\Gamma(\frac{5}{4})}\sqrt{\lambda} T^3 \ .
\label{qhatResult}
\end{equation}
The probability distribution (\ref{PkperpResult}) has a simple physical interpretation: the probability that the quark has gained transverse momentum $k_\perp$ is given by diffusion in transverse momentum space with a diffusion constant given by $\hat q L$.  This is indeed consistent with the physical expectation that transverse momentum broadening in a strongly coupled plasma is due to the accumulated effect of many small kicks by gluons from the medium: the quark performs Brownian motion in momentum space even though in coordinate space it remains on a light-like trajectory.  

If we attempt to plug RHIC-motivated numbers into the result (\ref{qhatResult}), taking $T=300$~MeV, $N_c=3$, $\alpha_{\rm SYM}=\frac{1}{2}$ and therefore $\lambda=6\pi$ yields 
$\hat q = 4.5$~GeV$^2$/fm, which turns out to be in the same ballpark as the values of $\hat q$ inferred from RHIC data on the suppression of high momentum partons in heavy ion 
collisions~\cite{Liu:2006ug,Liu:2006he}, as we reviewed around Fig.~\ref{fig:RAAphenix} 
in Section \ref{sec:JetQuenching}.   We can also write the result (\ref{qhatResult}) as
\begin{equation}
\hat q \simeq 57 \sqrt{ \alpha_{\rm SYM} \frac{N_c}{3} } \,T^3\ ,
\label{qhatResult2}
\end{equation}
which can be compared to the result (\ref{qhatKrelation},\ref{qhatKresult}) extracted via comparison to RHIC data in Ref.~\cite{Armesto:2009zi}. To make the comparison, we need to relate the QCD energy density $\varepsilon$ appearing in (\ref{qhatKrelation}) to $T$.  Lattice calculations of QCD thermodynamics indicate $\varepsilon \sim (9-11)\, T^4$ in the temperature regime that is relevant at RHIC~\cite{Borsanyi:2010cj}.  This then means that if in (\ref{qhatResult2}) 
we take $\alpha_{\rm SYM}$ within the range
$\alpha_{\rm SYM}= .66^{+.34}_{-.25}$,
the result (\ref{qhatResult2}) for the strongly 
coupled ${\cal N}=4$ SYM plasma is consistent with the result (\ref{qhatKresult}) obtained via comparing QCD jet quenching calculations to RHIC data.
The extraction of $\hat q$ by inference from  LHC data on several-hundred-GeV jets should be under better control, since then the separation of scales upon which the QCD calculations (and the basic assumption that energy loss is dominated by gluon radiation) will be more quantitatively reliable.

We have reviewed the ${\cal N}=4$ SYM calculation, but the jet quenching parameter can be calculated in any conformal theory with a gravity 
dual~\cite{Liu:2006he}.  In a large class of such theories in which the 
spacetime for the gravity dual is AdS$_5 \times$~M$_5$ for some internal manifold M$_5$ other than the five-sphere $S_5$ which gives ${\cal N}=4$ SYM theory~\cite{Liu:2006he},
\begin{equation}
\frac{\hat q_{\rm CFT}}{\hat q_{{\cal N}=4}}=
\sqrt{\frac{s_{\rm CFT}}{s_{{\cal N}=4}}}\ ,
\label{qhatversusentropy}
\end{equation}
with $s$ the entropy density.  This result makes a central qualitative lesson from (\ref{qhatResult}) clear: in a strongly coupled plasma, the jet quenching parameter is not proportional to the entropy density or to some number density of distinct scatterers.  This qualitative lesson is more robust than any attempt to make a quantitative comparison to QCD. But, we note that
if QCD were conformal, (\ref{qhatversusentropy}) would suggest 
\begin{equation}
\frac{\hat q_{\rm QCD}}{\hat q_{{\cal N}=4}} \approx 0.63\ .
\end{equation}
And, analysis of how $\hat q$ changes in a particular toy model in which nonconformality can be introduced by hand then suggests that introducing the degree of nonconformality seen in QCD thermodynamics may increase $\hat q$ by a few tens of percent~\cite{Liu:2008tz}.\footnote{$\hat q$ also increases with increasing nonconformality in strongly coupled ${\cal N}=2^*$ gauge theory~\cite{Gursoy:2009kk,HoyosBadajoz:2009pv}.}   Putting these two observations together, perhaps it is not surprising that the $\hat q$ for the strongly coupled plasma of ${\cal N}=4$ SYM theory is in the same ballpark as that extracted by comparison with RHIC data.

\section{Quenching a beam of strongly coupled synchrotron radiation}
\label{sec:Synchrotron}

In Sections \ref{sec:HQDrag}, \ref{sec:HQBroadening} and \ref{sec:AdSCFTDragWaves} we have gained insights into parton energy loss and jet quenching via studying how a single heavy quark slows and gets kicked as it moves through the strongly coupled plasma of ${\cal N}=4$ SYM theory, and how that single heavy quark excites the plasma through which it moves. In 
Section~\ref{sec:Stopping}, we have seen how a single light quark or gluon is stopped and thermalized by the strongly coupled $\N=4$ SYM plasma, assuming that all aspects of this physics are described at strong coupling.
In Section~\ref{sec:AdSCFTJetQuenching} we have analyzed jet quenching per se, but have done so via the strategy of working as far as possible within weakly coupled QCD and only using a holographic calculation within  ${\cal N}=4$ SYM theory for one small part of the story, namely the calculation of the jet quenching parameter $\hat q$ through which the physics of the strongly coupled medium enters the calculation.  This approach is justified in the high jet energy limit, where there is a clean separation of scales and where jet quenching is due to medium-induced gluon radiation.  But, the jets being studied at present at RHIC are not sufficiently energetic as to make one confident in the quantitative reliability of these approximations.  In addition to motivating the study of jet quenching in heavy ion collisions at the LHC, this
raises an obvious theoretical
question:  can we study jets themselves in ${\cal N}=4$ SYM theory?  Suppose we assume that the physics at all relevant scales is strongly coupled,  an assumption that is in a sense the opposite of that we made in Section~\ref{sec:AdSCFTJetQuenching}.  But, instead of considering just a single quark, can we analyze how an actual jet is modified by the strongly coupled plasma of ${\cal N}=4$ SYM theory, relative to how it would have developed in the vacuum of that theory? 
This question is pressing since one of the main goals 
of the ongoing high-$p_T$ measurements in heavy ion collisions at RHIC is full jet reconstruction, with experimentalists aiming to answer questions like how much wider in angle are jets in heavy ion collisions than in proton-proton collisions?  For example, the STAR collaboration is measuring the ratio of the energy within a cone of radius 0.2 radians to that within a cone of radius 0.4 radians, and comparing that ratio in gold-gold and proton-proton collisions.  It would be of value to get some insights into what to expect for such an observable in a strongly coupled plasma, via a holographic calculation in ${\cal N}=4$ SYM theory.

Unfortunately, the answer to the question as we have  just posed it  is negative.  In 
Ref.~\cite{Hofman:2008ar}, Hofman and Maldacena considered the following thought experiment.  Suppose you did electron-positron scattering in a world in which the electron and positron coupled to ${\cal N}=4$ SYM theory through a virtual photon, just as in the real world they couple to QCD.   What would happen in high energy scattering?  Would there be any ``jetty'' events?  They showed that the answer is no. Instead, the final state produced by a virtual photon in 
the conformal ${\cal N}=4$ SYM theory is a spherically symmetric outflow of energy.   Similar conclusions were also reached in Refs.~\cite{Hatta:2008tx,Chesler:2008wd}.   The bottom line is that there are no jets in strongly coupled ${\cal N}=4$ SYM theory, which would seem to rule out using this theory to study how jets are modified by propagating through the strongly coupled plasma of this theory.  

\begin{figure}[t] 
\begin{center}
\includegraphics[scale=0.55]{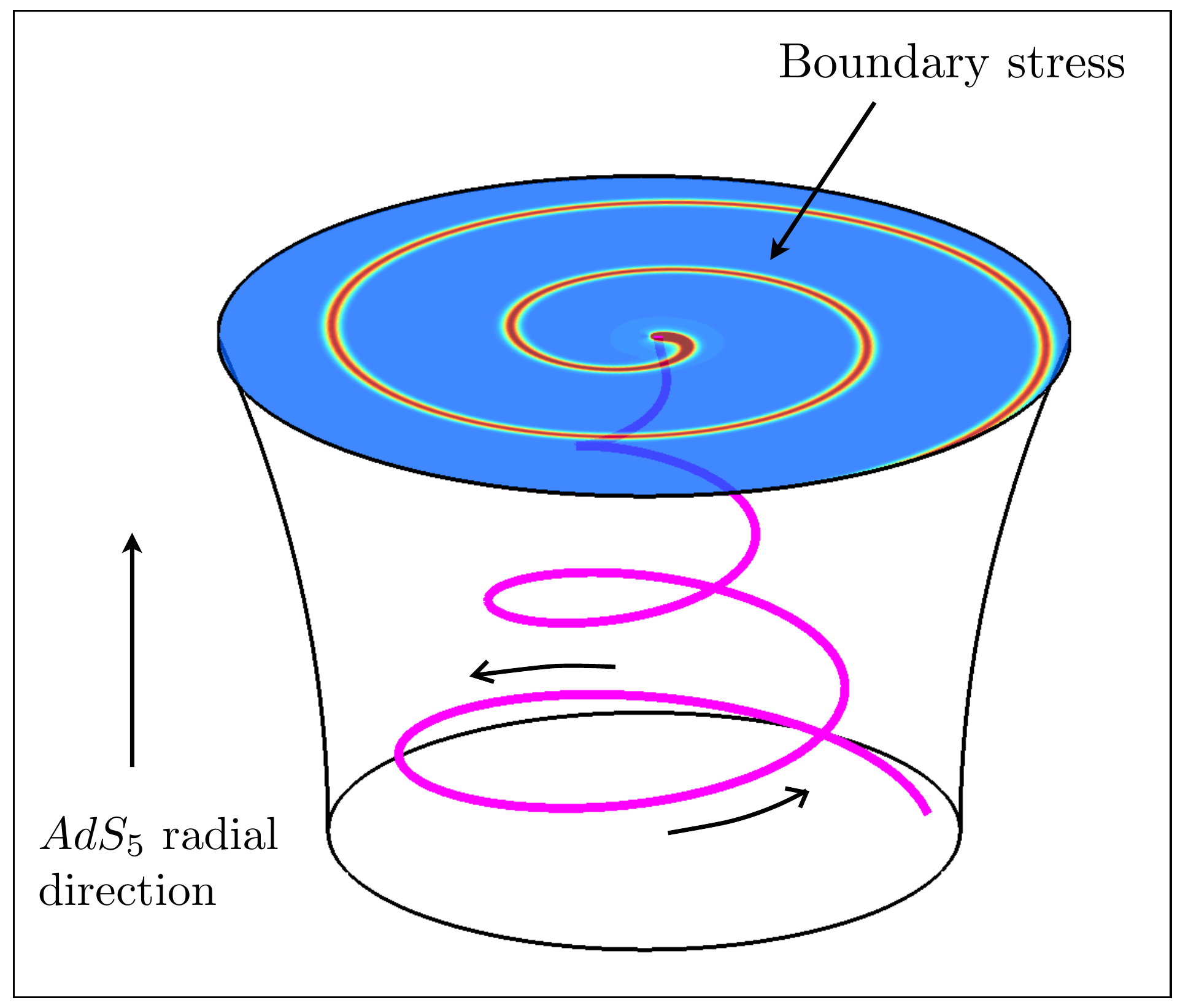}
\caption{\small 
\label{fig:SynchrotronCartoon}
Cartoon of the gravitational description of synchrotron radiation at strong coupling: the quark rotating at the boundary trails a rotating string behind it which hangs down into 
the bulk AdS$_5$ space.  This string acts as a source of gravitational waves in the bulk, 
and this gravitational radiation induces a stress tensor on the four-dimensional boundary.
By computing the bulk-to-boundary propagator one obtains the boundary-theory stress-tensor that describes the radiated energy.  The entire 
calculation can be done analytically~\cite{Athanasiou:2010pv}.
}
\end{center}
\end{figure}

Recent work~\cite{Athanasiou:2010pv} may provide a path forward.  These authors have found a way of creating a beam of gluons (and color-adjoint scalars) that is tightly collimated in angle and that propagates outwards forever in the ${\cal N}=4$ SYM theory vacuum without spreading in angle.  This beam is not literally a jet, since it is not produced far off shell.  But, we know from Hofman and Maldacena that a far off shell ``photon'' does not result in jets in this theory.  And, this beam of nonabelian radiation may yield a better cartoon of a jet than a single quark, heavy or light, or a single gluon, all of which have been analyzed previously.   This beam is created by synchrotron radiation from a test quark that is moving in a circle of radius $L$ with constant angular 
frequency $\omega$, as in Ref.~\cite{Fadafan:2008bq} but now, in Ref.~\cite{Athanasiou:2010pv}, in vacuum instead of in medium.    The way the calculation is done is 
sketched in Fig.~\ref{fig:SynchrotronCartoon}.  The logic of the calculation is as we described in Section~\ref{sec:AdSCFTDragWaves}, and so we shall not present it in detail. It turns out, however, that in vacuum the shape of the rotating string in the bulk and the corresponding form of the energy density of the outward-propagating radiation on the boundary can both be determined analytically~\cite{Athanasiou:2010pv}.

\begin{figure}
\begin{center}
\includegraphics[scale=0.27]{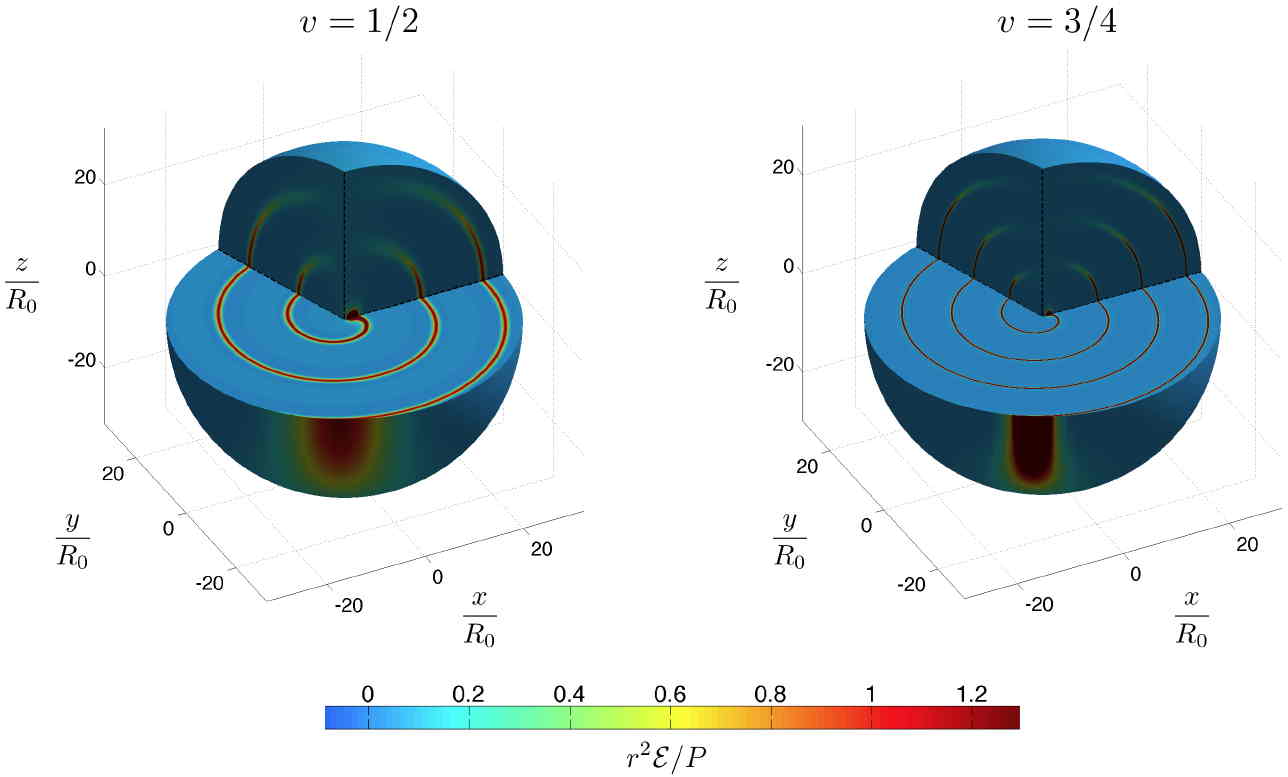}
\end{center}
\caption{\small 
\label{fig:SynchrotronCandy}
Cutaway plots of $r^2 {\cal E} /P$ for a test quark in circular motion with $v=1/2$ and $v=3/4$.  Here, ${\cal E}$ is the energy density and $P$ is the total power radiated per unit time.  We see a spiral of radiation, propagating radially outwards at the speed of light, without any spreading.  The spiral is localized about $\theta=\pi/2$ with a characteristic width $\delta\theta\propto1/\gamma$.  The radial thickness of the spiral is proportional to $1/\gamma^3$.
}
\end{figure}

Because the vacuum of ${\cal N}=4$ SYM theory is not similar to that of our world, studying the radiation of a test quark in circular motion does not have direct phenomenological motivation.  However, the angular distribution of the radiation turns out to be {\it very} similar to that of classic synchrotron radiation~\cite{Athanasiou:2010pv}:  when the quark is moving along its circle with an ultrarelativistic velocity $v$, the radiation is produced in a tightly collimated beam with an angular width that is proportional to $1/\gamma$, with $\gamma$ the Lorentz boost  factor associated with $v$, see Fig.~\ref{fig:SynchrotronCandy}.  The spiral pulse of radiation propagates outward without spreading, and the radial thickness  (or, equivalently, the duration in time) of the pulse is proportional to $1/\gamma^3$.
So, as a function of increasing $\gamma$ the beam, if Fourier analyzed, contains radiation at increasingly short wavelengths and high frequencies.   

Although it has recently been explained from the gravitational point of view in beautifully geometric terms~\cite{Hubeny:2010rd},
from the point of view of the non-abelian gauge theory it is surprising that the angular distribution of the radiation at strong coupling is so similar (see Ref.~\cite{Athanasiou:2010pv} for quantitative comparisons) to that at weak coupling, where what is radiated is a mixture of colored --- and therefore interacting --- gluons and scalars.   The fact that, even when the coupling is arbitrarily strong, as the pulses of radiation propagate outwards they do not spread at all and never isotropize indicates that intuition based upon parton branching~\cite{Hatta:2008tx} (namely that the non-abelian character of the radiation should result in energy flowing from short to long wavelengths as the pulses propagate outwards and should yield isotropization at large distances) is 
invalid in this context.\footnote{Very recent work~\cite{Hatta:2010dz} shows that isotropization via some analogue of parton branching is also not the correct picture for the radiation studied 
in Ref.~\cite{Hofman:2008ar} by Hofman and Maldacena:  this radiation also propagates outward as a pulse without any spreading, but this pulse is spherically symmetric at all radii.  So, in the case studied by Hofman and Maldacena there is no process of isotropization because the radiation is isotropic at all times while in the case illustrated in Fig.~\ref{fig:SynchrotronCandy} there is no isotropization because the radiation never becomes isotropic.}  

For our purposes, the result that the ``beam'' of radiated gluons (and scalars) propagates outward with a fixed angular width that we can select by picking $\gamma$ is fortuitous.  It means that via this roundabout method we have found a way of making a state that looks something like a jet.
It is not the same in all respects as a jet in QCD  in that it is not produced via the fragmentation of an initially far offshell parton.  But, it is a collimated beam of gluons of known, and controllable, angular width.  For this reason, the results obtained in the formal setting of a test quark moving in a circle open the way to new means of modeling jet quenching in heavy ion collisions.  The challenge is to repeat the calculation of Ref.~\cite{Athanasiou:2010pv} at nonzero temperature, in the presence of a horizon in the bulk spacetime.  Upon doing so, one could watch the initially tightly collimated beam of synchrotron radiation interact with the strongly coupled plasma that would then be present. 
As long as $\omega^2\gamma^3\gg\pi^2T^2$, we know from Ref.~\cite{Fadafan:2008bq} 
that the total power radiated is as if in vacuum~\cite{Mikhailov:2003er}.  In this regime, we therefore expect to see strongly coupled synchrotron radiation as in Fig.~\ref{fig:SynchrotronCandy} on length scales small compared to $1/T$.  But, as in Section~\ref{sec:AdSCFTDragWaves}, we expect that this beam of radiation should eventually thermalize locally, likely converting into an outgoing hydrodynamic wave moving at the speed of sound before ultimately isotropizing, dissipating and thermalizing completely. One could see by how much a beam prepared such that in vacuum it has a known, small, angular extent spreads in angle after propagating through $1/T$ (or $2/T$, or $3/T$) of strongly coupled plasma.  Such an analysis should provide interesting benchmarks against which to compare data from RHIC on how much wider in angle jets of a certain energy are in heavy ion collisions as compared to in proton-proton collisions. Early results on jets in heavy ion collisions at the LHC~\cite{Collaboration:2010bu} indicate that, for the high energy jets available in these collisions,  it will be possible to make calorimetric measurements of how the jets initiated by partons propagating in the medium produced in these collisions are broadened in angle.





\section{Velocity-scaling of the screening length and quarkonium suppression in a hot wind}
\label{sec:HotWind}

We saw in Section~\ref{quarkonium} that, because they are smaller than typical hadrons in QCD, heavy quarkonium mesons survive as bound states even at temperatures above the crossover from a hadron gas to quark-gluon plasma.  However, if the temperature of the quark-gluon plasma is high enough, they eventually dissociate.
An important physical mechanism underlying the dissociation is
the weakening attraction between the heavy quark and anti-quark in the bound state because the force between their color charges is screened by the medium.
The dissociation of charmonium and bottomonium bound states has
been proposed as a signal for the formation of a hot and
deconfined quark-gluon plasma in heavy ion collisions~\cite{Matsui:1986dk}, and as a means of gauging the temperatures reached during the collisions.

In the limit of large quark mass, 
the interaction between the quark and the anti-quark in a bound state in the thermal medium can be extracted from the thermal expectation value of the Wilson loop operator 
$\vev{W^F (\sC_{\rm static})}$, with $\sC_{\rm static}$  a rectangular loop 
with a short side of length $L$ in a spatial direction (say $x_1$) and a long side of length $\sT$ along the time direction.  This expectation value takes the form
 \begin{equation}
    \langle W^F(\sC_{\rm static}) \rangle = \exp \left[ - i\, {\cal T}\,
         E(L) \right] \ .
    \label{2.2}
\end{equation}
where $E(L)$ is the (renormalized) free energy of the quark-antiquark
pair with the self-energy of each quark subtracted. $E(L)$ defines an effective potential between the quark anti-quark pair.  The screening of the force between color charges due to the presence of the medium manifests itself in the flattening of $E(L)$ for $L$ greater than some characteristic length scale $L_s$ called the screening length.  In QCD, the flattening of the potential occurs smoothly, as seen in the lattice calculations illustrated in Fig.~\ref{fig:QQbarDissoc2} in Section~\ref{secmeson},
and one must make an operational definition of $L_s$.
For example, in the parametrization of 
(\ref{ScreenedPotentialParametrization}), $L_s$ can be set equal to $1/\mu$. 
$L_s$ decreases with increasing temperature and can be used to
estimate the scale of the dissociation temperature $T_{\rm diss}$ as
 \begin{equation} \label{dis} 
 L_s (T_{\rm diss}) \sim d
 \end{equation}
where $d$ is the size of a particular mesonic bound state at zero temperature.  The idea here is that once the temperature of the quark-gluon plasma is high enough that the potential between a quark and an antiquark separated by a distance corresponding to the size of a particular meson has been fully screened, that meson can no longer exist as a bound state in the plasma.  This means that larger quarkonium states dissociate at lower temperatures, and means that the ground-state bottomonium meson survives to the highest temperatures of all.  As we discussed at length in Section~\ref{quarkonium}, there are many important confounding effects that must be taken into account in order to realize the goal of using data on charmonium production in heavy ion collisions to provide evidence for this sequential pattern of quarkonium dissociation as a function of increasing temperature.  In this Section, we shall focus only on one of these physical effects, one on which calculations done via gauge/gravity duality have shed some 
light~\cite{Peeters:2006iu,Liu:2006nn,Chernicoff:2006hi,Caceres:2006ta,Argyres:2006vs,Avramis:2006em,Friess:2006rk,Talavera:2006tj,Chernicoff:2006yp,Liu:2006he,Avramis:2006nv,Mateos:2007vn,Natsuume:2007vc,Avramis:2007mv,Ejaz:2007hg,Athanasiou:2008pz,Liu:2008tz,Myers:2008cj,Faulkner:2008qk}.

In heavy ion collisions, quarkonium mesons are
produced moving with some velocity $\vec v$ with respect to the medium.
It is thus important to understand the effects of nonzero quarkonium velocity on the screening length and consequent dissociation of  bound states. 
To describe the interaction between a quark--anti-quark pair that is moving relative to the medium, it is convenient to boost into a frame in which the quark anti-quark pair is at rest, but feels a hot wind of QGP blowing past them.
 The effective quark potential can again be extracted from~\eqref{2.2}  evaluated in the boosted frame with $\sT$  now interpreted as the proper time of the dipole.  While much progress has been made in using lattice QCD calculations to extract the effective potential between 
 a quark--anti-quark pair at rest in the QGP, there are significant difficulties in using Euclidean lattice techniques to address the (dynamical as opposed to thermodynamic) problem of a quark--antiquark pair in a hot wind.  In the strongly coupled plasma of ${\cal N}=4$ SYM theory with large $N_c$, however, 
 the calculation can be done using gauge/gravity duality~\cite{Liu:2006nn,Liu:2006he,Chernicoff:2006hi}, 
 and requires only a modest extension of the standard methods reviewed in Section~\ref{sec:Wilson}.  Here, we sketch the derivation from Ref.~\cite{Liu:2006he}.

We 
start with a rectangular Wilson loop whose short transverse space-like side 
\begin{equation}
    \sigma = x_1 \in [-{L \ov 2}, {L \ov 2}] 
    \label{bound1}
\end{equation}
defines the separation $L$ between the quark--antiquark pair and whose long time-like sides extend along the $x_3=v\,t$ direction, describing a pair moving with speed $v$ in the $x_3$ direction.
In this frame, the plasma is at rest and the spacetime metric in the gravitational description is the familiar AdS black hole (\ref{AdSfiniteR}).  We then apply a Lorentz boost that rotates this Wilson loop
into the rest frame $(t',x_3')$ of the quark--anti-quark pair:
 \begin{eqnarray}
    dt &=& dt'\,  \cosh\eta - dx_3'\, \sinh\eta\, ,
    \label{boostcoord1}\\
    dx_3 &=& - dt'\,  \sinh\eta + dx_3'\, \cosh\eta\, ,
    \label{boostcoord2}
\end{eqnarray}
where the rapidity $\eta$ is given by $\tanh \eta = v$, meaning that $\cosh \eta=\gamma$.  After the AdS black hole metric has been transformed according to this boost, it describes the moving hot wind of plasma felt by the quark--anti-quark pair in its rest frame.  

In order to extract $E(L)$ it suffices to work in the limit in which the time-like extent of the Wilson loop ${\cal T}$ is much greater than its transverse extent $L$, meaning that the corresponding string worldsheet ``suspended'' from this Wilson loop and ``hanging down'' into the bulk is invariant under translations along the long direction of the Wilson loop.
Parametrizing
the two-dimensional world sheet with the coordinates $\sigma$ and $\tau=t$, the dependence
on $\tau$ is then trivial. The task is reduced to calculating the curve $r(\sigma)$ along 
which the worldsheet descends into the bulk
from positions on the boundary brane which we take to be located at $r=r_0\Lam$, with $\Lam$ a dimensionless UV cutoff that we shall take to infinity at the end of the calculation.  That is, the boundary conditions on $r(\sigma)$ are
\begin{equation}
r \left(\pm {L \ov 2}\right) = r_0 \Lam\, .
\end{equation}
It is then helpful to introduce dimensionless variables 
\begin{equation}
 r= r_0 y, \qquad \tilde\sigma = \sigma{r_0\over R^2}, \qquad
 l = {L r_0\over R^2} = \pi L T, \label{dimensionless}
\end{equation}
 where $T = {r_0 \ov \pi  R^2}$ is the temperature. Upon dropping the tilde, 
one is then seeking to determine the shape $y(\sigma)$ of the string world sheet 
satisfying the boundary conditions $y \le(\pm {l \ov 2}\ri) = \Lam$. 
From the boosted  AdS black hole metric, one finds that the 
Nambu-Goto action which must be extremized takes the form
 \begin{equation}
S({\cal C}) = -\sqrt{\lambda}\, {\cal T}\, T\, \int_0^{l/2} d\sigma\, {\cal L}\, ,
\label{peo}
\end{equation}
with a Lagrangian that reads ($y'= \p_\sig y$)
\begin{equation}
 {\cal L} = \sqrt{\left(y^4 - {\cosh^2 \eta } \right) \left(1 + {y'^2 \over y^4
 - 1} \right)}\, .
 \label{3.20}
 \end{equation}
We must now determine $y(\sigma)$ by extremizing (\ref{3.20}).  This can
be thought of as a classical mechanics problem, with $\sigma$ the analogue
of time. Since ${\cal L}$ does not depend on $\sig$ explicitly, the
corresponding Hamiltonian
\begin{equation}
    {\cal H} \equiv {\cal L} - y' \frac{\partial {\cal L}}{\partial y'}
    = {y^4-\cosh^2 \eta \over {\cal L}} = q
    \label{3.21}
\end{equation}
is a constant of the motion, which we denote by $q$.   In the calculation we are reviewing in this section, we take $\Lambda\rightarrow\infty$ at fixed, finite, rapidity $\eta$.  In this limit, the string worldsheet in the bulk is time-like, and $E(L)$ turns out to be real.  (The string world-sheet bounded by the rectangular Wilson loop that we are considering becomes space-like if $\sqrt{\cosh\eta}>\Lambda$.  In order to recover the light-like Wilson loop used in the calculation of the jet quenching parameter in Section~\ref{sec:AdSCFTJetQuenching}, one must first take $\eta\to\infty$ and only then take $\Lam\to\infty$.)


It follows from the Hamiltonian (\ref{3.21}) that solutions $y(\sigma)$ 
with $\Lambda>\sqrt{\cosh\eta}$ satisfy the equation of motion 
 \begin{equation}
 \label{3.34}
y' = {1 \over q} \sqrt{(y^4- 1) (y^4 - y_c^4)}
 \end{equation}
with
 \begin{equation}
  \label{3.35}
y_c^4 \equiv \cosh^2 \eta + q^2.
 \end{equation}
Note that $y_c^4 > \cosh^2 \eta\ge 1$.  The extremal string world sheet
begins at $\sigma=-\ell/2$ where $y=\Lambda$, and ``descends'' in $y$ until
it reaches a turning point, namely the largest value of $y$
at which $y'=0$.  It then ``ascends'' from the turning point to its
end point at $\sigma=+\ell/2$ where $y=\Lambda$.  By symmetry,
the turning point must occur at $\sigma=0$.  We see from (\ref{3.34})
that in this case, the turning point occurs at $y=y_c$ meaning that the
extremal surface stretches between $y_c$ and $\Lambda$.
The integration constant
$q$ can then be determined from the 
equation ${l \over 2} = \int^{l \over 2}_0 d \sigma$ which, upon using (\ref{3.34}), becomes
 \begin{equation}
 \label{3.36}
{l } =
 {2 q} \int_{y_c}^{\Lambda} dy \, {1 \over
\sqrt{(y^4- y_c^4) (y^4 - 1)} }\ .
 \end{equation}
The action for the extremal surface can be found by substituting
(\ref{3.34}) into (\ref{peo}) and (\ref{3.20}), yielding
 \begin{equation}
 \label{3.37}
S(l) =-{ \sqrt{ \lambda }{\cal T}  T}   \int_{y_c}^\Lambda dy \; {
y^4 - \cosh^2\eta \over \sqrt{(y^4 -1)(y^4-y_c^4)}} \ .
 \end{equation}

Equation (\ref{3.37}) contains not only the potential between
the quark-antiquark pair but also the static mass of the quark and antiquark
considered separately
in the moving medium. (Recall that we have boosted to the rest frame of
the quark and antiquark, meaning that the quark-gluon plasma is
moving.)
Since we are only
interested in the quark-antiquark potential, we need to subtract 
the action $S_0$ of two independent quarks from (\ref{3.37}) in order 
to obtain the quark--anti-quark potential in the dipole rest frame:
 \begin{equation}
    E(L) \sT = -S(l) + S_0\, .
    \label{3.38}
\end{equation}
The string configuration corresponding to a single quark
at rest in a moving
 $\N=4$ SYM plasma was obtained in Refs.~\cite{Herzog:2006gh,Gubser:2006bz},
 as we have reviewed 
 in Section~\ref{sec:HQDrag}.  From this configuration one finds that
  \begin{equation}
 S_0 = -\sqrt{\lambda}\,  {\cal T}\, T\,  \int_{1}^{\Lam} dy \, .
 \label{3.39}
 \end{equation}
To extract the quark-antiquark potential, we use (\ref{3.36}) to
solve for $q$ in terms of $l$ and then plug the corresponding $q(l)$
into (\ref{3.37}) and (\ref{3.38}) to obtain $E(L)$. Note that~\eqref{3.36} 
is manifestly finite as $\Lam \to \infty$ and the limit can be taken directly. 
\eqref{3.37} and~\eqref{3.39} are divergent separately when taking $\Lam \to \infty$, but the difference~\eqref{3.38} is finite.


We now describe general features of~\eqref{3.36} and~\eqref{3.38}. Denoting the RHS 
of~\eqref{3.36} (with $\Lam = \infty$) as function $l(q)$, one finds that for a given $\eta$, $l(q)$ has a maximum $l_{\rm max} (\eta)$ and equation (\ref{3.36}) has no solution when
$l > l_{\rm max} (\eta)$. Thus for $l > l_{\rm max}$, the only string worldsheet configuration is two disjoint strings and from~\eqref{3.38}, $E(L) = 0$, i.e. 
the quark and anti-quark are completely screened to the order of the approximation we are considering.
We can define the screening length 
as 
\begin{equation}
L_s \equiv  {l_{\rm max} (\eta) \ov \pi T}\ . 
\end{equation}
At $\eta =0$, i.e. the dipole at rest with the medium, one finds that
 \begin{equation}
 L_s (0) \approx {0.87 \ov \pi T}  \approx  \frac{0.28}{T}\ .
 \label{screL}
 \end{equation}
Similar criteria are used in the
definition of screening length in QCD~\cite{Karsch:2006sf}, although
in QCD there is no sharply defined length scale
at which screening sets in.   Lattice calculations of the
static potential between a heavy quark
and antiquark in QCD indicate a screening length $L_s \sim 0.5/T$ in hot
QCD with two flavors of light quarks~\cite{Kaczmarek:2005ui} and
$L_s\sim 0.7/T$ in hot QCD with no dynamical
quarks~\cite{Kaczmarek:2004gv}.   The fact that there {\it is} a sharply
defined $L_s$  is an artifact of the limit
in which we are working, in which $E(L)=0$ for $L>L_s$.\footnote{We are considering the contribution to $E(L)$ that is proportional to $\sqrt{\lambda}$.  For $l\gg l_{\rm max}$, the leading contribution to $E(L)$ is proportional to $\lambda^0$ and is determined by the exchange of the lightest supergravity mode between the two disjoint strings~\cite{Bak:2007fk}.}

\begin{figure}[t]
\begin{center}
\includegraphics[scale=0.55,angle=0]{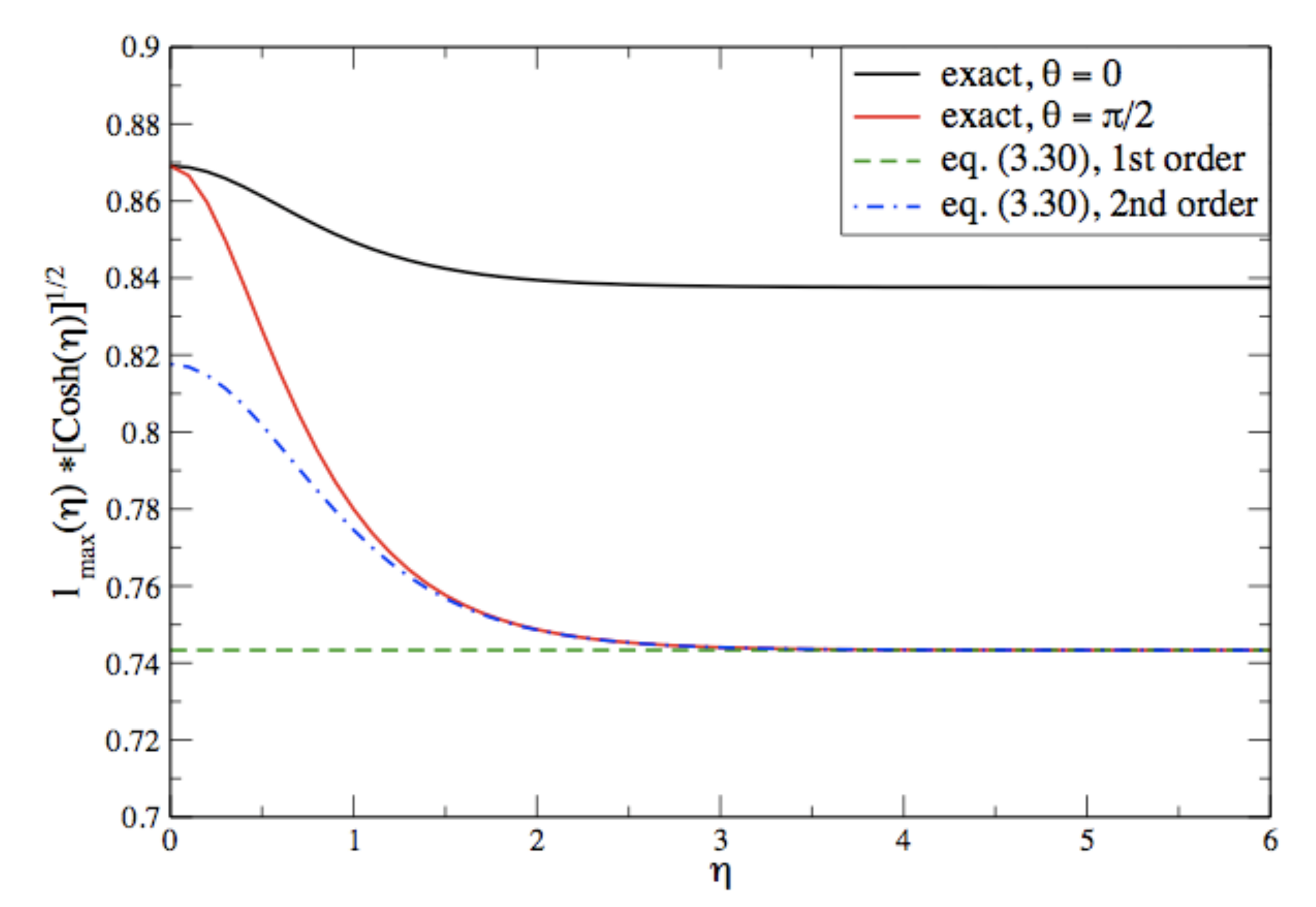}
\end{center}
\caption{\small The screening length $l_{\rm max}$ times its leading large-$\eta$
dependence $\sqrt{\cosh(\eta)}$. The exact results are given for dipoles oriented
perpendicular to the wind ($\theta=\pi/2$) and parallel to the wind ($\theta=0$).
The $\theta=\pi/2$ curve is compared to the analytical large-$\eta$ approximation
(\protect\ref{3.41}). Keeping only the first term in this analytical expression corresponds
to a horizontal line on the figure; including the term proportional to $(\cosh \eta)^{-5/2}$
improves the agreement with the exact result.
\label{fig6}
}\end{figure}

$L_s(\eta)$ can be obtained numerically, with results as illustrated in Fig.~\ref{fig6}.
One finds that $L_s(\eta)$ decreases with increasing velocity, indicating that quarkonia dissociate at a lower temperature when they are moving.
An analytical expression for $L_s$ can also be obtained in the limit of high velocity. 
Expanding (\ref{3.36}) in powers of $1/y_c^4$ and truncating the corresponding
expression at order $1/y_c^4$, one finds 
%
%
%
\bea
    l_{\rm max} &=& \frac{\sqrt{2\pi}\, \Gamma\left(\frac{3}{4}\right)}{
                             3^{3/4}\Gamma\left(\frac{1}{4}\right)}
                             \left( \frac{2}{\cosh^{1/2}\eta} + \frac{1}{5\cosh^{5/2}\eta}
                             + \cdots \right)
                            \nonumber \\
                         &=& 0.74333 \left( \frac{1}{\cosh^{1/2}\eta} + \frac{1}{10\, \cosh^{5/2}\eta}
                         +  \cdots \right)\ .
                          \label{3.41}
\eea
In Fig.~\ref{fig6}, exact results for $l_{\rm max}= \pi T L_s (\eta)$ as a function of $\eta$, obtained numerically,
are compared to the expression (\ref{3.41}) that is valid in the high velocity limit. One sees that the screening length decreases with increasing 
velocity to a good approximation according to the 
scaling~\cite{Peeters:2006iu,Liu:2006nn,Chernicoff:2006hi} 
\begin{equation}
L_s(v) \simeq \frac{L_s(0)}{\cosh^{1/2}\eta} =
\frac{L_s(0)}{\sqrt{\gamma}} ,
\label{velocityscaling}
\end{equation}
with $\gamma=1/\sqrt{1-v^2}$.
This velocity dependence suggests that
$L_s$ should be thought of
as, to a good approximation,  proportional to
(energy density)$^{-1/4}$, since the energy density increases like $\gamma^2$
as the wind  velocity is boosted.  
The velocity-scaling of $L_s$ has 
proved robust in the sense that it applies in various strongly
coupled plasmas other than
${\cal N}=4$ SYM~\cite{Avramis:2006em,Caceres:2006ta,Natsuume:2007vc,Liu:2008tz}  and
in the sense that it applies to baryons made of heavy quarks also~\cite{Athanasiou:2008pz}.


We have only described the calculation for the case in which the direction of the hot wind is perpendicular to the dipole.
With a little more effort, the calculation can be extended to general angles and 
has been analyzed in detail in~\cite{Liu:2006he}, to which we refer the readers for more details. The conclusion of~\cite{Liu:2006he} was that the dependences of $L_s$ and $E(L)$ on the angle between the dipole and the wind are very weak. For example, the black line in Fig.~\ref{fig6} gives 
the $\eta$ dependence of the screening length when the wind direction is parallel to the dipole. We see the difference from the perpendicular case is only about $12\%$. 

\begin{figure}[t]
\begin{center}
\includegraphics[scale=0.55,angle=0]{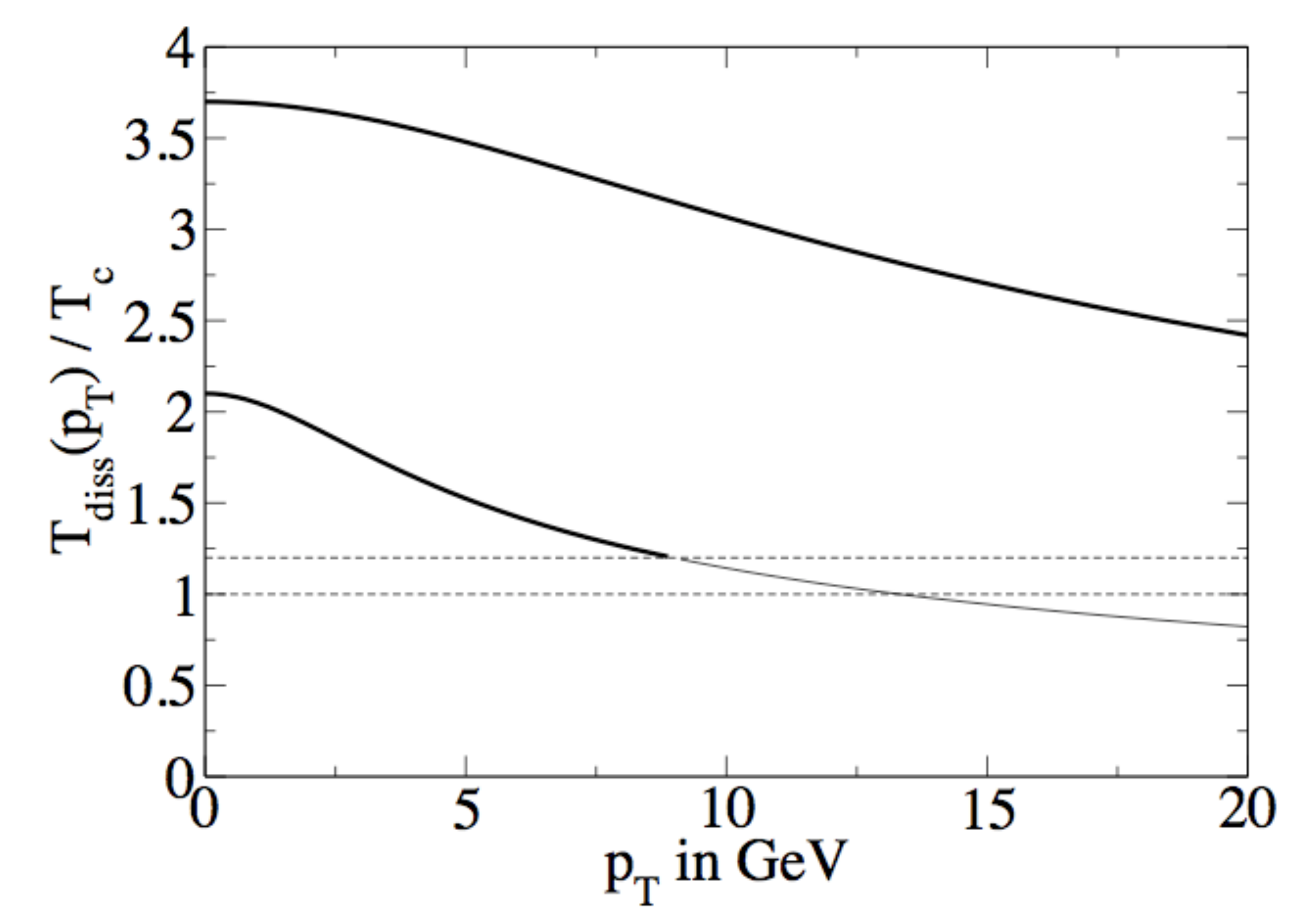}
\end{center}
\caption{\small A $1/\sqrt{\gamma}$-velocity scaling of the screening length
in QCD would imply a $J/\Psi$ dissociation temperature 
$T_{\rm diss}(p_T)$ that decreases significantly with $p_T$, while
that for the heavier $\Upsilon$ is affected less at a given $p_T$.  
The curves are schematic, in that we have arbitrarily taken $T_{\rm diss}(0)$ 
for the $J/\Psi$ to be $2.1\, T_c$ and 
we have increased $T_{\rm diss}(0)$ for the $\Upsilon$ over that for the $J/\psi$ 
by a factor corresponding to its smaller size in vacuum.  At a qualitative level, we expect to see fewer $J/\Psi$ ($\Upsilon$) mesons at $p_T$'s above that at which their dissociation temperature
is comparable to the temperatures reached in heavy ion collisions at RHIC (at the LHC).
}
\label{Fig:jpsidiss}
\vskip-0.05in
\end{figure}

If the velocity-scaling of $L_{s}$~(\ref{velocityscaling}) holds for QCD,
it will have qualitative consequences for quarkonium suppression
in heavy ion collisions~\cite{Liu:2006nn,Liu:2006he}.  From~\eqref{dis},
the dissociation temperature $T_{\rm diss}(v)$, defined as the temperature above which
$J/\Psi$ or $\Upsilon$ mesons with a given velocity do not exist, should scale with velocity as 
 \be \label{rro}
 T_{\rm diss} (v) \sim  T_{\rm diss} (v=0) (1-v^2)^{1/4} \ ,
  \ee
since $T_{\rm diss}(v)$ should be the temperature at which the screening length $L_s(v)$  is comparable to the size of the meson bound state. The scaling (\ref{rro}) indicates that slower mesons can exist up to higher temperatures than faster ones.
As illustrated schematically in Fig.~\ref{Fig:jpsidiss}, this scaling indicates that $J/\Psi$
suppression at RHIC (and $\Upsilon$ suppression at the LHC)
may increase markedly for $J/\Psi$'s ($\Upsilon$'s) with
transverse momentum $p_T$ above some threshold, on the assumption
that the temperatures reached in RHIC (LHC) collisions do 
not reach the dissociation temperature
of $J/\Psi$ ($\Upsilon$) mesons at zero
velocity~\cite{Karsch:2005nk,Satz:2005hx}.   Modelling this effect requires embedding results for quarkonium production in hard scatterings in nuclear collisions into a hydrodynamic code that describes the motion of the quark-gluon fluid produced in the collision, in order to evaluate the velocity of the hot wind felt by each putative quarkonium meson.  Such an analysis indicates that  
once $p_T$ is above the threshold at which $T_{\rm diss}(v)$ has dropped below the temperature reached in the collision, the decline in the $J/\Psi$ survival probability is significant, by more than a factor of four (two) in central (peripheral) collisions~\cite{Gunji:2007uy,Gunji:2007uyProceeding}, .
We should 
caution that, as we discussed in Section~\ref{quarkonium}, in modelling quarkonium production and suppression
versus $p_T$ in heavy ion collisions, various other effects like
secondary production or formation of $J/\Psi$ mesons outside the
hot medium at high $p_T$~\cite{Karsch:1987zw} remain to be
quantified.  The quantitative importance of these and other
effects may vary significantly, depending on details of their
model implementation. In contrast, Eq. (\ref{rro}) was obtained
directly from a field-theoretic calculation and its implementation
will not introduce additional model-dependent uncertainties.

The threshold $p_T$ above which the production of $J/\Psi$ mesons falls off due to their motion through the quark-gluon plasma
depends sensitively on the difference between $T({\rm diss}(v=0)$ and the temperature reached in the collision~\cite{Gunji:2007uyProceeding}, but should be somewhere around 5 GeV; perhaps a conservative range is 2 to 10 GeV.  Present data on the suppression of $J/\Psi$ production in CuCu collisions at RHIC show no signs of increased suppression above some $p_T$ 
threshold~\cite{Abelev:2009qaa,Atomssa:2009ek}, 
but the error bars are large above $p_T\sim 4$~GeV.
The kinematical range
in which this novel quarkonium suppression mechanism is
operational lies within experimental reach of near-future
high-luminosity AuAu runs at RHIC and will be studied thoroughly at the
LHC in both the $J/\Psi$ and $\Upsilon$ channels. 

The analysis of this section is built upon the calculation of the potential between a test quark and antiquark in the strongly coupled plasma of ${\cal N}=4$ SYM theory, a theory which in and of itself has no mesons.    Gaining insight into the physics of quarkonium mesons from calculations of the screening of the static quark-antiquark potential has a long history in QCD, as we have seen in 
Section~\ref{secmeson}.  But, we have also seen in that section that these approaches are gradually being superseded as lattice QCD calculations of quarkonium spectral functions themselves are becoming available.  In the present context also, we would like to move beyond drawing inferences about mesons from analyses of the potential $E(L)$ and the screening length $L_s$ to analyses of mesons themselves.  This is the subject of Section~\ref{mesons}, in which we shall carefully describe how once we have added heavy quarks to ${\cal N}=4$ SYM by adding a D7-brane in the gravity dual~\cite{Karch:2002sh},  as in Section~\ref{fundamental}, the fluctuations of the D7-brane then describe the quarkonium mesons of this 
theory.  We shall review the construction first in vacuum and then in the presence of the strongly coupled plasma at nonzero temperature.  We shall find that the results of this section prove robust, in that the velocity-scaling  (\ref{rro})
has also been obtained~\cite{Ejaz:2007hg}
by direct analysis of the dispersion relations
of mesons in
the plasma~\cite{Mateos:2007vn,Ejaz:2007hg}.
These mesons have a limiting velocity that is less than the speed of light and that decreases with increasing temperature~\cite{Mateos:2007vn}, and 
whose temperature dependence
is equivalent to (\ref{rro}) up to few percent corrections that have been
computed~\cite{Ejaz:2007hg} and that we shall show.    This is a key part of the story, with the velocity-dependent dissociation temperature of this section becoming a temperature dependent limiting velocity for explicitly constructed quarkonium mesons in Section~\ref{mesons}.  However, this cannot be the whole story since the dispersion relations seem to allow for mesons with arbitrarily large momentum even though they limit their velocity.  The final piece of the story is described in 
Section~\ref{sec:MesonWidths}, 
where we review the calculation of the leading contribution to the widths of these 
mesons~\cite{Faulkner:2008qk}, which was neglected in the earlier calculations of their dispersion relations.  Above some momentum, the width grows rapidly, increasing like $p_\perp^2$. And, the momentum above which this rapid growth of the meson width sets in is just the momentum at which the meson velocity first approaches its limiting value.  The physical picture that emerges is that at the momentum at which the mesons reach a velocity such that the hot wind they are feeling has a temperature sufficient to dissociate them, according to the analysis of this Section built upon the calculation of $L_s$, their widths in fact grow rapidly~\cite{Faulkner:2008qk}.

\chapter{Quarkonium mesons}
\label{mesons}

As discussed in section \ref{quarkonium}, heavy quarks and quarkonium  mesons, with masses such that $M/T \gg 1$, constitute valuable probes of the QGP. Since dynamical questions about these probes are very hard to answer from first principles, here we will study analogous questions in the strongly coupled ${\cal N}=4$ SYM plasma. In this case the gauge/string duality provides the tool that makes a theoretical treatment possible. Although for concreteness we will focus on the ${\cal N}=4$ plasma, many of the results that we will obtain are rather universal in the sense that, at least qualitatively, they hold for any strongly coupled gauge theory with a string dual.  Such results may give us insights relevant for the QCD quark-gluon plasma at temperatures at which it is reasonably strongly coupled.

The QGP only exists at temperatures $T > \tc$, so in QCD the condition 
$M/T \gg 1$ can only be realised by taking $M$ to be large. In contrast, ${\cal N}=4$ SYM is a conformal theory with no confining phase, so all temperatures are equivalent. In the presence of an additional scale, namely the quark or the meson mass, the physics only depends on the ratio $M/T$. This means that in the ${\cal N}=4$ theory the condition $M/T \gg 1$ can be realised by fixing $T$ and sending $M$ to infinity, or by fixing $M$ and sending $T \ra 0$; both limits are completely equivalent. In particular, the leading-order approximation to the heavy quark or quarkonium meson physics, in an expansion in $T/M$, may be obtained by setting $T=0$. For this reason, this is the limit that we will study first.

We will follow the nomenclature common in the QCD literature and refer to mesons made of two heavy quarks as `quarkonium mesons' or `quarkonia', as opposed to using the term `heavy mesons', which commonly encompasses mesons made of one heavy and one light quark.

\section{Adding quarks to ${\cal N}=4$ SYM}
\label{sec:AddingQuarks}
In section \ref{fundamental} we saw that $\nf$ flavours of fundamental matter can be added to ${\cal N}=4$ SYM by introducing $\nf$ D7-brane probes into the geometry sourced by the D3-branes, as indicated by the array \eqn{D3D7}, which we reproduce here (with the time direction included) for convenience: 
\be
\begin{array}{rccccccccccl}
\mbox{D3:}\,\,\, & 0 & 1 & 2 & 3 & \_ & \_ & \_ & \_ & \_ & \_ & \, \\
\mbox{D7:}\,\,\, & 0 & 1 & 2 & 3 & 4 & 5 & 6 & 7 & \_ & \_ & \,. 
\end{array}
\label{D3D7bis}
\ee

Before we proceed, let us clarify an important point. ${\cal N}=4$ SYM is a conformal theory, \ie its $\beta$-function vanishes exactly. Adding matter to it, even if the matter is massless, makes the quantum-mechanical $\beta$-function positive, at least perturbatively. This means that the theory develops a Landau pole in the UV and is therefore not well defined at arbitrarily-high energy scales.\footnote{Non-perturbatively, the possibility that a strongly coupled fixed point exists must be ruled out before reaching this conclusion. See \cite{D'Hoker:2010mk} for an argument in this direction based on supersymmetry.} However, since the beta-function (for the 't Hooft coupling 
$\lambda$) is proportional to $\nf/\nc$, the Landau pole occurs at a scale of order $e^{\nc/\nf}$. This is exponentially large in the limit of interest here, $\nf/\nc \ll 1$, and in fact the Landau pole disappears altogether in the strict probe limit $\nf/\nc \rightarrow 0$. On the string side, the potential pathology associated with a Landau pole manifests itself in the fact that a completely smooth solution that incorporates the backreaction of the D7-branes may not exist \cite{Aharony:1998xz,Grana:2001xn,Burrington:2004id,Kirsch:2005uy,Benini:2006hh,D'Hoker:2010mk}. In any case, the possible existence of a Landau pole at high energies will not be of concern for the applications reviewed here. In the gauge theory, it will not prevent us from extracting interesting infrared physics, just as the existence of a Landau pole in QED does not prevent one from calculating the conductivity of an electromagnetic plasma. In the string description, we will not go beyond the probe approximation, so the backreaction of the D7-branes will not be an issue.\footnote{For a review of `unquenched' models, \ie those in which the flavour backreaction is included, see \cite{Nunez:2010sf}.}  And finally, we note that we will work with the D3/D7 model because of its simplicity. We could work with a more sophisticated model with better UV properties, but this would make the calculations more involved while leaving the physics we are interested in essentially unchanged.

\begin{figure}
    \begin{center}
    	\includegraphics[width=0.37\textwidth]{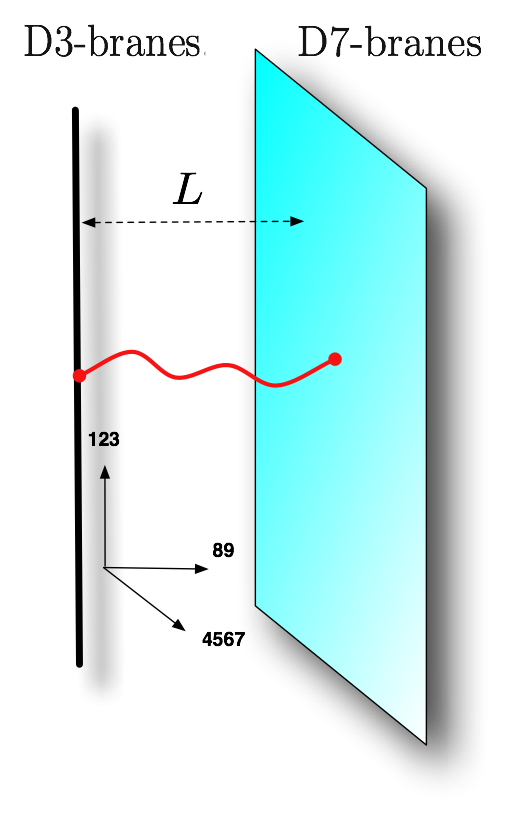}
    \end{center}
    \caption{\small D3/D7 system at weak coupling.}
\label{D3D7orientation}
\end{figure}

As illustrated in Fig.~\ref{D3D7orientation}, the D3-branes and the D7-branes can be separated a distance $L$ in the 89-directions.
This distance times the string tension, Eqn.~\eqn{tension}, is the minimum energy of a string stretching between the D3- and the D7-branes. Since the quarks arise as the lightest modes of these 3-7 strings, this energy is precisely the bare quark mass:
\be
M_\mt{q} = \frac{L}{2\pi \alpha'} \,.
\label{mq}
\ee
An important remark here is the fact that the branes in 
Fig.~\ref{D3D7orientation} are implicitly assumed to be embedded in flat spacetime. In Section~\ref{fundamental} this was referred to as the `first' or `open-string' description of the D3/D7 system, which is reliable in the regime $g_s \nc \ll 1$, in which the backreaction of the D3-branes on spacetime can be ignored. One of our main tasks in the following sections will be to understand how this picture is modified in the opposite regime, $g_s \nc \gg 1$, when the D3-branes are replaced by their backreaction on spacetime. In this regime the shape of the D7-branes may or may not be modified, but Eqn.~(\ref{mq}) will remain true provided the appropriate definition of $L$, to be given below, is used.

Although ${\cal N}=4$ SYM is a conformal theory, the addition of quarks with a nonzero mass introduces a scale and gives rise to a rich spectrum of quark-antiquark bound states, \ie mesons. In the following section we will study the meson spectrum in this theory at zero temperature in the regime of  strong `t Hooft coupling, $g_s \nc \gg 1$. On the gauge theory side this is inaccessible to conventional methods such as perturbation theory, but on the string side a classical  description in terms of D7-brane probes in a weakly curved $AdS_5 \times S^5$ applies. Our first task is thus to understand in more detail the way in which the D7-branes are embedded in this geometry. Since this is crucial for subsequent sections, we will in fact provide a fair amount of detail here.

\section{Zero temperature}
\label{zero}

\subsection{D7-brane embeddings}
\label{sec:D7atZeroT}

We begin by recalling that the coordinates in the $AdS_5 \times S^5$ metric \eqn{metric10D}, \eqn{AdSmetric} can be understood as follows. The four directions $t, x_i$ correspond to the 0123-directions in \eqn{D3D7bis}. The 456789-directions in the space transverse to the D3-branes give rise to the radial coordinate $r$ in $AdS_5$, defined through 
\be
r^2 = x_4^2 + \cdots + x_9^2 \,,
\ee
as well as five angles that parametrise the $S^5$. We emphasize that, once the gravitational effect of the D3-branes is taken into account, the 6-dimensional space transverse to the D3-branes is not flat, so the $x^4, \ldots, x^9$ coordinates are not Cartesian coordinates. However, they are still useful to label the different directions in this space.

The D7-branes share the 0123-directions with the D3-branes, so from now on we will mainly focus on the remaining directions. In the 6-dimensional space transverse to the D3-branes, the D7-branes span only a 4-dimensional subspace parametrised by $x_4, \ldots, x_7$. Since the D7-branes preserve the $SO(4)$ rotational symmetry in this space, it is convenient to introduce a radial coordinate $u$ such that 
\be
u^2 = x_4^2 + \cdots + x_7^2 \,,
\label{u}
\ee
as well as three spherical coordinates, denoted collectively by  $\Omega_3$, that parametrise an $S^3$. Similarly, it is useful to introduce a radial coordinate $U$ in the 89-plane through 
\be
U=x_8^2 + x_9^2 \,,
\label{U}
\ee
as well as a polar angle $\alpha$. In terms of these coordinates one has
\be
dx_4^2 + \cdots + dx_9^2 = du^2 + u^2 d\Omega_3^2 + dU^2 + U^2 d\alpha^2 \,.
\ee
Obviously, the overall radial coordinate $r$ satisfies $r^2 = u^2 + U^2$.

Since the D7-branes only span the 4567-directions, they only wrap an $S^3$ inside the $S^5$. The D7-brane worldvolume may thus be parametrized by the coordinates 
$\{t, x_i, u, \Omega_3\}$. In order to specify the D7-branes' embedding one must then specify the remaining spacetime coordinates, $U$ and $\alpha$, as functions of, in principle, all the worldvolume coordinates. However, translational symmetry in the $\{t, x_i\}$-directions and rotational symmetry in the $\{\Omega_3\}$-directions allow $U$ and $\alpha$ to depend only on $u$.  

\begin{figure}
    \begin{center}
    	\includegraphics[width=0.8\textwidth]{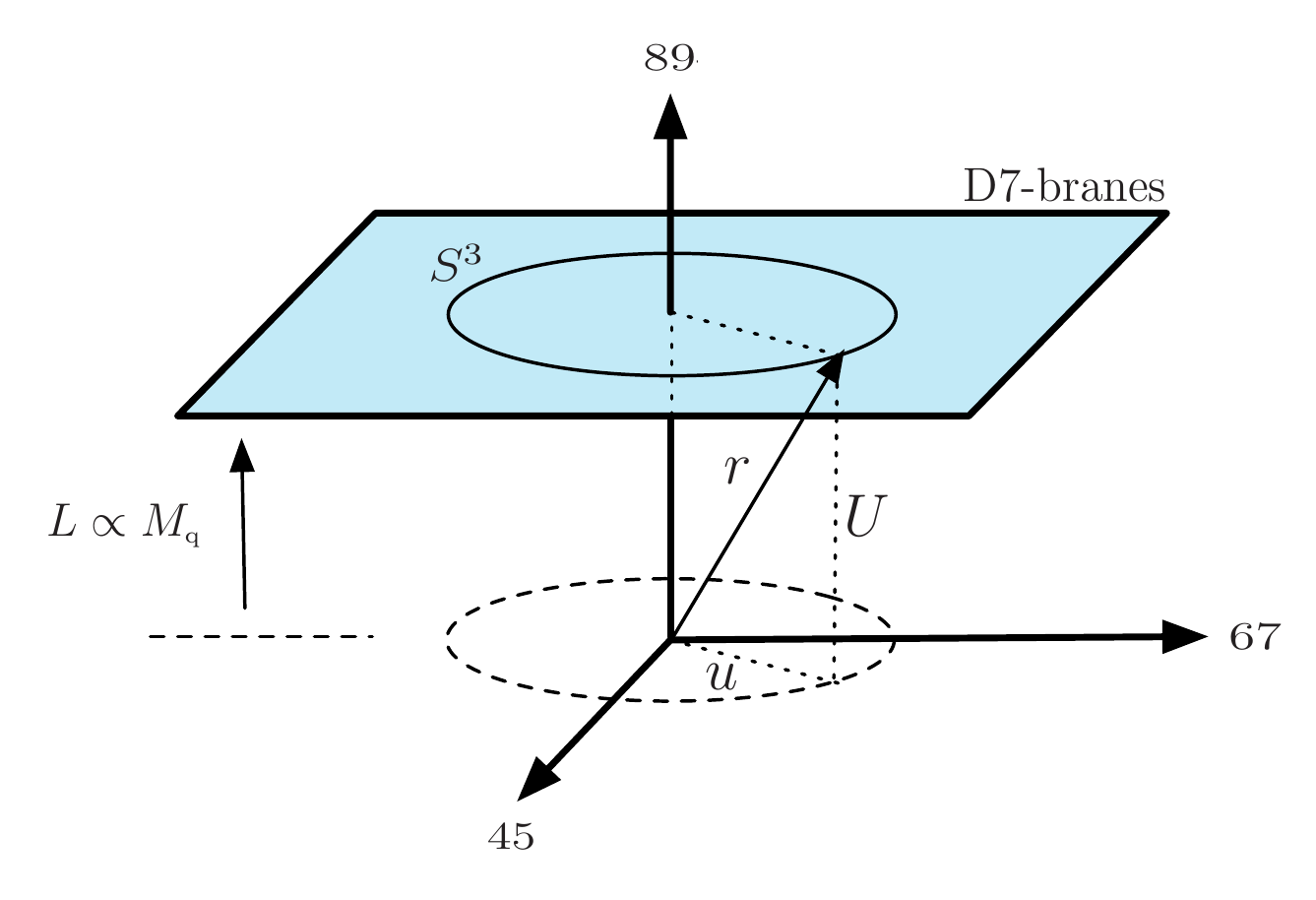}
    \end{center}
    \caption{\small Coordinates in the 6-dimensional space transverse to the D3-branes. Each axis actually represents two directions, $\ie$ a plane (or, equivalently, the radial direction in that plane). The asymptotic distance $L=U(u=\infty)$ is proportional to the quark mass, Eqn.~\eqn{mq}. We emphasize that the directions parallel to the D3-branes (the gauge theory directions $t, x_i$) are suppressed in this picture, and they should not be confused with the D7 directions shown in the figure, which lie entirely in the space transverse to the D3-branes.}
\label{Uur}
\end{figure}

In order to understand this dependence, consider first the case in which the spacetime curvature generated by the D3-branes is ignored. In this case, the D7-branes lie at a constant position in the 89-plane, see Fig.~\ref{Uur}. 
In other words, their embedding is given by $\alpha(u)=\alpha_0$ and $U(u)=L$, where  
$\alpha_0$ and $L$ are constants. The first equation can be understood as saying that, because of the $U(1)$ rotational symmetry in the 89-plane, the D7-branes can sit at any constant angular position; choosing $\alpha_0$ then breaks the symmetry. Since this $U(1)$ symmetry is respected by the D3-branes' backreaction (\ie since the $AdS_5 \times S^5$ metric is $U(1)$-invariant), it is easy to guess (correctly) that $\alpha(u)=\alpha_0$ is still a solution of the D7-branes' equation of motion in the presence of the D3-branes' backreaction. 

\begin{figure}
    \begin{center}
    	\includegraphics[width=0.7\textwidth]{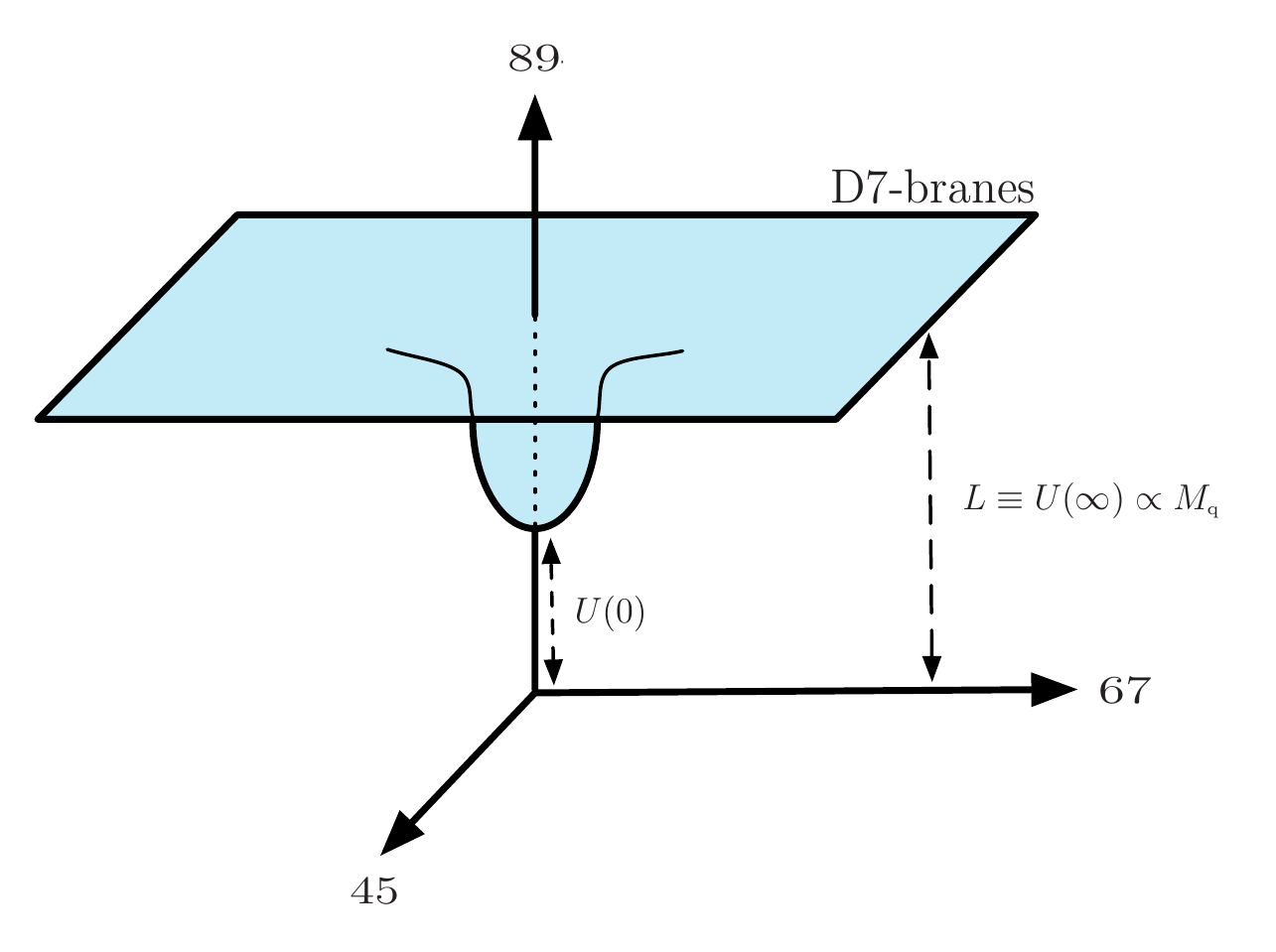}
    \end{center}
    \caption{\small Possible bending of the D7-branes at nonzero temperature. The asymptotic distance $L\equiv U(\infty)$ is proportional to the bare quark mass $\mq$, whereas the minimum distance $U(0)$ is related (albeit in a way more complicated than simple proportionality) to the quark thermal mass.}
\label{bending}
\end{figure}

The second equation, $U(u)=L$, says that the D7-branes lie at a constant distance from the D3-branes. In the absence of the D3-branes' backreaction this is easily understood: there is no force on the D7-branes and therefore they span a perfect 4-plane. In the presence of backreaction, one should generically expect that the spacetime curvature deforms the D7-branes as in Fig.~\ref{bending}, bending them towards the D3-branes at the origin. 
The reason that this does not happen for the D3/D7 system at zero temperature is that the underlying supersymmetry of the system guarantees an exact cancellation of forces on the D7-branes. In fact, it is easy to verify directly that $U(u)=L$ is still an exact solution of the D7-branes' equations of motion in the presence of the D3-branes' backreaction. The constant 
$L$ then determines the quark mass through Eqn.~\eqn{mq}. We will see below that the introduction of nonzero temperature breaks supersymmetry completely, and that consequently $U(u)$ becomes a non-constant function that one must solve for, and that this function contains information about the ground state of the theory in the presence of quarks. For example, its asymptotic behaviour encodes the value of the bare quark mass $\mq$ and the quark condensate $\qc$, whereas its value at $u=0$ is related to the quark thermal mass $\mth$. Since in this section we work at $T=0$, any nonzero quark mass corresponds to $\mq/T \to \infty$. In this sense one must think of the quarks in question as the analogue of heavy quarks in QCD, and of the  quark condensate  as the analogue of  $\langle \bar{c} c \rangle$ or 
$\langle \bar{b} b \rangle$. However, when we consider a nonzero temperature in subsequent sections, whether the holographic quarks described by the D7-branes are the analogues of heavy or light quarks in QCD will depend on how their mass (or, more precisely, the mass of the corresponding mesons) compares to the temperature. 

\begin{figure}
    \begin{center}
    	\includegraphics[width=1\textwidth]{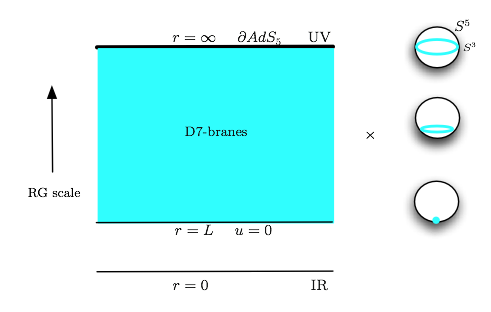}
    \end{center}
    \caption{\small D7-branes' embedding in $AdS_5 \times S^5$. At nonzero temperature this picture is slightly modified: First, a horizon appears at $r = r_0 >0$, and second, the D7-branes terminate at $r=U(0) < L$. This `termination point' corresponds to the tip of the branes in 
    Fig.~\ref{bending}.} 
\label{AdSpictureD7}
\end{figure}

We have concluded that, at zero temperature, the D7-branes lie at $U=L$ and are parametrised by $\{t, x_i, u, \Omega_3\}$. In terms of these coordinates, the induced metric on the D7-branes takes the form
\be
ds^2 = \fc{u^2+L^2}{R^2} \left( -dt^2 + dx_i^2 \right) +
\frac{R^2}{u^2+L^2} du^2 + 
\frac{R^2 u^2}{u^2+L^2} d\Omega_3^2 \,.
\label{indmetric}
\ee
We see that if $L=0$ then this metric is exactly that of $AdS_5 \times S^3$. The $AdS_5$ factor suggests that the dual gauge theory should still be conformally invariant. This is indeed the case in the limit under consideration: If $L=0$ the quarks are massless and the theory is classically conformal, and in the probe limit $\nf/\nc \rightarrow 0$ the quantum mechanical 
$\beta$-function, which is proportional to $\nf/\nc$, vanishes. If $L \neq 0$ then the metric above becomes $AdS_5 \times S^3$ only asymptotically, i.e.~for $u \gg L$, reflecting the fact that in the gauge theory conformal invariance is explicitly broken by the quark mass 
$\mq \propto L$, but is restored asymptotically at energies $E \gg \mq$. We also note that, if $L \neq 0$, then the radius of the three-sphere is not constant, as displayed in fig.~\ref{AdSpictureD7}; 
in particular, it shrinks to zero at $u=0$ (corresponding to $r=L$), at which point the D7-branes `terminate' from the viewpoint of the projection on $AdS_5$ \cite{Karch:2002sh}. In other words, if $L \neq 0$ then the D7-branes fill the $AdS_5$ factor of the metric only down to a minimum value of the radial direction proportional to the quark mass. As we anticipated above, at nonzero temperature one must distinguish between the bare and the thermal quark masses, related to $U(\infty)$ and $U(0)$ respectively. In this case the position in AdS at which the D7-branes terminate is $r=U(0) < L$, and therefore they fill the AdS space down to a radial position related to the thermal mass. Note also that at finite temperature a horizon is present at $r = r_0 >0$.

\subsection{Meson spectrum}
\label{zerospectrum}

We are now ready to compute the spectrum of low-spin mesons in the D3/D7 system following Ref.~\cite{Kruczenski:2003be}. The spectrum for more general D$p$/D$q$ systems was computed in \cite{Arean:2006pk,Myers:2006qr,Ramallo:2006et}.  Recall that mesons are described by open strings attached to the D7-branes. In particular, spin-zero and spin-one mesons correspond to the scalar and vector fields on the D7-branes. Large-spin mesons can be described as long, semi-classical strings  \cite{Kruczenski:2003be,Kruczenski:2004me,Paredes:2004is,Peeters:2005fq,Cotrone:2005fr,Peeters:2006iu,Bando:2006sp,Bigazzi:2006jt,Park:2009nb,AliAkbari:2009pf}, but we will not review them here.

For simplicity, we will focus on scalar mesons. Following Section \ref{sec:norma}, we know that in order to determine the spectrum of scalar mesons, we need to determine the spectrum of normalisable modes of small fluctuations of the scalar fields on the D7-branes. At this point we restrict ourselves to a single D7-brane, \ie we set $\nf = 1$, in which case  the dynamics is described by the DBI action \eqn{DBI}.
At leading order in the large-$\nc$ expansion, the spectrum for $\nf>1$ consists of $\nf^2$ identical copies of the single-flavour spectrum \cite{Kruczenski:2003uq}. 

We use the coordinates in Eqn.~\eqn{indmetric} as worldvolume coordinates for the D7-brane, which we collectively denote by $\sigma^\mu$. The physical scalar fields on the D7-brane are then $x^8 (\sigma^\mu)$, $x^9 (\sigma^\mu)$. By a rotation in the 89-plane we can assume that, in the absence of fluctuations, the D7-brane lies at $x^8 = 0,\, x^9 = L$. Then the fluctuations can be parametrised as
\be
x^8 = 0 +  \vp (\sigma^\mu) \ , \qquad 
x^9=L+ \tilde{\varphi} (\sigma^\mu)\ , 
\label{embed}
\ee
with $\vp$ and $\tilde{\varphi}$ the scalar fluctuations around the fiducial embedding. In order to determine the normalisable modes, it suffices to work to quadratic order in $\vp, \tilde{\varphi}$. Substituting \eqn{embed} in the DBI action \eqn{DBI} and expanding in $\vp, \tilde{\varphi}$ leads to a quadratic Lagrangian  whose corresponding  equation of motion is 
\be
\frac{R^4}{(u^2+L^2)^2} \Box \vp
+\frac{1}{u^3}\partial_u(u^3\partial_u\vp)
+\frac{1}{u^2}\nabla^2 \vp= 0\,, \label{mex}
\ee
where $\Box$ is the four-dimensional d'Alembertian associated with the Cartesian coordinates $t,x_i$, and $\nabla^2$ is the Laplacian on the three-sphere. The equation for 
$\tilde{\varphi}$ takes exactly the same form. Modes that transform non-trivially under rotations on the sphere correspond to mesons that carry nonzero R-charge. Since QCD does not possess an R-symmetry, we will restrict ourselves to R-neutral mesons, meaning that we will assume that $\vp$ does not depend on the coordinates of the sphere. We can use separation of variables to write these modes as
\be
\vp = \phi(u) e^{iq \cdot x} \,, \label{wavefunk}
\ee
where $x=(t,x_i)$. Each of these modes then corresponds to a physical meson state in the gauge theory with a well defined four-dimensional mass given by its eigenvalue under $\Box$, that is, $M^2= -q^2$. For each of these modes, Eqn.~\eqn{mex} results in an equation for $\phi(u)$ that, after introducing dimensionless variables through
\be
\bar{u} = \frac{u}{L}\ , \qquad \bar{M}^2 = -\frac{k^2R^4}{L^2}\ ,
\label{redefinition}
\ee
becomes
\be
\partial_{\bar{u}}^2\phi+\frac{3}{\bar{u}}\, \partial_{\bar{u}} \phi +\frac{\bar{M}^2}{(1+\bar{u}^2)^2}\,
\phi=0\ .
\label{meor}
\ee
This equation can be solved in terms of hypergeometric functions. The details can be found in Ref.~\cite{Kruczenski:2003be}, but we will not give them here because most of the relevant physics can be extracted as follows.

\begin{figure}
    \begin{center}
    	\includegraphics[width=0.4\textwidth]{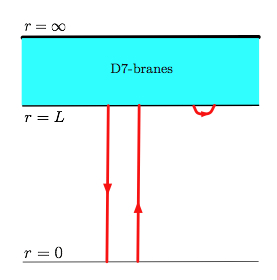}
    \end{center}
    \caption{\small String description of a quark, an antiquark and a meson. The string that describes the meson can be much shorter than those describing the quark and the antiquark.} 
\label{MesonFormation}
\end{figure}

Eqn.~\eqn{meor} is a second-order, ordinary differential equation with two independent solutions. The combination we seek must satisfy two conditions: It must be normalisable as 
$\bar{u} \rightarrow \infty$, and it must be regular as $\bar{u} \rightarrow 0$. For arbitrary values of $\bar{M}$, both conditions cannot be simultaneously satisfied. In other words, the values of $\bar{M}$ for which physically acceptable solutions exist are quantised. Since Eqn. \eqn{meor} contains no dimensionful parameters, the values of 
$\bar{M}$ must be pure numbers. These can be explicitly determined from the solutions of \eqn{meor} and they take the form \cite{Kruczenski:2003be}
\be
\bar{M}^2=4(n+1)(n+2) \sac n=0, 1, 2, \ldots \,.
\label{spectre}
\ee
Using this, and $M^2=-q^2= \bar{M}^2 L^2/R^4$, 
we derive the result that the four-dimensional mass spectrum of scalar mesons is
\be
M(n)=\frac{2L}{R^2}\sqrt{(n+1)(n+2)} 
= \frac{4\pi \mq}{\sqrt{\lambda}} \sqrt{(n+1)(n+2)} \,,
\label{spectrumEQN}
\ee
where in the last equality we have used the 
expressions $R^2/\alpha'=\sqrt{\lambda}$ 
and (\ref{mq}) to write $R$ and $L$ in terms of gauge theory parameters. We thus conclude that the spectrum consists of a discrete set of mesons with a mass gap given by the mass of the lightest meson:\footnote{In order to compare this and subsequent formulas with ref.~\cite{Kruczenski:2003be} and others, note that our definition \eqn{YM} of $\gym^2$ differs from the definition in some of those references by a factor of 2, for example
$g^2_\mt{[here]} = 2 g^2_\mt{\cite{Kruczenski:2003be}}$.}
\be
\mmes = 4 \pi \sqrt{2} \, \frac{\mq}{\sqrt{\lambda}} \,.
\label{gap}
\ee
Since this result is valid at large `t Hooft coupling, $\lambda \gg 1$, the mass of these mesons is much smaller than the mass of two constituent quarks. In other words, the mesons in this theory are very deeply bound. In fact, the binding energy 
\be  \label{benr}
E_{B} \equiv 2\mq - \mmes \lesssim 2 \mq  \sim \sqrt{\lam} M_{\rm mes} 
\ee
is so large that it almost cancels the rest energy of the quarks. This is clear from the gravity picture of `meson formation' (see Fig.~\ref{MesonFormation}), in which two strings of opposite orientation stretching from the D7-brane to $r=0$
(the quark-antiquark pair) join together to form an open string with both ends on the D7-brane (the meson). This resulting string is much shorter than the initial ones, and hence corresponds to a configuration with much lower energy.
This feature is an important difference with quarkonium mesons in QCD, such as charmonium or bottomonium, which are not deeply bound. Although this certainly means that caution must be exercised when trying to compare the physics of quarkonium mesons in holographic theories with the physics of quarkonium mesons in QCD, 
the success or failure of these comparisons cannot be assessed at this point. We will discuss this assessment in detail below, once we have learned more about the physics of holographic mesons. Suffice it to say here that some of this physics, such as the temperature or the velocity dependence of certain meson properties, turns out to be quite general and may yield insights into some of the challenges related to understanding the physics of quarkonia within the QCD quark-gluon plasma.

We close this section with a consistency check. The behaviour of the fluctuation modes at infinity is related to the high-energy properties of the theory. At high energy, we can ignore the effect of the mass of the quarks and the theory becomes conformal. 
The $u \rightarrow \infty$ behaviour is then related to the UV operator of the lowest conformal dimension, $\Delta$, that has the same quantum numbers as the meson 
\cite{Gubser:1998bc,Witten:1998qj}. Analysis of this behaviour for the solutions of Eqs.~\eqn{mex}, \eqn{meor} shows that $\Delta = 3$ \cite{Kruczenski:2003be}, as expected for a quark-bilinear operator.

\section{Nonzero temperature}
\label{nonzero}

\subsection{D7-brane embeddings}
\label{sec:D7atNonZeroT}

We now turn to the case of nonzero temperature, $T \neq 0$. This means that we must study the physics of a D7-brane in the black brane metric (cf. Eqn.~\eqn{AdSfiniteR}) 
\be
ds^2 = \frac{r^2}{R^2} \left( - f dt^2 + dx_1^2 + dx_2^2 + dx_3^2 \right) +
\frac{R^2}{r^2 f} dr^2 + R^2 d\Omega_5^2 \,,
\label{finiteT}
\ee
where
\be
f(r) = 1 - \frac{r_0^4}{r^4} \sac r_0 = \pi R^2 T \,.
\label{r0}
\ee
The study we must perform is conceptually analogous to that of the last few sections, but the equations are more involved and most of them must be solved numerically. These technical details are not very illuminating, and for this reason we will not dwell into them. Instead, we will focus on describing in detail the main conceptual points and results, as well as the physics behind them, which in fact can be understood in very simple and intuitive terms. 

\begin{figure}
	\begin{center}
		\begin{tabular}{c}
		\includegraphics[width= 0.70 \textwidth]{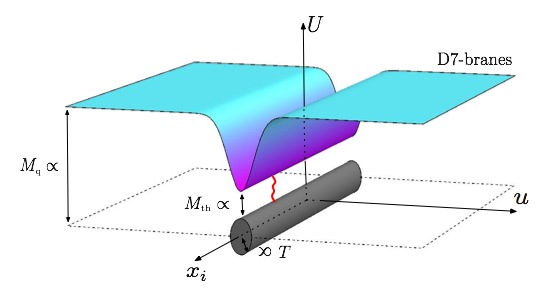} \\
		\includegraphics[width= 0.60 \textwidth]{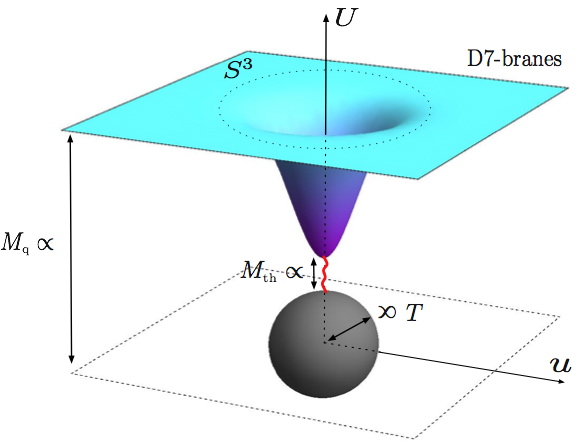}
		\end{tabular}
	\end{center}
\caption{\small D7-branes' embedding for small $T/\mq$. The branes bend towards the horizon, shown in dark grey. The radius of the horizon is proportional to its Hawking temperature, which is identified with the gauge theory temperature $T$ --- see Eqn.~\eqn{r0}. The asymptotic position of the D7-branes is proportional to the bare quark mass, $\mq$. The minimum distance between the branes and the horizon is related to the thermal quark mass, because this is the minimum length of a string (shown as a red wiggly line) stretching between the branes and the horizon. The top figure shows the two relevant radial directions in the space transverse to the D3-branes, $U$ and $u$ (introduced in Eqs.~(\ref{u}) and (\ref{U})), together with the gauge theory directions $x_i$ (time is suppressed). The  horizon has topology $\bbr{3} \times S^5$, where the first factor corresponds to the gauge-theory directions. This `cylinder-like' topology is manifest in the top figure. Instead, in the bottom figure the gauge theory directions are suppressed and the $S^3$ wrapped by the D7-branes in the space transverse to the D3-branes is shown, as in Figs.~\ref{Uur} and \ref{bending}. In this figure only the $S^5$ factor of the horizon is shown.}
\vspace{2mm}
\label{BendingWithBH}
\end{figure}

As mentioned above, at $T \neq 0$ all supersymmetry is broken. We therefore expect that the D7-branes will be deformed by the non-trivial geometry. In particular, the introduction of nonzero temperature corresponds, in the string description, to the introduction of a black brane in the background. Intuitively, we expect that the extra gravitational attraction will bend the D7-branes towards the black hole. This simple conclusion, which was anticipated in previous sections, has far-reaching consequences. At a qualitative level, most of the holographic physics of mesons in a strongly coupled plasma follows from this conclusion. An example of the D7-branes' embedding for a small value of $T/\mq$ is depicted in two slightly different ways in Fig.~\ref{BendingWithBH}.

\begin{figure}
    \begin{center}
    	\includegraphics[width=0.95\textwidth]{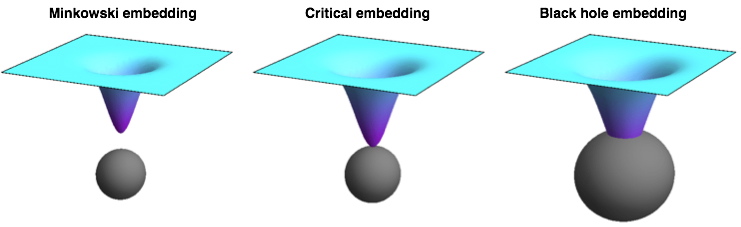}
    \end{center}
    \caption{\small Various D7-brane configurations in a black D3-brane
background with increasing temperature from left to right. At
low temperatures, the probe branes close off smoothly above
the horizon. At high temperatures, the branes fall through the
event horizon. In between, a critical solution exists in which the
branes just `touch' the horizon at a point. The critical configuration 
is never realized: a first-order phase transition occurs from a Minkowski to a 
black hole embedding (or vice versa) before the critical solution is reached.}
\vspace{2mm}
\label{transition}
\end{figure}

The qualitative physics of the D3/D7 system as a function of the dimensionless ratio $T/\mq$ is now easy to guess, and is captured by Fig.~\ref{transition}.
At zero temperature the horizon has zero size and the D7-branes span an exact hyperplane. At nonzero but sufficiently small $T/\mq$, the gravitational attraction from the black hole pulls the branes down but the branes' tension can still compensate for this. The embedding of the branes is thus deformed, but the branes remain entirely outside the horizon. Since in this case the induced metric on the D7-branes has no horizon, we will call this type of configuration a `Minkowski embedding'. In contrast, above a critical temperature $\td$,\footnote{The reason for the subscript will become clear shortly.} the gravitational force overcomes the tension of the branes and these are pulled into the horzion. In this case the induced metric on the branes possesses an event horizon, inherited from that of the spacetime metric. For this reason we will refer to such configurations as `black hole embeddings'. Between these two types of embeddings there exists a so-called critical embedding in which the branes just `touch the horizon at a point'. The existence of such an interpolating solution might lead one to suspect that the phase transition between Minkowski and black hole embeddings is continuous, i.e.~of second or higher order. However, as we will see in the next section, thermodynamic considerations reveal that a first-order phase transition occurs between a Minkowski and a black hole embedding. In other words, the critical embedding is skipped over by the phase transition, and near-critical embeddings turn out to be metastable or unstable.

As illustrated by the figures above, the fact that the branes  bend towards the horizon implies that the asymptotic distance between the two differs from their minimal distance. As we will see in Section \ref{sec:ThermoD7}, the asymptotic distance is proportional to the microscopic or `bare' quark mass, since it is determined by the non-normalizable mode of the field that describes the branes' bending. In contrast, the minimal distance between the branes and the horizon includes thermal (and quantum) effects, and for this reason we will refer to the mass of a string stretching between the bottom of the branes and the horizon (shown as a wiggly red curve in the figures) as the `thermal' quark mass. Note that this vanishes in the black hole phase. 

\begin{figure}
    \begin{center}
    	\includegraphics[width=0.85\textwidth]{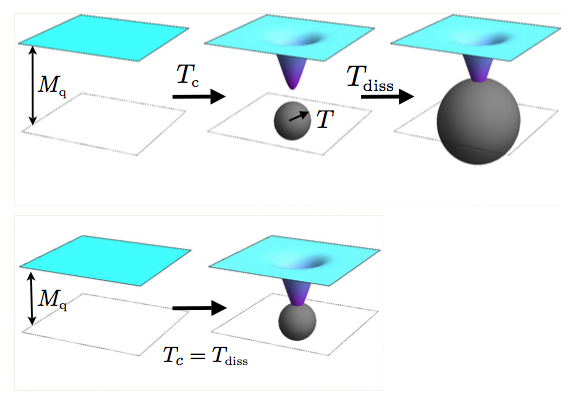}
    \end{center}
    \caption{\small Top: Sufficiently heavy quarkonium mesons remain bound in the deconfined phase (above $T_c$) and dissociate at $\td > T_c$. Bottom: In contrast, light mesons dissociate as soon as the deconfinement phase transition at $T=T_c$ takes place. This picture also applies to ${\cal N}=4$ SYM theory with $T_c=0$, as described in the main text. In $\N=4$ SYM theory, the top (bottom) panel applies when $M_q>0$ ($M_q=0$).}
\vspace{2mm}
\label{Tc-versus-Tdiss} 
\end{figure}

Although we will come back to this important point below, we wish to emphasize right from the start that {\it the phase transition under discussion is not a confinement-deconfinement phase transition}, since the presence of a black hole implies that both phases are deconfined. Instead, we will see that {\it the branes' phase transition corresponds to the dissociation of heavy quarkonium mesons}. In order to illustrate the difference most clearly, consider first a holographic model of a confining theory, as described in Section~\ref{sec:conf}; below we will come back to the case of ${\cal N}=4$ SYM. For all such confining models,  the difference between the deconfinement and the dissociation phase transitions is illustrated in Fig.~\ref{Tc-versus-Tdiss}. Below $T_c$, the theory is in a confining phase and therefore no black hole is present. At some $T_c$, a deconfinement transition takes place, which in the string description corresponds to the appearance of a black hole whose size is proportional to $T_c$. If the quark mass is sufficiently large compared to $T_c$ then the branes remain outside the horizon (top part of the figure); otherwise they fall through the horizon (bottom part of the figure). The first case corresponds to heavy quarkonium mesons that remain bound in the deconfined phase, and that eventually dissociate at some higher $\td > T_c$. The second case describes light mesons that dissociate as soon as the deconfinement transition takes place.  

Fig.~\ref{Tc-versus-Tdiss} also applies to ${\cal N}=4$ SYM theory with $T_c=0$ in the sense that, although the vacuum of the theory is not confining, there is no black hole at $T=0$. Note also that mesons only exist provided $\mq >0$, since otherwise the theory is conformal and there is no particle spectrum. This means that in ${\cal N}=4$ SYM theory any meson is a heavy quarkonium meson that remains bound for some range of temperatures above $T_c=0$, as described by the top part of Fig.~\ref{Tc-versus-Tdiss}. In the case $\mq=0$ we cannot properly speak of mesons, but we see that the situation is still described by the bottom part of the figure in the sense that in this case the branes fall through the horizon as soon as $T$ is raised above $T_c=0$.

The universal character of the meson dissociation transition was emphasized in Refs.~\cite{Mateos:2006nu,Mateos:2007vn}, which we will follow in our presentation. SpeciÞc examples  were originally seen in \cite{Babington:2003vm,Kruczenski:2003uq,Kirsch:2004km}, and aspects of these transitions in the D3/D7 system were studied independently in \cite{Albash:2006ew,Albash:2006bs,Filev:2007gb,Karch:2006bv}. Similar holographic transitions appeared in a slightly different framework in \cite{Aharony:2006da,Parnachev:2006dn,Gao:2006up,Antonyan:2006qy}. The D3/D7 system at nonzero temperature has been studied upon including the backreaction of the D7-branes in Ref.~\cite{Bigazzi:2009bk}.

\subsection{Thermodynamics of D7-branes}
\label{sec:ThermoD7}
In this subsection we shall show that the phase transition between Minkowski and black hole embeddings is a discontinuous, first-order phase transition.  The reader willing to accept this without proof can safely skip to Section~\ref{sec:ThermoMesons}.
Since we are working in the canonical ensemble (i.e.~at fixed temperature) we must compute the free energy density of the system per unit gauge theory three-volume, $F$,
and determine the configuration that minimizes it. In the gauge theory we know that this takes the form
\be
F = F_{{\cal N} =4} + F_\mt{flavour} \,,
\label{F}
\ee
where the first term is the $O(\nc^2)$ free energy of the ${\cal N} =4$ SYM theory in the absence of quarks, and the second term is the $O(\nc \nf)$ contribution due to the presence of quarks in the fundamental representation. Since the SYM theory without quarks is conformal, dimensional analysis completely fixes the first factor to be of the form $F_{{\cal N} =4} = C(\lambda) T^4$, where $C$ is a possibly coupling-dependent coefficient of order $\nc^2$. In contrast, in the presence of quarks of mass $\mq$ there is a dimensionless ratio $T/\mq$ on which the flavour contribution can depend non-trivially. Our purpose is to determine this contribution to leading order in the large-$\nc$, strong coupling limit. 

Our tool is of course the dual description of the ${\cal N} =4$ SYM theory with flavour as a system of $\nf$ D7-brane probes in the gravitational background of $\nc$ black D3-branes. As usual in finite-temperature physics, the free energy of the system may be computed through the identification
\be
\beta F = \se \,,
\ee
where $\beta=1/T$ and $\se$ is the Euclidean action of the system. In our case this takes the form
\be
\se = S_\mt{sugra} + S_\mt{D7} \,.
\label{se}
\ee
The first term is the contribution from the black hole gravitational background sourced by the D3-branes, and is computed by evaluating the Euclideanized supergravity action on this background. The second term is the contribution from the D7-branes, and is computed by evaluating the Euclidean version of the  DBI action \eqn{DBI} on a particular D7-brane configuration. The decomposition \eqn{se} is the dual version of that in \eqn{F}. The supergravity action scales as $1/g_s^2 \sim \nc^2$, and thus yields the free energy of the ${\cal N} =4$ SYM theory in the absence of quarks, i.e.~we identify 
\be
S_\mt{sugra} = \beta F_{{\cal N} =4} \,.
\ee
Similarly, the D7-brane action scales as $\nf / g_s \sim \nc \nf$, and represents the flavour contribution to the free energy:
\be
S_\mt{D7} = \beta F_\mt{D7} = \beta F_\mt{flavour} \,.
\label{D7free}
\ee
We therefore conclude that we must first find the solutions of the equations of motion of the D7-branes for any given values of $T$ and $\mq$, then evaluate their Euclidean actions, and finally use the identification above to compare their free energies and determine the thermodynamically preferred configuration. 

As explained above, in our case solving the D7-brane equations of motion just means finding the function $U(u)$, which is determined  by the condition that the D7-brane action be extremized. This leads to an ordinary, second-order, non-linear differential equation for $U(u)$. Its precise form can be found in e.g.~Ref.~\cite{Mateos:2007vn}, but is not very illuminating. However, it is easy to see that it implies the asymptotic, large-$u$ behaviour 
\be
U(u) = \frac{m \, r_0}{\sqrt{2}} + \frac{c \,r_0^3}{2\sqrt{2} \, u^2} + \cdots \,,
\labell{asympD7R}
\ee
where $m$ and $c$ are constants. The factors of $r_0$ have been introduced to make these constants dimensionless, whereas the numerical factors have been chosen to facilitate comparison with the literature. As usual (and, in particular, as in Section~\ref{sec:norma}), the leading and subleading terms correspond to the non-normalizable and to the normalizable modes, respectively. Their coefficients are therefore proportional to the source and the expectation values of the corresponding dual operator in the gauge theory. In this case, the position of the brane $U(u)$ is dual to the quark-mass operator 
${\cal O}_m \sim \bar{\psi} \psi$, so $m$ and $c$ are proportional to the quark mass and the quark condensate, respectively. The precise form of ${\cal O}_m$ can be found in Ref.~\cite{Kobayashi:2006sb}, where it is shown that the exact relation between $m, c$ and $\mq,  \langle {\cal O}_m \rangle$ is
\beqa 
\mq &=& \frac{r_0 m}{2^{3/2}\pi \ell_s^2} =
\frac{1}{2\sqrt{2}}\sqrt{\lambda}\,T\,m\,,\labell{mqD3D7}\\
\langle {\cal O}_m \rangle &=& - 2^{3/2}\pi^3 \ell_s^2 \nf T_\mt{D7}
r_0^3\, c = -\frac{1}{8\sqrt{2}}\sqrt{\lambda}\,\nf\,\nc\,T^3\,c\,.
\labell{cqD3D7} 
\eeqa
In particular, we recover the fact that the asymptotic value 
\be
L = \lim_{u\rightarrow \infty} U(u) = \frac{m r_0}{\sqrt{2}}
\ee
is related to the quark mass through Eqn.~\eqn{mq}, as anticipated in previous sections.

It is interesting to note that the dimensionless mass $m$ is given by the simple ratio 
\be
m=\frac{\bar{M}}{T} \,,
\label{littlem}
\ee
where 
\be 
\bar{M} = \frac{2 \sqrt{2} \mq}{\sqrt{\lambda}}  = \frac{M_\mt{mes}}{2\pi} 
\labell{mbarD3D7} 
\ee
is (up to a constant) precisely the meson gap at zero temperature, given in Eqn.~\eqn{gap}. As mentioned in Section \ref{sec:D7atZeroT}, and as we will elaborate upon in Section \ref{sec:RemarksConnectionQGP}, ${\cal O}_m$ must be thought of as the analogue of a heavy- or light-quark bilinear operator in QCD depending on whether the ratio $\mmes/T \sim m$ is large or small, respectively.  

\begin{figure}
\begin{center}
\begin{tabular}{cc}
\includegraphics[width=0.5 \textwidth]{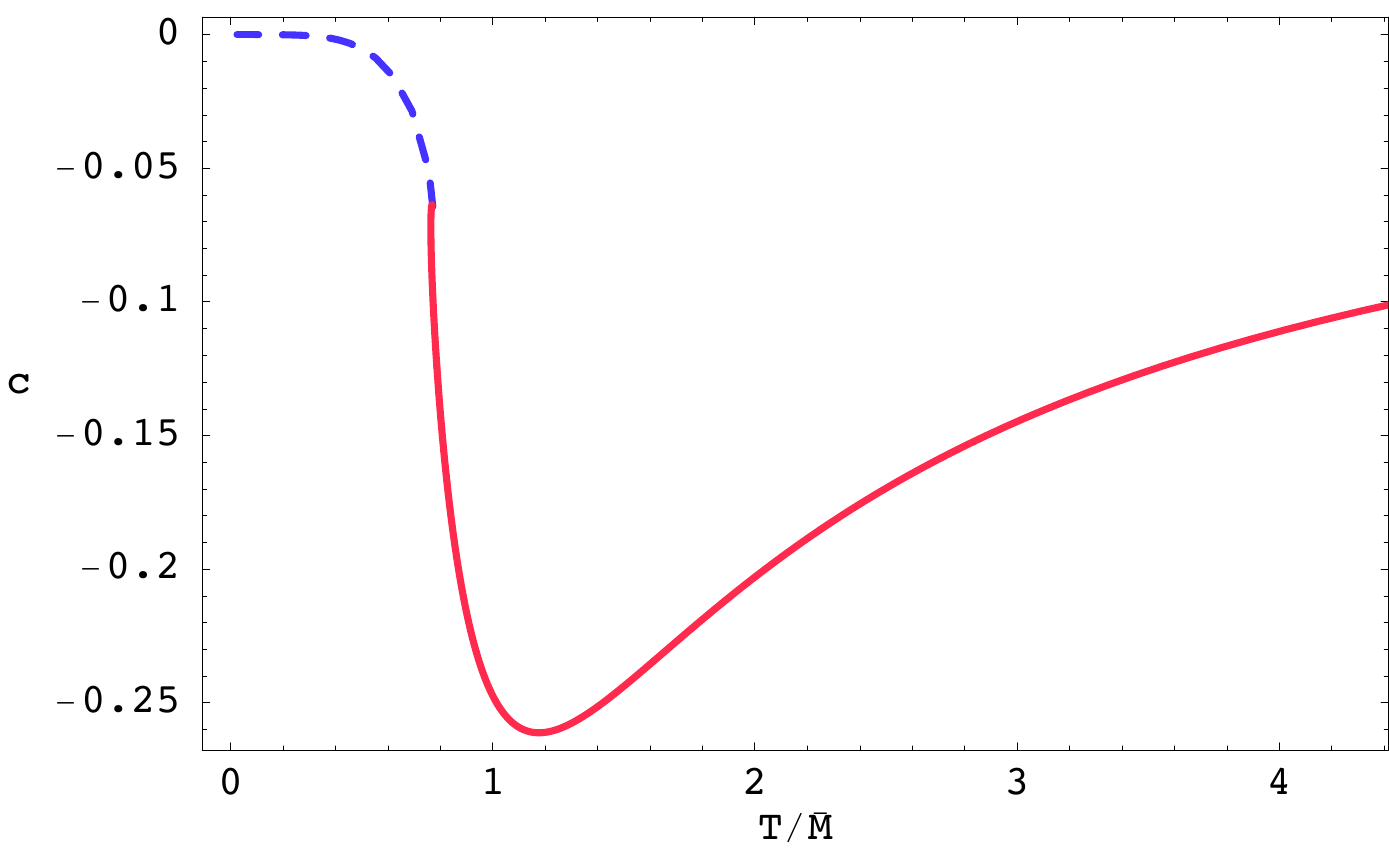} &
\includegraphics[width=0.5 \textwidth]{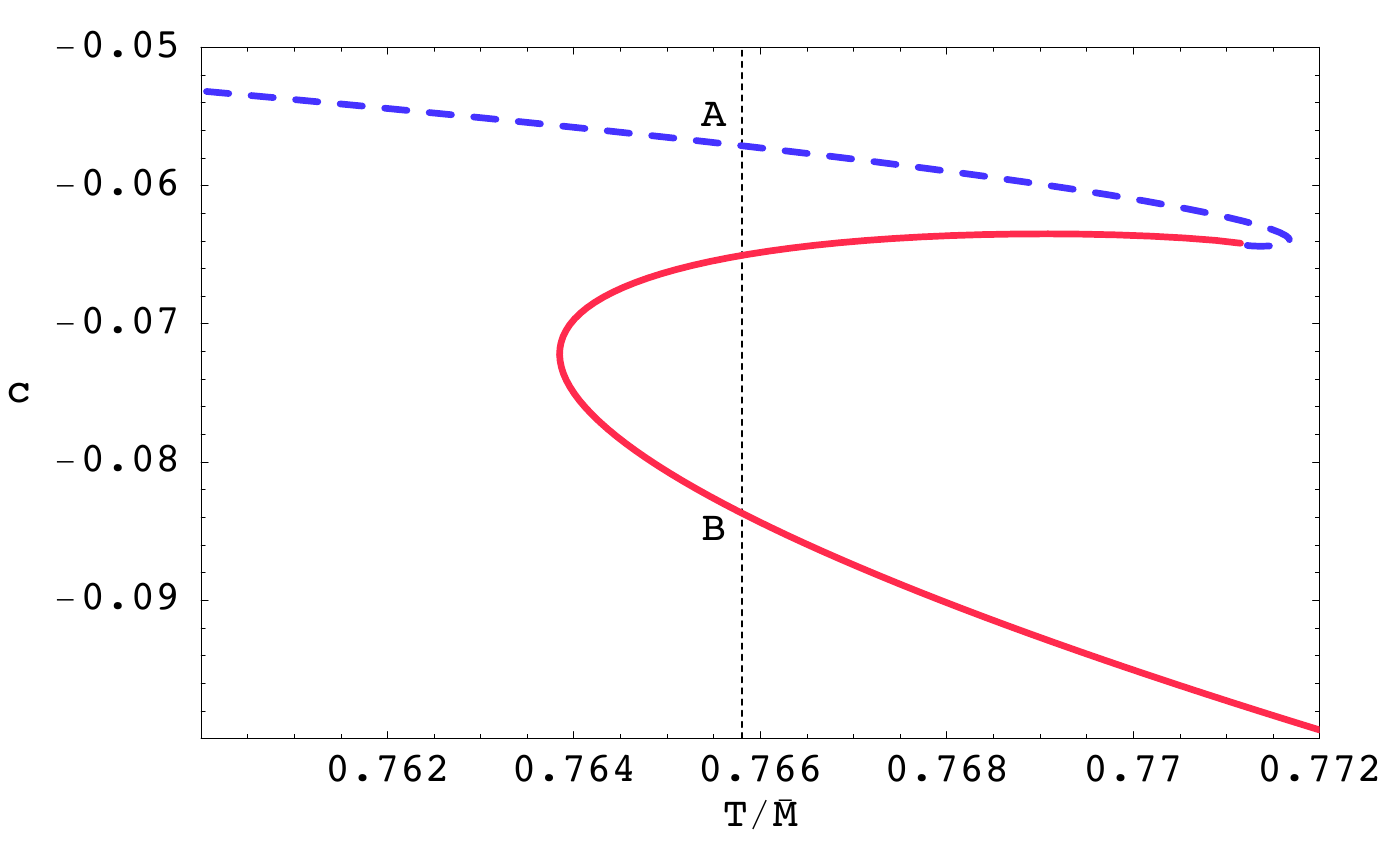} 
\end{tabular}
\end{center}
\caption{\small Quark condensate $c$ versus $T/\mbar = 1/m$. The blue dashed (red continuous) curves correspond to the Minkowski (black hole) embeddings.  The dotted vertical line indicates the precise temperature of the phase transition. The point where the two branches meet corresponds to the critical embedding.}
\label{mc}
\end{figure}

\begin{figure}
    \begin{center}
    	\includegraphics[width=0.55\textwidth]{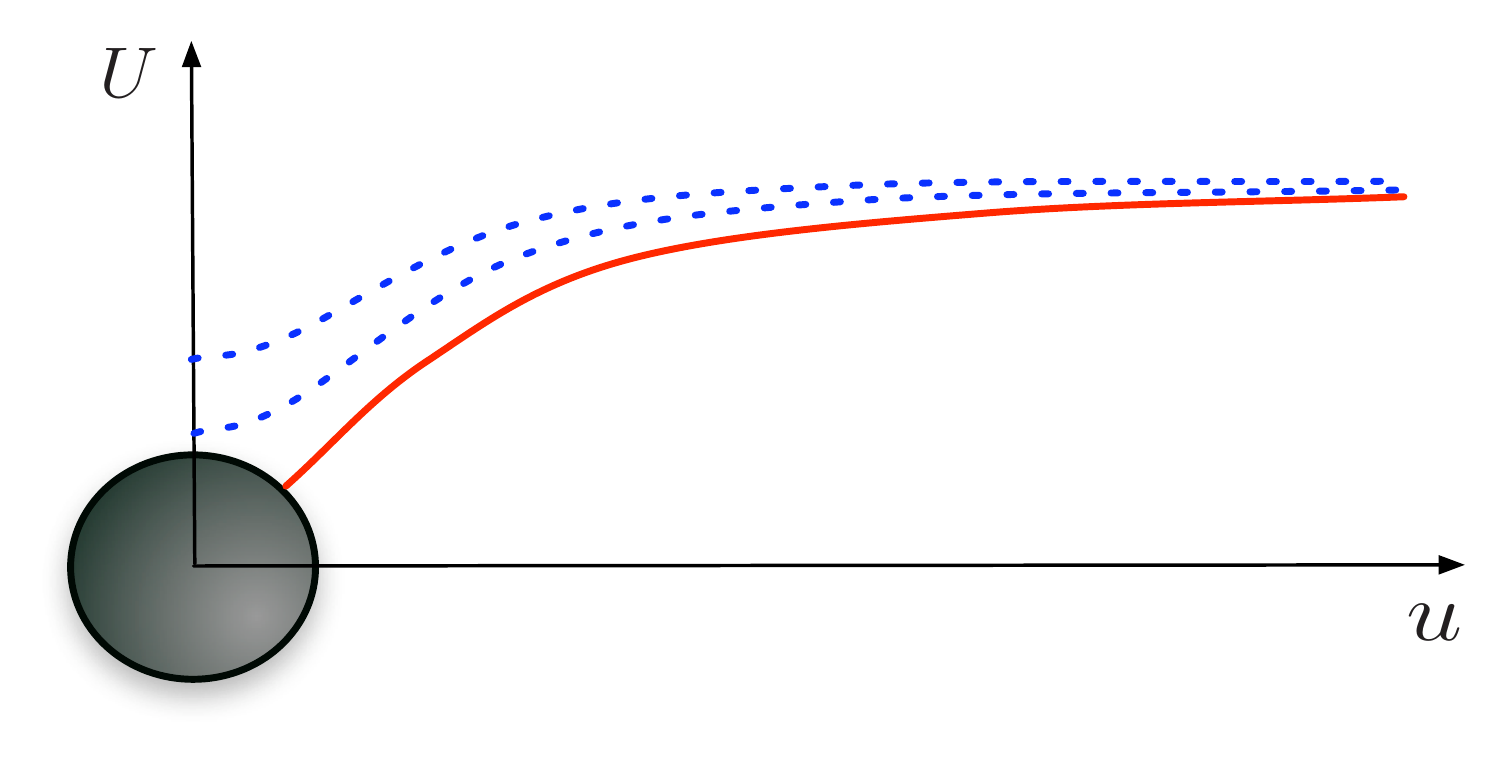}
    \end{center}
    \caption{\small Some representative D7-brane embeddings from the region 
in which $c$ is multi-valued. The three profiles correspond to the same  value of $m$ but differ in their value of $c$. Two of them, represented by blue, dashed curves, are of Minkowski type. The third one, represented by a red, continuous curve, is a black hole embedding.}
\label{multivalued}
\end{figure}

The constants $m$ and $c$ can be understood as the two integration constants that completely determine a solution of the second-order differential equation obeyed by $U(u)$. Mathematically, these two constants are independent, but the physical requirement that the solution be regular in the interior relates them to one another. The equation for $U(u)$ can be solved numerically (see, e.g.~Ref.~\cite{Mateos:2007vn}), and the resulting possible values of $c$ for each value of $m$ are plotted in Fig.~\ref{mc}.
We see from the `large-scale' plot on the left that $c$ is a single-valued function of $m$ for most values of the latter. However, the zoom-in plot on the right shows that, in a small region around 
$1/m = T/\mbar \simeq 0.766$, three values of $c$ are possible for a given value of $m$; a pictorial representation of a situation of this type is shown in Fig.~\ref{multivalued}. 
This multivaluedness is related to the existence of the phase transition which, as we will see, proceeds between points A and B through a discontinuous jump in the quark condensate and other physical quantities. The point in Fig.~\ref{mc} where the Minkowski and the black hole branches meet corresponds to the critical embedding. 

\begin{figure}
\begin{center}
\begin{tabular}{cc}
\includegraphics[width= 0.46 \textwidth]{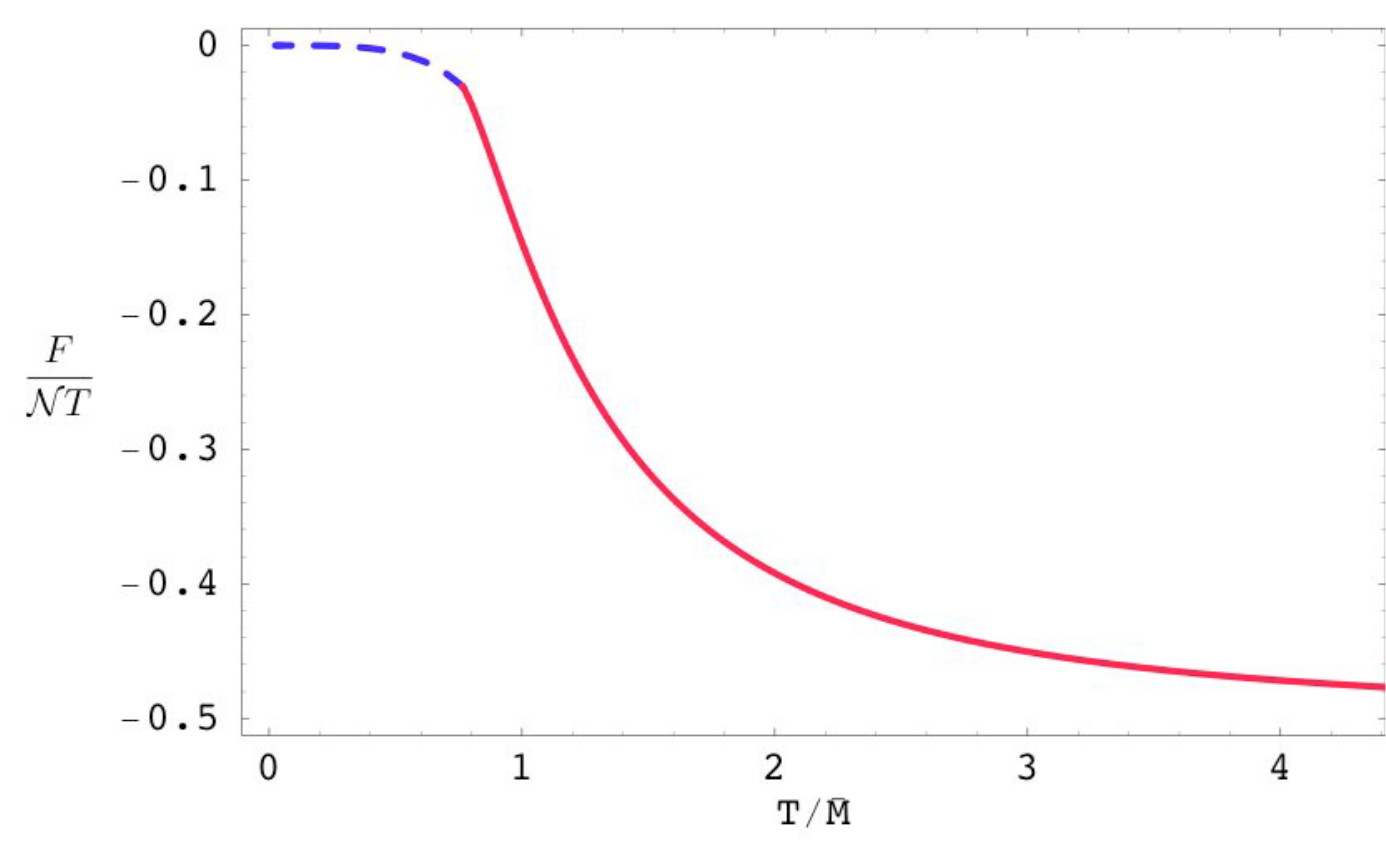} &
\includegraphics[width=0.48 \textwidth]{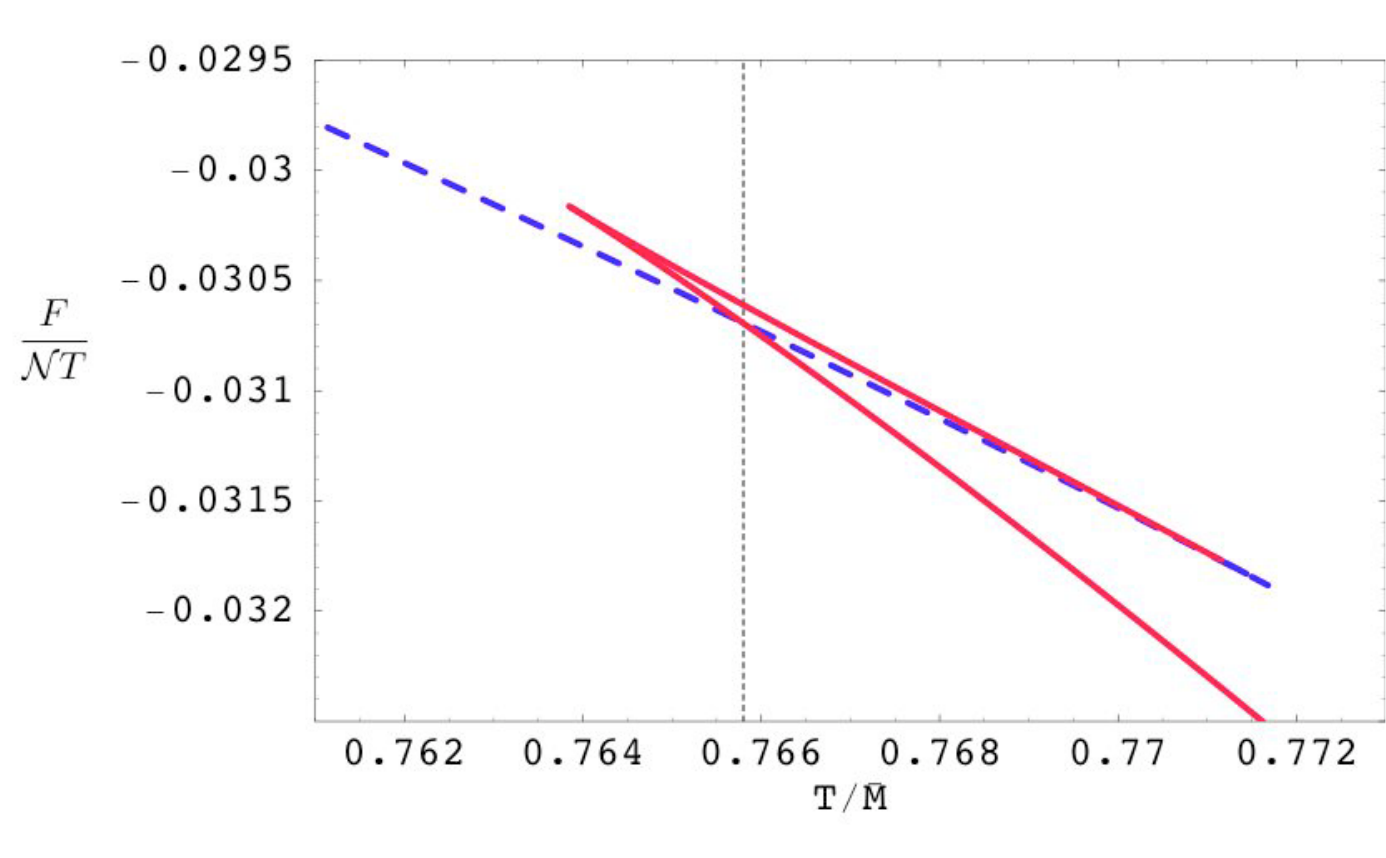} \\
\includegraphics[width=0.47 \textwidth]{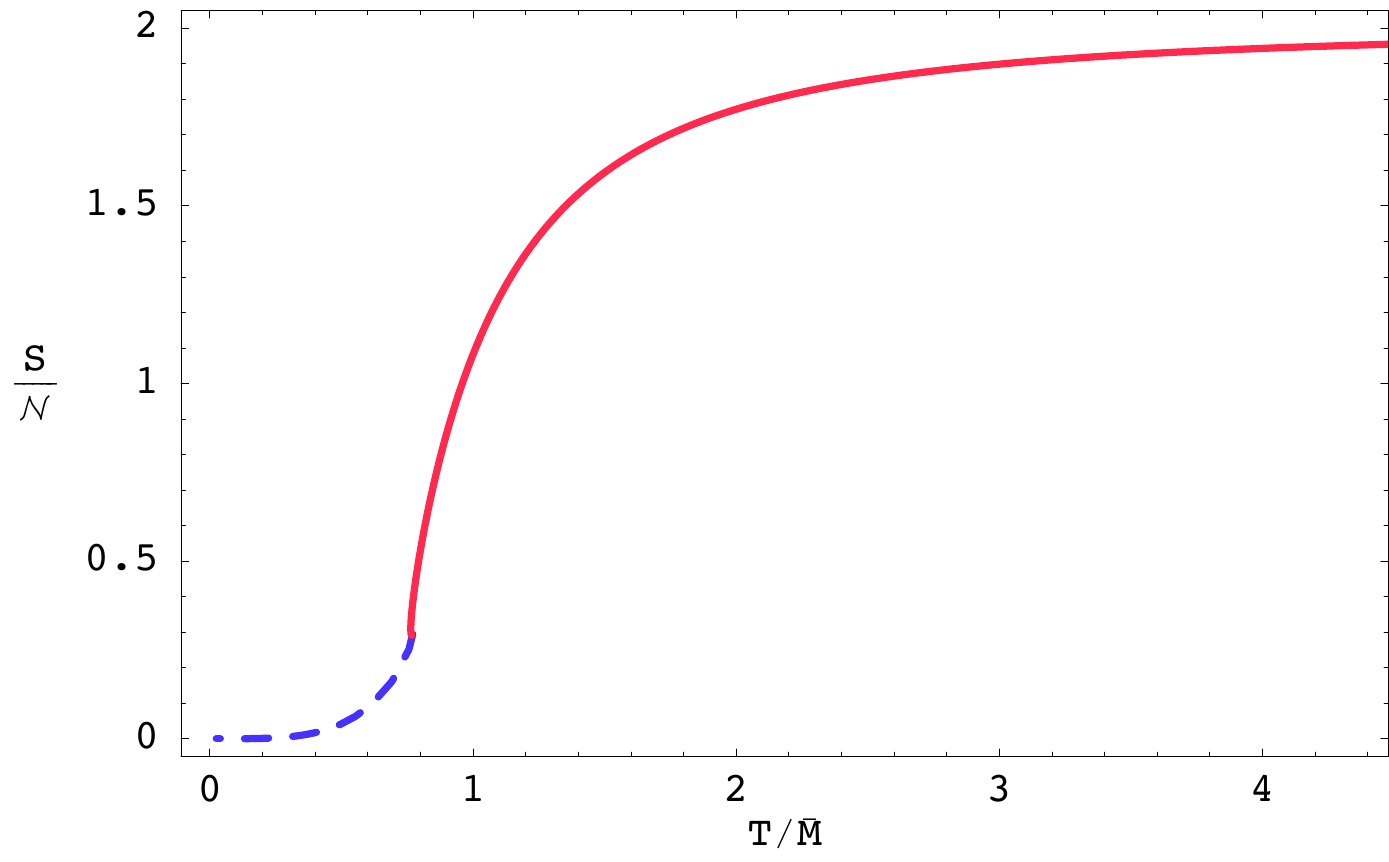}  &
\includegraphics[width=0.47 \textwidth]{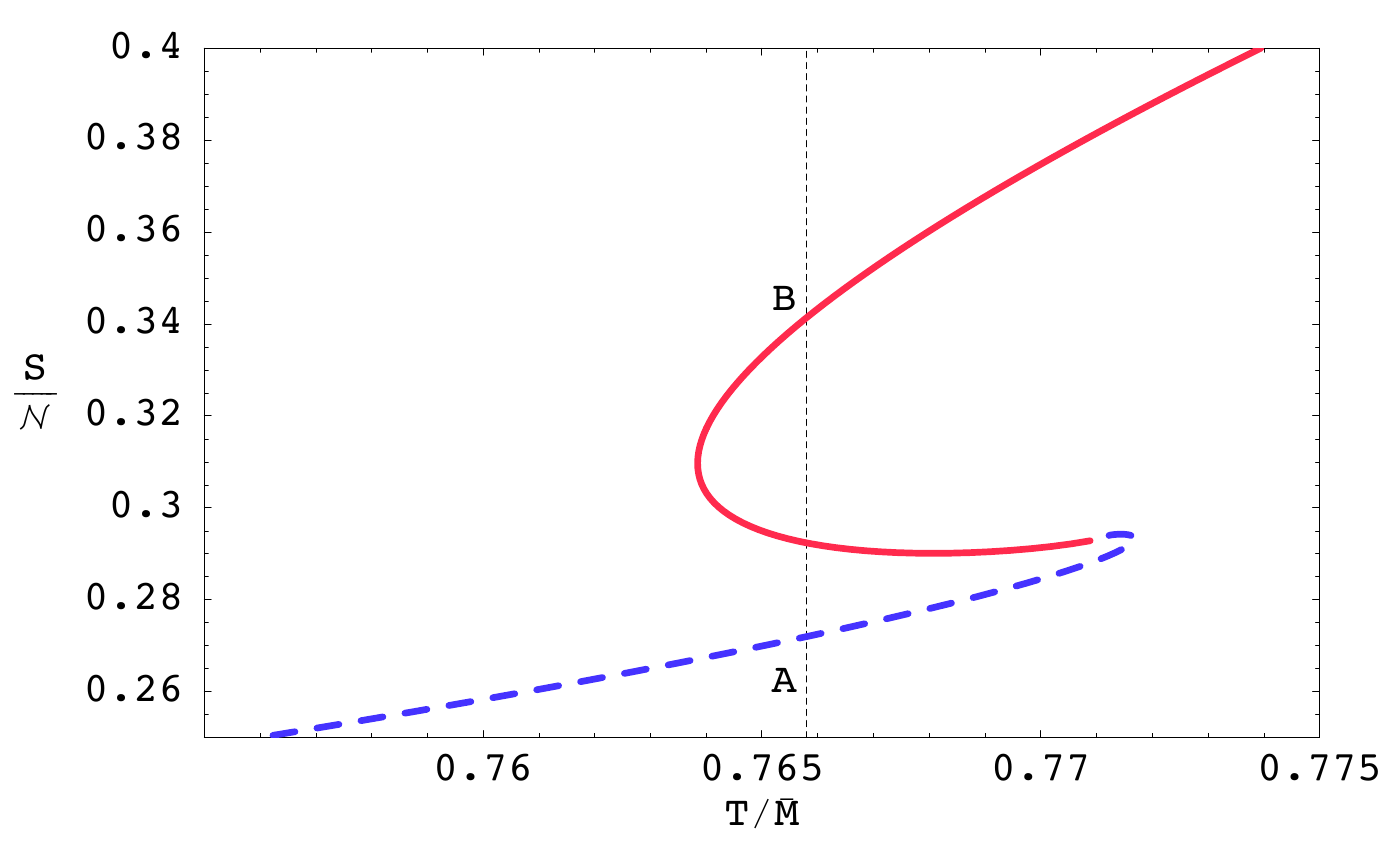} \\
\includegraphics[width=0.47 \textwidth]{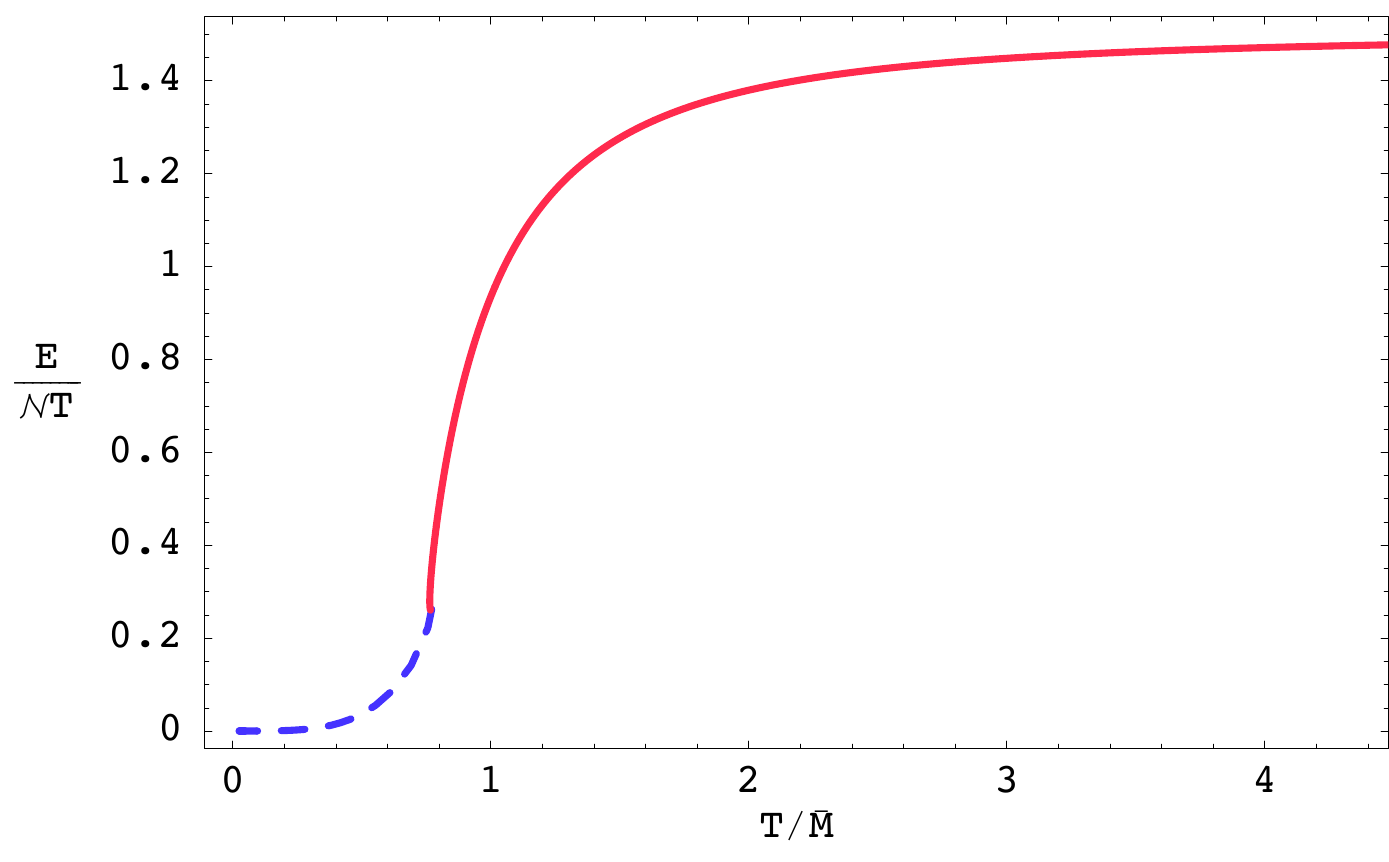} &
\includegraphics[width=0.47 \textwidth]{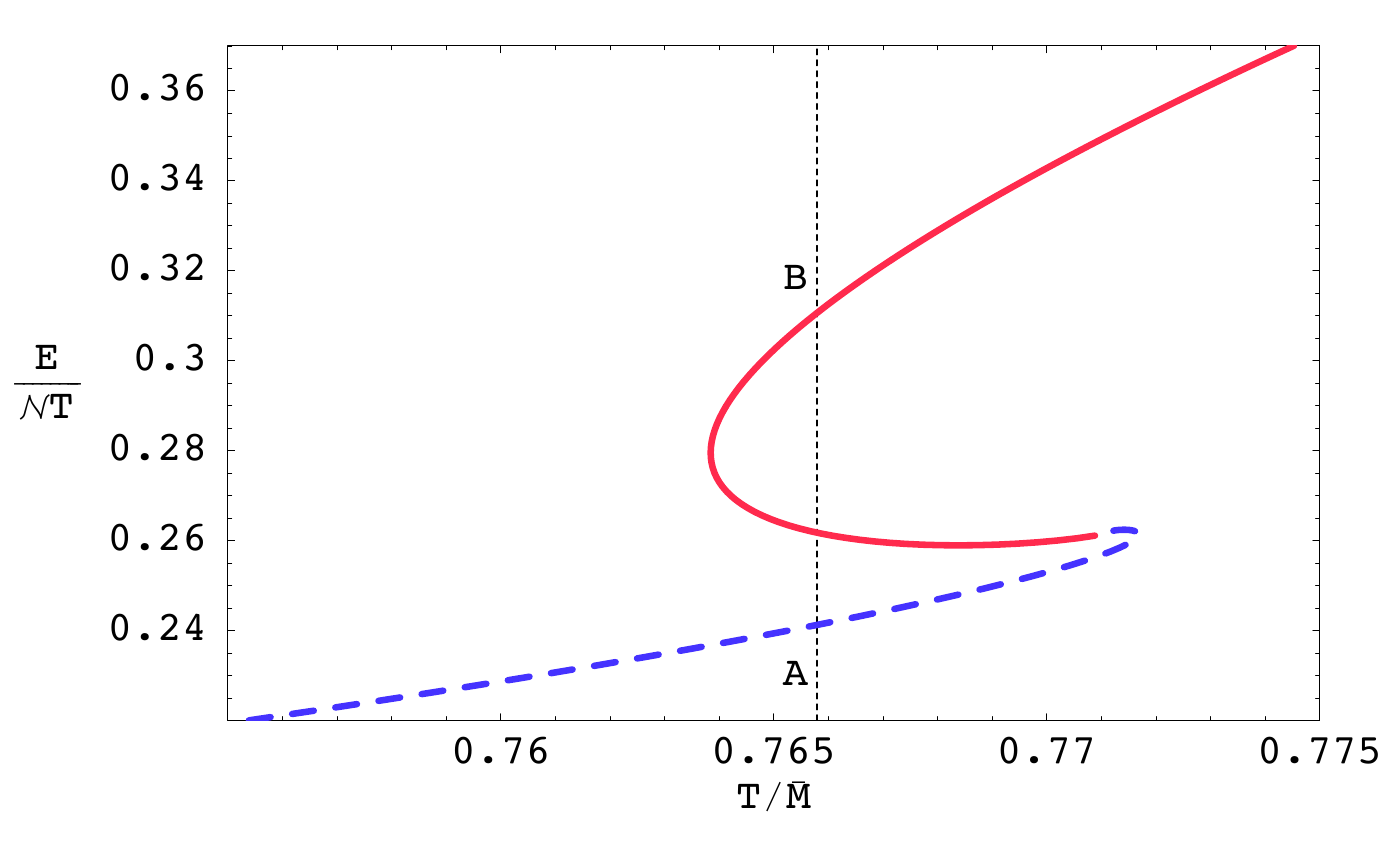}
\end{tabular}
\end{center}
\caption{\small Free energy, entropy and energy densities for a D7-brane in a black D3-brane background; note that 
$\N \propto T^3$. The blue dashed (red continuous) curves correspond to the Minkowski (black hole)
embeddings.  The dotted vertical line indicates the precise temperature of the phase transition.} 
\label{e}
\end{figure}

Having determined the regular D7-brane configurations, one must now compute their free energies and compare them in order to determine which one is preferred in the multivalued region. The result is shown in Fig.~\ref{e}(top), where the normalization constant is given by \cite{Mateos:2006nu,Mateos:2007vn}
\be 
{\cal N} = \frac{2\pi^2\nf T_\mt{D7} r_0^4}{4T}
=\frac{\lambda\nf\nc}{64}\,T^3 \,.
\label{ND3D7} 
\ee
The plot on the right shows the classic `swallow tail' form, typically associated with a first-order phase transition. As anticipated, Minkowski embeddings have the lowest free energy for temperatures $T < \td$, whereas the free energy is minimized by black hole embeddings for $T > \td$, with $\td \simeq 0.77 \mbar$  (i.e.~$m \simeq 1.3$). At $T=\td$ the Minkowski and the black hole branches meet and the 
thermodynamicaly-preferred embedding changes from one type to the other. The first-order nature of the phase transition follows from the fact that several physical quantities jump discontinuously across the transition. An example is provided by the quark condensate which, as illustrated in Fig.~\ref{mc}, makes a finite jump between the points labelled A and B. Similar discontinuities also appear in other physical quantities, like the entropy and energy density. These are easily obtained from the free energy through the usual thermodynamic relations
\be
S = -\frac{\partial F}{\partial T} \sac E=F+TS \,, 
\label{thermo}
\ee
and the results are shown in Fig.~\ref{e}. From the plots of the energy density one can immediately read off the qualitative behaviour of the specific heat $c_V = \prt E/ \prt T$. In particular, note that this slope must become negative as the curves approach the critical solution, indicating that the corresponding embeddings are thermodynamically unstable. Examining the fluctuation spectrum of the branes, we will show that a corresponding dynamical instability, manifested by a meson state becoming tachyonic, is present exactly for the same embeddings for which $c_V <0$. One may have thought that the phases near the critical point were metastable and thus accessible by `super-cooling' the system, but instead it turns out that over much of the relevant regime such phases are unstable.

We see from \eqn{ND3D7} that $\N \sim \lambda \nc \nf T^3$, which means that the
leading contribution of the D7-branes to all the various thermodynamic quantities will be order $\lambda \nc \nf$, in comparison to $\nc^2$ for the usual bulk gravitational contributions. The $\nc \nf$ dependence, anticipated below Eqn.~\eqn{F}, follows from large-$\nc$ counting. In contrast, as noted in Ref.~\cite{Mateos:2006nu,Mateos:2006yd}, the factor of $\lambda$ represents a strong-coupling enhancement over the
contribution of a simple free-field estimate for the $\nc \nf$ fundamental degrees of freedom. From the viewpoint of the string description, this enhancement is easy to understand by reexamining the relative normalization of the two terms in Eqn.~\eqn{se} more carefully than we did above. Ignoring only order-one, purely numerical factors, the supergravity action scales as $1/G$, with $G \sim g_s^2 \ls^8$ the ten-dimensional Newton's constant, whereas the D7-brane action scales as 
$\nf T_\mt{D7} \sim \nf / g_s \ls^8$. The ratio between the two normalizations  is therefore
\be
G \nf T_\mt{D7} \sim g_s \nf \sim \gym^2 \nf \sim \frac{\lambda \nf}{\nc} \,.
\label{notquite}
\ee
Thus the flavour contribution is suppressed with respect to the leading $O(\nc^2)$ contribution by $\lambda \nf /\nc$, i.e.~it is of order $\lambda \nc \nf$. We will come back to this point in the next subsection.

As the calculations above were all performed in the limit $\nc, \lambda \rightarrow \infty$ (with $\nf$ fixed), it is natural to ask how the detailed results depend on this approximation. Since the phase transition is first order, we expect that its qualitative features will remain unchanged within a finite radius of the $1/\nc, 1/\lambda$ expansions. Of course, finite-$\nc$ and finite-$\lambda$ corrections may eventually modify the behaviour described above. For example, at large but finite $\nc$ the black hole will emit Hawking radiation and each bit of the probe branes will experience a thermal bath at a temperature determined by the local acceleration. Similarly,  finite 't Hooft coupling corrections, which correspond to higher-derivative corrections both to the supergravity action and the D-brane action, will become important if the spacetime or the brane curvatures become large. It is certainly clear that both types of corrections will become  more and more important as the lower part of a Minkowski brane approaches the horizon, since as this happens the local temperature and the branes (intrinsic) curvature at their  tip increase. However, at the phase transition the minimum separation between the branes and the horizon is not parametrically small, and therefore the corrections above can be made arbitrarily small by taking $\nc$ and $\lambda$ sufficiently large but still finite. This confirms our expectation on general grounds that the qualitative aspects of the phase transition should be robust within a finite radius around the $1/\nc = 0$, $1/\lambda =0$ point. Of course, these considerations do not tell us whether the dissociation transition is first order or a crossover at $\nc=3$.

\subsection{Quarkonium thermodynamics}
\label{sec:ThermoMesons}

We have seen above that, in a large class of strongly coupled gauge theories with fundamental matter, this matter undergoes a first-order phase transition described on the gravity side by a change in the geometry of the probe D-branes. In this section we will elaborate on thermodynamical aspects of this transition from the gauge-theory viewpoint. Once we have learned more about the dynamics of holographic mesons in subsequent sections, in Section \ref{sec:RemarksConnectionQGP} we will return to the 
gauge-theory  viewpoint and discuss possible implications for the dynamics of quarkonium mesons in the QCD plasma. 

The temperature scale at which the phase transition takes place is set by the meson gap at zero temperature, $\td \sim \mmes$. As well as giving the mass gap in the meson spectrum, $1/\mmes$ is roughly the characteristic size of these bound states \cite{Hong:2003jm,Myers:2006qr}. The gluons and other adjoint fields are already in a deconfined phase at $\td$, so this new transition is not a confinement/deconfinement transition. Rather, the most striking feature of the new phase transition is the change in the meson spectrum, and so we refer to it as a `dissociation' or `melting' transition.

In the low-temperature phase, below the transition, stable mesons exist and their spectrum is discrete and gapped. This follows from the same general principles as in the zero-temperature case. The meson spectrum corresponds to the spectrum of normalizable fluctuations of the D7-branes around their fiducial embedding. For Minkowski embeddings the branes close off smoothly outside the black hole horizon and the admissible modes must also satisfy a regularity condition at the tip of the branes. On general grounds, we expect that 
the regular solution at the tip of the branes evolves precisely into  the normalizable solution at the boundary
only for a certain set of discrete values of the meson mass.
We will study the meson spectrum in detail in Section \ref{sec:mesdis}, and in Section \ref{sec:MesonWidths} we will see that mesons acquire finite decay widths at finite $\nc$ or finite coupling. Since the phase under consideration is not a confining phase, we can also introduce deconfined quarks into the system, represented by fundamental strings stretching between the D7-branes and the horizon. At a figurative level, in this phase we might describe quarks in the adjoint plasma as a `suspension'. That is, when quarks are added to this phase, they retain their individual identities. More technically, we may just say that quarks are well defined quasiparticles in the Minkowski phase.

In the high-temperature phase, at $T>\td$, no stable mesons exist.  Instead, as we will discuss in more detail in Section \ref{sec:BHQuasi}, the excitations of the fundamental fields in this phase are characterised by a discrete spectrum of quasinormal modes on the black hole embeddings \cite{Hoyos:2006gb,Myers:2007we}. The spectral function of some two-point meson
correlators in the holographic theory, of which we will see an example in Section \ref{sec:BHMesonSpectrum},
 still exhibits some broad peaks in a regime just above $\td$, which suggests that a few broad bound states persist just above the dissociation phase transition \cite{Myers:2007we,Mateos:2007yp}. This is quite analogous to the lattice approach, where similar spectral functions 
 are examined to verify the presence or absence of bound states. Hence, identifying $\td$ with the dissociation temperature should be seen as a (small) underestimate of the temperature at which mesons completely cease to exist. An appropriate figurative characterization of the quarks in this high-temperature phase would be as a `solution'. If one attempts to inject a localised quark charge into the system, it quickly falls through the horizon, i.e.~it spreads out across the entire plasma and its presence is reduced to diffuse disturbances of the supergravity and worldvolume fields, which are soon damped out \cite{Hoyos:2006gb,Myers:2007we}.  Technically speaking, we may just state that quarks are not well defined quasiparticles in the black hole phase.

The physics above is potentially interesting in connection with QCD, since evidence from several sources indicates that heavy quarkonium mesons remain bound in a range of temperatures above 
$T_c$. We will analyze this connection in more detail in Section \ref{sec:RemarksConnectionQGP}, once we have learned more about the properties of holographic mesons in subsequent sections. Here we would just like to point out one simple
physical parallel. The question of quarkonium bound states surviving in the quark-gluon plasma was first addressed by comparing the size of the bound states to the
screening length in the plasma \cite{Matsui:1986dk}. In the 
D3/D7 system, the size of the mesons can be inferred, for example, from the structure functions, and the relevant length scale that emerges 
is $d_\mt{mes} \sim \sqrt{\lambda}/\mq$ \cite{Hong:2003jm}. This can also be heuristically motivated as follows. As discussed in Section~\ref{sec:Wilson} (see Eqn.~\eqref{potB}) at zero temperature the potential between a quark-antiquark pair separated by a distance $\ell$ is given by
\be \label{eopr}
V \sim -{\sqrt{\lam} \ov \ell} \ .
\ee
We can then estimate the size $d_\mt{mes}$ of a meson by requiring $E_B \sim |V(d_\mt{mes})|$, where $E_B$ is the binding energy \eqn{benr}.\footnote{Eqn.~\eqn{benr} was derived at zero temperature, but as we will see in Section \ref{sec:mesdis} it is also parametrically correct at nonzero temperature.} This gives
\be 
\label{riri}
d_\mt{mes} \sim {\sqrt{\lam} \ov E_B } \sim {\sqrt{\lam} \ov \mq} \sim {1 \ov \mmes} \sim {R^2 \ov L} \,.
\ee
The last equality follows from 
Eqn.~\eqn{spectrumEQN} and is consistent with expectations based on the UV/IR correspondence \cite{Myers:2006qr}, since on the gravity side mesons are excitations near $r = L$. Just for comparison, we remind the reader that the weak-coupling formula for the size of quarkonium is 
$d_\mt{weak} \sim 1 /( \gym^2 \mq)$.

One intuitive way to understand why a meson has a very large size compared to its inverse binding energy or to the inverse quark mass is that, due to strong-coupling effects, the quarks themselves have an effective size of order $\dmes$. The effective size of a quark is defined as the largest of the following two scales: (i) its Compton wavelength, or (ii) the distance between a quark-antiquark pair at which their potential energy is large enough to 
pair-produce additional quarks and antiquarks. In a weakly coupled theory (i) is larger, whereas in a strongly coupled theory (ii) is larger. From Eqn.~\eqn{eopr} we see that this criterion gives an effective quark size of order 
$\sqrt{\lambda}/\mq$ instead of $1/\mq$. This heuristic estimate is supported by an explicit calculation of the size of the gluon cloud that dresses a 
quark~\cite{Hovdebo:2005hm}. These authors computed the expectation value $\langle \mbox{Tr} F^2 (x)\rangle$ sourced by a quark of mass $\mq$ and found that the characteristic size of the region in which this expectation value is nonzero is precisely $\sqrt{\lambda}/\mq$.

As reviewed in Section \ref{Loop-nonzero}, holographic studies of Wilson lines at nonzero temperature \cite{Rey:1998bq,Brandhuber:1998bs} reveal that the relevant screening length of the SYM plasma is of order $L_s \sim 1/T$ --- see Eqn.~\eqref{screL}. The argument
that the mesons should dissociate when the screening length is
shorter than the size of these bound states then yields
$\td \sim\mq/\sqrt{\lam} \sim \mmes$, in agreement with the results of the detailed calculations explained in previous sections. We thus see that the same physical reasoning which is used in QCD to estimate the dissociation temperature of, e.g.,~the $J/\psi$ meson can also be used to understand the dissociation of mesons in the the ${\cal N}=4$ SYM theory. This may still seem counterintuitive in view of the fact that the binding energy of these mesons is much larger that $\td$. In other words, one might have expected that the temperature required to break apart a meson would be of the order of the binding energy, $E_B \sim \mq$, instead of being parametrically smaller, 
\be
\td \sim \mmes \sim E_B / \sqrt{\lambda} \,.
\label{distt}  
\ee
However, this intuition relies on the expectation that the result of dissociating a meson is a quark-antiquark pair of mass $2\mq$. The gravity description makes it clear that this is not the case at strong coupling, since above $\td$ the branes fall through the horizon. Heuristically, one may say that this means that the `constituent' or `thermal' mass of the quarks becomes effectively zero. However, a more precise statement is simply that in the black hole phase quark-like quasiparticles simply do not exist, and therefore for the purpose of the present discussion it becomes meaningless to attribute a mass to them.

One point worth emphasizing is that there are two distinct processes that are occurring at $T\sim\mmes$. If we consider e.g.~the entropy density in Fig.~\ref{e}, we see that the phase transition occurs in the midst of a crossover signalled by a rise in $S/T^3$. We may write the contribution of the fundamental matter to the entropy density as
\be
S_\mt{flavour} = \frac{1}{8}\lambda\, \nf\,\nc\,T^3 \, H(x) \,,
\label{entro}
\ee
where $x=\lambda T^2 /\mq$ and $H(x)$ is the function shown in the plot of the free energy density in the top panels of Fig.~\ref{e}. $H$
rises from 0 at $x=0$ to 2 as $x\rightarrow\infty$, but the most
dramatic part of this rise occurs in the vicinity of $x=1$. Hence
it seems that new degrees of freedom, i.e.~the fundamental
quarks, are becoming `thermally activated' at $T\sim\mmes$. We note that the phase transition produces a discontinuous jump in which $H$ only increases by about 0.07, i.e.~the jump at the phase
transition only accounts for about 3.5\% of the total entropy
increase. Thus the phase transition seems to play an small role in
this crossover and produces relatively small changes in the thermal properties of the fundamental matter, such as the energy and entropy densities.

As $\mmes$ sets the scale of the mass gap in the meson spectrum, it is tempting to associate the crossover above with the thermal excitation of mesonic degrees of freedom. However, the pre-factor $\lambda\, \nf\,\nc$ in \reef{entro} indicates that this reasoning is incorrect: if mesons provided the relevant degrees of
freedom, we should have $S_\mt{flavour}\propto\nf^2$. Such a contribution can be obtained either by a one-loop calculation of the fluctuation determinant around the classical D7-brane configuration, or by taking into consideration the D7-branes backreaction to second order in the $\nf/\nc$ expansion as in \cite{Cherkis:2002ir,GomezReino:2004pw,Erdmenger:2004dk,Bigazzi:2009bk}. One can make an analogy here with the entropy of a confining theory (cf.~Section \ref{sec:conf}). In the low-temperature, confining phase the absence of a black hole horizon implies that the classical-gravity saddle point yields zero entropy, which means that the entropy is zero at order $\nc^2$.  One must look at the fluctuation determinant to see the entropy contributed by the
supergravity modes, i.e.~by the gauge-singlet glueballs, which is of order $\nc^0$.  

We thus see that the factor of $\nf\nc$ in $S_\mt{flavour}$ is naturally interpreted as counting the number of degrees of freedom associated with deconfined quarks, with the factor of $\lambda$ demonstrating that the contribution of the quarks is enhanced at strong coupling. A complementary interpretation of \reef{entro} comes from reorganizing the pre-factor as
\be
\lambda\,\nf\,\nc=(\gym^2 \nf)\,\nc^2\,.
\label{reorg}
\ee
The latter expression suggests that the result corresponds to the
first-order correction of the adjoint entropy due to quark loops. As explained at the end of Section \ref{fundec}, we are working in a `not quite' quenched approximation, in that contributions of the D7-branes represent the leading order contribution in an expansion in $\nf/\nc$, and so quark loops are suppressed but not completely. In view of the discussion below Eqn.~\eqn{notquite}, it is clear that the expansion for the classical gravitational back-reaction of the D7-branes is controlled by  $\lambda\nf/\nc=\gym^2 \nf$. Hence this
expansion corresponds to precisely the expansion in 
quark loops on the gauge theory side.

We conclude that the strongly coupled theory brings together these two otherwise distinct processes. That is, because the ${\cal N}=4$ SYM theory is strongly coupled at all energy scales, the dissociation of the quarkonium bound states and the thermal activation of the quarks happen at essentially the same temperature. Note that this implies that the phase transition should not be thought of as \emph{exclusively} associated with a discontinuous change in the properties of mesons --- despite the fact that this is the aspect that is more commonly emphasized. The phase transition is also associated with a discontinuous change in the properties of quarks since, as explained above, these exist as well defined quasiparticles in the Minkowski phase but not in the black hole phase. In fact, as the discussion around Eqn.~\eqn{reorg} makes clear, in 
the ${\cal O}(\nf \nc)\,$-$\,$approximation considered here the observed discontinuous jump in the thermodynamic functions comes \emph{entirely} from the discontinuous change in the properties of quarks. In this approximation, the discontinuous jump in the thermodynamic functions associated with the discontinuous change in the properties of mesons simply cannot be detected, since it is of order 
$\nf^2$ and its determination would require a one-loop calculation. Fortunately, however, the change in the mesons' properties can be inferred, e.g.~from the comparison of their spectra above and below $\td$.

It is instructive to contrast this behavior with that which is expected to occur at weak coupling.
In this regime, one expects that the dissociation of the quarkonium mesons may well be just a crossover rather
than a (first-order) transition.  Moreover, since the weakly bound mesons are much larger than $1/M_{\rm mes} \sim 1/(2 M_q)$, their
dissociation transition will occur at a $T_{\rm diss}$ that is much lower than $M_q$.  On the other hand, the quarks would not be thermally activated until the temperature $T_{\rm activ}\sim M_q$, above which the number densities of unbound quarks and antiquarks are no longer Boltzmann-suppressed.  Presumably, the thermal activation would again correspond to a crossover
 rather than a phase transition. The key point is that these two temperatures are widely separated at weak coupling.
\begin{figure}
    \begin{center}
    	\includegraphics[width=0.65\textwidth]{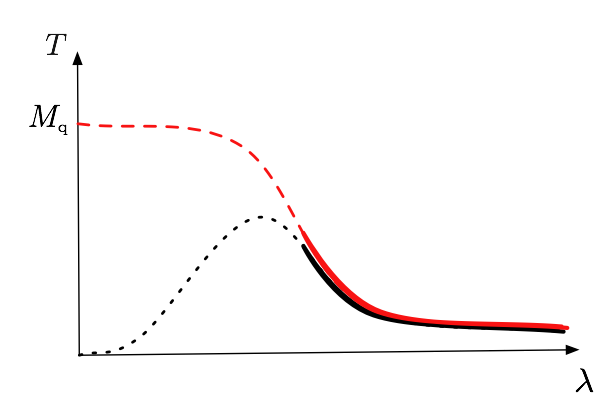}
    \end{center}
    \caption{\small A qualitative representation of the simplest possibility interpolating between the weak- and the strong-coupling regimes in $\N=4$ SYM theory. The solid and the dotted black curves correspond to $T=\td$. At strong coupling this corresponds to a first-order phase transition (solid black curve), whereas at weak coupling it corresponds to a crossover (dotted black curve). 
The dashed red  curve corresponds to $T=T_\mt{activ}$. At strong coupling this takes place immediately after the phase transition, whereas at weak coupling it is widely separated from $\td$.} 
    \label{weak-strong}
\end{figure}
Fig.~\ref{weak-strong} is an `artistic' representation of the
simplest behaviour which would interpolate between strong and weak
coupling. One might expect that the dissociation point and the thermal
activation are very close for $\lambda \gg1$. The line of first-order
phase transitions must end somewhere and so one might expect that it terminates at a critical point around 
$\lambda \sim1$. Below this point, both processes would only represent crossovers and their respective temperatures would diverge from one another, approaching the weak-coupling behaviour described above.

We close with a comment about a possible comparison to QCD. Although it would be interesting to look for signs of a crossover or a phase transition associated with quarkonium dissociation, for example in lattice QCD, the above discussion makes it clear that much caution must be exercised in trying to compare with the holographic results described here.  The differences can be traced back to the fact that, unlike the holographic theory considered here, QCD is not strongly coupled at the scale set by the mass of the heavy quark or of the corresponding heavy quarkonium meson. 
For this reason, in QCD the binding energy of a quarkonium meson is $E_B \ll \mmes \lesssim 2 \mq$ and, since one expects that $\td \sim E_B$, this implies that at the dissociation temperature the quarkonium contribution to (say)  the total entropy density would be Boltzmann suppressed, i.e.~it would be of order $S_\mt{flavour} \sim \nf^2 \exp ( -\mmes / \td) \ll 1$. In contrast, in the holographic setup there is no exponential suppression because $\td \sim \mmes$. Note also that the quarkonium contribution should scale as $\nf^2$, and therefore
the exponential suppression is a further suppression on top of the already small one-loop contribution discussed in the paragraph above Eqn.~\eqn{reorg}. That is, there are two sources of suppression relative to the leading 
$O(\nf \nc)\,$-$\,$contribution in the holographic theory.    Although $\nf/\nc$ is not small in QCD, the Boltzmann suppression is substantial and will likely make the thermodynamic effects of any quarkonium dissociation transition quite a challenge to identify.

\section{Quarkonium mesons in motion and in decay}
\label{sec:MinkSpectrum}

In previous sections, we examined the thermodynamics of the phase transition between Minkowski and black hole embeddings, and we argued that from the gauge-theory viewpoint it corresponds to a meson-dissociation transition. In particular, we argued that quarkonium 
bound states exist on Minkowski embeddings, i.e.~at $T<\td$, that they are absolutely stable in the large-$\nc$, strong-coupling limit, and that their spectrum is discrete and gapped. We will begin Section~\ref{sec:mesdis} by studying this spectrum quantitatively, which will allow us to understand how the meson spectrum is modified with respect to that at zero temperature, described in 
Section~\ref{zerospectrum}. The spectrum on black hole embeddings will be considered in Section~\ref{sec:BHQuasi}.

After describing the spectrum of quarkonium mesons at rest, we will determine their dispersion relations. This  will allow us to study mesons in motion with respect to the plasma and, in particular, to determine how the dissociation temperature depends on the meson velocity.  As discussed in Section~\ref{quarkonium}, one of the hallmarks of a quark-gluon plasma is the screening of colored objects. Heavy quarkonia provide an important probe of this effect since the existence (or absence) of quark-antiquark bound states and their properties 
are sensitive to the screening properties of the medium in which they are embedded. In Section~\ref{sec:HotWind} we studied this issue via computing the potential between an external quark-antiquark pair, at rest in the plasma or moving through it with velocity $v$.
In particular, we found that the dissociation temperature scales with $v$ as 
\be \label{temsc}
\td (v) \simeq \td (v=0) (1-v^2)^{1/4} \,,
\ee
which could have important implications for the phenomenon of quarkonium suppression in heavy ion collisions. By studying dynamical mesons in a thermal medium, we will be able to reexamine this issue in a more `realistic' context.

We will show in Section~\ref{sec:MesonWidths} that both finite-$\nc$ and finite-coupling corrections 
generate nonzero meson decay widths, as one would expect in a thermal medium. We shall find that the dependence of the widths on the meson momentum yields further understanding of how (\ref{temsc}) arises.

We will close in Section~\ref{sec:RemarksConnectionQGP} 
with a discussion of the potential connections between the properties of quarkonium mesons in motion in a holographic plasma and those of quarkonium mesons in motion in the QCD plasma.

\subsection{Spectrum and dispersion relations} \label{sec:mesdis}

In order to determine the meson spectrum on Minkowski embeddings, we proceed as in Section~\ref{zerospectrum}. For simplicity we will focus on fluctuations of the position of the branes $U(u)$ with no angular momentum on the $S^3$, i.e.~we write 
\be
\delta U = \U (u) \, e^{-i \omega t} e^{i{\bf q} \cdot {\bf x}} \,.
\ee
The main difference between this equation and its zero-temperature counterpart \eqn{wavefunk} is that in the latter case Lorentz invariance implies the usual relation $\omega^2 - q^2 = M^2$ between the energy $\omega$, the spatial three-momentum $q$, and the mass $M$ of the meson. At nonzero temperature, boost invariance is broken because the plasma defines a preferred frame in which it is at rest and the mesons develop a non-trivial dispersion relation 
$\omega(q)$. In the string description this is determined by requiring normalizability and regularity of  $\U(u)$: For each value of $q$, these two requirements are mutually compatible only for a discrete set of values $\omega_n(q)$, where different values of $n$ label different excitation levels of the meson. We define the `rest mass' of a meson as the energy 
$\omega(0)$ at vanishing three-momentum, $q=0$, in the rest-frame of the plasma. 

\begin{figure}
    \begin{center}
    	\includegraphics[width=0.85\textwidth]{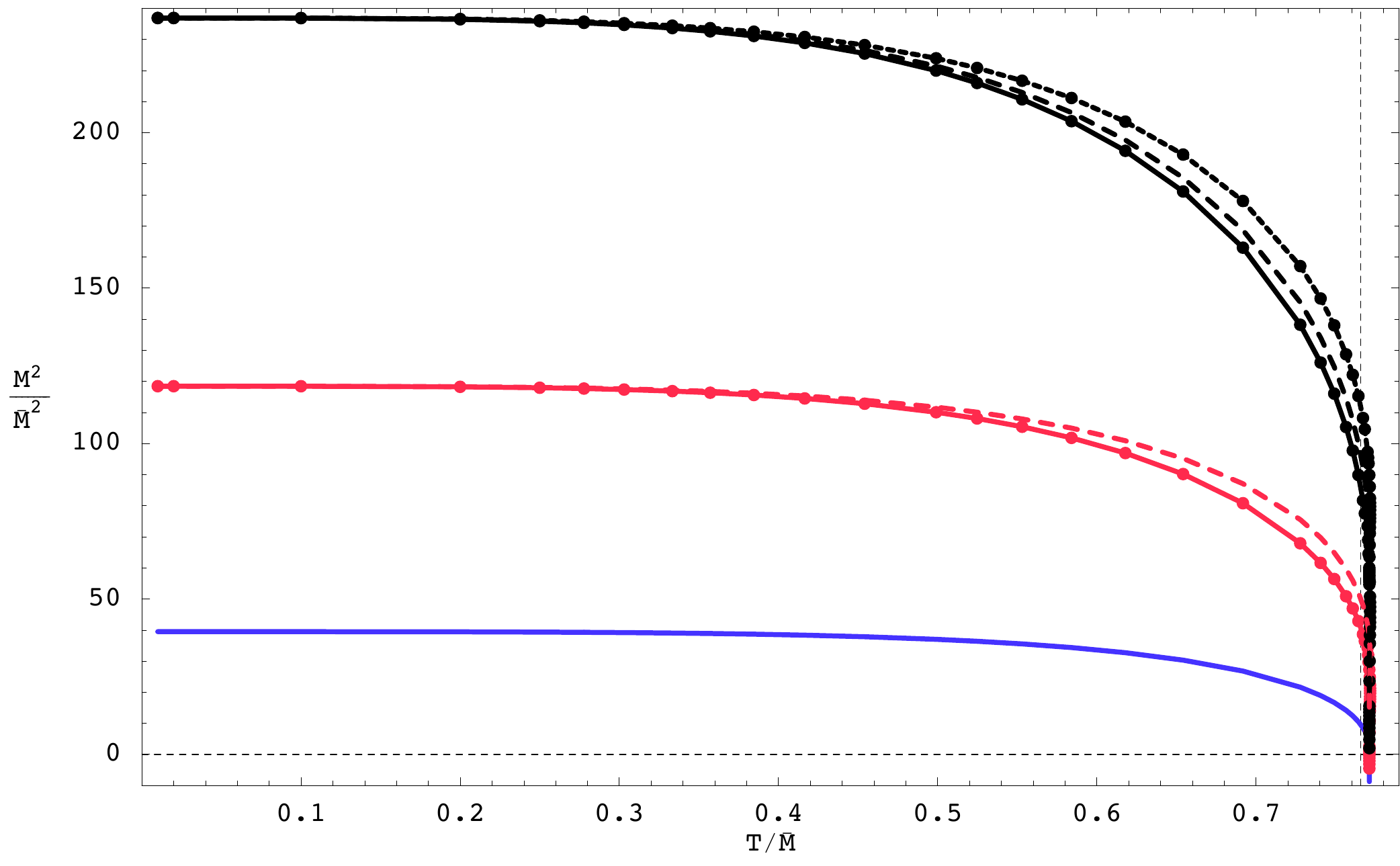}
    \end{center}
\caption{\small Meson mass spectrum $M^2=\omega^2|_{q=0}$ for Minkowski embeddings in the D3/D7 system. Continuous curves correspond to radially excited mesons with radial quantum number $n=0,1,2$ from bottom to top, respectively. Dashed lines correspond to mesons with angular momentum on the $S^3$. The dashed vertical line indicates the temperature of the phase transition.
Note that modes become tachyonic slightly beyond this temperature.} 
\label{mesonMassesRW} 
\end{figure}

Fig.~\ref{mesonMassesRW}, taken from Ref.~\cite{Mateos:2007vn}, shows the rest mass of the mesons as a function of temperature and quark mass. 
Note that in the zero-temperature limit, the spectrum coincides with the zero-temperature spectrum \eqn{spectrumEQN}. In particular, the lightest meson has a mass squared matching Eqn.~\eqn{mbarD3D7}: $M_\mt{mes}^2= 4\pi^2 \mbar^2\simeq 39.5\,\mbar^2$. 

The meson masses decrease as the temperature increases. Heuristically, this can be understood in geometrical terms from Fig.~\ref{BendingWithBH}, which shows that the thermal quark mass $M_\mt{th}$ decreases as the temperature increases and the tip of the D7-branes gets closer to the black hole horizon.
The thermal shift in the meson masses becomes more significant at the phase transition, and slightly beyond this point some modes actually become tachyonic. This happens precisely in the same region in which Minkowski embeddings become thermodynamically unstable because $c_V <0$. In other words, Minkowski embeddings develop thermodynamic and dynamic instabilities 
at exactly the same $T/{\bar M}$, just beyond that at which the first order dissociation transition occurs.


\begin{figure}[t]
\begin{center}
\includegraphics[width= \textwidth]{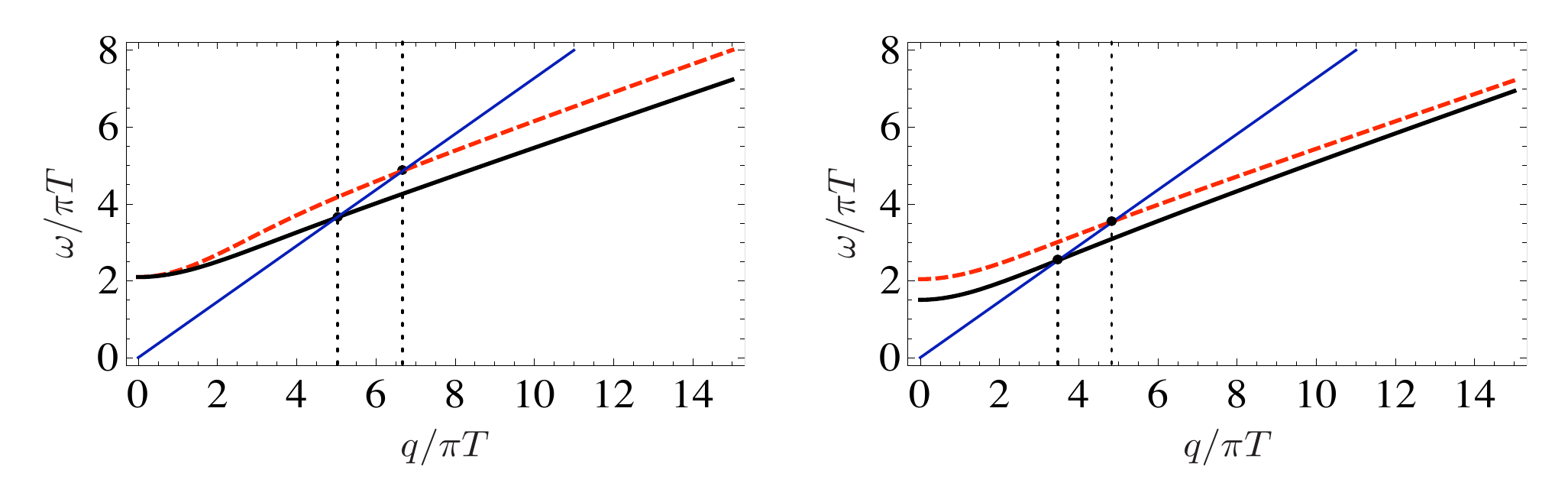}  
\end{center}
\caption{\small Left: Dispersion relation for the transverse (black, continuous curve) and longitudinal (red, dashed curve) $n=0$ modes of a heavy vector meson with $\vlim = 0.35$ in the ${\cal N}=4$ SYM plasma. The dual D7-brane has $m=1.3$, corresponding to a temperature just below $T_{\rm diss}$. 
Right: Analogous curves for a scalar (black, continuous curve) and pseudoscalar (red, dashed curve) meson. In both plots the blue, continuous straight lines correspond to $\omega = v q$ for some $v$ such that $\vlim<v \leq 1$. The black, dotted, vertical lines mark the crossing points between the meson dispersion relations and the blue lines. 
} 
\label{dispersion}
\end{figure}

\begin{figure}[t]
\begin{center}
\includegraphics[width=0.7 \textwidth]{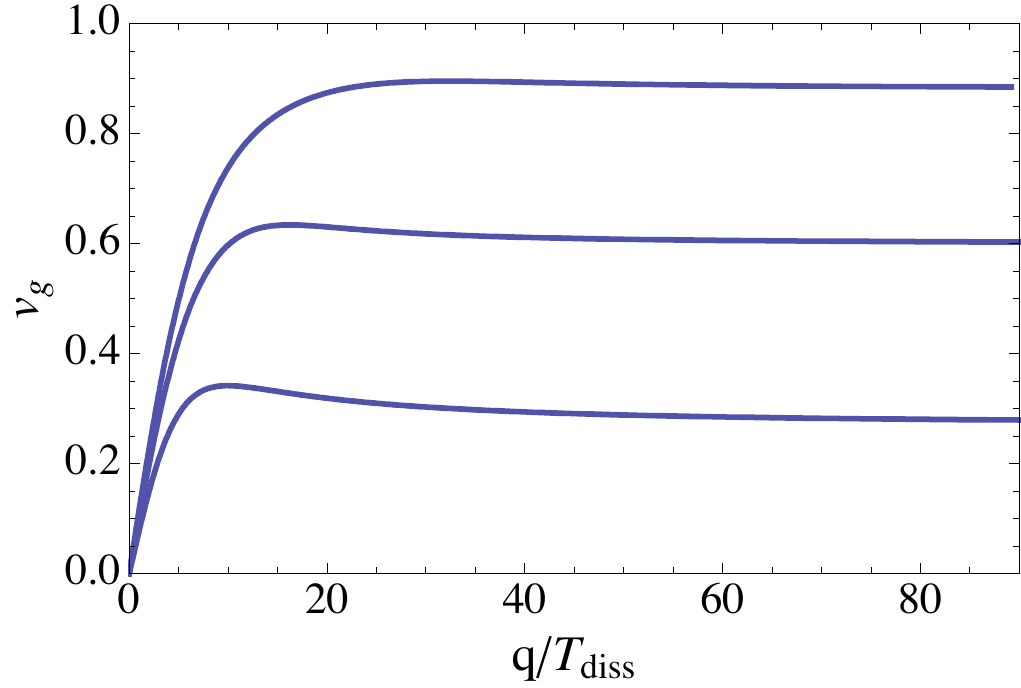}  
\end{center}
\caption{\small Group velocities $v_g $ for $n=0$ scalar meson modes with $T / \td \approx 0.65, 0.92$ and 1 from top to bottom. We see that $\vlim < 1$ decreases with increasing temperature. ($\vlim$ would approach zero if we included the unstable Minkowski embeddings with $T > \td$.) The group velocity approaches its large-$q$ value $\vlim$ from above, i.e. $v_g$ reaches a maximum before settling into the limiting velocity $\vlim$. The maximum exists also for the top curve even though it is less clearly visible. We will refer to the momentum at which $v_g$ reaches the maximum as $q_m$. Clearly $q_m$ decreases with temperature.}
\label{fig:group}
\end{figure}

We now turn to quarkonium 
mesons moving through the plasma, that is to modes with $q \neq 0$. The dispersion relation for scalar mesons was first computed in Ref.~\cite{Mateos:2007vn} and then revisited in Ref.~\cite{Ejaz:2007hg}. The dispersion relation for (transverse) vector mesons appeared in Ref.~\cite{CasalderreySolana:2009ch}. An exhaustive discussion of the dispersion relations for all these cases can be found in Ref.~\cite{CasalderreySolana:2010xh}. The result for the lowest-lying ($n=0$) vector, scalar and pseudoscalar quarkonia is shown in 
Fig.~\ref{dispersion}, taken from Ref.~\cite{CasalderreySolana:2010xh}. Fig.~\ref{fig:group}, taken from Ref.~\cite{Ejaz:2007hg}, shows the group velocity $v_g = d \om / d q$ for the $n=0$ scalar mesons at three different temperatures.

An important  feature of these plots is their behavior at large momentum. In this regime we find that $\omega$ grows linearly with $q$. Naively, one might expect that the constant of proportionality should be one. However, one finds instead that
\beq 
\omega = \vlim \, q \,,
\labell{fast}
\eeq
where $\vlim <1$ and where $\vlim$ depends on $m={\bar M}/T$ but at a given temperature is the same for all quarkonium modes. In the particular case of $m=1.3$, illustrated in 
Fig.~\ref{dispersion}, one has $\vlim \simeq 0.35$.
 In other words, there is a subluminal limiting velocity for quarkonium 
 mesons moving through the plasma.
 And, as illustrated in Fig.~\ref{fig:group}, one finds that the limiting velocity decreases with 
 increasing temperature.  Fig.~\ref{fig:group} 
also illustrates another generic feature of the dispersion relations, namely that 
the maximal group velocity  is attained at some $q_m < \infty$ and as $q$ is increased further
the group velocity approaches $\vlim$ from above.  Since $v_g$ at $q_m$ is not much greater than $\vlim$, we will not always distinguish between these two velocities.
We will come back to the physical interpretation of $q_m$ at the end of this subsection.
  
The existence of a subluminal limiting velocity, which was discovered in \cite{Mateos:2007vn} and subsequently elaborated upon in \cite{Ejaz:2007hg}, is easily understood from the perspective of the dual gravity description \cite{Mateos:2007vn,Ejaz:2007hg}. Recall that  mesonic states have wave functions supported on the D7-branes. Since highly energetic mesons are strongly attracted by the gravitational pull of the black hole, their wave-function is very concentrated at the bottom of the branes (see Fig.~\ref{BendingWithBH}). Consequently, their velocity is limited by the local speed of light at that point.  As seen by an observer at the boundary, this limiting velocity is
\beq 
\vlim =\left. \sqrt{- g_{tt}/g_{xx}}\right|_{\mt{tip}} \,,
\labell{seer}
\eeq
where $g$ is the induced metric on the D7-branes. 
Because of the black hole redshift, $\vlim$ is lower than the speed of light at infinity (i.e.~at the boundary), which is normalized to unity.  Note that, as the temperature increases, the bottom of the brane gets closer to the horizon and the redshift becomes larger, thus further 
reducing $\vlim$; this explains the temperature dependence in  Fig.~\ref{fig:group}.
In the gauge theory, the above translates into the statement that $\vlim$ is lower than the speed of light in the vacuum. The reason for this interpretation is that the absence of a medium in the gauge theory corresponds to the absence of a black hole on the gravity side, in which case $\vlim=1$ everywhere. Eqn.~\eqn{seer} yields $\vlim \simeq 0.35$ at $m=1.3$, in agreement with the numerical results displayed in Fig.~\ref{dispersion}.

\begin{figure}
\begin{tabular}{cc}
\includegraphics[width=0.45 \textwidth]{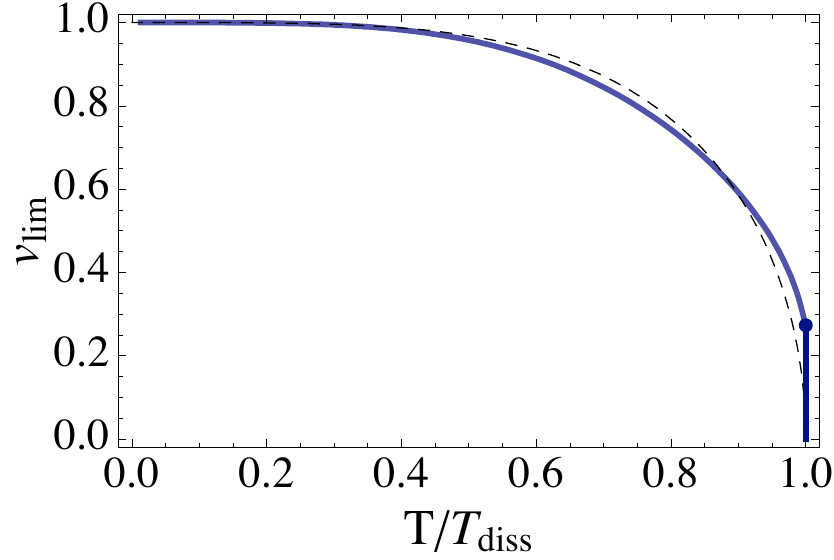}  
& \,\,\,\,
\includegraphics[width=0.5 \textwidth]{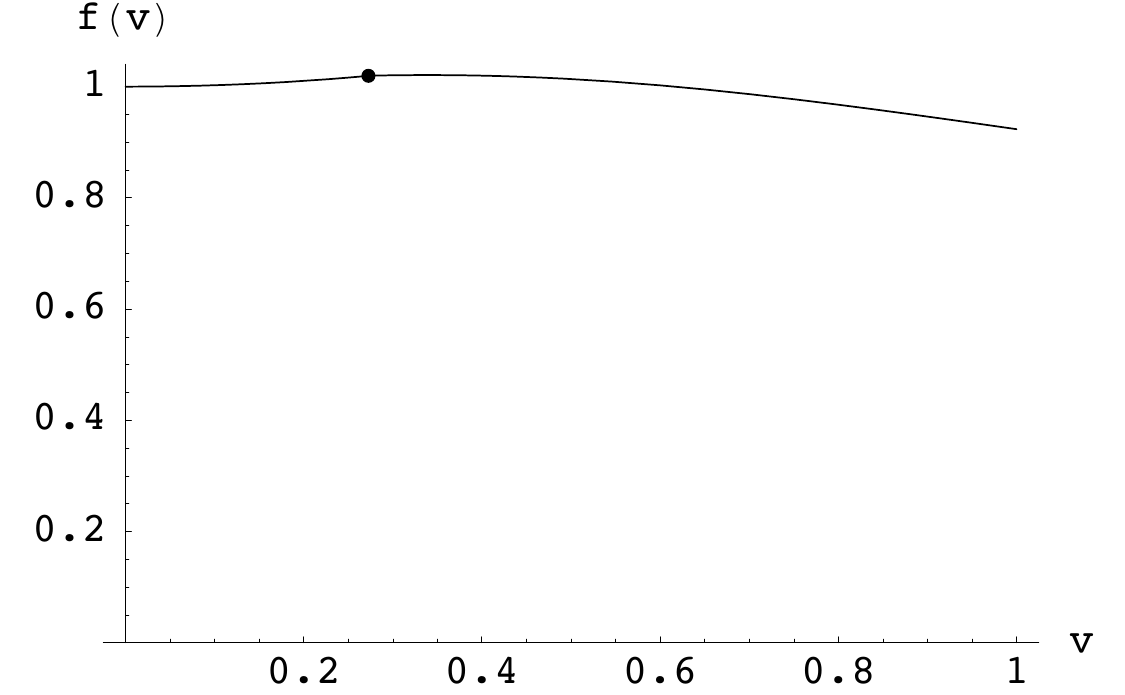} 
\end{tabular}
\caption{\small Left panel: The solid curve is the limiting velocity $\vlim$ as a function of $T/\td$, where $\td$ is the temperature of the dissociation transition at zero velocity. The dissociation transition occurs at the dot, where $v_{\rm lim}=0.27$.
The dashed curve is the approximation obtained by setting $f(v) = 1$ in Eqn.~\eqref{parafv}. Right panel: $f(v)$, the ratio of the solid and dashed curves in the left panel at a given $v$. We see that $f(v)$
is within a few percent of 1 at all velocities.} 
\label{fig:dis}
\end{figure}

It is also instructive to plot $\vlim$ as a function of $T/\td$, as done in Fig.~\ref{fig:dis}, taken from~\cite{Ejaz:2007hg}.
Although this curve was derived as a limiting meson velocity at a given temperature, it can also be read (by asking where it cuts horizontal lines rather than vertical ones) as giving $\td (v)$, the temperature below which mesons with a given velocity $v$ are found and above which no mesons with that velocity exist. In order to compare this result for $\td$ at all velocities to~\eqref{temsc}, one can parametrize the curve in Fig.~\ref{fig:dis} as
\be  \label{parafv}
\td(v) = f(v) (1-v^2)^{1/4} \, \td (0) \,.
\ee
In the left panel of Fig.~\ref{fig:dis}, the dashed line is obtained by setting $f(v) =1$, which is of course just~\eqref{temsc}. In the right panel, $f(v)$ is shown to be close to $1$ for all velocities, varying between 1.021 at its maximum and 0.924 at $v = 1$. Recall that the 
scaling~\eqref{temsc} was first obtained via the analysis of the potential between a moving test quark and antiquark, as described in Section~\ref{sec:HotWind}. The weakness of the dependence of $f(v)$ on $v$ is a measure of the robustness with which that simple scaling describes the velocity dependence of the dissociation temperature for quarkonium mesons in a fully dynamical calculation. In other words, to a good approximation $\vlim(T)$ can be determined by setting $v=\vlim$ on the right-hand side of  \eqref{temsc}, yielding 
\begin{equation}\label{temscInverse}
\vlim(T) \simeq \sqrt{1-\left(\frac{T}{T_{\rm diss}(v=0)}\right)^4} \ .
\end{equation}

Thus we reach a rather satisfactory picture that the subluminal limiting velocity~\eqref{fast} is in fact a manifestation in the physics of dynamical mesons of the 
velocity-enhanced screening of Section~\ref{sec:HotWind}. 
However, in the case of the low-spin mesons whose dynamics we are considering in this Section, there is an important addition to our earlier picture: Although the quarkonium mesons have a limiting velocity, they can nevertheless manage to remain bound at arbitrarily large momenta thanks to their modified dispersion relations. The latter allow the group velocity to remain less than $\vlim$, and consequently $\td (v)$ as given in~\eqref{temsc} to remains higher than $T$, all the way out to arbitrarily large momenta.
In other words, there exist meson bound states of arbitrarily large spatial momentum, but no matter how large the momentum the group velocity never exceeds $\vlim$. In this sense, low-spin mesons realize the first of two simple 
possibilities by which mesons may avoid exceeding 
$\vlim$.  A second possibility, more closely related to the analysis of Section~\ref{sec:HotWind},  is that meson states with momentum larger than a certain value simply cease to exist. This possibility is realized in the case of high-spin mesons. Provided $J \gg 1$, these mesons can be reliably described as long, semiclassical strings whose ends are attached to the bottom of the D7-branes. The fact that the endpoints do not fall on top of one another is of course due  to the fact that they are rotating around one another in such a way that the total angular momentum of the string is $J$. These type of mesons were first studied~\cite{Kruczenski:2003be} at zero temperature, and subsequently considered at nonzero temperature in Ref.~\cite{Peeters:2006iu}. These authors also studied the possibility that, at the same time that the endpoints of the string rotate around one another in a given plane, they also move with a certain velocity in the direction orthogonal to that plane. The result of the analysis was that, for a fixed spin $J$, string solutions exist only up to a maximum velocity $\vlim < 1$.

As we saw in Fig.~\ref{fig:group}, the group velocity of quarkonium mesons reaches a {\it maximum} at some value of the momentum $q=q_m$ before approaching the limiting value $\vlim$.  
There is a simple intuitive explanation for the location of $q_m$: it can be checked numerically 
that $q_m$ is always close to the `limiting momentum' $q_\mt{lim}$ that would follow
from~\eqref{temsc} if one assumes the {\it standard} dispersion relation for the meson. Thus, 
$q_m$ can  be thought of as a characteristic momentum scale where the velocity-enhanced screening effect  starts to be important. For the curves in Fig.~\ref{fig:group}, to the left of the maximum one finds approximately standard dispersion relations with a thermally corrected meson mass. To the right of the maximum, the dispersion relations approach the limiting 
behavior~\eqref{fast}, with $v_g$ approaching $\vlim$, as a consequence of the enhanced screening.

\subsection{Decay widths} 
\label{sec:MesonWidths}

We saw above that at $T < \td$ (Minkowski embeddings) there is a discrete and gapped spectrum of  absolutely stable quarkonium mesons, i.e.~the mesons have zero width.
The reason is that in this phase the D-branes do not touch the black hole horizon. Since the mesons'  wave-functions are supported on the branes, this means that the mesons cannot fall into the black hole. In the gauge theory this translates into the statement that the mesons cannot disappear into the plasma, which implies that the meson widths are strictly zero in the limit $\nc,\lam \to \infty$. This conclusion only depends on the topology of the Minkowski embedding. In particular, it applies even when higher-order perturbative corrections in $\apr$ are included, which implies that the widths of mesons should remain zero to all orders in the perturbative 
$1 / \sqrt{\lam}$ expansion.  In contrast, in the black hole phase the D-branes fall into the black hole and a meson has a nonzero probability of disappearing through the horizon, that is, into the plasma. As a consequence, we expect the mesons to develop thermal widths in the black hole phase, even in the limit $\nc,\lam \to \infty$. In fact, as we will see in Section~\ref{sec:BHQuasi}, the widths are generically comparable to the energies of the mesons, and hence the mesons can no longer be interpreted as quasiparticles. 

We thus encounter a somewhat unusual situation: the quarkonium mesons are absolutely stable for $T < \td$, but completely disappear for $T > \td$.  The former is counterintuitive because, on general grounds, we expect that any bound states should always have a nonzero width when immersed in 
a  medium with $T > 0$.
In the case of these mesons,  we expect that they can decay and acquire a width through the following channels: 

\ben 

\item  Decay to gauge singlets such as glueballs, lighter mesons, etc;  \label{en:1}

\begin{figure}[t]
\begin{center}
\includegraphics[width=0.8 \textwidth]{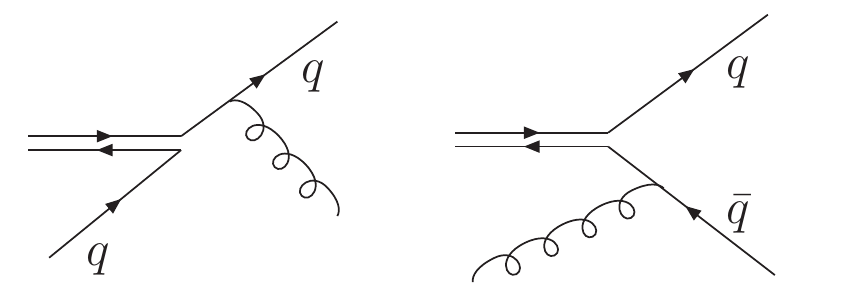}  
\end{center}
\caption{\small Sketches taken from Ref.~\cite{Faulkner:2008qk}  showing the relevant
thermal processes contributing to the meson width. $q$ ($\bar q$) denotes a quark (antiquark). The left diagram corresponds to the breakup of 
a meson by a quark from the  thermal medium, while the right diagram corresponds to breakup of a meson by an energetic gluon.
For large $\lambda$ the first process is dominant,
coming from the single instanton sector. }
\label{fig:feyn}
\end{figure}

\item  Breakup by high-energy gluons (right diagram in Fig.~\ref{fig:feyn}); \label{en:2}

\item  Breakup by thermal medium quarks (left diagram in Fig.~\ref{fig:feyn}). \label{en:3}
\een
Process \eqref{en:1} is suppressed by $1/\nc^2$ (glueballs) or $1/\nc$ (mesons), while~\eqref{en:2} and~\eqref{en:3} are unsuppressed in the large-$\nc$ limit.  Since \eqref{en:1} is also present in the vacuum, we will focus on~\eqref{en:2} and~\eqref{en:3}, which are medium effects. They are shown schematically  in Fig.~\ref{fig:feyn}.

The width due to~\eqref{en:2} is proportional to a Boltzmann factor $e^{-\beta E_B}$ for creating a gluon that is energetic enough to break up the bound state, while that due to~\eqref{en:3} is proportional to a Boltzmann factor $e^{-\beta M_\mt{th}}$ for creating a thermal quark, where $M_\mt{th}$ is the thermal mass of the quark --- see Fig.~\ref{BendingWithBH}. Given Eqn.~\eqref{distt} and the fact that in the Minkowski phase $T < \td$, both Boltzmann factors are suppressed by 
$e^{-\sqrt{\lam}} \sim 
e^{-R^2 / \apr}$, so we recover the result that these mesons are stable in the infinite-$\lam$ limit.  In particular, there is no width at any perturbative order in the $1/\sqrt{\lam}\,$--$\,$expansion,  consistent with the conclusion 
from the string theory side. Furthermore, since the binding energy is $E_B \approx 2 M_\mt{th}$, in the regime where $\lam$ is large (but not infinite), the width from process~\eqref{en:3} will dominate over that from  process~\eqref{en:2}.  

We now review the result from the string theory calculation of the meson widths in 
Ref.~\cite{Faulkner:2008qk}. As discussed above, the width is non-perturbative in $1 / \sqrt{\lam} \sim \apr / R^2$, and thus should correspond to some instanton effect on the string worldsheet. 
The basic idea is very simple:  even though in a Minkowski embedding the brane is separated from the black hole horizon and classically a meson living on the brane cannot 
fall into the black hole, quantum mechanically (from the viewpoint of the string worldsheet)
it has a nonzero probability of tunneling into the black hole and the meson therefore develops a width. At leading order, the instanton describing this tunneling process is given by 
a (Euclidean) string worldsheet stretching between the tip of the D7-brane to the black hole horizon (see Fig.~\ref{BendingWithBH}) and winding around the Euclidean time direction. 
Heuristically, such a worldsheet creates a small tunnel between the brane and the black hole through which 
mesons can fall into the black hole. The instanton action is $\beta M_\mt{th}$, as can be read off immediately from the geometric picture just described, and its exponential gives rise to the Boltzmann factor expected from process~\eqref{en:3}. From the gauge theory perspective, such an instanton can be interpreted as creating a thermal quark from the medium, and a meson can disappear into the medium via interaction with it as shown in the left diagram of Fig.~\ref{fig:feyn}. 

The explicit expression for the meson width due to such instantons is somewhat complicated, so we refer the reader to the original literature~\cite{Faulkner:2008qk}. Although the width appears to be highly model dependent and is exponentially small in the regime of a large but finite $\lam$ under consideration, remarkably its momentum dependence has some universal features at large momentum in~\cite{Faulkner:2008qk}. Specifically, one finds that
 \be \label{eq:ratio}
 {\Ga (q) \ov \Ga (0)} = {|\psi ({\rm tip}; \vec q)|^2 \ov |\psi ({\rm tip}; \vec q=0)|^2} \,,
 \ee
 where $\Ga (q)$ denotes the width of a meson with spatial momentum $q$ and $\psi ({\rm tip}; q)$ 
 its wave-function evaluated at the tip of the 
 D7-branes (i.e.~where it is closest to the black hole). This result is intuitively obvious because  a meson tunnels into the black hole from the tip of the branes. In particular, as discussed in detail  
 in Ref.~\cite{Ejaz:2007hg}, at large momentum $q$ the wave-function becomes localized around the tip of the brane and can be approximated by that of a spherical harmonic oscillator with a potential proportional to $q^2 z^2$, where $z$ is the proper distance from the tip of the branes.\footnote{Note that there are four transverse directions along the D7-brane as we move away from the tip (not including the other $(3+1)$ dimensions parallel to the boundary). Thus this is a four-dimensional harmonic oscillator.}
 It then immediately follows that for large $q$
the width (\ref{eq:ratio}) scales as $q^2$.  Furthermore for temperatures $T\ll \mmes$ and $q \gg \mmes^3 / T^2$, one finds the closed-form expression
\be
{\Ga_n (q) \ov \Ga_n (0)} \approx \frac{ 2 (4 \pi)^4}{(n+2) (n+3/2)} \frac{ T^4 q^2}{\mmes^6} \,,
\ee 
where $n$ labels different mesonic excitations (see~\eqref{spectre}). 

\begin{figure}
\begin{center}
\includegraphics[width=0.7 \textwidth]{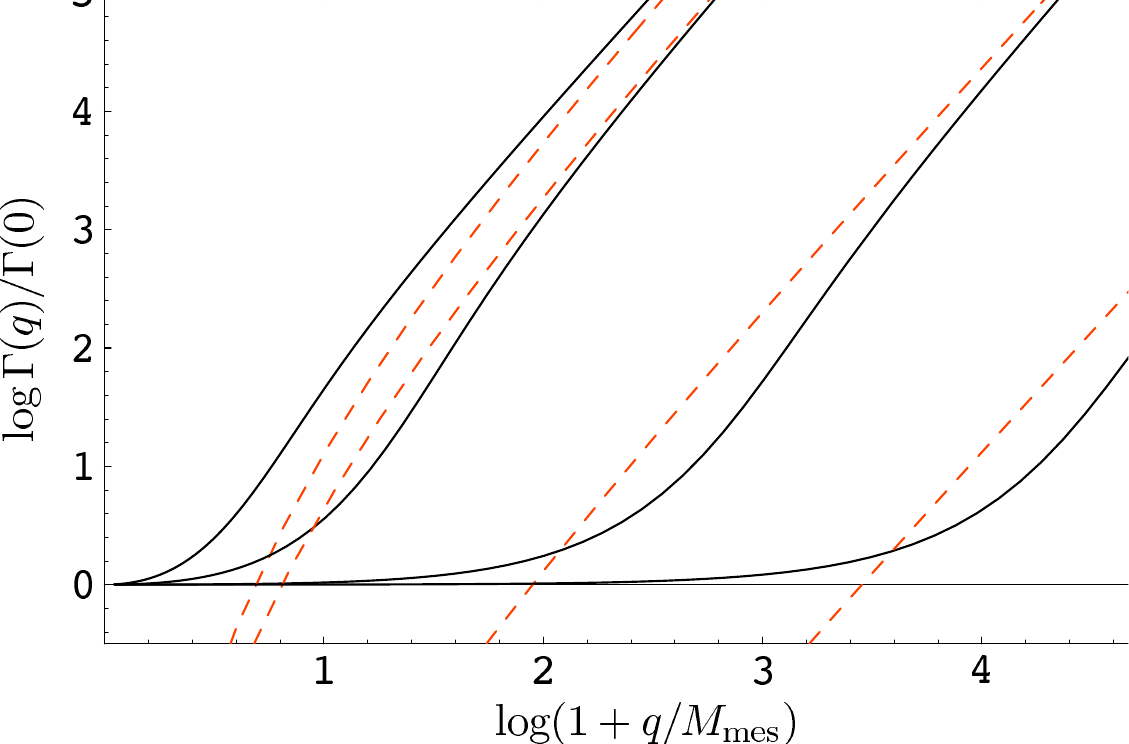}  
\end{center}
\caption{\small The behavior of the width as a function
of $q$ for $T/\td=0.99,\, 0.71,\, 0.3,\, 0.13$
from left to right. The dashed lines are analytic results for large momenta.
} 
\label{fig:width}
\end{figure}

It is also instructive to plot the full $q$-dependence of~\eqref{eq:ratio} obtained numerically, 
as done in 
Fig.~\ref{fig:width} (taken from~\cite{Faulkner:2008qk}) for $n=0$ mesons at various  temperatures.  Fig.~\ref{fig:width} has the interesting feature that the width is roughly constant for small $q$, but turns up quadratically
around $q/\mmes \approx 0.52 (\td/T)^2$. This is roughly the momentum $q_m \sim q_\mt{lim}$ at which the group velocity of a meson achieves its maximum in Fig.~\ref{fig:group} which,   
as discussed in Section~\ref{sec:mesdis}, can be considered as the characteristic momentum scale where velocity-enhanced screening becomes significant. 
This dramatic increase of meson widths beyond $q_m$ can also be understood intuitively:  when velocity-enhanced screening becomes significant, interaction between the quark and antiquark in a meson becomes further weakened, which makes it easier for a thermal medium quark or gluon to break it apart. 

We now briefly comment on the gravity description of process~\eqref{en:2} mentioned earlier, i.e.~the right diagram in Fig.~\ref{fig:feyn}. For such a process to happen the gluon should have an energy above the binding energy of the meson. The density of such gluons is thus suppressed by $e^{-2 \beta M_\mt{th}}$ and should be described by
an instanton and anti-instanton. We expect that contributions from such processes are also controlled by the the value of the meson wave-function at the tip of the branes, and thus likely have similar growth with momentum.

Finally, we note that, as $T$ increases, $M_\mt{th}$ decreases and thus the meson width increases quickly with temperature, but remains exponentially suppressed until $\td$ is reached, after which we are in the black hole phase. As will be discussed in Section~\ref{sec:BHQuasi}, in this phase quarkonium quasiparticles no longer exist.

\subsection{Connection with the quark-gluon plasma}
\label{sec:RemarksConnectionQGP}

Let us now recapitulate the main qualitative features regarding heavy mesons in a strongly coupled plasma:
\ben

\item They survive deconfinement. \label{en:11}

\item Their dispersion relations have  a subluminal limiting velocity at large momentum. 
The limiting velocity decreases with increasing temperature and 
as a result the motion of a meson with large momentum dramatically slows down near $\td$.
\label{en:22}

\item At large momenta, meson widths increase dramatically with momentum. 
\label{en:33}

\item The limiting velocity is reached and the increase in widths applies
when $q \gg q_\mt{lim}$, 
where $q_\mt{lim}$ is the `limiting' momentum following from~\eqref{temsc} if one 
assumes the {\it standard} dispersion relation.
\label{en:44}

\een

Properties~\eqref{en:11}--\eqref{en:33} are universal in the sense that they apply to the deconfined phase of any gauge theory with a string dual in the large-$\nc$, strong-coupling limit. The reason for this is that they are simple consequences of general geometric features 
following from two universal aspects of the gauge/string duality: (i) the fact that the deconfined phase of the gauge theory is described on the gravity side by a black hole geometry  \cite{Witten:1998zw}, and (ii) the fact that a finite number $\nf$ of quark flavours is described by $\nf$ D-brane probes \cite{Karch:2000gx,Karch:2002sh}.  Property~\eqref{en:44} was established by explicit numerical calculations in specific models. However, given that $q_\mt{lim}$ can be motived in a model-independent way from~\eqref{temsc}, it is likely to also be universal even though though this was not manifest in our discussion above.

We have seen that  properties~\eqref{en:22} and \eqref{en:33}  can be considered direct consequences of velocity-enhanced screening, which as discussed in Section~\ref{sec:HotWind} can have important implications for quarkonium suppression in heavy ion collisions.


It is interesting that properties~\eqref{en:11} and to some degree \eqref{en:22} can be independently motivated in QCD whether or not a string dual of QCD exists.
The original argument \cite{Matsui:1986dk} for~\eqref{en:11}  is simply that the heavier the quarkonium meson, the smaller its size. 
And, it is reasonable to expect a meson to remain bound until the screening length in the plasma becomes comparable to the meson size, and for sufficiently heavy quarkonia this happens at $\td > \tc$.  
As we have discussed in Sections~\ref{quarkonium} and \ref{secmeson}, this conclusion is supported by calculations of both the static quark-antiquark potential 
and of Minkowski-space spectral functions in lattice-regularized QCD. 
The ballpark estimate for the dissociation temperature of heavy mesons suggested by the above studies roughly agrees with that from the D3/D7 system. For example, for the $J/\psi$ meson the former estimate is 
$\tc \lesssim \td \lesssim 2 \tc$. Allowing for a certain range in the precise value of $150 \,\mbox{MeV} \lesssim \tc \lesssim 190 \, \mbox{MeV}$, this translates into 
$300 \, \mbox{MeV} \lesssim \td \lesssim 380 \, \mbox{MeV}$.  In the D3/D7 model, we see from 
Fig.~\ref{e} that meson states melt at 
$\td \simeq 0.766 \mbar$.
The scale $\mbar$ is related to the mass $M_\mt{mes}$ of the lightest meson in the theory at zero temperature through Eqn.~\eqn{mbarD3D7}. Therefore we have \mbox{$\td (M_\mt{mes}) \simeq 0.122 M_\mt{mes}$}. For the $J/\psi$, taking $M_\mt{mes} \simeq 3$ GeV gives $\td (J/\psi) \simeq 366$ MeV. Although it is gratifying that this comparison leads to qualitative agreement, it must be taken with some caution because meson bound states in the D3/D7 system are deeply bound, i.e.~$M_\mt{mes} \ll 2M_q$, whereas the binding energy of charmonium states in QCD is a small fraction of the charm mass, 
i.e.~$M_{c\bar{c}} \simeq 2 M_c$. An additional difference comes from the fact that
in QCD the dissociation of charmonium states is expected to happen sequentially, with excited states (that are larger) dissociating first, whereas in the D3/D7 system all meson states are comparable in size and dissociate at the same temperature. Presumably, in the D3/D7 system this is an artifact of the large-$\nc$, strong-coupling approximation under consideration, and thus corrections away from this limit should make holographic mesons dissociate sequentially too.

There is a simple (but incomplete) argument for property~\eqref{en:22} that applies to QCD just as well 
as to $\N=4$ SYM theory~\cite{Liu:2006nn,Peeters:2006iu,Chernicoff:2006hi,Ejaz:2007hg}: a meson moving through the plasma with velocity $v$ experiences a higher energy density, boosted by a factor of 
$\gamma^2$. Since energy density is proportional to $T^4$, this can be thought of as if the meson sees an effective temperature that is boosted by a factor of $\sqrt{\gamma}$, meaning $T_\mt{eff}(v) = (1-v^2)^{-1/4} T$. A velocity-dependent dissociation temperature scaling like (\ref{temsc}) follows immediately and from this a subluminal limiting 
velocity (\ref{temscInverse}) can be inferred.  Although this argument is seductive, it can be seen in several ways that it is incomplete.  For example, we would have reached a different $T_\mt{eff}(v)$ had we started by observing that the entropy density $s$ is boosted by a factor of $\gamma$ and is proportional to $T^3$.  And, furthermore, there really is no single effective temperature seen by the moving quarkonium meson.
The earliest analysis of quarkonia moving through a weakly coupled QCD plasma
with some velocity $v$ showed that the meson sees a blue-shifted temperature in some
directions and a red-shifted temperature in others~\cite{Chu:1988wh}.
Although the simple argument does not work by itself, it does mean that all we need from the calculations done via gauge/string duality is the result that $T_{\rm diss}(v)$ behaves as if it is controlled by the boosted energy density  --- \ie we need the full calculation only for the purpose of justifying the use of the particular simple argument that works.  This suggests 
that property~\eqref{en:22}, and in particular the scaling in Eqs.~(\ref{temsc}) and (\ref{temscInverse}), are general enough that they may apply to the quark-gluon plasma of QCD whether or not it has a gravity dual.



As explained towards the end of Section \ref{sec:mesdis}, there are at least two simple ways in which a limiting velocity for quarkonia may be implemented. It may happen that meson states with momentum above a certain $q_\mt{lim}$ simply do not exist, in which case one expects that 
$v_\mt{lim}=v(q_\mt{lim})$. The second possibility is that the dispersion relation of mesons may become dramatically modified beyond a certain $q_\mt{lim}$ in such a way that, although meson states of arbitrarily high momentum exist, their group velocity never exceeds a certain $v_\mt{lim}$. It is remarkable that both possibilities are realized in gauge theories with a string dual, the former by high-spin mesons and the latter by low-spin mesons. However, note that even in the case of 
low-spin mesons, $q_\mt{lim}$ remains the important momentum scale beyond which we expect more significant quarkonium suppression for two reasons. First, meson widths increase significantly for $q > q_\mt{lim}$. Therefore, although it is an overstatement to say that these mesons also cease to exist above $q_\mt{lim}$, their existence becomes more and more transient at higher and 
higher $q$.
Second, due to the modified dispersion relation mesons with $q > q_\mt{lim}$ slow down and they spend a longer time in the medium, giving the absorptive imaginary part more time to cause dissociation.
It will be very interesting to see whether future measurements at RHIC or the LHC will show the suppression of $J/\Psi$ or $\Upsilon$ production increasing markedly above some threshold transverse momentum $p_T$, as we described in 
Section~\ref{sec:HotWind}.

In practice, our ability to rigorously verify the properties~\eqref{en:11}--\eqref{en:44} in QCD is limited due to the lack of tools that are well-suited for this purpose. It is therefore reassuring that they hold for all strongly coupled, large-$\nc$ plasmas with a gravity dual, for which the gravity description provides just such a tool.


\section{Black hole embeddings}
\label{sec:BHQuasi}

We now consider the phase $T > \td$, which is described by a D7-brane with a black hole embedding.  We will give a qualitative argument that in this regime the system {\it generically} contains no quarkonium quasiparticles.  We have emphasized the word `generically' because exceptions arise when certain large ratios of physical scales are introduced `by hand', as we will see later. We will illustrate the absence of quasiparticles in detail by computing a spectral function of two electromagnetic currents in the next subsection.

\subsection{Absence of quasiparticles}
\label{absence}
In the gravity description of physics at $T>T_{\rm diss}$, the meson widths may be seen by studying so-called quasi-normal modes. These are analogues of the fluctuations we studied in the case of Minkowski embeddings in that a normalizable fall-off is imposed at the boundary. However, the regularity condition at the tip of the branes is replaced by the so-called in-falling boundary condition at the horizon. Physically, this is the requirement that energy can flow into the horizon but cannot come out of it (classically). Mathematically, it is easy to see that this boundary condition forces the frequency of the mode to acquire a negative imaginary part, and thus corresponds to a nonzero meson width. The meson in question may then be considered a quasiparticle if and only if this width is much smaller than the real part of the frequency. In the case at hand, the meson widths increase as the area of the induced horizon on the branes increases, and go to zero only when the horizon shrinks to zero size. This is of course to be expected, since it is the presence of the induced horizon that causes the widths to be nonzero in the first place. We are thus led to the suggestion that meson-like quasiparticles will be present  in the black hole phase only when the size of the induced horizon on the branes can be made parametrically small. This expectation can be directly verified by explicit calculation of the quasi-normal modes on the branes~\cite{Hoyos:2006gb,Evans:2008tv,Kaminski:2009ce,Kaminski:2009dh}, and we will confirm it indirectly below by examining the spectral function of two electromagnetic currents. For the moment, let us just note that this condition is not met in the system under consideration because, as soon as the phase transition at $T = \td$ takes place, the area of the induced horizon on the brane is an order-one fraction of the area of the background black hole. This can be easily seen from Fig.~\ref{e} by comparing the entropy density (which is a measure of the horizon area) at the phase transition to the entropy density at 
asymptotically high temperatures: 
\be
\frac{s_\mt{transition}}{s_\mt{high\, T}} \approx \frac{0.3}{2}\approx 15 \% \,.
\ee
This indicates that there is no \emph{parametric} reason to expect quasiparticles with narrow widths above the transition.  We shall confirm by explicit calculation in the next subsection that there are no quasiparticle excitations in the black hole phase.

\subsection{Meson spectrum from a spectral function}
\label{sec:BHMesonSpectrum}
Here we will illustrate some of the general expectations discussed above by examining the in-medium spectral function of two electromagnetic currents in the black hole phase. We choose this particular correlator because it is related to thermal photon production, which we will discuss in the next Section. We will see that no narrow peaks exist for stable black hole embeddings, indicating the absence of long-lived quasiparticles. These peaks will appear, however, as we artificially push the system into the unstable region close to the critical embedding (see Fig.~\ref{transition}), thus confirming our expectation that quasiparticles should appear as the area of the induced horizon on the branes shrinks to zero size.

\begin{figure}
    \begin{center}
    	\includegraphics[width=0.75\textwidth]{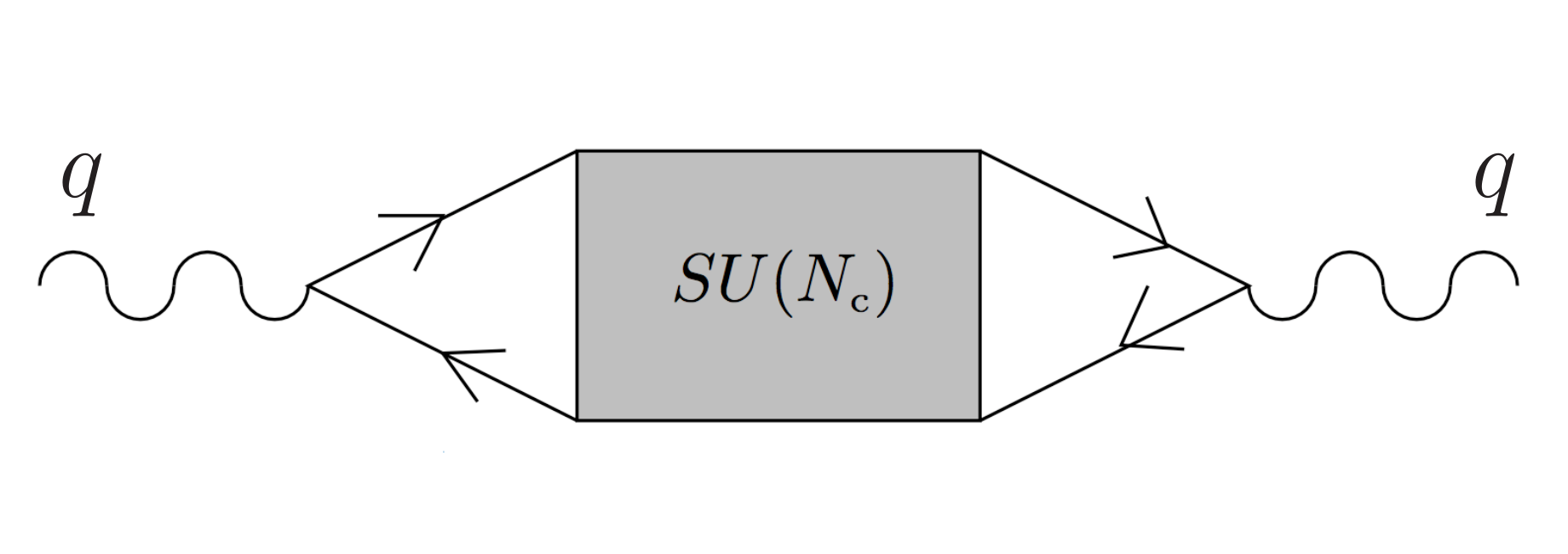}
    \end{center}
\caption{\small Diagrams contributing to the two-point function of electromagnetic currents. The external line corresponds to a photon of momentum $q$. As explained in the text, to leading order in the electromagnetic coupling constant only $SU(\nc)$ fields `run' in the loops represented by the shaded blob.} 
\label{loops} 
\end{figure}

${\cal N} =4$ SYM coupled to $\nf$ flavours of equal-mass quarks is an $SU(\nc)$ gauge theory with a global $U(\nf)$ symmetry. In order to couple this theory to electromagnetism we should gauge a 
$\uem$ subgroup of $U(\nf)$ by adding a dynamical photon $A_\mu$ to the theory; for simplicity we will assume that all quarks have equal electric charge, in which case $\uem$ is the diagonal subgroup of $U(\nf)$. In this extended theory we could then compute correlation functions of the conserved current $\jem_\mu$ that couples to the $\uem$ gauge field. The string dual of this $SU(\nc) \times \uem$ gauge theory is unknown, so we cannot perform this calculation holographically. However, as noted in \cite{CaronHuot:2006te}, we can perform it to leading order in the electromagnetic coupling constant $e$, because at this order correlation functions of electromagnetic currents in the gauged and in the ungauged theories are identical. This is very simple to understand diagramatically, as illustrated for the two-point function in Fig.~\ref{loops}. In the ungauged theory only $SU(\nc)$ fields `run' in the loops, represented by the shaded blob. The gauged theory contains additional diagrams in which the photon also runs in the loops, but these necessarily involve more photon vertices and therefore contribute at higher orders in $e$.  Thus one can use the holographic description to compute the `$SU(\nc)$ blob' and obtain the result for the correlator to leading order in $e$.  

Using this observation, the authors of  Ref.~\cite{CaronHuot:2006te} first did a holographic computation of  the spectral density of two R-symmetry currents in ${\cal N}=4$ SYM theory, 
to which finite-coupling corrections were computed in \cite{Hassanain:2009xw,Hassanain:2010fv}. 
The result for the R-charge spectral density is identical, up to an overall constant, with the spectral density of two electromagnetic currents in ${\cal N}=4$ SYM theory coupled to massless quarks. This, and the extension to nonzero quark mass, were obtained in Ref.~\cite{Mateos:2007yp}, which we now follow. 

The relevant spectral function is defined as
\be
\chi_{\mu\nu} (k) = 2 \, \mbox{Im} \, G^\mt{R}_{\mu\nu} (k) \,,
\label{spectral}
\ee
where $k^\mu=(\omega, q)$ is the photon null momentum (i.e.~$\omega^2= q^2$) and 
\be
G^\mt{R}_{\mu \nu} (k) =  i \int d^{d+1} x \, e^{- i k_\mu x^\mu} \, \Theta (t) \langle [ \jem_\mu (x), \jem_\nu (0) ] \rangle 
\label{green}
\ee
is the retarded correlator of two electromagnetic currents. The key point in this calculation is to identify the field in the string description that is dual to the operator of interest here, namely the conserved current $\jem_\mu$. We know from the discussion in Section \ref{AdS/CFT} that conserved currents are dual to gauge fields on the string side. Moreover, since $\jem_\mu$ is constructed out of fields in the fundamental representation, we expect its dual field to live on the D7-branes. The natural (and correct) candidate turns out to be the $U(1)$ gauge field associated with the diagonal subgroup of the $U(\nf)$ gauge group living on the worldvolume of the $\nf$ D7-branes. Once this is established, one must just follow the general prescription explained in Section \ref{AdS/CFT}. The technical details of the calculation can be found in Ref.~\cite{Mateos:2007yp}, so here we will only describe the results and their interpretation. In addition, we will concentrate on the trace of the spectral function, $\chi^\mu_{\,\, \mu} (k) \equiv \eta^{\mu\nu} \chi_{\mu\nu} (k)$, since this is the quantity that determines the thermal photon production by the plasma (see next section). 

\begin{figure}
    \begin{center}
    	\includegraphics[width=0.75\textwidth]{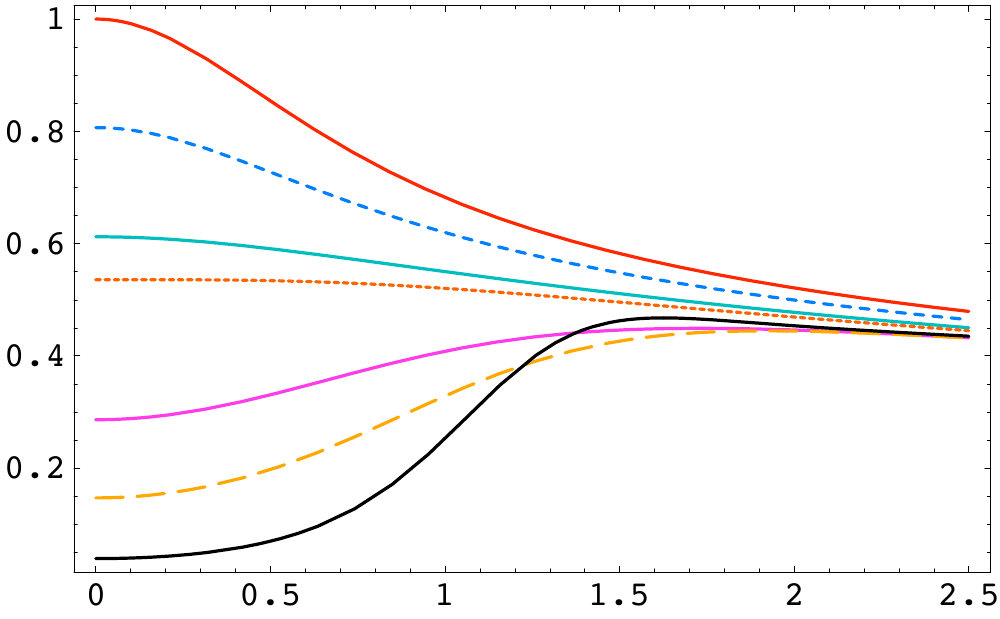}
	\put(-175,-15){$\bar{\omega}=\omega / 2 \pi T$}
	\put(-375,110){\large{$\frac{\chi^\mu_{\,\,\mu}(\bar{\omega})}{8 \tilde{\cal N}_\mt{D7} \bar{\omega}}$}}
    \end{center}
\caption{\small Trace of the spectral function as a function of the dimensionless frequency $\bar{\omega}=\omega/2\pi T$ for (from top to bottom on the left-hand side) $m=\{0, 0.6, 0.85, 0.93, 1.15, 1.25, 1.32 \}$. The last value corresponds to that at which the phase transition from a black hole to a Minkowski embedding takes place. Recall that 
$\tilde{\cal N}_\mt{D7}=  \nf \nc T^2/4$.} 
\label{D3D7wplot}
\end{figure}

The trace of the spectral function for stable black hole embeddings is shown in Fig.~\ref{D3D7wplot} for several values of the quark mass $m$. 
Note that this is a function of only one variable, since for an on-shell photon $\omega=q$. The normalisation constant that sets the scale on the vertical axis is 
$\tilde{\cal N}_\mt{D7} = \nf \nc T^2 /4$.  The $\nf \nc$ scaling of the spectral function reflects the number of electrically charged degrees of freedom in the plasma; in the case of two R-symmetry currents, $\nf \nc$ would be replaced by 
$\nc^2$ \cite{CaronHuot:2006te}. All curves decay as 
$\omega^{-1/3}$ for large frequencies. Note that $\chi \sim \omega$ as $\omega \rightarrow 0$. This is consistent with the fact that the value at the origin of each of the curves yields the electric conductivity of the plasma at the corresponding quark mass, namely
\be
\sigma = \frac{e^2}{4} \lim_{\omega \rightarrow 0} \frac{1}{\omega} 
\eta^{\mu\nu} \chi_{\mu\nu} (\omega=q)\,.
\label{null}
\ee
This formula is equivalent to the 
perhaps-more-familiar expression in terms of the zero-frequency limit of the spectral function at vanishing spatial momentum (see Appendix~\ref{sec:Green-Kubo}):
\be
\sigma = \frac{e^2}{6} \lim_{\omega \rightarrow 0} \frac{1}{\omega} 
\delta^{ij} \chi_{ij} (\omega, q=0) = 
\frac{e^2}{6} \lim_{\omega \rightarrow 0} \frac{1}{\omega} 
\eta^{\mu\nu} \chi_{\mu\nu} (\omega, q=0)\,,
\label{normal}
\ee
where in the last equality we used the fact that $\chi_{00}(\omega \neq 0,q=0)=0$, as implied by the Ward identity $k^\mu \chi_{\mu\nu} (k)= 0$. To see that the two expressions \eqn{null} and \eqn{normal} are equivalent, suppose that $q$ points along the 1-direction. Then the Ward identity, together with the symmetry of the spectral function under the exchange of its spacetime indices, imply that $\omega^2  \chi_{00} = q^2 \chi_{11}$. For null momentum this yields $- \chi_{00} + \chi_{11} = 0$, so we see that  Eqn.~\eqn{null} reduces to
\be
\sigma = \frac{e^2}{4} \lim_{\omega \rightarrow 0} \frac{1}{\omega} 
\Big[ \chi_{22}(\omega=q) + \chi_{33}(\omega=q) \Big] \,.
\label{null2}
\ee
The diffusive nature of the hydrodynamic pole of the correlator implies that at low frequency and momentum the spatial part of the spectral function behaves as
\be
\chi_{ij}(\omega, q)  \sim \frac{\omega^3}{\omega^2 + D^2 q^4} \,,
\ee
where $D$ is the diffusion constant for electric charge. This means that we can replace $q=\omega$ by $q=0$ in Eqn.~\eqn{null2}, thus arriving at expression \eqn{normal}.

\begin{figure}
    \begin{center}
    	\includegraphics[width=0.75\textwidth]{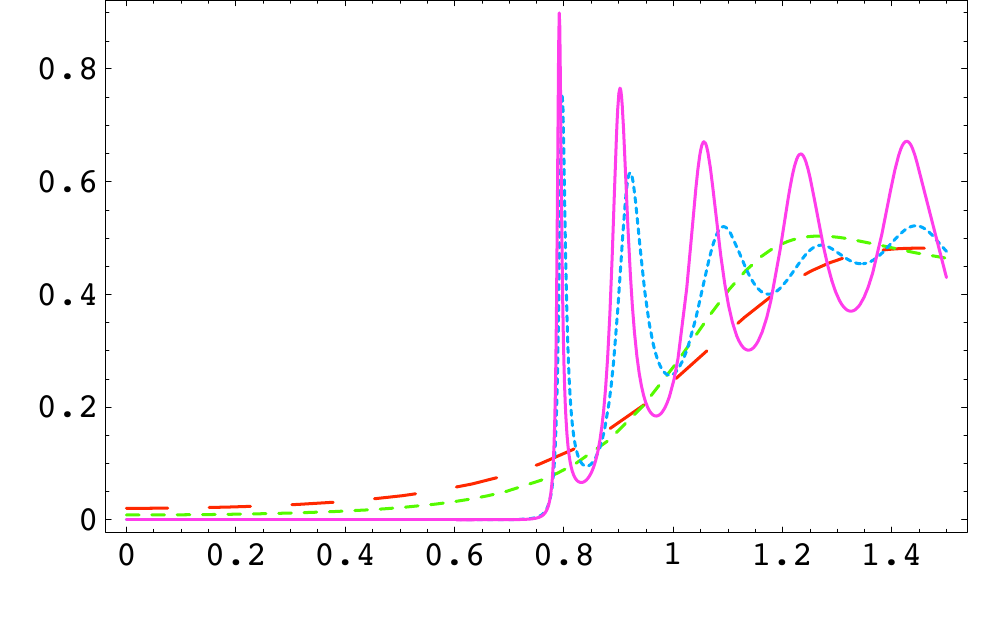}
	\put(-180,-1){$\bar{\omega}=\omega / 2 \pi T$}
	\put(-365,110){\large{$\frac{\chi^\mu_{\,\,\mu}(\bar{\omega})}{8 \tilde{\cal N}_\mt{D7} \bar{\omega}}$}}
       \end{center}
\caption{\small Trace of the spectral function as a function of the dimensionless frequency $\bar{\omega}=\omega/2\pi T$ for non-stable black hole embeddings. 
Curves with higher, narrower peaks correspond to embeddings that are closer to the critical embedding.} 
\label{D3D7wplotbis}
\end{figure}

For the purpose of our discussion, the most remarkable feature of the spectral functions displayed in Fig.~\ref{D3D7wplot} is the absence of any kind of high, narrow peaks that may be associated with a quasiparticle excitation in the plasma. This feature is shared by thermal spectral functions of other operators on stable black hole embeddings. We thus confirm our expectation that no quasiparticles exist in this phase. In order to make contact with the physics of the Minkowski phase, in which we do expect the presence of quarkonium quasiparticles, the authors of Ref.~\cite{Mateos:2007yp} computed the spectral function for black hole embeddings beyond the phase transition, i.e.~in the region below $T_{\rm diss}$  in which these embeddings are metastable or unstable. The results for the spectral function are shown in Fig.~\ref{D3D7wplotbis}. 
The most important feature of these plots is the appearance of well defined peaks in the spectral function, which become higher and narrower, seemingly approaching delta-functions, as the embedding approaches the critical embedding (see Fig.~\ref{transition}). Thus the form of the spectral function appears to approach the form we expect for Minkowski embeddings,\footnote{An analogous result was found in \cite{Myers:2007we} for time-like momenta.} namely an infinite sum of delta-functions supported at a discrete set of energies $\omega^2= q^2$.  (However, a precise map between the peaks in Fig.~\ref{D3D7wplotbis} and the meson spectrum in a Minkowski embedding is not easy to establish \cite{Paredes:2008nf}.)
 Each of these delta-functions is associated with a meson mode on the D7-branes with {\it null} four-momentum. The fact that the momentum is null may seem surprising in view of the fact that, as explained above, the meson spectrum in the Minkowski phase possesses a mass gap, but in fact it follows from the dispersion relation for these mesons displayed in Fig.~\ref{dispersion}. To see this, consider the dispersion relation $\omega (q)$ for a given meson in the Minkowski phase. The fact that there is a mass gap means that $\omega > 0$ at $q =0$. On the other hand, in the limit of infinite spatial momentum, $q \rightarrow \infty$, the dispersion relation takes the form $\omega \simeq \vlim q$ with $\vlim < 1$. Continuity then implies that there must exist a value of $q$ such that $\omega (q) = q$. This is illustrated in Fig.~\ref{dispersion} by the fact that the dispersion relations intersect the blue lines. 
Since in the Minkowski phase the mesons are absolutely stable in the large-$\nc$, strong-coupling limit under consideration, we see that each of them gives rise to a delta-function-like (i.e.~zero-width) peak in the spectral function of electromagnetic currents at null momentum. Below we will see some potential implications of this result for heavy-ion collisions.

\section{Two universal predictions}
\label{sec:Peak}

We have just seen that the fact that heavy mesons remain bound in the plasma, and the fact that their limiting velocity is subluminal, imply  that the dispersion relation of a heavy meson must cross the light-cone, defined by $\omega=q$, at some energy $\omega=\opeak$ indicated by the vertical line in Fig.~\ref{dispersion}.  In this section we will see that this simple observation leads to two universal consequences. Implications for deep inelastic scattering have been studied in \cite{Iancu:2009py} but will not be reviewed here.

\subsection{A meson peak in the thermal photon spectrum}

\begin{figure}[t]
\begin{center}
\includegraphics[scale=.45]{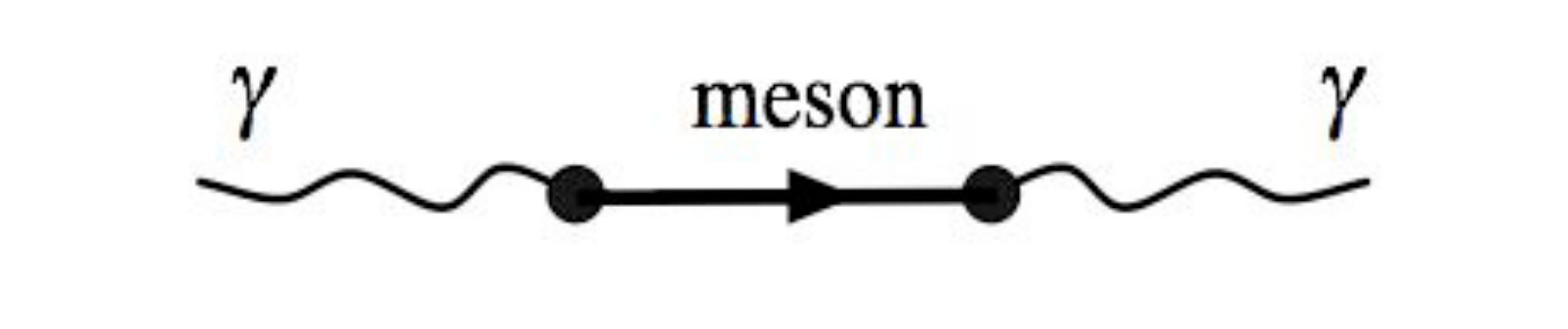}
\end{center}
\caption{\small In-medium vector meson--photon mixing. The imaginary part of this diagram yields the meson decay width into photons.}
\label{decay}
\end{figure}

At the crossing point between the meson dispersion relation and the light-cone, the meson four-momentum is null, that is $\omega_\mt{meson}^2 =q_\mt{meson}^2$. If the meson is flavourless and has spin one, then at this point its quantum numbers are the same as those of a photon. Such a meson can then decay into an on-shell photon, as depicted in Fig.~\ref{decay}. Note that, in the vacuum, only the decay into a virtual photon would be allowed by kinematics. In the medium, the decay can take place because of the modified dispersion relation of the meson. Also, note that the decay will take place unless the photon-meson coupling vanishes for some reason (e.g.~a symmetry). No such reason is known in QCD. 

The decay process of Fig.~\ref{decay} contributes a resonance peak, at a position $\omega=\opeak$, to the in-medium spectral function of two electromagnetic currents \eqn{green} evaluated at null-momentum $\omega = q$. This in turn produces a peak in the spectrum of thermal photons emitted by the plasma,
\be
\frac{dN_\gamma}{d\omega} \sim e^{-\omega/T} \, \chi^\mu_{\,\,\,\mu}  (\omega,T) \,.
\ee
The width of this peak is the width of the meson in the plasma. 

The analysis above applies to an infinitely-extended plasma at constant temperature.  Assuming that these results can be extrapolated to QCD, a crucial question is whether a peak in the photon spectrum could be observed in a heavy ion collision experiment. Natural heavy vector mesons to consider are the $J/\psi$ and the $\Upsilon$, since these are expected to survive deconfinement. We wish to compare the number of photons coming from these mesons to the number of photons coming from other sources. Accurately calculating the meson contribution would require a precise theoretical understanding of the dynamics of these mesons in the quark-gluon plasma, which at present is not available. Our goal will therefore be to estimate the order of magnitude of this effect with a simple recombination model. The details can be found in  Ref.~\cite{CasalderreySolana:2008ne}, so here we will only describe the result for heavy ion collisions at LHC energies. 

\begin{figure}
\begin{center}
\includegraphics[scale=0.35]{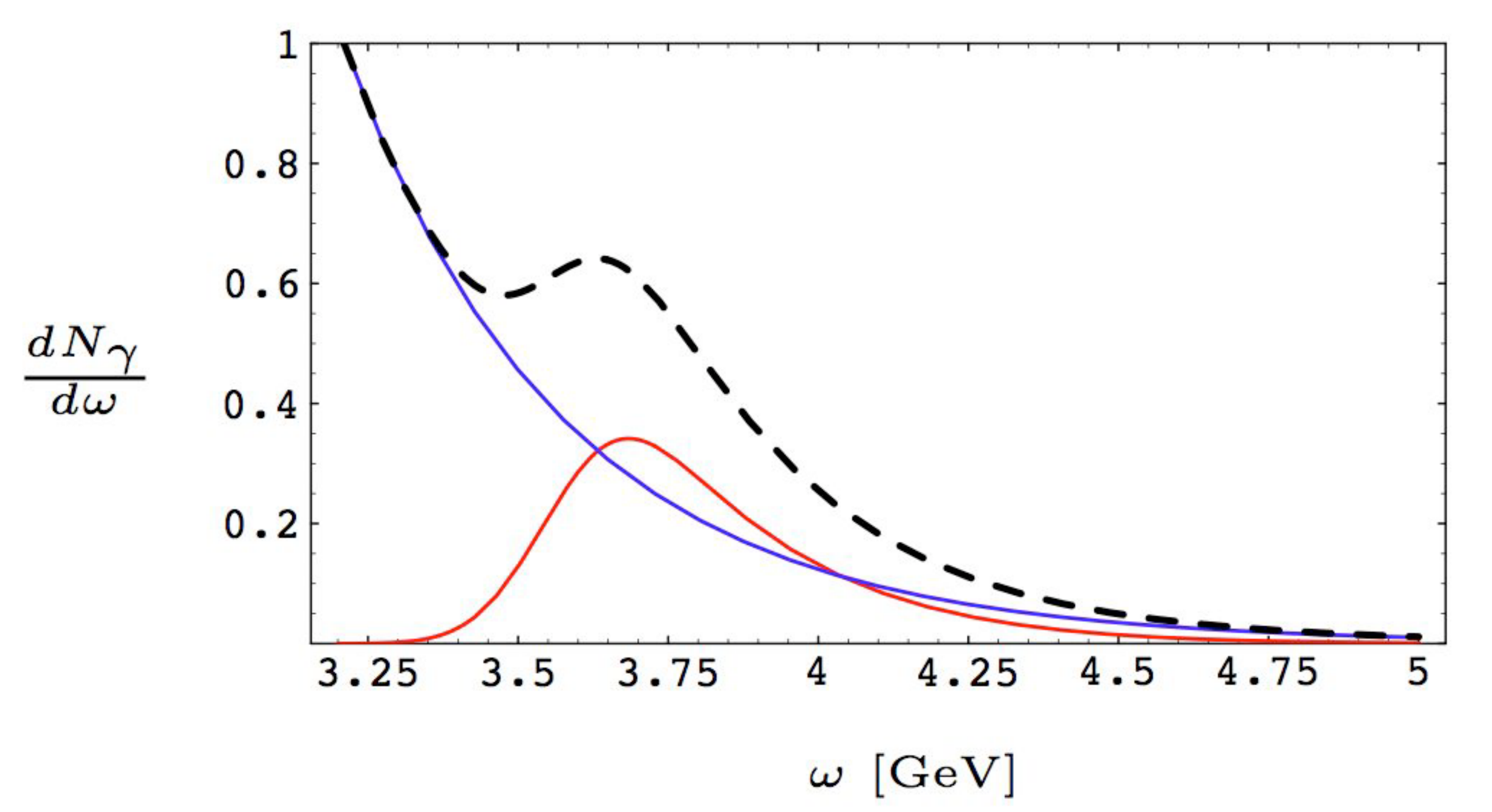} 
\end{center}
\caption{\small Thermal photon spectrum for LHC energies, $\td=1.25\, T_c$ and $M_\mt{charm}=1.7$~GeV. The (arbitrary) normalization is the same for all curves. The continuous, monotonically decreasing, blue curve is the background from light quarks. The continuous, red curve is the signal from $\jpsi$ mesons. The dashed, black curve is the sum of the two. See Ref.~\cite{CasalderreySolana:2008ne} for details.}
\label{spectrumFIG}
\end{figure}

The result is summarized in Fig.~\ref{spectrumFIG}, which shows the thermal photon spectrum coming from light quarks, the contribution from $\jpsi$ mesons, and the sum of the two, for a thermal charm mass $M_\mt{charm}=1.7$ GeV and a $\jpsi$  dissociation temperature $\td=1.25 T_c$.
Although the value of $M_\mt{charm}$ is relatively high, the values of $M_\mt{charm}$ and 
$\td$ are within the range commonly considered in the literature. For the charm mass a typical range is $1.3 \leq M_\mt{charm} \leq 1.7$ GeV because of a substantial thermal contribution --- see 
e.g. Ref.~\cite{Mocsy:2005qw} and Refs.~therein. The value of $\td$ is far from settled, but a typical range is $\tdec \leq \td \leq 2 \tdec$ \cite{Kaczmarek:2004gv,Petreczky:2004pz,Kaczmarek:2005ui,Mocsy:2007yj,Mocsy:2008eg, Asakawa:2003re,Datta:2003ww,Hatsuda:2005nw,Karsch:2005nk,Satz:2006kba,Asakawa:2002xj}. We have chosen these values for illustrative purposes, since they lead to an order-one enhancement in the spectrum. We emphasize, however, that whether this photon excess manifests itself as a peak, or only as an enhancement smoothly distributed over a broader range of frequencies, depends sensitively on these and other parameters. Qualitatively, the dependence on the main ones is as follows. Decreasing the quark mass decreases the magnitude of the $\jpsi$ contribution. Perhaps surprisingly, higher values of $\td$ make the peak less sharp. The in-medium width of the $\jpsi$ used in 
Fig.~\ref{spectrumFIG} was 100 MeV. Increasing this by a factor of two turns the peak into an enhancement. 
Crucially, the $\jpsi$ contribution depends quadratically on the $\ccbar$ cross-section. Since at RHIC energies this is believed to be ten times smaller than at LHC energies, the enhancement discussed above is presumably unobservable at RHIC. 

These considerations show that a precise determination of the enhancement is not possible without a very detailed understanding of the in-medium dynamics of the $\jpsi$. On the other hand, they also illustrate that there exist reasonable values of the parameters for which this effect yields an order-one enhancement, or even a peak, in the spectrum of thermal photons produced by the quark-gluon plasma. This thermal excess is concentrated at photon energies roughly between 3 and 5 GeV. In this range the number of thermal photons in heavy ion collisions at the LHC is expected to be comparable to or larger than that of photons produced in initial partonic collisions that can be described using perturbative QCD~\cite{Arleo:2007ms}. Thus, we expect the thermal excess above to be observable even in the presence of the pQCD background. 

The authors of Ref.~\cite{CasalderreySolana:2008ne} also examined the possibility of an analogous effect associated with the $\Upsilon$ meson, in which case $\opeak \sim 10$ GeV. At these energies the number of thermal photons is very much smaller than that coming from initial partonic collisions, so an observable effect is not expected.

\subsection{A new mechanism of quark energy loss: Cherenkov emission of mesons}
\label{sec:Cherenkov}
\begin{figure}
\begin{center}
\includegraphics[scale=.65]{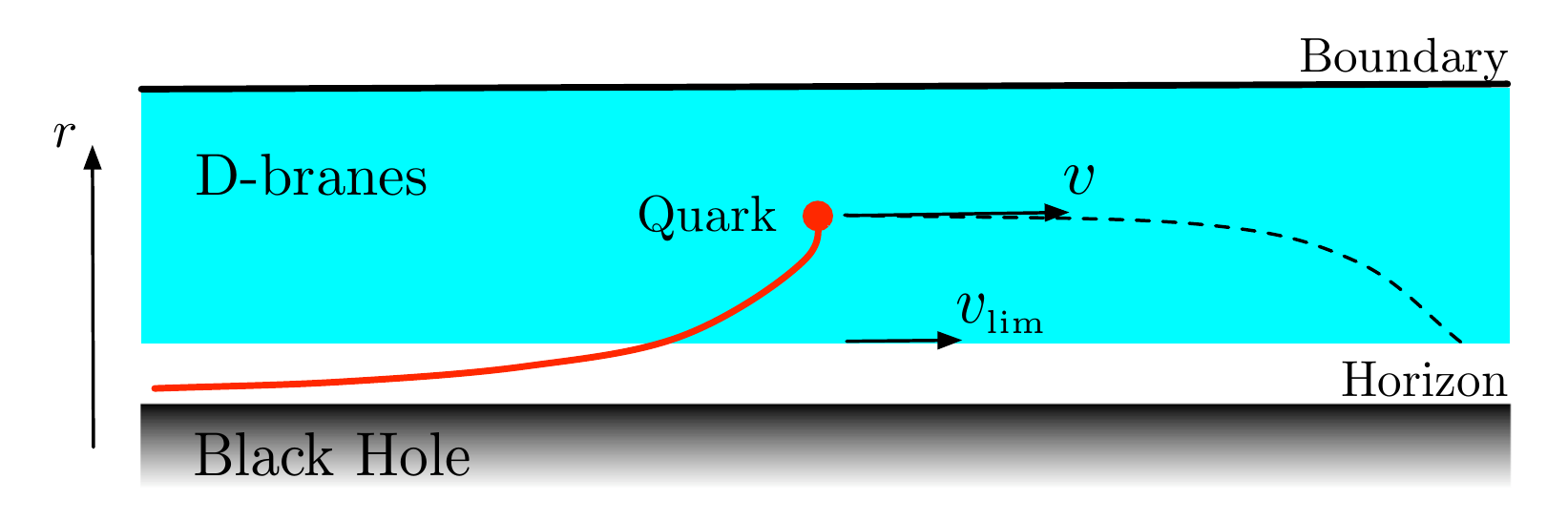}
\end{center}
\caption{\small D-branes and open string in a black brane geometry.}
\label{cherenkov-with-quark}
\end{figure}

We now turn to another universal prediction that follows from the existence of a subluminal limiting velocity for mesons in the plasma. Consider a highly energetic quark moving through the plasma. In order to model this we consider a string whose endpoint moves with an arbitrary velocity $v$ at an arbitrary radial position $r_q$ --- see Fig.~\ref{cherenkov-with-quark}. Roughly speaking, the interpretation of $r_q$ in the gauge theory is that of the inverse size of the gluon cloud that dresses the quark. This can be seen, for example, by holographically computing the profile of $\langle \mbox{Tr} F^2(x) \rangle$ around a static quark source dual to a string whose endpoint sits at $r=r_q$ \cite{Hovdebo:2005hm}.

\begin{figure}
\begin{center}
\includegraphics[scale=.55]{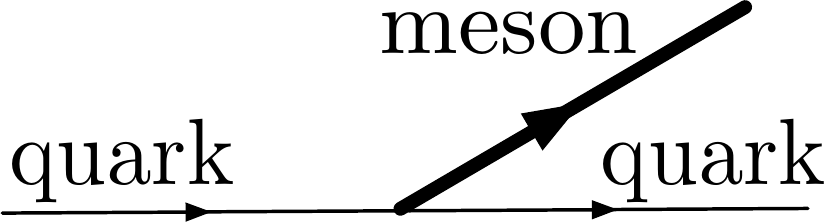}
\end{center}
\caption{\small Effective quark-meson coupling.} 
\label{quark-meson-coupling}
\end{figure}

Two simple observations now lead to the effect that we are interested in. The first one is that the string endpoint is charged under the scalar and vector fields on the branes. In the gauge theory, this corresponds to an effective quark-meson coupling (see Fig.~\ref{quark-meson-coupling}) of order $e \sim 1/\sqrt{\nc}$. Physically, this can be understood very simply. The fields on the branes describe fluctuations around the equilibrium configuration. The string endpoint pulls on the branes and therefore excites (i.e.~it is charged under) these fields. The branes' tension is of order $1/g_s \sim \nc$, where $g_s$ is the string coupling constant, whereas the string tension is $\nc$-independent. This means that the deformation of the branes caused by the string is of order $e^2 \sim 1/\nc$. We thus conclude that the dynamics of the `branes+string endpoint' system is (a generalization of) that of classical electrodynamics in a medium in the presence of a fast-moving charge.

The second observation is that the velocity of the quark may exceed the limiting velocity of the mesons, since the redshift at the position of the string endpoint is smaller than that at the bottom of the branes. As in ordinary electrodynamics, if this happens then the string endpoint loses energy by Cherenkov-radiating into the fields on the branes. In the gauge theory, this translates into the quark losing energy by Cherenkov-radiating scalar and vector quarkonium mesons. The rate of energy loss is set by the square of the coupling, and is therefore of order $1/\nc$.

\begin{figure}
\begin{center}
\includegraphics[scale=.5]{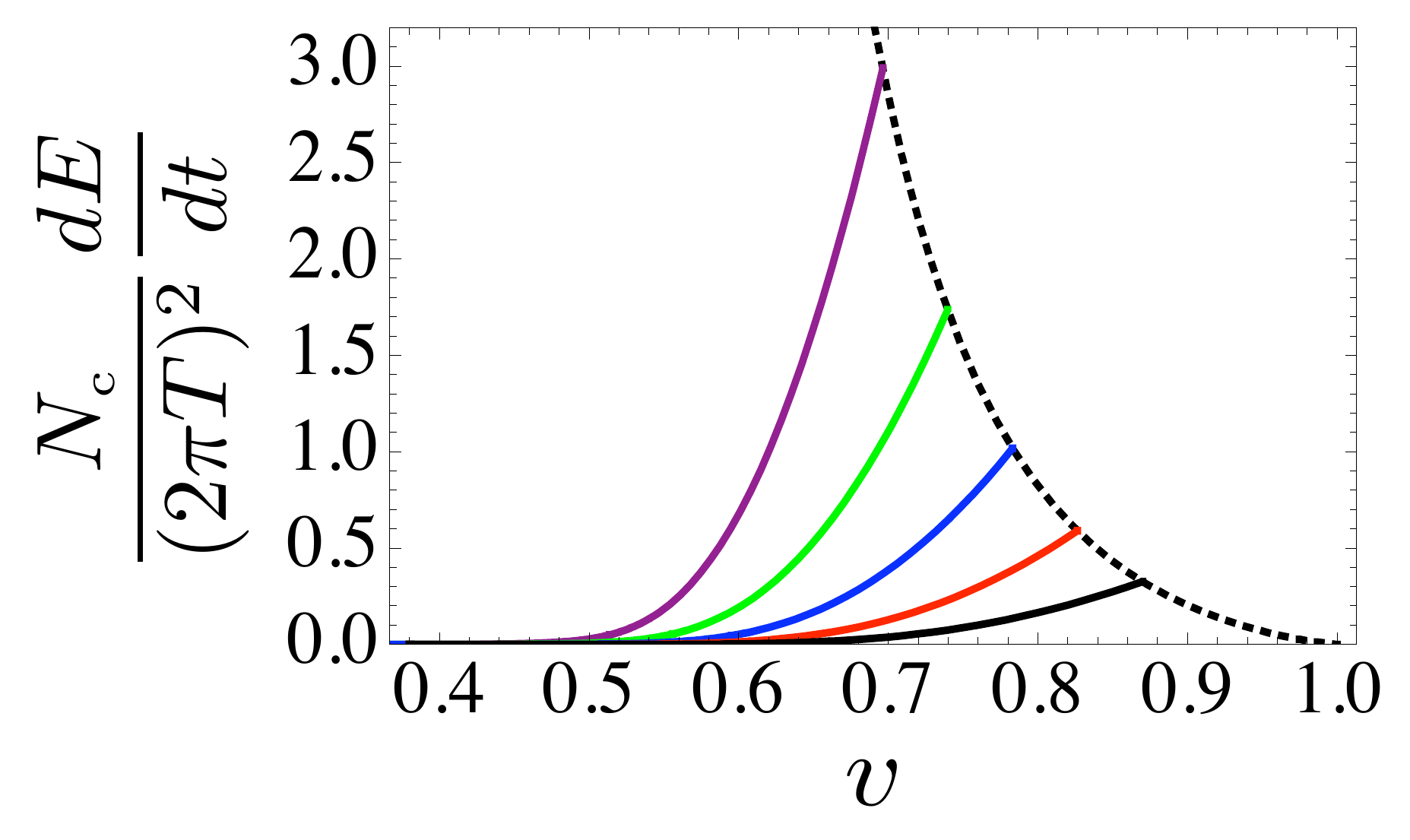}
\end{center}
\caption{\small Cherenkov energy loss into the transverse mode of vector quarkonium mesons. The continuous curves correspond to increasing values of $r_q$ from left to right. The dotted curve is defined by the endpoints of the constant-$r_q$ curves. See Refs.~\cite{CasalderreySolana:2010xh,CasalderreySolana:2009ch} for details.} 
\label{result}
\end{figure}

The quantitative details of the energy lost to Cherenkov radiation of quarkonium mesons by a quark propagating through the ${\cal N}=4$ plasma can be found in \cite{CasalderreySolana:2010xh,CasalderreySolana:2009ch}, so here we will only describe the result. For simplicity, we will assume that the quark moves with constant velocity along a straight line at a constant radial position. In reality, $r_q$ and $v$ will of course decrease with time because of the black hole gravitational pull and the energy loss. However, we will concentrate on the initial part of the trajectory (which is long provided the initial quark energy is large), for which $r_q$ and $v$ are approximately constant \cite{Chesler:2008uy} 
--- see Fig.~\ref{cherenkov-with-quark}. Finally, for illustrative purposes we will focus on the energy radiated into the transverse modes of vector mesons.
The result is depicted in Fig.~\ref{result}, and its main qualitative features are as follows.

As expected, we see that the quark only radiates into meson modes with phase velocity lower than $v$ --- those to the right of the dashed, vertical lines in Fig.~\ref{dispersion}. For fixed $r_q$, the energy loss increases monotonically with $v$ up to the maximum allowed value of $v$ --- the local speed of light at $r_q$. As $r_q$ decreases, the characteristic momentum $q_\mt{char}$ of the modes into which the energy is deposited increases.  These modes become increasingly peaked near the bottom of the branes, and the energy loss diverges. However, this mathematical divergence is removed by physical effects we have not taken into account. For example, for sufficiently large $q$ the mesons' wave functions become concentrated on a region whose size is of order the string length, and hence stringy effects become important \cite{Ejaz:2007hg}. Also, as we saw in Section \ref{sec:MesonWidths},  mesons acquire widths $\Gamma \propto q^2$ at large $q$ \cite{Faulkner:2008qk} and can no longer be treated as well defined quasiparticles. Finally, the approximation of a constant-$v$, constant-$r_q$ trajectory ceases to be valid whenever the energy loss rate becomes large. 

The Cherenkov radiation of quarkonium mesons by quarks depends only on the qualitative features of the dispersion relation of Fig.~\ref{dispersion}, which are universal for all gauge theory plasmas with a dual gravity description. Moreover, as we explained in Section~\ref{sec:RemarksConnectionQGP}, it is conceivable that they may also hold for QCD mesons such as the $J/\psi$ or the $\Upsilon$ whether or not a string dual of QCD exists. Here we will examine some qualitative consequences of this assumption for heavy ion collision experiments. Since the heavier the meson the more perturbative its properties become, we expect that our conclusions are more likely to be applicable to the charmonium sector than to the bottomonium sector. 

An interesting feature of energy loss by Cherenkov radiation of quarkonia is that, unlike other energy-loss mechanisms, it is largely independent of the details of the quark excited state, such as the precise features of the gluon cloud around the quark, etc. In the gravity description these details would be encoded in the precise profile of the entire string, but the Cherenkov emission only depends on the trajectory of the string endpoint. This leads to a dramatic simplification which, with the further approximation of rectilinear uniform motion, reduces the parameters controlling the energy loss to two simple ones: the string endpoint velocity $v$ and its radial position $r_q$. In order to obtain a ballpark estimate of the magnitude of the energy loss, we will assume that in a typical collision quarks are produced with order-one values of $r_q$ (in units of $R^2 T$). Under these circumstances the energy loss is of order unity in units of $(2\pi T)^2/N_c$, which for a temperature range of $T=200-400\,$ MeV and $N_c=3$ leads to $dE/dx \approx 2-8\,$ GeV/fm. This is is of the same order of magnitude as other mechanisms of energy loss in the plasma; for example, the BDMPS radiative energy loss $dE/dx=\alpha_s  C_F \hat q  L/2$ yields values of $dE/dx=7-40$ GeV/fm for $\hat q=1-5$ GeV$\,^2/$fm, $\alpha_s=0.3$ and $L\approx 6$ fm. Since our gravity calculation is strictly valid only in the infinite-quark-energy limit (because of the linear trajectory approximation), we expect that our estimate is more likely to be applicable to highly energetic quarks at the LHC than to those at RHIC. 

Even if in the quark-gluon plasma the magnitude of Cherenkov energy loss turns out to be subdominant with respect to other mechanisms, its velocity dependence and its geometric features may still make it identifiable. Indeed, this mechanism would only operate for quarks moving at velocities $v > \vlim$, with $\vlim$ the limiting velocity of the corresponding quarkonium meson in the plasma. The presence of such a velocity threshold is the defining characteristic of Cherenkov energy loss. The precise velocity at which the mechanism starts to operate may actually be higher than $\vlim$ in some cases, since the additional requirement that the energy of the quark be equal or larger than the in-medium mass of the quarkonium meson must also be met. 

Cherenkov mesons would  be radiated at a characteristic angle $\cos \theta_c=v_\mt{lim}/v$ with respect to the emitting quark, where $v$ is the velocity of the quark. Taking the gravity result as guidance, $\vlim$ could be as low as $\vlim=0.35$ at the quarkonium dissociation 
temperature~\cite{Mateos:2007vn}, corresponding to an angle as large as 
$\theta_\mt{c} \approx 1.21$ rad. This would result in an excess of heavy quarkonium associated with high-energy quarks passing through the plasma. Our estimate of the energy loss suggests that the number of emitted $J/\psi$'s, for example, could range from one to three per fm. This emission pattern is similar to the emission of sound waves by an energetic 
parton~\cite{CasalderreySolana:2004qm}  that we have reviewed in Section~\ref{sec:AdSCFTDragWaves}, in that both effects lead to a non-trivial angular structure. One important difference, however, is that the radiated quarkonium mesons would not thermalize and hence would not be part of a hydrodynamic shock wave. As in the Mach cone case, the meson emission pattern could be reflected in azimuthal dihadron correlations triggered by a high-$p_T$ hadron. Due to surface bias, the energetic parton in the triggered direction is hardly modified, while the one propagating in the opposite direction moves through a significant amount of medium, emitting quarkonium mesons. Thus, under the above assumptions, the dihadron distribution with an  associated $J/\psi$ would have a ring-like structure  peaked at an angle  $\theta \approx \pi-\theta_\mt{c}$.
Even if this angular structure were to prove hard to discern, the simpler correlation that in events with a high-$p_T$ hadron there are more $J/\Psi$ mesons than in typical events may suffice as a distinctive signature, although further phenomenological modelling is required to establish this.

A final observation is that Cherenkov energy loss also has a non-trivial temperature dependence, since it requires that there are meson-like states in the plasma, and therefore it does not take place at temperatures above the meson dissociation temperature. Similarly, it is reasonable to assume that it does not occur at temperatures below $T_c$, since in this case we do not expect the meson dispersion relations to become spacelike.\footnote{This assumption is certainly correct for plasmas with a gravity dual, since the corresponding geometry does not include a black hole horizon if $T< T_c$.} 
Under these circumstances, the Cherenkov mechanism is only effective over a limited range of temperatures $T_c<T< T_\mt{diss}$ which, if $T_\mt{diss} \gtrsim 1.2 T_c$ as in Ref.~\cite{Mocsy:2007yj}, is a narrow interval. As was pointed out in Ref.~\cite{Liao:2008dk}, a mechanism of energy loss which is confined to a narrow range of temperatures in the vicinity of $T_c$ concentrates the emission of energetic partons to a narrow layer within the collision geometry and is able to explain $v_2$-data at high $p_T$  at RHIC~\cite{Adler:2006bw,Adams:2004wz}. Provided that the meson dissociation temperature $T_\mt{diss}$ is not much larger than $T_c$, the Cherenkov radiation of quarkonium mesons is one such mechanism.

\chapter{Concluding remarks and outlook}

Since the purpose of heavy ion collisions is to study the properties of Quantum Chromodynamics at extreme temperature and energy density, any successful phenomenology  must ultimately be based on QCD. However, as discussed in Section~\ref{intro}, heavy ion phenomenology requires 
strong-coupling techniques not only for bulk thermodynamic quantities 
like the QCD equation of state, but also for many
dynamical quantities at nonzero 
temperature, such as transport coefficients, relaxation times and quantities accessed by probes 
propagating through a plasma. By now, lattice-regularized
QCD calculations provide 
well controlled results for the former, but progress on all the latter quantities is likely to
be slow since one needs to overcome both conceptual limitations and limitations in computing power. Alternative strong-coupling tools are therefore desirable. The gauge/string duality provides one such tool for performing non-perturbative calculations for a wide class of non-abelian plasmas. 

In this review we have mostly focussed on results obtained within one particular example of a gauge/string duality, namely the case in which the gauge theory is $\mathcal N=4$ SYM or a small deformation thereof. One reason for this is pedagogical: $\N=4$ SYM is arguably the simplest and best understood case of a gauge/string  duality.  By now many examples are known of more sophisticated string duals of non-supersymmetric, non-conformal QCD-like theories that exhibit  confinement, spontaneous chiral symmetry breaking, thermal phase transitions, etc. However, many of these features become unimportant in the deconfined phase. For this reason, for the purpose of studying the QCD quark-gluon plasma, the restriction to $\mathcal N=4$ SYM not only yields a gain in simplicity, but also does not imply a significant loss of generality, at least at the qualitative level. Moreover, none of these `more realistic' theories can be considered in any sense a controlled approximation to QCD. Indeed, many differences remain including the presence of adjoint fermions and scalar fields, the lack of asymptotic freedom, the large-$\nc$ approximation, etc. Some of these differences may be overcome if string theory in asymptotically AdS spacetimes becomes better understood. However,   in the supergravity (plus classical strings and branes) limit currently accessible these caveats  remain important to bear in mind when comparing to QCD. 

In this context, it is clearly questionable to assess the interplay between heavy ion phenomenology and the gauge/string duality correspondence solely on the basis of testing the numerical agreement between theory and experiment.  Rather, one should view this interplay in light of the standard scientific strategy that to gain significant insight into problems that cannot be addressed within the current state of the art, it is useful 
to find a closely related theory within which such problems can be addressed with known technology and which encompasses the essential features of interest.  
For many dynamical features of phenomenological interest in heavy ion physics, controlled 
strong-coupling calculations in QCD are indeed not in immediate reach with the current state of the art. In contrast,  within the gauge/string correspondence, it has been possible to formulate and solve the same problems in the strongly coupled plasmas of a large class of non-abelian quantum field theories. Among these, strongly coupled $\N=4$ SYM theory at large-$N_c$ turns out to provide the simplest model for the strongly coupled plasma being produced and probed in heavy ion collisions. 
Very often in the past, when theoretical physicists have introduced some model for the purpose of capturing the essence of some phenomenon or phenomena involving strongly coupled dynamics, the analysis of that model has then required further uncontrolled approximations. (Examples abound: Nambu--Jona-Lasinio models in which the QCD interaction is first replaced by a four-fermi coupling but one then still has to make a mean-field approximation;  linear sigma models, again followed by a mean-field approximation; bag models; $\ldots$)  A great advantage of using a quantum field theory with a gravity dual as a model is that once we have picked such a theory, the calculations needed to address the problems of interest can be done rigorously at strong coupling, without requiring any further compromise. 
In some cases, the mere formulation of the problem in a way suitable for a gravitational dual calculation can lead to new results within QCD~\cite{Baier:2007ix,Laine:2009dd,CaronHuot:2009uh}. In many others, as we have seen, the existence of these solutions allows one to examine and understand the physics responsible for the processes of interest.    The most important output of a successful model is understanding. Controlled quantitative calculations come later.  Understanding how the dynamics works, what is important and what is extraneous, what the right picture is that helps you to think about the physics in a way that is both insightful and predictive, these 
must all come first.

At the least, the successes to date of the applications of gauge/string duality to problems arising from heavy ion collisions indicate that it provides us with a successful model, in the sense of the previous paragraph.  However, there are many indications that it provides more.
In solving these problems, some regularities have emerged in the form of universal properties, by which we mean properties common to all strongly coupled theories with gravity duals in the large-$\nc$, strong-coupling limit. These include both quantitative observables, such as the 
ratio of the thermodynamic potentials at strong and weak coupling (Section~\ref{sec:BulkDynamicalProperties}) and the
value $\eta/s=1/4\pi$ at strong coupling (Section~\ref{sec:TransportProperties}), and qualitative features, such as the familiar fact that heavy quarkonium mesons remain bound in the plasma as well as the
 discovery that the dissociation temperature for quarkonium mesons drops with increasing meson velocity $v$ like $(1-v^2)^{1/4}$ (Sections~\ref{sec:HotWind} 
and \ref{sec:MinkSpectrum}) and that
high-momentum dispersion relations become space-like (Section~\ref{sec:MinkSpectrum}). The discovery of these generic properties is important in order to extract lessons for QCD. Indeed, the fact that some  properties are valid in a class of gauge-theory plasmas so broad as to include theories in different numbers of dimensions, with different field content, with or without chemical potentials, with or without confinement and chiral symmetry breaking, etc.~leads one to suspect that such properties 
might be universal across the plasmas in a class of theories that includes the QCD  --- whether or not a string dual of QCD itself exists.
The domain of applicability of this putative universality is at present unknown, both in the sense that we do not know to what theories it may apply and in the sense that we cannot say {\it a priori} which observables and phenomena are universal and which others are theory-specific details. One guess as to  a possible characterization in theory space could be that these universal features may be common across all gauge-theory plasmas that have no quasiparticle description
(Section \ref{sec:Quasi-particlesSpectralFunctions}).

Even results obtained via the gauge/string duality that are not universal may provide guidance for our understanding of QCD at nonzero temperature and for heavy ion phenomenology. In many cases, these are strong-coupling results that differ parametrically from the corresponding weak-coupling results and therefore deliver valuable qualitative messages for the modeling of heavy ion collisions. In particular, the small values of the ratio $\eta/s$, of the relaxation time $\tau_\pi$ 
(Section~\ref{sec:TransportProperties}) and of the 
heavy-quark diffusion constant (Section~\ref{sec:HQBroadening}) showed that such small values can be realized in a gauge-theory plasma. In addition, the result for $\eta/s$ in $\mathcal N=4$ theory, combined with the results for the entropy density, pressure and energy density (Section~\ref{sec5.1}) have taught us that a theory can have almost identical thermodynamics at zero and infinite coupling and yet have radically different hydrodynamics. The lesson that this provides for QCD is that the thermodynamic observables, although they are available from lattice simulations, are not good indicators of whether the quark-gluon plasma is weakly or strongly coupled, whereas the transport coefficients are. 

Important lessons have also been extracted from the strong-coupling calculations of  the jet quenching parameter $\hat q$ and the heavy-quark drag coefficient $\eta_D$ and momentum broadening $\kappa$ (Sections~\ref{sec:HQDrag}, \ref{sec:HQBroadening} and~\ref{sec:AdSCFTJetQuenching}). These showed not only that these quantities can attain values significantly larger than indicated by perturbative estimates but also that while in perturbation theory  both $\hat q$ and  $\kappa$ are proportional to the entropy density, this is not the case at strong coupling, where both these quantities and $\eta_D$ 
scale with the square root of the number of degrees of freedom. 
This result, which is valid for a large class of theories,  corrected a 
na\"ive physical expectation that was supported by perturbation theory. 

Perhaps even more fundamentally, the availability of rigorous, reliable results for any strongly coupled plasma (leave apart for a large class of them) 
can alter the very intuition we use to think about the dynamics of the quark-gluon plasma. In perturbation theory, one thinks of the plasma as being made of quark and gluon quasiparticles.
However, gravity calculations of correlation functions valid at strong coupling show no evidence of the existence of any quasiparticle excitations composed from gluons and light quarks
(Section \ref{sec:Quasi-particlesSpectralFunctions}).  (Heavy, small, quarkonium mesons do survive as quasiparticles up to some dissociation temperature
(Sections \ref{sec:BHMesonSpectrum} and \ref{sec:Peak}).)   The presence or absence of quasiparticles is a 
 major qualitative difference between the weak- and 
 strong-coupling pictures of the plasma  which is largely independent of  the caveats associated with the use of gauge/string duality that we have described above.

Finally, two important roles played by the gauge/string duality are 
 that of a testing ground of existing ideas and models relevant for heavy ion collisions, and that of a source of new ones in a regime in which guidance and inspiration from perturbation theory may be inapplicable or misleading. In its first role, the duality provides a rigorous field-theoretical framework within 
 which to verify our intuition about the plasma. For example, explicit calculations gave support to 
 previously suggested ideas about the hydrodynamical response of the plasma to high-energy particles (Section~\ref{sec:AdSCFTDragWaves}) or the possibility that heavy quarkonium mesons survive deconfinement (Section~\ref{sec:D7atNonZeroT}). In its second role, the duality has generated qualitatively new ideas which could not have been guessed from perturbation theory. Examples of these are the non-trivial velocity dependence of screening lengths (Section~\ref{sec:HotWind}), the in-medium conversion of mesons into photons (Section~\ref{sec:Peak}), the energy loss of heavy quarks via Cherenkov emission of mesons (Section~\ref{sec:Cherenkov}), and the appearance of a phase transition associated with 
 the dissociation of heavy quarkonium bound states (Section~\ref{sec:ThermoD7}).   

In summary, while it is true that caution and a critical mind must be exercised  when trying to extract lessons from any gauge/gravity calculation, paying particular attention to its limitations and range of applicability, it is also undeniable that over the last few years a broad suite of qualitatively novel insights relevant
for heavy ion phenomenology have emerged from detailed and quantitative calculations in the gravity duals of non-abelian field theories.
 As the phenomenology of heavy ion collisions moves to new, more
quantitative, and more detailed, studies in the RHIC program, and as it moves to novel challenges at the LHC, we have every reason to 
expect that experimental information about additional properties of hot QCD matter will come into theoretical 
focus. Understanding properties of the QCD plasma, as well as its response to and its effects on probes, at strong 
coupling will therefore remain key issues in future analyses. We expect that the gauge/string correspondence will continue
to play an important role in making progress on these issues. 

\section*{Acknowledgements}
We gratefully acknowledge our many collaborators with whom and from whom we have learned the subjects that we have reviewed. We also acknowledge T. Faulkner for helping us with some of the figures in this review. 
The work of HL was supported in part by a DOE Outstanding Junior Investigator grant. This research was supported in part by the DOE Offices of Nuclear and High Energy 
Physics under grants $\#$DE-FG02-94ER40818 and $\#$DE- 
FG02-05ER41360. DM is supported in part by grants 2009-SGR-168, MEC FPA 2007-66665-C02, MEC FPA 2007-66665-C01 and CPAN CSD2007-00042 Consolider-Ingenio 2010. 
JCS is supported by a  Marie Curie Intra-European Fellowship 
PIEF-GA-2008-220207.

\appendix

\chapter{Green-Kubo formula for transport coefficients}
\label{sec:Green-Kubo}

Transport coefficients of a gauge theory plasma, such as the shear
viscosity $\eta$,  can be extracted from correlation functions of the 
gauge theory via a relation known as the Green-Kubo formula. Here, we derive this relation for the case of the shear viscosity.
Let us consider a system in  equilibrium and let us work
in the fluid rest frame, meaning that
$u^{\mu}=(1,{\bf 0})$. Deviations from equilibrium are studied by introducing
a small external source of the type
\be
S=S_0+\frac{1}{2}\int d^4x\,  T^{\mu\nu} h_{\mu\nu}\, ,
\ee
where $S_0$ ($S$) is the action of the theory in the absence (presence) of the perturbation
$h_{\mu\nu}$.
To leading order in the 
perturbation, the expectation value of the stress tensor is 
\be
\label{eq:lrsp}
\llangle T^{\mu \nu}(x)\rrangle=\llangle T^{\mu \nu} (x)\rrangle_0 
                               -\frac{1}{2}\int d^4y\, G^{\mu \nu, \alpha\beta}_R (x-y) h_{\alpha \beta} (y)
\ee
where the subscript $0$ indicates the unperturbed expectation value and the retarded correlator is
given by
\be
\label{grTT}
i  G^{\mu \nu, \alpha\beta}_R (x-y)\equiv \theta \left(x^0-y^0\right)
                                    \llangle \left[T^{\mu\nu}(x),T^{\alpha \beta}(y)\right] \rrangle\ .
\ee

To extract the shear viscosity, we concentrate on an external  perturbation of the form
\be
h_{xy}(t,z) \, .
\ee
Upon Fourier transforming, this is equivalent
to using rotational invariance to choose the wave vector of the 
perturbation, ${\bf k}$, along the $\hat{z}$ direction.
The off-diagonal components of the stress tensor are then given by
\be
\label{eq:lrsplw}
\llangle T^{xy}(\omega,k)\rrangle=-G_R^{xy,xy}(\omega,k) h_{xy}(\omega,k) \ ,
\ee
to linear order in the perturbation.
In the long wavelength limit, in which the typical variation of the 
perturbed metric is large compared with any correlation length, we obtain
\be 
\llangle T^{xy}\rrangle (t,z) =-\int \frac{d\omega}{2\pi} e^{-i\omega t} G_R^{xy,xy}(\omega,k=0) h_{xy}(\omega,z)\ .
\ee
This long wavelength expression may be compared to the hydrodynamic approximation 
by studying the reaction of the system to the source within the effective theory. 
The source $h_{\mu\nu}$
can be interpreted as a modification on the metric,
\be
g_{\mu\nu}\rightarrow g_{\mu\nu} + h_{\mu \nu}\ .
\ee
To leading order in the perturbation, the shear tensor defined in \Eq{eq:shear} is given by
\be
\sigma_{xy}=2\,\Gamma^0_{xy}=\del_0 h_{xy} \,,
\ee
where $\Gamma^{\mu}_{\nu\rho}$ are the Christoffel symbols.
The hydrodynamic approximation is valid in the long time limit, when
all microscopic processes have relaxed. In this limit, we can compare 
the linear response expression (\ref{eq:lrsplw}) to the expression obtained upon making the hydrodynamic approximation, namely
(\ref{eq:pilinear}). We conclude that
\be\label{etaco}
\eta=-\lim_{\omega \rightarrow 0} \frac{1}{\omega} \lim_{k\rightarrow 0}{\rm Im} G_R^{xy,xy}\left(\omega,{\bf k}\right) \,.
\ee
This result is known as the Green-Kubo formula for the shear viscosity.

The above discussion for the stress tensor can also be generalized to other conserved currents.
In general, the low frequency limit of $G_R (\om, \vec k)$ for a conserved current operator  $O$ defines a transport coefficient $\chi$ 
\be \label{trco}
\chi = -\lim_{\om \to 0} \lim_{\vec k \to 0} {1 \ov \om} {\rm Im} G_R (\om, {\bf k}) \, ,
\ee
where the retarded correlator is defined analogously to \Eq{grTT}
\be
i  G_R (x-y)\equiv \theta \left(x^0-y^0\right)
                                    \llangle \left[O(x),O(y)\right] \rrangle\ .
\ee

\chapter{Hawking temperature of a general black brane metric} 
\label{app:HawT}

Here we calculate the Hawking temperature for a general class of black brane metrics of the form
\be
\label{mp1}
ds^2 = g(r) \Big[ -f(r) dt^2 + d\vec x^2 \Big] + {1 \ov h (r)} dr^2 \,.
\ee
where we assume that $f(r) $ and $h(r)$ have a first-order zero at the horizon $r=r_0$, whereas $g(r)$ is non-vanishing there.
We follow the standard method~\cite{Gibbons:1976ue} and demand that the Euclidean continuation of the metric~\eqn{mp1},
\be
\label{mp2}
ds^2 = g(r) \Big[ f(r) d\te^2 + d\vec x^2 \Big] + {1 \ov h (r)} dr^2 \,,
\ee
obtained by the replacement $t \ra - i \te$, be regular at the horizon.
Expanding \eqn{mp2} near $r=r_0$ one finds
\be
\label{apme}
ds^2 \approx \rho^2 d \th^2 + d \rho^2 + g(r_0) \, d \vec x^2 \,,
\ee
where we have introduced new variables $\rho, \theta$ defined as
\be \label{cheor}
\rho = 2 \sqrt{r-r_0 \ov h'(r_0)} \sac
\th = {\te \ov 2} \sqrt{g(r_0) f'(r_0) h'(r_0)} \,.
\ee
The first two terms in the metric (\ref{apme}) describe a plane in polar coordinates, so in order to avoid a conical singularity at $\rho=0$ we must require $\th$ to have period $2 \pi$. From (\ref{cheor}) we then see that the period $\beta=1/T$ of the Euclidean time must be
 \be \label{HWT}
 \beta = \frac{1}{T} = {4 \pi \ov \sqrt{g(r_0) f'(r_0) h'(r_0)}}  \,.
 \ee

\chapter{Holographic renormalization, one-point functions and an example of a Euclidean two-point function} 
\label{app:A}

Here we will illustrate the general prescription 
of Section \ref{sec:eucl} for computing Euclidean correlators. 
We will begin by giving a derivation of Eq.~\eqref{imeq} at linear order in the external source, although we note that Eq.~\eqref{imeq} is in fact valid at the nonlinear level~\cite{Klebanov:1999tb}, as follows from a simple generalization of the discussion that we shall present. Then, we will calculate the two-point function of a scalar operator $\sO(x)$ in $\sN =4$ SYM at zero temperature. Although our main interest is in four-dimensional boundary theories, for the sake of generality we will present the formulas for a general dimension $d$.

Let $\Phi$ be the scalar field in AdS dual to $\sO$. The Euclidean two-point function of $\sO$ is then given by the right-hand side of Eq.~(\ref{eepp}) with $n=2$. In order to evaluate this, we first need to solve the classical equation of motion for $\Phi$ subject to the boundary condition \eqn{bc}, and then evaluate the action on that solution. Since in order to obtain the two-point function we only need to take two functional derivatives of the action, it suffices to keep only the terms in the action that are quadratic in $\Phi$, ignoring all interaction terms. At this level, the action is given by Eq.~\eqn{quaDcBIS}, except without  the minus sign inside 
$\sqrt{-g}$, as appropriate for Euclidean signature:
\be \label{quaDc}
S =   - {1 \ov 2 } \int dz \, d^{d} x \, \sqrt{g} \,  \left[  g^{MN} \p_M \Phi \p_N \Phi + m^2 \Phi^2 \right] + \cdots \, .
 \ee
Note that we have adopted an overall sign convention for the Euclidean action appropriate for~\eqref{equi}.  The metric is that of pure Euclidean AdS and takes the form 
\be \label{EADSA}
ds^2 = {R^2 \ov z^2} \le(dz^2 + \de_{\mu \nu} dx^\mu dx^\nu \ri) \,.
\ee
We will work in momentum space along the boundary directions.
The equation of motion for $\Phi(z,k)$ then takes the form \eqn{eom1A}, which we reproduce here for convenience  
\be \label{eom1a1}
z^{d+1} \p_z \le(z^{1-d} \p_z \phi \ri) - k^2 z^2 \Phi - m^2 R^2 \Phi = 0 \,,
\ee
with $k^2=\delta_{\mu\nu} k^\mu k^\nu$ as appropriate in Euclidean signature. 

A simple integration by parts shows that, when evaluated on a solution $\Phi_c$, $S$ reduces to the boundary term
\be
S [\Phi_c] = -\ha  \lim_{\ep \to 0} \int_{z = \ep} {d^d k \ov (2 \pi)^d} \, \Pi_c (-k)\Phi_c (k) \,,
\ee
where  $\Pi_c$ is the canonical momentum associated with the $z$-foliation,
\be 
\Pi = - \sqrt{-g} \, g^{zz} \p_z \Phi \,,
\ee
evaluated at the solution $\Phi_c$. 
Since $z=0$ is a regular singular point of \Eq{eom1a1}, it is possible to choose a basis for $\Phi_{1,2}$ 
given by
\be 
\Phi_1 \to z^{d-\De} , \qquad \Phi_2 \to z^{\De}, \qquad \mbox{as } z \to 0\ ,
\ee 
with the corresponding canonical momentum $\Pi_{1,2} (z,k)$ behaving as  
\be
\Pi_1 \to  - \left(d-\De \right) z^{-\De}, \qquad \Pi_2 \to -\De z^{-\left(d-\De\right)},  
\quad  \mbox{as } z \to 0 \ ,
\ee
where
\be
\De = {d \ov 2} + \nu, \qquad \nu = \sqrt{{d^2 \ov 4} + m^2 R^2}  \ .
\ee
Then $\Phi_c$ and its canonical momentum can be expanded as 
\begin{eqnarray}
\Phi_c(z,k) &=& A(k) \Phi_1 (z,k) + B (k) \Phi_2 (z,k) \,, \nn
\Pi_c (z,k) &=& A(k) \Pi_1 (z,k) + B (k) \Pi_2 (z,k) \,,
\label{solexA}
\end{eqnarray}
as in~\eqref{asumA}, and  the classical on-shell action becomes   
\bea \label{clas}
S [\Phi_c]& =&- \ha  \lim_{\ep \to 0} \int_{z=\ep} {d^d k \ov (2 \pi)^d}\, \le[(A(-k)A(k) \Pi_1(-k) \Phi_1(k) + B(-k) B(k) \Pi_2(-k) \Phi_2(k) \right. \nonumber
\\
& & \qquad \qquad \qquad \qquad\left. + \, A(-k)B(k) (\Pi_1(-k) \Phi_2 (-k)+ \Phi_1 (-k)\Pi_2(k) )\ri] \,.
\eea
Note that given $\nu > 0$, in the $\ep \to 0$ limit the first term on the RHS of~\eqref{clas} contains divergences and thus $S$ requires renormalization. These divergences can be interpreted as dual to UV divergences of the boundary gauge theory. A local counter-term action $S_{ct}$ defined on the cutoff surface $z = \ep$ can be introduced to cancel the divergences. From~\eqref{clas} we need to 
choose \footnote{We assume that $2\nu$ is not an integer. If that is the case,
then extra logarithmic terms arise. See discussion in \cite{Skenderis:2002wp}.} 
\be \label{ctac}
S_{ct} = \ha \int_{z= \ep} {d^d k \ov (2 \pi)^d}\,  { \Pi_1 (-k) \ov \Phi_1 (k)} \, \Phi(-k) \Phi (k)  \ .
\ee
 Note that in this expression $\Pi_1(-k)/\Phi_1(k)$ is a function of $z$ which in the $z\to \ep$ limit is independent of $k$. The renormalized on-shell action is then given by
\be
S^{\rm (ren)} [\Phi_c] \equiv S[\Phi_c] + S_{ct} [\Phi_c] =  \ha   \int {d^d k \ov (2 \pi)^d}\, 2 \nu \, A(-k)B(k)  \ ,
\ee
where we have dropped terms which vanish in the $\ep \to 0$ limit; the action is now finite. 

We now impose the (momentum space version of the) boundary condition~\eqref{bc} on $\Phi_c$, i.e. 
\be 
\Phi_c (\ep,k) = \ep^{d-\De} \phi (k) \qquad \mbox{as } \ep \to 0 \ ,
\ee
which from equation~\eqref{solexA} gives 
\be 
A(k) = \phi (k) + \mbox{terms that vanish as } \ep \to 0 \,.
\ee 
We also need to impose the condition that $\Phi_c$ be regular everywhere in the interior. This extra condition then fixes the solution of  \eqn{eom1a1} completely, which in turn determines the ratio $\chi \equiv B / A$ in terms of which $B = \chi \phi$. The renormalized action can now be written as 
\be 
S^{\rm (ren)} [\Phi_c] = \ha  \int {d^d k \ov (2 \pi)^d}\, 2 \nu \, \chi \phi(-k) \phi(k) \,.
 \ee
It follows that the one-point function is given by
 \be 
 \vev{\sO (k)}_\phi = {\delta S^{\rm (ren)} [\Phi_c] \ov \delta \phi (-k)}  = 2 \nu \chi \phi(k)  = 2 \nu B(k) \,,
 \ee
 which is the momentum space version of~\eqref{imeq}. The two point function is then 
 \be 
 G_E (k) = {\vev{\sO (k)}_\phi \ov \phi(k)} = 2 \nu {B(k) \ov A(k)} \ .
 \ee

Note that the above discussion only uses the form of equation~\eqref{eom1a1} near $z=0$ and thus applies to generic asymptotically-AdS geometries.  

For pure AdS, equation~\eqref{eom1a1} can in fact be solved exactly with 
$\Phi_{1,2}$  given by
\be  \label{asol1}
\Phi_2 = \Ga (1 + \nu) \le({k \ov 2}\ri)^{-\nu} z^{d \ov 2} I_{ \nu} (kz), \qquad
\Phi_1 = \Ga (1 - \nu) \le({k \ov 2}\ri)^{\nu} z^{d \ov 2} I_{ - \nu} (kz) \,,
\ee
where $I(x)$ is the modified Bessel function of the first kind. 
Requiring $\Phi_c$ to be regular at $z \to \infty$ determines the solution up to an overall multiplicative constant:
 \be \label{asol2}
 \Phi_c = z^{d \ov 2} K_\nu (kz) \,,
 \ee
where $K(x)$ is the modified Bessel function of the second kind.  From~\eqref{solexA}, \eqref{asol1} and \eqref{asol2} we then find that
 \be
 {B \ov A} = {\Ga (-\nu) \ov \Ga (\nu)} \le({k \ov 2} \ri)^{2 \nu}
 \ee
 and thus 
 \be
 G_E (k) = 2 \nu {\Ga (-\nu) \ov \Ga (\nu)} \le({k \ov 2} \ri)^{2 \nu} \ .
 \ee




\bibliography{bib_jorge.bib,bib_hong.bib,bib_david.bib,bib_krishna.bib,bib_urs.bib}

\end{document}